\documentclass[french,12pt,a4paper]{book}
\usepackage[utf8]{inputenc}
\usepackage[T1]{fontenc}
\usepackage[english]{babel}
\usepackage[default,oldstyle, scale=.95]{opensans} 
\usepackage{caption}

\makeatletter
\newcommand{\enableopenany}{%
  \@openrightfalse%
}
\newcommand{\disableopenany}{%
  \@openrighttrue%
}
\makeatother

\usepackage{amsmath, dsfont}
\usepackage{amssymb}
\usepackage{epsfig}
\usepackage{calc}
\usepackage{amsfonts}
\usepackage{enumerate}
\usepackage{graphicx}
\usepackage{cancel}
\usepackage{array}
\usepackage{inputenc}
\usepackage{multirow}
\usepackage{subcaption}
\usepackage{tensor}

\usepackage{leftindex}
\usepackage[dvipsnames]{xcolor}
\usepackage[numbers]{natbib}
\usepackage{hyperref}
\usepackage{comment}
\usepackage{soul}
\usepackage{framed}
\usepackage{bibentry}
\usepackage[normalem]{ulem}
\usepackage{emptypage}
\nobibliography*

\definecolor{Prune}{RGB}{99,0,60} 
\definecolor{B1}{RGB}{49,62,72} 
\definecolor{C1}{RGB}{124,135,143}
\definecolor{D1}{RGB}{213,218,223}
\definecolor{A2}{RGB}{198,11,70}
\definecolor{B2}{RGB}{237,20,91}
\definecolor{C2}{RGB}{238,52,35}
\definecolor{D2}{RGB}{243,115,32}
\definecolor{A3}{RGB}{124,42,144}
\definecolor{B3}{RGB}{125,106,175}
\definecolor{C3}{RGB}{198,103,29}
\definecolor{D3}{RGB}{254,188,24}
\definecolor{A4}{RGB}{0,78,125}
\definecolor{B4}{RGB}{14,135,201}
\definecolor{C4}{RGB}{0,148,181}
\definecolor{D4}{RGB}{70,195,210}
\definecolor{A5}{RGB}{0,128,122}
\definecolor{B5}{RGB}{64,183,105}
\definecolor{C5}{RGB}{140,198,62}
\definecolor{D5}{RGB}{213,223,61}

\hypersetup{
  colorlinks,
  citecolor=A4}

\usepackage{mdframed}
\usepackage{multirow} 
\usepackage{multicol} 
\usepackage{tikz}
\usepackage{graphicx}
\usepackage[absolute]{textpos} 
\usepackage{colortbl}
\usepackage{array}
\usepackage{geometry}
\usepackage{titlesec}
\usepackage{hyperref}
\usepackage{ragged2e}
\hypersetup{ 
    colorlinks=true,
    linkcolor=black,
    urlcolor=A5}

\makeatletter
\renewcommand{\bibentry}[1]{\nocite{#1}{\frenchspacing\@nameuse{BR@r@#1\@extra@b@citeb}}}
\makeatother

\makeatletter
\newcommand*{\rom}[1]{\expandafter\@slowromancap\romannumeral #1@}
\makeatother

\begin{document}

\begin{titlepage}

\newgeometry{left=6cm,bottom=2cm, top=1cm, right=1cm}

\tikz[remember picture,overlay] \node[opacity=1,inner sep=0pt] at (-13mm,-135mm){\includegraphics{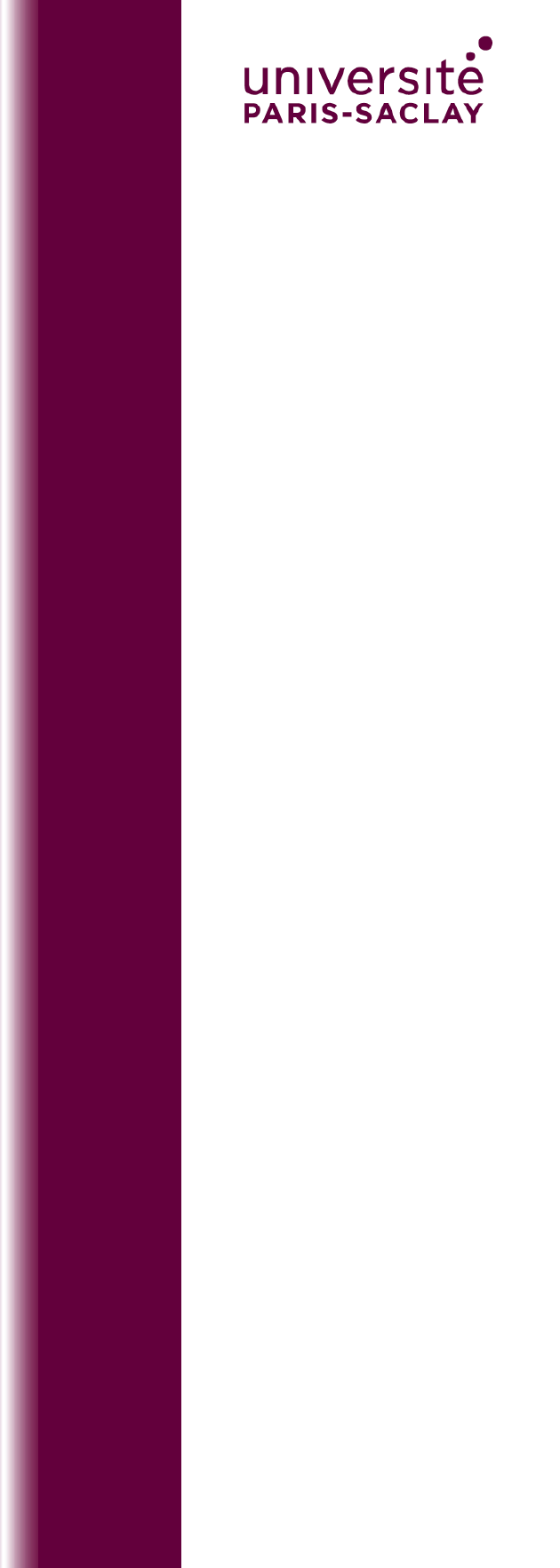}};


\color{white}

\begin{picture}(0,0)
\put(-152,-743){\rotatebox{90}{\Large \textsc{THESE DE DOCTORAT}}} \\
\put(-120,-743){\rotatebox{90}{NNT : 2024UPASP036}}
\end{picture}
 
\vspace{20mm} 


\flushright
\vspace{10mm} 
\color{Prune}

\fontsize{22}{26}\selectfont
  \Huge Exact black hole solutions in scalar-tensor theories \\

\normalsize
\color{black}
\Large{\textit{Solutions exactes de trous noirs en théories scalaire-tenseur}} \\

\fontsize{8}{12}\selectfont

\vspace{1.5cm}

\normalsize
\textbf{Thèse de doctorat de l'université Paris-Saclay} \\

\vspace{6mm}

\small École doctorale n$^{\circ}$ 564, Physique en Île-de-France (PIF)\\
\small Spécialité de doctorat : Physique\\
\small Graduate School : Physique. Référent : Faculté des sciences d’Orsay \\
\vspace{6mm}

\footnotesize Thèse préparée dans l'unité de recherche \textbf{IJCLab} (Université Paris-Saclay, CNRS), sous la direction de \textbf{Christos CHARMOUSIS}, Directeur de recherche, et la co-direction de \textbf{Eugeny BABICHEV}, Directeur de recherche \\
\vspace{15mm}

\textbf{Thèse soutenue à Paris-Saclay, le 14 juin 2024, par}\\
\bigskip
\Large {\color{Prune} \textbf{Nicolas LEC\OE UR}} 

\vspace{\fill} 

\bigskip

\flushleft
\small {\color{Prune} \textbf{Composition du jury}}\\
{\color{Prune} \scriptsize {Membres du jury avec voix délibérative}} \\
\vspace{2mm}
\scriptsize
\begin{tabular}{|p{7cm}l}
\arrayrulecolor{Prune}
\textbf{Karim NOUI} &   Président\\ 
Professeur, IJCLab, Université Paris-Saclay & \\
\textbf{Blaise GOUT\'ERAUX} &  Rapporteur \& Examinateur \\ 
Professeur, CPHT, Ecole Polytechnique   &   \\ 
\textbf{David LANGLOIS} &  Rapporteur \& Examinateur\\
Directeur de recherche, APC, Université Paris Cité &   \\ 
\textbf{Laura BERNARD} &  Examinatrice \\ 
Chargée de recherche, LUTH, Universités PSL et Paris Cité &   \\ 
\textbf{Mokhtar HASSAINE} &  Examinateur \\ 
Professeur, Instituto de Matemática y Física, Universidad de Talca &   \\ 
\textbf{Francesco NITTI} &  Examinateur \\ 
Professeur, APC, Université Paris Cité   &   \\
\textbf{Danièle STEER} &  Examinatrice \\ 
Professeure, APC, Université Paris Cité  &   \\

\end{tabular} 

\end{titlepage}

\thispagestyle{empty}
\newgeometry{top=1.5cm, bottom=1.25cm, left=2cm, right=2cm}

\noindent 
\includegraphics[height=2.45cm]{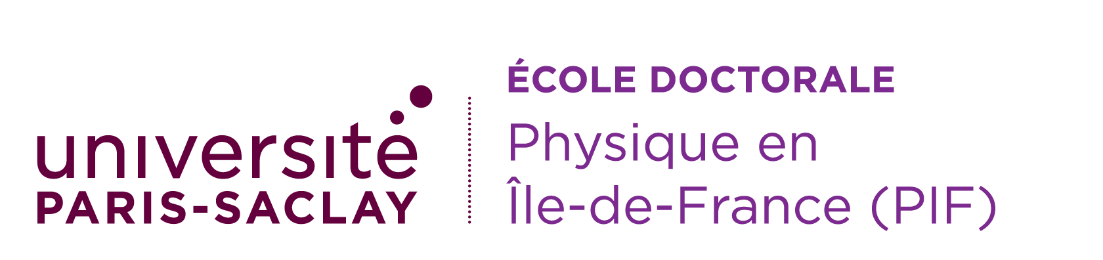}
\vspace{1cm}

\small

\begin{mdframed}[linecolor=Prune,linewidth=1]

\textbf{Titre :} Solutions exactes de trous noirs en théories scalaire-tenseur 

\noindent \textbf{Mots clés :} Trous noirs chevelus, Scalaire-tenseur, Réduction dimensionnelle, Symétries

\vspace{-.5cm}
\begin{multicols}{2}
\noindent \textbf{Résumé :} La Relativité Générale admet une unique solution de trou noir, caractérisée par sa masse $M$, son moment angulaire $J$, et sa charge électrique $Q$. On dit donc que les trous noirs en Relativité Générale n'ont pas de cheveux, c'est-à-dire pas d'autre quantité physique indépendante (théorème de calvitie). \\

\noindent Malgré les innombrables succès de la Relativité Générale, des problèmes subsistent, comme celui de la singularité au centre des trous noirs, où la courbure de l'espace-temps devient infinie. Les théories de gravité modifiée tentent de résoudre ces limitations. \\

\noindent Cette thèse teste le théorème de calvitie dans une modification populaire de la gravitation, appelée théories scalaire-tenseur, où un unique degré de liberté (un champ scalaire) est ajouté à l'habituelle métrique de l'espace-temps de la Relativité Générale. En exploitant diverses symétries, de nouveaux trous noirs, dits chevelus, sont obtenus. Certains contournent véritablement le théorème de calvitie, en étant caractérisés par une nouvelle quantité, distincte de $M$, $J$ ou $Q$. Un progrès intéressant est également réalisé, puisque dans certains cas, la singularité disparaît : la courbure de l'espace-temps demeure finie même au c\oe ur du trou noir.\\

\noindent Des liens théoriques sont établis entre les théories scalaire-tenseur (qui prennent place dans les quatre dimensions usuelles de l'espace-temps), et les théories de gravité en dimensions supérieures. Enfin, des propriétés propres aux théories scalaire-tenseur permettent de transformer des solutions initiales de trous noirs en d'autres solutions de géométrie très différente, comme des trous de ver.
\end{multicols}

\end{mdframed}

\vspace{8mm}

\begin{mdframed}[linecolor=Prune,linewidth=1]

\textbf{Title:} Exact black hole solutions in scalar-tensor theories


\begin{multicols}{2}
\noindent \textbf{Abstract:} General Relativity allows for a unique black hole solution, characterized by its mass $M$, angular momentum $J$, and electric charge $Q$. Black holes in General Relativity are thus said to have no hair, that is, no other independent physical quantity (no-hair theorem).
\\

\noindent Despite the numerous successes of General Relativity, some limitations remain, like the central singularity possessed by black holes, where the curvature of spacetime becomes infinite. Modified theories of gravity try to solve some of these shortcomings.\\

\noindent This thesis tests the no-hair theorem in a popular modification of gravity, called scalar-tensor theories, where a unique degree of freedom (a scalar field) is added on top of the usual metric of spacetime of General Relativity. Using various symmetries, new black holes, called hairy black holes, are obtained. Some of them evade strongly the no-hair theorem, being characterized by a new quantity, distinct from $M$, $J$ or $Q$. An interesting progress is also achieved, since in certain cases, the usual singularity disappears: the curvature of spacetime remains bounded even at the core of the black hole.\\

\noindent Moreover, theoretical links are established between scalar-tensor theories (which take place in the usual four dimensions of spacetime), and theories of gravity in higher dimensions. Finally, certain particular properties of scalar-tensor theories enable to transform initial black hole solutions into other solutions with very distinct geometry, like wormholes.
\end{multicols}
\end{mdframed}

\newpage
\thispagestyle{empty}

\newgeometry{left=4cm,bottom=4cm, top=5.5cm, right=4cm}
 


\vspace{100mm}
\flushright
\color{Prune}

\fontsize{22}{26}\selectfont
  \hspace{-6mm}\Huge \mbox{Thèse préparée dans le laboratoire} \\
  \Large \color{black} Laboratoire de physique des deux \\ infinis Irène Joliot-Curie (IJCLab) \\
  
\flushright \includegraphics[scale=0.165]{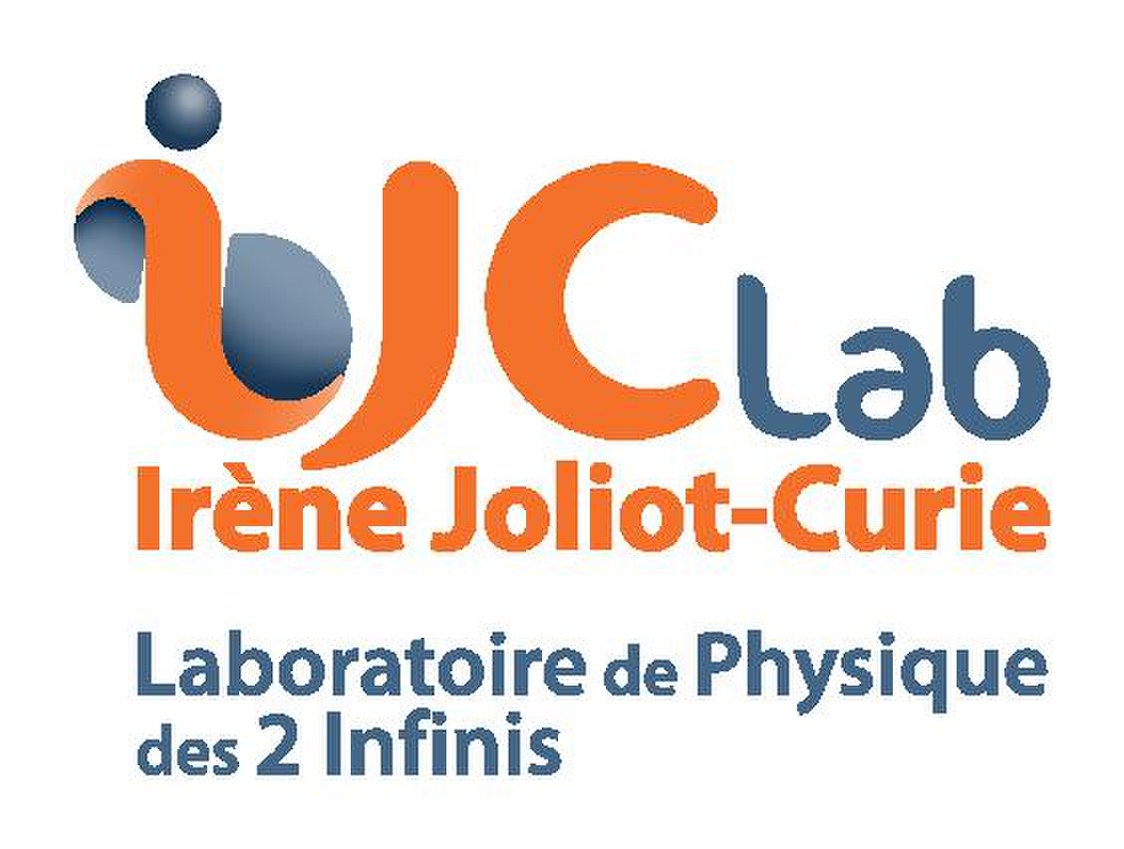}

\vspace{25mm} 
\color{Prune}

\fontsize{22}{26}\selectfont
  \Huge Contrat doctoral financé par \\
  \Large \color{black} Contrat Doctoral Spécifique Normalien \\ \'Ecole Normale Supérieure de Lyon \\

\flushright \includegraphics[scale=0.24]{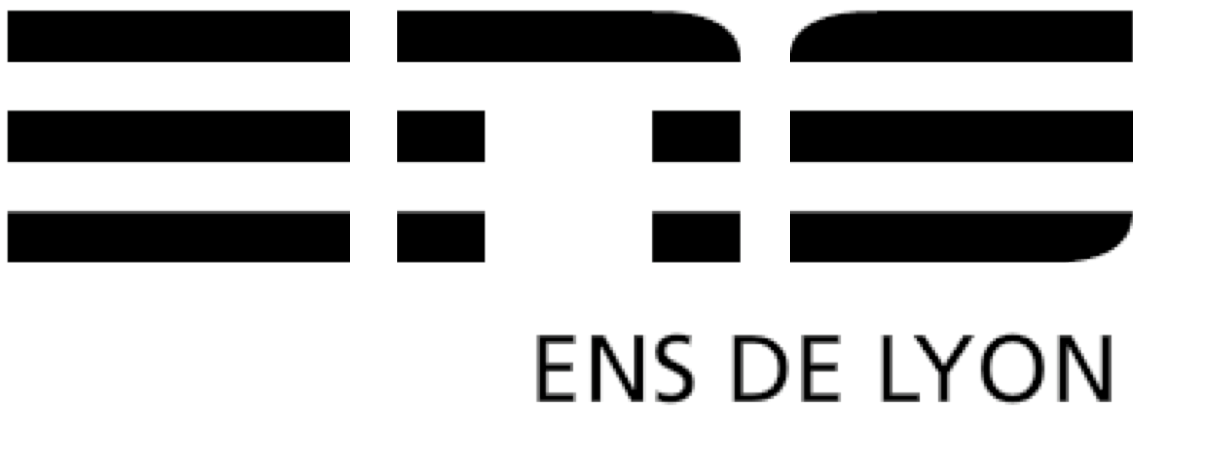}

\normalsize \justifying

\titleformat{\chapter}[hang]{\bfseries\Huge\color{Prune}\raggedright\justifying}{\thechapter\ -}{.1ex}
{\vspace{0.1ex}
}
[\vspace{1ex}]
\titlespacing{\chapter}{0pc}{0ex}{0.5pc}

\titleformat{\section}[hang]{\raggedright\justifying\bfseries\huge}{\thesection\ .}{0.5pt}
{\vspace{0.1ex}
}
[\vspace{0.1ex}]
\titlespacing{\section}{1.5pc}{4ex plus .1ex minus .2ex}{.8pc}

\titleformat{\subsection}[hang]{\bfseries\LARGE\raggedright\justifying}{\thesubsection\ .}{1pt}
{\vspace{0.1ex}
}
[\vspace{0.1ex}]
\titlespacing{\subsection}{3pc}{2ex plus .1ex minus .2ex}{.1pc}

\titleformat{\subsubsection}[hang]{\bfseries\large\raggedright\justifying}{\thesubsubsection\ .}{1pt}
{\vspace{0.1ex}
}
[\vspace{0.1ex}]
\titlespacing{\subsection}{3pc}{2ex plus .1ex minus .2ex}{.1pc}

\renewcommand{\chaptermark}[1]{\markboth{#1}{}}
\renewcommand{\sectionmark}[1]{\gdef\rightmark{#1}}

\newpage

\newgeometry{top=3.2cm,left=2.8cm,right=2.8cm,bottom=3cm}
\enableopenany
\pagenumbering{roman}
\chapter*{Remerciements}
\label{chap:rem}
\addcontentsline{toc}{chapter}{\nameref{chap:rem}}
\chaptermark{REMERCIEMENTS}
\sectionmark{REMERCIEMENTS}
La thèse est une expérience extrêmement enrichissante, mais également un travail de longue haleine, qu'il vaut mieux entreprendre entouré des bonnes personnes. J'aimerais profiter de ces quelques lignes pour les remercier.
\\

Tout d'abord, il est difficile de remercier suffisamment mes deux maîtres de thèse, Christos et Eugeny. Durant ces trois années, vous m'avez apporté énormément, sur le plan scientifique tout d'abord. C'est un privilège que de découvrir le monde des trous noirs guidé par de tels spécialistes. Nous avons beaucoup discuté autour du tableau noir, et je crois que cela nous a permis de fournir du bon travail. J'ai également apprécié tous nos débats sur des sujets moins scientifiques mais tout aussi sérieux, comme le football, sur lequel la formation d'Eugeny reste néanmoins à parfaire. Je suis heureux d'avoir pu vous connaître dans d'autres cadres que le Club, notamment en Grèce, ou chez Christos et Héléna.\\

Nos discussions incluaient bien sûr d'autres personnes, au premier rang desquels je souhaite remercier les deux chercheurs que j'ai le plus côtoyés hormis Christos et Eugeny, à savoir Karim et Mokhtar. Karim, en plus de tout le reste, je te suis véritablement reconnaissant de m'avoir orienté vers Christos et Eugeny lorsque je cherchais une thèse. Mokhtar, merci infiniment pour tous nos échanges, que ce soit à l'Institut Pascal ou par Zoom, et pour ton soutien lors de toutes mes démarches de cette dernière année.\\

Bien entendu, je remercie sincèrement David Langlois et Blaise Goutéraux, membres du jury et rapporteurs de cette thèse, pour avoir accepté de relire en détail ce manuscrit et pour leurs commentaires enrichissants. Merci à Laura Bernard, Francesco Nitti, Danièle Steer, et de nouveau Karim et Mokhtar, de prendre part à mon jury de thèse. Merci également à Adam Falkowski et à \'Eric Gourgoulhon pour les conseils prodigués, non seulement dans le cadre du comité de suivi de thèse, mais aussi au fil de nos autres rencontres. \\

J'ai pu côtoyer de nombreuses autres personnes dans le laboratoire, chercheurs établis ou (post-)doctorants, qui toutes ont rendu la vie quotidienne de thésard plus agréable. Kilian et Simon, je suis content que nous ayons effectué nos thèses simultanément. Je dois remercier particulièrement Kilian pour l'organisation des séminaires de jeunes chercheurs. J'espère que tu as fini par me convaincre d'y participer, mais je n'ai pas la réponse à l'heure d'écrire ces lignes. Merci à Jay notamment pour les bons moments passés à Madrid. Merci plus généralement à Donald, Gioacchino, Jacopo, Mathieu, Méril, Michele, Panagiotis, Salva, Teseo, Valentine, à qui je souhaite une bonne continuation. Je tiens aussi à remercier Yann, pour raconter tant d'histoires intéressantes et pour s'être gentiment préoccupé de mon avenir ces derniers temps.
\\

Par ailleurs, j'ai eu la chance de partager mon bureau avec différents jeunes chercheurs, que je suis très heureux d'avoir rencontrés et dont j'apprécie grandement la compagnie. Aimeric, Hugo, Joffrey, Simon, je vous souhaite à tous une belle carrière. Dans un autre contexte, je remercie tout particulièrement Thanasis, Theodoros et Panagiota. Nous avons bien travaillé ensemble, certes surtout par Zoom, mais chacune de nos rencontres m'a beaucoup plu. Panagiota, merci beaucoup pour ton aide dans les démarches de ces derniers mois. Merci également à tous les chercheurs qui m'ont accueilli au cours de mes différents voyages, en particulier Sabir Ramazanov, Tony Padilla et Thomas Sotiriou. 
\\

Enfin, je remercie chaleureusement tous ceux qui m'ont entouré de leur affection, depuis bien avant cette thèse. Alexandre(s), Amaury, Antoine, Maeva, Mario, Philippine, Vincent, votre amitié compte beaucoup pour moi. Maman, Papa, Antoine, merci d'avoir toujours voulu le meilleur pour moi. Cette thèse vous doit beaucoup.\\

\newpage
\thispagestyle{empty}
\begin{figure}
\centering
\includegraphics[width=0.7\linewidth]{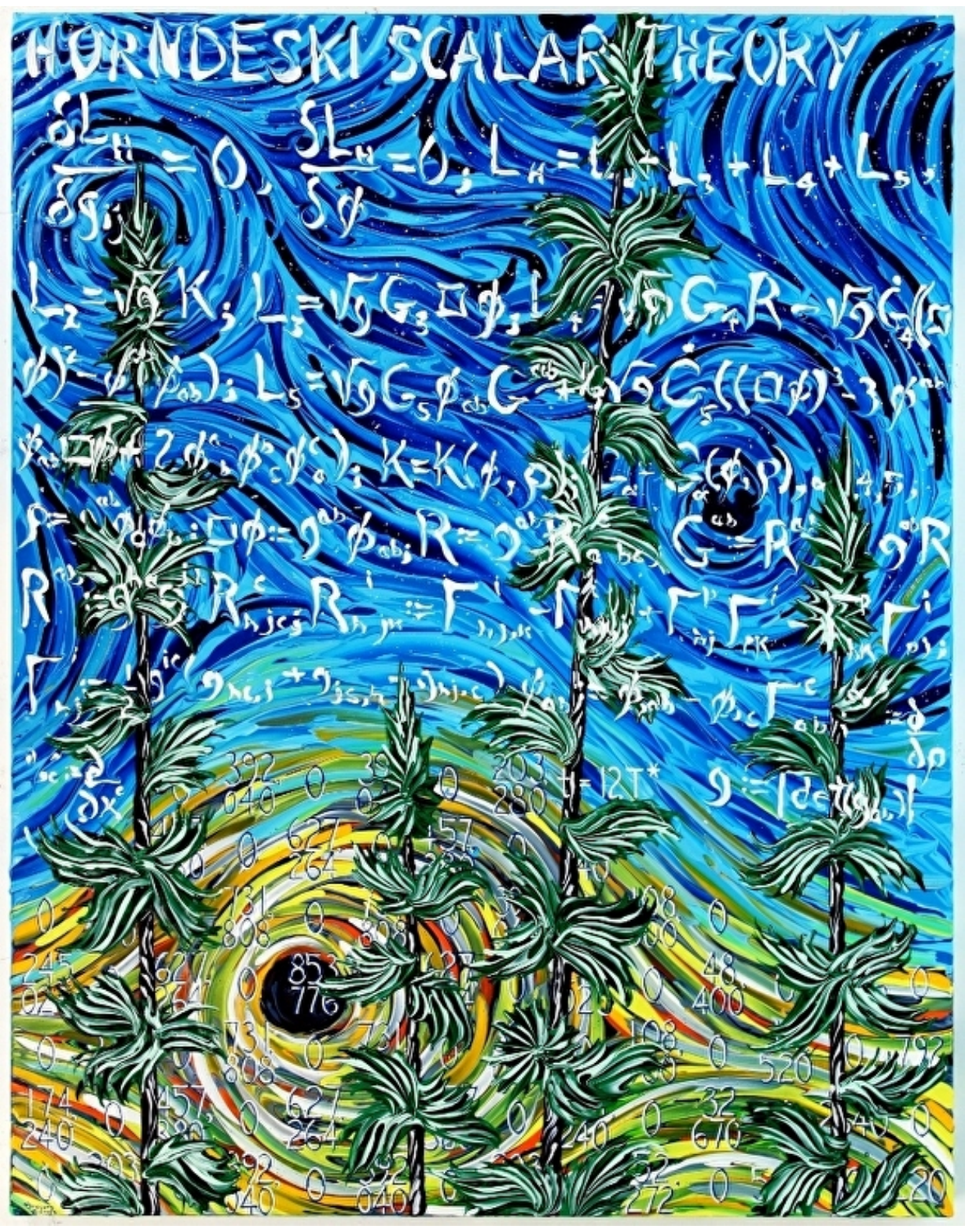}
\captionsetup{labelformat=empty,justification=centering,margin=1.5cm}\caption{\textit{Horndeski Scalar Theory, Past, Present and Future}, Gregory Horndeski, Propri\'et\'e de l'Universit\'e de Leyde} \label{fig:coverfig}
\end{figure}

\chapter*{Résumé de la thèse}
\label{chap:reslongfin}
\addcontentsline{toc}{chapter}{\nameref{chap:reslongfin}}
\chaptermark{RÉSUMÉ DE LA THÈSE}
\sectionmark{RÉSUMÉ DE LA THÈSE}
L'interaction gravitationnelle est l'une des quatre interactions élémentaires, les trois autres étant les interactions forte, faible et électromagnétique. La théorie standard de la gravitation est la Relativité Générale, formulée par Einstein en 1915. À cette époque, la gravitation était décrite par la théorie de Newton, datant du \rom{17}$^\text{e}$ siècle, et rendant compte très précisément des nombreux phénomènes impliquant la gravitation, à une exception notable près.\\

En effet, l'astronome français Urbain Le Verrier, célèbre pour avoir prédit, grâce à la théorie newtonienne, l'existence de Neptune avant son observation expérimentale, constata en 1859 le problème dit de l'avance du périhélie de Mercure. L'évolution, au fil des années, des paramètres de la trajectoire elliptique de Mercure autour du Soleil, ne peut en effet pas être expliquée par la mécanique newtonienne. Par ailleurs, la théorie de Newton se confronte, au début du \rom{20}$^\text{e}$ siècle, à un autre problème : en 1905, Einstein présente sa théorie de la relativité restreinte, qui conduit à l'introduction d'un espace-temps à quatre dimensions (trois d'espace, une de temps), et selon laquelle aucune information physique ne peut voyager plus vite que la vitesse de la lumière. Or, dans la théorie newtonienne, la force de gravitation se propage instantanément. 
\\

Ces deux problèmes sont résolus par la Relativité Générale qu'Einstein présente en 1915. Dans cette théorie, l'énergie de la matière déforme l'espace-temps, et la gravitation est la manifestation de cette courbure de l'espace-temps. En plus de rendre compte de l'avance du périhélie de Mercure, la Relativité Générale a été vérifiée par une multitude d'autres tests expérimentaux qui en font une théorie physique extrêmement bien établie. Les équations d'Einstein, qui remplacent les équations de Newton, impliquent également l'existence de \textbf{trous noirs} et d'\textbf{ondes gravitationnelles}. Leur réalité physique a été entérinée pour de bon avec la première détection directe par LIGO, en 2015, d'ondes gravitationnelles résultant de la fusion de deux trous noirs, et les premières images directes d'un trou noir publiées en 2019 par l'Event Horizon Telescope. 
\\

En dépit de ces innombrables succès, il subsiste certaines limitations à la Relativité Générale. De même que pour la théorie de Newton, ces problèmes existent tant sur le plan théorique qu'expérimental. Du point de vue théorique, la gravitation est la seule des quatre interactions élémentaires à n'avoir pas été incorporée dans le formalisme de la physique quantique, plus précisément de la théorique quantique des champs. Cela implique qu'il doit exister une théorie, plus fondamentale que la Relativité Générale, de gravité quantique. De tels effets quantiques jouent un rôle important aux très petites distances, où d'ailleurs la Relativité Générale ne fonctionne plus. En effet, la Relativité Générale elle-même prédit l'existence de \textbf{singularités d'espace-temps}, que ce soit au centre des trous noirs ou bien au moment du Big Bang. 
\\

La Relativité Générale est également confrontée au \textbf{problème de l'énergie noire}. L'énergie noire est une forme d'énergie, représentant près de 70\% du contenu énergétique de l'Univers, et responsable de son expansion accélérée, démontrée expérimentalement en 1998 par la High-Z Supernova Search Team. Dans le modèle usuel de cosmologie, l'énergie noire est représentée par une quantité dont l'énergie est uniforme, et la pression est opposée à l'énergie. Cette quantité, appelée constante cosmologique, permet de rendre compte des observations, mais son origine et son énergie, en comparaison avec les prédictions quantiques, demeurent complètement inexpliquées. 
\\

En un mot, de même que la théorie de Newton est très précise pour de très nombreuses expériences du quotidien, mais doit être remplacée par la Relativité Générale en raison de certains limitations, de même, la Relativité Générale, bien qu'accumulant de nombreuses réussites sur de très nombreux aspects, doit être modifiée pour tenir compte d'autres aspects. Certaines approches, très ambitieuses, proposent une théorie quantique de la gravitation (gravité quantique à boucles, théorie des cordes). D'autres approches, plus pragmatiques, visent à étudier des théories qui diffèrent de façon contrôlée de la Relativité Générale, et qui pourraient résoudre certains de ses problèmes mentionnés ci-dessus, sans pour autant prétendre être des théories ultimes de la gravitation. On parle alors de \textbf{théories de gravité modifiée}.
\\

Le panorama des théories de gravité modifiée est vaste, et cette thèse se concentre sur les théories dites \textbf{scalaire-tenseur}. "Tenseur" réfère à la \textbf{métrique de l'espace-temps}. Cette métrique est le champ physique usuel de la Relativité Générale, elle contient l'information concernant la courbure de l'espace-temps. C'est un champ dit tensoriel car il permet de définir la géométrie de l'espace-temps. En sus de ce champ métrique, habituel, est introduit un \textbf{champ scalaire}, c'est-à-dire simplement un champ physique qui, en tout point de l'espace-temps, prend une valeur donnée\footnote{Dans des situations physiques plus familières, la température est un champ scalaire (elle prend une certaine valeur numérique en tout point du globe); la vitesse du vent est un champ vectoriel (elle a une certaine direction et intensité). Un exemple de champ tensoriel, dans le contexte d'un matériau que l'on déforme, est le "tenseur des déformations", qui décrit l'état de déformation de ce matériau.}. On comprend que l'ajout d'un champ scalaire est l'une des façons les plus simples de modifier la gravité. Or, on s'attend à ce que les déviations à la Relativité Générale apparaissent petit à petit lorsque l'on sort de son domaine de validité : le champ scalaire devrait donc permettre de décrire ces premiers écarts à la Relativité Générale.
\\

L'une des prédictions les plus fameuses, déjà mentionnée, de la Relativité Générale, est l'existence de trous noirs, régions de l'espace-temps où le champ gravitationnel est si intense que rien, pas même la lumière, ne peut s'en échapper. La région dont rien ne peut s'échapper est délimitée par ce que l'on appelle l'\textbf{horizon des événements}. Il existe une unique solution de trou noir aux équations d'Einstein de la Relativité Générale, à savoir le trou noir de Kerr-Newman. Il est complètement caractérisé par sa masse $M$, son moment angulaire $J$ et sa charge électrique $Q$ : deux trous noirs avec même $M$, $J$ et $Q$ sont identiques. On dit que les trous noirs sont chauves, ou qu'ils n'ont pas de \textbf{cheveux}, c'est-à-dire pas d'autre quantité physique indépendante (\textbf{théorème de calvitie}).\\

En conséquence, le terme de "\textbf{trou noir chevelu}" désigne un trou noir qui diffère de celui de la Relativité Générale. On distingue \textbf{cheveu primaire} et \textbf{cheveu secondaire}. Un trou noir possède un cheveu primaire lorsqu'il est caractérisé par une quantité différente de $M$, $J$ et $Q$. C'est cette nouvelle quantité que l'on appelle cheveu primaire. Un trou noir possède un cheveu secondaire s'il demeure déterminé par $M$, $J$ et $Q$, mais qu'il diffère de la Relativité Générale en étant accompagné par d'autres champs physiques (dans le présent contexte, le champ scalaire des théories scalaire-tenseur). L'étude des trous noirs est bien plus simple lorsqu'ils n'ont pas de charge électrique $Q$, et surtout, pas de moment angulaire $J$. Dans ce cas, le trou noir de Kerr-Newman devient un trou noir \textbf{à symétrie sphérique}, dit de \textbf{Schwarzschild} (1916), paramétré seulement par sa masse $M$. \\

Cette thèse vise à \textbf{tester le théorème de calvitie dans les théories scalaire-tenseur}. En d'autres termes, elle étudie les solutions de trous noirs de ces théories, afin d'analyser dans quelle mesure ces trous noirs diffèrent de ceux de la Relativité Générale. Si des différences sont constatées, elles pourraient induire des modifications, par exemple dans le signal d'ondes gravitationnelles résultant de la fusion de deux trous noirs. Ces considérations sont d'autant plus importantes que des détecteurs bien plus précis (LISA et Einstein Telescope) sont en cours de construction.\\

Les équations des théories scalaire-tenseur sont bien plus compliquées que celles de la Relativité Générale, qui déjà ne sont pas simples. La plupart des solutions à ces équations sont donc obtenues avec des simulations numériques ou des approximations. Toutefois, cette thèse se concentre sur \textbf{les solutions ayant une expression de forme fermée}, c'est-à-dire pour lesquelles la métrique du trou noir peut être écrite explicitement, avec des fonctions mathématiques usuelles. Par souci de concision, le terme "solution exacte" est souvent employé, bien qu'il soit imprécis puisqu'une solution numérique peut être exacte. \\

Les solutions de forme fermée ne représentent pas seulement un défi sur le plan mathématique, elles comportent des avantages certains. Elles sont plus pratiques que les solutions numériques, qui nécessitent l'écriture et l'exécution d'un code numérique, et dont l'étude des propriétés peut être laborieuse. Les propriétés d'une solution de forme fermée peuvent au contraire être étudiées bien plus aisément. Enfin, ces solutions peuvent être facilement partagées, et n'importe qui peut les étudier sans nécessiter de code numérique. \\

Ce manuscrit comporte six chapitres. Les trois premiers consistent pour l'essentiel de résultats connus avant cette thèse, rassemblés et organisés de façon à introduire les notions nécessaires à la compréhension des chapitres 4 à 6, qui eux présentent les nouveaux résultats de cette thèse.\\

Le premier chapitre détaille les motivations pour modifier la Relativité Générale, puis présente le \textbf{théorème de Lovelock} (1971). Pour rappel, la Relativité Générale est une théorie de gravitation où l'unique champ gravitationnel est la métrique (on peut parler de théorie \textbf{purement métrique}), dans un espace-temps à quatre dimensions. Le théorème de Lovelock se place dans un espace-temps de dimension arbitraire (quatre, cinq, six, etc.) et détermine la théorie de gravitation purement métrique la plus générale. Le résultat est une somme de termes, appelés \textbf{invariants de courbure de Lovelock}. En quatre dimensions d'espace-temps, la théorie obtenue est la Relativité Générale ! Cela implique que, pour modifier la Relativité Générale, il faut par exemple introduire de nouveaux champs en plus de la métrique, comme le champ scalaire des théories scalaire-tenseur.
\\

Les aspects génériques de ces théories scalaire-tenseur sont étudiés au chapitre 2. Depuis les travaux pionniers de \textbf{Horndeski} (1974), les théories scalaire-tenseur ont été classifiées en différentes catégories : de la moins générale à la plus générale, il existe les théories Horndeski, beyond Horndeski, et DHOST. Au-delà de ce vocabulaire, l'important est qu'il existe des transformations de la métrique, appelées \textbf{transformations conformes-difformes}, qui permettent de passer d'une théorie scalaire-tenseur à une autre. Par ailleurs, tout problème physique se trouve simplifié par l'existence de \textbf{symétries}, par exemple, en mécanique classique, l'invariance par rotation ou translation. La situation est la même avec les théories scalaire-tenseur, et le chapitre 2 décrit et analyse en détail ces symétries.\\

Un lien intéressant est d'autre part établi entre les théories de Lovelock mentionnées précédemment, et les théories scalaire-tenseur. En effet, alors que la théorie purement métrique de Lovelock dans un espace-temps à quatre dimensions coïncide avec la Relativité Générale, ce n'est plus le cas à partir de cinq dimensions, où la théorie, appelée théorie d'\textbf{Einstein-Gauss-Bonnet}, est plus générale. Il est possible de considérer une théorie purement métrique d'Einstein-Gauss-Bonnet en dimension supérieure, puis de décomposer la métrique de cet espace-temps afin d'obtenir une théorie scalaire-tenseur dans un espace-temps à quatre dimensions (on parle de \textbf{réduction dimensionnelle}). La théorie ainsi obtenue reproduit des propriétés de la théorie d'Einstein-Gauss-Bonnet, et est donc appelée \textbf{théorie d'Einstein-Gauss-Bonnet à quatre dimensions, ou 4DEGB}.\\

Le chapitre 3 rentre dans le c\oe ur du sujet, en dressant un état de l'art des différents trous noirs chevelus en théories scalaire-tenseur qui existaient dans la littérature avant le début de cette thèse. Pour certains de ces trous noirs, la métrique est la même qu'en Relativité Générale, mais accompagnée d'un champ scalaire non nul. On dit que le champ scalaire est furtif, et par extension, on parle de solutions de \textbf{trous noirs furtifs}. Pour d'autres, la métrique diffère du cas de la Relativité Générale, auquel cas on parle de \textbf{trous noirs non furtifs}. Par exemple, de tels trous noirs ont été obtenus dans la théorie 4DEGB décrite plus haut. Toutes ces solutions chevelues n'ont qu'un \textbf{cheveu secondaire}. Cela signifie, dans le cas de la symétrie sphérique par exemple, que leur métrique peut différer (dans le cas non furtif) ou pas (dans le cas furtif) de la métrique de Schwarzschild de la Relativité Générale, elle reste dans tous les cas complètement caractérisée par un unique paramètre physique, la masse $M$. \\

Les nouveaux résultats de cette thèse sont présentés dans les chapitres 4, 5 et 6. Le chapitre 4 étudie en toute généralité une vaste classe de théories scalaire-tenseur, dans le cas de la symétrie sphérique. Bien que les équations à résoudre semblent extrêmement compliquées, une réécriture systématique et compacte de ces équations est entreprise, et permet d'obtenir de nombreuses nouvelles solutions de trous noirs. Par exemple, un trou noir furtif est obtenu, mais pour lequel le champ scalaire possède des propriétés assez distinctes par rapport aux trois noirs furtifs usuels. Un résultat remarquable est l'obtention de trous noirs avec \textbf{cheveu primaire} : ces trous noirs sont caractérisés non seulement par leur masse $M$, mais également par une nouvelle quantité, le \textbf{cheveu scalaire primaire} $q$, qui n'a pas d'équivalent en Relativité Générale. De plus, lorsque $q$ et $M$ sont reliés d'une certaine façon, \textbf{la singularité centrale de ces trous noirs disparaît} : la courbure de l'espace-temps demeure finie, même en plein centre de ces trous noirs.
\\

Le chapitre 5 s'intéresse à diverses généralisations de la théorie 4DEGB. Tout d'abord, l'on considère des théories qui ont une forme assez générique, mais similaire à 4DEGB. En étudiant les équations, il est possible de restreindre la forme générique de ces théories et d'obtenir de nouvelles solutions. Ensuite, rappelons que la théorie 4DEGB a été obtenue par réduction dimensionnelle de la théorie Einstein-Gauss-Bonnet, qui est la plus simple des théories de Lovelock (hormis la Relativité Générale). Le reste du chapitre 5 s'intéresse donc à la réduction dimensionnelle, non plus de la théorie Einstein-Gauss-Bonnet, mais de \textbf{la théorie de Lovelock la plus générale}. Des liens intéressants sont ainsi établis entre théories génériques de Lovelock, et théories scalaire-tenseur à quatre dimensions, et une nouvelle solution de trou noir est obtenue dans ce contexte. Certains résultats peuvent être démontrés rigoureusement, tandis que d'autres restent à l'état de conjectures. 
\\

Enfin, il a été évoqué précédemment l'existence de transformations conformes-difformes, permettant de naviguer entre différentes théories scalaire-tenseur. En fait, ces transformations permettent également d'\textbf{engendrer de nouvelles solutions} : en commençant avec une certaine théorie scalaire-tenseur admettant une solution connue, et en appliquant une transformation conforme-difforme à cette solution, la métrique obtenue est solution des équations d'une nouvelle théorie scalaire-tenseur. \\

Ce principe est utilisé à deux reprises dans le chapitre 6. En transformant des solutions de trous noirs, de nouvelles solutions, de géométries très différentes, sont obtenues : d'une part, un \textbf{trou noir en expansion}, d'autre part, un \textbf{trou de ver}. Un trou de ver est un objet astrophysique qui connecte deux régions distinctes de l'espace-temps. En Relativité Générale, un trou de ver ne peut rester ouvert assez longtemps pour permettre à un humain de le traverser que s'il est maintenu en place par de la matière exotique, dont l'énergie est négative, ce qui remet en cause la possibilité de l'existence de tels trous de ver. Dans notre cas, grâce aux modifications à la gravitation apportées par le champ scalaire, le trou de ver obtenu ne nécessite la présence d'aucune matière. Cela complète le panorama des nombreuses solutions de forme fermée pouvant être construites en théories scalaire-tenseur.

\chapter*{List of abbreviations}
\label{chap:abb}
\addcontentsline{toc}{chapter}{\nameref{chap:abb}}
\noindent\textbf{4DEGB} Four-dimensional Einstein-Gauss-Bonnet\vspace{0.25cm}\\
\noindent\textbf{ADM} Arnowitt-Deser-Misner\vspace{0.25cm}\\
\noindent\textbf{adS} Anti-de Sitter\vspace{0.25cm}\\
\noindent\textbf{BBMB} Bocharova–Bronnikov–Melnikov–Bekenstein\vspace{0.25cm}\\
\noindent\textbf{BCL} Babichev-Charmousis-Lehébel\vspace{0.25cm}\\
\noindent\textbf{BH} Black hole\vspace{0.25cm}\\
\noindent\textbf{DGP} Dvali-Gabadadze-Porrati\vspace{0.25cm}\\
\noindent\textbf{DHOST} Degenerate Higher-Order Scalar-Tensor\vspace{0.25cm}\\
\noindent\textbf{dS} De Sitter\vspace{0.25cm}\\
\noindent\textbf{EFT} Effective field theory\vspace{0.25cm}\\
\noindent\textbf{EGB} Einstein-Gauss-Bonnet\vspace{0.25cm}\\
\noindent\textbf{EHT} Event Horizon Telescope\vspace{0.25cm}\\
\noindent\textbf{FLRW} Friedmann-Lemaître-Robertson-Walker\vspace{0.25cm}\\
\noindent\textbf{GB} Gauss-Bonnet\vspace{0.25cm}\\
\noindent\textbf{GLPV} Gleyzes-Langlois-Piazza-Vernizzi\vspace{0.25cm}\\
\noindent\textbf{GR} General Relativity\vspace{0.25cm}\\
\noindent\textbf{GW} Gravitational wave\vspace{0.25cm}\\
\noindent\textbf{KK} Kaluza-Klein\vspace{0.25cm}\\
\noindent\textbf{LISA} Laser Interferometer Space Antenna\vspace{0.25cm}\\
\noindent\textbf{MTZ} Martinez-Troncoso-Zanelli\vspace{0.25cm}\\
\noindent\textbf{QFT} Quantum field theory\vspace{0.25cm}\\
\noindent\textbf{QNM} Quasi-normal mode\\

\newpage

\chapter*{List of symbols}
\label{chap:sym}
\addcontentsline{toc}{chapter}{\nameref{chap:sym}}
\noindent $g_{\mu\nu}$ Metric tensor field\vspace{0.15cm}\\
\noindent $\mathrm{d}s^2$ Associated line element\vspace{0.15cm}\\
\noindent $\phi$ Scalar field\vspace{0.15cm}\\
\noindent $g$ Determinant of $g_{\mu\nu}$\vspace{0.15cm}\\
\noindent $\mathcal{L}_V$ Lie derivative along the vector field $V$\vspace{0.15cm}\\
\noindent $\nabla$ Covariant derivative\vspace{0.15cm}\\
\noindent $\Gamma^\mu_{\nu\rho}$ Connection components\vspace{0.15cm}\\
\noindent $\epsilon_{\mu\nu\rho\sigma}$ Levi-Civita tensor\vspace{0.15cm}\\
\noindent $R_{\mu\nu\rho\sigma}$ Riemann tensor\vspace{0.15cm}\\
\noindent $R_{\mu\nu}$ Ricci tensor\vspace{0.15cm}\\
\noindent $R$ Ricci scalar\vspace{0.15cm}\\
\noindent $\mathcal{G}$ Gauss-Bonnet scalar\vspace{0.15cm}\\
\noindent $\mathcal{R}^{(k)}$ Lovelock curvature invariant of order $k$\vspace{0.15cm}\\
\noindent $\Lambda$ Cosmological constant\vspace{0.15cm}\\
\noindent $T_{\mu\nu}$ Energy-momentum tensor\vspace{0.15cm}\\
\noindent $(t,r,\theta,\varphi)$ Spherically-symmetric coordinates or Boyer-Lindquist coordinates\vspace{0.15cm}\\
\noindent $\mathrm{d}\Omega^2$ Metric of the unit two-sphere\vspace{0.15cm}\\
\noindent $M$ ADM mass (or an integration constant proportional to the ADM mass)\vspace{0.15cm}\\
\noindent $a=J/M$ Angular momentum per unit mass\vspace{0.15cm}\\
\noindent $a(t)$ or $A(t)$ FLRW scale factor\vspace{0.15cm}\\
\noindent $\phi_\mu=\partial_\mu\phi=\nabla_\mu\phi$\vspace{0.15cm}\\
\noindent $\phi_{\mu\nu}=\nabla_\mu\nabla_\nu\phi$\vspace{0.15cm}\\
\noindent $\Box\phi=g^{\mu\nu}\nabla_\mu\nabla_\nu\phi$\vspace{0.15cm}\\
\noindent $X=-\phi^\mu\,\phi_\mu/2$ for (beyond) Horndeski theories\vspace{0.15cm}\\
\noindent $X=\phi^\mu\,\phi_\mu$ for DHOST theories\vspace{0.15cm}\\
\noindent $G_{2,3,4,5}(\phi,X)$ Horndeski / Generalized Galileon functions\vspace{0.15cm}\\
\noindent $F_{4,5}(\phi,X)$ Beyond Horndeski functions\vspace{0.15cm}\\
\noindent $A_{1,2,3,4,5}(\phi,X)$ Quadratic DHOST functions\vspace{0.15cm}\\
\noindent $\mathcal{J}^\mu_\text{shift}$ Noether current associated to shift symmetry\vspace{0.15cm}\\
\noindent $F'$ Derivative of the function $F$ with respect to the radial coordinate $r$\vspace{0.15cm}\\
\noindent $F_\phi$ Derivative of the function $F$ with respect to $\phi$\vspace{0.15cm}\\
\noindent $F_X$ Derivative of the function $F$ with respect to $X$
\newpage 
\chaptermark{CONTENTS}
\sectionmark{CONTENTS}
\tableofcontents 





\disableopenany
\newpage
\pagenumbering{arabic}

\chapter*{Introduction}
\label{chap:intro}
\addcontentsline{toc}{chapter}{\nameref{chap:intro}}
\chaptermark{INTRODUCTION}
\sectionmark{INTRODUCTION}
The standard theory of gravitation is General Relativity (GR), formulated in 1915 by Albert Einstein~\cite{1915SPAW.......778E}. At that time, modifying the established, Newtonian theory of gravitation, was motivated from both experimental and theoretical considerations. 
\\

On the experimental side, Newton's theory of gravitation was not able to predict the correct precession of the perihelion of Mercury~\cite{verrier1}. With Newton's law, the main effect which explains precession of perihelion is the presence of other planets - the fact that a planet undergoes not only the attraction from the Sun, but also from other planets. An hypothetical planet, Vulcan, was even introduced to account for the behaviour of Mercury's orbit~\cite{baum2003search}. No experimental evidence of Vulcan was ever found: it turned out that one needed to modify, not the celestial content of our solar system, but the theory of gravitation itself. Switching from Newton's equations to Einstein's equations solved Mercury's perihelion's problem~\cite{Einstein:1916vd}.
\\

On the theoretical side, Newton's law predicts that, in the gravitational interaction between two masses $m_1$ and $m_2$, a change in $m_1$ implies an instantaneous change of the force exerted on $m_2$. This is incompatible with the relativistic framework developed by Einstein in his special theory of relativity in 1905~\cite{Einstein:1905:EBK}, which predicts that no signal can travel with a speed greater than the speed of light $c$.
\\

Apart from these shortcomings, Newton's theory was very much valid and accurate - and still is nowadays in situations of everyday's life. Of course, the development of GR led to many further experimental predictions, from bending of light~\cite{1920RSPTA.220..291D} to black holes (BHs) and gravitational waves (GWs), all of which have been successfully verified, see the review~\cite{Will:2014kxa}, affirming each time more the extraordinary firmness of GR. It remains however possible to point out limitations of GR, both experimental and theoretical, thus drawing a tempting parallel with the situation of the Newtonian theory 110 years ago.
\\

From the theoretical point of view, gravity, as described by GR, is the only one among the four fundamental forces of nature\footnote{The remaining ones being electromagnetism, weak and strong force.} which has not been accommodated within the framework of quantum field theory (QFT). In particular, GR is a non-renormalizable QFT~\cite{Feynman:1996kb,Hamber:2009zz}, and thus merely stands as an effective field theory (EFT) of a more fundamental theory of quantum gravity~\cite{Donoghue:1994dn}. Quantum effects are expected to play a fundamental role at distances close to the Planck scale. As is stands, GR is breaking down at those distances: there exist singularities of spacetime both at the center of BHs and at the Big Bang~\cite{Hawking:1973uf,Hawking:1976ra}. 
\\

From the experimental point of view, GR is challenged by the observed accelerated expansion of the universe~\cite{SupernovaSearchTeam:1998fmf,SDSS:2005xqv,Astier:2012ba}. On the one hand, this accelerated expansion is in fact perfectly accounted for in GR by the cosmological constant, but on the other hand, the origin and value of this cosmological constant remain a complete mystery~\cite{Weinberg:1988cp,Polchinski:2006gy,Bousso:2007gp,Padilla:2015aaa}.
\\

Therefore, just as Newton's gravity is an extremely useful and accurate theory in a certain domain of validity, but must be modified into GR to make it consistent with a bunch of observations and other theories, GR is an impressive theory regarding many aspects of gravitation, but must be modified to account for other aspects. One may try to formulate the theory of quantum gravity mentioned above, either by quantizing the gravitational field without unifying it to the other forces (e.g. loop quantum gravity~\cite{Ashtekar:2004eh}), or by proposing a unified framework for all forces (e.g. string theory and M-theory~\cite{Becker:2006dvp}). Another less ambitious, more pragmatic approach consists in studying theories which differ from GR in a controlled manner, and may solve some of the shortcomings encountered by GR (such as the dark energy/cosmological constant problem), but do not pretend to be an ultimate theory of gravitation, for instance to possess a better quantum behaviour. These latter theories are generically called \textbf{modified gravity theories}~\cite{Clifton:2011jh}.
\\

Among the large landscape of modified gravity theories, \textbf{scalar-tensor theories}~\cite{Fujii:2003pa,Kobayashi:2019hrl} will be the focus of this thesis. 'Tensor' refers to the metric tensor $g_{\mu\nu}$, which is the usual gravitational field of GR, and possesses two degrees of freedom. At low energies, a modified theory of gravity could naturally include effectively a single additional degree of freedom as compared to GR. This is the 'scalar' part, namely a scalar field $\phi$, coming on top of the metric field and bringing the number of degrees of freedom from two to three. From this brief description, it is quite clear that scalar-tensor theories are both simple and robust modifications of gravity. 
\\

One of the most famous successful predictions of GR is the existence of BHs~\cite{Schwarzschild:1916uq,Oppenheimer:1939ue,Finkelstein:1958zz}, regions of spacetime where the gravitational field is so strong that nothing, not even light, can escape from it. The region from which nothing can escape is delimited by the so-called \textbf{event horizon}. BHs are the result of gravitational collapse~\cite{Joshi:2011rlc}, as first demonstrated by Oppenheimer and Snyder~\cite{Oppenheimer:1939ue}. GR (coupled to electromagnetism) allows for a unique BH solution~\cite{Bekenstein:1971,Teitelboim:1972,Mazur:2000pn,Chrusciel:2012jk}, namely the Kerr-Newman BH\footnote{and its extensions~\cite{CARTER1968399} with positive cosmological constant (Kerr-Newman-de Sitter) and negative one (Kerr-Newman-anti de Sitter).}~\cite{Newman:1965my}. The Kerr-Newman BH is parameterized by three integration constants: the mass $M$, the angular momentum $J$ and the electric charge $Q$. BHs in GR are thus said to have \textbf{no hair}, i.e., no other independent, externally observable physical quantity~\cite{Misner:1973prb}.\\

Accordingly, the expression '\textbf{hairy BH}' refers to a BH possessing one of the following two kinds of hair~\cite{Herdeiro:2015waa}: \textbf{primary hair}, which is a global charge distinct from mass, angular momentum or electric charge; and \textbf{secondary hair}, where the BH metric is dressed with non-trivial additional fields (i.e. other than electromagnetic) but remains entirely determined by $M$, $J$ and $Q$. The Kerr-Newman solution reduces to the Kerr BH~\cite{Kerr:1963ud} when there is no Maxwell field ($Q=0$), which itself reduces to the spherically-symmetric Schwarzschild BH~\cite{Schwarzschild:1916uq} when there is no rotation ($J=0$).
\\

As mentioned before, on the experimental side, GR is above all challenged by the cosmological problem of the accelerated expansion of the universe. On the other hand, GR remains unchallenged regarding its predictions about strong field gravity and thus BHs. The first detection of GWs by LIGO in 2015~\cite{article_ligo}, coming from the merger of two BHs, confirmed once and for all the existence of BHs. The various detected signals agree with the expectations from GR. Notably, after the merger, the system relaxes to a stationary BH, emitting GWs which are a superposition of signals with a frequency depending on the properties of the BH, called quasi-normal modes (QNMs)~\cite{Kokkotas:1999bd}. QNMs predicted by GR are in agreement with all observations~\cite{LIGOScientific:2016lio,Ghosh:2021mrv,LIGOScientific:2021sio}. Future GW detectors, such as the Laser Interferometer Space Antenna (LISA)~\cite{LISA:2017pwj} or the Einstein Telescope~\cite{Maggiore:2019uih} will provide more and more accurate insights into these astronomical events. Other striking evidence of the reality of BHs is of course the first direct image of a BH, namely the supermassive BH M87$^\star$, published in 2019 following observations made by the Event Horizon Telescope (EHT)~\cite{EventHorizonTelescope:2019dse}.
\\

The most unpleasant feature concerning BHs in GR is the existence of a singularity, where curvature of spacetime becomes infinite. In any case, scalar-tensor theories are well-motivated modifications of gravity, and one must therefore study their BH solutions and see to what extent they differ or coincide with the BHs of GR. If there is a discrepancy, is it only quantitative (e.g., modification of the value of the gravitational field at a given point of spacetime) or qualitative (e.g., removal of the spacetime singularity)? Such discrepancies could in turn be observed, for instance in the GW signal of binary BH mergers or in the orbit of stars around a BH.
\\

The field equations of scalar-tensor theories are much more intricate than those of GR, which already are far from simple. Most solutions of these field equations are therefore obtained with numerical simulations, or sometimes by an expansion in small parameters, typically a coupling constant. This thesis nevertheless aims at describing \textbf{exact, closed-form BH solutions in scalar-tensor theories}. The important word here is 'closed-form', it means that the metric of the BH can be written down explicitly (if possible with simple, usual mathematical functions). The word 'exact' mostly stands for the fact that no approximation in small parameters is used. Just talking about 'exact solutions' would not be accurate, since a numerical solving may be exact; however, now that it has been explained, this abuse of terminology is largely used in this thesis, starting with its title. The relevance of closed-form solutions may be motivated from different aspects.
\\

First, they are more convenient than a numerical solution. A numerical solution relies on a solving algorithm, presented by a particular researcher, and other researchers willing to reproduce the solution must in turn implement the same (or a similar) code on their own machine. The properties of numerical BHs are not necessarily obvious, typically, if the theory is parameterized by two coupling constants and the BH by its mass, one must make a number of different simulations to try to infer approximately which values of these different parameters lead to no horizon (naked singularity), one horizon, two horizons. A closed-form BH, on the contrary, is more explicit: its properties can in general be inferred from basic maths; and more 'democratic': anyone can take the proposed solution, plug it into the field equations, and verify that it is indeed a solution, and anyone can study himself its physical properties.
\\

Second, closed-form solutions are quite elegant and attractive. They argue for some kind of simplicity and naturalness of the theory under consideration. It is probably better for the scope of GR that the Schwarzschild and Kerr BHs admit a closed form. Of course, if the fundamental laws of nature are such that the most complete theory does not allow for exact solutions, this is how it is and one must accept it. Nevertheless, even complicated theories like string theory admit some simple closed-form results which serve as starting points towards more complicated, approximate or numerical results. These closed-form results of simple situations often contain many hints about what one can expect from numerical results of more complicated situations. 
\\

The plan of this thesis unfolds as follows:
\\

\textbf{Chapter~\ref{chap:modifGR}, Modifying General Relativity: from Lovelock's theorem to scalar-tensor theories}. This chapter starts with a brief overview of GR and its shortcomings, then presents Lovelock's theorem, which constrains the possibilities when it comes to modifying gravity. Adding a scalar degree of freedom and moving into scalar-tensor theories is not the only one, but certain other approaches are in fact equivalent, thus arguing for the robustness of scalar-tensor gravity.
\\

\textbf{Chapter~\ref{chap:st}, Generic aspects of scalar-tensor theories}. This chapter reviews the different classes of scalar-tensor theories (Horndeski, beyond Horndeski and DHOST), their relations through conformal-disformal transformations, and the symmetries (shift and conformal) they may be endowed with.
\\

\textbf{Chapter~\ref{sec:state}, State-of-the-art of closed-form black hole solutions in scalar-tensor theories}. This chapter describes the various usual closed-form BH solutions in scalar-tensor theories; by usual, it is meant all solutions which existed before this thesis started.
\\

\textbf{Chapter~\ref{chap:4}, Static, spherically-symmetric black holes in shift-symmetric beyond Horndeski theories}. This chapter studies the scalar-tensor theories of the shift-symmetric, beyond Horndeski class, under the assumption that the BH metric is static and spherically-symmetric. This enables a compact reformulation of the field equations, and their integration in a number of cases. Among the new BH solutions, some have secondary hair (they are fully characterized by a unique parameter, their mass $M$), while other have primary hair (they are parameterized by another integration constant, in addition to the mass).
\\

\textbf{Chapter~\ref{chap:5}, Lovelock invariants and conformally-coupled scalar field}. This chapter gathers results on couplings between the scalar field and Lovelock invariants. New BH solutions are found in four-dimensional scalar-tensor theories with no shift nor conformal symmetry. The chapter continues with higher-dimensional considerations, namely a scalar field coupled to higher-order Lovelock invariants in a conformally-invariant way. These considerations pave the way towards compactifying higher-dimensional gravity down to lower (for instance four) dimensions.
\\

\textbf{Chapter~\ref{chap:6}, Generation of solutions with conformal and disformal transformations}. This chapter presents two closed-form solutions obtained through disformal and conformal transformations. One of them is a wormhole, the other one a rotating BH embedded in an expanding universe.
\\

Chapters 1 to 3 consist in their vast majority of previously known results, gathered and organized by the author in order to introduce the subject, and prepare the reader for the remaining chapters. Chapters 4 to 6 present results which were obtained in the course of this thesis. Only the results of section~\ref{sec:diagkk} are completely new, that is, unpublished. The other results of chapters 4 to 6 are based on the eight following works (six publications in peer-reviewed journals, one conference proceedings and one preprint) realized during this thesis:
\\

\noindent\cite{Bakopoulos:2022csr}~\bibentry{Bakopoulos:2022csr}
\\

\noindent\cite{Babichev:2022awg}~\bibentry{Babichev:2022awg}
\\

\noindent\cite{Bakopoulos:2022gdv}~\bibentry{Bakopoulos:2022gdv}
\\

\noindent\cite{Babichev:2023rhn}~\bibentry{Babichev:2023rhn}
\\

\noindent\cite{Babichev:2023dhs}~\bibentry{Babichev:2023dhs}
\\

\noindent\cite{Babichev:2023mgk}~\bibentry{Babichev:2023mgk}
\\

\noindent\cite{Babichev:2023psy}~\bibentry{Babichev:2023psy}
\\

\noindent\cite{Bakopoulos:2023fmv}~\bibentry{Bakopoulos:2023fmv}

\chapter[Modifying General Relativity: from Lovelock's theorem to scalar-tensor theories]{Modifying General Relativity: from Lovelock's theorem to scalar-tensor theories}\label{chap:modifGR}
\chaptermark{Chap.~1~~Modifying GR: from Lovelock's theorem to scalar-tensor theories}
This chapter starts with a brief overview of GR, Sec.~\ref{sec:reminder_gr}, first presenting its mathematical framework, notations and equations, then some of its results and shortcomings, which constitute motivations for modifying GR. In a second time, Sec.~\ref{sec:lovelock_th} presents \textbf{Lovelock's theorem}, which sets constraints on possible modified theories of gravity. The possibilities for modifying gravity are enumerated, and \textbf{scalar-tensor theories} emerge as a simple and natural modification of gravity, possessing links or equivalences with other modified theories of gravity. The link with higher-dimensional gravity, which is of importance in the rest of this manuscript, is highlighted in Sec.~\ref{sec:compact}. Finally, Sec.~\ref{sec:higher_order} presents the \textbf{Ostrogradsky instability} and how to evade it. On the one hand, this instability plagues some other natural modifications of gravity, like $f(R)$ theories; on the other hand, the evasion of this instability is at the core of the correct construction of scalar-tensor theories, described in Chap.~\ref{chap:st}.
\section[General Relativity: overview, results, shortcomings]{General Relativity: overview, results, shortcomings}\label{sec:reminder_gr}
\sectionmark{1.1~~GR: overview, results, shortcomings}
In this section, the definitions, standard results, notations and conventions are those of the usual notebooks where the interested reader can find any details on GR~\cite{Hawking:1973uf,Misner:1973prb,Wald:1984rg,Carroll:2004st,Poisson:2009pwt}.
\subsection{Mathematical framework, notations and equations}\label{subsec:mathFram}
This extremely brief reminder on Einstein's GR enables to set notations and, more importantly, to have in mind the various assumptions of GR: modifying gravity will indeed amount to modifying these assumptions. Spacetime is a four-dimensional manifold $\mathcal{M}$ endowed with a metric tensor $g_{\mu\nu}$, i.e. a symmetric covariant two-tensor field. The metric is Lorentzian, i.e. has signature $(-,+,+,+)$, so its determinant $g\equiv \det(g)$ is negative. More generally, given a coordinate chart $\left(x^0,x^1,x^2,x^3\right)$, the components of a tensor field $V$ of type $(m,n)$ ($m$-contravariant and $n$-covariant) are written $V^{\mu_1\cdots\mu_m}_{\qquad\,\,\,\nu_1\cdots\nu_n}$, such that the tensor field is
\begin{equation}
V = V^{\mu_1\cdots\mu_m}_{\qquad\,\,\,\nu_1\cdots\nu_n}\partial_{\mu_1}\otimes\cdots\otimes\partial_{\mu_m}\otimes\mathrm{d}x^{\nu_1}\otimes\cdots\otimes\mathrm{d}x^{\nu_n},
\end{equation}
where $\partial_\mu$ is the vector field $\frac{\partial}{\partial x^\mu}$ and $\mathrm{d}x^\nu$ is the dual one-form, $\mathrm{d}x^\nu\left(\partial_\mu\right)=\delta^\nu_\mu$. Indices are lowered and raised with respectively the metric $g_{\mu\nu}$ and its inverse $g^{\mu\nu}$. The symmetrization and antisymmetrization of indices are denoted by parenthesis and brackets respectively and defined by
\begin{align}
V_{(\mu\nu)} \equiv \frac{1}{2}\left(V_{\mu\nu}+V_{\nu\mu}\right), \quad V_{(\mu\nu\rho)} \equiv \frac{1}{6}\left(V_{\mu\nu\rho}+V_{\nu\rho\mu}+V_{\rho\mu\nu}+V_{\nu\mu\rho}+V_{\mu\rho\nu}+V_{\rho\nu\mu}\right),\\
V_{[\mu\nu]} \equiv \frac{1}{2}\left(V_{\mu\nu}-V_{\nu\mu}\right), \quad V_{[\mu\nu\rho]} \equiv \frac{1}{6}\left(V_{\mu\nu\rho}+V_{\nu\rho\mu}+V_{\rho\mu\nu}-V_{\nu\mu\rho}-V_{\mu\rho\nu}-V_{\rho\nu\mu}\right),
\end{align}
and so on, with a factor $1/n!$ where $n$ is the number of (anti)symmetrized indices, and a minus sign in front of odd permutations for antisymmetrization. The Levi-Civita tensor is
\begin{equation}
\epsilon_{\mu\nu\rho\sigma}\equiv \sqrt{-g}\,\varepsilon_{\mu\nu\rho\sigma}, 
\end{equation}
where $\varepsilon_{\mu\nu\rho\sigma}$ is the completely antisymmetric symbol with $\varepsilon_{0123}=1$. A tensor field of type $(1,0)$ is a \textbf{vector field} $V^\mu$. Its \textbf{covariant derivative} is defined as
\begin{equation}
\nabla_\nu V^\mu \equiv \partial_\nu V^\mu +\Gamma^\mu_{\nu\rho}V^\rho,
\end{equation}
where $\Gamma^\mu_{\nu\rho}$ are the components of the \textbf{connection}. On the other hand, a \textbf{scalar field} is simply a function on spacetime, $\phi:\mathcal{M}\to \mathbb{R}$ (other cases include for instance complex-valued scalar fields, but this thesis focuses on real scalar fields). The covariant derivative is extended to any type of tensors by requiring Leibniz (product) rule, commutation with contractions, and that it reduces to partial derivative when acting on scalar fields. The \textbf{Riemann tensor} $R^{\mu}_{\,\,\,\nu\rho\sigma}$ measures the failure of covariant derivatives to commute,
\begin{equation}
\left(\nabla_\rho\nabla_\sigma-\nabla_\sigma\nabla_\rho\right)V^\mu = R^{\mu}_{\,\,\,\nu\rho\sigma}V^\nu-T^\lambda_{\rho\sigma}\nabla_\lambda V^\mu,
\end{equation}
where the \textbf{torsion tensor} appears,
\begin{equation}
T^\lambda_{\rho\sigma}\equiv 2\Gamma^\lambda_{[\rho\sigma]}.\label{eq:torsion}
\end{equation}
The Riemann tensor has components
\begin{equation}
R^{\mu}_{\,\,\,\nu\rho\sigma}=\partial_\rho\Gamma^{\mu}_{\sigma\nu}-\partial_\sigma\Gamma^{\mu}_{\rho\nu}+\Gamma^\mu_{\rho\lambda}\Gamma^\lambda_{\sigma\nu}-\Gamma^\mu_{\sigma\lambda}\Gamma^\lambda_{\rho\nu}.
\end{equation}
The \textbf{Ricci tensor and Ricci scalar} are respectively
\begin{equation}
R_{\mu\nu}\equiv R^{\rho}_{\,\,\,\mu\rho\nu},\quad R \equiv R^\mu_\mu.
\end{equation}
In GR, the connection is assumed to be \textbf{without torsion} and \textbf{compatible with the metric},
\begin{equation}
T^\lambda_{\rho\sigma}=0,\quad Q_{\mu\nu\rho}=0,
\end{equation} 
where appears the \textbf{non-metricity tensor}~\cite{Lazkoz:2019sjl},
\begin{equation}
Q_{\mu\nu\rho}\equiv\nabla_\mu g_{\nu\rho}=\partial_\mu g_{\nu\rho} - \Gamma^\lambda_{\mu\nu}g_{\lambda\rho}-\Gamma^\lambda_{\mu\rho}g_{\nu\lambda}.\label{eq:nonmetr}
\end{equation}
This uniquely defines the connection as being the \textbf{Levi-Civita connection}, with components called the \textbf{Christoffel symbols},
\begin{equation}
\Gamma^\mu_{\nu\rho}=\frac{1}{2}g^{\mu\lambda}\left(\partial_{\nu}g_{\rho\lambda}+\partial_\rho g_{\nu\lambda}-\partial_\lambda g_{\nu\rho}\right).\label{eq:christ}
\end{equation}
The action functional $S$ of GR is the \textbf{Einstein-Hilbert action} (with a \textbf{cosmological constant} $\Lambda$),
\begin{equation}
S_{\text{EH}}\left[g_{\mu\nu}\right]=\frac{1}{2\kappa}\int \mathrm{d}^4x\sqrt{-g} \left(R-2\Lambda\right).\label{eq:ehaction}
\end{equation}
In front of the integral appears $\kappa = 8\pi G c^{-4}$, where $G$ is Newton's constant and $c$ the speed of light. Adding a matter action $S_{\text{m}}\left[g_{\mu\nu},\Psi_{\text{m}}\right]$ depending on matter fields $\Psi_{\text{m}}$, the total action reads
\begin{equation}
S\left[g_{\mu\nu},\Psi_{\text{m}}\right]=S_{\text{EH}}\left[g_{\mu\nu}\right] + S_{\text{m}}\left[g_{\mu\nu},\Psi_{\text{m}}\right].
\end{equation}
Varying with respect to $g_{\mu\nu}$ leads to the field equations of GR, or \textbf{Einstein's field equations}, 
\begin{equation}
G_{\mu\nu}+\Lambda g_{\mu\nu} = \kappa T_{\mu\nu},\label{eq:einstein_eq}
\end{equation}
where $G_{\mu\nu}$ is the Einstein tensor and $T_{\mu\nu}$ is the energy-momentum tensor of the matter fields,
\begin{equation}
G_{\mu\nu} \equiv R_{\mu\nu} - \frac{1}{2}Rg_{\mu\nu},\quad T_{\mu\nu} \equiv \frac{-2}{\sqrt{-g}}\frac{\delta S_{\text{m}}}{\delta g^{\mu\nu}}.
\end{equation}
Using the definition of the Ricci tensor, one can prove by simple computation the (contracted) \textbf{Bianchi identity},
\begin{equation}
\nabla^\mu G_{\mu\nu} = 0,
\end{equation}
which implies that matter fields verify energy-momentum conservation,
\begin{equation}
\nabla^\mu T_{\mu\nu}=0.
\end{equation}
Finding solutions to Einstein's equations is made easier in situations displaying symmetries. Symmetries of spacetime are associated to the existence of a \textbf{Killing vector field} $K=K^\mu\partial_\mu$, obeying the Killing equation
\begin{equation}
\nabla_{(\mu} K_{\nu)}=0.
\end{equation}
Indeed, this equation is equivalent to the vanishing of the \textbf{Lie derivative} of the metric along $K$,
\begin{align}
\mathcal{L}_K g_{\mu\nu}{}&{}\equiv K^\rho\partial_\rho g_{\mu\nu}+g_{\rho\nu}\partial_\mu K^\rho+g_{\mu\rho}\partial_\nu K^\rho\label{eq:lieintro1}\\
{}&{}=K^\rho\nabla_\rho g_{\mu\nu}+g_{\rho\nu}\nabla_\mu K^\rho+g_{\mu\rho}\nabla_\nu K^\rho\label{eq:lieintro2}\\
{}&{}=g_{\rho\nu}\nabla_\mu K^\rho+g_{\mu\rho}\nabla_\nu K^\rho.
\end{align}
Just a for the covariant derivative, the definition (\ref{eq:lieintro1}) is extended to analogous expressions for any type of tensor, by using the Leibniz rule.
\subsection{Black holes and singularities; cosmology and dark energy problem}\label{subsec:bhsing}
From now on, units are chosen such that $c=G=1$. Einstein's equations (\ref{eq:einstein_eq}), although non-linear, admit a number of solutions. The three most important solutions are probably BHs, isotropic and homogeneous universe, and GWs (which are in general obtained by linearizing Einstein's equations around a flat background, but can also be a full nonlinear solution~\cite{Kundt:1962svr,Robinson:1962zz,Robinson:1960zzb}). All these solutions have a number of consequences which can be confronted with experiments, and the theory of GR appears as extremely well-tested~\cite{Will:2014kxa}. However, as regards BHs and cosmology, their overall success comes with a small number of shortcomings, which justify the need to modify gravity.
\subsubsection{Black holes in General Relativity: Schwarzschild and Kerr metrics}
The unique~\cite{Bekenstein:1971,Teitelboim:1972,Mazur:2000pn,Chrusciel:2012jk} stationary\footnote{see definition below.}, asymptotically flat BH of GR in vacuum (no Maxwell term) is the Kerr metric~\cite{Kerr:1963ud}, which has the following line element $\mathrm{d}s^2\equiv g_{\mu\nu}\mathrm{d}x^\mu\mathrm{d}x^\nu$ in Boyer-Lindquist coordinates~\cite{Boyer:1966qh} $(t,r,\theta,\varphi)$:
\begin{align}
\mathrm{d}s^2={}&{}-\left(1-\frac{2Mr}{\Sigma}\right)\mathrm{d}t^2+\frac{\Sigma}{\Delta}\mathrm{d}r^2+\Sigma\mathrm{d}\theta^2-\frac{4Mar\sin^2\theta}{\Sigma}\mathrm{d}t\mathrm{d}\varphi+\frac{\sin^2\theta}{\Sigma}\Upsilon\mathrm{d}\varphi^2.\label{eq:kerr_intro}
\end{align}
The Kerr BH reduces to the Schwarzschild BH~\cite{Schwarzschild:1916uq} if there is no rotation ($a=0$),
\begin{equation}
\mathrm{d}s^2=-\left(1-\frac{2M}{r}\right)\mathrm{d}t^2+\frac{\mathrm{d}r^2}{\left(1-\frac{2M}{r}\right)}+r^2\mathrm{d}\Omega^2.
\end{equation}
$M$ is the mass of the BH, $a$ the angular momentum per unit mass~\cite{Arnowitt:1961zz,Gibbons:1976ue}, and
\begin{equation}
\Sigma = r^2+a^2\cos^2\theta,\quad \Delta = r^2+a^2-2Mr,\quad \Upsilon=\left(r^2+a^2\right)^2-a^2\Delta\sin^2\theta.\label{eq:delta_def_intro}
\end{equation}
Also, $\mathrm{d}\Omega^2\equiv \mathrm{d}\theta^2+\sin^2\theta\,\mathrm{d}\varphi^2$ is the metric on the unit 2-sphere. The Kerr and Schwarzschild BHs are solutions of Einstein's equations (\ref{eq:einstein_eq}) in the absence of cosmological constant, $\Lambda=0$. The Schwarzschild and Kerr solutions were found respectively in 1916 and 1963. If one includes a non-vanishing $\Lambda$, these metrics are generalized to the Kerr-de Sitter (dS) or Kerr-anti-de Sitter (adS) BHs~\cite{CARTER1968399}. 
\\

The Schwarzschild and Kerr spacetimes are \textbf{stationary}, i.e. they admit an asymptotically timelike Killing vector field, namely $\partial_t$. Schwarzschild metric is even said to be \textbf{static}, because the line element is unchanged under $t\to -t$. Both spacetimes are also \textbf{axisymmetric}, that is, they have a spacelike Killing vector, $\partial_\varphi$. In fact, Schwarzschild metric is \textbf{spherically-symmetric}, because, along with $\partial_\varphi$, the vectors $-\sin\varphi\,\partial_\theta-\cot\theta\cos\varphi\,\partial_\varphi$ and $\cos\varphi\,\partial_\theta-\cot\theta\sin\varphi\,\partial_\varphi$ form an $SO(3)$ algebra\footnote{for the bracket between vector fields $\left[X,Y\right]^\mu\equiv X^\nu\partial_\nu Y^\mu-Y^\nu\partial_\nu X^\mu$.}. A static and spherically-symmetric spacetime (like Schwarzschild) has the generic form
\begin{equation}
\mathrm{d}s^2=-N(r)^2f(r)\mathrm{d}t^2+\mathrm{d}r^2/f(r)+r^2\mathrm{d}\Omega^2.\label{eq:static_intro}
\end{equation}
A \textbf{causal curve} is a parameterized curve $x^\mu(\lambda)$ on spacetime ($\lambda$ is the parameter), such that the velocity $u^\mu=\mathrm{d}x^\mu/\mathrm{d}\lambda$ is timelike, $u^\mu u_\mu<0$, or null, $u^\mu u_\mu=0$. The Schwarzschild and Kerr spacetimes are called 'BHs' because they admit a region such that no causal curve can escape from it. The boundary of such a region is called the \textbf{event horizon}. Being a BH, and admitting an event horizon, is one and the same thing. The event horizon is a null surface, i.e. its normal is a null vector field.
\\

Finding the location of the event horizon is quite easy for Schwarzschild, and more generally for a static and spherically-symmetric spacetime (\ref{eq:static_intro}): the horizon $r_h$ is root of $g^{rr}$, i.e. $f\left(r_h\right)=0$. In particular, the Schwarzschild radius is $r_\text{Sch}=2M$. For a non-spherically-symmetric spacetime like Kerr (\ref{eq:kerr_intro}), the causal structure is more complicated:
\begin{itemize}
\item Static observers (with velocity $u\propto \partial_t$) cease to exist when $\partial_t$ ceases to be timelike. This happens when $g_{tt}$ becomes positive, i.e. at $r=M+\sqrt{M^2-a^2\cos^2\theta}$ for Kerr. This surface $r=r(\theta)$ is called the \textbf{static limit or ergosphere}.
\item Stationary observers (with velocity $u=u^t\partial_t+u^\varphi\partial_\varphi$) cease to exist when $g_{tt}g_{\varphi\varphi}-g_{t\varphi}^2$ becomes negative, i.e. at $r=M+\sqrt{M^2-a^2}$ for Kerr. This surface is called the \textbf{stationary limit}. 
\item For Kerr, this stationary limit is a null surface and is in fact the event horizon of the Kerr BH. Note however that the stationary limit has no reason to coincide with the event horizon for a generic stationary spacetime. The fact that this Kerr event horizon corresponds to $g^{rr}=0$ is also due to the specific form of Kerr, but does not hold true in general. It holds true in general only for static, spherically-symmetric spacetimes (\ref{eq:static_intro}). 
\end{itemize}
\subsubsection{Singularity at the core of black holes}
One may be tempted to quantify the amount of spacetime curvature for the above Schwarzschild and Kerr metrics, by computing a Lorentz scalar associated with curvature. The trace of the vacuum Einstein's equations implies that the Ricci scalar $R$ of both metrics vanish. Another relevant Lorentz scalar is the \textbf{Kretschmann scalar},
\begin{equation}
K\equiv R_{\mu\nu\rho\sigma}R^{\mu\nu\rho\sigma}.
\end{equation}
Its value for the Kerr and Schwarzschild BHs is
\begin{equation}
K_\text{Kerr}=\frac{48M^2}{\Sigma^6}\Bigl[r^6-15r^4a^2\cos^2\theta+15r^2a^4\cos^4\theta-a^6\cos^6\theta\Bigr],\quad K_\text{Sch}=\frac{48M^2}{r^6}.
\end{equation}
One sees that it diverges at $\Sigma=0$ (i.e. $r=0$ and $\theta=\pi/2$) for Kerr, and at the center $r=0$ for Schwarzschild. These are \textbf{curvature singularities}: the curvature inside the BHs of GR diverges and becomes infinite, indicating a breakdown of the theory. For precise definitions of singularities and their appearance as endstate of gravitational collapse, the reader is referred to~\cite{Hawking:1973uf,Hawking:1976ra}. GR cannot be trusted anymore near the singularity, and a new theory is needed, which would modify the geometry of spacetime at these scales and erase the singularity.
\subsubsection{Homogeneous, isotropic universe in General Relativity: FLRW metric}
Cosmology usually assumes that the universe is homogeneous and isotropic, which is the case at sufficiently large scales and enables to describe the universe with a very good accuracy. Under this assumption\footnote{and the assumption that sections of constant time are flat and not spherical nor ellipsoidal, which we assume here for brevity of the discussion.}, one can endow spacetime with coordinates $(t,x,y,z)$, where $(x,y,z)$ are the usual Cartesian coordinates, such that the metric reads
\begin{equation}
\mathrm{d}s^2=-\mathrm{d}t^2+a(t)^2\left(\mathrm{d}x^2+\mathrm{d}y^2+\mathrm{d}z^2\right),\label{eq:firstflrw}
\end{equation}
and the energy-momentum tensor of the matter inside the universe reads
\begin{equation}
T^\mu_{\nu}\partial_\mu\otimes\mathrm{d}x^\nu=-\rho(t)\,\partial_t\otimes\mathrm{d}t+p(t)\left(\partial_x\otimes\mathrm{d}x+\partial_y\otimes\mathrm{d}y+\partial_z\otimes\mathrm{d}z\right).
\end{equation}
In the metric, which is called Friedmann-Lema\^itre-Robertson-Walker (FLRW) metric, $a(t)$ is the \textbf{scale factor}, each section of given time $t=t_0$ is seen to be the flat space metric [multiplied by a constant factor $a(t_0)$]. In $T_{\mu\nu}$, $\rho=\rho(t)$ is the energy density and $p=p(t)$ the pressure. Note that the cosmological constant $\Lambda$ can be taken into account as a part of $T_{\mu\nu}$ with $\rho_\Lambda=-p_\Lambda=\Lambda/\kappa$. Then, Einstein's equations (\ref{eq:einstein_eq}) determine the evolution of the scale factor:
\begin{equation}
\frac{\dot{a}^2}{a^2}=\frac{\kappa\rho}{3},\quad 2\frac{\ddot{a}}{a} +\frac{\dot{a}^2}{a^2}=-\kappa p.\label{eq:Fried}
\end{equation}
The universe currently undergoes a phase of accelerating expansion. This fact comes most famously from supernova observations~\cite{SupernovaSearchTeam:1998fmf}, but also from e.g. baryon acoustic oscillations~\cite{SDSS:2005xqv}. We refer the interested reader to the concise review~\cite{Astier:2012ba}. In the framework of GR and FLRW cosmology, this can be explained if the energy content of the universe is dominated by the cosmological constant: $\rho\approx\rho_\Lambda$, $p\approx p_\Lambda$. Eqs. (\ref{eq:Fried}) then lead to
\begin{equation}
a(t)\propto \exp\left(t\sqrt{\frac{\Lambda}{3}}\right).
\end{equation}
This form of the scale factor is indeed totally consistent with the current accelerating expansion of the universe provided the cosmological constant has value~\cite{Planck:2018nkj}
\begin{equation}
\Lambda_\text{exp}=(1.11\pm 0.02)\times 10^{-52}\,\text{m}^{-2}.
\end{equation}
This value implies that the cosmological constant represents a fraction of the total energy density of the universe of around 68\%~\cite{Planck:2018vyg}. 
\subsubsection{Cosmological constant problem}
However, trying to give a theoretical interpretation of this value leads to a \textbf{fine-tuning problem}. Let us describe this problem without entering in all subtleties, which can be found in~\cite{Weinberg:1988cp,Polchinski:2006gy,Bousso:2007gp,Padilla:2015aaa} and references therein. It comes from the confrontation between QFT and GR. According to QFT, the vacuum has an energy, 
\begin{equation}
V_{\text{vac}}\sim\sum \mathcal{O}(1)m^4,
\end{equation}
where the sum bears on all particles and $m$ is their respective mass. QFT and the equivalence principle also imply that the vacuum gravitates as a cosmological constant, thus giving a contribution $\Lambda_\text{QFT}=8\pi GV_\text{vac}c^{-4}$. There is also the contribution $\Lambda$ from the cosmological constant appearing in the GR action (\ref{eq:ehaction}). Therefore, the total cosmological constant is $\Lambda+\Lambda_\text{QFT}$, and this should coincide with the experimental value supporting the current accelerated expansion of the universe:
\begin{equation}
\Lambda_\text{exp}\stackrel{?}{\sim}\Lambda+\Lambda_\text{QFT}=\Lambda+8\pi GV_\text{vac}c^{-4}.\label{eq:finetune}
\end{equation}
However, $V_\text{vac}\gtrsim (\text{TeV})^4$ as regards the finite contributions of the Standard Model, which includes particles up to the TeV mass scale. Therefore, rewriting (\ref{eq:finetune}) in $\text{m}^{-2}$ gives:
\begin{equation}
10^{-52} \sim \Lambda+10^6.
\end{equation}
Thus, there must be a fine-tuning of around 60 digits between the non-diverging parts\footnote{Again, a more detailed explanation can be found for exemple in~\cite{Padilla:2015aaa}.} of $\Lambda$ and $\Lambda_\text{QFT}$. This is the \textbf{cosmological constant problem}: the current era of accelerated expansion of the universe can be accounted for in GR by an FLRW universe dominated by the cosmological constant, but the value of the cosmological constant remains mysterious. More generally, the cosmological constant problem may be called the \textbf{dark energy problem}. Dark energy is defined as the unknown source of energy which drives the accelerated expansion of the universe. In this context, the cosmological constant is only one of the various candidates for dark energy (its main competitor being scalar fields), but, as illustrated by the cosmological constant problem, up to now, none of the candidates offers a fully satisfactory answer to the question of dark energy. For a detailed account on different approaches to dark energy, see~\cite{Peebles:2002gy,Copeland:2006wr}.
\\

Apart from the two that we have detailed (BH singularities and dark energy problem), other good reasons exist to modify gravity, like the absence of consistent quantum description of the gravitational interaction, see e.g.~\cite{tHooft:1974toh,Deser:1974cz}. The problem of quantum gravity rather leads to new theoretical frameworks, like loop quantum gravity~\cite{Ashtekar:2004eh} or string theory/M-theory~\cite{Becker:2006dvp}. On the other hand, the dark energy problem typically leads, as already mentioned, to the introduction of scalar fields as dark energy candidates. Another approach is to modify GR, leading to the \textbf{modified theories of gravity}, which do not aim at solving all the shortcomings of GR (like its quantum behaviour) but only part of them, by introducing controlled deviations from GR. Many ways exist to modify gravity. We will briefly present some of them below (paragraph~\ref{subsec:various}), but we refer the reader to the comprehensive review~\cite{Clifton:2011jh} for details on all existing possibilities and their consequences.
\\

It is not possible to modify GR arbitrarily. There exists a number of constraints, as illustrated by the theorem demonstrated by David Lovelock in 1971. 
\section[Constraints on possible modifications: Lovelock's theorem]{Constraints on possible modifications: Lovelock's theorem}\label{sec:lovelock_th}
\sectionmark{1.2~~Constraints on possible modifications: Lovelock's theorem}
\subsection{Lovelock's theorem}
Lovelock's theorem~\cite{Lovelock:1971yv} greatly helps one to understand how to, or how not to, modify GR. The assumptions of Lovelock's theorem are the following:
\begin{enumerate}
\item Spacetime is a $D$-dimensional Lorentzian manifold (in GR, $D=4$),
\item Gravity is mediated by a unique metric tensor field $g_{\mu\nu}$, so the action functional (without matter fields) is of the form $S\left[g_{\mu\nu}\right]$, 
\item The metric $g_{\mu\nu}$ defines a covariant derivative $\nabla$ through the Levi-Civita connection, with Christoffel symbols (\ref{eq:christ}),
\item The theory is diffeomorphism-invariant,
\item The field equations are of second order (i.e. second order in the derivatives of the fields, here the metric $g_{\mu\nu}$).
\end{enumerate}
Also, the field equations are assumed to be conserved: $\nabla^\mu\mathcal{E}_{\mu\nu}=0$, where $\mathcal{E}_{\mu\nu}\equiv 2/\sqrt{-g}\,\delta S/\delta g^{\mu\nu}$ (this ensures energy-momentum conservation when including matter fields). Under these assumptions, Lovelock proved that the action functional reads
\begin{equation}
S\left[g_{\mu\nu}\right] = \int\mathrm{d}^Dx\sqrt{-g}\sum_{k=0}^{\lfloor\frac{D-1}{2}\rfloor}\alpha_k\mathcal{R}^{\left(k\right)},\label{eq:lov}
\end{equation} 
where the upper limit of the sum is the floor part of $\left(D-1\right)/2$. The $\alpha_k$'s are coupling constants, while $\mathcal{R}^{\left(k\right)}$ is a curvature invariant, of order $k$ in powers of the Riemann tensor, given by
\begin{equation}
\mathcal{R}^{\left(k\right)} \equiv \frac{1}{2^k}\delta^{\mu_1\nu_1\cdots\mu_k\nu_k}_{\alpha_1\beta_1\cdots\alpha_k\beta_k}\prod_{i=1}^{k} R^{\alpha_i\beta_i}_{\quad\,\mu_i\nu_i}.\label{eq:lov_def}
\end{equation}
The generalized Kronecker delta is
\begin{equation}
\delta^{\rho_1\cdots\rho_p}_{\sigma_1\cdots\sigma_p}\equiv p!\,\delta^{\rho_1}_{[\sigma_1}\cdots\delta^{\rho_p}_{\sigma_p]}.
\end{equation}
Notably, $\mathcal{R}^{\left(0\right)}=1$ and $\mathcal{R}^{\left(1\right)}=R$, the Ricci scalar. One can normalize the couplings so that $\alpha_1=1$, and call $\alpha_0=-2\Lambda$, then the sum up to $k=2$ reads
\begin{equation}
\sum_{k=0}^{2}\alpha_k\mathcal{R}^{\left(k\right)} = -2\Lambda + R + \alpha_2\mathcal{G},\label{eq:egb}
\end{equation}
This is the Einstein-Hilbert Lagrangian with an additional term,
\begin{equation}
\mathcal{G}\equiv \mathcal{R}^{\left(2\right)}=R^2-4R_{\mu\nu}R^{\mu\nu}+R_{\mu\nu\rho\sigma}R^{\mu\nu\rho\sigma},
\end{equation}
called the Gauss-Bonnet (GB) invariant. The following term in the sum (cubic Lovelock invariant, $k=3$), which we will need later, is still rather compact,
\begin{align}
\mathcal{R}^{\left(3\right)} ={}&{} R^3-12RR_{\mu\nu}R^{\mu\nu}+3RR_{\mu\nu\rho\sigma}R^{\mu\nu\rho\sigma}+16R\indices{_\mu^\nu}R^{\mu\rho}R_{\rho\nu}+24R^{\mu\nu}R^{\rho\sigma}R_{\mu\rho\nu\sigma} \nonumber\\{}&{}- 24 R^{\mu\nu}R\indices{_\mu^\rho^\sigma^\lambda}R_{\nu\rho\sigma\lambda}-8R\indices{_\mu^\nu_\rho^\sigma}R^{\mu\lambda\rho\tau}R_{\lambda\sigma\tau\nu} + 2 R\indices{_\mu_\nu^\rho^\sigma}R^{\mu\nu\lambda\tau}R_{\lambda\tau\rho\sigma}.
\end{align}
By varying the Lovelock action, one obtains the field equations,
\begin{equation}
\mathcal{E}_{\mu\nu}\equiv \frac{2}{\sqrt{-g}}\frac{\delta S}{\delta g^{\mu\nu}}=2\sum_{k=0}^{\lfloor\frac{D-1}{2}\rfloor}\alpha_k \mathcal{H}_{\mu\nu}^{(k)}, \label{eq:lov_eq}
\end{equation}
where $\mathcal{H}_{\mu\nu}^{(k)}$, which comes from the variation of $\mathcal{R}^{(k)}$, generalizes the Einstein tensor:
\begin{align}
\mathcal{H}_{\quad\nu}^{(k)\mu}\equiv{}&{}\frac{-1}{2^{k+1}}\delta^{\mu\,\mu_1\nu_1\cdots\mu_k\nu_k}_{\nu\,\alpha_1\beta_1\cdots\alpha_k\beta_k}\prod_{i=1}^{k} R^{\alpha_i\beta_i}_{\quad\,\mu_i\nu_i},\quad \mathcal{H}_{\quad\mu}^{(k)\mu}=\frac{2k-D}{2}\mathcal{R}^{(k)},\quad \nabla^\mu\mathcal{H}^{(k)}_{\mu\nu}=0,\label{eq:lov_trace}\\
\mathcal{H}_{\mu\nu}^{(0)}={}&{}\frac{-1}{2}g_{\mu\nu},\quad \mathcal{H}_{\mu\nu}^{(1)}=G_{\mu\nu}=R_{\mu\nu}-\frac{1}{2}Rg_{\mu\nu},\\
\mathcal{H}_{\mu\nu}^{(2)}={}&{}2\left(RR_{\mu\nu}+R_{\mu\rho\sigma\lambda}R_\nu^{\,\,\,\rho\sigma\lambda}-2R_{\mu\rho}R^{\rho}_\nu-2R_{\mu\rho\nu\sigma}R^{\rho\sigma}\right)-\frac{1}{2}\mathcal{G}g_{\mu\nu},\quad \cdots
\end{align}
Why does the sum of Lovelock invariants stop at $\lfloor (D-1)/2\rfloor$? The $k$-th Lovelock invariant (\ref{eq:lov_def}) is defined by an antisymmetrization over $2k$ indices, so vanishes in $D$ dimensions for $k>D/2$. So the sum is \textit{a priori} non-trivial up to $k=\lfloor D/2\rfloor$, which, for odd $D$, coincides indeed with $\lfloor (D-1)/2\rfloor$. However, for even $D$, it might seem that the sum forgets the Lovelock invariant of order $k=D/2$. This invariant (which is the GB invariant $\mathcal{G}=\mathcal{R}^{\left(2\right)}$ for $D=4$, the cubic invariant $\mathcal{R}^{\left(3\right)}$ for $D=6$, etc.), is in fact a boundary term, due to a non-trivial result, the Chern-Gauss-Bonnet theorem~\cite{chern}: for a manifold $\mathcal{M}$ of even dimension $D$ with volume form $\mathrm{d}\mathcal{V}$,
\begin{equation}
\int_\mathcal{M}\mathcal{R}^{\left(D/2\right)}\mathrm{d}\mathcal{V} =32\pi^2\chi\left(\mathcal{M}\right),
\end{equation}
where the constant $\chi\left(\mathcal{M}\right)$ is the Euler characteristic of $\mathcal{M}$. So $\mathcal{R}^{\left(D/2\right)}$, although non-zero in general, does not contribute to the field equations when varying the action (\ref{eq:lov}), and can be forgotten. In particular, in dimension $D=4$, the GB term is a boundary term, and all further Lovelock invariants vanish. Consequently, under the assumptions of Lovelock, the only possible gravitational action functional in $D=4$ is the Einstein-Hilbert one (with cosmological constant)~\cite{Lovelock:1972vz}. 
\subsection{Various ways of modifying gravity}\label{subsec:various}
Accordingly, any modification of GR can be traced back to a corresponding modification of Lovelock's assumptions:
\begin{enumerate}
\item Changing the number of dimensions: spacetime is no more four-dimensional but has dimension $D>4$.
\item Adding extra fields mediating gravity (scalar $\phi$, vector $A^\mu$, etc.) so that the vacuum action functional acquires the form $S\left[g_{\mu\nu},\phi,A^\mu,\cdots\right]$ (one can also include a second metric field, leading to bigravity theories~\cite{Schmidt-May:2015vnx}). We insist that this is the action \textbf{without matter fields}: the extra fields $\phi$, $A^\mu$, etc., are not at all matter fields, on the contrary, they are responsible for the gravitational interaction just as the metric $g_{\mu\nu}$ is.
\item Relaxing geometric assumptions on the covariant derivative structure, by allowing non-vanishing torsion or non-metricity, see Eqs. (\ref{eq:torsion}) and (\ref{eq:nonmetr}), which then contribute into the action functional.
\item Breaking diffeomorphism invariance.
\item Allowing for field equations of order $>2$ in derivatives.
\end{enumerate}
\subsubsection{Equivalence between $f(R)$ theories (point 5) and scalar-tensor theories (point 2)}
The generality of this classification is very useful. For instance, the most obvious way of modifying gravity may well be to keep the exact same assumptions as GR, but postulating that the form of the action must be modified. Rather than being the Einstein-Hilbert action $S_\text{EH}\left[g_{\mu\nu}\right]$, the action is more general:
\begin{equation}
S_\text{EH}\left[g_{\mu\nu}\right]=\int\mathrm{d}^4x\sqrt{-g}R\to S\left[g_{\mu\nu}\right]=\int\mathrm{d}^4x\sqrt{-g}f(R).\label{eq:fofr}
\end{equation}
The Lagrangian of the modified theory is now an arbitrary function $f(R)$ of the Ricci scalar, and GR corresponds to the choice $f(R)=R$. These theories are called $f(R)$ theories~\cite{Sotiriou:2008rp,DeFelice:2010aj}. The new action satisfies assumptions 1 to 4 of Lovelock's theorem. Consequently, it is breaking assumption 5, so leads to field equations of order higher than two. Of course, one can check this by direct computation of the field equations. Unfortunately, higher-order field equations lead in general to the propagation of unphysical degrees of freedom, called Ostrogradsky ghosts, with a resulting \textbf{Ostrogradsky instability}~\cite{Woodard:2015zca}. We will detail this point, which will be important later in the manuscript, in Sec.~\ref{sec:higher_order}.
\\

As regards point 2 above, the most simple field that one can add on top of the usual metric tensor field $g_{\mu\nu}$ is a scalar field $\phi$. This leads to \textbf{scalar-tensor theories}~\cite{Fujii:2003pa,Kobayashi:2019hrl}, described by an action $S\left[g_{\mu\nu},\phi\right]$. Interestingly, the $f(R)$ action (\ref{eq:fofr}) can be rewritten~\cite{Clifton:2011jh} as a scalar-tensor theory, provided the function $f$ is such that its second derivative has a constant sign: $f''>0$ or $f''<0$ for all values of its argument. Indeed, consider the scalar-tensor action
\begin{equation}
S\left[g_{\mu\nu},\phi\right]=\int\mathrm{d}^4x\sqrt{-g}\Bigl[f(\phi)+f'(\phi)(R-\phi)\Bigr].\label{eq:eqfr}
\end{equation}
The equivalence of (\ref{eq:eqfr}) with $f(R)$ theory is seen as follows. Varying the action with respect to $\phi$ leads to the following field equation,
\begin{equation}
f''(\phi)(R-\phi)=0.
\end{equation}
Because of the assumption on $f''$, this leads to $\phi=R$. Replacing $\phi$ with $R$ in action (\ref{eq:eqfr}) shows the equivalence with the $f(R)$ action. In the literature, it is quite usual to go further and rewrite the scalar-tensor action (\ref{eq:eqfr}) under a more usual form, in terms of a new scalar field $\psi$ which is coupled to the Ricci scalar with a term $\psi R$. To this aim, one needs to define $\psi$ as
\begin{equation}
\psi= f'(\phi).
\end{equation}
Because of the assumption on $f''$, $f'$ is invertible, and one can note $F$ its inverse function. Therefore $\phi=F(\psi)$. Introducing a potential
\begin{equation}
V(\psi)= f(F(\psi))-\psi F(\psi),
\end{equation}
the scalar-tensor action $S\left[g_{\mu\nu},\phi\right]$ is rewritten as the following scalar-tensor action $S\left[g_{\mu\nu},\psi\right]$,
\begin{equation}
S\left[g_{\mu\nu},\psi\right]=\int\mathrm{d}^4x\sqrt{-g}\Bigl[\psi R+V(\psi)\Bigr].\label{eq:eqfinalstfofr}
\end{equation}
In a word, the $f(R)$ action with $f''>0$ or $f''<0$ is equivalent to the scalar-tensor action (\ref{eq:eqfinalstfofr}). Although the original $f(R)$ action leads to higher-order field equations, this is no more the case of the scalar-tensor action (\ref{eq:eqfinalstfofr}), which gives second-order field equations. Point 5 (higher-order field equations) and point 2 (extra fields) above thus naturally lead to considering, respectively, $f(R)$ theories and scalar-tensor theories, which are seen to be tightly linked.
\subsubsection{Extra dimensions (point 1) and Kaluza-Klein compactification}
As regards point 1, models with extra dimensions are often constructed so as to cure problems coming rather from high energy physics, like the hierarchy problem, e.g. by adding large extra dimensions (ADD model by Arkani-Hamed, Dimopoulos, Dvali~\cite{Arkani-Hamed:1998jmv}) or warped extra dimensions (Randall-Sundrum model~\cite{Randall:1999ee}). Another model, the DGP (Dvali, Gabadadze, Porrati) theory~\cite{Dvali:2000hr}, aims at reproducing the cosmic acceleration of dark energy without needing any cosmological constant. In this model, the action is the sum of a four-dimensional and a five-dimensional Einstein-Hilbert actions.\\

Concerning the purely gravitational aspects, we have seen with Lovelock's theorem that allowing for extra dimensions enables to add new terms (the Lovelock invariants) in a gravitational action still depending only on the metric. One must then find what to do with these extra dimensions, taking into account the physical relevance of the usual four-dimensional spacetime.\\

A process of interest is the \textbf{Kaluza-Klein (KK) compactification} of these extra dimensions. This is named after the KK theory, developed in the 1920s. In Kaluza's model~\cite{Kaluza:1921tu}, a five-dimensional metric $g_{AB}$, $A,B=0,\cdots, 4$ obeys the dynamics of the five-dimensional Einstein-Hilbert action. Grossly speaking, the components $g_{AB}$ are split into a four-dimensional metric $g_{\mu\nu}$, $\mu,\nu=0,\cdots, 3$, a scalar field $\phi\sim g_{44}$ and a four-dimensional one-form field $A_\mu\sim g_{\mu 4}$. Klein then gave the theory a quantum interpretation, proposing that the additional fifth dimension is \textbf{compact}, being a circle whose radius is a multiple of the Planck length~\cite{Klein:1926fj}. KK theory leads to field equations for the four-dimensional metric and vector field, and thus stands as a unified field theory of gravitation and electromagnetism.
\\

In this manuscript, KK compactification will be used in relation to scalar-tensor theories. This process will be illustrated below, first in Sec.~\ref{sec:compact}, then in Sec.~\ref{sec:fourd}, in the case of the compactification of the Einstein-Gauss-Bonnet theory down to a four-dimensional scalar-tensor theory. Later on, KK compactifications of more general Lovelock theories will be performed, in Sec.~\ref{sec:diagkk}. The obtained four-dimensional scalar-tensor theory may retain some of the features of the higher-dimensional theory, thus drawing an interesting link between point 1 (extra dimensions) and scalar-tensor theories.
\subsubsection{Other modifications of gravity (points 3 and 4)}
As regards point 4, several mechanisms can break diffeomorphism invariance, like giving a preferred time direction (Horava-Lifshitz gravity~\cite{Horava:2009uw}) or endowing the graviton with a mass (massive gravity). Note however that, by adding auxiliary fields, there exist healthy versions of massive gravity which do not break diffeomorphism invariance~\cite{deRham:2010kj}, see~\cite{deRham:2014zqa} for a review on massive gravity. Again, there exists interesting links between different modifications of gravity. The five-dimensional DGP model mentioned above can be shown to be equivalent to ghost-free massive gravity~\cite{deRham:2014zqa}. Both models reduce to a scalar-tensor theory in the so-called decoupling limit. For massive gravity, this limit corresponds to sending the mass of graviton $m\to 0$, the Planck mass to infinity, while keeping the product of both finite. The scalar is then the helicity-$0$ mode~\cite{deRham:2010ik}. For DGP model, this limit is reached at distances smaller than the ratio between the four and five-dimensional Planck masses~\cite{Luty:2003vm}. 
\\

On the other hand, authorizing torsion or non-metricity (point 3) allows for many new possibilities, like $f(T)$ and $f(Q)$ theories. The interested reader is referred to the respective reviews~\cite{Cai:2015emx} and~\cite{BeltranJimenez:2019tme}. 
\subsubsection{Scalar-tensor theories: adding a unique degree of freedom}
To cut a long story short, various ways exist which enable to modify GR. The present thesis focuses on scalar-tensor theories, which we just defined above as theories with an action functional $S\left[g_{\mu\nu},\phi\right]$ depending on the usual metric field $g_{\mu\nu}$, and on a new field, the scalar $\phi$. More precisely, the attention will be restricted to \textbf{scalar-tensor theories for which there exists a unique scalar degree of freedom on top of the usual two metric degrees of freedom of GR}. This is not so obvious as it may seem: an action $S\left[g_{\mu\nu},\phi\right]$ may possess more than three dynamical degrees of freedom in total. This is in general the case if the field equations are of order higher than two, and is linked to the existence of unstable, unphysical degrees of freedom, the Ostrogradsky ghosts~\cite{Woodard:2015zca}. This question will thus be of importance for the correct formulation of scalar-tensor theories, and is therefore dealt with in the last section of this chapter, Sec.~\ref{sec:higher_order}.
\\

With this definition as theories where a unique degree of freedom is added thanks to a scalar field, scalar-tensor theories appear as a very natural way of modifying GR. For instance, from the EFT point of view, one may admit that GR describes accurately the physics up to some energy scale, but that some additional degrees of freedom (the first of which being $\phi$) are excited when the energy increases. This simplicity as adding a single degree of freedom also implies the robustness of scalar-tensor theories: we have seen that another natural modification of gravity, the $f(R)$ theories, are equivalent (under certain mathematical assumptions) to scalar-tensor theories. We have also briefly mentioned the link between scalar-tensor theories and other modifications of gravity, like the higher-dimensional theories which, after certain KK compactifications, lead to scalar-tensor theories. Let us now explore the details of this claim by performing an explicit compactification.
\section[Diagonal Kaluza-Klein compactification of Einstein-Gauss-Bonnet gravity]{Diagonal Kaluza-Klein compactification of Einstein-Gauss-Bonnet gravity}\label{sec:compact}
\sectionmark{1.3~~Diagonal Kaluza-Klein compactification of Einstein-Gauss-Bonnet gravity}
Our journey towards scalar-tensor gravity started with Lovelock's theorem, stating that a purely metric theory of gravity in $\mathfrak{D}$ dimensions (the use of the notation $\mathfrak{D}$ is for latter convenience), without torsion nor non-metricity, and yielding second-order and conserved field equations, has action
\begin{equation}
S\left[g_{\mu\nu}\right] = \int\mathrm{d}^\mathfrak{D}x\sqrt{-g}\sum_{k=0}^{\lfloor\frac{\mathfrak{D}-1}{2}\rfloor}\alpha_k\mathcal{R}^{\left(k\right)}=\int\mathrm{d}^\mathfrak{D}x\sqrt{-g}\left(R-2\Lambda+\bar{\alpha}\mathcal{G}+\cdots\right),
\end{equation} 
where Eq. (\ref{eq:lov_def}) recalls the definition of $\mathcal{R}^{\left(k\right)}$, and we have set the couplings $\alpha_1=1,$ $\alpha_0=-2\Lambda$ and $\alpha_2=\bar{\alpha}$. $\mathcal{G}=R^2-4R_{\mu\nu}R^{\mu\nu}+R_{\mu\nu\rho\sigma}R^{\mu\nu\rho\sigma}$ is the GB invariant. Let us consider in detail the \textbf{Einstein-Gauss-Bonnet} (EGB) action without cosmological constant, designing explicitly $\mathfrak{D}$-dimensional quantities with a subscript $(\mathfrak{D})$,
\begin{equation}
S_{(\mathfrak{D}),\text{EGB}}\left[g_{(\mathfrak{D})AB}\right]\equiv \int\mathrm{d}^\mathfrak{D}x\sqrt{-g_{(\mathfrak{D})}}\left(R_{(\mathfrak{D})}+\bar{\alpha}\mathcal{G}_{(\mathfrak{D})}\right),\label{eq:degb}
\end{equation}
which stands as the first non-trivial modification to the Einstein-Hilbert action in $\mathfrak{D}>4$, but reduces to GR in $\mathfrak{D}=4$, where $\mathcal{G}$ is a boundary term. This result implied that modifications of gravity had to be looked for, typically, with extra dimensions or extra fields. Let us now show that, in fact, these two approaches may coincide when one performs a \textbf{KK dimensional reduction}. Split the $\mathfrak{D}$-dimensional Lorentzian line element into a $D$-dimensional Lorentzian line element, plus an $n$-dimensional Euclidean line element with $n=\mathfrak{D}-D$,
\begin{equation}
g_{(\mathfrak{D})AB}(x^\rho)\mathrm{d}x^A\mathrm{d}x^B=g_{(D)\mu\nu}(x^\rho)\mathrm{d}x^\mu\mathrm{d}x^\nu+\mathrm{e}^{-2\phi(x^\rho)}\widetilde{g}_{(n)ab}\left(x^c\right)\mathrm{d}x^a\mathrm{d}x^b,\label{eq:diag_dec}
\end{equation}
where the original spacetime with $\mathfrak{D}$ dimensions corresponds to indices $A,B,\cdots$, the $D$-dimensional \textbf{target} spacetime to $\mu,\nu,\cdots$ and the $n$-dimensional \textbf{internal} space to $a,b,\cdots$. This decomposition is said to be \textbf{diagonal}, since there are no cross-terms $\mathrm{d}x^\mu\mathrm{d}x^a$. \\

Moreover, the $\mathfrak{D}$ and $D$-dimensional metrics do not depend on the internal space coordinates $x^a$, while in the case of the internal space, the dependence on the target spacetime coordinates $x^\mu$ is completely encoded into the conformal factor $\mathrm{e}^{-2\phi(x^\mu)}$, where $\phi$ is a scalar field. Obviously, this choice of decomposition is related to the will of relating the $\mathfrak{D}$-dimensional, purely metric EGB theory (\ref{eq:degb}), to a $D$-dimensional scalar-tensor theory. Had we included cross-terms $\mathrm{d}x^\mu\mathrm{d}x^a$, this would have given rise to one-form fields $A_\mu\sim g_{\mu a}$, like in the original KK model. This decomposition for the EGB theory was performed in~\cite{Charmousis:2012dw}, see also~\cite{VanAcoleyen:2011mj} for similar considerations, leading to the following results. With straightforward calculations\footnote{The aim at this point is of course not to enter in any technical details. These will be rather developed later on, in Sec.~\ref{sec:diagkk}.}, Eq. (\ref{eq:diag_dec}) yields
\begin{align}
\sqrt{-g_{(\mathfrak{D})}} ={}&{} \sqrt{-\widetilde{g}_{(n)}}\sqrt{-g_{(D)}}\mathrm{e}^{-n\phi},\\
R_{(\mathfrak{D})}={}&{}R_{(D)}+\widetilde{R}_{(n)}\mathrm{e}^{2\phi}+n\left(n-1\right)\left(\partial\phi\right)^2,\\
\mathcal{G}_{(\mathfrak{D})}={}&{}\mathcal{G}_{(D)}+\widetilde{\mathcal{G}}_{(n)}\mathrm{e}^{4\phi}+2\widetilde{R}_{(n)}\mathrm{e}^{2\phi}\Bigl[R_{(D)}+\left(n-2\right)\left(n-3\right)\left(\partial\phi\right)^2\Bigr]\nonumber\\ {}&{}-4n\left(n-1\right)G_{(D)}^{\mu\nu}\partial_\mu\phi\,\partial_\nu\phi+2n\left(n-1\right)\left(n-2\right)\Box\phi\left(\partial\phi\right)^2\nonumber\\ {}&{}-n\left(n-1\right)^2\left(n-2\right)\left(\partial\phi\right)^4.
\end{align}
In these equations, quantities with $\sim$ are those associated to the metric of the internal space $\widetilde{g}_{(n)ab}$. Since $\phi$ depends only on the target spacetime, all terms involving its derivatives are $D$-dimensional terms,
\begin{equation}
\left(\partial\phi\right)^2 = g_{(D)}^{\mu\nu}\partial_\mu\phi\,\partial_\nu\phi,\quad \Box\phi = g_{(D)}^{\mu\nu}\nabla_{(D)\mu}\nabla_{(D)\nu}\phi.
\end{equation}
Assume now that the curvature invariants $\widetilde{R}_{(n)},\widetilde{\mathcal{G}}_{(n)}$ of the internal space are constants, for instance if the internal space is maximally-symmetric or a product of spheres. Then, 
\begin{equation}
S_{(\mathfrak{D}),\text{EGB}}\left[g_{(\mathfrak{D})\mu\nu}\right] = \mathcal{V}_{(n)}\,S_{(D),n,\text{EGB}}\left[g_{(D)\mu\nu},\phi\right].
\end{equation}
The constant proportionality factor $\mathcal{V}_{(n)}$ is the volume of the internal space,
\begin{equation}
\mathcal{V}_{(n)}=\int\mathrm{d}^nx\sqrt{-\widetilde{g}_{(n)}},
\end{equation}
while $S_{(D),n,\text{EGB}}\left[g_{(D)\mu\nu},\phi\right]$ is a $D$-dimensional scalar-tensor action. It reads
\begin{align}
S_{(D),n,\text{EGB}}\left[g_{(D)\mu\nu},\phi\right] ={}&{} \int\mathrm{d}^Dx\sqrt{-g_{(D)}}\,\mathrm{e}^{-n\phi}\Bigl\{R_{(D)}+\widetilde{R}_{(n)}\mathrm{e}^{2\phi}+n\left(n-1\right)\left(\partial\phi\right)^2\nonumber\\{}&{}+\bar{\alpha}\Bigl[\mathcal{G}_{(D)}+\widetilde{\mathcal{G}}_{(n)}\mathrm{e}^{4\phi}+2\widetilde{R}_{(n)}\mathrm{e}^{2\phi}\Bigl(R_{(D)}+\left(n-2\right)\left(n-3\right)\left(\partial\phi\right)^2\Bigr)\nonumber\\{}&{}-4n\left(n-1\right)G_{(D)}^{\mu\nu}\partial_\mu\phi\,\partial_\nu\phi+2n\left(n-1\right)\left(n-2\right)\Box\phi\left(\partial\phi\right)^2\nonumber\\ {}&{}-n\left(n-1\right)^2\left(n-2\right)\left(\partial\phi\right)^4\Bigr]\Bigr\}.\label{eq:egbkk}
\end{align}
In fact, $S_{(D),n,\text{EGB}}$ depends not only on the internal space dimension $n$, but also on its geometry, since its Ricci and GB scalars, $\widetilde{R}_{(n)}$ and $\widetilde{\mathcal{G}}_{(n)}$, appear in the action. These geometrical terms of the internal space are mere coupling constants for the resulting $D$-dimensional theory. Thus, as announced, modifying gravity by adding extra dimensions can translate into the inclusion of extra fields to a $D$-dimensional theory following a KK procedure, so in particular to a four-dimensional theory by taking $D=4$. \\

For clarity, we emphasize that, although the GB invariant is a boundary term in four dimensions, it obviously contributes non-trivially to the field equations when coupled to a scalar field $\phi$ as in action (\ref{eq:egbkk}). In fact, beyond Lovelock's mathematical result, the GB invariant is physically motivated by many considerations from string theory, where for instance, 1-loop corrected heterotic string effective action presents terms as $\mathrm{e}^\phi\mathcal{G}$, where $\phi$ is the dilaton~\cite{Zwiebach:1985uq,Nepomechie:1985us,Callan:1986jb,Candelas:1985en,Gross:1986mw}.\\

It will be seen in Sec.~\ref{sec:fourd} that the most relevant interpretation of action (\ref{eq:egbkk}) is obtained by taking a singular limit $\bar{\alpha}\to\infty$, $n\to 0$ and $\bar{\alpha}\,n=\text{constant}$. For the moment however, let us conclude this first chapter by a study of the possible pathologies brought about by higher-order field equations, as in $f(R)$ theories. The associated Ostrogradsky instability, and its possible circumvention, will constrain the admissible scalar-tensor theories.
\section[Higher-order field equations: Ostrogradsky instability, and evading it by degeneracy of the kinetic matrix]{Higher-order field equations: Ostrogradsky instability, and evading it by degeneracy of the kinetic matrix}\label{sec:higher_order}
\sectionmark{1.4~~Degeneracy of the kinetic matrix: evading the Ostrogradsky instability}
\subsection{A toy model}
The issue of higher-order field equations is often presented by using the following toy model of analytical mechanics, see~\cite{Langlois:2015cwa}. Let us use the same toy model and briefly repeat the analysis of this latter reference. The variables are $\phi(t)$ and $q^i(t)$ with $i=1,\cdots,n$, and a dot means time derivative. The action reads
\begin{equation}
S\left[\phi,q^i\right]=\int\mathrm{d}t\left[\frac{1}{2}a\ddot{\phi}^2+\frac{1}{2}k_{ij}\dot{q}^i\dot{q}^j+b_i\ddot{\phi}\,\dot{q}^i+c_i\dot{\phi}\,\dot{q}^i\right].
\end{equation}
The field equations are
\begin{align}
0={}&{}a \ddddot{\phi}+b_i\dddot{q}^i-c_i\ddot{q}^i,\\
0={}&{}k_{ij}\ddot{q}^j+b_i\dddot{\phi}+c_i\ddot{\phi}.
\end{align}
To compute the number of degrees of freedom, one replaces $\dot{\phi}$ by $Q$, and imposes that $\dot{\phi}=Q$ by using a Lagrange multiplier $\lambda$, yielding
\begin{equation}
S\left[\phi,q^i,Q,\lambda\right]=\int\mathrm{d}t\left[\frac{1}{2}a\dot{Q}^2+\frac{1}{2}k_{ij}\dot{q}^i\dot{q}^j+b_i\dot{Q}\,\dot{q}^i+c_i Q\,\dot{q}^i-\lambda\left(Q-\dot{\phi}\right)\right].
\end{equation}
The new field equations are
\begin{align}
c_i\,\dot{q}^i-\lambda={}&{}a\ddot{Q}+b_i\ddot{q}^i,\label{eq:degen_1}\\
-c_i\,\dot{q}^i={}&{} b_i\ddot{Q}+k_{ij}\ddot{q}^j,\label{eq:degen_2}\\
\dot{\phi}={}&{} Q,\quad \dot{\lambda}=0\label{eq:degen_3},
\end{align}
and one can check that they are equivalent to the ones of the first formulation.
\subsection{Invertible kinetic matrix: Ostrogradsky ghost instability}
From the above system of equations, one can read the kinetic matrix $M$ of the system,
\begin{equation}
M = \left(\begin{array}{cc}
a & b_i \\ b_i & k_{ij}
\end{array}\right).
\end{equation}
If $M$ is invertible, the above system of differential equations (\ref{eq:degen_1}-\ref{eq:degen_3}) requires initial conditions for $Q$, $\dot{Q}$, $q^i$, $\dot{q}^i$, $\lambda$ and $\phi$, so $2(n+2)$ initial conditions, describing therefore $n+2$ degrees of freedom. However, the variables of the initial action, $S\left[\phi,q^i\right]$, seemed to imply only $n+1$ degrees of freedom. The $(n+2)$-th degree of freedom is an unphysical degree of freedom, the Ostrogradsky ghost. Its appearance is linked to the presence of derivatives of order higher than two in the original field equations. It can be identified more rigorously in the framework of Hamiltonian analysis. In this framework, the kinetic term of the Ostrogradsky degree of freedom is shown to have the wrong sign. The Hamiltonian is unbounded from below: this is the Ostrogradsky instability.
\subsection{Degenerate kinetic matrix: evading the Ostrogradsky instability}
Let us now present the case where the kinetic matrix $M$ is non-invertible, or in other words \textbf{degenerate}. For the purposes of the argument, we assume that the kinetic matrix $k_{ij}$ of the variables $q^i$ is non-degenerate. Also, for the higher-order derivatives to be present, we assume $a\neq 0$ and $b_i\neq 0$. Then, the determinant of the total kinetic matrix $M$ is 
\begin{equation}
\det(M)=\det(k)\left(a-b_i\left(k^{-1}\right)^{ij}b_j\right),
\end{equation}
so the degeneracy of $M$ is equivalent to $a-b_i\left(k^{-1}\right)^{ij}b_j=0$. The kernel of $M$ is then generated by the vector $v$,
\begin{equation}
v \equiv \left(\begin{array}{c} v^0\\ v^i
\end{array}\right)
 \equiv \left(\begin{array}{c} -1 \\ \left(k^{-1}\right)^{ij}b_j
 \end{array}\right).
\end{equation}
Projecting (\ref{eq:degen_1}-\ref{eq:degen_2}) along $v$ yields
\begin{equation}
c_i\left(\dot{q}^i+v^i\dot{Q}\right)=\lambda.
\end{equation}
Introducing a new variable $x^i\equiv \dot{q}^i+v^i\dot{Q}$, Eqs. (\ref{eq:degen_1}-\ref{eq:degen_2}) are thus equivalent to
\begin{align}
\lambda={}&{}c_i\,\dot{x}^i,\label{eq:degen_4}\\
0={}&{} k_{ij}\ddot{x}^j+c_i\dot{Q}.
\end{align}
Taking the time derivative of (\ref{eq:degen_4}) and using (\ref{eq:degen_3}) to replace $Q$ by $\dot{\phi}$ leads to the equivalent system 
\begin{align}
0={}&{}c_i\,\ddot{x}^i,\\
0={}&{} c_i\,\ddot{\phi}+k_{ij}\ddot{x}^j.
\end{align}
This is a second order system for the $n+1$ variables $\phi$ and $x^i$. As a conclusion, the degeneracy of the kinetic matrix $M$ is seen to imply that the $(n+2)$-th degree of freedom (the Ostrogradsky ghost) does not appear. 
\subsection{Consequences for $f(R)$ and scalar-tensor theories}
The previous analysis extends very generically beyond the simple toy model. Therefore, the $f(R)$ theories, which have field equations of order four, suffer from Ostrogradsky instabilities. While $f(R)$ theories might have seemed, at first sight, very straightforward modifications of gravity (since they depend only on the metric field, just as GR), their higher-order equations of motion justifies, among other arguments given previously, that this thesis focuses on another natural modification of gravity, namely scalar-tensor theories.
\\

The Ostrogradsky instability explains that one may at first restrict their attention to field equations of order two, as is most common in physics. This retrospectively explains one of the assumption of Lovelock's theorem: 'The field equations are of second order'. The construction of scalar-tensor theories followed the same principle\footnote{Historically, the first major proposal of a scalar-tensor theory is the Brans-Dicke theory~\cite{Brans:1961sx}, building upon a proposal of Jordan~\cite{Jordan1959ZumGS}. Brans-Dicke theory will be briefly presented at the end of Sec.~\ref{sec:horngal}.}: \textbf{Gregory Horndeski} constructed as early as in 1974 the most general action $S\left[g_{\mu\nu},\phi\right]$ with second-order field equations~\cite{Horndeski:1974wa}. Later on, more general, healthy actions $S\left[g_{\mu\nu},\phi\right]$ were constructed~\cite{Gleyzes:2014dya,Langlois:2015cwa}, with higher-order field equations, but where degeneracy of the kinetic matrix is required in order for the theory to propagate a unique additional degree of freedom (on top of the usual two metric degrees of freedom). These are the \textbf{Degenerate Higher-Order Scalar-Tensor (DHOST)} theories~\cite{Langlois:2015cwa}. The next chapter presents the Horndeski and DHOST theories, and their properties. 
\chapter{Generic aspects of scalar-tensor theories}\label{chap:st}
\chaptermark{Chap.~2~~Generic aspects of scalar-tensor theories}
This second chapter presents the generic aspects of scalar-tensor theories, which have been defined in Chap.~\ref{chap:modifGR} as theories, described by an action functional $S\left[g_{\mu\nu},\phi\right]$, propagating two metric and one scalar degrees of freedom. The understanding of Ostrogradsky instability, Sec.~\ref{sec:higher_order}, implies that the most straightforward way of constructing such scalar-tensor theories is to impose second-order field equations. This leads to the scalar-tensor theories of \textbf{Horndeski}, now also called \textbf{generalized Galileons}, and presented in Sec.~\ref{sec:horngal}. In a second time, Sec.~\ref{sec:disf} defines and analyzes a particular redefinition of the metric field, called \textbf{disformal transformation}. Indeed, this naturally paves the way towards the introduction of scalar-tensor theories with field equations of order higher than two, but evading the Ostrogradsky instability. These are the \textbf{beyond Horndeski} and \textbf{DHOST} theories, Sec.~\ref{sec:bhdhost}. \\

Then, Sec.~\ref{sec:symm} presents the various \textbf{symmetries} of scalar-tensor theories which may simplify the analysis of their field equations. One of these symmetries is local conformal invariance, and is tightly linked to the formulation of a \textbf{four-dimensional Einstein-Gauss-Bonnet (4DEGB)} theory. This is a scalar-tensor theory which reproduces the features of the higher-dimensional EGB theory as formulated by Lovelock. Sec.~\ref{sec:fourd} gives all details about this 4DEGB theory.
\section{Scalar-tensor theories of Horndeski}\label{sec:horngal}
\sectionmark{2.1~~Scalar-tensor theories of Horndeski}
The Horndeski action is the most general four-dimensional scalar-tensor action $S\left[g_{\mu\nu},\phi\right]$ with a variational principle leading to field equations of second order. Explicitly,
\begin{equation}
\mathcal{E}_{\mu\nu}\equiv \frac{2}{\sqrt{-g}}\frac{\delta S}{\delta g^{\mu\nu}},\quad \mathcal{E}_\phi\equiv\frac{1}{\sqrt{-g}}\frac{\delta S}{\delta\phi},\label{eq:field_eq}
\end{equation}
contain at most second-order derivatives of the metric $g_{\mu\nu}$ and the scalar field $\phi$. This action was constructed as early as in 1974 by Gregory Horndeski~\cite{Horndeski:1974wa}. A review of Horndeski theories can be found in~\cite{Kobayashi:2019hrl}. The following paragraph briefly sketches the original construction by Horndeski, while a more detailed account of this construction can be found in appendix~\ref{chap:appHornProof}.
\subsection{Original derivation by Horndeski}\label{subsec:horndeski}
Horndeski started from a generic four-dimensional action with diffeomorphism invariance $S\left[g_{\mu\nu},\phi\right]=\int\mathrm{d}^4x\sqrt{-g}\mathcal{L}$. An infinitesimal diffeomorphism is parameterized by a vector field $\delta\xi^\mu$ under which the fields are modified according to their Lie derivatives along $-\delta\xi^\mu$,
\begin{equation}
\delta g^{\mu\nu}=\nabla^\mu\delta\xi^\nu+\nabla^\nu\delta\xi^\mu,\quad \delta\phi = -\delta\xi^\mu\partial_\mu\phi.
\end{equation}
Diffeomorphism invariance of $S$ then implies, after integrating by parts and using the symmetry of $\mathcal{E}_{\mu\nu}$,
\begin{equation}
0=\delta S=\int\mathrm{d}^4x\sqrt{-g}\,\delta\xi^\mu\Bigl\{-\nabla^\nu\left(\frac{2}{\sqrt{-g}}\frac{\delta S}{\delta g^{\mu\nu}}\right)-\frac{1}{\sqrt{-g}}\frac{\delta S}{\delta\phi}\partial_\mu\phi\Bigr\}.\label{eq:diffeo}
\end{equation}
Since $\delta\xi^\mu$ is arbitrary, one gets the '\textbf{Bianchi identity}' for scalar-tensor theories,
\begin{equation}
\nabla^\nu\mathcal{E}_{\mu\nu}=-\mathcal{E}_\phi\,\partial_\mu\phi.\label{eq:bianchi}
\end{equation}
Because $\mathcal{E}_\phi$ is second-order in derivatives, $\nabla^\nu\mathcal{E}_{\mu\nu}$ is also second-order, and not third-order as naively expected. Horndeski thus starts by finding \textbf{the most general symmetric tensor of the form $A_{\mu\nu}=A_{\mu\nu}\left(g_{\mu\nu},\partial g_{\mu\nu},\partial\partial g_{\mu\nu},\phi,\partial\phi,\partial\partial\phi\right)$ which is such that $\nabla^\nu A_{\mu\nu}$ is at most of second order in the derivatives of both $g_{\mu\nu}$ and $\phi$}. Then, because of (\ref{eq:bianchi}), Horndeski imposes in addition that this $A_{\mu\nu}$ satisfy $\nabla^\nu A_{\mu\nu}=\partial_\mu\phi\,B\left(g_{\mu\nu},\partial g_{\mu\nu},\partial\partial g_{\mu\nu},\phi,\partial\phi,\partial\partial\phi\right)$ for some Lorentz scalar $B$. Such an $A_{\mu\nu}$ then corresponds to the metric field equations $\mathcal{E}_{\mu\nu}$. Finally, Horndeski is able to find the Lagrangian which gives rise to the field equations $\mathcal{E}_{\mu\nu}$ and $\mathcal{E}_\phi$ by inspection of $g^{\mu\nu}\mathcal{E}_{\mu\nu}$. With this procedure, the Horndeski action reads
\begin{align}
S=\int{}&{}\mathrm{d}^4x\sqrt{-g}\Bigl\{\delta^{\alpha\beta\gamma}_{\mu\nu\sigma}\Bigl[K_1\phi^\mu_\alpha R_{\beta\gamma}^{\,\,\,\,\,\,\nu\sigma}+\frac{2}{3}K_{1X}\phi^\mu_\alpha\phi^\nu_\beta\phi^\sigma_\gamma+K_3\phi_\alpha\phi^\mu R_{\beta\gamma}^{\,\,\,\,\,\,\nu\sigma}\nonumber\\{}&{}+2K_{3X}\phi_\alpha\phi^\mu\phi^\nu_\beta\phi^\sigma_\gamma\Bigr]+\delta^{\alpha\beta}_{\mu\nu}\Bigl[FR_{\alpha\beta}^{\,\,\,\,\,\,\mu\nu}+2F_X\phi^\mu_\alpha\phi^\nu_\beta+2K_8\phi_\alpha\phi^\mu\phi^\nu_\beta\Bigr]\nonumber\\{}&{}+6\left(XK_8-F_\phi\right)\Box\phi+K_9\Bigr\}.\label{eq:original}
\end{align}
It is parameterized by four arbitrary functions $K_i\left(\phi,X\right)$, $i=1,3,8,9$, of the scalar field $\phi$ and of its kinetic term $X$,
\begin{equation}
X \equiv -\frac{1}{2}g^{\mu\nu}\partial_\mu\phi\partial_\nu\phi=-\frac{1}{2}\phi^\mu\phi_\mu,
\end{equation}
and the following abbreviations are used,
\begin{equation}
\phi_\mu\equiv\partial_\mu\phi=\nabla_\mu\phi,\quad\phi_{\mu\nu}\equiv\nabla_\mu\nabla_\nu\phi=\phi_{\nu\mu},\quad f_X \equiv\frac{\partial f}{\partial X},\quad f_\phi\equiv\frac{\partial f}{\partial\phi}.\label{eq:notations}
\end{equation}
Finally, the function $F\equiv F\left(\phi,X\right)$ appearing in the action (\ref{eq:original}) must verify
\begin{equation}
F_X = 2\left(K_3+2XK_{3X}-K_{1\phi}\right).
\end{equation}
\subsection{Equivalence with the generalized Galileon}
The original form (\ref{eq:original}) of the action is no more used nowadays. The current form can be called the \textbf{generalized Galileon form of Horndeski theories}, and is obtained in the following way. One starts from the \textbf{Galileon} action~\cite{Nicolis:2008in}. This is the most general four-dimensional scalar field action within a fixed Minkowski background, invariant under the Galilean transformation for the scalar field, $\phi\to\phi+b_\mu x^\mu+c$ with constant $b_\mu$ and $c$, and leading to second-order field equations. The Galileon action reads
\begin{align}
S\left[\phi\right] =\int\mathrm{d}^4 x\Bigl\{c_1\phi{}&{}+c_2X-c_3X\Box\phi+c_4X\left[\left(\Box\phi\right)^2-\partial_\mu\partial_\nu\phi\,\partial^\mu\partial^\nu\phi\right]\nonumber\\{}&{}-\frac{c_5}{3}X\Bigl[\left(\Box\phi\right)^3-3\Box\phi\partial_\mu\partial_\nu\phi\,\partial^\mu\partial^\nu\phi+2\partial_\mu\partial_\nu\phi\,\partial^\nu\partial^\rho\phi\,\partial_\rho\partial^\mu\phi\Bigr]\Bigr\},\label{eq:galileon_action_first}
\end{align}
where $c_1$, $c_2$, $c_3$, $c_4$, $c_5$ are constants, and of course, the kinetic term is appropriate to the fixed Minkowski metric $\eta_{\mu\nu}$, $X=-\eta^{\mu\nu}\partial_\mu\phi\,\partial_\nu\phi/2$. Eq. (\ref{eq:galileon_action_first}) is a pure scalar action, and one can transform it into a scalar-tensor action by making the replacements $\eta_{\mu\nu}\to g_{\mu\nu}$, $\partial_\mu\to\nabla_\mu$, and multiplying the Lagrangian by $\sqrt{-g}$. However, in addition to this usual procedure, additional terms must be included in order to maintain second-order field equations for the metric $g_{\mu\nu}$. These additional terms are emphasized in red below. The \textbf{covariant Galileon} action thus obtained~\cite{Deffayet:2009wt,Deffayet:2009mn} is
\begin{align}
S\left[g_{\mu\nu},\phi\right] =\int\mathrm{d}^4 x{}&{}\sqrt{-g}\Bigl\{c_1\phi+c_2X-c_3X\Box\phi+{\color{red}\frac{c_4}{2}X^2R}+ c_4X\left[\left(\Box\phi\right)^2-\phi_{\mu\nu}\phi^{\mu\nu}\right]\nonumber\\{}&{}+{\color{red}c_5X^2G^{\mu\nu}\phi_{\mu\nu}}-\frac{c_5}{3}X\Bigl[\left(\Box\phi\right)^3-3\Box\phi\phi_{\mu\nu}\phi^{\mu\nu}+2\phi_{\mu\nu}\phi^{\nu\rho}\phi_\rho^\mu\Bigr]\Bigr\}.\label{eq:galileon_action_second}
\end{align}
We recall that $X=-g^{\mu\nu}\phi_\mu\phi_\nu/2$, $\phi_\mu=\nabla_\mu\phi$, $\phi_{\mu\nu}=\nabla_\mu\nabla_\nu\phi$. Note that the covariant Galileon action (\ref{eq:galileon_action_second}) is called 'Galileon' only by analogy with the original Galileon action (\ref{eq:galileon_action_first}): it does not enjoy anymore the Galilean symmetry. One can thus push the analogy further by constructing the \textbf{generalized Galileon} action: this time, one starts by constructing a four-dimensional scalar field action $S\left[\phi\right]$ in a fixed Minkowski background, with the only requirement that the field equations be of second order. There is no more reference to the Galilean symmetry, and one finds an action which generalizes the original Galileon action (\ref{eq:galileon_action_first}). Then, one promotes this scalar field action $S\left[\phi\right]$ to a scalar-tensor action $S\left[g_{\mu\nu},\phi\right]$ in the same way as one promoted the original Galileon action (\ref{eq:galileon_action_first}) to the covariant Galileon action (\ref{eq:galileon_action_second}). The obtained theory is the generalized Galileon~\cite{Deffayet:2011gz},
\begin{align}
S\left[g_{\mu\nu},\phi\right] = \int{}&{}\mathrm{d}^4 x\sqrt{-g}\Bigl\{G_2\left(\phi,X\right)-G_3\left(\phi,X\right)\Box\phi+G_4\left(\phi,X\right)R\nonumber\\{}&{}+G_{4X}\left[\left(\Box\phi\right)^2-\phi_{\mu\nu}\phi^{\mu\nu}\right]+G_5\left(\phi,X\right)G^{\mu\nu}\phi_{\mu\nu}\nonumber\\{}&{}-\frac{G_{5X}}{6}\Bigl[\left(\Box\phi\right)^3-3\Box\phi\phi_{\mu\nu}\phi^{\mu\nu}+2\phi_{\mu\nu}\phi^{\nu\rho}\phi_\rho^\mu\Bigr]\Bigr\}.\label{eq:galileon_action}
\end{align}
It is parameterized by four arbitrary functions $G_2\left(\phi,X\right)$, $G_3\left(\phi,X\right)$, $G_4\left(\phi,X\right)$ and $G_5\left(\phi,X\right)$ of the scalar field $\phi$ and its kinetic term $X$. In fact, this action (\ref{eq:galileon_action}) is the \textbf{generalized Galileon form of Horndeski theories}, since it is equivalent to the original Horndeski action (\ref{eq:original}). Indeed, it can be obtained from (\ref{eq:original}) through integration by parts and identification of the four functions $G_2$, $G_3$, $G_4$ and $G_5$ as~\cite{Kobayashi:2011nu}
\begin{align}
G_2={}&{}K_9+4X\int\left(K_{8\phi}-2K_{3\phi\phi}\right)\mathrm{d}X,\quad G_5 = -4K_1,\nonumber\\ G_3={}&{} 6F_\phi-2XK_8-8XK_{3\phi}+2\int\left(K_8-2K_{3\phi}\right)\mathrm{d}X,\quad G_4=2F-4XK_3.\label{eq:change_gal}
\end{align}
The advantage of the Galileon approach is that it yields a nicely written and concise action (\ref{eq:galileon_action}) in a not so difficult way. However, its equivalence with the Horndeski action (\ref{eq:original}) is paramount, since only the Horndeski approach guarantees that the obtained action is the most general four-dimensional scalar-tensor action with second order field equations. This being said, Horndeski theories are nowadays exclusively dealt with under their generalized Galileon form (\ref{eq:galileon_action}).
\subsection{Common metric-scalar couplings in the Galileon formalism}\label{subsec:common}
Importantly, the Horndeski, or generalized Galileon, action, is defined through the requirement of second-order field equations. Therefore, the form (\ref{eq:galileon_action}) is the most general leading to such field equations \textbf{only up to integration by parts}. Notably, some usual terms of scalar-tensor theories are commonly written in a compact way which is not necessarily the one of the generalized Galileon notations. However, if one tries such a Lagrangian and finds that it leads to second-order field equations, then there necessarily exists a way to rewrite it under the Galileon notations with integration by parts. 
\subsubsection{Fab Four theory}
This is in particular the case for the Fab Four Lagrangians~\cite{Charmousis:2011bf,Charmousis:2011ea}, which were introduced so as to allow self-tuning mechanisms on FLRW backgrounds,
\begin{align}
\mathcal{L}_\text{John}\equiv{}&{} V_\text{J}(\phi)G^{\mu\nu}\phi_\mu\phi_\nu,\\
\mathcal{L}_\text{Ringo}\equiv{}&{} V_\text{R}(\phi)\mathcal{G},\\
\mathcal{L}_\text{Paul}\equiv{}&{}V_\text{P}(\phi)P^{\mu\rho\nu\sigma}\phi_\mu\phi_\nu\phi_{\rho\sigma}.\label{eq:paulterm}
\end{align}
We recall that $G_{\mu\nu}$ is the Einstein tensor and $\mathcal{G}$ the GB scalar. The tensor $P_{\mu\rho\nu\sigma}$ is
\begin{equation}
P_{\mu\rho\nu\sigma} \equiv R_{\mu\rho\nu\sigma}+g_{\mu\sigma}R_{\rho\nu}+g_{\rho\nu}R_{\mu\sigma}-g_{\mu\nu}R_{\rho\sigma}-g_{\rho\sigma}R_{\mu\nu}+\frac{1}{2}\left(g_{\mu\nu}g_{\rho\sigma}-g_{\mu\sigma}g_{\rho\nu}\right)R.\label{eq:doubledula_first}
\end{equation}
This definition could be used in any spacetime dimension, and in four dimensions, it is equivalent to $P^{\mu\rho\nu\sigma}=-\epsilon^{\mu\rho\alpha\beta}R_{\alpha\beta\lambda\tau}\epsilon^{\nu\sigma\lambda\tau}/4$, i.e. the double dual of the Riemann tensor~\cite{Misner:1973prb}. The functions $V_{\cdots} (\phi)$ are arbitrary functions of $\phi$. We did not write the George term which is just the $G_4$ term of Horndeski. However, for the three terms we have written, the equivalence is not straightforward. For the coupling to GB, the Galileon correspondence is
\begin{align}
G_2 ={}&{} 8V_{\text{R}\phi\phi\phi\phi}X^2\left(3-\ln X\right),\quad G_3 = 4V_{\text{R}\phi\phi\phi}X(7-3\ln X),\nonumber\\ G_4={}&{}4V_{\text{R}\phi\phi}X(2-\ln X),\quad G_5=-4V_{\text{R}\phi}\ln X.\label{eq:gbhorn}
\end{align}
This was demonstrated at the level of the field equations by~\cite{Kobayashi:2011nu}, and more recently shown directly at the level of the action by~\cite{Langlois:2022eta}, using the expression of $\mathcal{G}$ as a total derivative (in four dimensions) found by~\cite{Colleaux:2019ckh}.
\\

As concerns the coupling to the Einstein tensor, it is a $G_5$ term,
\begin{equation}
G_5 = -\int V_\text{J}(\phi)\mathrm{d}\phi,
\end{equation}
but if $V_\text{J}(\phi)=1$, it can also be written as a $G_4$ term,
\begin{equation}
G_4 = X.
\end{equation}
Finally, let us take the Paul term with $V_\text{P}(\phi)=1$ for simplicity (it is the only case which will be needed in this manuscript). It corresponds to~\cite{Narikawa:2013pjr}
\begin{equation}
G_5 = 3X.\label{eq:paultermnari}
\end{equation}
\subsubsection{(Generalized) Brans-Dicke theory}
One of the first and most well-known scalar-tensor theory was introduced by Brans and Dicke in 1961~\cite{Brans:1961sx}. It was further generalized by Bergmann~\cite{Bergmann:1968ve} and Wagoner~\cite{Wagoner:1970vr}, leading to the generalized Brans-Dicke theory:
\begin{equation}
S=\int\mathrm{d}^4x\sqrt{-g}\left[\phi R-\frac{\omega(\phi)}{\phi}\left(\partial\phi\right)^2-V(\phi)\right].\label{eq:action_bd}
\end{equation}
This belongs to Horndeski with
\begin{equation}
G_2=-V(\phi)+2\frac{\omega(\phi)}{\phi},\quad G_4=\phi.
\end{equation}
The particular case of Brans-Dicke corresponds to $\omega(\phi)=\text{constant}$ and $V(\phi)=0$. We mostly introduce this action for its simplicity and historical importance (it was first constructed to get a varying Newton's constant, replaced by $1/\phi$), but it will not be of great interest for us. Indeed, it allows for no asymptotically flat hairy BH solution but one, the BBMB BH~\cite{Bocharova:1970skc,Bekenstein:1974sf}, as will be explained in Sec.~\ref{sec:nohair} dedicated to no-hair theorems. This somehow explains the necessity of going beyond the simple Brans-Dicke theory, and considering rather the Horndeski theory, in order to get new BH phenomenology. 
\section{Disformal transformations of Horndeski theories}\label{sec:disf}
\sectionmark{2.2~~Disformal transformations of Horndeski theories}
Now that the modern Horndeski action (\ref{eq:galileon_action}) is constructed, one can naturally wonder about its behaviour under redefinitions of the two fields upon which it depends, the metric field $g_{\mu\nu}$ and the scalar field $\phi$. In particular, since $g_{\mu\nu}$ is a symmetric covariant two-tensor, it is tempting to construct a symmetric covariant two-tensor from the scalar field $\phi$, and to add it to the metric. An infinity of possibilities exist: with the notations of (\ref{eq:notations}), $\phi_\mu\phi_\nu$, $\phi_{\mu\nu}$, $\phi_{\mu\rho}\phi^\rho_\nu$... The simplest possibility is the one which yields derivatives of not too high order: if one acts upon $\phi_\mu\phi_\nu$ with $\partial$ or $\nabla$, one gets terms with at most $\nabla\nabla\phi$, while with any other possibility, one would get terms with at least $\nabla\nabla\nabla\phi$. One thus defines a \textbf{disformal transformation}~\cite{Bekenstein:1992pj} as the following redefinition of the fields:
\begin{equation}
g_{\mu\nu}\mapsto \tilde{g}_{\mu\nu} \equiv g_{\mu\nu}+D\left(\phi,X\right)\phi_\mu\phi_\nu,\quad \phi\mapsto\tilde{\phi}\equiv\phi.\label{eq:disf_first}
\end{equation}
The disformal transformation is seen to act only upon the metric. The transformation is parameterized by a function $D\left(\phi,X\right)$ of the scalar field and its kinetic term. Again, if one were considering more general functions of other scalar quantities constructed from $\phi$, like $D\left(\phi,X,\Box\phi\right)$, derivatives of high order in $\phi$ would appear more rapidly. Even with the simple disformal transformation (\ref{eq:disf_first}), it not obvious that a Horndeski action $S\left[g_{\mu\nu},\phi\right]$ leads to a sensible action in terms of $\tilde{g}_{\mu\nu}$. For instance, the term $G^{\mu\nu}\phi_{\mu\nu}$, which appears in the Horndeski action (\ref{eq:galileon_action}), contains terms in $\left(\partial\partial g_{\mu\nu}\right)\left(\partial\partial\phi\right)$, which, once rewritten in terms of $\tilde{g}_{\mu\nu}$ and $\phi$, give terms in $\left(\partial\partial\partial\phi\right)\left(\partial\partial\phi\right)$. 
\subsection{Transformation of the action}
However, using integration by parts, the scalar-tensor action written in terms of the disformal metric $\tilde{g}_{\mu\nu}$,
\begin{equation}
\tilde{S}\left[\tilde{g}_{\mu\nu},\phi\right]\equiv S\left[g_{\mu\nu},\phi\right],\label{eq:stilde}
\end{equation}
can be recast in a relatively simple form, see Eq. (\ref{eq:calc_stilde}) below~\cite{Bettoni:2013diz,Babichev:2022awg}. To this aim, an important expression is the one for the inverse of the disformal metric,
\begin{equation}
\tilde{g}^{\mu\nu}=g^{\mu\nu}-\frac{D}{1-2DX}\phi^\mu\phi^\nu,\label{eq:gtilde_inv}
\end{equation}
where it is seen that the disformal transformation is well-defined as mapping invertible symmetric tensors to invertible symmetric tensors if and only if $1-2DX\neq 0$. This is confirmed by the computation of the determinant of the disformal metric. Understanding this computation will be useful later, see Eq. (\ref{eq:det_big}), so we briefly explain it. First write
\begin{equation}
\tilde{g}_{\mu\nu}=g_{\mu\rho}\left(\delta^\rho_\nu+Dg^{\rho\sigma}\phi_\sigma\phi_\nu\right),
\end{equation}
then take the determinant of both sides (we write here the $\det$ in order to distinguish clearly matrices from determinants),
\begin{equation}
\det\left(\tilde{g}\right)=\det\left(g\right)\times\det\left(I+Dg^{-1}\mathcal{P}\right),\label{eq:dets}
\end{equation}
where $I$ is the identity matrix and $\mathcal{P}$ is the rank one matrix $\mathcal{P}_{\mu\nu}\equiv\phi_\mu\phi_\nu$. Thus $g^{-1}\mathcal{P}$ is also of rank one, and its unique nonvanishing eigenvalue is its trace, $\left(g^{-1}\mathcal{P}\right)^\mu_\mu=g^{\mu\nu}\phi_\mu\phi_\nu=-2X$. So the matrix $I+Dg^{-1}\mathcal{P}$ has a single eigenvalue different from $1$ and equal to $1-2DX$, yielding the final expression
\begin{equation}
\det\left(\tilde{g}\right)=\det\left(g\right)\left(1-2DX\right),\label{eq:dets2}
\end{equation}
or, dropping as usual the notation $\det$,
\begin{equation}
\tilde{g}=g\left(1-2DX\right).\label{eq:detdisf}
\end{equation}
These ingredients, as well as other detailed in appendix~\ref{chap:appdisf}, enable to write the scalar-tensor action $\tilde{S}\left[\tilde{g}_{\mu\nu},\phi\right]= S\left[g_{\mu\nu},\phi\right]$ as~\cite{Gleyzes:2014dya}
\begin{align}
\tilde{S}\left[\tilde{g}_{\mu\nu},\phi\right] =\int\mathrm{d}^4 x\sqrt{-\tilde{g}}\Bigl\{{}&{}\widetilde{G_2}-\widetilde{G_3}\widetilde{\Box}\phi+\widetilde{G_4}\tilde{R}+\widetilde{G_{4}}_{\tilde{X}}\left[\left(\widetilde{\Box}\phi\right)^2-\widetilde{\phi_{\mu\nu}}\widetilde{\phi^{\mu\nu}}\right]\nonumber\\{}&{}+\widetilde{G_5}\tilde{G}^{\mu\nu}\widetilde{\phi_{\mu\nu}}-\frac{\widetilde{G_{5}}_{\tilde{X}}}{6}\Bigl[\left(\widetilde{\Box}\phi\right)^3-3\widetilde{\Box}\phi\widetilde{\phi_{\mu\nu}}\widetilde{\phi^{\mu\nu}}+2\widetilde{\phi_{\mu\nu}}\widetilde{\phi^{\nu\rho}}\widetilde{\phi_\rho^\mu}\Bigr]\nonumber\\{}&{}+\widetilde{\mathcal{L}}_{4b}+\widetilde{\mathcal{L}}_{5b}\Bigr\}.\label{eq:calc_stilde}
\end{align}
On the one hand, there are the usual terms of the Horndeski action (\ref{eq:galileon_action}), but of course now in terms of the disformal metric $\tilde{g}_{\mu\nu}$ and with modified Horndeski functions $\widetilde{G_2}$, $\widetilde{G_3}$, $\widetilde{G_4}$, $\widetilde{G_5}$ which are determined by the initial Horndeski functions and the disformal factor $D\left(\phi,X\right)$, and are given in appendix~\ref{chap:appdisf}. On the other hand, the Lagrangian contains two additional contributions $\widetilde{\mathcal{L}}_{4b}$ and $\widetilde{\mathcal{L}}_{5b}$. They read
\begin{align}
\widetilde{\mathcal{L}}_{4b} ={}&{} 2\widetilde{F_4}\left\lbrace 
\tilde{X}\left[\left(\widetilde{\Box}\phi\right)^2-\widetilde{\phi_{\mu\nu}}\widetilde{\phi^{\mu\nu}}\right]+
\widetilde{\Box}\phi\widetilde{\phi^\mu}\widetilde{\phi_{\mu\nu}}\widetilde{\phi^\nu}-\widetilde{\phi_\mu}\widetilde{\phi^{\mu\nu}}\widetilde{\phi_{\nu\rho}}\widetilde{\phi^\rho}
\right\rbrace,  \label{eq:l4b} \\
\widetilde{\mathcal{L}}_{5b}
={}&{}\widetilde{F_5}\Bigl\{
2\tilde{X}\left[\left(\widetilde{\Box}\phi\right)^3 - 3\widetilde{\Box}\phi
\widetilde{\phi_{\mu\nu}}\widetilde{\phi^{\mu\nu}} +2\widetilde{\phi_{\mu\nu}}\widetilde{\phi^{\nu\rho}}\widetilde{\phi^\mu_\rho}\right]+3\Bigl[\left(\widetilde{\Box}\phi\right)^2\widetilde{\phi^\mu}\widetilde{\phi_{\mu\nu}}\widetilde{\phi^\nu}\nonumber\\
{}&{}-2\widetilde{\Box}\phi
\widetilde{\phi_\mu}\widetilde{\phi^{\mu\nu}}\widetilde{\phi_{\nu\rho}}\widetilde{\phi^\rho}-\widetilde{\phi_{\mu\nu}}\widetilde{\phi^{\mu\nu}}
\widetilde{\phi^\rho}\widetilde{\phi_{\rho\sigma}}\widetilde{\phi^\sigma}+2\widetilde{\phi_\mu}\widetilde{\phi^{\mu\nu}}\widetilde{\phi_{\nu\rho}}\widetilde{\phi^{\rho\sigma}}\widetilde{\phi_\sigma}\Bigr]\Bigr\}.\label{eq:l5b}
\end{align}
The two additional functions $\widetilde{F_4}$ and $\widetilde{F_5}$, which, as the Horndeski functions, depend on the scalar field and its kinetic term, are determined by the initial Horndeski functions and the disformal factor $D\left(\phi,X\right)$, see appendix~\ref{chap:appdisf}. It can be seen, in particular, that $\widetilde{F_4}$ vanishes if $D$ does not depend on $X$ and either there is no $G_5$ term in the initial action or $G_5$ and $D$ do not depend on $\phi$. On the other hand, $\widetilde{F_5}$ vanishes if there is no $G_5$ in the initial action or $D$ does not depend on $X$. These subcases, where $\widetilde{F_4}$ and $\widetilde{F_5}$ vanish, determine when the Horndeski class is \textbf{stable} under the disformal transformation (\ref{eq:disf_first}), i.e. mapped to another Horndeski action. Generically however, the Horndeski action is not stable, and the disformal action belongs to a class of theories which can be written as
\begin{align}
S\left[g_{\mu\nu},\phi\right] = \int\mathrm{d}^4 x{}&{}\sqrt{-g}\Bigl\{G_2-G_3\Box\phi+G_4R+G_{4X}\left[\left(\Box\phi\right)^2-\phi_{\mu\nu}\phi^{\mu\nu}\right]\nonumber\\{}&{}+G_5G^{\mu\nu}\phi_{\mu\nu}-\frac{G_{5X}}{6}\Bigl[\left(\Box\phi\right)^3-3\Box\phi\phi_{\mu\nu}\phi^{\mu\nu}+2\phi_{\mu\nu}\phi^{\nu\rho}\phi_\rho^\mu\Bigr]\nonumber\\{}&{}+ F_4 \epsilon^{\mu\nu\rho\sigma}\epsilon^{\alpha\beta\gamma}_{\,\,\,\,\,\,\,\,\,\,\sigma}\phi_\mu\phi_\alpha\phi_{\nu\beta}\phi_{\rho\gamma}+F_5\epsilon^{\mu\nu\rho\sigma}\epsilon^{\alpha\beta\gamma\delta}\phi_\mu\phi_\alpha\phi_{\nu\beta}\phi_{\rho\gamma}\phi_{\sigma\delta}\Bigr\}.\label{eq:beyond_action_first}
\end{align}
This new class of theories can be considered for itself, and no more as coming from a disformal transformation of a Horndeski action, so there are no more $\sim$ (disformal) quantities. All functions $G_2$, $G_3$, $G_4$, $G_5$, $F_4$, $F_5$ depend as usual on $\phi$ and $X$. Also, the new terms (\ref{eq:l4b}-\ref{eq:l5b}) are written in an equivalent, more concise but less explicit way, involving the Levi-Civita tensor $\epsilon_{\mu\nu\rho\sigma}$. This class of actions (\ref{eq:beyond_action_first}) will be studied below, paragraph~\ref{subsec:beyond_degen}, in more detail as the \textbf{beyond Horndeski theories}, see~\cite{Gleyzes:2014qga,Gleyzes:2014dya} and also~\cite{Crisostomi:2016tcp}. Before that, let us go back to the framework of the disformal transformation, and wonder if the transformed action $\tilde{S}\left[\tilde{g}_{\mu\nu},\phi\right]$ is equivalent to the original action $S\left[g_{\mu\nu},\phi\right]$.
\subsection{(In)equivalence of the transformed action}\label{subsec:ineq}
Indeed, a fundamental question is whether the disformal transformation (\ref{eq:disf_first}), which maps the action $S\left[g_{\mu\nu},\phi\right]$ to the new action $\tilde{S}\left[\tilde{g}_{\mu\nu},\phi\right]$, amounts to a mere redefinition of the fields and thus to the description of the same physics, or whether the two theories $S\left[g_{\mu\nu},\phi\right]$ and $\tilde{S}\left[\tilde{g}_{\mu\nu},\phi\right]$ are inequivalent~\cite{Zumalacarregui:2013pma}. This question can be answered from two points of view. First, the distinction between inequivalent theories becomes clear when it comes to defining the \textbf{physical metric} to which matter fields couple minimally. Assume for example that the physical metric is the initial metric $g_{\mu\nu}$. This means that the total action including matter fields, collectively denoted as $\Psi_{\text{m}}$, reads
\begin{equation}
S_{\text{tot}}\left[g_{\mu\nu},\phi,\Psi_{\text{m}}\right]=S\left[g_{\mu\nu},\phi\right]+S_{\text{m}}\left[g_{\mu\nu},\Psi_{\text{m}}\right].\label{eq:stot}
\end{equation} 
$S_{\text{m}}\left[g_{\mu\nu},\Psi_{\text{m}}\right]$ is the matter action, and minimal coupling of the matter to the metric $g_{\mu\nu}$ means that $S_{\text{m}}$ does not depend on the scalar field $\phi$. Then, if the disformal transformation (\ref{eq:disf_first}) mapping $g_{\mu\nu}$ to $\tilde{g}_{\mu\nu}$ is invertible, one gets an equivalent action
\begin{equation}
\tilde{S}_{\text{tot}}\left[\tilde{g}_{\mu\nu},\phi,\Psi_{\text{m}}\right]\equiv S_{\text{tot}}\left[g_{\mu\nu},\phi,\Psi_{\text{m}}\right]
\end{equation}
and this action is explicitly obtained as
\begin{equation}
\tilde{S}_{\text{tot}}\left[\tilde{g}_{\mu\nu},\phi,\Psi_{\text{m}}\right]=\tilde{S}\left[\tilde{g}_{\mu\nu},\phi\right]+S_{\text{m}}\left[\tilde{g}_{\mu\nu}-D\,\phi_\mu\phi_\nu,\Psi_{\text{m}}\right],\label{eq:stildetot}
\end{equation}
where the part $\tilde{S}\left[\tilde{g}_{\mu\nu},\phi\right]$ is the beyond Horndeski action described above and defined by (\ref{eq:stilde}). The matter fields $\Psi_{\text{m}}$ are seen to be non-minimally coupled in the frame $\tilde{g}_{\mu\nu}$, i.e., they do not couple only to the metric $\tilde{g}_{\mu\nu}$ but also to the scalar field $\phi$. While both actions $\tilde{S}_{\text{tot}}\left[\tilde{g}_{\mu\nu},\phi,\Psi_{\text{m}}\right]$ and $S_{\text{tot}}\left[g_{\mu\nu},\phi,\Psi_{\text{m}}\right]$ are equivalent, the action $S_{\text{tot}}\left[g_{\mu\nu},\phi,\Psi_{\text{m}}\right]$ is said to be written in terms of the physical metric $g_{\mu\nu}$ to which matter fields couple minimally. Once this notion of physical metric is understood, it becomes clear how the disformal transformation (\ref{eq:disf_first}) can be used to define two physically inequivalent theories. One can indeed perform the disformal transformation only on the vacuum, pure scalar-tensor part of the action, Eq. (\ref{eq:stilde}). If the disformal transformation is invertible, this leads to two equivalent theories in vacuum, but described in different frames: $S\left[g_{\mu\nu},\phi\right]$ and $\tilde{S}\left[\tilde{g}_{\mu\nu},\phi\right]$. From this, one can \textbf{define} a theory including matter and for which the disformal metric $\tilde{g}_{\mu\nu}$ is \textbf{defined} to be the physical metric, thus leading to a total action of the form
\begin{equation}
\tilde{S}_{\text{tot}}^{\neq}\left[\tilde{g}_{\mu\nu},\phi,\Psi_{\text{m}}\right] = \tilde{S}\left[\tilde{g}_{\mu\nu},\phi\right]+\tilde{S}_{\text{m}}^{\neq}\left[\tilde{g}_{\mu\nu},\Psi_{\text{m}}\right],\label{eq:sdiff}
\end{equation}
where the superscript '$\neq$' emphasizes the difference of the corresponding actions with the actions appearing in (\ref{eq:stildetot}). Now the matter action $\tilde{S}_{\text{m}}^{\neq}\left[\tilde{g}_{\mu\nu},\Psi_{\text{m}}\right]$ is \textbf{defined} so that the matter fields $\Psi_{\text{m}}$ are minimally coupled in the frame $\tilde{g}_{\mu\nu}$. This is the first way of getting inequivalent theories, namely (\ref{eq:stot}) and (\ref{eq:sdiff}), from an invertible disformal transformation (\ref{eq:disf_first}): while the vacuum part (\ref{eq:stilde}) of both actions are equivalent, the equivalence is broken by the \textbf{definition} of the matter action as being, for instance, minimally coupled as illustrated. 
\\

In the previous argument, one uses the matter action to get inequivalent theories from disformal transformations because the disformal transformation is assumed invertible. Therefore, the second way which can lead to inequivalent actions $S\left[g_{\mu\nu},\phi\right]$ and $\tilde{S}\left[\tilde{g}_{\mu\nu},\phi\right]$ is if the disformal transformation is \textbf{not} invertible. The Jacobian of the transformation (\ref{eq:disf_first}) is the 'matrix'
\begin{equation}
\begin{pmatrix}
\frac{\delta \tilde{g}_{\mu\nu}\left(x\right)}{\delta g_{\rho\sigma}\left(y\right)} & \frac{\delta \tilde{g}_{\mu\nu}\left(x\right)}{\delta \phi\left(y\right)} \\
\frac{\delta \tilde{\phi}\left(x\right)}{\delta g_{\rho\sigma}\left(y\right)} & \frac{\delta \tilde{\phi}\left(x\right)}{\delta \phi\left(y\right)}
\end{pmatrix} =
\begin{pmatrix}
\frac{1}{2}\left(\delta^\rho_\mu\delta^\sigma_\nu+\delta_\mu^\sigma\delta^\rho_\nu+D_X\phi^\rho\phi^\sigma\phi_\mu\phi_\nu\right)\delta\left(x,y\right) & \frac{\delta \tilde{g}_{\mu\nu}\left(x\right)}{\delta \phi\left(y\right)} \\
0 & \delta\left(x,y\right)
\end{pmatrix},
\end{equation}
where $\delta\left(x,y\right)$ is the usual Dirac distribution. Since it is upper triangular, there is no need to compute $\delta \tilde{g}_{\mu\nu}\left(x\right)/\delta \phi\left(y\right)$: the disformal transformation is invertible if and only if the matrix~\cite{Zumalacarregui:2013pma,Jirousek:2022rym}
\begin{equation}
J_{\mu\nu}^{\rho\sigma} \equiv \frac{1}{2}\left(\delta^\rho_\mu\delta^\sigma_\nu+\delta_\mu^\sigma\delta^\rho_\nu+D_X\phi^\rho\phi^\sigma\phi_\mu\phi_\nu\right)\equiv I_{\mu\nu}^{\rho\sigma}+\frac{1}{2}D_X\phi^\rho\phi^\sigma\phi_\mu\phi_\nu
\end{equation}
is invertible. This is effectively a $10\times 10$ matrix, with indices being a symmetric pair $\mu\nu$ and a symmetric pair $\rho\sigma$. $I_{\mu\nu}^{\rho\sigma}=\left(\delta^\rho_\mu\delta^\sigma_\nu+\delta_\mu^\sigma\delta^\rho_\nu\right)/2$ is the identity matrix, as underlined by the fact that $I_{\mu\nu}^{\rho\sigma}=\delta g_{\mu\nu}/\delta g_{\rho\sigma}$. One thus sees that the determinant of $J_{\mu\nu}^{\rho\sigma}$ can be determined in an analogous way as the determinant $\det\left(I+Dg^{-1}\mathcal{P}\right)$ was computed between Eqs. (\ref{eq:dets}) and (\ref{eq:dets2}). Indeed, the matrix $\phi^\rho\phi^\sigma\phi_\mu\phi_\nu$ is the analogous of the rank one matrix with two indices $\mathcal{P}_{\mu\nu}=\phi_\mu\phi_\nu$. Therefore the determinant of $J_{\mu\nu}^{\rho\sigma}$ is one, plus the trace of this part of rank one,
\begin{equation}
\det\left(J_{\mu\nu}^{\rho\sigma}\right) = 1+\frac{1}{2}D_X \phi^\mu\phi^\nu\phi_\mu\phi_\nu = 1+2X^2D_X.\label{eq:det_big}
\end{equation}
As a consequence, the disformal transformation is a well-defined and invertible map between invertible symmetric tensors if and only if\footnote{Note that $1-2DX$ must in any case not be negative. Otherwise, the disformed metric would not be Lorentzian, as shown by the relation (\ref{eq:detdisf}) between the determinants, $\tilde{g}=g(1-2DX)$.}
\begin{equation}
1-2DX\not\in\left\lbrace 0,\infty\right\rbrace, \quad 1+2X^2D_X \text{ keeps a constant sign, }>0 \text{ or } <0.
\end{equation}
The first condition comes from the invertibility of $\tilde{g}_{\mu\nu}$, Eqs. (\ref{eq:gtilde_inv}) and (\ref{eq:detdisf}). In a nutshell, \textbf{two ways enable to get physically inequivalent theories with a disformal transformation (\ref{eq:disf_first}). If the disformal transformation is not invertible, i.e. $1-2DX\in\left\lbrace 0,\infty\right\rbrace$ or $1+2X^2D_X$ changes sign, then the vacuum actions $S\left[g_{\mu\nu},\phi\right]$ and $\tilde{S}\left[\tilde{g}_{\mu\nu},\phi\right]$ themselves are not equivalent. If on the contrary the disformal transformation is invertible, then the vacuum actions are physically equivalent, but the total actions (\ref{eq:stot}) and (\ref{eq:sdiff}), obtained by coupling matter minimally to the metrics $g_{\mu\nu}$ and $\tilde{g}_{\mu\nu}$ respectively, are inequivalent.}
\newpage
\section[Beyond Horndeski theories and Degenerate Higher-Order Scalar-Tensor (DHOST) theories]{Beyond Horndeski theories and Degenerate Higher-Order Scalar-Tensor (DHOST) theories}\label{sec:bhdhost}
\sectionmark{2.3~~Beyond Horndeski and DHOST theories}
\subsection{Beyond Horndeski theories}\label{subsec:beyond_degen}
We have just explained that a disformal transformation maps a Horndeski theory into a new class of theory, of the form
\begin{align}
S\left[g_{\mu\nu},\phi\right] = \int\mathrm{d}^4 x{}&{}\sqrt{-g}\Bigl\{G_2-G_3\Box\phi+G_4R+G_{4X}\left[\left(\Box\phi\right)^2-\phi_{\mu\nu}\phi^{\mu\nu}\right]\nonumber\\{}&{}+G_5G^{\mu\nu}\phi_{\mu\nu}-\frac{G_{5X}}{6}\Bigl[\left(\Box\phi\right)^3-3\Box\phi\phi_{\mu\nu}\phi^{\mu\nu}+2\phi_{\mu\nu}\phi^{\nu\rho}\phi_\rho^\mu\Bigr]\nonumber\\{}&{}+ F_4 \epsilon^{\mu\nu\rho\sigma}\epsilon^{\alpha\beta\gamma}_{\,\,\,\,\,\,\,\,\,\,\sigma}\phi_\mu\phi_\alpha\phi_{\nu\beta}\phi_{\rho\gamma}+F_5\epsilon^{\mu\nu\rho\sigma}\epsilon^{\alpha\beta\gamma\delta}\phi_\mu\phi_\alpha\phi_{\nu\beta}\phi_{\rho\gamma}\phi_{\sigma\delta}\Bigr\}.\label{eq:beyond_action_bis}
\end{align}
This class of theory is called \textbf{beyond Horndeski theories} or GLPV (Gleyzes, Langlois, Piazza, Vernizzi) theories~\cite{Gleyzes:2014dya}. They are really distinct from Horndeski theories, i.e. cannot be rewritten with integration by parts in a Horndeski form. Indeed, the two additional Lagrangians, in $F_4(\phi,X)$ and $F_5(\phi,X)$, lead to field equations of order three. As we saw in Sec.~\ref{sec:higher_order}, in general, field equations of order higher than two lead to the propagation of Ostrogradsky ghosts, i.e. additional degrees of freedom associated to a Hamiltonian unbounded from below. We also explained that this Ostrogradsky instability is circumvented if the kinetic matrix of the system is degenerate. In the present case, the analysis of the degrees of freedom is of course more intricate than in the toy model of Sec.~\ref{sec:higher_order}, requiring for instance to perform an Arnowitt-Deser-Misner (ADM)~\cite{PhysRev.116.1322} decomposition. The detailed study can be found in~\cite{Gleyzes:2014dya}, and reveals that \textbf{the beyond Horndeski theory propagates the usual two metric and one scalar degrees of freedom, provided the following relation holds:}
\begin{equation}
3F_5\left(G_4-2XG_{4X}\right)=XF_4G_{5X}.\label{eq:degen_bh}
\end{equation}
In other words, beyond Horndeski theories, verifying condition (\ref{eq:degen_bh}), do not suffer from Ostrogradsky instability. As a consequence, the intuitive origin of beyond Horndeski theories as disformal transformations of Horndeski theories, although useful, is no more essential: action (\ref{eq:beyond_action_bis}) can be considered in itself as a perfectly valid scalar-tensor action, and, among other things, its BH solutions may be studied. The field equations of beyond Horndeski theories are presented in appendix~\ref{sec:eq_horn}.
\subsection{DHOST theories}
Beyond Horndeski theories, although leading to field equations of order higher than two, do not suffer from Ostrogradsky instability thanks to the degeneracy of the kinetic matrix. One may therefore ask the following question: can one generalize further the beyond Horndeski theories, and find the most general scalar-tensor theory evading the Ostrogradsky instability thanks to the degeneracy of the kinetic matrix? This program was initiated by David Langlois and Karim Noui in 2015~\cite{Langlois:2015cwa,Langlois:2015skt}, note in addition the important work~\cite{Crisostomi:2016czh}. This then led to further generalizations~\cite{BenAchour:2016fzp,BenAchour:2016cay}, see also the review~\cite{Langlois:2018dxi}. They classify the \textbf{Degenerate Higher-Order Scalar-Tensor (or DHOST)} theories, that is, the scalar-tensor theories where the degeneracy of the kinetic matrix prevents the existence of Ostrogradsky ghosts, no matter the order of the field equations\footnote{This approach is now being performed also for $U(1)$ vector-tensor theories, see~\cite{Colleaux:2023cqu}.}. 
\\

Again, the reader interested in the construction of these theories is referred to the original articles. DHOST theories were first constructed at \textbf{quadratic order}~\cite{Langlois:2015cwa}, then at \textbf{cubic order}~\cite{BenAchour:2016fzp}, while further orders remain to be constructed. These orders refer to the order in second derivatives of the scalar field. For instance, in the beyond Horndeski action (\ref{eq:beyond_action_bis}), $G_4$ and $F_4$ are quadratic terms, while $G_5$ and $F_5$ are cubic terms (on the other hand, $G_2$ and $G_3$ are linear terms which do not play any role in the kinetic matrix). In this manuscript, we will mostly use the formalism of Horndeski and beyond Horndeski theories, but some of the solutions we will review or construct are based on the DHOST notations. Specifically, at no time will we need the cubic DHOST action. Let us thus present only the quadratic DHOST action, which reads
\begin{align}
S=\int\mathrm{d}^4x\sqrt{-g}\Bigl\{F\,R+P+Q\,\Box\phi{}&{}+A_1\phi_{\mu\nu}\phi^{\mu\nu}+A_2\left(\Box\phi\right)^2+A_3\,\Box\phi\phi^\mu\phi_{\mu\nu}\phi^\nu\nonumber \\ {}&{} +A_4\,\phi^\mu\phi_{\mu\nu}\phi^{\nu\rho}\phi_\rho+A_5\left(\phi^\mu\phi_{\mu\nu}\phi^\nu\right)^2\Bigr\}.\label{eq:dhost_general}
\end{align}
As usual, the functions $F$, $P$, $Q$, $A_{1,2,3,4,5}$ are functions of $\phi$ and its kinetic term $X$. \textbf{Traditionally, in the context of DHOST, the kinetic term is understood to be $X\equiv \phi^\mu\phi_\mu$, while we recall that $X=-\phi^\mu\phi_\mu/2$ for Horndeski and beyond Horndeski theories. In this manuscript, rather than trying to unify the notations, we have decided to keep the traditional conventions. Therefore, each time a DHOST action is considered, the kinetic term is understood to be $X= \phi^\mu\phi_\mu$, while each time a (beyond) Horndeski action is considered, $X=-\phi^\mu\phi_\mu/2$. We will repeat this caveat when necessary.}
\\

There are obviously common terms between the quadratic DHOST Lagrangian (\ref{eq:dhost_general}) and the beyond Horndeski Lagrangian (\ref{eq:beyond_action_bis}). $F$, in front of the Ricci scalar, is like $G_4$, while $P\sim G_2$ and $Q\sim G_3$. The terms in $A_1$ and $A_2$ appear in $G_{4X}$ but also in $F_4$, while $A_3$ and $A_4$ appear in $F_4$ [see the explicit expression of the $F_4$ Lagrangian, Eq. (\ref{eq:l4b})]. The $A_5$ Lagrangian is completely new. As explained, the (beyond) Horndeski Lagrangians $G_5$ and $F_5$ are cubic, so one would need to write down the cubic DHOST Lagrangian to find them appear. \textbf{In any case, DHOST is the most general framework of scalar-tensor theories evading Ostrogradsky instability, which means that (beyond) Horndeski theories are particular cases of DHOST theories}.
\\

The quadratic DHOST action (\ref{eq:dhost_general}) is in fact not yet a DHOST action: it is indeed a quadratic, higher-order, scalar-tensor theory, but one must impose the degeneracy conditions of the kinetic matrix as computed in~\cite{Langlois:2015cwa}. Very briefly, these degeneracy conditions lead to distinct classes I, II and III, decomposing themselves into subclasses, for instance Ia and Ib. Among all these classes and subclasses, it was shown that all subclasses but the subclass Ia suffer from an instability, when it comes to studying the quadratic action for perturbations around a cosmological background~\cite{Langlois:2017mxy,Langlois:2017dyl,Langlois:2018dxi}. Therefore, the only phenomenologically viable theory is DHOST Ia, for which the degeneracy conditions are:
\begin{align}
A_1={}&{}-A_2,\label{eq:a1dhost}\\
F\neq{}&{} -XA_2,\\
A_4={}&{}\frac{1}{8(F+XA_2)^2}\Bigl[16XA_2^3+4(3F+16XF_X)A_2^2-X^2FA_3^2\nonumber\\{}&{}+(16X^2F_X-12XF)A_3A_2+16F_X(3F+4XF_X)A_2\nonumber\\{}&{}+8F(XF_X-F)A_3+48FF_X^2\Bigr],\label{eq:a4dhost}\\
A_5={}&{} \frac{(4F_X+2A_2+XA_3)(-2A_2^2+3XA_2A_3-4F_XA_2+4FA_3)}{8(F+XA_2)^2}\label{eq:a5dhost}.
\end{align}
DHOST Ia theories thus depend on five arbitrary functions of $\phi$ and $X$: $P$, $Q$, $F$, $A_2$ and $A_3$. Finally, note that the condition (\ref{eq:degen_bh}), given above and ensuring that the generic beyond Horndeski action (\ref{eq:beyond_action_bis}) does not suffer from Ostrogradsky instability, is a particular case of the most general degeneracy conditions found when one combines quadratic and cubic DHOST actions~\cite{BenAchour:2016fzp}. The field equations of quadratic DHOST are presented in appendix~\ref{sec:eq_dhost}.
\subsection{Interpretation in terms of conformal-disformal transformations}\label{subsec:intconfdisf}
Sec.~\ref{sec:disf} presented the disformal transformation of the metric,
\begin{equation}
g_{\mu\nu}\to\tilde{g}_{\mu\nu}=g_{\mu\nu}+D(\phi,X)\phi_\mu\phi_\nu.
\end{equation}
Starting with a Horndeski action $S\left[g_{\mu\nu},\phi\right]$, it was shown that one obtains a new action
\begin{equation}
\tilde{S}\left[\tilde{g}_{\mu\nu},\phi\right]\equiv S\left[g_{\mu\nu},\phi\right],\label{eq:sstildenew}
\end{equation}
and that the new action $\tilde{S}\left[\tilde{g}_{\mu\nu},\phi\right]$ belongs to the beyond Horndeski class, with an action of the form (\ref{eq:beyond_action_bis}). There exists a similar interpretation of DHOST theories~\cite{BenAchour:2016cay}, through this time \textbf{conformal-disformal transformations},
\begin{equation}
g_{\mu\nu}\to\tilde{g}_{\mu\nu}=C(\phi,X)g_{\mu\nu}+D(\phi,X)\phi_\mu\phi_\nu,\label{eq:confdisf}
\end{equation}
where, in addition to the disformal factor $D(\phi,X)$, now appears a conformal factor $C(\phi,X)$. Under such a conformal-disformal transformation, if the original scalar-tensor action $S\left[g_{\mu\nu},\phi\right]$ belongs to the (beyond) Horndeski class, then the resulting action $\tilde{S}\left[\tilde{g}_{\mu\nu},\phi\right]$ as defined by (\ref{eq:sstildenew}) belongs to the DHOST Ia class. We have purposedly detailed previously the case of a pure disformal transformation, in Sec.~\ref{sec:disf} and appendix~\ref{chap:appdisf}. This permitted to see, among others, the invertibility conditions for such a transformation, as well as the equivalence or not between the initial and transformed actions. Analogous considerations can of course be applied to the generic conformal-disformal transformation, and the interested reader is referred to~\cite{BenAchour:2016cay}. The important point in exiting beyond Horndeski and entering more general DHOST is that the conformal factor $C$ depend on $X$. If $C=C(\phi)$, then a beyond Horndeski action is merely transformed into another beyond Horndeski action. The viable DHOST Ia class is itself stable under a generic transformation (\ref{eq:confdisf}).
\\

This stability enables \textbf{generation of solutions}~\cite{BenAchour:2020wiw} in the following way. If one starts from a \textbf{seed solution} $\left(g_{\mu\nu},\phi\right)$ of the variational principle of the initial action $S$, and the transformation (\ref{eq:confdisf}) is invertible, then the image of the seed solution, $\left(\tilde{g}_{\mu\nu},\phi\right)$, is solution of the variational principle of the new action $\tilde{S}$. This useful tool has been widely used, to generate among others: novel rotating BH solutions~\cite{Anson:2020trg,BenAchour:2020fgy}, see paragraph~\ref{subsec:disf_kerr}; wormhole solutions~\cite{Bakopoulos:2021liw,Babichev:2022awg}, see Sec.~\ref{sec:wormhole}; rotating BHs embedded in an expanding universe~\cite{Babichev:2023mgk}, see Sec.~\ref{sec:kerrflrw}; and nonlinear GW solutions~\cite{BenAchour:2024zzk,BenAchour:2024tqt}.
\subsection{Observational constraints and GW170817}
From Horndeski to DHOST, an enormous amount of scalar-tensor theories is now at the disposal of the researcher willing to modify gravity. These new actions can be studied for themselves, and this is the point of view which is mostly adopted in this thesis. Namely, we will consider scalar-tensor actions $S\left[g_{\mu\nu},\phi\right]$, study their mathematical properties and possible simplifying symmetries, see Sec.~\ref{sec:symm}, analyze their field equations, and find their closed-form BH solutions.
\\

However, the ultimate aim of the physicist is to come with theories which explain the observations. As briefly reviewed in Chap.~\ref{chap:modifGR}, modifications of gravity are typically introduced to overcome the shortcomings of GR, such as BH singularities or the dark energy problem. Beyond the rather mathematical and calculational point of view which is adopted in this thesis, it is therefore paramount that, at least, the new scalar-tensor theories stick to the existing observations, just as GR does~\cite{Will:2014kxa}.
\\

If a particular scalar-tensor theory is to be considered seriously as a modification of GR, it must therefore predict correctly the various following tests, see~\cite{Clifton:2020xhc} for a short but detailed presentation of them and of others:
\begin{itemize}
\item advance of periapsis: perihelion precession of Mercury~\cite{2011CeMDA.111..363F}; LAGEOS satellites precession~\cite{Lucchesi:2010zzb}; precession of the star S2 around Sagittarius A$^\star$, the central BH of the Milky Way, as measured by the GRAVITY collaboration~\cite{GRAVITY:2020gka}; precession in binary pulsars~\cite{Lorimer:2008se};
\item gravitational lensing of light, e.g. by the Sun as measured using Very Long Baseline Interferometry~\cite{PhysRevLett.92.121101};
\item Shapiro time delay~\cite{Shapiro:1964uw}, measured with the help of the Cassini spacecraft~\cite{Bertotti:2003rm};
\item BH shadows, which were measured for the supermassive BH M87$^\star$ by the EHT~\cite{EventHorizonTelescope:2019dse,EventHorizonTelescope:2019ggy};
\item \textbf{speed of GWs}.
\end{itemize}
This ultimate point may bring tight constraints on the admissible scalar-tensor theories. It is however sometimes hastily claimed that this constraint rules out entire possibilities for scalar-tensor theories. Let us therefore elaborate a bit on this point.
\\

After their theoretical prediction by Einstein in 1916~\cite{1916SPAW.......688E,1918SPAW.......154E}, GWs received a first indirect experimental evidence in 1974, from the observed orbital decay of the Hulse-Taylor binary pulsar~\cite{1975ApJ...195L..51H}. Indeed, this decay matched the one predicted by GR, as energy is transmitted from the binary trajectory into gravitational radiation~\cite{1979Natur.277..437T,1982ApJ...253..908T}. However, the first direct detection of GWs was made in 2015 by the LIGO detectors, which observed the gravitational radiation generated by the merger of two BHs~\cite{article_ligo}. The sixth GW event, GW170817, detected by LIGO and Virgo in 2017~\cite{LIGOScientific:2017ync,LIGOScientific:2017vwq}, is up to date the only one with an electromagnetic counterpart, starting with a short gamma-ray burst discovered by the Fermi Gamma-ray Space Telescope. Comparing the times of detection implies the following constraint on the speed of GWs $c_T$ as compared to the speed of light $c$~\cite{LIGOScientific:2017zic},
\begin{equation}
-3\times 10^{-15}\leq \frac{c_T}{c}-1\leq 7\times 10^{-16},\label{eq:constraint_serious}
\end{equation} 
essentially implying the \textbf{equality between the speed of GWs and the speed of light}. The speed of GWs can be easily computed in scalar-tensor theories by writing down the quadratic action for linear cosmological perturbations. We refer the reader interested in the precise computation to the references which appear in the following lines. For instance, for Horndeski theories, the speed of GWs is~\cite{Kobayashi:2019hrl,Kobayashi:2011nu}
\begin{equation}
c_T^2 = \frac{G_4-X\left(\ddot{\phi}\,G_{5X}+G_{5\phi}\right)}{G_4-2XG_{4X}-X\left(H\dot{\phi}\,G_{5X}-G_{5\phi}\right)},\label{eq:cthorn}
\end{equation}
with $H\equiv \dot{a}/a$ the Hubble rate of the FLRW expansion with scale factor $a$. If $c_T$ is to be equal to the speed of light, no matter the background cosmological evolution, this imposes
\begin{equation}
G_{4X}=0,\quad G_5=0.
\end{equation}
As regards the quadratic DHOST action (\ref{eq:dhost_general}), submitted to the degeneracy conditions (\ref{eq:a1dhost}-\ref{eq:a5dhost}) of the viable class Ia, the equality between $c_T$ and $c$ requires~\cite{Langlois:2017dyl}
\begin{equation}
A_2=0.\label{eq:dhostct1}
\end{equation}
In fact, some scalar-tensor theories predict the rapid decay of GWs into scalar waves~\cite{Creminelli:2018xsv}. This would be in contradiction with the observation on Earth of GWs emitted by binaries located one billion lightyears away, as for GW170817. For DHOST Ia, evading this decay implies the additional constraint $A_3=0$. To appreciate the drastic reduction of admissible DHOST theories, note that one goes from action (\ref{eq:dhost_general}), with degeneracy conditions (\ref{eq:a1dhost}-\ref{eq:a5dhost}), to the following action,
\begin{align}
S=\int\mathrm{d}^4x\sqrt{-g}\Bigl\{F\,R+P+Q\,\Box\phi+\frac{6F_X^2}{F}\phi^\mu\phi_{\mu\nu}\phi^{\nu\rho}\phi_\rho\Bigr\}.
\end{align}
While the original DHOST Ia action is parameterized by five independent functions $P$, $Q$, $F$, $A_1$ and $A_3$, the surviving one depends only on $P$, $Q$ and $F$.\\

GW170817 thus arguably reduces considerably the freedom of the theoretical physicists when it comes to considering scalar-tensor theories. As regards the work realized during this thesis, the main results of which are presented in Chaps.~\ref{chap:4},~\ref{chap:5} and~\ref{chap:6}, all novel closed-form BH (and wormhole) solutions are found in theories which do not satisfy $c_T=1$ according to the previous criteria, since they all have $G_{4X}\neq 0$ in the Horndeski framework. Does this mean that these new solutions are physically irrelevant? In fact, there exist two arguments relativising the necessity for a scalar-tensor theory to satisfy the above constraints.\\

The first argument was pointed out by~\cite{deRham:2018red} and has to do with the status of scalar-tensor theories as EFTs for dark energy~\cite{Gubitosi:2012hu}. The parameters of an EFT depend \textit{a priori} on the energy/frequency scale $k$. In particular, the speed of GW may generically have such dependence, $c_T=c_T(k)$. However, the LIGO bound coming from GW170817 applies to GWs at a frequency between around $10$ and $100$ Hz, while the EFT for dark energy stands as an EFT for describing cosmology on energy scales which are smaller by about $20$ orders of magnitude. Ref.~\cite{deRham:2018red} analyzes the precise example of a Horndeski theory where $c_T\neq c$ as predicted by the formula (\ref{eq:cthorn}), but where the operators appearing at the cutoff scale end affecting $c_T$, bringing it back to $c_T=c$ at the energy scales probed by LIGO.\\

The second argument comes from considering the brute observational constraint on $c_T$, Eq. (\ref{eq:constraint_serious}), seriously. This means that $c_T$ is indeed very close to $c$, but that this does not force one to impose strictly $c_T=c$. In this case, the bounds (\ref{eq:constraint_serious}) are merely used to put constraints on, e.g., the coupling constants of a particular scalar-tensor theory. An interesting example of this point of view is~\cite{Clifton:2020xhc}. In this article, the authors consider the four-dimensional Einstein-Gauss-Bonnet theory, to be introduced in Sec.~\ref{sec:fourd} of the present manuscript. This theory has all Horndeski terms $G_{2,3,4,5}$, so $c_T\neq c$. \\

More precisely, they consider the case where this theory is parameterized by a unique coupling constant $\alpha$, that they constrain using experimental data. The perihelion precession for instance leads to $\left\lvert\alpha\right\rvert\leq 10^{10}\,\text{m}^2$. On the other hand, the FLRW metric is a solution of the field equations of this theory, with a homogeneous scalar field $\phi$. One can then use the expression for $G_4$, $G_5$ and $\phi$ into the expression for the speed of GWs in Horndeski, Eq. (\ref{eq:cthorn}). Forcing the obtained $c_T$ to lie within the experimental bounds (\ref{eq:constraint_serious}) gives a much lighter constraint on $\alpha$, namely $\left\lvert\alpha\right\rvert\leq 10^{36}\,\text{m}^2$.\\

The former argument is more general, while the second rather requires individual checks for a given theory with its particular FLRW solution. Although the constraints brought about by GW170817 represent a great progress in understanding and restricting scalar-tensor theories, we hope that these two arguments will convince the reader of the following fact: \textbf{it is not because a Horndeski theory possesses a non-zero $G_{4X}$ or $G_5$ that this theory is a priori irrelevant (and equivalently for DHOST)}. 
\\

Now that this important parenthesis regarding experimental constraints bearing on scalar-tensor theories has been precised, we go back to the generic presentation of scalar-tensor theories and their properties. In paragraph~\ref{subsec:intconfdisf}, we mentioned the possibility of generating new solutions from existing ones. Even though this is a handy technique, it still requires to start from a seed solution, so one must find ways to build scalar-tensor solutions from scratch, before having the possibility to apply a conformal-disformal transformation on them. As is common and fundamental in physics~\cite{10.1093/0198248601.001.0001}, the use of \textbf{symmetries} is at the core of the construction of many solutions in scalar-tensor theories. We now turn on to the description of the most commonly used such symmetries. 
\section{Symmetries of scalar-tensor theories}\label{sec:symm}
\sectionmark{2.4~~Symmetries of scalar-tensor theories}
Given a scalar-tensor action $S\left[g_{\mu\nu},\phi\right]$, its invariance under a continuous symmetry implies the existence of a conserved Noether current which should simplify the solving of the field equations. This section first studies \textbf{shift symmetry}, where the invariance transformation bears only on the scalar field. As will be seen, shift symmetry leads to great simplifications, particularly for static and spherically-symmetric spacetimes, and thus stands as a common assumption. The analysis then turns to \textbf{(global) conformal symmetry}, which involves both the metric and the scalar field. This naturally leads in a third time to \textbf{local conformal symmetry}. 
\subsection{Shift (and parity) symmetry}\label{subsec:shift}
The previous general introduction to scalar-tensor theories showed that they are all parameterized by functions of the scalar field $\phi$ and its kinetic term $X$: $G_{2,3,4,5}\left(\phi,X\right)$ for Horndeski theories, with in addition $F_{4,5}\left(\phi,X\right)$ for beyond Horndeski theories, and analogously for DHOST theories. Each of these functions multiply terms which depend on the scalar field \textbf{only through its derivatives}. Consequently, if the functions parameterizing the theory themselves depend only on $X=-\phi^\mu\phi_\mu/2$ and not on $\phi$, the entire action depends on the scalar field \textbf{only through its derivatives}. The scalar-tensor action $S\left[g_{\mu\nu},\phi\right]$ then acquires \textbf{shift symmetry} under shifts of the scalar field by a constant value,
\begin{equation}
\phi\to\phi+\text{constant}.\label{eq:shift_sym}
\end{equation}
In particular, the value of the scalar field is determined only up to an additive constant. Remembering the definition (\ref{eq:field_eq}) of the scalar field equation of motion $\mathcal{E}_\phi$, one can write the variation $\delta S$ of the action under a generic change $\delta \phi$ of the scalar field as
\begin{align}
\int\mathrm{d}^4x\sqrt{-g}\,\delta\phi\,\mathcal{E}_\phi=\delta S={}&{}\int\mathrm{d}^4x\sqrt{-g}\,\delta\left(\partial_\mu\phi\right)\frac{1}{\sqrt{-g}}\frac{\delta S}{\delta\left(\partial_\mu\phi\right)},\nonumber\\
{}={}&{}-\int\mathrm{d}^4x\sqrt{-g}\,\delta\phi\,\nabla_\mu \mathcal{J}^\mu_\text{shift}
\end{align}
where the last equality of the first line comes from the fact that $S$ depends only on the derivatives of the scalar field, and the last equality is obtained through integration by parts and introduction of the current
\begin{equation}
\mathcal{J}^\mu_\text{shift} \equiv \frac{1}{\sqrt{-g}}\frac{\delta S}{\delta\left(\partial_\mu\phi\right)}.\label{eq:shift_current}
\end{equation}
Because $\delta\phi$ is arbitrary, this current is seen to provide an integration of the scalar field equation of motion,
\begin{equation}
\mathcal{E}_\phi=-\nabla_\mu \mathcal{J}^\mu_\text{shift}.\label{eq:ephij}
\end{equation}
This Noether current $\mathcal{J}^\mu_\text{shift}$ is given in appendix~\ref{chap:appfieldeq}. Shift symmetry will therefore simplify the solving of the field equations, and this will be even more the case for static and spherically-symmetric spacetimes.
\subsubsection{Shift symmetry in static, spherically-symmetric spacetimes}
Indeed, assume a metric ansatz of the form
\begin{equation}
\mathrm{d}s^2=-h\left(r\right)\mathrm{d}t^2+\mathrm{d}r^2/f\left(r\right)+r^2\mathrm{d}\Omega^2,\label{eq:metric_ansatz_1}
\end{equation}
where $\left(t,r,\theta,\varphi\right)$ are the usual spherically-symmetric coordinates, $\mathrm{d}\Omega^2=\mathrm{d}\theta^2+\sin^2\theta\mathrm{d}\varphi^2$ is the metric of the two-sphere with unit radius, and $h\left(r\right)$ and $f\left(r\right)$ can be called respectively the \textbf{redshift} and \textbf{shape} functions, to be determined by the field equations. A scalar field compatible with the symmetries of the metric should not introduce any $t$-dependent, $\theta$-dependent or $\varphi$-dependent terms in the field equations. However, the action depends on the scalar field only through its derivatives. As a consequence, it matters only that the \textbf{derivatives} of the scalar field do not depend on $t$, $\theta$ nor $\varphi$. This allows for a linear dependence of the scalar field $\phi$ on these coordinates. However, a linear dependence on the angles $\theta$ or $\varphi$ would yield a multi-valued scalar field. The ansatz for the scalar field $\phi$, compatible with the symmetries of the metric ansatz (\ref{eq:metric_static_ansatz}), can therefore be taken as~\cite{Babichev:2010kj,Babichev:2012re,Babichev:2013cya}
\begin{equation}
\phi=qt+\psi\left(r\right),\label{eq:scalar_ansatz_1}
\end{equation}
where $q$ is a constant. Since the scalar field is dimensionless, and $t$ has\footnote{In all the manuscript, a physical quantity is said to have dimension $n$ if it is homogeneous to a mass to the power $n$, taking into account the units $c=G=1$.} dimension $1$, $q$ has dimension $-1$. Use now the diffeomorphism invariance of the action under a diffeomorphism $\delta\xi^\mu$, Eq. (\ref{eq:diffeo}), along with (\ref{eq:ephij}). This leads to
\begin{align}
0 = \delta S ={}&{} \int\mathrm{d}^4x\sqrt{-g}\,\delta\xi^\mu \left[ -\nabla_\nu\mathcal{E}_{\mu}^\nu+\nabla_\nu \mathcal{J}^\nu_\text{shift}\,\phi_\mu\right]
\nonumber\\ {}={}&{}\int\mathrm{d}^4x\sqrt{-g}\left[\left(\mathcal{E}^\nu_\mu-J^\nu\phi_\mu\right)\nabla_\nu\delta\xi^\mu-\mathcal{J}^\nu_\text{shift}\phi_{\nu\mu}\delta\xi^\mu\right],
\end{align}
where one integrates by parts to get the final line. Remarkably, choosing the diffeomorphism vector field to be $\delta\xi^\mu=\left(\delta\xi^t\left(r\right),0,0,0\right)$, taking into account the ansatz (\ref{eq:metric_ansatz_1}-\ref{eq:scalar_ansatz_1}) which enables to compute explicitly the covariant derivatives $\nabla_\nu\delta\xi^\mu$ and $\phi_{\nu\mu}$, this last integral simplifies to
\begin{equation}
0 = \int\mathrm{d}^4x\sqrt{-g}\,\left(\delta\xi^t\right)'\left(\mathcal{E}_t^r-q\mathcal{J}^r_\text{shift}\right).
\end{equation}
Therefore, given the arbitrariness of $\delta\xi^t\left(r\right)$,
\begin{equation}
\mathcal{E}^r_t=q\mathcal{J}^r_\text{shift}.\label{eq:super_important}
\end{equation}
This identity, found by~\cite{Babichev:2015rva}, implies that, if the scalar field has a linear time dependence, $q\neq 0$, then the radial component $\mathcal{J}^r_\text{shift}$ of the current vanishes \textbf{on shell} (i.e. when the field equations are verified). This result is much stronger than the result (\ref{eq:ephij}), which implied only that $\nabla_\mu \mathcal{J}^\mu_\text{shift}$ was vanishing on shell. This result does not hold \textit{a priori} for a purely radial scalar field, $q=0$. In this case, when the field equations are verified,
\begin{equation}
0=\nabla_\mu \mathcal{J}^\mu_\text{shift}=\frac{1}{\sqrt{-g}}\partial_\mu\left(\sqrt{-g}\mathcal{J}^\mu_\text{shift}\right)=\frac{1}{\sqrt{-g}}\partial_r\left(r^2\sin\theta\sqrt{\frac{h}{f}} \mathcal{J}^r_\text{shift}\right),
\end{equation}
because with this ansatz, $\mathcal{J}^r_\text{shift}$ is the only non-identically vanishing component of the current and depends only on $r$. Therefore there exists a constant $\mathfrak{c}$ such that~\cite{Hui:2012qt}
\begin{equation}
\mathcal{J}^r_\text{shift}=\frac{\mathfrak{c}}{r^2}\sqrt{\frac{f}{h}}.
\end{equation} 
As announced, $\mathcal{J}^r_\text{shift}$ is no more vanishing, unless $\mathfrak{c}=0$. However, consider the case where the metric (\ref{eq:metric_ansatz_1}) is a BH with horizon $r_h$, 
\begin{equation}
h\left(r_h\right)=f\left(r_h\right)=0.
\end{equation}
Then, it is possible to impose $\mathfrak{c}=0$ by assuming that the norm $\mathcal{J}^\mu_\text{shift} \mathcal{J}_{\text{shift}\,\mu}$ of the current is finite in the whole spacetime. Indeed,
\begin{equation}
\mathcal{J}^\mu_\text{shift} \mathcal{J}_{\text{shift}\,\mu} = \frac{\mathfrak{c}^2}{r^4h},
\end{equation}
so this norm diverges at the horizon unless $\mathfrak{c}=0$.
\subsubsection*{Parity symmetry}
Going back to the general case of a shift-symmetric scalar-tensor action, each term in the action now has a definite behaviour under the \textbf{parity transformation} $\phi\to -\phi$. Indeed, each function parameterizing the action ($G_{2,3,4,5}$, $F_{4,5}$, etc.) now depends only on $X$, which is parity-symmetric. Therefore, if the expression mutliplied by this function is itself parity-symmetric, then the corresponding term also enjoys this symmetry. For instance, for the beyond Horndeski action (\ref{eq:beyond_action_bis}), if shift symmetry is assumed, one gets parity symmetry of the action under $\phi\to -\phi$ by keeping only the terms $G_2\left(X\right)$, $G_4\left(X\right)$ and $F_4\left(X\right)$. Parity symmetry is discrete, so does not seem to imply as many simplifications as shift symmetry. However, it will be seen in Chap.~\ref{chap:4} and more precisely in Sec.~\ref{sec:parity} that, when studying explicitly the field equations, parity symmetry helps a lot to get exact solutions.
\subsection{Global conformal symmetry}\label{subsec:globalconffinal}
In curved spacetime, scale invariance of a theory, or \textbf{conformal symmetry}, involves a Weyl rescaling of the metric field, $g_{\mu\nu}\to \omega^2 g_{\mu\nu}$, where the rescaling parameter $\omega$ is here assumed to be a constant (the case of non-constant $\omega$ corresponds to \textbf{local} conformal symmetry and will be dealt with just after). This implies the following rescalings: $\sqrt{-g}\to\omega^4\sqrt{-g}$, $R\to R/\omega^2$, so the Einstein-Hilbert action is of course not invariant. Thanks to the scalar field, it is however possible to construct scalar-tensor actions enjoying the global conformal symmetry, provided one imposes, in addition to the Weyl rescaling of the metric, a rescaling of the scalar\footnote{It would be more accurate to speak of Weyl symmetry rather than conformal symmetry; however, we use this abuse of terminology, since it is ubiquitous in the literature.}
\begin{equation}
g_{\mu\nu}\to \omega^2 g_{\mu\nu},\quad \phi\to\frac{\phi}{\omega}.\label{eq:conf_inv}
\end{equation}
Then for the kinetic term, $X\to X/\omega^4$, and one identifies the invariant combination 
\begin{equation}
Y\equiv \frac{X}{\phi^4},\quad Y\to Y.
\end{equation}
One then easily understands the forms of, for example, the Horndeski functions $G_{2,3,4,5}$ guaranteeing conformal invariance: they have a part in $\phi^n$ where the power $n$ is chosen so as to cancel all possible $\omega$ factors, multiplying an arbitrary function of the invariant combination $Y$. The most general Horndeski action possessing the global conformal invariance under (\ref{eq:conf_inv}) is thus~\cite{Padilla:2013jza}
\begin{align}
S\left[g_{\mu\nu},\phi\right] = {}&{}\int\mathrm{d}^4 x\sqrt{-g}\Bigl\{\phi^4a_2\left(Y\right)-\phi a_3\left(Y\right)\Box\phi+\phi^2a_4\left(Y\right)R+\frac{a_{4Y}}{\phi^2}\left[\left(\Box\phi\right)^2-\phi_{\mu\nu}\phi^{\mu\nu}\right]\nonumber\\{}&{}+\frac{a_5\left(Y\right)}{\phi}G^{\mu\nu}\phi_{\mu\nu}-\frac{a_{5Y}}{6\phi^5}\Bigl[\left(\Box\phi\right)^3-3\Box\phi\phi_{\mu\nu}\phi^{\mu\nu}+2\phi_{\mu\nu}\phi^{\nu\rho}\phi_\rho^\mu\Bigr]\Bigr\},\label{eq:global_conf_ac}
\end{align}
where $a_2$, $a_3$, $a_4$, $a_5$ are arbitrary functions of $Y$, and a subscript $Y$ means derivation with respect to $Y$. While the use of the Noether current of shift symmetry is highly common, the Noether current of such conformally-invariant theories has never been used up to date to construct new solutions. In fact, this latter Noether current had never been written down in the literature. Therefore, we computed it, and present it in appendix~\ref{chap:appconfnoeth} for all practical purposes: one might end finding it useful for the discovery of new BH solutions in theories enjoying global conformal symmetry. 
\subsection{Local conformal symmetry and generalized conformal symmetry}\label{subsec:local}
Global conformal symmetry thus leads to a plethora of possible actions, Eq. (\ref{eq:global_conf_ac}). The number of candidate actions may be grealty reduced by rather considering \textbf{local conformal symmetry}, where the invariance transformation is now under
\begin{equation}
g_{\mu\nu}\to\Omega^2(x)g_{\mu\nu},\quad\phi\to\frac{\phi}{\Omega(x)},\label{eq:local_conf_Omega}
\end{equation}
where the previous constant parameter $\omega$ is promoted to a function on spacetime, $\Omega=\Omega(x)$. In fact, for latter convenience, we will rather work with the following local conformal transformation,
\begin{equation}
g_{\mu\nu}\to \mathrm{e}^{2\sigma(x)}g_{\mu\nu},\quad\phi\to\phi-\sigma(x),\label{eq:local_conf}
\end{equation}
which is the same as above provided $\sigma=\ln\Omega$ and the scalar is redefined as $\phi\to\ln\phi$. In this paragraph~\ref{subsec:local}, we are first going to present formal aspects of local conformal symmetry and generalized conformal symmetry, which is a related invariance to be defined below. Only at the end of the paragraph will the explicit form of the corresponding actions be presented. The following presentation elaborates on results of~\cite{Bocharova:1970skc,Bekenstein:1974sf,Martinez:2002ru,Jackiw:2005su,Padilla:2013jza,Fernandes:2021dsb}, trying to organize them in a coherent and unified way.
\subsubsection{Equivalence between local conformal symmetry and $g^{\mu\nu}\mathcal{E}_{\mu\nu}+\mathcal{E}_\phi=0$}
Let us prove the following result: \textbf{a scalar-tensor action $S\left[g_{\mu\nu},\phi\right]$ has local conformal invariance under (\ref{eq:local_conf}) if and only if the trace of its metric field equations, and its scalar field equation, obey the following off-shell relation:
\begin{equation}
g^{\mu\nu}\mathcal{E}_{\mu\nu}+\mathcal{E}_\phi=0.\label{eq:geomegal0}
\end{equation}}The proof is straightforward. Consider an arbitrary scalar-tensor action $S\left[g_{\mu\nu},\phi\right]$ and an infinitesimal transformation (\ref{eq:local_conf}), $\sigma(x)=-\epsilon(x)$. Introduce the following notations for the transformed metric and scalar fields,
\begin{equation}
g^\epsilon_{\mu\nu}=\left(1-2\epsilon\right)g_{\mu\nu},\quad \phi^\epsilon\equiv \phi+\epsilon.\label{eq:geps}
\end{equation}
Accordingly, the action is transformed into
\begin{equation}
S^\epsilon\left[g_{\mu\nu},\phi\right]\equiv S\left[g^\epsilon_{\mu\nu},\phi^\epsilon\right].\label{eq:geps2}
\end{equation}
Using the definition (\ref{eq:field_eq}) of the field equations, the variation of the action is then
\begin{equation}
S^\epsilon\left[g_{\mu\nu},\phi\right]-S\left[g_{\mu\nu},\phi\right]=\int\mathrm{d}^4x\sqrt{-g}\,\epsilon(x)\left(g^{\mu\nu}\mathcal{E}_{\mu\nu}+\mathcal{E}_\phi\right).\label{eq:geps3}
\end{equation}
The action $S$ has local conformal invariance if and only if the left hand side vanishes, which is seen to be indeed equivalent to $g^{\mu\nu}\mathcal{E}_{\mu\nu}+\mathcal{E}_\phi=0$.
\subsubsection{Generalized conformal symmetry}
There exists another notion of conformal symmetry, dubbed \textbf{generalized conformal symmetry}, which is not properly speaking a usual symmetry of the action, but will turn out to appear in many situations. \textbf{A scalar-tensor action $S\left[g_{\mu\nu},\phi\right]$ is said to have generalized conformal symmetry if the Lorentz scalar density $\delta S/\delta \phi\left[g_{\mu\nu},\phi\right]$ has local conformal invariance under the transformation (\ref{eq:local_conf})}. It is sometimes said that the scalar field equation of motion is conformally invariant, however, we rather defined the scalar field equation to be the Lorentz scalar $\mathcal{E}_\phi=(-g)^{-1/2}\delta S/\delta\phi$, so this is a slight abuse of terminology.
\\

This notion of generalized conformal symmetry is in fact the correct way of getting a practical use of the notion of local conformal symmetry, as we are going to explain in the next few lines. Indeed, local conformal symmetry (\ref{eq:local_conf}) is a gauge symmetry which thus removes a degree of freedom. In (\ref{eq:local_conf}), one may choose the transformation parameter $\sigma(x)$ to be the scalar field $\phi$, thus setting the scalar field to zero and highlighting the removal of a degree of freedom. In a word, local conformal invariance destroys the additional degree of freedom that we purposedly introduced by considering scalar-tensor theories. 
\\

This is why this local conformal symmetry should be \textbf{broken}. Take an action $S_\text{loc}\left[g_{\mu\nu},\phi\right]$ with local conformal invariance, add to it another action $S_\text{\st{loc}}\left[g_{\mu\nu},\phi\right]$ without the invariance, then the total action $S\left[g_{\mu\nu},\phi\right]=S_\text{\st{loc}}\left[g_{\mu\nu},\phi\right]+S_\text{loc}\left[g_{\mu\nu},\phi\right]$ does not have local conformal invariance. There is thus an infinity of ways of breaking the invariance, as many as possible choices for $S_\text{\st{loc}}\left[g_{\mu\nu},\phi\right]$. 
\\

Nevertheless, a natural option for $S_\text{\st{loc}}\left[g_{\mu\nu},\phi\right]$ is to take an action which does not depend on the scalar field, $S_\text{met}\left[g_{\mu\nu}\right]$. Of course, in four dimensions, the only possible choice is the Einstein-Hilbert action with cosmological constant, according to Lovelock's theorem, but we keep the notation $S_\text{met}\left[g_{\mu\nu}\right]$ to highlight the generality of the present considerations (which will be used in dimension higher than four in Chap.~\ref{chap:5}). Thus, the local conformal symmetry is broken by coupling the invariant action to a pure metric action. The total action is then
\begin{equation}
S\left[g_{\mu\nu},\phi\right]=S_\text{met}\left[g_{\mu\nu}\right]+S_\text{loc}\left[g_{\mu\nu},\phi\right].\label{eq:smetsloc}
\end{equation}
As a consequence,
\begin{equation}
\frac{\delta S}{\delta\phi}\left[g_{\mu\nu},\phi\right] = \frac{\delta S_\text{loc}}{\delta\phi}\left[g_{\mu\nu},\phi\right].\label{eq:deltasdeltasloc}
\end{equation}
$S_\text{loc}$ has local conformal invariance, so $\delta S_\text{loc}/\delta\phi$ also, so $\delta S/\delta\phi$ also, according to (\ref{eq:deltasdeltasloc}). In other words, $S$ displays generalized conformal invariance. One thus understands that \textbf{generalized conformal invariance naturally arises when one breaks the local conformal invariance by an action $S_\text{met}\left[g_{\mu\nu}\right]$ independent of the scalar field}. This explains the relevance of the notion of generalized conformal invariance, which might have seemed obscure at first sight.
\\

However, Eq. (\ref{eq:smetsloc}) is not the most general form of an action with generalized conformal invariance: their may exist Lagrangian terms which are not purely metric (as in $S_\text{met}$) nor locally conformally invariant (as in $S_\text{loc}$), and still lead to a locally conformally invariant $\delta/\delta \phi$. The explicit form of such terms will be seen later, for the moment, let us pursue with abstract characterizations of generalized conformal invariance.
\subsubsection{Equivalence between generalized conformal symmetry and $g^{\mu\nu}\mathcal{E}_{\mu\nu}+\mathcal{E}_\phi=\mathfrak{R}\left(g_{\mu\nu}\right)$}
An interesting characterization of generalized conformal invariance is the following: \textbf{a scalar-tensor action $S\left[g_{\mu\nu},\phi\right]$ has generalized conformal invariance if and only if there exists a Lorentz scalar $\mathfrak{R}=\mathfrak{R}\left(g_{\mu\nu}\right)$, built out of the metric $g_{\mu\nu}$ (and independent of $\phi$), such that off shell, 
\begin{equation}
g^{\mu\nu}\mathcal{E}_{\mu\nu}+\mathcal{E}_\phi=\mathfrak{R}\left(g_{\mu\nu}\right).\label{eq:puregeomfisrtoc}
\end{equation}}In other words, such scalar-tensor action satisfies on shell a \textbf{purely geometric equation}, namely $\mathfrak{R}=0$. \\

Note that the characterization of \textbf{local} conformal invariance, see Eq. (\ref{eq:geomegal0}), can be reformulated as the fact that the purely geometric quantity is in this case vanishing off shell: $\mathfrak{R}=0$.\\
 
The proof unfolds as follows. Consider an arbitrary scalar-tensor action $S\left[g_{\mu\nu},\phi\right]$. The variation of the action under an infinitesimal conformal transformation is again (\ref{eq:geps3}), with the notations of Eqs. (\ref{eq:geps}-\ref{eq:geps2}). Because $g^\epsilon_{\mu\nu}$ does not depend on $\phi$, one has
\begin{equation}
\frac{\delta S^\epsilon}{\delta \phi}\left[g_{\mu\nu},\phi\right]=\frac{\delta S}{\delta\phi}\left[g^\epsilon_{\mu\nu},\phi^\epsilon\right].\label{eq:deltadeltaeps}
\end{equation}
Combining (\ref{eq:deltadeltaeps}) with (\ref{eq:geps3}) leads to
\begin{equation}
\frac{\delta S}{\delta\phi}\left[g^\epsilon_{\mu\nu},\phi^\epsilon\right]-\frac{\delta S}{\delta\phi}\left[g_{\mu\nu},\phi\right]=\int\mathrm{d}^4x\sqrt{-g}\,\epsilon(x)\frac{\delta}{\delta\phi}\left(g^{\mu\nu}\mathcal{E}_{\mu\nu}+\mathcal{E}_\phi\right).
\end{equation}
By definition, the action $S$ enjoys generalized conformal invariance if and only if the left hand side vanishes, which is seen to be equivalent to 
\begin{equation}
\frac{\delta}{\delta\phi}\left(g^{\mu\nu}\mathcal{E}_{\mu\nu}+\mathcal{E}_\phi\right)=0.
\end{equation}
This is turn is equivalent to the existence of a Lorentz scalar $\mathfrak{R}\left(g_{\mu\nu}\right)$ (i.e. a purely geometric quantity) such that $g^{\mu\nu}\mathcal{E}_{\mu\nu}+\mathcal{E}_\phi=\mathfrak{R}$, as announced.
\\

We have thus defined local conformal symmetry and generalized conformal symmetry (which includes local conformal symmetry but is more general), and seen that they are equivalent to the existence of an $\mathfrak{R}=\mathfrak{R}\left(g_{\mu\nu}\right)$ such that $g^{\mu\nu}\mathcal{E}_{\mu\nu}+\mathcal{E}_\phi=\mathfrak{R}$ off shell, with in particular $\mathfrak{R}=0$ in the case of local conformal symmetry. Let us now briefly describe the construction of the most general scalar-tensor action with these symmetries. The construction is done in four dimensions, and will be generalized for later purposes to arbitrary dimensions in Chap.~\ref{chap:5}.
\subsubsection{Most general action with local conformal invariance (broken by an Einstein-Hilbert term): the BBMB/MTZ action}
A straightforward way to construct the most general action enjoying local conformal invariance consists in identifying a metric $\hat{g}_{\mu\nu}$, depending on both $g_{\mu\nu}$ and $\phi$, $\hat{g}_{\mu\nu}=\hat{g}_{\mu\nu}\left(g_{\rho\sigma},\phi\right)$, and invariant under the transformation (\ref{eq:local_conf}). One easily finds
\begin{equation}
\hat{g}_{\mu\nu} = \mathrm{e}^{2\phi}g_{\mu\nu}.
\end{equation}
The only scalar density of unit weight built out of $\hat{g}_{\mu\nu}$ is $\sqrt{-\hat{g}}F(\hat{g}_{\mu\nu})$, where $F(\hat{g}_{\mu\nu})$ is an arbitrary Lorentz scalar constructed out of $\hat{g}_{\mu\nu}$. So the most general four-dimensional action with local conformal invariance is
\begin{equation}
S_{\text{loc}}\left[g_{\mu\nu},\phi\right]=\int\mathrm{d}^4x\sqrt{-\hat{g}}F(\hat{g}_{\mu\nu}).
\end{equation}
Importantly, it is a scalar-tensor action for $g_{\mu\nu}$ and $\phi$: the hat quantities on the right-hand side must be understood as functions of $g_{\mu\nu}$ and $\phi$. Now, by requiring second-order field equations, one is obviously led to the usual Einstein-Hilbert Lagrangian for $\hat{g}_{\mu\nu}$, $F(\hat{g}_{\mu\nu})=-\beta\hat{R}-2\lambda$, where $\beta$ and $\lambda$ are coupling constants, with $\lambda$ appearing as a 'cosmological constant' for the $\hat{g}_{\mu\nu}$ Lagrangian. So
\begin{align}
S_{\text{loc}}\left[g_{\mu\nu},\phi\right]{}&{}=\int\mathrm{d}^4x\sqrt{-\hat{g}}\left(-\beta\hat{R}-2\lambda\right)\nonumber\\{}&{}=\int\mathrm{d}^4x\sqrt{-g}\Bigl[-\beta\mathrm{e}^{2\phi}\left(R+6\left(\partial\phi\right)^2\right)-2\lambda\mathrm{e}^{4\phi}\Bigr].\label{eq:slocpara}
\end{align}
One gets the last form by using~\cite{Carneiro:2004rt,Dabrowski:2008kx}
\begin{equation}
\sqrt{-\hat{g}}=\mathrm{e}^{4\phi}\sqrt{-g},\quad \hat{R}=\mathrm{e}^{-2\phi}\left(R-6\Box\phi-6\left(\partial\phi\right)^2\right),
\end{equation}
and integrating by parts. Remember that an action with local conformal invariance is not of particular interest in itself, since the additional scalar field degree of freedom is removed by the gauge symmetry. As explained earlier, it is therefore usual to consider the action $S_\text{loc}$ coupled to the Einstein-Hilbert action,
\begin{equation}
S_{\text{BBMB/MTZ}}\left[g_{\mu\nu},\phi\right]\equiv \int\mathrm{d}^4x\sqrt{-g}\Bigl[R-2\Lambda-\beta\mathrm{e}^{2\phi}\left(R+6\left(\partial\phi\right)^2\right)-2\lambda\mathrm{e}^{4\phi}\Bigr].\label{eq:bbmbmtz}
\end{equation}
This action is called \textbf{BBMB/MTZ} for the following reason. If $\Lambda=0=\lambda$, Bocharova, Bronnikov, Melnikov~\cite{Bocharova:1970skc} and Bekenstein~\cite{Bekenstein:1974sf} studied the BH solutions of the action. On the other hand, Martinez, Troncoso, Zanelli~\cite{Martinez:2002ru} studied the case of non-zero $\Lambda$ and $\lambda$. Of course, the BBMB/MTZ action does not enjoy anymore local conformal invariance but generalized conformal invariance, so there exists a geometric equation $\mathfrak{R}=g^{\mu\nu}\mathcal{E}_{\mu\nu}+\mathcal{E}_\phi$. First, $\mathfrak{R}_\text{loc}=0$ for the part with local conformal invariance. Second, $\mathfrak{R}_\text{EH}=2g^{\mu\nu}\left(G_{\mu\nu}+\Lambda g_{\mu\nu}\right)=8\Lambda-2R$ for the Einstein-Hilbert part. So the geometric equation verified by the BBMB/MTZ action is the following,
\begin{equation}
\mathfrak{R}_\text{BBMB/MTZ}=8\Lambda-2R.\label{eq:geombbmb}
\end{equation}
For instance, if $\Lambda=0$, any scalar-tensor solution of the BBMB action must have vanishing scalar curvature, $R=0$. Chap.~\ref{sec:state} reviews the already existing exact BH solutions, and the solutions of the BBMB/MTZ action will be given more precisely in paragraph~\ref{subsec:bbmbbh}. For the moment, let us stress on the fact that the BBMB/MTZ action is just the sum, in four dimensions, of the most general pure metric action (Einstein-Hilbert) with the most general action with local conformal invariance, $S_\text{loc}$. Does this give the most general action with generalized conformal invariance? The answer is no, as shown by the following lines.
\subsubsection{Most general action with generalized conformal invariance: the Fernandes action}
The construction of the most general action with generalized conformal invariance is quite similar in the idea to the above construction, and mostly requires an additional technical step. Since $\delta S/\delta\phi$ must have local conformal invariance, one must have
\begin{equation}
\frac{\delta S}{\delta\phi}=\sqrt{-\hat{g}}F(\hat{g}_{\mu\nu}).\label{eq:ferncons1}
\end{equation}
As before, $\hat{g}_{\mu\nu}=\mathrm{e}^{2\phi}g_{\mu\nu}$ is the metric with local conformal invariance, and $F(\hat{g}_{\mu\nu})$ is a Lorentz scalar built out of this metric. For the construction of an action with local conformal invariance, $F(\hat{g}_{\mu\nu})$ was simply given by Lovelock's theorem as the Einstein-Hilbert Lagrangian for $\hat{g}_{\mu\nu}$, $F(\hat{g}_{\mu\nu})=-\beta\hat{R}-2\lambda$ (provided one restricts to second-order field equations, which we again do here). \\

In fact, the most general Lorentz scalar that one can build from the four-dimensional metric $\hat{g}_{\mu\nu}$, and consistent with second-order field equations, is more general than $F(\hat{g}_{\mu\nu})=-\beta\hat{R}-2\lambda$. Indeed, due to Lovelock's theorem, one can write down more generally
\begin{equation}
F(\hat{g}_{\mu\nu})=-\beta\hat{R}-2\lambda-\alpha\hat{\mathcal{G}}.
\end{equation}
There is now a contribution from the GB scalar of $\hat{g}_{\mu\nu}$. Of course, included inside a Lagrangian, this contribution is a boundary term, this is why it was not taken into account when constructing $S_\text{loc}$ above. However, it must now be taken into account, yielding
\begin{equation}
\frac{\delta S}{\delta\phi}=\sqrt{-\hat{g}}\left(-2\beta\hat{R}-8\lambda-\alpha\hat{\mathcal{G}}\right)\label{eq:deltasbetalambdalpha}
\end{equation}
(the coefficients in front of $\beta$ and $\lambda$ are not taken as before because $F(\hat{g}_{\mu\nu})$ now appears in $\delta S/\delta\phi$ and no more in $S$ itself). All what remains is a technical part: how to reconstruct $S$ from the knowledge of $\delta S/\delta \phi$? We will not explain this reconstruction explicitly: the interested reader may read the original article~\cite{Fernandes:2021dsb}, which uses methods presented in~\cite{soper2008classical}. The important points are the following. First, the $\beta$ and $\lambda$ terms in (\ref{eq:deltasbetalambdalpha}), once 'integrated', give the same contributions as in the BBMB/MTZ action (\ref{eq:bbmbmtz}). Second, the $\alpha$ part in (\ref{eq:deltasbetalambdalpha}) gives a new contribution as compared to the BBMB/MTZ action. This new contribution involves both $g_{\mu\nu}$ and $\phi$, does not have local conformal invariance (not even up to integration by parts), although its variation $\delta/\delta\phi$ has local conformal invariance. Third, and as for the BBMB/MTZ case, one can freely add the Einstein-Hilbert action, which preserves both the generalized conformal invariance and the second-order field equations. At the end of the day, the resulting action, as constructed by Fernandes~\cite{Fernandes:2021dsb}, is
\begin{align}
S_{\text{Fern}}\left[g_{\mu\nu},\phi\right] =\int\mathrm{d}^4x\sqrt{-g}{}&{}\Bigl\{R-2\Lambda-2\lambda\mathrm{e}^{4\phi}-\beta\mathrm{e}^{2\phi}\Bigl[R+6\left(\partial\phi\right)^2\Bigr]\nonumber\\ {}&{}+\alpha\Bigl[-\phi\mathcal{G}+4G^{\mu\nu}\phi_\mu\phi_\nu+4\Box\phi\left(\partial\phi\right)^2+2\left(\partial\phi\right)^4\Bigr]\Bigr\}.\label{eq:fern_first}
\end{align}
We repeat that the associated BH solutions are presented in Chap.~\ref{sec:state}, more precisely in paragraph~\ref{subsec:bhegb}. Remember that \textbf{the Fernandes action} is the most general one with generalized conformal invariance and second-order field equations; more general actions are obtained if one allows higher-order field equations~\cite{Ayon-Beato:2023bzp}. The Fernandes action thus belongs to the Horndeski class. The associated Horndeski functions are
\begin{align}
G_2 ={}&{} -2\lambda\mathrm{e}^{4\phi}+12\beta\mathrm{e}^{2\phi}X+8\alpha X^2,\quad G_3 = 8\alpha X,\nonumber\\ G_4 ={}&{} 1-\beta\mathrm{e}^{2\phi}+4\alpha X,\quad G_5 = 4\alpha\ln\left\lvert X\right\rvert,\label{eq:horn_conf}
\end{align}
see Eq. (\ref{eq:gbhorn}) to understand the $G_5$ term. By construction also and because of its generalized conformal invariance, the Fernandes action satisfies a geometric equation, computed to be
\begin{equation}
\mathfrak{R}_\text{Fern}=g^{\mu\nu}\mathcal{E}_{\mu\nu}+\mathcal{E}_\phi = 8\Lambda-2R-\alpha\mathcal{G}.\label{eq:rfern}
\end{equation}
Here comes a striking fact. Consider the Lovelock's EGB action in dimension $D\geq 5$, $S=\int\mathrm{d}^Dx\sqrt{-g}(R-2\Lambda+\bar{\alpha}\mathcal{G})$. The field equations, as given by (\ref{eq:lov_eq}-\ref{eq:lov_trace}), have the following trace:
\begin{equation}
g^{\mu\nu}\mathcal{E}_{\mu\nu}=2D\Lambda+(2-D)R+\bar{\alpha}(4-D)\mathcal{G}.\label{eq:traceegb}
\end{equation}
One sees that the first two terms give the same contribution (\ref{eq:rfern}) as for the Fernandes action upon taking $D=4$. The last term $\bar{\alpha}(4-D)\mathcal{G}$ vanishes, but, if one defines a new coupling $\alpha\equiv \bar{\alpha}(D-4)$, one can impose the following limit: send the dimension $D\to 4$, while simultaneously sending the coupling $\bar{\alpha}\to \infty$, such that $\alpha$ remains a constant. Then the last term $\bar{\alpha}(4-D)\mathcal{G}$ gives the same contribution $-\alpha\mathcal{G}$ as in (\ref{eq:rfern}) in the limit $D\to 4$. It thus seems that, up to a rescaling of the GB coupling constant $\bar{\alpha}$, the trace of the EGB field equations is reproduced by the Fernandes theory. The next section explains in detail this link and why the Fernandes action can be called the \textbf{four-dimensional Einstein-Gauss-Bonnet (4DEGB) theory}.
\section{Four-dimensional Einstein-Gauss-Bonnet theory}\label{sec:fourd}
\sectionmark{2.5~~Four-dimensional Einstein-Gauss-Bonnet theory}
The obtention of the 4DEGB theory has been the core of an intense research activity between 2019 and 2021. This quest was initiated by Glavan and Lin~\cite{Glavan:2019inb}. Several aspects of their work were incorrect, but stirred up the discussion and led to better proposals. An excellent and very detailed review on the 4DEGB gravity can be found in~\cite{Fernandes:2022zrq}. Here, we rather adopt the unified picture of KK regularization, introduced in~\cite{Babichev:2023psy} by the present author, among others.\\

Sec.~\ref{sec:compact} presented the diagonal KK compactification of the $\mathfrak{D}$-dimensional EGB theory,
\begin{equation}
S_{(\mathfrak{D}),\text{EGB}}\left[g_{(\mathfrak{D})AB}\right]= \int\mathrm{d}^\mathfrak{D}x\sqrt{-g_{(\mathfrak{D})}}\left(R_{(\mathfrak{D})}+\bar{\alpha}\mathcal{G}_{(\mathfrak{D})}\right).\label{eq:degb_bis}
\end{equation}
This compactification is based on a diagonal decomposition of the $\mathfrak{D}$-dimensional metric:
\begin{equation}
g_{(\mathfrak{D})AB}\mathrm{d}x^A\mathrm{d}x^B=g_{(D)\mu\nu}\mathrm{d}x^\mu\mathrm{d}x^\nu+\mathrm{e}^{-2\phi}\widetilde{g}_{(n)ab}\mathrm{d}x^a\mathrm{d}x^b.\label{eq:diagansatz}
\end{equation}
A scalar field $\phi$, which depends only on the target spacetime coordinates $x^\mu$, appears as a conformal factor of the internal space metric $\widetilde{g}_{(n)ab}$. The target spacetime has dimension $D$, and the internal space dimension $n=\mathfrak{D}-D$. Let us focus on the case where $D=4$, while the case of general $D$ is in Sec.~\ref{sec:compact}. If the internal space is such that its curvature invariants $\widetilde{R}_{(n)}$ and $\widetilde{\mathcal{G}}_{(n)}$ are constants, the original action is compactified down to a four-dimensional scalar-tensor action,
\begin{align}
S_{n,\text{EGB}}\left[g_{\mu\nu},\phi\right] ={}&{} \int\mathrm{d}^4x\sqrt{-g}\,\mathrm{e}^{-n\phi}\Bigl\{R+\widetilde{R}_{(n)}\mathrm{e}^{2\phi}+n\left(n-1\right)\left(\partial\phi\right)^2+\bar{\alpha}\Bigl[\mathcal{G}+\widetilde{\mathcal{G}}_{(n)}\mathrm{e}^{4\phi}\nonumber\\{}&{}+2\widetilde{R}_{(n)}\mathrm{e}^{2\phi}\Bigl(R+\left(n-2\right)\left(n-3\right)\left(\partial\phi\right)^2\Bigr)-4n\left(n-1\right)G^{\mu\nu}\partial_\mu\phi\,\partial_\nu\phi\nonumber\\{}&{}+2n\left(n-1\right)\left(n-2\right)\Box\phi\left(\partial\phi\right)^2-n\left(n-1\right)^2\left(n-2\right)\left(\partial\phi\right)^4\Bigr]\Bigr\}.\label{eq:egbkkchristos_bis}
\end{align}
The subscripts $D=4$ are dropped for brevity. The Ricci and GB scalars of the $n$-dimensional internal space, $\widetilde{R}_{(n)}$ and $\widetilde{\mathcal{G}}_{(n)}$, are seen to play the role of coupling constants for the four-dimensional theory. Now that the action is written down, the dimension $n$ can just as well be taken as a parameter of the theory, taking its values in $\mathbb{R}$. 
\\

Action (\ref{eq:egbkkchristos_bis}) was first introduced by~\cite{Charmousis:2012dw}. In this same article, the authors took the internal space to be a product of $n/2$ two-spheres of same radius $\rho$, giving
\begin{equation}
\widetilde{R}_{(n)} = n/\rho^2,\quad \widetilde{\mathcal{G}}_{(n)} = n\left(n-2\right)/\rho^4.\label{eq:prodspheres}
\end{equation}
They took a spherically-symmetric ansatz
\begin{equation}
\mathrm{d}s^2=-f(r)\mathrm{d}t^2+\mathrm{d}r^2/f(r)+r^2\mathrm{d}\Omega^2,\quad \phi=\phi(r),
\end{equation}
and obtained exact solutions for any $n$. Nevertheless, the metric function behaves asymptotically as $f\left(r\right)\approx 1/\left(n+1\right)\left(1-2M/r^{n+1}\right)$. This is not a Schwarzschild-like behaviour, unless $n=0$, but in this case (\ref{eq:egbkkchristos_bis}) reduces to pure GR. \\

The way out of this deadlock was first described by Lu and Pang~\cite{Lu:2020iav}, who defined the following limit:
\begin{equation}
n\to 0,\quad \bar{\alpha}\to \infty,\quad n\bar{\alpha}=\text{constant}\equiv \alpha,\label{eq:limitlupang}
\end{equation}
where $\alpha$ is a new coupling constant. Lu and Pang applied this procedure to a maximally-symmetric internal space of curvature $\gamma$. The Riemann tensor of such an internal space is $\widetilde{R}_{(n)abcd} = \gamma\left(\widetilde{g}_{(n)ac}\widetilde{g}_{(n)bd}-\widetilde{g}_{(n)ad}\widetilde{g}_{(n)bc}\right)$. This yields
\begin{equation}
\widetilde{R}_{(n)} = \gamma n\left(n-1\right),\quad \widetilde{\mathcal{G}}_{(n)} = \gamma^2n\left(n-1\right)\left(n-2\right)\left(n-3\right).\label{eq:maxsym}
\end{equation}
Nevertheless, their prescription in fact applies to any internal space such that $\widetilde{R}_{(n)}\propto n$ and $\widetilde{\mathcal{G}}_{(n)}\propto n$, like the product of two-spheres (\ref{eq:prodspheres}). Indeed, $\widetilde{R}_{(n)}$ and $\widetilde{\mathcal{G}}_{(n)}$ are multiplied by $\bar{\alpha}$ in the action (\ref{eq:egbkkchristos_bis}), so they possess a regular limit under the prescription (\ref{eq:limitlupang}). The only term of (\ref{eq:egbkkchristos_bis}) which remains ambiguous in this limit is $\mathrm{e}^{-n\phi}\bar{\alpha}\mathcal{G}$. It is schematically regularized by expanding the exponential as
\begin{equation}
\mathrm{e}^{-n\phi}\bar{\alpha}\mathcal{G} = \underbrace{\cancel{\bar{\alpha}\mathcal{G}}}_\text{BT} - n\bar{\alpha}\phi\mathcal{G} +\bar{\alpha}\,\mathcal{O}\left(n^2\right) \underset{(\ref{eq:limitlupang})}{\longrightarrow} -\alpha\phi\mathcal{G}. 
\end{equation}
In the intermediate step, $\bar{\alpha}\mathcal{G}$ drops out since it is a boundary term (BT) in four dimensions. Very generally, the procedure of Lu and Pang can thus be applied for any internal space for which the following \textbf{regularized curvature invariants} are well-defined,
\begin{equation}
\widetilde{R}_\text{reg} \equiv \lim_{n\to 0}\frac{\widetilde{R}_{(n)}}{n},\quad \widetilde{\mathcal{G}}_\text{reg} \equiv \lim_{n\to 0}\frac{\widetilde{\mathcal{G}}_{(n)}}{n}.\label{eq:curv_reg}
\end{equation}
With these notations, applying the limit (\ref{eq:limitlupang}) to the KK action (\ref{eq:egbkkchristos_bis}) brings about the \textbf{regularized KK action}~\cite{Babichev:2023psy}:
\begin{align}
S\left[g_{\mu\nu},\phi\right] = \int\mathrm{d}^4x\sqrt{-g}\Bigl\{R-2\Lambda+\alpha\Bigl[{}&{}-\phi\mathcal{G}+\widetilde{\mathcal{G}}_\text{reg}\mathrm{e}^{4\phi}+2\widetilde{R}_\text{reg}\mathrm{e}^{2\phi}\Bigl(R+6\left(\partial\phi\right)^2\Bigr)\nonumber\\ {}&{}+4G^{\mu\nu}\phi_\mu\phi_\nu+4\Box\phi\left(\partial\phi\right)^2+2\left(\partial\phi\right)^4\Bigr]\Bigr\}.\label{eq:kkreg}
\end{align}
We introduced a cosmological constant $\Lambda$ which could have been taken into account since the beginning. Shortly after the publication of Lu and Pang, two other articles, Refs.~\cite{Fernandes:2020nbq} and~\cite{Hennigar:2020lsl}, came in almost simultaneously with another proposal for regularizing EGB gravity in four dimensions. In this case, the GB piece in four dimensions is defined as follows. First, consider the $D$-dimensional piece $\int\mathrm{d}^Dx\sqrt{-g}\hat{\alpha}\mathcal{G}$. Second, substract the identical action but for the conformally-related metric $\hat{g}_{\mu\nu}=\mathrm{e}^{2\phi}g_{\mu\nu}$. Third, redefine the coupling $\hat{\alpha}=\alpha/\left(D-4\right)$. Finally, compute the limit when $D\to 4$. The obtained action is nothing but (\ref{eq:kkreg}) with $\widetilde{R}_\text{reg}=\widetilde{\mathcal{G}}_\text{reg}=0$: it can thus be seen as a regularized KK along a flat internal space.
\\

The Fernandes action (\ref{eq:fern_first}) was introduced a few months later by Fernandes~\cite{Fernandes:2021dsb}, as the most general action in four dimensions with generalized conformal invariance (and second-order field equations), as we explained above. Comparison of the Fernandes action with the regularized KK action (\ref{eq:kkreg}) shows that both actions are identical, under the identification:
\begin{equation}
2\lambda = -\alpha\widetilde{\mathcal{G}}_\text{reg},\quad \beta = -2\alpha\widetilde{R}_\text{reg}.\label{eq:link}
\end{equation}
In other words, \textbf{the diagonal KK regularization of the EGB theory leads to the most general scalar-tensor action with generalized conformal invariance as constructed by Fernandes}. One now understands better the link between the trace of the EGB field equations (\ref{eq:traceegb}) and the geometric equation (\ref{eq:rfern}), implied by the generalized conformal invariance and which we recall,
\begin{equation}
\mathfrak{R}=g^{\mu\nu}\mathcal{E}_{\mu\nu}+\mathcal{E}_\phi=8\Lambda-2R-\alpha\mathcal{G}.
\end{equation} 
The Fernandes/regularized KK action are seen to extend naturally a geometric equation of the higher-dimensional theory down to four dimensions. In addition, because of the Fernandes construction, which we explained briefly between (\ref{eq:ferncons1}) and (\ref{eq:fern_first}), this action is the most general which reproduces this equation. At the end of the day, one can thus consider this action to be the \textbf{4DEGB theory},
\begin{align}
S_{\text{4DEGB}}\left[g_{\mu\nu},\phi\right] =\int\mathrm{d}^4x\sqrt{-g}{}&{}\Bigl\{R-2\Lambda-2\lambda\mathrm{e}^{4\phi}-\beta\mathrm{e}^{2\phi}\Bigl[R+6\left(\partial\phi\right)^2\Bigr]\nonumber\\ {}&{}+\alpha\Bigl[-\phi\mathcal{G}+4G^{\mu\nu}\phi_\mu\phi_\nu+4\Box\phi\left(\partial\phi\right)^2+2\left(\partial\phi\right)^4\Bigr]\Bigr\}.\label{eq:4degb}
\end{align}
Remember that it belongs to the Horndeski class with Horndeski functions (\ref{eq:horn_conf}). The theory acquires shift-symmetry when $\lambda=\beta=0$: the $\phi\mathcal{G}$ term is shift-symmetric because $\mathcal{G}$ is a boundary term in four dimensions. Note also that the $\mathrm{e}^{2\phi}X$ term in $G_2$ can be transformed into a canonical kinetic term for the scalar field $\Phi\equiv \mathrm{e}^\phi$, but we will continue working with $\phi$. This appears to be important, since scalar field perturbations can be affected by the absence of the lowest order term in the action. 
\\

The static, spherically-symmetric BH solutions of the 4DEGB action will be reviewed in paragraph~\ref{subsec:bhegb}. Note that they will be obtained for different respective values of the couplings~\cite{Fernandes:2021dsb}:
\begin{align}
(1)\quad {}&{}\lambda=\frac{3\beta^2}{4\alpha},\label{eq:linkmax1}\\ 
(2)\quad {}&{}\lambda=\beta=0,\label{eq:linkflat1}\\
(3)\quad {}&{}\lambda=\frac{\beta^2}{4\alpha}.\label{eq:linkprod1}
\end{align}
Using the relation between the coupling constants and the Ricci and GB invariants of the internal space, Eq. (\ref{eq:link}), as well as the value of these curvature invariants for a product of two-spheres (\ref{eq:prodspheres}) and for a maximally-symmetric space (\ref{eq:maxsym}), one can see that these values of the couplings correspond respectively to
\begin{align}
(1)\quad {}&{}\text{Maximally-symmetric internal space},\label{eq:linkmax2}\\ 
(2)\quad {}&{}\text{Flat internal space},\label{eq:linkflat2}\\
(3)\quad{}&{}\text{Internal space product of two-spheres}.\label{eq:linkprod2}
\end{align}
This presentation of the 4DEGB theory closes Chap.~\ref{chap:st} by demonstrating a powerful interpretation of scalar-tensor theories, at least from a theoretical point of view: this particular scalar-tensor theory reproduces features of the lowest order modification to GR in higher dimensions as proved by Lovelock's theorem. In fact, even some of the BH solutions of this 4DEGB theory appear as continuations in $D=4$ of BH solutions existing in the EGB theory in dimension $D\geq 5$. These solutions, among other exact solutions in scalar-tensor theories, are presented in Chap.~\ref{sec:state}.
\chapter[State-of-the-art of closed-form black hole solutions in scalar-tensor theories]{State-of-the-art of closed-form black hole solutions in scalar-tensor theories}\label{sec:state}
\chaptermark{Chap.~3~~State-of-the-art of closed-form BH solutions in scalar-tensor theories}
The framework of scalar-tensor theories and their symmetries has been clearly set up in Chap.~\ref{chap:st}. It is now time to describe the various BH solutions which were encountered thus far in scalar-tensor theories. By 'thus far', we mean that this third chapter will present the solutions which were obtained before this thesis started, in Fall 2021. We allow for a cosmological constant $\Lambda$ in the action, but not for Maxwell terms or any kind of matter: we are interested in vacuum solutions. \\

Sec.~\ref{sec:nohair} recaps essential features on \textbf{no-hair theorems}. The main result is the following: apart from the \textbf{BBMB solution}~\cite{Bocharova:1970skc,Bekenstein:1974sf}, scalar-tensor theories of the historical (generalized) Brans-Dicke form (\ref{eq:action_bd}) do not allow for asymptotically flat BHs. On the contrary, Horndeski theories and beyond enable to get a lot of hairy BH solutions. On the one hand, Sec.~\ref{sec:stealth} focuses on so-called \textbf{stealth BHs}, where the metric coincides with a GR metric, but is dressed with a non-trivial scalar field. On the other hand, Sec.~\ref{sec:nonstealth} describes \textbf{non-stealth solutions}, where the metric itself does not possess any equivalent in GR.
\section[No-hair theorems in scalar-tensor theories and the BBMB black hole]{No-hair theorems in scalar-tensor theories and the BBMB black hole}\label{sec:nohair}
\sectionmark{3.1~~No-scalar-hair theorems and the BBMB black hole}
\subsection{Challenge represented by asymptotically flat BHs}\label{subsec:challenge}
There exists a very instructive review by Herdeiro and Radu about \textit{Asymptotically flat black holes with scalar hair}~\cite{Herdeiro:2015waa}. They first review the existing no-scalar-hair theorems and their assumptions, in order to present in a second time scalar-tensor BH solutions obtained by violating some of these assumptions. The title of the review is interesting, highlighting that these no-scalar-hair theorems bear on \textbf{asymptotically flat BHs}.\\

This implies the following rule of thumb: \textbf{there exists less restrictions for the obtention of BHs with dS or adS asymptotics than for the obtention of asymptotically flat BHs}. Concretely, there exists scalar-tensor theories which allow for BHs with dS or adS asymptotics, but not for asymptotically flat BHs. In general, in such theories, the effective cosmological constant appearing in the metric is constructed out of the various coupling constants appearing in the Lagrangian terms coupling the scalar field with the metric. As a precise instance of these considerations, an early example was given by Rinaldi in 2012~\cite{Rinaldi:2012vy} (see also~\cite{Anabalon:2013oea,Minamitsuji:2013ura}). He considered the following action, with one coupling constant $z$,
\begin{equation}
S = \int\mathrm{d}^4x\sqrt{-g}\Bigl[\frac{R}{2}-\frac{1}{2}\left(g^{\mu\nu}-zG^{\mu\nu}\right)\partial_\mu\phi\,\partial_\nu\phi\Bigr].\label{eq:actionrinaldi}
\end{equation}
His ansatz is static and spherically-symmetric,
\begin{equation}
\mathrm{d}s^2=-h(r)\mathrm{d}t^2+\mathrm{d}r^2/f(r)+r^2\mathrm{d}\Omega^2.
\end{equation}
A solution to the field equations is given by
\begin{align}
h(r)={}&{}\frac{3}{4}+\frac{r^2}{12z}-\frac{2M}{r}+\frac{\sqrt{z}}{4r}\arctan\left(\frac{r}{\sqrt{z}}\right),\\
f(r)={}&{}\frac{4\left(r^2+z\right)^2h(r)}{\left(r^2+2z\right)^2},\\
\left[\phi'(r)\right]^2={}&{}-\frac{r^2\left(r^2+2z\right)^2}{4z\left(r^2+z\right)^3h(r)},
\end{align}
where a prime stands for derivative with respect to $r$. $M$ is an integration constant. The behaviour as $r\to\infty$ of the metric is
\begin{align}
h(r)={}&{}\frac{3}{4}+\frac{r^2}{12z}-\frac{2M-\pi\sqrt{z}/8}{r}+\mathcal{O}\left(\frac{1}{r^2}\right),\\
f(r)={}&{}\frac{7}{3}+\frac{r^2}{3z}-\frac{8M-\pi\sqrt{z}/2}{r}+\mathcal{O}\left(\frac{1}{r^2}\right).
\end{align}
$M$ is seen to appear as a mass term (term $1/r$), and since it is an integration constant, it would be possible to absorb the contribution of $\pi\sqrt{z}$ by a redefinition of $M$. The behaviour of $h(r)$ and $f(r)$ is very similar to the one of Schwarzschild-adS: as announced, the BH solution allows for an effective cosmological constant term, which can be removed only if $z\to \infty$, but the scalar-tensor action (\ref{eq:actionrinaldi}) makes no sense in this limit.
\\

On the contrary, we will see many examples where, once an asymptotically flat scalar-tensor BH is obtained, its (a)dS generalization is straightforward: one includes the usual cosmological constant term $-2\Lambda$ in the action, and the metric function typically just acquires the usual term\footnote{either exactly or in the $r\to\infty$ limit.} $-\Lambda r^2/3$. We have thus understood the challenge represented by asymptotically flat BHs, that is, the fact that many non-minimal scalar-tensor couplings imply an effective non-zero cosmological constant and thus asymptotic flatness is more difficult to achieve. We now go back to the presentation of no-hair theorems~\cite{Herdeiro:2015waa} and their consequences.
\subsection{Minimally-coupled scalar field}\label{subsec:mincoupl}
First of all, for an action
\begin{equation}
S=\int\mathrm{d}^4x\sqrt{-g}\Bigl[R-\frac{1}{2}\partial^\mu\phi\,\partial_\mu\phi-V(\phi)\Bigr],\label{eq:mincoupl}
\end{equation}
the scalar field is said to be \textbf{minimally-coupled}. In this case, the scalar field stands in fact as a matter field, entering the usual Einstein's equations of GR (\ref{eq:einstein_eq}) only through its energy-momentum tensor. Action (\ref{eq:mincoupl}) should therefore not be considered as a modification of gravity/scalar-tensor theory, as opposed, for instance, to the Brans-Dicke theory\footnote{and of course the Horndeski and DHOST theories.} (\ref{eq:action_bd}) where the scalar field is \textbf{non-minimally coupled} to the metric through the term $\phi R$, leading to a varying Newton's constant. Still, this minimally-coupled case deserves some consideration. \\

It was shown by Chase~\cite{Chase:1970omy}, see also~\cite{PhysRev.174.1578}, that 'every zero-mass scalar field which is gravitationally coupled, static and asymptotically flat, becomes singular at a simply-connected event horizon'. In other words, he proved a no-scalar-hair theorem for $V(\phi)=0$ in action (\ref{eq:mincoupl}). Bekenstein then generalized it~\cite{Bekenstein:1972ny} to the case of massive scalar, $V(\phi)=\mu^2\phi^2/2$.\\

As regards more general potentials $V(\phi)$, Bekenstein again established a no-hair theorem in the case of spherical symmetry~\cite{Bekenstein:1995un}, which can in fact be generalized~\cite{Herdeiro:2015waa} to stationary, axisymmetric asymptotically flat BH spacetime. Under the assumption that the scalar field itself is stationary and axisymmetric, and that the potential is such that\footnote{for instance if $V(\phi)$ is an even power of $\phi$.} $\phi V_\phi\geq 0$, then the scalar field must be identically zero and the BH solutions of action (\ref{eq:mincoupl}) are the ones of GR. 
\\

Many solutions with minimally-coupled scalar field were found by violating the assumption bearing on the potential. They are obtained in spherical symmetry, by 'reverse engineering' the scalar field potential $V(\phi)$: one first imposes the BH solution, then finds the potential supporting it. The first example was given by~\cite{Bechmann:1995sa}, followed by many others~\cite{Dennhardt:1996cz,Bronnikov:2005gm,Anabalon:2012ih,Cadoni:2015gfa,Bakopoulos:2023hkh,Karakasis:2023hni}. Note that some of them assume in fact a so-called phantom scalar field, i.e., the kinetic term in action (\ref{eq:mincoupl}) has rather a wrong sign, $+\partial^\mu\phi\,\partial_\mu\phi/2$, bringing usually about stability problems. 
\subsection{(Generalized) Brans-Dicke scalar field}
We recall the generalized Brans-Dicke action of paragraph~\ref{subsec:common},
\begin{equation}
S=\int\mathrm{d}^4x\sqrt{-g}\left[\phi R-\frac{\omega(\phi)}{\phi}g^{\mu\nu}\partial_\mu\phi\,\partial_\nu\phi-V(\phi)\right],\label{eq:bdbis}
\end{equation}
where the original Brans-Dicke theory corresponds to $\omega(\phi)=\omega=\text{constant}$ and $V(\phi)=0$. Hawking~\cite{Hawking:1972qk} established a no-scalar-hair theorem for the original Brans-Dicke theory, which was more recently generalized by Sotiriou and Faraoni~\cite{Sotiriou:2011dz} for the generalized Brans-Dicke theory. The argument is similar and lies on the use of a conformal transformation and a redefinition of the scalar field:
\begin{equation}
\hat{g}_{\mu\nu}\equiv\phi g_{\mu\nu},\quad \hat{\phi}\equiv \int\sqrt{2\omega(\phi)+3}\frac{\mathrm{d}\phi}{\phi}.\label{eq:confhawking}
\end{equation}
The action is transformed into
\begin{equation}
S=\int\mathrm{d}^4x\sqrt{-\hat{g}}\Bigl[\hat{R}-\frac{1}{2}\hat{g}^{\mu\nu}\partial_\mu\hat{\phi}\,\partial_\nu\hat{\phi}-\hat{V}(\hat{\phi})\Bigr],
\end{equation}
where $\hat{V}(\hat{\phi})=V(\phi)/\phi^2$. One can thus apply the no-hair theorem for a minimally-coupled scalar field, provided the potential satisfies the conditions mentioned above.
\subsection{Evading no-hair theorem in Brans-Dicke: the BBMB black hole}\label{subsec:bbmbbh}
There however exists a loophole in the previous argument: what does happen if the conformal transformation (\ref{eq:confhawking}) becomes singular, i.e. if $\phi$ reaches zero? This enables to understand the possibility for the \textbf{BBMB BH}~\cite{Bocharova:1970skc,Bekenstein:1974sf}. As a reminder, the BBMB/MTZ action (\ref{eq:bbmbmtz}) was constructed as the most general four-dimensional action with local conformal invariance (broken by a mere Einstein-Hilbert term with cosmological constant),
\begin{equation}
S_{\text{BBMB/MTZ}}\left[g_{\mu\nu},\phi\right]\equiv \int\mathrm{d}^4x\sqrt{-g}\Bigl[R-2\Lambda-\beta\mathrm{e}^{2\phi}\left(R+6\left(\partial\phi\right)^2\right)-2\lambda\mathrm{e}^{4\phi}\Bigr].\label{eq:bbmb_bis}
\end{equation}
The BBMB action corresponds to $\Lambda=\lambda=0$, otherwise it is the MTZ action. Because of the generalized conformal invariance, we recall, see (\ref{eq:geombbmb}), that any solution to the BBMB/MTZ action must satisfy a pure geometric equation, $\mathfrak{R}_\text{BBMB/MTZ}=0$, where
 \begin{equation}
\mathfrak{R}_\text{BBMB/MTZ}=g^{\mu\nu}\mathcal{E}_{\mu\nu}+\mathcal{E}_\phi=8\Lambda-2R.
\end{equation}
This geometric equation can be easily solved for a static, spherically-symmetric ansatz of the form
\begin{equation}
\mathrm{d}s^2=-f(r)\mathrm{d}t^2+\mathrm{d}r^2/f(r)+r^2\mathrm{d}\Omega^2,
\end{equation}
giving
\begin{equation}
f(r)=1-\frac{\Lambda r^2}{3}-\frac{2M}{r}+\frac{\mu}{r^2}.\label{eq:riccizerosol}
\end{equation}
There are two integration constants, the mass $M$, and $\mu$. On the other hand, with a static scalar field $\phi=\phi(r)$, the following combination of field equations is very simple,
\begin{equation}
\mathcal{E}^t_t-\mathcal{E}^r_r=4\mathrm{e}^{2\phi}\beta f\left[\left(\phi'\right)^2-\phi''\right],
\end{equation}
giving $\phi=\ln\left[\alpha/(r+\beta)\right]$ with two integration constants $\alpha$ and $\beta$. If $\alpha=0$, $\phi$ vanishes and the metric is Schwarzschild(-(a)dS). If $\alpha\neq 0$, plugging everything into the remaining field equations fixes the values of $\mu$, $\alpha$, $\beta$ and of the coupling constant $\lambda$, giving finally
\begin{equation}
f(r)=\left(1-\frac{M}{r}\right)^2-\frac{\Lambda r^2}{3},\quad \phi = \ln\left[\frac{M}{\sqrt{\beta}(r-M)}\right],\quad \lambda=-\Lambda\beta^2.\label{eq:bbmbsol}
\end{equation}
If $\Lambda=0$, the self-interacting potential $-2\lambda\mathrm{e}^{4\phi}$ is not allowed, and the metric is asymptotically flat, with mass $M$. This BBMB BH is extremal with a double horizon at $r_h=M$, where the scalar field $\phi$ diverges. The MTZ spacetime~\cite{Martinez:2002ru}, with $\Lambda\neq 0$, requires a non-vanishing self-interacting potential. If $\Lambda<0$, the spacetime is a naked singularity. If $\Lambda>0$, the spacetime is asymptotically dS, with a cosmological horizon. There exists an event horizon $r_h$ if and only if $\Lambda<\frac{3}{16M^2}$,
\begin{equation}
r_h = \frac{\sqrt{3}-\sqrt{3-4M\sqrt{3\Lambda}}}{2\sqrt{\Lambda}},
\end{equation}
and $r_h>M$, so the scalar field divergence is hidden behind the horizon in this case. \\

The existence of the MTZ solution is of course not trivial, but, because of its dS behaviour, it does not fall within the range of the no-scalar-hair theorems. From this point of view, the BBMB BH is more puzzling. Indeed, the BBMB action, Eq. (\ref{eq:bbmb_bis}) with $\Lambda=\lambda=0$, can be recast under the generalized Brans-Dicke form (\ref{eq:bdbis}) with the following identification between the scalar fields,
\begin{equation}
\phi_\text{BD}=1-\beta\exp\left(2\phi_\text{BBMB}\right).
\end{equation}
Using the solution (\ref{eq:bbmbsol}) for $\phi_\text{BBMB}$, this yields on shell
\begin{equation}
\phi_\text{BD}=\frac{r(r-2M)}{\left(r-M\right)^2}.
\end{equation}
Thus, $\phi_\text{BD}=0$ at $r=2M$, and the conformal transformation of Hawking (\ref{eq:confhawking}) becomes singular at that point, thus preventing the identification between the BBMB theory and a minimally-coupled scalar field in another metric frame. The example of this loophole in the no-hair argument may be thought to pave the way for other BH solutions in Brans-Dicke theory. However, up to date, \textbf{the BBMB BH remains the only known hairy BH solution in Brans-Dicke theory}. If other solutions exist, they remain to be discovered, more than 50 years after the proposal of the BBMB BH. \\

To cut a long story short, scalar-tensor BHs in \textbf{second-order Lagrangians} are highly constrained by no-scalar-hair theorems. 'Second-order' here means that, for both minimally-coupled and (generalized) Brans-Dicke scalar fields, the sum of the order of derivatives of both the metric and the scalar does not exceed two in any term of the action. As seen above, BHs with minimally-coupled scalar exist, but these theories are not modified theories of gravity, the scalar is just a matter field in the framework of GR. Regarding Brans-Dicke, it allows up to date for a unique asymptotically flat BH, namely the BBMB BH, which itself is plagued by a divergence of the scalar field at the extremal horizon of the BH, see Eq. (\ref{eq:bbmbsol}). 
\\

This justifies the relevance of \textbf{higher-order Lagrangians} when it comes to constructing hairy BHs. 'Higher-order' here refers to the structure of the Horndeski, beyond Horndeski and DHOST Lagrangians constructed in Chap.~\ref{chap:st}: they display terms where the sum of derivatives is strictly higher than two, for instance $G_4(\phi,X)R$. As regards such higher-order Lagrangians, there exists one most famous no-hair theorem.
\subsection{No-hair theorem for shift-symmetric scalar-tensor theories}\label{subsec:huinic}
This theorem was established by Hui and Nicolis~\cite{Hui:2012qt}. A clear and precise discussion of its assumptions can be found e.g. in~\cite{Lehebel:2018zga}, which we briefly follow. For definiteness, we formulate it in the case of a Horndeski theory, but it can be readily extended to any DHOST theory. Consider thus a Horndeski theory parameterized by the functions $G_{2,3,4,5}$. The assumptions are the following:
\begin{enumerate}
\item The theory has shift symmetry (see paragraph~\ref{subsec:shift}): $G_{2,3,4,5}$ are all functions of the kinetic term $X$ only. This implies the existence of the Noether current $\mathcal{J}^\mu_\text{shift}$, Eq. (\ref{eq:shift_current}),
\item Spacetime is spherically symmetric, static, and asymptotically flat,
\item The scalar field respects these symmetries: $\phi=\phi(r)$ with the usual radial coordinate $r$, and the gradient of $\phi$ vanishes at spatial infinity,
\item The norm of the current, $\mathcal{J}_\text{shift}^2=\mathcal{J}^\mu_\text{shift} \mathcal{J}_{\text{shift}\,\mu}$, is finite on the horizon,
\item The action contains a canonical kinetic term $X\subset G_2$,
\item All Horndeski functions $G_i(X)$ are analytic at $X=0$.
\end{enumerate}
Under these assumptions, the scalar field must be trivial (i.e. equal to any constant value, because of shift symmetry) and the BH solutions are the ones of GR. The theorem is quite easy to establish and we refer the interested reader to~\cite{Hui:2012qt,Lehebel:2018zga}: more interesting for our purposes are the consequences of this theorem. The respective violations of these assumptions are:
\begin{enumerate}
\item By going beyond shift symmetry and considering a generic theory $G_{2,3,4,5}(\phi,X)$, there exists no no-scalar-hair theorem.
\item The assumption of spherically-symmetric and static spacetime is in general maintained, since going beyond this case is very intricate. Also, the hypothesis of asymptotic flatness is kept: we are mostly interested in asymptotically flat solutions, as explained in paragraph~\ref{subsec:challenge}, and evading a no-hair theorem regarding asymptotically flat solutions by finding asymptotically (a)dS solutions would not be relevant.
\item The assumption $\phi=\phi(r)$ can be relaxed. As explained in paragraph~\ref{subsec:shift}, see Eq. (\ref{eq:scalar_ansatz_1}), the ansatz $\phi(t,r)=qt+\psi(r)$ with $q$ a constant is compatible with the symmetries of the spacetime. Indeed, shift symmetry means that only the derivatives of the scalar field enter in the action and in the field equations, so the linear time dependence is compatible with the static assumption.
\item The norm of the current is in general kept finite.
\item The assumption of canonical kinetic term is attracting, since its absence may bring about perturbative problems or strong coupling issues. Removing the canonical kinetic term therefore appears as double-edged: it may enable to evade the no-hair theorem, but the obtained BHs can suffer from pathologies.
\item Violating the assumption of analyticity of the Horndeski functions is an easy way of obtaining BH solutions. For instance, a coupling of the scalar field with the GB invariant yields logarithmic Horndeski functions, see Eq. (\ref{eq:gbhorn}). Consequently, there has existed for a long time many solutions in theories involving GB couplings. For the recent 4DEGB theory that we presented in Sec.~\ref{sec:fourd}, these are closed-form BH solutions, and we will detail them in paragraph~\ref{subsec:bhegb}. Nevertheless, no-scalar-hair theorems make of course no distinction between closed-form solutions and other, since what matters is the existence or not of a BH metric, not the fact that we are able to write it down with usual mathematical functions. \\

Therefore, although this thesis focuses on closed-form solutions, it is worth mentioning the other solutions, either numerical or requiring perturbative expansions, involving GB couplings. Indeed, they prove that relaxing the analyticity assumption of the no-hair theorem opens up many possibilities regarding BH solutions in scalar-tensor theories. The historical example (1995) is the one by Kanti, Mavromatos, Rizos, Tamvakis and Winstanley~\cite{Kanti:1995vq}. Other examples include static solutions~\cite{Sotiriou:2013qea,Sotiriou:2014pfa,Antoniou:2017hxj,Antoniou:2017acq,Saravani:2019xwx,Bakopoulos:2020dfg,Babichev:2022djd} and spinning solutions~\cite{Kleihaus:2011tg,Ayzenberg:2014aka,Maselli:2015tta,Kleihaus:2015aje,Collodel:2019kkx,Delgado:2020rev}. GB couplings are also famous for allowing \textbf{BH scalarization}, i.e. the existence of GR solutions which, under certain conditions, become unstable and acquire a non-trivial scalar field dressing a non-GR metric. See~\cite{Doneva:2017bvd,Silva:2017uqg,Antoniou:2021zoy} for static examples and~\cite{Cunha:2019dwb,Dima:2020yac,Herdeiro:2020wei,Berti:2020kgk} for rotating ones. For recent reviews on scalarization, see~\cite{Antoniou:2023gwd,Doneva:2022ewd}.
\end{enumerate}
\begin{table}
\begin{center}
\begin{tabular}{| m{2.4cm} || m{2.5cm} | m{3.4cm} | m{2.6cm} | m{2.8cm} |} 
 \hline
$\phi=\cdots$ & $G_{2,4}\left(X\right)$ (shift + parity sym. Horndeski) & \raggedright\arraybackslash $G_{2,3,4,5}\left(\phi,X\right)$ (Generic Horndeski) & $G_{2,4}\left(X\right)$ and $F_4\left(X\right)$ (shift + parity sym. beyond Horndeski) & \begin{center}DHOST\end{center}\\ [0.5ex] 
 \hline\hline
 $\phi\left(r\right)$ & BCL~\cite{Babichev:2017guv} & 4DEGB~\cite{Lu:2020iav,Fernandes:2021dsb} & Other BCL~\cite{Babichev:2017guv} & Stealth Schwarzschild \cite{Motohashi:2019sen};\newline \raggedright\arraybackslash Regular BHs (Kerr-Schild construction)~\cite{Babichev:2020qpr,Baake:2021jzv} \\ 
 \hline
 $qt+\psi\left(r\right)$ and $X=\text{cst}$ & (Historical) stealth Schwarzschild \cite{Babichev:2013cya}; \newline Other stealth Schwarzschild \cite{Kobayashi:2014eva} &  Stealth Schwarzschild~\cite{Minamitsuji:2018vuw} & 
  Stealth Schwarzschild \cite{Babichev:2016kdt}  &  Stealth Schwarzschild \cite{Motohashi:2019sen,Minamitsuji:2019shy,BenAchour:2018dap} and Kerr~\cite{Charmousis:2019vnf}; Disformed Kerr~\cite{Anson:2020trg,BenAchour:2020fgy}  \\
 \hline
 $qt+\psi\left(r\right)$ and $X\neq \text{cst}$ &  \centering $\emptyset$ & Shift-symmetric 4DEGB~\cite{Hennigar:2020lsl} & \centering $\emptyset$ &  \begin{center} $\emptyset$  \end{center} \\ [1ex] 
 \hline
\end{tabular}
\caption{State-of-the-art of scalar-tensor BHs before this thesis, i.e. in Fall 2021.\label{table:1}}
\end{center}
\end{table}
This presentation of no-scalar-hair theorems is now complete, and enables to move on to the core subject of this Chap.~\ref{sec:state}, namely existing closed-form BH solutions in Horndeski, beyond Horndeski and DHOST theories. Table~\ref{table:1} summarizes this state-of-the-art that we are going to detail. \textbf{We limit our interest to BHs which display satisfactory asymptotic behaviour, i.e. coincide with Schwarzschild/Kerr at leading order far away from the core of the BH, or with their (a)dS generalizations. We also consider only BHs for which the scalar field is not divergent outside and at the horizon}. Beyond the brief review presented here, we refer the reader to~\cite{Babichev:2016rlq} and to our recent review~\cite{Babichev:2023psy}.
\section[General Relativity metric: stealth black holes, and their disformal transformations]{General Relativity metric: stealth black holes, and their disformal transformations}\label{sec:stealth}
\sectionmark{3.2~~GR metric: stealth black holes, and their disformal transformations}
When looking for BHs in scalar-tensor gravity, it may of course be useful to keep in mind the situation in non-modified gravity, that is, in GR. The unique~\cite{Bekenstein:1971,Teitelboim:1972,Mazur:2000pn,Chrusciel:2012jk} stationary, asymptotically flat BH of GR with Maxwell field is the Kerr-Newman BH~\cite{Newman:1965my}, which, in absence of electric charge, reduces to the Kerr BH~\cite{Kerr:1963ud},
\begin{align}
\mathrm{d}s^2={}&{}-\left(1-\frac{2Mr}{\Sigma}\right)\mathrm{d}t^2+\frac{\Sigma}{\Delta}\mathrm{d}r^2+\Sigma\mathrm{d}\theta^2-\frac{4Mar\sin^2\theta}{\Sigma}\mathrm{d}t\mathrm{d}\varphi+\frac{\sin^2\theta}{\Sigma}\Upsilon.\label{eq:kerr}
\end{align}
The Kerr BH itself reduces to the Schwarzschild BH~\cite{Schwarzschild:1916uq} if there is no rotation ($a=0$),
\begin{equation}
\mathrm{d}s^2=-\left(1-\frac{2M}{r}\right)\mathrm{d}t^2+\frac{\mathrm{d}r^2}{\left(1-\frac{2M}{r}\right)}+r^2\mathrm{d}\Omega^2.
\end{equation}
$M$ is the mass of the BH, $a$ the angular momentum per unit mass, and
\begin{equation}
\Sigma = r^2+a^2\cos^2\theta,\quad \Delta = r^2+a^2-2Mr,\quad \Upsilon=\left(r^2+a^2\right)^2-a^2\Delta\sin^2\theta.\label{eq:delta_def}
\end{equation}
If one includes a cosmological constant $\Lambda$, the Kerr metric is generalized to the Kerr-(a)dS BH~\cite{CARTER1968399}, written in the following compact form,
\begin{align}
\mathrm{d}s^2=-\frac{\Delta_r}{\Xi^2\Sigma}\left[\mathrm{d}t-a\sin^2\theta\,\mathrm{d}\varphi\right]^2+{}&{}\Sigma\left(\frac{\mathrm{d}r^2}{\Delta_r}+\frac{\mathrm{d}\theta^2}{\Delta_\theta}\right)\nonumber\\{}&{}+\frac{\Delta_\theta\sin^2\theta}{\Xi^2\Sigma}\left[a\mathrm{d}t-\left(r^2+a^2\right)\mathrm{d}\varphi\right]^2,\label{eq:kerr-ds}
\end{align}
with
\begin{equation}
\Delta_r=\left(1-\frac{\Lambda r^2}{3}\right)\left(r^2+a^2\right)-2Mr,\quad \Xi=1+\frac{\Lambda a^2}{3},\quad \Delta_\theta = 1+\frac{\Lambda a^2}{3}\cos^2\theta,
\end{equation}
or the Schwarzschild-(a)dS BH,
\begin{equation}
\mathrm{d}s^2=-\left(1-\frac{2M}{r}-\frac{\Lambda r^2}{3}\right)\mathrm{d}t^2+\frac{\mathrm{d}r^2}{\left(1-\frac{2M}{r}-\frac{\Lambda r^2}{3}\right)}+r^2\mathrm{d}\Omega^2.\label{eq:sdsfirst}
\end{equation}
A possible starting point in the quest for scalar-tensor BHs is to assume that the metric is like in GR, thus letting only the scalar field unknown. Of course, the metric field equations $\mathcal{E}_{\mu\nu}=0$ can always be rewritten as
\begin{equation}
G_{\mu\nu}+\Lambda g_{\mu\nu}=T^\phi_{\mu\nu},
\end{equation}
by isolating the Einstein tensor on the left, and considering the right hand side as an effective energy-momentum tensor of the scalar field, $T^\phi_{\mu\nu}$, by analogy with the field equations of GR in the presence of matter. Here, $\Lambda$ is the usual cosmological constant included in the action with a $-2\Lambda$ term. If the metric is a vacuum BH of GR, then it satisfies $G_{\mu\nu}+\Lambda g_{\mu\nu}=0$, so $T^\phi_{\mu\nu}=0$. If $\phi$ is trivial (typically, $\phi=0$, but also $\phi=\text{constant}$ if the theory is shift-symmetric), then this BH is just a GR BH. If $\phi$ is not trivial, it is said to be \textbf{stealth} (since its energy-momentum tensor vanishes), and the BH is a hairy BH since it is dressed with a non-trivial field. Such a hairy BH is itself called a \textbf{stealth BH}: the metric is identical to GR, but supported by a non-trivial scalar field.
\\

In fact, we will also say that the BH is stealth if the energy-momentum tensor of $\phi$ is proportional to the metric, $T^\phi_{\mu\nu}=-\Lambda_\phi g_{\mu\nu}$ with $\Lambda_\phi$ a constant, thus supporting a GR BH with an effective cosmological constant,
\begin{equation}
\Lambda_{\text{eff}}\equiv\Lambda+\Lambda_\phi.
\end{equation}
This phenomenon, where the total cosmological constant of spacetime includes a contribution from the scalar field, is known as \textbf{self-tuning}~\cite{Charmousis:2015aya,Babichev:2016kdt,Babichev:2017lmw}. In the following, $\Lambda_{\text{eff}}$ refers to the cosmological constant which appears in the metric, while $\Lambda$ is the 'bare' cosmological constant included in the action. We sometimes drop the suffix (a)dS for brevity.
\subsection{Stealth Schwarzschild black holes}\label{subsec:stealthsch}
The first stealth (Schwarzschild) BH was constructed by Babichev and Charmousis~\cite{Babichev:2013cya}. In fact, they had not considered from a start a Schwarzschild-dS metric, but rather a static, spherically-symmetric ansatz,
\begin{equation}
\mathrm{d}s^2=-h\left(r\right)\mathrm{d}t^2+\mathrm{d}r^2/f\left(r\right)+r^2\mathrm{d}\Omega^2,\label{eq:ansatz_stealth_metric}
\end{equation}
in the context of a particular shift-symmetric theory,
\begin{equation}
S = \int\mathrm{d}^4x\sqrt{-g}\left(R-2\Lambda-\eta\left(\partial\phi\right)^2+\beta G^{\mu\nu}\partial_\mu\phi\,\partial_\nu\phi\right),\label{eq:action_bc}
\end{equation}
which belongs to the Horndeski class with $G_2=2\eta X-2\Lambda$ and $G_4=1+\beta X$, $\eta$ and $\beta$ being coupling constants. They chose to allow for a time-dependent scalar field, finding that the form
\begin{equation}
\phi=qt+\psi\left(r\right)\label{eq:ansatz_stealth_scalar}
\end{equation}
was consistent with the symmetries of the metric because of the shift symmetry of the action, as explained in paragraph~\ref{subsec:shift}. This amounts to breaking assumption 3 of the Hui-Nicolis no-hair theorem of paragraph~\ref{subsec:huinic}. Only by inspection of the field equations did they find that this setup could lead to a stealth Schwarzschild solution, provided the scalar field kinetic term was a constant, 
\begin{equation}
X=X_0.
\end{equation}
This finding paved the way towards many examples of stealth Schwarzschild solutions in other theories than (\ref{eq:action_bc}), by this time assuming from a start a constant kinetic term. The situation was thus studied in shift-symmetric Horndeski~\cite{Kobayashi:2014eva}, beyond Horndeski~\cite{Babichev:2016kdt}, quadratic DHOST~\cite{Motohashi:2019sen} and cubic DHOST~\cite{Minamitsuji:2019shy} theories, and in theories without shitft symmetry, both Horndeski~\cite{Minamitsuji:2018vuw} and quadratic DHOST~\cite{BenAchour:2018dap}. Let us thus take this point of view and show, for instance, the ubiquity of stealth Schwarzschild solutions in the quadratic DHOST Ia theory with shift and parity symmetry,
\begin{align}
S=\int\mathrm{d}^4x\sqrt{-g}\Bigl\{F\left(X\right)R+{}&{}P\left(X\right)+A_2\left(X\right)\left[\left(\Box\phi\right)^2-\phi_{\mu\nu}\phi^{\mu\nu}\right]+A_3\left(X\right)\Box\phi\phi^\mu\phi_{\mu\nu}\phi^\nu\nonumber \\ {}&{} +A_4\left(X\right)\phi^\mu\phi_{\mu\nu}\phi^{\nu\rho}\phi_\rho+A_5\left(X\right)\left(\phi^\mu\phi_{\mu\nu}\phi^\nu\right)^2\Bigr\},\label{eq:dhost_shift_par}
\end{align}
where $F\left(X\right)$, $P\left(X\right)$, $A_2\left(X\right)$ and $A_3\left(X\right)$ are free, while $A_4\left(X\right)$ and $A_5\left(X\right)$ are determined by the degeneracy conditions (\ref{eq:a4dhost}-\ref{eq:a5dhost}). With the usual notations of DHOST, $X=\phi^\mu\,\phi_\mu$. Assume the ansatz (\ref{eq:ansatz_stealth_metric}) for the metric, with
\begin{equation}
f=h=1-2M/r-\Lambda_{\text{eff}}\, r^2/3
\end{equation}
the Schwarzschild(-(a)dS) function, and (\ref{eq:ansatz_stealth_scalar}) for the scalar, with $X=X_0=\text{const}$. The assumption that the spacetime is stealth, with total cosmological constant $\Lambda_{\text{eff}}$, gives
\begin{equation}
G_{\mu\nu}=-\Lambda_{\text{eff}}\,g_{\mu\nu},\quad R=4\Lambda_{\text{eff}},\quad R_{\mu\nu} = \Lambda_{\text{eff}}\,g_{\mu\nu}.
\end{equation}
Taking also into account $X=X_0$,
the metric field equations $\mathcal{E}_{\mu\nu}$ of appendix~\ref{sec:eq_dhost} greatly simplify,
\begin{align}
\mathcal{E}_{\mu\nu} ={}&{} \left\lbrace A_2(X_0)\mathfrak{P}-P(X_0)-2\Lambda_{\text{eff}}\Bigl[F+XA_2\Bigr]_{X=X_0}\right\rbrace g_{\mu\nu}\nonumber\\{}&{}+\left\lbrace 2P_X(X_0)+\mathfrak{P}\Bigl[2A_{2X}-A_3\Bigr]_{X=X_0}+\Lambda_{\text{eff}}\Bigl[8F_X+4A_2+XA_3\Bigr]_{X=X_0}\right\rbrace\phi_\mu\phi_\nu\nonumber\\{}&{} +2A_2(X_0)\Bigl[R_{\mu\rho\nu\sigma}\phi^\rho\phi^\sigma+\phi_{\mu\rho}\phi^{\rho}_\nu-\Box\phi\phi_{\mu\nu}\Bigr],\label{eq:metric_eq_stealth}
\end{align}
where for brevity, the following notation is introduced,
\begin{equation}
\mathfrak{P}\equiv\left(\Box\phi\right)^2-\phi_{\rho\sigma}\phi^{\rho\sigma}.\label{eq:frakE}
\end{equation}
In particular, the $A_4$ and $A_5$ Lagrangians do not contribute (this is true for any spacetime, as long as $X$ is constant). Since $X$ is constant, then the scalar field must have a true linear time dependence, i.e. $q\neq 0$, otherwise $\phi$ would be constant, that is to say trivial for a shift-symmetric theory. Remember the important result (\ref{eq:super_important}): for a static, spherically-symmetric spacetime where the scalar field has a linear time dependence, the radial component of the shift-symmetric current $\mathcal{J}_{\text{shift}}^\mu$ must vanish:
\begin{equation}
\mathcal{J}^r_{\text{shift}}=0.
\end{equation}
This current, see appendix~\ref{sec:eq_dhost}, acquires the following simplified form when $X=X_0$ and $G_{\mu\nu}=-\Lambda_{\text{eff}}\,g_{\mu\nu}$,
\begin{align}
\mathcal{J}^\mu_{\text{shift}}=\Bigl\{ 2P_X(X_0)+\mathfrak{P}{}&{}\Bigl[2A_{2X}-A_3\Bigr]_{X=X_0}\nonumber\\{}&{}+\Lambda_{\text{eff}}\Bigl[8F_X+2A_2+XA_3\Bigr]_{X=X_0}\Bigr\}\phi^\mu.\label{eq:current_stealth}
\end{align}
$\phi'$ cannot vanish identically, otherwise the scalar field would be proportional to the time coordinate $t$, which diverges to infinity at the horizon. Thus, $\mathcal{J}^r_{\text{shift}}=0$ implies that the expression between braces vanishes, so $\mathcal{J}^\mu_{\text{shift}}$ as a whole vanishes. Now, the expression between braces only displays constant terms but one, the term $\mathfrak{P}$ of Eq. (\ref{eq:frakE}), which is multiplied by $(2A_{2X}-A_3)_{X=X_0}$. Thus, either one imposes this latter combination of theory functions to vanish at $X=X_0$, or one can let the theory functions free, in which case the scalar field must be such that $\mathfrak{P}$ be constant. Let us proceed further in this direction. First, one can solve the equation $X=X_0$,
\begin{equation}
X_0=X=\partial_\mu\phi\,\partial^\mu\phi = f\left(\psi'\right)^2-\frac{q^2}{f},
\end{equation}
where $f=1-2M/r-\Lambda_{\text{eff}} \,r^2/3$. This gives 
\begin{equation}
\psi\left(r\right)=\pm \int\frac{\sqrt{q^2+X_0 f(r)}}{f(r)}\mathrm{d}r,
\end{equation}
which leads to
\begin{equation}
\mathfrak{P}=\frac{2}{r^2}\left(q^2+X_0\right)-2\Lambda_{\text{eff}}X_0.
\end{equation}
Thus $\mathfrak{P}$ is constant, equal to $2q^2\Lambda_{\text{eff}}$, if and only if
\begin{equation}
X=X_0=-q^2.
\end{equation}
The vanishing of $\mathcal{J}^\mu_{\text{shift}}$, Eq. (\ref{eq:current_stealth}), is then equivalent to
\begin{equation}
0=\Biggl\{\Lambda_{\text{eff}}\Bigl[q^2\left(4A_{2X}-3A_3\right)+8F_X+2A_2\Bigr]+2P_X\Biggr\}_{X=X_0=-q^2}.\label{eq:stealthcond1}
\end{equation}
Using all this, one finally finds that the metric field equations (\ref{eq:metric_eq_stealth}) vanish if and only if
\begin{equation}
0=\Bigl[P+2\Lambda_{\text{eff}}\left(F-q^2A_2\right)\Bigr]_{X=X_0=-q^2}.\label{eq:stealthcond2}
\end{equation}
To summarize, the quadratic DHOST theory with shift and parity symmetry (\ref{eq:dhost_shift_par}) accomodates a stealth Schwarzschild(-(a)dS) solution with cosmological constant $\Lambda_{\text{eff}}$,
\begin{align}
\mathrm{d}s^2={}&{}-f\mathrm{d}t^2+\frac{\mathrm{d}r^2}{f}+r^2\mathrm{d}\Omega^2,\quad f=1-\frac{2M}{r}-\frac{\Lambda_{\text{eff}} r^2}{3},\nonumber\\ \phi ={}&{} q\left(t\pm\int\frac{\sqrt{1-f}}{f}\mathrm{d}r\right),\label{eq:stealth_final}
\end{align}
provided both conditions (\ref{eq:stealthcond1}-\ref{eq:stealthcond2}) are satisfied. Note that the situation is quite distinct depending on whether or not $\Lambda_{\text{eff}}$ vanishes. If it vanishes, then the scalar field kinetic or k-essence term $P\left(X\right)$ has a double root at $X=X_0=-q^2$, and the value of $q^2$ is unfixed: $q$ is a free integration constant of the solution. If $\Lambda_{\text{eff}}\neq 0$, then, given a set of theory functions $F$, $P$, $A_2$, $A_3$, conditions (\ref{eq:stealthcond1}-\ref{eq:stealthcond2}) constitute a system of two equations for two unknowns $\Lambda_{\text{eff}}$ and $q$, so, in general, the values of both the total cosmological constant $\Lambda_{\text{eff}}$ and the scalar field kinetic term $q^2$ are completely and uniquely fixed by the theory, namely:
\begin{align}
\Lambda_\text{eff} ={}&{} \left[\frac{3PA_3-4\left(PA_2\right)_X}{8FA_{2X}+16F_XA_2+4A_2^2-6FA_3}\right]_{X=X_0=-q^2},\\  q^2={}&{}\left[\frac{4FP_X-8F_XP-2PA_2}{4\left(PA_2\right)_X-3PA_3}\right]_{X=X_0=-q^2}.
\end{align}
Consider for instance the case of the action mentioned at the beginning of this study~\cite{Babichev:2013cya},
\begin{equation}
S = \int\mathrm{d}^4x\sqrt{-g}\left(R-2\Lambda-\eta\left(\partial\phi\right)^2+\beta G^{\mu\nu}\partial_\mu\phi\,\partial_\nu\phi\right).
\end{equation}
It corresponds in the DHOST vocabulary to $F=1-\beta X/2$, $P=-\eta X-2\Lambda$, $A_2=\beta$ and $A_3=0$. Using the previous equations in this situation, one immediately finds the expressions for $\Lambda_\text{eff}$ and for $q^2$, and in particular that if $\Lambda_\text{eff}=0$, then $\eta=0$ and the scalar field kinetic term $-\eta\left(\partial\phi\right)^2$ disappears from the action:
\begin{equation}
\Lambda_{\text{eff}} = \frac{-\eta}{\beta},\quad \eta q^2=\Lambda-\Lambda_{\text{eff}}.
\end{equation}
There therefore appears to be an important difference between asymptotically flat and asymptotically (a)dS cases. In the asymptotically flat case, the double root of the scalar field k-essence term $P(X)$ at $X_0$ brings about perturbative problems, namely non-hyperbolic equations for the scalar field perturbations~\cite{Babichev:2018uiw}. In the (a)dS case, this problem is not encountered thanks to the presence of a kinetic term for the scalar field perturbations~\cite{Babichev:2017lmw}. As regards the stealth Schwarzschild BH, problems arising from the study of linear perturbations and QNMs are also highlighted in~\cite{Langlois:2021aji,Roussille:2022vfa}.\\

Importantly, for a Schwarzschild-adS metric, $\Lambda_{\text{eff}}<0$, the scalar field (\ref{eq:stealth_final}) is seen to become imaginary when $r$ is sufficiently large, so such solutions are not admissible: only $\Lambda_{\text{eff}}\geq 0$ is possible. Regarding the apparent divergence of the radial part of $\phi$ at the horizon $f\left(r_h\right)=0$, one must not forget that the time coordinate $t$ also diverges at the horizon. Therefore, \textbf{one must change to horizon-crossing coordinates}~\cite{Carroll:2004st} $\left(u,r,\theta,\varphi\right)$ or $\left(v,r,\theta,\varphi\right)$, where $u$ and $v$ are respectively the advanced and retarded null times,
\begin{equation}
u\equiv t-r^\star,\quad v\equiv t+r^\star,\quad r^\star\equiv\int\frac{\mathrm{d}r}{f\left(r\right)}\quad\text{(tortoise coordinate).}
\end{equation}
The solution is regular at the future event horizon and past cosmological horizon (where $v$ is finite) if the $+$ sign is chosen in the scalar field (\ref{eq:stealth_final}); and at the past event horizon and future cosmological horizon (where $u$ is finite) if the $-$ sign is chosen. Indeed, one then has respectively
\begin{align}
\mathrm{d}s^2={}&{}-f\mathrm{d}v^2+2\mathrm{d}v\mathrm{d}r+r^2\mathrm{d}\Omega^2,\quad \phi = q\left(v+\int\frac{\sqrt{1-f}-1}{f}\mathrm{d}r\right),\\ \mathrm{d}s^2={}&{}-f\mathrm{d}u^2-2\mathrm{d}u\mathrm{d}r+r^2\mathrm{d}\Omega^2,\quad \phi = q\left(u-\int\frac{\sqrt{1-f}-1}{f}\mathrm{d}r\right),
\end{align}
and the scalar field is well-behaved at the corresponding horizon. However, $v$ diverges at the past event or future cosmological horizons, and vice-versa for $u$. \textbf{In a word:\begin{enumerate}
\item For $\Lambda_{\text{eff}}> 0$, the scalar field cannot be made regular at both (event and cosmological) future horizons or both (event and cosmological) past horizons. 
\item Moreover, and this also occurs for $\Lambda_{\text{eff}}=0$, the scalar field cannot be made regular at both future and past event horizons.
\end{enumerate}}
In particular, there is no maximally-extended solution for which the scalar field is finite everywhere. The next paragraph presents the inclusion of rotation. The additional intricacy brought about by rotation compels one to fix more theory functions than in the spherically-symmetric case. This may lead on the one hand to certain issues that we discuss below, but on the other hand, this enables to cure the first problem mentioned above, namely to get regularity at, e.g., both future horizons. However, the second point remains unsolved.
\subsection{Stealth Kerr black holes}\label{subsec:stealthkerr}
The generalization of the previous construction to a rotating BH was first performed by Charmousis, Crisostomi, Gregory and Stergioulas~\cite{Charmousis:2019vnf}. Due to the recent GW event GW170817~\cite{LIGOScientific:2017ync,LIGOScientific:2017vwq}, the considered quadratic DHOST action (\ref{eq:dhost_shift_par}) was limited to its pieces ensuring a speed of GWs equal to the speed of light, that is to say
\begin{equation}
A_2\equiv 0,
\end{equation}
see Eq. (\ref{eq:dhostct1}). We shall also make this assumption to keep the discussion more simple, although, given the lines of the proof, the reader will see that one could easily include $A_2$. This time, the metric is assumed to have a Kerr-((a)dS) form (\ref{eq:kerr-ds}) with total cosmological constant $\Lambda_{\text{eff}}$, and for the moment, the only assumption made on the scalar field $\phi=\phi\left(t,r,\theta,\varphi\right)$ is that its kinetic term is constant,
\begin{equation}
X=X_0,
\end{equation}
by taking inspiration from what works in the Schwarzschild case. The previous simplification of field equations, Eq. (\ref{eq:metric_eq_stealth}), only used the fact that $X=X_0$ and $G_{\mu\nu}=-\Lambda_{\text{eff}}\,g_{\mu\nu}$. This equation is thus still valid, with less terms because now $A_2=0$,
\begin{align}
\mathcal{E}_{\mu\nu} ={}&{}\phi_\mu\phi_\nu\left[ 2P_X+\Lambda_{\text{eff}}\left(8F_X+XA_3\right)-A_3\mathfrak{P}\right]_{X=X_0} - g_{\mu\nu}\left(P+2\Lambda_{\text{eff}}F\right)_{X=X_0},
\end{align}
where $\mathfrak{P}$ is still defined by (\ref{eq:frakE}). With the simple symmetries of Schwarzschild, $\mathfrak{P}$ could be made a constant, and this enabled to proceed without fixing the theory functions. This is no more the case with Kerr. Consequently, the factor multiplying $\mathfrak{P}$ must be forced to vanish, by imposing $A_3(X_0)=0$. Then, the factors of $\phi_\mu\phi_\nu$ and of $g_{\mu\nu}$ must vanish separately, giving
\begin{equation}
2P_X(X_0)+8\Lambda_{\text{eff}}F_X(X_0)=0,\quad P(X_0)+2\Lambda_{\text{eff}}F(X_0)=0,\quad A_3(X_0)=0,\label{eq:kerr_cond}
\end{equation}
where we repeated the condition on $A_3$. \\

In a word, the DHOST Ia action (\ref{eq:dhost_shift_par}) with $A_2\equiv 0$ and $A_4$, $A_5$ given by the degeneracy conditions (\ref{eq:a4dhost}-\ref{eq:a5dhost}), admits a stealth Kerr-((a)dS) solution with cosmological constant $\Lambda_{\text{eff}}$, provided the theory functions $F$, $P$ and $A_3$ satisfy (\ref{eq:kerr_cond}), and provided the scalar field $\phi$ is such that $X=X_0=\text{constant}$: as soon as this latter requirement is dealt with, the proof of existence of Kerr solution will be complete. By analogy with the Schwarzschild case, let us note $X_0=-q^2$, 
\begin{equation}
-q^2=X_0=g^{\mu\nu}\partial_\mu\phi\,\partial_\nu\phi\equiv g^{\mu\nu}p_\mu p_\nu,
\end{equation}
where we introduced in the last line the one-form $p_\mu\mathrm{d}x^\mu\equiv \mathrm{d}\phi=\partial_\mu\phi\,\mathrm{d}x^\mu$, which is seen to correspond to a vector field $p^\mu$ with constant four-norm $-q^2$. In other words, \textbf{$p^\mu$ is the tangent vector to the geodesic of a test particle with mass $q$}, and the scalar field $\phi$ is called a \textbf{Hamilton-Jacobi function for the geodesic trajectory}, which means $\mathrm{d}\phi=p_\mu\mathrm{d}x^\mu$. Fortunately, the integration of Kerr-((a)dS) geodesics was carried out by Carter in 1968~\cite{Carter:1968rr}. Using his results, one gets the following form for the Hamilton-Jacobi scalar field,
\begin{equation}
\phi = -Et + L_z\varphi + S_r(r) +S_\theta(\theta),
\end{equation}
where $E$ and $L_z$ are the conserved energy and angular momentum (along the Kerr axis of rotation) for the corresponding geodesic trajectory, while the precise form of the functions $S_r(r)$ and $S_\theta(\theta)$ can be found in~\cite{Charmousis:2019vnf}. Rather than entering technical details, let us give the main conclusions of~\cite{Charmousis:2019vnf}:
\begin{enumerate}
\item \textbf{A stealth Kerr-adS solution ($\Lambda_{\text{eff}}<0$) is not possible} (at least with this construction).
\item \textbf{Stealth Kerr-dS solutions ($\Lambda_{\text{eff}}>0$) exist and can be made regular at both (event and cosmological) future horizons, or both past horizons}. When the rotation parameter $a$ is taken to zero, this leads to a stealth Schwarzschild-dS solution for which the scalar field has thus more regularity than for the one presented in the previous paragraph~\ref{subsec:stealthsch}. The necessary trade-off to this gain in regularity lies in conditions (\ref{eq:kerr_cond}): the presently considered construction requires to fix the theory functions, which was not the case for the former construction.
\item \textbf{Stealth Kerr solutions ($\Lambda_{\text{eff}}=0$) exist and the associated Hamilton-Jacobi scalar field has no $\theta$-dependence, $S_\theta(\theta)=0$. In the Boyer-Lindquist coordinates of the Kerr metric (\ref{eq:kerr}), it reads
\begin{equation}
\phi = q\left(t\pm\int\frac{\sqrt{2Mr\left(r^2+a^2\right)}}{\Delta}\mathrm{d}r\right).\label{eq:hj}
\end{equation}
This reduces to the Schwarzschild scalar field (\ref{eq:stealth_final}) when $a=0$. Just as for Schwarzschild, by changing to horizon-crossing coordinates, one can check the regularity at the future event horizon (for $+$ sign) or at the past event horizon (for $-$ sign), but regularity at both future and past event horizons is not achieved}.
\end{enumerate}
Around this first stealth Kerr solution, other interesting works include Ref.~\cite{Takahashi:2020hso}, which undertook a systematic study of the most general quadratic DHOST theory in order to determine conditions for it to admit the stealth Kerr solution\footnote{and, more generally, the Kerr-Newman-dS solution.}. The linear metric perturbations of the stealth Kerr solution were studied in~\cite{Charmousis:2019fre} and they were found to have a modified Teukolsky form.\\

Regarding the stability of stealth solutions in DHOST theories, odd-parity perturbations around static stealth solutions were discussed in~\cite{Takahashi:2019oxz}. On the other hand, Refs.~\cite{deRham:2019gha} and~\cite{Takahashi:2021bml} extended the analysis to even-parity perturbations, stating that they were strongly coupled. In fact, this strong coupling problem also exists for the stealth Kerr solution~\cite{deRham:2019gha}, and is most probably due to the fact that the scalar field kinetic term is absent for this solution. This entails the non-hyperbolic character of the scalar perturbation equation. Discussions on the significance of strong coupling in stealth solutions, and ways to circumvent the problem, have been recently presented in~\cite{DeFelice:2022xvq}.
\\

With all these caveats in mind, the stealth Kerr solution remains an interesting illustration of the possibility to go beyond spherical symmetry in scalar-tensor theories. It may be taken as a starting point to construct other solutions. Indeed, remember paragraph~\ref{subsec:intconfdisf}, where we explained the generation of solutions. Starting from a seed solution of a seed scalar-tensor theory, one can perform a conformal-disformal transformation,
\begin{equation}
g_{\mu\nu}\to\tilde{g}_{\mu\nu}=C(\phi,X)g_{\mu\nu}+D(\phi,X)\phi_\mu\phi_\nu,\label{eq:confdisfdisfkerr}
\end{equation}
and $\tilde{g}_{\mu\nu}$, along with an unchanged scalar field, is solution to a new scalar-tensor action. This fact is now used in the next paragraph.
\subsection{Disformal Kerr black hole}\label{subsec:disf_kerr}
The stealth Kerr solution paves the way towards a \textbf{non-stealth} rotating solution, i.e., a rotating solution with metric different from Kerr. This solution is obtained for free thanks to a disformal transformation: the \textbf{disformal Kerr metric} is defined by
\begin{equation}
\tilde{g}_{\mu\nu} = g_{\mu\nu}^{\text{stealth Kerr}} - \frac{D}{q^2}\partial_\mu\phi\,\partial_\nu\phi.
\end{equation}
This metric was constructed in~\cite{Anson:2020trg,BenAchour:2020fgy}, where all details and justifications of the following lines can be found. Comparing with the generic disformal transformation (\ref{eq:confdisfdisfkerr}), this corresponds to a constant conformal factor $C(\phi,X)=1$ and a constant disformal factor $D(\phi,X)=-D/q^2$, where $D$ is a constant, and $q$ is the constant appearing in the Hamilton-Jacobi scalar field (\ref{eq:hj}). After rescaling the time coordinate as $t\left(1+D\right)\to t$, the resulting line element reads
\begin{align}
\mathrm{d}\tilde{s}^2={}&{}-\left(1-\frac{2\tilde{M}r}{\Sigma}\right)\mathrm{d}t^2+\Sigma\mathrm{d}\theta^2+\frac{\sin^2\theta}{\Sigma}\Upsilon\mathrm{d}\varphi^2-\frac{4\sqrt{1+D}\tilde{M}ar\sin^2\theta}{\Sigma}\mathrm{d}t\mathrm{d}\varphi\nonumber\\{}&{}+\frac{\Sigma\Delta-2D\left(1+D\right)\tilde{M}r\left(r^2+a^2\right)}{\Delta^2}\mathrm{d}r^2-2D\frac{\sqrt{2\tilde{M}r\left(r^2+a^2\right)}}{\Delta}\mathrm{d}t\mathrm{d}r.\label{eq:disf_kerr}
\end{align}
This metric, along with the unchanged scalar field (\ref{eq:hj}), is solution of the variational principle of a new DHOST action, which can be found in~\cite{Anson:2021yli}. \\

In (\ref{eq:disf_kerr}), $\tilde{M}\equiv M/\left(1+D\right)$, while $\Sigma$, $\Delta$ and $\Upsilon$ are the same as in the seed Kerr metric, see (\ref{eq:delta_def}), in particular, $\Delta$ still features the parameter $M$ and not $\tilde{M}$. There is a ring singularity at $\Sigma=0$, just like for Kerr.  The disformed Kerr metric is a stationary, axisymmetric, asymptotically flat spacetime. Regarding the conserved charges, the disformed Kerr metric has mass $\tilde{M}$, while its angular momentum per unit mass is $\tilde{a}\equiv a\sqrt{1+D}$. The disformal transformation thus rescales the mass and angular momentum, $J = aM\to\tilde{J} = J/\sqrt{1+D}$. \\

Importantly, for vanishing rotation $a=0$, i.e. if the seed metric is stealth Schwarzschild, then a change of coordinates removes the $\mathrm{d}t\mathrm{d}r$ term of (\ref{eq:disf_kerr}), and the disformed metric is again Schwarzschild, with a mere rescaling of the mass, see~\cite{Babichev:2017lmw,BenAchour:2020wiw}. In other words, the disformal transformation maps stealth Schwarzschild to stealth Schwarzschild. On the other hand, the disformal transformation turns out to be non-trivial when $a\neq 0$: the Ricci tensor of (\ref{eq:disf_kerr}) is not zero, so the disformed Kerr metric is really different from the usual Kerr metric. \\

A major difference between the disformed Kerr metric and the Kerr metric is the following. The Kerr metric (\ref{eq:kerr}) possesses a property of \textbf{circularity}, i.e. invariance under $\left(t,\varphi\right)\to\left(-t,-\varphi\right)$. The disformed Kerr metric as written in (\ref{eq:disf_kerr}) does not possess this invariance because of the $\mathrm{d}t\mathrm{d}r$ term. This statement is not due to a bad choice of coordinates: the authors of~\cite{Anson:2020trg} refer to the more fundamental definition of circularity, which means integrability of the 2-submanifold orthogonal to the two Killing vector fields $\partial_t$ and $\partial_\varphi$ of an axisymmetric spacetime, to prove that the disformed Kerr metric really is \textbf{non-circular}. The non-circularity is partly linked to the fact that the disformed Kerr metric is not an Einstein space, i.e. does not satisfy $R_{\mu\nu}\propto g_{\mu\nu}$. Indeed, Einstein metrics belonging to the class of stationary and axisymmetric spacetimes are known to be circular spacetimes~\cite{Frolov:1998wf}.
\\

Non-circular spacetimes have a richer causal structure than circular ones~\cite{Johannsen:2013rqa}. As recalled in paragraph~\ref{subsec:bhsing}, for Kerr, there is an ergosphere or static limit, where $\partial_t$ is null. Inside the ergosphere, static observers, i.e. observers with constant $r$, $\theta$ and $\varphi$, cease to exist. There is also a stationary limit, which is defined as the hypersurface inside which stationary observers cease to exist. Stationary observers are observers with constant $r$ and $\theta$. For circular spacetimes like Kerr, this stationary limit coincides with the event horizon (this is a particular case of the rigidity theorem, see e.g.~\cite{Hawking:1973uf}). 
\\

However, when the spacetime is non-circular, the stationary limit is no more the event horizon, which must be looked for as a null hypersurface inside the stationary limit. As a summary, for a generic stationary, axisymmetric but non-circular spacetime, there are three remarkable hypersurfaces: ergosphere, stationary limit and event horizon. This is a property of circular spacetimes, like Kerr, that the stationary limit and the event horizon coincide.
\\

Concerning the particular case of the disformed Kerr metric, these surfaces were studied in~\cite{Anson:2020trg}. The static limit or ergosphere lies at
\begin{equation}
r\left(\theta\right) = \tilde{M}+\sqrt{\tilde{M}^2-a^2\cos^2\theta},
\end{equation}
which is the same expression as for Kerr, up to the rescaling of the mass. On the other hand, the stationary limit is located at $r=R_0\left(\theta\right)$ such that
\begin{equation}
P\left(R_0\left(\theta\right),\theta\right)=0
\end{equation}
where 
\begin{equation}
P\left(r,\theta\right)\equiv r^2+a^2-2\tilde{M}r+\frac{2D\tilde{M}ra^2\sin^2\theta}{\Sigma}.
\end{equation}
This yields a fourth-order algebraic equation for $R_0\left(\theta\right)$. Finally, the event horizon $r=R\left(\theta\right)$ is given by a nonlinear ordinary differential equation,
\begin{equation}
\left(\frac{\mathrm{d}R\left(\theta\right)}{\mathrm{d}\theta}\right)^2+P\left(R\left(\theta\right),\theta\right)=0.
\end{equation}
Ref.~\cite{Anson:2020trg} investigated numerically the parameter space allowing solution to this equation. It was found in particular that for a nonvanishing disformal parameter $D$, this candidate event horizon exists only for an angular momentum per unit mass $\tilde{a}<\tilde{a}_c<\tilde{M}$, with the upper bound depending on the value of $D$, $\tilde{a}_c = \tilde{a}_c\left(D\right)$.\\

The disformal parameter $D$ in the disformed Kerr metric marks its departure from the usual Kerr geometry. This parameter can be constrained by observations, since it deforms the shadow with respect to the one of a Kerr BH~\cite{Long:2020wqj} or modifies the orbit of stars around the BH~\cite{Anson:2021yli}. It was recently shown~\cite{Babichev:2024hjf} that even a small $D$ can have a non-negligible effect on the globally accumulated phase of the gravitational waveform of an Extreme Mass Ratio Inspiral, and that this effect could in principle be detected by LISA.

\section[Metric different from General Relativity: non-stealth black holes]{Metric different from General Relativity: non-stealth black holes}\label{sec:nonstealth}
\sectionmark{3.3~~Metric different from GR: non-stealth black holes}
The previous section was concluded with a \textbf{non-stealth} scalar-tensor BH, i.e., for which the metric is not an Einstein metric, unlike the GR BHs. This non-stealth BH was generated as a disformal transformation of a stealth Kerr BH. The present section now focuses on more general non-stealth BHs, that is, which can be directly obtained by solving the scalar-tensor field equations, and not as a transformation of a GR BH. A first example of such a genuine non-stealth BH is the BBMB BH, presented in paragraph~\ref{subsec:bbmbbh} as the (up to date) unique example of hairy BH in second-order scalar-tensor gravity. By going to higher-order scalar-tensor theories (Horndeski and beyond), many more examples can be obtained.
\subsection{Black holes in (beyond) Horndeski theory with shift and parity symmetry}\label{subsec:bcl}
The first non-stealth BH in a higher-order scalar-tensor theory, and with correct asymptotic behaviour, was found by Babichev, Charmousis and Leh\'ebel~\cite{Babichev:2017guv}, and is thus sometimes dubbed BCL. Consider first the generic shift-symmetric beyond Horndeski action,
\begin{align}
S\left[g_{\mu\nu},\phi\right] = \int\mathrm{d}^4 x{}&{}\sqrt{-g}\Bigl\{G_2-G_3\Box\phi+G_4R+G_{4X}\left[\left(\Box\phi\right)^2-\phi_{\mu\nu}\phi^{\mu\nu}\right]\nonumber\\{}&{}+G_5G^{\mu\nu}\phi_{\mu\nu}-\frac{G_{5X}}{6}\Bigl[\left(\Box\phi\right)^3-3\Box\phi\phi_{\mu\nu}\phi^{\mu\nu}+2\phi_{\mu\nu}\phi^{\nu\rho}\phi_\rho^\mu\Bigr]\nonumber\\{}&{}+ F_4 \epsilon^{\mu\nu\rho\sigma}\epsilon^{\alpha\beta\gamma}_{\,\,\,\,\,\,\,\,\,\,\sigma}\phi_\mu\phi_\alpha\phi_{\nu\beta}\phi_{\rho\gamma}+F_5\epsilon^{\mu\nu\rho\sigma}\epsilon^{\alpha\beta\gamma\delta}\phi_\mu\phi_\alpha\phi_{\nu\beta}\phi_{\rho\gamma}\phi_{\sigma\delta}\Bigr\},\label{eq:beyond_action_ter}
\end{align}
where shift symmetry means that all $G_{2,3,4,5}$ and $F_{4,5}$ depend on the kinetic term $X$ only. Take a static and spherically-symmetric ansatz,
\begin{equation}
\mathrm{d}s^2=-h(r)\mathrm{d}t^2+\mathrm{d}r^2/f(r)+r^2\mathrm{d}\Omega^2,\quad \phi=\phi(r),\label{eq:ansatz_bcl}
\end{equation} 
giving a kinetic term
\begin{equation}
X=-\frac{1}{2}\partial_\mu\phi\,\partial^\mu\phi=-\frac{1}{2}f\left(\phi'\right)^2.
\end{equation}
The Noether current associated to shift symmetry, $\mathcal{J}^\mu_{\text{shift}}$, has a unique component, $\mathcal{J}^r_{\text{shift}}$. As explained in the paragraph~\ref{subsec:shift} about shift symmetry, if one makes the natural assumption that the four-norm $\mathcal{J}^\mu_\text{shift} \mathcal{J}_{\text{shift}\,\mu}$ does not diverge, then $\mathcal{J}^r_{\text{shift}}=0$. The covariant expression for the current can be found in appendix~\ref{sec:eq_horn}, and it leads with the present symmetries to
\begin{align}
\mathcal{J}^r_{\text{shift}}={}&{}-G_{2X}f\phi'-XG_{3X}f\frac{4h+rh'}{rh}-\frac{4X}{r^2h\phi'}\Bigl\{ G_{4X}\left[h\left(f-1\right)+rfh'\right]\nonumber\\{}&{}+2XG_{4XX}f\left(rh\right)'\Bigr\}+\frac{Xfh'}{r^2h}\left[G_{5X}\left(3f-1\right)+2XG_{5XX}f\right]\nonumber\\{}&{}-\frac{16X^2f\left(rh\right)'}{r^2h\phi'}\left(2F_4+XF_{4X}\right)-\frac{12X^2f^2h'}{r^2h}\left(5F_5+2XF_{5X}\right).\label{eq:jrleh}
\end{align}
One can then identify potentially interesting forms of the functions $G_k$ and $F_k$, as the ones which remove explicit dependencies in $\phi'$ and can thus provide source terms for the scalar field in the equation $\mathcal{J}^r_{\text{shift}}=0$. \\

For instance, $G_2\propto \sqrt{\lvert X\rvert}$ gives $G_{2X}\propto 1/\sqrt{\lvert X\rvert}\propto 1/\phi'$, which is exactly what compensates the factor $\phi'$ multiplying $G_{2X}$ in the current (\ref{eq:jrleh}). Regarding $G_4$, one must have that both $XG_{4X}$ and $X^2G_{4XX}$ be proportional to $\sqrt{\lvert X\rvert}\propto \phi'$ in order to compensate the factor $1/\phi'$. Therefore, $G_4\propto \sqrt{\lvert X\rvert}$. It is easy to do it for every term and to see that the following choices give source terms for $\phi$ in the equation $\mathcal{J}^r_{\text{shift}}=0$,
\begin{align}
{}&{}G_2\supset \sqrt{\lvert X\rvert},\quad G_3\supset \ln \lvert X\rvert,\quad G_4\supset\sqrt{\lvert X\rvert},\nonumber\\{}&{} G_5\supset\ln \lvert X\rvert,\quad F_4\supset \lvert X\rvert^{-3/2},\quad F_5\supset X^{-2}.\label{eq:sourcecur}
\end{align}
With this reasoning, the BCL BH comes from the choice $G_2\propto X$ (canonical kinetic term), while $G_4$ has a canonical Einstein-Hilbert term ($G_4\supset 1$), and a non-minimal coupling $G_4\supset \sqrt{\lvert X\rvert}$ which provides a source term for the scalar field in the equation $\mathcal{J}^r_{\text{shift}}=0$. The action thus reads
\begin{equation}
S = \int\mathrm{d}^4x\sqrt{-g}\left\lbrace \left(1+\beta\sqrt{\left(\partial\phi\right)^2/2}\right) R -\frac{\eta}{2}\left(\partial\phi\right)^2-\frac{\beta}{\sqrt{2\left(\partial\phi\right)^2}}\left[\left(\Box\phi\right)^2-\phi_{\mu\nu}\phi^{\mu\nu}\right]\right\rbrace.
\end{equation}
The Horndeski functions are precisely $G_2=\eta X$ and $G_4=1+\beta\sqrt{-X}$ with two coupling constants $\eta$ and $\beta$. By construction, the equation $\mathcal{J}^r_{\text{shift}}=0$ admits a solution if and only if $\beta$ and $\eta$ have the same sign, and this solution reads
\begin{equation}
\phi(r) = \pm\frac{\sqrt{2}\beta}{\eta}\int\frac{\mathrm{d}r}{r^2\sqrt{f(r)}}.\label{eq:phi_bcl}
\end{equation}
The metric field equation along $tt$ then gives the profile of $f(r)$, while the one along $rr$ proves that $h(r)=f(r)$. At the end of the day, the BH solution is metric (\ref{eq:ansatz_bcl}) with
\begin{equation}
h=f=1-\frac{2M}{r}-\frac{\beta^2}{2\eta r^2},
\end{equation}
where the integration constant $M$ is the ADM mass~\cite{Arnowitt:1961zz} of the BH. Note that $f$ is of the general form (\ref{eq:riccizerosol}), so the Ricci scalar vanishes, $R=0$. However, the Ricci tensor does not vanish and
\begin{equation}
R_{\mu\nu}R^{\mu\nu}=\frac{\beta^4}{\eta^2 r^8},
\end{equation}
so $r=0$ is a curvature singularity. The horizon structure is the following: 
\begin{enumerate}
\item If $\eta>0$, there is a unique horizon.
\item If $\eta<0$ and $M<-\beta/\sqrt{-2\eta}$, there are no horizons (naked singularity).
\item If $\eta<0$ and $M\geq -\beta/\sqrt{-2\eta}$, there are two horizons (a double horizon in case of equality).
\end{enumerate}
Surprisingly, $M=0$ is a spacetime with vanishing mass, but not flat: it is a BH if $\eta>0$ and a naked singularity if $\eta <0$. Regarding now the scalar field (\ref{eq:phi_bcl}), it is real and finite for any $r\geq r_h$, while it becomes imaginary when $f\left(r\right)<0$. The scalar field kinetic term is well-defined apart from the central singularity,
\begin{equation}
X = -\frac{\beta^2}{\eta^2 r^4}.
\end{equation}
\textbf{Importantly, one can compare the present setup with the assumptions of the Hui-Nicolis no-hair theorem of paragraph~\ref{subsec:huinic}. The only broken assumption is the sixth one, because the form of $G_4$ is not analytic at $X=0$. This explains the obtention of a hairy BH, while remaining consistent with the no-hair theorem}.
\\

The linear perturbations and QNMs of the BCL BH have been studied in~\cite{Langlois:2021aji,Roussille:2022vfa}, where some pathologies are reported.
\\

Note that the same Ref.~\cite{Babichev:2017guv} also studied a beyond Horndeski case, with $G_4=1$, $G_2\propto X$ and $F_4\propto \left\lvert X\right\rvert^{-3/2}$. The $F_4$ term thus provides a source for the scalar field in the equation $\mathcal{J}^r_\text{shift}=0$ according to (\ref{eq:sourcecur}). An asymptotically flat BH with Schwarzchild asymptotic behaviour is also obtained in this case. Again, the non-analyticity of $F_4$ explains the evasion from the no-hair theorem.\\

Both these closed-form solutions were thus found in the context of (beyond) Horndeski theories with shift symmetry and parity symmetry, i.e. symmetry under $\phi\to -\phi$. Breaking parity symmetry, e.g. introducing $G_3$ and $G_5$ terms in Horndeski, prevented for some years the obtention of non-stealth closed-form solutions. Only solutions with numerical analysis or perturbative expansions could be obtained, as the numerous ones with a GB term that we mentioned in paragraph~\ref{subsec:huinic}, or the ones obtained in the cubic Galileon theory (i.e. with $G_3=X$)~\cite{Babichev:2016fbg}. In this case, a linear time dependence of the scalar field ensures the evasion of the no-hair theorem. These numerical cubic Galileon solutions were extended to the rotating case, first approximately~\cite{VanAelst:2019kku} then exactly~\cite{Grandclement:2023xwq}.\\

In the same vein, theories without shift symmetry seemed reluctant to closed-form non-stealth solutions. This resistance of scalar-tensor theories without parity or shift symmetry was broken in the wake of the formulation of the 4DEGB scalar-tensor theory of gravity. The reader can refer back to Sec.~\ref{sec:fourd}, where it is explained that this four-dimensional theory is obtained by a regularized KK compactification of the higher-dimensional, Lovelock EGB theory of gravity, and that is reproduces features of this theory. In addition to its intereseting theoretical properties, this Horndeski theory with all $G_2$, $G_3$, $G_4$ and $G_5$ terms allows for closed-form solutions, which are described in the next paragraph.\newpage
\subsection{Black holes in the four-dimensional Einstein-Gauss-Bonnet theory}\label{subsec:bhegb}
Let us now present the various exact, asymptotically flat BH solutions to the 4DEGB action,
\begin{align}
S_{\text{4DEGB}}\left[g_{\mu\nu},\phi\right] =\int\mathrm{d}^4x\sqrt{-g}{}&{}\Bigl\{R-2\lambda\mathrm{e}^{4\phi}-\beta\mathrm{e}^{2\phi}\Bigl[R+6\left(\partial\phi\right)^2\Bigr]\nonumber\\ {}&{}+\alpha\Bigl[-\phi\mathcal{G}+4G^{\mu\nu}\phi_\mu\phi_\nu+4\Box\phi\left(\partial\phi\right)^2+2\left(\partial\phi\right)^4\Bigr]\Bigr\}.\label{eq:4degb_bis}
\end{align}
As announced, it is a Horndeski theory, corresponding to
\begin{align}
G_2 ={}&{} -2\lambda\mathrm{e}^{4\phi}+12\beta\mathrm{e}^{2\phi}X+8\alpha X^2,\quad G_3 = 8\alpha X,\nonumber\\ G_4 ={}&{} 1-\beta\mathrm{e}^{2\phi}+4\alpha X,\quad G_5 = 4\alpha\ln\left\lvert X\right\rvert.\label{eq:horn_conf_bis}
\end{align}
It is seen to acquire shift symmetry when $\lambda=\beta=0$, however, it never has parity symmetry. We do not include the cosmological constant term $-2\Lambda$ to keep the discussion more simple, but all the following asymptotically flat solutions admit direct generalizations, which can be found in the various references encountered in this paragraph, when including the cosmological constant.\\

The considered solutions are of the static, spherically symmetric, homogeneous form (\textbf{in the BH context, by homogeneous, we mean $g_{tt}=-g^{rr}$}),
\begin{equation}
\mathrm{d}s^2 = -f\left(r\right)\mathrm{d}t^2+\mathrm{d}r^2/f\left(r\right)+r^2\mathrm{d}\Omega^2.\label{eq:hom_ans}
\end{equation}
As explained in Secs.~\ref{sec:symm} and~\ref{sec:fourd}, the 4DEGB action coincides with the most general action with generalized conformal symmetry. This implies the existence of a geometric equation $\mathfrak{R}=0$, see Eq. (\ref{eq:rfern}), where
\begin{equation}
\mathfrak{R}=g^{\mu\nu}\mathcal{E}_{\mu\nu}+\mathcal{E}_\phi=-2R-\alpha\mathcal{G}.
\end{equation}
For ansatz (\ref{eq:hom_ans}), this geometric equation directly implies
\begin{equation}
f = 1+\frac{r^2}{2\alpha}\left(1\pm\sqrt{1+\frac{8\alpha M}{r^3}+\frac{\mu}{r^4}}\right).
\end{equation}
There are two integration constants, $M$ and $\mu$. The $+$ branch gives inconsistent asymptotic behaviour, with a (a)dS behaviour and a function $f(r)$ which does not reduce to Schwarzschild in the limit of vanishing coupling $\alpha\to 0$. It is therefore common to discard it and to consider only the $-$ branch. The integration constant $M$ can then be seen to be the ADM mass of the spacetime, since $f=1-2M/r+o(1/r)$ when $r\to\infty$. The remaining field equations then constrain the value of the second integration constant $\mu$, the scalar field profile, and the respective values of the coupling constants $\lambda$, $\beta$ and $\alpha$ appearing in action (\ref{eq:4degb_bis}). In particular, the scalar field profile is obtained by considering the combination of metric field equations $\mathcal{E}^t_t-\mathcal{E}^r_r$.
\\

We will present these solutions according to the geometric characteristics of the internal space, as it appears in the regularized KK picture (\ref{eq:kkreg}). Importantly, they all correspond to different relative values of the couplings $\lambda$, $\beta$ and $\alpha$, thus they are solutions to different theories. 
\subsubsection{Maximally-symmetric internal space}
The first solution was obtained by~\cite{Lu:2020iav}, following their KK regularization procedure along a maximally-symmetric internal space. This corresponds to $\lambda = 3\beta^2/\left(4\alpha\right)$, see Eqs. (\ref{eq:linkmax1},\ref{eq:linkmax2}). The metric is
\begin{equation}
f\left(r\right) = 1+\frac{r^2}{2\alpha}\left(1-\sqrt{1+\frac{8\alpha M}{r^3}}\right),\label{eq:met_lu_pang}
\end{equation}
i.e. the second integration constant is $\mu=0$. Very interestingly, this profile can be seen as the continuation to $D\to 4$ of the spherically-symmetric Boulware-Deser solution of pure metric EGB gravity in $D\geq 5$~\cite{Boulware:1985wk}, see also~\cite{Glavan:2019inb}. The integration constant $M$ is the mass of the spacetime, which behaves asymptotically as
\begin{equation}
f\left(r\right) = 1-\frac{2M}{r}+\frac{4\alpha M^2}{r^4}+\mathcal{O}\left(\frac{1}{r^7}\right).\label{eq:asymp_lu_pang}
\end{equation}
The scalar field is radial and reads\footnote{The divergence of $\phi\left(r\right)$ as $r\to\infty$ is irrelevant as it can be eliminated by field redefinition. Indeed, the redefined scalar $\Phi=\mathrm{e}^\phi$ vanishes as $r\to\infty$. Action (\ref{eq:4degb_bis}) displays a canonical kinetic term for this scalar field $\Phi$, as already mentioned below Eq. (\ref{eq:4degb}).\label{f:redef_scal}}
\begin{equation}
\phi\left(r\right) = \ln\left(\frac{\sqrt{-2\alpha/\beta}}{r}\right)-\ln\left(\sigma\left(c\pm\int\frac{\mathrm{d}r}{r\sqrt{\left\lvert f\left(r\right)\right\rvert}}\right)\right),\label{eq:phi_lu_pang}
\end{equation}
where $c$ is an arbitrary integration constant and $\sigma$ is a function with an expression depending on the sign of $f\left(r\right)$: if $f\left(r\right)\geq 0$, $\sigma = \cosh$, while if $f\left(r\right)<0$, $\sigma = \left\lvert\cos\right\rvert$. Therefore the scalar field is well-defined when $f\left(r\right)\geq 0$ (in particular outside the event horizon), and even when $f\left(r\right)<0$ apart on a set of null measure. The horizons of (\ref{eq:met_lu_pang}) depend on the sign of $\alpha$. If $\alpha>0$, there are two possible horizons at
\begin{equation}
r_{\pm} = M\pm\sqrt{M^2-\alpha},\label{eq:rpm}
\end{equation}
in other words, the spacetime has no horizons if $M< \sqrt{\alpha}$, and has two horizons otherwise. For $\alpha<0$, the square root in (\ref{eq:met_lu_pang}) becomes ill-defined before reaching $r_+$ if the mass is too small (naked singularity), while if the mass is sufficiently large, there is indeed a horizon at $r_+$, but no inner horizon, since the square root in (\ref{eq:met_lu_pang}) becomes ill-defined. \\

Ref.~\cite{Charmousis:2021npl} proved that the expression (\ref{eq:rpm}) for the horizon would strongly constrain the negative values of $\alpha$ if there were a Birkhoff-like theorem~\cite{Israel:1967wq} for the considered 4DEGB theory. In this case, an atomic nucleus of radius $R$ would produce the gravitational field (\ref{eq:met_lu_pang}). But a nucleus is not a BH, so the horizon must be hidden below the atomic radius: $r_+<R$. This yields $-10^{-30}\,\text{m}^2<\alpha<0$. On the other hand, when $\alpha>0$, the metric is well-defined up to $r=0$, where it behaves as
\begin{equation}
f\left(r\right)=1-\sqrt{\frac{2Mr}{\alpha}}+\frac{r^2}{2\alpha}+\mathcal{O}\left(r^{7/2}\right).\label{eq:lu_pang_zero}
\end{equation}
Although $f\left(r\right)$ is well-defined at $r=0$, the curvature invariants diverge there. It is indeed known~\cite{Burzilla:2020utr} that the curvature invariants do not diverge only if $f\left(r\right)=1+\mathcal{O}\left(r^2\right)$. There is therefore a curvature singularity at $r=0$.

\subsubsection{Flat internal space}
The second solution was presented by~\cite{Hennigar:2020lsl}, after their regularization which corresponds to a flat internal space and therefore to $\lambda=\beta=0$, see Eqs. (\ref{eq:linkflat1},\ref{eq:linkflat2}). The metric potential $f\left(r\right)$ is exactly the same as above, Eq. (\ref{eq:met_lu_pang}), while the radial scalar field is now given by
\begin{equation}
\phi\left(r\right)=\int\frac{\pm 1-\sqrt{f\left(r\right)}}{r\sqrt{f\left(r\right)}}\mathrm{d}r.\label{eq:phi_mann}
\end{equation}
This holds up to a global additive constant since this particular theory is shift-symmetric. The scalar field is well-defined outside and at the horizon, $r\geq r_+$, but becomes imaginary when $f\left(r\right)<0$. The kinetic term itself is imaginary for $f\left(r\right)<0$,
\begin{equation}
X = -\frac{\left(\pm 1-\sqrt{f\left(r\right)}\right)^2}{2r^2}.
\end{equation}
This unpleasant feature was cured in~\cite{Charmousis:2021npl}, by adding a linear time dependence to the scalar field. The obtained scalar field reads
\begin{equation}
\phi(t,r) = qt + \int\frac{\pm\sqrt{q^2r^2+f\left(r\right)}-f\left(r\right)}{rf\left(r\right)}\mathrm{d}r.\label{eq:cured}
\end{equation}
The metric remains unchanged: it does not feel the influence of the scalar hair $q$, which is an arbitrary integration constant. As explained in Sec.~\ref{sec:stealth} regarding stealth solutions, the regularity of the scalar field at the horizon $r_+$ is manifest in horizon-crossing coordinates, since $\phi\left(v,r=r_+\right)=qv+\text{const.}$ or $\phi\left(u,r=r_+\right)=qu+\text{const.}$ depending on the sign of $q$ and the choice of $\pm$ in (\ref{eq:cured}). We recall that $v=t+r^\star$, $u=t-r^\star$ and $r^\star = \int\mathrm{d}r/f\left(r\right)$. Moreover, since $f\left(r\right)$ as given by (\ref{eq:met_lu_pang}) is bounded from below, there always exists a range of the integration constant $q$ such that the scalar field is well-defined in the whole spacetime. The same then holds for the kinetic term,
\begin{equation}
X = \frac{\pm 2\sqrt{q^2r^2+f\left(r\right)}-1-f\left(r\right)}{2r^2},
\end{equation}
apart from the central singularity at $r=0$. Note that $X$ here is not constant, as opposed to the case of stealth solutions studied in Sec.~\ref{sec:stealth}. On another hand, the linear perturbations of the BH solution in the shift-symmetric 4DEGB theory, and the associated QNMs, were analyzed in~\cite{Langlois:2022eta,Roussille:2022vfa}, showing the existence of pathologies. The QNMs and shadow of the BH were also presented in~\cite{Konoplya:2020bxa}, while~\cite{Guo:2020zmf} considered, in addition to the shadow, the properties of orbits around this BH.\\

Finally, since these solutions (both with $\phi=\phi(r)$ and $\phi=qt+\psi(r)$) are obtained in a theory with shift symmetry, it should be verified how the no-hair theorem of paragraph~\ref{subsec:huinic} is violated. In the case where $\phi=qt+\psi(r)$, it is of course due to the form of the scalar field. In the case $\phi=\phi(r)$, the theorem is circumvented thanks to the absence of a canonical kinetic term $X\subset G_2$. 
\subsubsection{Internal space as a product of two spheres}
The third solution was described by~\cite{Fernandes:2021dsb}, and in fact corresponds to the KK regularization for an internal space which is a product of two-spheres of identical radius. This corresponds to $\lambda = \beta^2/\left(4\alpha\right)$, see Eqs. (\ref{eq:linkprod1},\ref{eq:linkprod2}). The obtained metric is this time different, with a non-vanishing second integration constant $\mu$,
\begin{equation}
f\left(r\right) = 1+\frac{r^2}{2\alpha}\left(1-\sqrt{1+\frac{8\alpha M}{r^3}+\frac{8\alpha^2}{r^4}}\right).\label{eq:prod_sphere_sol}
\end{equation}
The asymptotic departure from Schwarzschild gets larger than in (\ref{eq:asymp_lu_pang}), since now,
\begin{equation}
f\left(r\right)=1-\frac{2M}{r}-\frac{2\alpha}{r^2}+\mathcal{O}\left(\frac{1}{r^4}\right).
\end{equation}
In fact, this solution directly stems from a higher-dimensional solution~\cite{Dotti:2005rc} through the diagonal KK reduction presented in Secs.~\ref{sec:compact} and~\ref{sec:fourd}~\cite{Charmousis:2012dw}. The expression for the scalar field is
\begin{equation}
\phi\left(r\right)=\ln\left(\frac{\sqrt{-2\alpha/\beta}}{r}\right).\label{eq:log_scal}
\end{equation}
The candidate event horizon is now located at
\begin{equation}
r_+ = M+\sqrt{M^2+\alpha}.
\end{equation}
For $\alpha>0$, there is always a unique horizon at $r_+$, while for $\alpha<0$, the square root in (\ref{eq:prod_sphere_sol}) might become ill-defined. It turns out that there is no horizon for small masses, and a unique horizon at $r_+$ for sufficiently large masses. This was studied in detail in~\cite{Babichev:2022awg}. Also, the same kind of Birkhoff conjecture as explained between (\ref{eq:rpm}) and (\ref{eq:lu_pang_zero}) now severly constrains the positive values of $\alpha$ for the present theory, $0<\alpha<10^{-30}\,\text{m}^2$. When the metric reaches $r=0$ with no problems in the square root, it behaves there as
\begin{equation}
f\left(r\right) = 1-\text{sgn}\left(\alpha\right)\sqrt{2}-\frac{Mr}{\sqrt{2}\left\lvert\alpha\right\rvert}+\mathcal{O}\left(r^2\right),
\end{equation}
where $\text{sgn}$ is the sign function, $\text{sgn}(\gtrless 0)=\pm 1$. Again, $f\left(r\right)$ is well-defined at $r=0$, but not sufficiently regular, so there is a curvature singularity at $r=0$. A major difference with the previous spacetime (\ref{eq:met_lu_pang}) is that (\ref{eq:prod_sphere_sol}) does not reduce to flat spacetime when $M=0$, but is rather a naked singularity at $r=0$. This is due to the $8\alpha^2/r^4$ term in (\ref{eq:prod_sphere_sol}), which ultimately comes from the higher-dimensional origin of the solution with a horizon of non-trivial topology (product of two-spheres)~\cite{Charmousis:2012dw,Dotti:2005rc}. Note finally that a more detailed study of this solution is performed in paragraph~\ref{sec:seedbhsol}.
\\

As a summary, the 4DEGB action (\ref{eq:4degb_bis}) admits three distinct scalar-tensor BH solutions, for three distinct theories with respectively $\lambda=3\beta^2/\left(4\alpha\right)$, $\lambda=\beta=0$ and $\lambda=\beta^2/\left(4\alpha\right)$. It was in fact shown in~\cite{Fernandes:2021dsb} that these possibilities are the only ones\footnote{More precisely, Ref.~\cite{Fernandes:2021dsb} also finds a solution with constant scalar field, but we do not mention it because the field equations become partly degenerate at this constant value (strong coupling). This constant scalar even allows for a non-perturbative rotating solution~\cite{Fernandes:2023vux}. However, degeneracy is manifest since this solution is parameterized by two arbitrary functions of $\theta$, which are not fixed by the field equations.} with the ansatz (\ref{eq:hom_ans}), which assumes $g_{tt}=-g^{rr}$. Furthermore, for the shift-symmetric case $\lambda=\beta=0$, unicity of the static BH solutions was proved in~\cite{Fernandes:2021ysi}. In other words, BHs with $g_{tt}\neq -g^{rr}$ do not exist in this case.\\

Each of the obtained theories can be mapped to the regularized KK action (\ref{eq:kkreg}) with different internal space (respectively maximally-symmetric, flat, and product of two-spheres). Nevertheless, as regards the solutions, only the case of the product of two-spheres directly descends from a solution to EGB gravity in $\mathfrak{D}$ dimensions through the diagonal KK reduction (\ref{eq:diagansatz}). This means that both the metric (\ref{eq:prod_sphere_sol}) and the scalar (\ref{eq:log_scal}) are clearly mapped to the higher-dimensional solution of~\cite{Dotti:2005rc} in the diagonal KK ansatz (\ref{eq:diagansatz}). For the other two solutions, a hypothetical KK origin remains unclear. Indeed, their metric function (\ref{eq:met_lu_pang}) generalizes the higher-dimensional Boulware-Deser solution~\cite{Boulware:1985wk}, but their scalar field (\ref{eq:phi_lu_pang}) or (\ref{eq:phi_mann}) does not fit into the diagonal KK reduction picture (\ref{eq:diagansatz}).
\\

This completes the state-of-the-art of scalar-tensor, closed-form BH solutions, with sensible asymptotic behaviour and regular scalar field outside and at the horizon, which existed before this thesis started. For conciseness, we have not detailed some interesting regular BHs which were constructed in DHOST theories, and rather refer the reader to~\cite{Babichev:2020qpr,Baake:2021jzv}. \\

To summarize, as regards higher-order scalar-tensor theories (from Horndeski to DHOST), hairy BH solutions can be obtained by violating one of the assumptions of the Hui-Nicolis no-hair theorem of paragraph~\ref{subsec:huinic}. For instance, a scalar field with a linear time dependence (which is allowed in shift-symmetric theories) enables to obtain many stealth solutions, i.e., the metric is the same as in GR although the scalar field is non-trivial. All these stealth solutions exist provided the kinetic term $X$ of the scalar field is constant, $X=X_0$. This is not a problem in itself, but it turns out that, for the asymptotically flat stealth solutions, the kinetic or k-essence term of the scalar field must have a double root at $X_0$, and this implies pathologies as regards their perturbative aspects.
\\

Non-stealth solutions, for which the metric is different from GR, also exist, in theories with or without shift symmetry. For shift-symmetric theories, the Hui-Nicolis theorem can still be circumvented for instance by considering Horndeski functions which are non-analytic functions of $X$. This yields the BCL BH, a very simple example of non-stealth BH. Going beyond the shift-symmetric case could in principle offer many possibilities for hairy BHs, since it automatically evades to no-hair theorem. Despite this encouraging fact, as regards non-stealth, closed-form solutions beyond shift symmetry, only two existed before this thesis, namely (\ref{eq:met_lu_pang}) and (\ref{eq:prod_sphere_sol}). They are obtained in the context of the 4DEGB theory. Another non-stealth solution exists in the shift-symmetric 4DEGB theory.  
\\

It is now time to present, in Chaps.~\ref{chap:4} and~\ref{chap:5}, the new exact BH solutions obtained during this thesis. Chap.~\ref{chap:4} proposes a systematic study of the shift-symmetric beyond Horndeski field equations, which will lead to many\footnote{More precisely, infinitely many, since certain infinite families of theories, and their corresponding solutions, are identified.} new solutions. Among them, a stealth Schwarzschild solution where, as opposed to the usual ones, the kinetic term $X$ is not constant. Also, all non-stealth solutions mentioned above have secondary hair. This means that, in spherical symmetry, although their metric differs from Schwarzschild, they remain characterized by a unique parameter, their mass $M$. In Chap.~\ref{chap:4}, a solution with primary hair, i.e. parameterized by an additional integration constant $q$ distinct from $M$, is obtained. For a particular relation between $q$ and $M$, the central curvature singularity even disappears.
\\

From a very different approach, non-stealth solutions are also obtained in Chap.~\ref{chap:5}. This time, shift symmetry is not assumed anymore. Inspiration is therefore taken from the theory which we saw allowed closed-form solutions in this context, the 4DEGB theory. This 4DEGB theory has two distinctive features: its link with the higher-dimensional EGB theory through KK compactification, and its generalized conformal invariance. Both these ideas are used in Chap.~\ref{chap:5} and yield new relevant theories and solutions.
\\

Table~\ref{table:new} enables to appreciate at a glance these new contributions. 
\begin{table}
\begin{center}
\begin{tabular}{| m{2.4cm} || m{2.5cm} | m{3.3cm} | m{2.8cm} | m{2.8cm} |} 
 \hline
$\phi=\cdots$ & $G_{2,4}\left(X\right)$ (shift + parity sym. Horndeski) & \raggedright\arraybackslash $G_{2,3,4,5}\left(\phi,X\right)$ (Generic Horndeski) & $G_{2,4}\left(X\right)$ and $F_4\left(X\right)$ (shift + parity sym. beyond Horndeski) & \begin{center}DHOST\end{center}\\ [0.5ex] 
 \hline\hline
 $\phi\left(r\right)$ & BCL~\cite{Babichev:2017guv} & 4DEGB~\cite{Lu:2020iav,Fernandes:2021dsb};\newline \raggedright\arraybackslash{\color{red}Extensions without conformal invariance \cite{Babichev:2023dhs}, Sec.~\ref{sec:selecting};} \newline {\color{red}Other higher-order Lovelock compactifications \cite{Bakopoulos:2022csr}, Sec.~\ref{sec:nonpar} and paragraph~\ref{subsec:lastconj}} & Other BCL~\cite{Babichev:2017guv};\newline \raggedright\arraybackslash{\color{red} New secondary hair solutions \cite{Bakopoulos:2022csr}, Sec.~\ref{sec:purebey}} & Stealth Schwarzschild \cite{Motohashi:2019sen};\newline \raggedright\arraybackslash Regular BHs (Kerr-Schild construction)~\cite{Babichev:2020qpr,Baake:2021jzv} \\ 
 \hline
 $qt+\psi\left(r\right)$ and $X=\text{cst}$ & (Historical) stealth Schwarzschild \cite{Babichev:2013cya}; \newline Other stealth Schwarzschild \cite{Kobayashi:2014eva} &  Stealth Schwarzschild~\cite{Minamitsuji:2018vuw} & 
  Stealth Schwarzschild \cite{Babichev:2016kdt}  &  Stealth Schwarzschild \cite{Motohashi:2019sen,Minamitsuji:2019shy,BenAchour:2018dap} and Kerr~\cite{Charmousis:2019vnf}; Disformed Kerr~\cite{Anson:2020trg,BenAchour:2020fgy}  \\
 \hline
 $qt+\psi\left(r\right)$ and $X\neq \text{cst}$ &  {\color{red}New Stealth Schwarzschild \cite{Bakopoulos:2023fmv}, Sec.~\ref{sec:pure}} & Shift-symmetric 4DEGB~\cite{Hennigar:2020lsl} & \raggedright\arraybackslash{\color{red}Primary hair solutions and regular BHs \cite{Bakopoulos:2023fmv}, Sec.~\ref{sec:primary}} & \raggedright\arraybackslash{\color{red}Conformal Kerr \cite{Babichev:2023mgk}, Sec.~\ref{sec:kerrflrw}} \\ [1ex] 
 \hline
\end{tabular}
\caption{State-of-the-art of scalar-tensor BHs after this thesis, in Fall 2023. In red, the new solutions constructed during this thesis, and presented in the following chapters.\label{table:new}}
\end{center}
\end{table}

\chapter[Static, spherically-symmetric black holes in shift-symmetric beyond Horndeski theories]{Static, spherically-symmetric black holes in shift-symmetric beyond Horndeski theories}\label{chap:4}
\chaptermark{Chap.~4~~Static, spherical BHs in shift-symmetric beyond Horndeski theories}
This chapter studies a particular class of scalar-tensor theories, namely beyond Horndeski theories with shift symmetry. By restricting to static and spherical symmetry, one can rewrite the field equations in a compact way, Sec.~\ref{sec:rewrite}, which facilitates their solving. Integration of the field equations can then be performed in quite generic cases where all beyond Horndeski functions $G_{2,3,4,5}$ and $F_{4,5}$ are present, Sec.~\ref{sec:nonpar}. The obtained theories are parameterized by a parameter $n$, and generalize the shift-symmetric 4DEGB theory studied in paragraph~\ref{subsec:bhegb}, which corresponds to $n=1$.
\\

Still, the greatest simplification to the field equations occurs by restricting to parity symmetry, under $\phi\to -\phi$. In pure Horndeski theories, Sec.~\ref{sec:pure}, this leads notably to a \textbf{new stealth Schwarzschild solution} which does not suffer \textit{a priori} from the same perturbative problems as the usual stealth Schwarzschild of paragraph~\ref{subsec:stealthsch}. In beyond Horndeski theories with a radial scalar field, Sec.~\ref{sec:purebey}, new non-stealth BHs are obtained, qualitatively similar to the BCL BH of paragraph~\ref{subsec:bcl}. Finally, beyond Horndeski theories with a time dependent scalar field, $\phi=qt+\psi(r)$, allow for BHs with \textbf{primary scalar hair}, that is, the metric is parameterized not only by its mass $M$, but also by $q$, which is an arbitrary integration constant. Also, when a certain relation holds between $M$ and $q$, the curvature singularity disappears. This is presented in Sec.~\ref{sec:primary}.
\section{The setup: theory, ansatz, and system of equations}
\sectionmark{4.1~~The setup: theory, ansatz, and system of equations}
This chapter focuses on \textbf{beyond Horndeski theories with shift symmetry}. Beyond Horndeski theories were presented in Chap.~\ref{chap:st}, see specifically Eq. (\ref{eq:beyond_action_bis}). They are in general parameterized by functions $G_{2,3,4,5}(\phi,X)$ and $F_{4,5}(\phi,X)$ of the scalar field $\phi$ and its kinetic term $X$. Shift symmetry, that is, symmetry under constant shifts $\phi\to\phi+\text{constant}$, was described in paragraph~\ref{subsec:shift}. A beyond Horndeski action acquires shift symmetry when all of its coupling functions depend only on $X$, but not on $\phi$. Thus, the action under consideration in this chapter reads:
\begin{align}
S\left[g_{\mu\nu},\phi\right] {}&{}= \int\mathrm{d}^4 x\sqrt{-g}\Bigl\{G_2\left(X\right)-G_3\left(X\right)\Box\phi+G_4\left(X\right)R+G_{4X}\left[\left(\Box\phi\right)^2-\phi_{\mu\nu}\phi^{\mu\nu}\right]\nonumber\\{}&{}+G_5\left(X\right)G^{\mu\nu}\phi_{\mu\nu}-\frac{G_{5X}}{6}\Bigl[\left(\Box\phi\right)^3-3\Box\phi\phi_{\mu\nu}\phi^{\mu\nu}+2\phi_{\mu\nu}\phi^{\nu\rho}\phi_\rho^\mu\Bigr]\nonumber\\{}&{}+ F_4\left(X\right)\epsilon^{\mu\nu\rho\sigma}\epsilon^{\alpha\beta\gamma}_{\hspace{0.5cm}\sigma}\phi_\mu\phi_\alpha\phi_{\nu\beta}\phi_{\rho\gamma}+F_5\left(X\right)\epsilon^{\mu\nu\rho\sigma}\epsilon^{\alpha\beta\gamma\delta}\phi_\mu\phi_\alpha\phi_{\nu\beta}\phi_{\rho\gamma}\phi_{\sigma\delta}\Bigr\}.\label{eq:shift_action}
\end{align}
In addition and as explained in paragraph~\ref{subsec:beyond_degen}, the following relation must hold,
\begin{equation}
3F_5\left(G_4-2XG_{4X}\right)=XF_4G_{5X},\label{eq:beyond_degen_bis}
\end{equation}
so as to evade the appearance of an Ostrogradsky ghost degree of freedom.\\

This chapter describes the study and solving of the field equations coming from the variational principle of an action of the form (\ref{eq:shift_action}), for the case of a static, spherically-symmetric metric ansatz,
\begin{equation}
\mathrm{d}s^2 = -h\left(r\right)\mathrm{d}t^2+\mathrm{d}r^2/f\left(r\right)+r^2\mathrm{d}\Omega^2. \label{eq:metric_static_ansatz}
\end{equation}
The metric is said to be \textbf{homogeneous} if $h=f$, this is for instance the case of the usual Schwarzschild BH, $h=f=1-2M/r$, and \textbf{non-homogeneous} otherwise. As usual, the metric (\ref{eq:metric_static_ansatz}) is a BH if it admits a horizon, that is to say, a radius $r=r_h$ such that
\begin{equation}
h\left(r_h\right) = f\left(r_h\right) = 0.\label{eq:horizon_def}
\end{equation} 
As explained in paragraph~\ref{subsec:shift}, the ansatz for the scalar field $\phi$, compatible with the symmetries of the metric ansatz (\ref{eq:metric_static_ansatz}), can be taken as
\begin{equation}
\phi=qt+\psi\left(r\right),\label{eq:scalar_ansatz}
\end{equation}
where $q$ is a constant. Indeed, since the action (and hence the field equations) depend only on the gradient of the scalar field, the term linear in time does not introduce any time dependence into the field equations. Also, due to shift symmetry, $\phi$ is defined up to an irrelevant additive constant. With this ansatz, the kinetic term is
\begin{equation}
X = -\frac{1}{2}\partial_\mu\,\phi\partial^\mu\phi = \frac{1}{2}\left(\frac{q^2}{h}-f\psi'^2\right).\label{eq:x_shift}
\end{equation}
Again in paragraph~\ref{subsec:shift}, the Noether current associated to shift symmetry,
\begin{equation}
\mathcal{J}^\mu_\text{shift} = \frac{1}{\sqrt{-g}}\frac{\delta S}{\delta (\partial_\mu\phi)},
\end{equation}
was defined and studied. We recalled in particular that the scalar field equation is $\mathcal{E}_\phi=-\nabla_\mu \mathcal{J}^\mu_\text{shift}$, but that the ansatz (\ref{eq:metric_static_ansatz}-\ref{eq:scalar_ansatz}) implies in fact that $\mathcal{J}_\text{shift}^r=0$. This is a consequence of diffeomorphism invariance if the scalar has a true linear time dependence ($q\neq 0$)~\cite{Babichev:2015rva}, while if $q=0$, this comes from imposing that the norm $\mathcal{J}^\mu_\text{shift} \mathcal{J}_{\text{shift}\,\mu}$ be finite at the BH horizon. 
\\

In fact, diffeomorphism invariance also relates the divergence of the metric field equations to the scalar field equation through a 'Bianchi identity',
\begin{equation}
\nabla^\nu\mathcal{E}_{\mu\nu}=-\mathcal{E}_\phi\,\nabla_\mu \phi,
\end{equation}
see Eq. (\ref{eq:bianchi}). If $\phi$ is not constant, the scalar field equation $\mathcal{E}_\phi$, equivalent to $\mathcal{J}_\text{shift}^r=0$, is thus implied by the metric field equations $\mathcal{E}_{\mu\nu}$. With the spherically-symmetric ansatz (\ref{eq:metric_static_ansatz}-\ref{eq:scalar_ansatz}), the non-vanishing components of $\mathcal{E}_{\mu\nu}$ are $tt$, $rr$, $\theta\theta$, $\varphi\varphi$, and also $tr$ if $q\neq 0$. However, spherical symmetry implies $\mathcal{E}_{\varphi\varphi}=\sin^2\theta\,\mathcal{E}_{\theta\theta}$, and if $q\neq 0$, $\mathcal{J}_\text{shift}^r$ is proportional to $\mathcal{E}_{tr}$ by (\ref{eq:super_important}). 
\\

At the end of the day, \textbf{the field equations for a shift-symmetric scalar-tensor action\footnote{Note indeed that the above reasoning did not make use of the particular beyond Horndeski class (\ref{eq:shift_action}), so it remains true for any shift-symmetric theory with the same metric and scalar ansatz.}, with static metric ansatz (\ref{eq:metric_static_ansatz}) and non-constant scalar field $\phi$ of the form (\ref{eq:scalar_ansatz}) reduce, in the usual spherical coordinates $\left(t,r,\theta,\varphi\right)$, to a system of three coupled differential equations,
\begin{equation}
\mathcal{E}_{tt}=0,\quad \mathcal{E}_{rr}=0,\quad \mathcal{J}_\text{shift}^r=0,\label{eq:field_eq_4}
\end{equation}
for the three unknown functions $h\left(r\right)=-g_{tt}$, $f\left(r\right)=g^{rr}$ and $\psi\left(r\right)$ [with $\phi=qt+\psi\left(r\right)$].}
\\

The analysis of this chapter gathers the results of two articles,~\cite{Bakopoulos:2022csr} and~\cite{Bakopoulos:2023fmv}, of which the present author is one of the authors. The former article studied the case of $q=0$, while the latter article the case of $q\neq 0$.
\section{Rewriting the field equations}\label{sec:rewrite}
\sectionmark{4.2~~Rewriting the field equations}
The action under consideration is the generic beyond Horndeski action with shift symmetry, Eq. (\ref{eq:shift_action}). The covariant form of the field equations is given in appendix~\ref{sec:eq_horn}. Naively specifying to the ansatz (\ref{eq:metric_static_ansatz},\ref{eq:scalar_ansatz}) for the field equations of interest (\ref{eq:field_eq_4}), they still remain very intricate. Nevertheless, a careful rewriting of the field equations will enable to integrate them exactly in a number of cases. To this aim, one needs to define the following quantities,
\begingroup
\allowdisplaybreaks
\begin{align}
Z\left(X\right)\equiv{}&{}2XG_{4X}-G_4+4X^2F_4,\label{eq:z}\\
Y\left(X\right)\equiv{}&{}12X^2F_5-XG_{5X},\label{eq:y}\\
W\left(X\right)\equiv{}&{}G_{4X}+2XF_4 = \left(Z+G_4\right)/\left(2X\right),\\
\mathcal{A} \equiv{}&{} 4rZ_X-\frac{2X}{f\psi'}\left(r^2G_{3X}+G_{5X}\right)+2f\psi'Y_X-\frac{2}{\psi'}Y+\frac{12q^2X}{h\psi'}F_5\label{eq:cala}\\
={}&{} 4rZ_X-\frac{2X}{f\psi'}\left(r^2G_{3X}+G_{5X}\right)+\frac{q^2}{h \psi'}G_{5X}+\frac{f\psi'}{X}(Y+2Y_X X),\\
\mathcal{B} \equiv{}&{}rZ+f\psi'Y,\\
\mathcal{C} \equiv{}&{}rF_4+3F_5f\psi'.\label{eq:c}
\end{align}
\endgroup
The functions $Z\left(X\right)$ and $Y\left(X\right)$ enable to replace the beyond Horndeski functions, going from the pair $\left(G_4,F_4\right)$ to $\left(G_4,Z\right)$, and from the pair $\left(G_5,F_5\right)$ to $\left(G_5,Y\right)$. The relevance of the combination $G_{4X}+2XF_4$, appearing both in $Z\left(X\right)$ and $W\left(X\right)$, and of the combination $12XF_5-G_{5X}$ appearing in $Y\left(X\right)$, is not surprising, since in the action, $G_{4X}$ and $2XF_4$ multiply the same term $\left[\left(\Box\phi\right)^2-\phi_{\mu\nu}\phi^{\mu\nu}\right]$, while $-G_{5X}/6$ and $2XF_5$ multiply the same term $\Bigl[\left(\Box\phi\right)^3-3\Box\phi\phi_{\mu\nu}\phi^{\mu\nu}+2\phi_{\mu\nu}\phi^{\nu\rho}\phi_\rho^\mu\Bigr]$ [see the explicit form of the $F_4$ and $F_5$ Lagrangians, Eqs. (\ref{eq:l4b}-\ref{eq:l5b})].
\\

The interest lies in solutions where the radial part of the scalar field, $\psi\left(r\right)$, is not constant\footnote{Otherwise $\phi=qt+\text{cst.}$ diverges at the horizon, where $t$ itself diverges.} $\psi'\neq 0$. For convenience, we can thus rescale the independent field equations, from $\left(\mathcal{J}_\text{shift}^r,\mathcal{E}_{rr},\mathcal{E}_{tt}\right)$ to $\left(\mathcal{E}_J,\mathcal{E}_r,\mathcal{E}_t\right)$, where
\begin{equation}
\mathcal{E}_J=-\frac{r^2}{f\psi'}\mathcal{J}_\text{shift}^r,\quad \mathcal{E}_r = r^2f\mathcal{E}_{rr},\quad \mathcal{E}_t = \frac{r^2}{hf}\mathcal{E}_{tt}.
\end{equation}
In terms of the quantities introduced above, the independent field equations $\left(\mathcal{E}_J,\mathcal{E}_r,\mathcal{E}_t\right)$ can be written, after inspection, as:
\begin{align}
\mathcal{E}_J={}&{} -\frac{fh'}{2h}\mathcal{A}+r^2G_{2X}+2G_{4X}-2rf\psi'G_{3X}-2fZ_X\nonumber\\
{}&{} +\frac{2q^2f}{h}\left(W_X+2F_4\right)+\frac{2q^2f}{h}\mathcal{C}\left(\frac{f'}{f}-\frac{h'}{h}\right),\\
\mathcal{E}_r={}&{} -\frac{2fh'}{h}\mathcal{B}-r^2G_2-2G_4-2fZ+\frac{2q^2f}{h}W-\frac{4q^2f}{h}\mathcal{C}X'-f\left(\psi'\right)^2\mathcal{E}_J,\\
\mathcal{E}_t={}&{} X'\mathcal{A}+2\left(\frac{f'}{f}-\frac{h'}{h}\right)\mathcal{B}-\frac{\mathcal{E}_r}{f}-\left(\left(\psi'\right)^2+\frac{q^2}{fh}\right)\mathcal{E}_J.
\end{align}
This rewriting is an identity, i.e. holds true even when the field equations are not satisfied. Now, for a solution of the field equations, $\mathcal{E}_J=\mathcal{E}_r=\mathcal{E}_t=0$, leading to the following system:
\begin{align}
X'\mathcal{A} ={}&{} 2\left(\frac{h'}{h}-\frac{f'}{f}\right)\mathcal{B}, \label{eq1}\\
\frac{fh'}{2h}\mathcal{A} = {}&{} r^2G_{2X}+2G_{4X}-2rf\psi'G_{3X}-2fZ_X\nonumber\\{}&{}+\frac{2q^2f}{h}\left(W_X+2F_4\right)+\frac{2q^2f}{h}\mathcal{C}\left(\frac{f'}{f}-\frac{h'}{h}\right),\label{eq2}\\
\frac{2fh'}{h}\mathcal{B}={}&{}-r^2G_2-2G_4-2fZ+\frac{2q^2f}{h}W-\frac{4q^2f}{h}\mathcal{C}X'\label{eq3}.
\end{align}\newpage
\section[Compatibility conditions for the generic case with $q=0$, $\phi=\phi(r)$]{Compatibility conditions for the generic case with $q=0$, $\phi=\phi(r)$}\label{sec:nonpar}
\sectionmark{4.3~~Compatibility conditions for the generic case with $q=0$, $\phi=\phi(r)$}
One is now faced with a system of three equations~(\ref{eq1}-\ref{eq3}). The next section~\ref{sec:parity} will be interested in the great simplification which occurs in the case of parity symmetry (invariance under $\phi\to -\phi$), which corresponds to $G_3=G_5=F_5=0$. But, as long as the full system~(\ref{eq1}-\ref{eq3}) is fresh in the mind of the reader, let us begin by the generic case, where all beyond Horndeski functions are non-vanishing. This case seems impossible to treat in all generality, but \textbf{much can be said if one is looking for a homogeneous BH, i.e. $h=f$, and a static scalar field, $q=0$}. Then, Eq.~(\ref{eq1}) implies that $X$ is constant, or $\mathcal{A}=0$. Constant $X$ is in general associated with stealth solutions, and we rather decide to look at the case $\mathcal{A}=0$.    
\\

The second equation, (\ref{eq2}), then has a vanishing left hand side, so the right hand side must also vanish. One can choose to make the right hand side vanish by imposing it to be proportional to $\mathcal{A}$, in the following sense: there exists a function of $X$, denoted $Q(X)$, such that
\begin{equation}
r^2G_{2X}+2G_{4X}-2rf\psi'G_{3X}-2fZ_X=f\psi' Q\mathcal{A}.\label{eq:choice}
\end{equation} 
Note that this condition is sufficient for Eq.~(\ref{eq2}) to be satisfied, but is by no means necessary. It will just enable to select a number of \textbf{compatibility conditions} for a generic beyond Horndeski theory to admit closed-form solutions, but \textit{a priori}, other theories could admit such solutions without satisfying these compatibility conditions. Using the definition (\ref{eq:cala}) of $\mathcal{A}$, (\ref{eq:choice}) becomes
\begin{align}
r^2G_{2X}+2G_{4X}{}&{}-2rf\psi'G_{3X}-2fZ_X\nonumber \\{}={}&{}4f\psi'QrZ_X-2XQ\left(r^2G_{3X}+G_{5X}\right)-2fQY-4fXQY_X.\label{eq:compq0}
\end{align}
Therefore, a set of sufficient (but again, not necessary) compatibility conditions for Eq.~(\ref{eq2}) to be verified is
\begin{align}
G_{2X}={}&{}-2XQG_{3X}=4XQ^2Z_X,\label{eq:compq1}\\ 
G_{4X}={}&{}-XQG_{5X},\label{eq:compq2}\\ 
Z_X={}&{}\left(Y+2XY_X\right)Q.\label{eq:compq3}
\end{align}
These conditions are obtained by identifying terms in (\ref{eq:compq0}) with the same powers in $r$, $f$ and $\psi'$. On the other hand, the degeneracy condition (\ref{eq:beyond_degen_bis}) ensuring the absence of ghost degree of freedom, can be rewritten in terms of $Z$ and $Y$ as
\begin{equation}
Y = \frac{XG_{5X}Z}{G_4-2XG_{4X}}.
\end{equation}
Combining this with condition (\ref{eq:compq2}) yields
\begin{equation}
Z=QY\left(2X-\frac{G_4}{G_{4X}}\right).\label{eq:compq4}
\end{equation}
Finally, conditions (\ref{eq:compq3}) and (\ref{eq:compq4}) combine to give
\begin{equation}
2XQ_XY+QY=\left(QY\frac{G_4}{G_{4X}}\right)_X.\label{eq:compq5}
\end{equation}
The five conditions (\ref{eq:compq1}-\ref{eq:compq3}) and (\ref{eq:compq5}) relate the seven  functions $G_{2,3,4,5}$, $Q$, $Z$ and $Y$ (remember that $F_4$ is known from $G_4$ and $Z$, while $F_5$ from $G_5$ and $Y$). In this framework, only two, out of these seven functions of $X$, are therefore independent. 
\\

To summarize, when the compatibility conditions (\ref{eq:compq1}-\ref{eq:compq3}) are satisfied, the original system of equations~(\ref{eq1}-\ref{eq3}), with $q=0$, reduces to
\begin{align}
X'\mathcal{A} ={}&{} 2\left(\frac{h'}{h}-\frac{f'}{f}\right)\mathcal{B},\label{eq1comp} \\
0={}&{}\left(\frac{h'}{2h}-\psi'Q\right)\mathcal{A},\label{eq2comp}\\
\frac{2fh'}{h}\mathcal{B}={}&{}-r^2G_2-2G_4-2fZ,\label{eq3comp}
\end{align}
where the rewriting~(\ref{eq2comp}) comes from the definition of $Q$, Eq. (\ref{eq:compq0}). Moreover, taking into account the compatibility conditions (\ref{eq:compq1}-\ref{eq:compq3}), the original expression of $\mathcal{A}$, Eq. (\ref{eq:cala}), simplifies to
\begin{equation}
\mathcal{A}=-\frac{2Z_X}{Q\sqrt{f}\sqrt{-2X}}\Bigl[\left(\sqrt{f}-rQ\sqrt{-2X}\right)^2-\frac{G_{4X}}{Z_X}\Bigr].\label{calarew}
\end{equation}
The above compatibility conditions were motivated by the case $f=h$, and therefore the system of equations~(\ref{eq1comp}-\ref{eq3comp}) naturally simplifies in this case. The first equation~(\ref{eq1comp}) leads to $\mathcal{A}=0$ (if we assume $X$ not constant). Then, thanks to the rewriting~(\ref{calarew}), $\mathcal{A}=0$ can be directly solved for $X$, or equivalently for the scalar field $\phi=\psi(r)$, in terms of $f(r)$. Also, $\mathcal{A}=0$ ensures that the second equation~(\ref{eq2comp}) is verified. Finally, the third equation~(\ref{eq3comp}) is a first order differential equation for $f(r)$. 
\\

As a consequence, all what remains to do is to fix two functions among $G_{2,3,4,5}$, $Q$, $Z$ and $Y$, then compute the remaining ones by virtue of the conditions (\ref{eq:compq1}-\ref{eq:compq3}) and (\ref{eq:compq5}), and find the associated homogeneous solution, that is, $\phi=\psi(r)$ and $f=h$. The most interesting case, as presented in our original article~\cite{Bakopoulos:2022csr}, is motivated as follows. In the state-of-the-art of previously existing BH solutions, Chap.~\ref{sec:state}, the only shift-symmetric Horndeski theory without parity symmetry, and allowing closed-form solutions, is the shift-symmetric 4DEGB theory, see more precisely paragraph~\ref{subsec:bhegb}. The shift symmetry is associated to a KK reduction along a flat internal space. The Horndeski functions are given by Eq. (\ref{eq:horn_conf_bis}) with $\beta=\lambda=0$, so there is a unique coupling constant $\alpha$,
\begin{equation}
G_2 = 8\alpha X^2,\quad G_3 = 8\alpha X,\quad G_4 = 1+4\alpha X,\quad G_5 = 4\alpha\ln\left\lvert X\right\rvert.\label{eq:4degbreminder}
\end{equation}
Very interestingly, these functions satisfy the compatibility conditions (\ref{eq:compq1}-\ref{eq:compq3}) and (\ref{eq:compq5}), with a constant function $Q(X)$. Let us generalize this case by fixing from now on a constant value for $Q$,
\begin{equation}
Q=-\gamma_1,
\end{equation}
where $\gamma_1$ is an arbitrary dimensionless constant. Eq. (\ref{eq:compq5}) leads to
\begin{equation}
Y=\gamma_2 G_{4X},\label{eq:yval}
\end{equation}
where $\gamma_2$ is an integration constant. Eq. (\ref{eq:compq3}) then gives
\begin{equation}
Z = \gamma_1\gamma_2\left(G_4-2XG_{4X}\right).\label{eq:zval}
\end{equation}
We have fixed $Q=-\gamma_1$, so there remains a unique function to fix in order to determine completely the theory. Since $Z$ and $Y$ are seen to be determined by $G_4$, we fix $G_4$. We choose it to be similar to the 4DEGB case (\ref{eq:4degbreminder}), namely a constant term $1$ corresponding to a canonical Einstein-Hilbert term, plus a power of $X$:
\begin{equation}
G_4=1-2\alpha\left(-2X\right)^n.
\end{equation}
The power $n$ is \textit{a priori} an arbitrary real number, and $n=1$ corresponds to the shift-symmetric 4DEGB theory (\ref{eq:4degbreminder}). Then, the remaining equations (\ref{eq:compq1}-\ref{eq:compq2}) give
\begin{align}
G_{2}={}&{}-4\alpha\gamma_1^3\gamma_2n(2n-1)\frac{(-2X)^{n+1}}{n+1},\\ G_{3}={}&{}4\alpha\gamma_1^2\gamma_2 (2n-1)(-2X)^{n},\\
G_4 ={}&{} 1-2\alpha(-2X)^n,\label{eq:g4comp}\\ 
G_{5}={}&{}\frac{4\alpha n}{\gamma_1}\frac{(-2X)^{n-1}}{n-1}.\label{eq:g5compfirst}
\end{align}
Note however that for particular values of $n$ (those for which the denominators vanish), the integration rather leads to logarithms, for example $G_5$ with $n=1$, consistently with (\ref{eq:4degbreminder}). Note also that one could include a cosmological constant term $-2\Lambda$ in $G_2$. We have recalled the form of $G_4$ in order to have all the Horndeski functions at a glance. It remains however to compute the beyond Horndeski functions $F_4$ and $F_5$. Indeed, combining the definitions (\ref{eq:z}-\ref{eq:y}) of $Z$ and $Y$, their current value (\ref{eq:yval}-\ref{eq:zval}), and the form (\ref{eq:g4comp}-\ref{eq:g5compfirst}) of $G_4$ and $G_5$, one gets
\begin{align}
F_4 ={}&{} \frac{1+\gamma_1\gamma_2}{4X^2}\left(1+2(2n-1)\alpha(-2X)^n\right),\\ F_5={}&{}\frac{4\left(1+\gamma_1\gamma_2\right)}{3\gamma_1}\alpha n(-2X)^{n-3}.
\end{align}
The role of the constants $\gamma_1$ and $\gamma_2$ thus becomes clear: the theory under consideration belongs to the Horndeski class if and only if $\gamma_1\gamma_2=-1$, otherwise, it is of the beyond Horndeski type. In particular, for $\gamma_1\gamma_2=-1$ and $n=1$, one retrieves the 4DEGB case (\ref{eq:4degbreminder}).  
\\

One can now turn to the integration of the field equations~(\ref{eq1comp}-\ref{eq3comp}), in the case $f=h$. Using~(\ref{calarew}), $\mathcal{A}=0$ gives the scalar field $\phi=\psi(r)$,
\begin{equation}
\phi(r)=\int\frac{1-\sqrt{\gamma_1\gamma_2(1-2n)f(r)}}{r\gamma_1\sqrt{\gamma_1\gamma_2(1-2n)f(r)}}\mathrm{d}r.\label{eq:scalar_gen}
\end{equation}  
We will comment on the regularity of the scalar field later, for the moment, let us finish the integration of the field equations by solving~(\ref{eq3comp}). Because of the form of $\phi(r)$, one is tempted to introduce a function
\begin{equation}
F(r)^2\equiv \gamma_1\gamma_2(1-2n)f(r).\label{eq:defF}
\end{equation}
Then, the last field equation~(\ref{eq3comp}) can be integrated once with respect to the radial coordinate r. This integration yields the following equation:
\begin{equation}
    (n+1)\left[\gamma_1^3 \gamma_2 (1-2 n)\right]^n r^{2n}\left(1-2n+F^2\right)-2\alpha\left(1-F\right)^{2n}\left(1+2nF+F^2\right)-\mu r^{2n-1}=0, \label{eqalgsqrt}
\end{equation}
where $\mu$ is an integration constant. The above is an algebraic equation with degree $2(n+1)$ in $F$. For $n\geq 1$, integer or half-integer, it becomes a polynomial equation in $F$. Due to the link (\ref{eq:defF}) between $F(r)$ and the metric function $f(r)$, the equation is properly defined only when $f(r)\geq 0$, however, for the case $n=1$, appear only $F^0$, $F^2$ and $F^4$, and one gets a proper polynomial in $f(r)$ at any point of spacetime. 
The explicit solution reads
\begin{equation}
    f(r) = -\frac{1}{\gamma_1 \gamma_2}+\frac{r^2\gamma_1^2}{2\alpha}\left(1\pm\sqrt{1-\frac{2\alpha\mu}{r^3\gamma_1^6\gamma_2^2}}\right).
\end{equation}
Remember that the theory belongs to the Horndeski class (rather than beyond Horndeski) if and only if $\gamma_1\gamma_2=-1$. This is why the above $f(r)$ profile reduces to the BH solution of the shift-symmetric 4DEGB theory, Eq. (\ref{eq:met_lu_pang}), when $\gamma_1 \gamma_2=-1$. Note that the extension of the solution in beyond Horndeski, $\gamma_1\gamma_2\neq -1$, picks up a solid angle deficit: $f(r)\to -1/(\gamma_1\gamma_2)$ when $r\to\infty$ (as explained in paragraph~\ref{subsec:bhegb}, we are interested in the $-$ branch, since the $+$ branch presents a (a)dS behaviour).
\\

Let us now move on to the general $n$ case. Equation (\ref{eqalgsqrt}) implies the following asymptotic behaviour for the metric function:
\begin{equation}
    f(r)=-\frac{1}{\gamma_1\gamma_2}+\frac{\mu}{(n+1)\gamma_1\gamma_2 (1-2n)\left[\gamma_1^3\gamma_2(1-2n)\right]^n}\frac{1}{r} +\mathcal{O}\left(\frac{1}{r^{2n}}\right). \label{eqasymplupanggeneral}
\end{equation}
Hence the scalar field (\ref{eq:scalar_gen}) behaves at infinity as
\begin{equation}
    \phi(r)=\frac{1-\sqrt{2n-1}}{\gamma_1\sqrt{2n-1}}\ln(r)+\mathcal{O}\left(\frac{1}{r}\right).
\end{equation}
The scalar field diverges like $\ln(r)$ at infinity, except for $n=1$ where $\phi=\mathcal{O}\left(\frac{1}{r}\right)$. However, the kinetic term $X=-f (\phi')^2/2$ always vanishes at infinity. As regards the metric, it reaches asymptotically Minkowski spacetime only for the Horndeski case, $\gamma_1 \gamma_2=-1$. If $\gamma_1 \gamma_2\neq -1$, the asymptotic metric is only locally asymptotically flat: there is a global deficit angle as in the case of the gravitating monopole solution in GR~\cite{Barriola:1989hx}. In what follows, we thus focus on the asymptotically flat, Horndeski case, $\gamma_1 \gamma_2=-1$. Without loss of generality, one can set $\gamma_1=1$ and $\gamma_2=-1$, since $\gamma_1$ can always be absorbed by a redefinition of the coupling constant $\alpha$. The asymptotic expansion~(\ref{eqasymplupanggeneral}) shows that the integration constant $\mu$ is related to the ADM mass $M$ by
\begin{equation}
    \mu = -2M(n+1)(2n-1)^{n+1}.
\end{equation}
\textbf{Let us summarize: the Horndeski theory
\begin{align}
G_{2}={}&{}4\alpha n(2n-1)\frac{(-2X)^{n+1}}{n+1},\quad G_{3}=-4\alpha (2n-1)(-2X)^{n},\nonumber\\ G_4 ={}&{} 1-2\alpha(-2X)^n,\quad G_{5}=4\alpha n\frac{(-2X)^{n-1}}{n-1},\label{eq:g5comp}
\end{align}
admits a homogeneous ($f=h$) BH solution, given as the root of an algebraic equation
\begin{align}
    {}&{} 0=(n+1)\left(2n-1\right)^n r^{2n-1}\left[(2n-1)(2M-r)+rF^2\right]-2\alpha\left(1-F\right)^{2n}\left(1+2nF+F^2\right),\nonumber\\ {}&{}F(r)^2\equiv (2n-1)f(r). \label{eqsimplen}
\end{align}
The scalar field supporting this BH is
\begin{equation}
\phi(r) = \int\frac{1-\sqrt{(2n-1)f(r)}}{r\sqrt{(2n-1)f(r)}}\mathrm{d}r,\label{eq:phiprimenonpar}
\end{equation}
showing that one needs $n>1/2$}. Note that the existence of these BH solutions is compatible with the Hui-Nicolis no-hair theorem (paragraph~\ref{subsec:huinic}) because of the absence of canonical kinetic term $X\subset G_2$. Indeed, this would correspond in (\ref{eq:g5comp}) to $n=0$, which in fact removes all Horndeski functions.
\\ 

The algebraic equation~(\ref{eqsimplen}) cannot be solved analytically for $n\neq 1$: the BH solutions are therefore known implicitly. This is one of the rare exceptions to the declared aim of this thesis, namely finding closed-form solutions. Still, using~(\ref{eqsimplen}), much can be said about the behaviour of the solution near $r=0$ or for small coupling $\alpha$, and about the horizon structure. The interested reader is referred to~\cite{Bakopoulos:2022csr} as regards the behaviour for small $r$ or $\alpha$. Notably, the main conclusion concerning the $r\to 0$ region is that, although $f(r)$ is finite at $r=0$, there still exists a curvature singularity there because $f(r)$ is not sufficiently regular near $r=0$, very much like in the 4DEGB case. We rather decide to focus here on the horizons. 
\\

The existence of a horizon is indicated by the vanishing of the metric function $f(r)$. Thus, setting $F(r)=0$ in~(\ref{eqsimplen}) defines the horizon radius as the value $r_h$ which satisfies the equation:
\begin{equation}
    (n+1)\left(2n-1\right)^{n+1} r_h^{2n-1}\left(2M-r_h\right)=2\alpha\,. \label{eqhorizonn}
\end{equation}
As expected, the coupling parameter $\alpha$ induces a deviation from the Schwarzschild radius $r_\text{Sch}=2M$. For clarity and in order to compare with the GR limit, we restrict from now on to the case $M>0$. The presence of the coupling parameter $\alpha$ in~\eqref{eqhorizonn} also determines the number of roots of that equation. Given the form of Eq.~\eqref{eqhorizonn} as a polynomial of order $2n$ in $r_h$, much can be said about its real, positive roots. To this end, we define the quantities:
\begin{equation}
    r_n\equiv\left(1-\frac{1}{2n}\right)2M<2M,\qquad \alpha_n\equiv\frac{1}{2}(n+1)\left(2n-1\right)^{3n} \left(\frac{M}{n}\right)^{2n}>0.
\end{equation}
The following results are then easy to prove. If $\alpha<0$, there is a unique horizon, with  $r_h>2M$. If $0<\alpha<\alpha_n$, there are exactly two horizons, with $r_{h-}<r_n<r_{h+}<2M$. One has that $r_{h+}\to \left(2M\right)^-$ when $n\to\infty$, and $r_{h-}\to 0^+$ when $\alpha\to 0^+$. If $\alpha>\alpha_n$, the spacetime has no horizons. These results are consistently illustrated by the 4DEGB case, which corresponds to $n=1$, where
\begin{equation}
    r_n =M,\quad \alpha_n = M^2,\quad r_{h\pm} = M\pm\sqrt{M^2-\alpha}.
\end{equation}
Before moving on to the parity-symmetric case, let us announce a result which will be obtained in paragraph~\ref{subsec:lastconj}. There, a regularized KK reduction of the cubic Lovelock invariant $\mathcal{R}^{(3)}$ is performed down to four dimensions along a flat internal space, and the obtained theory belongs to the theories identified above, Eq. (\ref{eq:g5comp}), with $n=2$. Therefore, as regards KK reduction along a flat internal space, the reduction of the quadratic Lovelock invariant, that is, the GB scalar, leads to (\ref{eq:g5comp}) with $n=1$ (4DEGB), while the reduction of the cubic Lovelock invariant leads to (\ref{eq:g5comp}) with $n=2$. One can thus propose the following conjecture: \textbf{the KK reduction, along a flat internal space, of the Lovelock action of order $k$, corresponds to the Horndeski theory (\ref{eq:g5comp}) with $n=k-1$}. This remains a conjecture for the moment, although strongly motivated by the considerations developed in Sec.~\ref{sec:diagkk}. 
\section{Parity-symmetric simplification}\label{sec:parity}
\sectionmark{4.4~~Parity-symmetric simplification}
The system of equations~(\ref{eq1}-\ref{eq3}) has thus been shown to admit solutions when all the beyond Horndeski functions are non-vanishing, and although these solutions are implicit, much can be said about their horizon structure. However, the form of the system~(\ref{eq1}-\ref{eq3}), and the new quantities (\ref{eq:z}-\ref{eq:c}), show that the solving is much easier when $G_3=G_5=F_5=0$, i.e. when the theory acquires parity symmetry under $\phi\to-\phi$. 
\\

Indeed, in this case, $Y\left(X\right)=0$, $\mathcal{A}=4rZ_X$, $\mathcal{B}=rZ$, such that the radial part of the scalar, $\psi'\left(r\right)$, disappears completely: only the kinetic term $X$ remains, and solving for $X$ will determine the scalar field through (\ref{eq:x_shift}). Of course, even when $G_3$, $G_5$ and $F_5$ are present, one can in principle replace every $\psi'$ in terms of $X$ thanks to (\ref{eq:x_shift}) again, but this would introduce complicated terms since $\psi'=\pm\sqrt{q^2/h-2X}/f$. Also, and as we will see in a moment, only when $G_3=G_5=F_5=0$ does equation~(\ref{eq1}) integrate directly. For all these reasons, we now restrict to theories with $G_3=G_5=F_5=0$. The theories considered from now on in this section therefore have the following action,
\begin{align}
S\left[g_{\mu\nu},\phi\right] = \int\mathrm{d}^4 x\sqrt{-g}\Bigl\{G_2\left(X\right)+G_4\left(X\right){}&{}R+G_{4X}\left[\left(\Box\phi\right)^2-\phi_{\mu\nu}\phi^{\mu\nu}\right]\nonumber\\{}&{}+F_4\left(X\right)\epsilon^{\mu\nu\rho\sigma}\epsilon^{\alpha\beta\gamma}_{\hspace{0.5cm}\sigma}\phi_\mu\phi_\alpha\phi_{\nu\beta}\phi_{\rho\gamma}\Bigr\},\label{eq:shift_parity_beyond_action}
\end{align}
which depends on three functions $G_2\left(X\right)$, $G_4\left(X\right)$ and $F_4\left(X\right)$, and, with the ansatz (\ref{eq:metric_static_ansatz}-\ref{eq:scalar_ansatz}), the independent field equations~(\ref{eq1}-\ref{eq3}) reduce to:
\begin{align}
2X'Z_X ={}&{} \left(\frac{h'}{h}-\frac{f'}{f}\right)Z, \label{eq1bis}\\
\frac{2fh'}{h}rZ_X = {}&{} r^2G_{2X}+2G_{4X}-2fZ_X+\frac{2q^2f}{h}\left(W_X+2F_4\right)+\frac{2q^2f}{h}\mathcal{C}\left(\frac{f'}{f}-\frac{h'}{h}\right),\label{eq2bis}\\
\frac{2fh'}{h}rZ={}&{}-r^2G_2-2G_4-2fZ+\frac{2q^2f}{h}W-\frac{4q^2f}{h}\mathcal{C}X'\label{eq3bis}.
\end{align}
In the following, we again restrict to the case of non-constant kinetic term $X$: $X'\neq 0$. 
\section{Solving the field equations and the method of the function $\mathfrak{G}$}
\sectionmark{4.5~~Solving the field equations and the method of the function $\mathfrak{G}$}
Eq.~(\ref{eq1bis}) integrates directly to
\begin{equation}
\frac{f}{h}=\frac{\gamma^2}{Z^2},\label{eq:simple}
\end{equation}
where $\gamma\neq 0$ is a constant of integration. A homogeneous BH ($f=h$) corresponds to $Z=\gamma$ (more generally if $Z=\text{const.}$, this constant can be rescaled to $\gamma$, by rescaling $h$, by rescaling the time coordinate $t$), while a non-homogeneous BH ($f\neq h$) is obtained as soon as $Z$ depends non-trivially on $X$. The combination $Z\times(\ref{eq2bis})-Z_X\times(\ref{eq3bis})$ then gives
\begin{equation}
r^2\left(G_2Z\right)_X+2\left(G_4Z\right)_X\left(1-\frac{q^2\gamma^2}{2Z^2X}\right)=0.\label{eq:simple2}
\end{equation}
Since the metric functions $f$ and $h$ do not appear in this equation, this equation determines $X=X\left(r\right)$, and thus the scalar field $\phi=qt+\psi\left(r\right)$ through (\ref{eq:x_shift}).  $X$ is determined as a function of $r$, $f$ is related to $h$ by (\ref{eq:simple}), and finally, $h$ is determined by~(\ref{eq3bis}), which simplifies as:
\begin{align}
2\gamma^2 \left(h\left(r\right) r-\frac{q^2 r}{2 X}\right)'+r^2 G_2 Z+2G_4 Z={}&{}\frac{q^2\gamma^2}{ZX^2}\left[XG_4+rX'\left(2XG_{4X}-G_4\right)\right]\nonumber\\ {}={}&{} \frac{2q^2\gamma^2}{Z} \sqrt{\frac{r}{X}}\left(\sqrt{\frac{r}{X}}G_4 \right)'.\label{eq:simple3}
\end{align}
\textbf{The whole system of field equations for a static, spherically-symmetric metric (\ref{eq:metric_static_ansatz}) and scalar field with linear time dependence (\ref{eq:scalar_ansatz}), for the generic beyond Horndeski action with shift and parity symmetry (\ref{eq:shift_parity_beyond_action}), has therefore been rephrased as three short equations (\ref{eq:simple}-\ref{eq:simple3}), where the beyond Horndeski function $F_4\left(X\right)$ has been replaced by the function $Z\left(X\right)$ defined by (\ref{eq:z}).}
\\

In order to integrate Eq. (\ref{eq:simple2}), introduce a function\footnote{This function is denoted $\mathcal{G}$ in the original article~\cite{Bakopoulos:2022csr}. Here, $\mathfrak{G}$ is used to avoid any confusion with the GB curvature scalar.} $\mathfrak{G}=\mathfrak{G}\left(X\right)$ such that, when the field equations are satisfied, its derivative $\mathfrak{G}_X$ with respect to $X$ can be expressed in terms of $X$ and $r$ as
\begin{equation}
\mathfrak{G}_X=\frac{\alpha r^2+\beta\left(1-\frac{q^2\gamma^2}{2Z^2X}\right)}{\epsilon r^2+\delta\left(1-\frac{q^2\gamma^2}{2Z^2X}\right)},\label{eq:gx}
\end{equation}
where $\alpha$, $\beta$, $\epsilon$ and $\delta$ are constants, with $\delta\neq 0$. It is always possible to define such a $\mathfrak{G}$ at least locally: since we assume that $X\left(r\right)$ is not constant, there exist points where the inverse function $r\left(X\right)$ is defined. Then, one can easily check that Eq. (\ref{eq:simple2}) is verified provided the functions of $X$, $G_2Z$ and $G_4Z$, which appear in this equation, are related to the function $\mathfrak{G}$ as
\begin{align}
    G_2Z={}&{}\epsilon\, \mathfrak{G} -\alpha X+ C,\label{eq:g2z}\\
    2G_4Z={}&{}\delta\, \mathfrak{G}-\beta X+ D,\label{eq:g4z}
\end{align}
where $C$ and $D$ are constants. Importantly, \textbf{conditions (\ref{eq:gx}-\ref{eq:g4z}) are sufficient conditions for integrating the field equations, but not necessary}: we will give an example below where these conditions are not verified, but a BH solution is obtained. Still, they remain interesting conditions, facilitating the solving, and we refer to it as \textbf{the method of the function $\mathfrak{G}$}. The field equations are then completely integrated, and both the theory functions $G_2$, $G_4$ and $F_4$, and the corresponding solution [that is to say, the functions $h\left(r\right)$, $f\left(r\right)$ and $X\left(r\right)$], are found by following the algorithmic steps:
\begin{enumerate}
\item Among the four functions $\mathfrak{G}\left(X\right)$, $Z\left(X\right)$, $G_2\left(X\right)$ and $G_4\left(X\right)$, \textbf{choose} the form of two of them.
\item The remaining two functions are determined by (\ref{eq:g2z}) and (\ref{eq:g4z}). One then knows $\mathfrak{G}$, $Z$, $G_2$ and $G_4$.
\item $F_4\left(X\right)$ is determined by the definition (\ref{eq:z}) of $Z$.
\item The kinetic term of the scalar field, $X$, is determined as a function of $r$ by (\ref{eq:gx}).
\item The metric function $h\left(r\right)=-g_{tt}$ is given by the first order differential equation (\ref{eq:simple3}).
\item The metric function $f\left(r\right)=g^{rr}$ is given by (\ref{eq:simple}).
\end{enumerate}
These steps were followed in our original articles~\cite{Bakopoulos:2022csr,Bakopoulos:2023fmv}, and yielded solutions that we are going to present in a moment. Of course, not all expressions of $\mathfrak{G}$ and $Z$ enable to get a closed-form solution for $X\left(r\right)$ via (\ref{eq:gx}) or for the metric functions via (\ref{eq:simple3}). In addition, the algorithmic solving just presented, although necessary in a first time, forces one to work with many coupling constants [at least the $\gamma$, $\alpha$, $\beta$, $\epsilon$ and $\delta$ of Eqs. (\ref{eq:simple}) and (\ref{eq:gx})], while it is possible in a second time to remove a lot of these coupling constants, either because they are redundant or because they must be fixed to a certain value to give a relevant asymptotic behaviour. As a consequence, we will detail the algorithmic procedure only in the next section (where in fact the step 3 above is absent because we are looking into the pure Horndeski case $F_4=0$), in order to make this procedure clear, but then we will proceed with a more simple presentation of the most relevant results. 
\section{Pure Horndeski, homogeneous black holes}\label{sec:pure}
\sectionmark{4.6~~Pure Horndeski, homogeneous black holes}
Let us start by the case of Horndeski theory, by removing the beyond Horndeski function: $F_4=0$. Also, the metric is assumed homogeneous: $f=h$, which implies $Z=\gamma$, see (\ref{eq:simple}). Since $F_4=0$, Eq. (\ref{eq:z}) yields
\begin{equation}
G_4=-\gamma+\eta\sqrt{\left\lvert X\right\rvert},\label{eq:g4horn}
\end{equation} 
with $\eta$ a coupling. One then computes $\mathfrak{G}$, which yields on the one hand $G_2$,
\begin{equation}
G_2=\frac{\zeta}{\gamma}X+2\kappa\eta\sqrt{\left\lvert X\right\rvert}-2\Lambda,\label{eq:g2horn}
\end{equation}
and on the other hand $X$, which is determined by the algebraic equation
\begin{equation}
\pm 2\zeta r^2X\sqrt{\pm X}+2X\gamma\eta\left(1+\kappa r^2\right)=q^2\gamma\eta.\label{eq:alg_x_horn}
\end{equation}
In these last equations, we have defined the new coupling constants:
\begin{equation}
\kappa \equiv \frac{\epsilon}{\delta},\quad \zeta \equiv \beta\kappa-\alpha,\quad \Lambda \equiv \frac{\epsilon\left(D+2\gamma^2\right)-C\delta}{2\gamma\delta},
\end{equation}
where $\Lambda$ is seen to appear in $G_2$ as a usual, bare cosmological constant. 
\subsection{Case $q=0$, $\phi=\phi\left(r\right)$}
We first assume $\phi=\phi\left(r\right)$, i.e. $q=0$ in $\phi=qt+\psi\left(r\right)$, and $\psi\left(r\right)$ then coincides with $\phi\left(r\right)$. Then, the kinetic term reduces to $X=-f\left(\phi'\right)^2/2$. Therefore, in order for the scalar field to be real at least outside the horizon (where $f=g^{rr}>0$), $X$ must be negative. It can then be inferred from Eq. (\ref{eq:alg_x_horn}) that $\gamma\eta\zeta\left(1+r^2\kappa\right)$ must be positive, and that
\begin{equation}
X = -\gamma^2\eta^2\frac{\left(1+r^2\kappa\right)^2}{\zeta^2\, r^4}.\label{eq:x_radial}
\end{equation}
The metric functions $h=f$ are obtained thanks to Eq. (\ref{eq:simple3}). One finds
\begin{equation}
h\left(r\right)=1-\frac{\eta^2\kappa}{\zeta}-\frac{2M}{r}+r^2\left[\frac{\Lambda}{3\gamma}-\frac{\eta^2\kappa^2}{6\zeta}\right]+\frac{\eta^2}{2\zeta r^2}.
\end{equation}
In the case where the coupling $\kappa$ is not zero, the constant term in the metric function is not equal to $1$, but rather differs from $1$ by a term determined by the coupling constants, $-\eta^2\kappa/\zeta$. Also, the (a)dS term, proportional to $r^2$, does not vanish if the bare cosmological constant $\Lambda=0$: only a nonzero $\Lambda$, adjusted with respect to the couplings, can remove the (a)dS term. On the other hand, the case $\kappa=0$ does not possess these unpleasant features: the metric function then reads
\begin{equation}
h\left(r\right) = 1-\frac{2M}{r}+r^2\frac{\Lambda}{3\gamma}+\frac{\eta^2}{2\zeta r^2}.
\end{equation}
In order to get the usual expression for the cosmological term, one must fix $\gamma=-1$: by inspection of (\ref{eq:g4horn}-\ref{eq:g2horn}), this is in any case equivalent to a mere rescaling of the couplings.
\\

\textbf{To sum up, with the method of the functon $\mathfrak{G}$, the only pure Horndeski action with shift and parity symmetry, i.e. parameterized by two functions $G_2\left(X\right)$ and $G_4\left(X\right)$, admitting as a solution a static, spherically-symmetric, homogeneous BH dressed with a scalar field $\phi=\phi\left(r\right)$ and displaying the usual asymptotics, is
\begin{equation}
G_2=-\zeta X-2\Lambda,\quad G_4=1+\eta\sqrt{-X},
\end{equation}
where the product $\eta\zeta$ must be negative, and the associated BH is $\mathrm{d}s^2=-h\mathrm{d}t^2+\mathrm{d}r^2/h+r^2\mathrm{d}\Omega^2$ with
\begin{equation}
h\left(r\right)=1-\frac{2M}{r}-r^2\frac{\Lambda}{3}+\frac{\eta^2}{2\zeta r^2}.\label{eq:bcl_bis_0}
\end{equation}
The corresponding scalar field and its kinetic term, coming from (\ref{eq:x_radial}), read
\begin{equation}
\phi\left(r\right)=\pm\frac{\eta\sqrt{2}}{\zeta}\int\frac{\mathrm{d}r}{r^2\sqrt{h\left(r\right)}},\quad X=-\frac{\eta^2}{\zeta^2r^4}.\label{eq:bcl_bis}
\end{equation}}Interestingly, this is nothing but the BCL BH~\cite{Babichev:2017guv} presented in paragraph~\ref{subsec:bcl}. The method of the function $\mathfrak{G}$ thus enables to find what remains, up to date, the only non-stealth homogeneous BH in a Horndeski theory with only $G_2$ and $G_4$ and a static scalar field ($q=0$). It should be stressed upon again that the method of the function $\mathfrak{G}$ is a sufficient condition to integrate the field equations, but not necessary. As a consequence, there may \textit{a priori} exist other BH solutions in such Horndeski theories, but they remain unknown up to date.
\subsection{Case $q\neq 0$, $\phi=qt+\psi\left(r\right)$}
In the other case where the scalar field has a linear time dependence, the $\gamma$ coupling can in the same way be taken to be $-1$. The theory functions and the kinetic term, Eqs. (\ref{eq:g4horn}-\ref{eq:alg_x_horn}), are given by
\begin{align*}
G_4={}&{}1+\eta\sqrt{\left\lvert X\right\rvert},\quad\hspace{-0.1cm} G_2=-\zeta X+2\kappa\eta\sqrt{\left\lvert X\right\rvert}-2\Lambda,\\ 0={}&{}\pm 2\zeta r^2X\sqrt{\pm X}-2X\eta\left(1+\kappa r^2\right)+q^2\eta.
\end{align*}
Taking into account this algebraic equation verified by $X\left(r\right)$, Eq. (\ref{eq:simple3}) integrates to
\begin{equation}
h\left(r\right)=1-\frac{2M}{r}-r^2\frac{\Lambda}{3}+\frac{\zeta}{2r}\int r^2X\left(r\right)\mathrm{d}r.\label{eq:480}
\end{equation}
The solution therefore has a Schwarzschild(-(a)dS) behaviour, modified by the last term. On the one hand, if $\zeta\neq 0$, the algebraic equation can be solved in the limit $r\to\infty$, yielding for this last term
\begin{equation}
\frac{\zeta}{2r}\int r^2X\left(r\right)\mathrm{d}r=\left\lbrace\begin{split}& \frac{q^2\zeta}{4\kappa}\pm\frac{q^3\zeta^2\ln r}{4\sqrt{2}\eta\kappa^{5/2}r}+\mathcal{O}\left(\frac{1}{r^2}\right)\text{ if }\kappa\neq 0\\ &\pm\frac{r^2\eta^2\kappa^2}{6\zeta}\pm\frac{\eta^2\kappa}{\zeta}-\frac{q^2\zeta}{2\kappa}+\mathcal{O}\left(\frac{1}{r^2}\right)\text{ if }\kappa\neq 0\text{ (other branches)}\\ & \pm r^{2/3}\frac{3\zeta/10}{2^{2/3}}\left(\frac{-q^2\eta}{\zeta}\right)^{2/3}+\mathcal{O}\left(\frac{1}{r^{2/3}}\right)\text{ if }\kappa= 0\\ \end{split}\right.
\end{equation}
This adds very unusual terms: $r^{2/3}$ diverges but not like a cosmological constant term, $\ln r/r$ vanishes as $r\to\infty$ but dominates the mass term. In the second line, an effective cosmological constant term, depending on the couplings, appears, as well as a non-vanishing constant term. On the other hand, for $\zeta=0$, the metric (\ref{eq:480}) reduces to Schwarzschild and the kinetic term acquires a closed-form expression permitting to compute the scalar field. 
\\

At the end of the day, \textbf{with the method of the function $\mathfrak{G}$, the only Horndeski action with shift and parity symmetry, admitting as a solution a static, spherically-symmetric, homogeneous BH dressed with a scalar field $\phi=qt+\psi\left(r\right)$ and displaying the usual asymptotics is
\begin{equation}
G_2=2\kappa\eta\sqrt{\left\lvert X\right\rvert}-2\Lambda,\quad G_4=1+\eta\sqrt{\left\lvert X\right\rvert},\label{eq:g2g4newst}
\end{equation}
and the associated BH is Schwarzschild(-dS): $f=h=1-2M/r-\Lambda r^2/3$, while the scalar field and its kinetic term read
\begin{equation}
\phi=q\left(t\pm\int\frac{\sqrt{1-\frac{h\left(r\right)}{1+\kappa r^2}}}{h\left(r\right)}\mathrm{d}r\right),\quad X=\frac{q^2/2}{1+\kappa r^2}.\label{eq:phinewst}
\end{equation}}As explained e.g. in paragraph~\ref{subsec:stealthsch}, the regularity of the scalar field at the horizon can be verified by changing to horizon-crossing coordinates, with advanced time $v$ or retarded time $u$. Indeed, if a $+$ sign is chosen in $\phi$, then $\phi$ behaves like $qv$ at the horizon, so is finite at the future event horizon. If a $-$ sign is chosen, then $\phi$ is rather finite at the past event horizon. However, it is not possible to achieve regularity of the scalar field at both the future and past event horizons.\\

The existence of this hairy \textbf{new stealth Schwarzschild BH} is compatible with the Hui-Nicolis no-hair theorem of paragrapgh~\ref{subsec:huinic}. Indeed, the scalar field has a linear time dependence, and $G_2$ and $G_4$ are not analytic, so two assumptions are violated. However, the norm $\mathcal{J}^\mu_\text{shift}\mathcal{J}_{\text{shift}\,\mu}$ of the current is everywhere finite. This is not trivial \textit{a priori}: although $\mathcal{J}^r_\text{shift}=0$, the current has a non-trivial $\mathcal{J}^t_\text{shift}$ component because of the time dependence of $\phi$. This norm has value
\begin{equation}
\mathcal{J}^\mu_\text{shift}\mathcal{J}_{\text{shift}\,\mu}=-\frac{8\eta^2\kappa^2h(r)}{1+\kappa r^2}.
\end{equation}
It is worth appreciating the difference between this new stealth Schwarzschild BH, and the usual stealth Schwarzschild BHs presented in paragraph~\ref{subsec:stealthsch}. For the usual stealth solutions, the kinetic term is a constant, $X=X_0$. This is not a problem in itself, however, it was seen in paragraph~\ref{subsec:stealthsch} that for the asymptotically flat case, the kinetic/k-essence term of the scalar field had to admit a double root at $X_0$: $G_2(X_0)=G_{2X}(X_0)=0$. This leads to perturbative problems, that is, non-hyperbolic equations for the scalar field perturbations~\cite{Babichev:2018uiw}. \\

Here, for the new stealth Schwarzschild, there is a proper $G_2$ term (\ref{eq:g2g4newst}), even if $\Lambda=0$, and $X$ is not constant. Given these differences with the usual stealth solutions, it would be interesting to study the perturbative aspects of the new stealth Schwarzschild, in order to see if they possess a better behaviour than the one of the usual stealth Schwarzschild solutions. 
\\

The case of homogeneous BHs in Horndeski theories, $F_4=0$, has thus been treated with the method of the function $\mathfrak{G}$. Again, this method is not exhaustive, and other solutions might exist which do not fit into this framework. Still, the results are useful. For a radial scalar field ($q=0$), it enabled to retrieve the BCL BH. For a scalar field with linear time dependence ($q\neq 0$), it led to a new stealth Schwarzschild BH with different characteristics as compared to the usual stealth solutions. The next section now goes back to beyond Horndeski theories, $F_4\neq 0$, starting with a purely radial scalar field $\phi=\phi\left(r\right)$.
\section{Beyond Horndeski black holes with $q=0$, $\phi=\phi\left(r\right)$}\label{sec:purebey}
\sectionmark{4.7~~Beyond Horndeski black holes with $q=0$, $\phi=\phi\left(r\right)$}
\subsection{Homogeneous black hole with an $X^2$ k-essence term}
Again, homogeneity is assumed ($f=h$) as a start. Then $Z$ is constant, $Z=\gamma$, and exactly as before, the computations turn out to show that one can impose $Z=\gamma=-1$ without loss of generality. The algorithmic procedure is thus completely determined if we fix one of the three functions $\mathfrak{G}\left(X\right)$, $G_2\left(X\right)$ or $G_4\left(X\right)$. A possibility is to impose an $X^2$ k-essence term with a usual bare cosmological constant,
\begin{equation}
G_2\left(X\right)=\eta X^2-2\Lambda,\label{eq:g2bhbh}
\end{equation}
where $\eta$ is a coupling constant. From (\ref{eq:g2z}), $\mathfrak{G}$ is obtained, and from (\ref{eq:g4z}), $G_4$ is found to be
\begin{equation}
G_4\left(X\right)=1+\zeta X+\kappa\eta X^2,\label{eq:g4bhbh}
\end{equation}
where, as compared to (\ref{eq:g2z}-\ref{eq:g4z}), we have defined
\begin{equation}
\kappa = \frac{\delta}{2\epsilon},\quad \zeta = \frac{\beta}{2}-\lambda\alpha,
\end{equation}
while $C$ and $D$ have been fixed to give the usual $-2\Lambda$ in $G_2$ and the Einstein-Hilbert term $1\subset G_4$. The theory now belongs to the beyond Horndeski class: using this $G_4$ and the fact that $Z=-1$ into the definition (\ref{eq:z}) of $Z$, the beyond Horndeski function is computed to be
\begin{equation}
F_4\left(X\right)=-\frac{\zeta}{4X}-\frac{3\eta\kappa}{4}.\label{eq:f4first}
\end{equation}
With these notations, and using Eq. (\ref{eq:gx}), $X\left(r\right)$ reads
\begin{equation}
X\left(r\right) = -\frac{\zeta/\eta}{r^2+2\kappa}.\label{eq:xphirfirst}
\end{equation}
Then, (\ref{eq:simple3}) gives the metric function $f=h$ to be
\begin{equation}
h\left(r\right)=1-\frac{2M}{r}-r^2\frac{\Lambda}{3}+\frac{\zeta^2}{2\eta\sqrt{2\kappa}}\frac{\pi/2-\arctan\left(\frac{r}{\sqrt{2\kappa}}\right)}{r}.
\end{equation}
This is seen to possess a Schwarzschild behaviour (or Scwharzschild-(a)dS if $\Lambda\neq 0$), with ADM mass $M$, at leading order, while the $\arctan$ term induces subleading order corrections: as $r\to\infty$,
\begin{equation}
h\left(r\right)=1-\frac{2M}{r}-r^2\frac{\Lambda}{3}+\frac{\zeta^2}{2\eta r^2}+\mathcal{O}\left(\frac{1}{r^4}\right).
\end{equation}
Note that the $\arctan$ profile is quite similar to the solution obtained first in the Horndeski theory of~\cite{Rinaldi:2012vy} and that we described in paragraph~\ref{subsec:challenge}. The important difference is that we manage here to reach asymptotic flatness for $\Lambda=0$, whereas~\cite{Rinaldi:2012vy} has an always non-zero effective cosmological constant. Going to beyond Horndeski theories seems to be the key point in order to attain asymptotic flatness with such a profile.
\\

On the other hand, the scalar field $\phi=\phi\left(r\right)$ is directly deduced from (\ref{eq:xphirfirst}),
\begin{equation}
\phi\left(r\right)=\pm\sqrt{\frac{2\zeta}{\eta}}\int\frac{\mathrm{d}r}{\sqrt{\left(r^2+2\kappa\right)h\left(r\right)}}.
\end{equation}
\subsection{Rewriting with a canonical kinetic term}
Interestingly, one can rewrite the previous solution with a canonical kinetic term $X\subset G_2$. The pattern found above, see (\ref{eq:g2bhbh}-\ref{eq:g4bhbh}), is schematically
\begin{equation}
G_2 \sim X^2,\quad G_4 \sim 1+X+X^2.
\end{equation}
It is therefore tempting to make the replacement $X\to\sqrt{-X}$ in order to get a canonical kinetic term in $G_2$. There is no precise way to make such a replacement, since such a procedure does not correspond to a covariant redefinition of the scalar field. Using the brute force method of trial and error, one however manages to find the following theory, 
\begin{equation}
G_2=\frac{8\eta\beta^2}{\lambda^2}X-2\Lambda,\quad G_4=1+4\eta\beta\left(\sqrt{-X}+\beta X\right),\quad F_4 = -\frac{\eta\beta^2}{X}.\label{eq:rewriting_cano}
\end{equation}
This theory is parameterized by three couplings $\eta$, $\beta$, $\lambda$, with $\beta>0$ and $\lambda>0$. It admits as a solution the same BH metric as before, up to the changes in the couplings,
\begin{equation}
h\left(r\right)=1-\frac{2M}{r}-r^2\frac{\Lambda}{3}+\eta\frac{\arctan\left(r/\lambda\right)-\pi/2}{r/\lambda},\label{eq:hrcanon}
\end{equation}
while the scalar field and its kinetic term are found from (\ref{eq:simple2}),
\begin{equation}
\phi\left(r\right)=\pm\frac{1}{\sqrt{2}\beta}\int\frac{\mathrm{d}r}{\left(1+\left(r/\lambda\right)^2\right)\sqrt{h\left(r\right)}},\quad X\left(r\right)=-\frac{\lambda^4}{4\beta^2\left(r^2+\lambda^2\right)^2}.\label{eq:associated_x}
\end{equation}
The replacement $X\to\sqrt{-X}$ seems quite faithful as regards $G_2$ and $G_4$, $F_4$ however looses its constant part as compared to (\ref{eq:f4first}). While the method of the function $\mathfrak{G}$ enabled to find the BH with an $X^2$ k-essence term, and then to guess the form of $G_2$ and $G_4$ for the BH with canonical kinetic term, it is thus unclear how the two actions are related - if they are related at all, beyond the superficial resemblance. \textbf{Still, the virtue of this rewriting with a canonical kinetic term is to prove that the method of the function $\mathfrak{G}$ is not at all general as regards the solving of the field equations}. Indeed, the theory functions (\ref{eq:rewriting_cano}) are inconsistent with the equations (\ref{eq:g2z}-\ref{eq:g4z}) yielding $G_2$ and $G_4$ in terms of a function $\mathfrak{G}$.\\

These beyond Horndeski BHs (the one with $G_2\sim X^2$ and the one with $G_2\sim X$) are compatible with the Hui-Nicolis no-hair theorem of paragraph~\ref{subsec:huinic} by violation of assumption 6. Indeed, $F_4$ is not analytic. In the case $G_2\sim X^2$, assumption 5 (presence of a canonical kinetic term) is also violated.\\

Let us comment on the horizons implied by the metric function $h(r)$ of Eq. (\ref{eq:hrcanon}), focusing for simplicity on the asymptotically flat case, $\Lambda=0$. The presence or absence of horizons can be understood by looking at the behaviour of the metric near $r=0$,
\begin{equation}
h\left(r\right)=1+\eta-2\frac{M-M_0}{r}-\frac{\eta r^2}{3\lambda^2}+\mathcal{O}\left(r^4\right).
\end{equation}
There appears a threshold mass $M_0$, defined by
\begin{equation}
M_0\equiv -\frac{\pi\eta\lambda}{4}.
\end{equation}
For $M>M_0$, there is a unique horizon (this is in particular the case for $\eta>0$). For $M<M_0$,  there are two horizons if $M$ is not too small, and zero horizon (naked singularity) for small masses. For the limit case $M=M_0$, there is zero or one horizon according to $\eta\gtrless -1$, and $h\left(r\right)$ does not diverge at $r=0$ but the spacetime remains singular there because $h\left(0\right)\neq 1$.
\\

This beyond Horndeski BH is seen to be quite similar to the BCL BH of Eqs. (\ref{eq:bcl_bis_0}-\ref{eq:bcl_bis}), which was obtained in a pure Horndeski theory. For instance, in both cases, the spacetime does not reduce to Minkowski for vanishing mass $M$. The scalar field is well-defined outside and at the horizon, but becomes imaginary inside the horizon. The main qualitative difference with respect to the BCL BH is that the kinetic term $X$ diverges at the curvature singularity $r=0$ for the BCL BH, whereas for the beyond Horndeski BH under consideration, $X$ remains finite, see (\ref{eq:xphirfirst}) or (\ref{eq:associated_x}).
\subsection{Associated non-homogeneous black hole}
Interestingly, once a homogeneous ($f=h$) beyond Horndeski BH with radial scalar field ($q=0$) is obtained, it becomes very easy to generate non-homogeneous BHs ($f\neq h$). Indeed, for $q=0$, the final field equations (\ref{eq:simple}-\ref{eq:simple3}) are seen to depend only on the products $G_2Z$ and $G_4Z$. Therefore, one can
\begin{enumerate}
\item Start from a homogeneous BH solution, with functions $G_2^\text{hom}$, $G_4^\text{hom}$, $Z^\text{hom}=-1$, $h\left(r\right)$ determined by the differential equation (\ref{eq:simple3}), and $X\left(r\right)$ determined by (\ref{eq:simple2}).
\item Choose $Z^\text{non-hom}\left(X\right)$ so as to fix the relation between the two metric functions $f$ and $h$, see Eq. (\ref{eq:simple}).
\item Choose new functions $G_2^\text{non-hom}$ and $G_4^\text{non-hom}$ so that the products $G_2Z$ and $G_4Z$ remain unchanged as compared to the homogeneous solution, i.e. 
\begin{equation}
G_2^\text{non-hom}Z^\text{non-hom}=-G_2^\text{hom},\quad G_4^\text{non-hom}Z^\text{non-hom}=-G_4^\text{hom}.
\end{equation}
On the other hand, $F_4$ is obtained from the definition (\ref{eq:z}) of $Z\left(X\right)$.
\item The form of Eqs. (\ref{eq:simple2}-\ref{eq:simple3}) then implies that $h\left(r\right)$ and $X\left(r\right)$ are unchanged as compared to the homogeneous case.
\end{enumerate}
Let us apply this procedure to the BH with canonical kinetic term just found, with theory functions (\ref{eq:rewriting_cano}). $X\left(r\right)$, as given by (\ref{eq:associated_x}), then vanishes as $\mathcal{O}\left(1/r^4\right)$ when $r\to\infty$. So, in order for $f=h/Z^2$ [see (\ref{eq:simple})] to display the same asymptotics as $h$, one can choose for instance $Z=-1+\xi^2 X$ with $\xi$ a new coupling. Step $3$ above then leads to the new theory functions:
\begin{align}
G_2 ={}&{} \frac{8\eta\beta^2X/\lambda^2-2\Lambda}{1-\xi^2 X},\quad G_4=\frac{1+4\eta\beta\left(\sqrt{-X}+\beta X\right)}{1-\xi^2 X}\nonumber\\F_4 ={}&{} \frac{\xi^6X^2-\left(3\xi^2+4\eta\beta^2\right)\xi^2 X-8\eta\beta\xi^2\sqrt{-X}-4\eta\beta^2}{4X\left(1-\xi^2 X\right)^2}.
\end{align}
Then, according to step $4$, the corresponding BH solution is
\begin{equation}
h\left(r\right)=1-\frac{2M}{r}-r^2\frac{\Lambda}{3}+\eta\frac{\arctan\left(r/\lambda\right)-\pi/2}{r/\lambda},\quad f\left(r\right)=\frac{h\left(r\right)}{\left(1+\frac{\xi^2\lambda^4}{4\beta^2\left(r^2+\lambda^2\right)^2}\right)^2},\label{eq:nonhombh}
\end{equation}
dressed with a scalar field
\begin{equation}
\phi\left(r\right)=\pm\frac{1}{\sqrt{2}\beta}\int\frac{1+\frac{\xi^2\lambda^4}{4\beta^2\left(r^2+\lambda^2\right)^2}}{\left(1+\left(r/\lambda\right)^2\right)\sqrt{h\left(r\right)}}\mathrm{d}r,\quad X\left(r\right)=-\frac{\lambda^4}{4\beta^2\left(r^2+\lambda^2\right)^2}.
\end{equation}
Obviously, this procedure could be generalized to arbitrary choices of $Z\left(X\right)$ preserving the good asymptotics of $f\left(r\right)$, typically $Z=-1+\xi^2 X^n$ with $n\geq 1$. One thus sees that \textbf{in beyond Horndeski theories with shift and parity symmetry and radial scalar field $\phi=\phi\left(r\right)$, each homogeneous BH solution in a given theory gives rise to an infinity of non-homogeneous BHs in distinct theories.}
\section{Beyond Horndeski black holes with $q\neq 0$, $\phi=qt+\psi\left(r\right)$}\label{sec:primary}
\sectionmark{4.8~~Beyond Horndeski black holes with $q\neq 0$, $\phi=qt+\psi\left(r\right)$}
\subsection{Black holes with primary scalar hair}
Let us now study the case where the scalar field admits a linear time dependence, $\phi=qt+\psi\left(r\right)$, for a homogeneous BH, i.e. $Z=\gamma=-1$. Eq. (\ref{eq:simple2}), which determines $X$, becomes
\begin{equation}
r^2G_{2X}+2G_{4X}\left(1-\frac{q^2}{2X}\right)=0.\label{eq:new_x_eq}
\end{equation}
Rather than solving it again step-by-step with the method of the function $\mathfrak{G}$, we present directly the results that this procedure yields. If $G_2$ and $G_4$ are related as
\begin{equation}
G_2=\frac{2}{\lambda^2}\left(G_4-1\right)-2\Lambda,\label{eq:g2g4}
\end{equation}
then (\ref{eq:new_x_eq}) implies
\begin{equation}
X = \frac{q^2/2}{1+r^2/\lambda^2},
\end{equation}
which in turn determines the scalar field,
\begin{equation}
\phi=q\left(t\pm\int\frac{\sqrt{1-\frac{h\left(r\right)}{1+r^2/\lambda^2}}}{h\left(r\right)}\mathrm{d}r\right).\label{eq:phiprimary}
\end{equation}
A natural choice for $G_4$ is a usual Einstein-Hilbert term, with an additional monomial term in $X$. This also determines $G_2$ through (\ref{eq:g2g4}), and $F_4$ through the definition (\ref{eq:z}) of $Z\left(X\right)$:
\begin{equation}
G_4 = 1-\frac{2^{n}\eta}{2n-1}X^n,\quad G_2=-\frac{2^{n+1}\eta}{\left(2n-1\right)\lambda^2}X^n-2\Lambda,\quad F_4 = \eta \left(2X\right)^{n-2}.\label{eq:beyondprimary}
\end{equation}
The metric function $f=h$ is then obtained from (\ref{eq:simple3}),
\begin{align}
h\left(r\right)=1-\frac{2M}{r}-r^2\frac{\Lambda}{3}-\frac{2\lambda}{3r}\eta q^{2n}\Biggl[\frac{r^3/\lambda^3}{\left(1+r^2/\lambda^2\right)^{n-1}}{}&{}\leftindex_2F_1\left(1,\frac{5}{2}-n,\frac{5}{2},-\frac{r^2}{\lambda^2}\right)\nonumber\\{}&{}-\frac{3\sqrt{\pi}\,\Gamma\left(n-3/2\right)}{4\Gamma\left(n\right)}\Biggr],\label{eq:gamma}
\end{align}
where $\Gamma$ is the Euler Gamma function, and
\begin{equation}
\leftindex_2F_1\left(a,b,c,z\right)\equiv\sum_{n=0}^\infty\frac{\left(a\right)_n\left(b\right)_n}{\left(c\right)_n}\frac{z^n}{n!},\quad \left(a\right)_n\equiv a\left(a+1\right)\cdots\left(a+n-1\right),
\end{equation}
is the ordinary hypergeometric function. The term between brackets on the second line of (\ref{eq:gamma}) has a leading order behaviour as $\mathcal{O}\left(r^{3-2n}\right)$ when $r\to\infty$. Since the brackets are multiplied by $1/r$, this term is subdominant with respect to the usual Schwarzschild term $-2M/r$ provided $n>3/2$. The asymptotic expansion is then\footnote{The limiting case $n=3/2$ is not included, since it gives terms $\ln\left(r\right)/r$ which are not subdominant with respect to the Schwarzschild term.}
\begin{equation}
h\left(r\right) = 1-\frac{2M}{r}-r^2\frac{\Lambda}{3}+\frac{2\eta q^{2n}}{2n-3}\left(\frac{\lambda}{r}\right)^{2n-2}+\mathcal{O}\left(\frac{1}{r^{2n}}\right),\label{eq:primary_asym}
\end{equation}
and $M$ is the ADM mass of the solution. Very interestingly, $M$ is not the unique integration constant parameterizing the solution: there exists a \textbf{primary scalar hair} $q$, which can take \textit{a priori} any real value, and appears as an independent integration constant of the solution, although we had first introduced it in our scalar field ansatz $\phi=qt+\psi\left(r\right)$. When $q=0$, the scalar hair disappears and one gets back a Schwarzschild solution with a trivial scalar. When both $q$ and $M$ vanish, the solution reduces to flat spacetime. When $M$ vanishes but not $q$, the spacetime has zero mass, but does not reduce to flat spacetime and remains freely parameterized by $q$. 
\\

All BHs that we have presented previously had \textbf{secondary hair}. This means that, just as in GR, they are parameterized only by their mass $M$ and angular momentum $J$ (in the spherical case, only by $M$, and if one were including Maxwell terms, there would also be the charge $Q$). These secondary hair BHs are still said to be hairy, since they are dressed by a non-trivial scalar field. Their metric is either the same as in GR (these are the stealth BHs) or different from GR (these are the non-stealth BHs). However, this is the first time that a scalar-tensor BH with primary hair is constructed: there exists an integration constant $q$, parameterizing the BH, and which is distinct from $M$, $J$ and $Q$.
\\

To be perfectly precise, there exist interesting examples of BHs with primary hair for a minimally-coupled complex scalar field~\cite{Herdeiro:2014goa,Herdeiro:2015gia}. There, and as we stated in paragraph~\ref{subsec:mincoupl}, the minimally-coupled scalar is part of the matter energy-momentum tensor in GR rather than a modification of gravity. Also, a primary hair BH was constructed in a bi-scalar extension of Horndeski theory~\cite{Charmousis:2014zaa}. However, as explained in Chap.~\ref{chap:st}, the framework of this thesis is scalar-tensor theory with a single real scalar field. In this framework, such BH with primary hair is a novelty in the literature. For a review on the difference between primary and secondary hair, see~\cite{Herdeiro:2015waa}. Importantly, note that our original article~\cite{Bakopoulos:2023fmv} had only found a first example of such primary scalar hair, namely the case $n=2$ in Eqs. (\ref{eq:beyondprimary}-\ref{eq:gamma}). Generalizations, in particular the general $n$ case, were then performed in~\cite{Baake:2023zsq,Bakopoulos:2023sdm}. Observational implications of the primary hair were also studied, like its gravitational lensing properties~\cite{Mushtaq:2024qse}.
\\

The horizon structure of these BHs will be studied below, by specifying to particular values of $n$ for simplicity. For the moment, let us remark that the scalar field (\ref{eq:phiprimary}) is regular either at the future event horizon or past event horizon, as can be checked by changing to horizon-crossing coordinates, see the detailed explanation in paragraph~\ref{subsec:stealthsch}. However, no regularity at both (future and past) horizons can be achieved. \\

Note also that the beyond Horndeski functions (\ref{eq:beyondprimary}) are analytic (for $n$ integer $\geq 2$). The no-hair theorem of parapraph~\ref{subsec:huinic} is thus violated thanks to the linear time dependence of the scalar field, and the absence of canonical kinetic term. It is worth noticing that, due to the linear time dependence of $\phi$, there remains a non-vanishing component $\mathcal{J}^t_\text{shift}$ of the Noether current, but the norm $\mathcal{J}^\mu_\text{shift}\mathcal{J}_{\text{shift}\,\mu}$ is everywhere finite, including at $r=0$. The expression for generic $n$ is cumbersome and not enlightening, but for instance, the norm in the case $n=2$ is
\begin{equation}
\mathcal{J}^\mu_\text{shift}\mathcal{J}_{\text{shift}\,\mu}=-\frac{4\eta^2q^6\lambda^4\left(11r^2-\lambda^2\right)^2h(r)}{9\left(r^2+\lambda^2\right)^6}.
\end{equation}
\subsection{Regular black holes}
The behaviour of (\ref{eq:gamma}) near $r=0$ brings valuable information:
\begin{equation}
h\left(r\right)=1-\frac{4M-\lambda\eta q^{2n}\sqrt{\pi}\,\Gamma\left(n-3/2\right)/\Gamma\left(n\right)}{2r}-\frac{r^2}{3}\left(\Lambda+\frac{2\eta q^{2n}}{\lambda^2}\right)+\mathcal{O}\left(r^4\right).
\end{equation}
This shows that there exists a threshold mass,
\begin{equation}
M_0 = \frac{\lambda\eta q^{2n}\sqrt{\pi}\,\Gamma\left(n-3/2\right)}{4\Gamma\left(n\right)},
\end{equation}
such that, when $r\to 0$, $h\left(r\right)\to-\infty$ if $M>M_0$ or $h\left(r\right)\to+\infty$ if $M<M_0$. When $M=M_0$, in other words when the primary scalar hair $q$ is related to the mass as
\begin{equation}
q = \left(\frac{4M\Gamma\left(n\right)}{\lambda\eta\sqrt{\pi}\,\Gamma\left(n-3/2\right)}\right)^{\frac{1}{2n}},\label{eq:qm}
\end{equation}
then the singular term proportional to $1/r$ disappears near $r=0$. More generally, for generic $r$, the metric then takes the form
\begin{equation}
h\left(r\right)=1-r^2\frac{\Lambda}{3}-\frac{8M\Gamma\left(n\right)r^2}{3\lambda^3\sqrt{\pi}\,\Gamma\left(n-3/2\right)\left(1+r^2/\lambda^2\right)^{n-1}}\leftindex_2F_1\left(1,\frac{5}{2}-n,\frac{5}{2},-\frac{r^2}{\lambda^2}\right).\label{eq:reg}
\end{equation}
Obviously, the mass $M$ is still present and the expansion of the hypergeometric function as $r\to\infty$ gives the same Schwarzschild behaviour at leading order as in (\ref{eq:primary_asym}). The striking feature is that $h\left(r\right)$ becomes an even function of $r$, and this is known~\cite{Burzilla:2020utr} to imply regularity of all curvature invariants and of their derivatives, that is, in addition to the regularity of the Ricci scalar $R$ or the Kretschmann scalar $K$, one has, for example, regularity of $\Box^p R$ or $\Box^p K$ for arbitrary $p$.
\\

In a word, \textbf{beyond Horndeski theories (\ref{eq:beyondprimary}) enable the obtention of BHs with primary scalar hair (\ref{eq:gamma}), parameterized by two integration constants: the mass $M$, and the scalar hair $q$. When $q$ and $M$ are related as in (\ref{eq:qm}), the BH with primary hair reduces to a BH with secondary hair (\ref{eq:reg}), parameterized only by its mass $M$. For this BH with secondary hair, there is no curvature singularity: its curvature invariants are infinitely regular}.
\\

Strictly speaking, the fact that these BHs have infinitely regular curvature invariants is the only thing that one can affirm for the moment. Indeed, going further and claiming that these BHs are \textbf{regular BHs} (as is done in the title of this paragraph) is a short way of presenting our results, but remains not precise for the moment and the lack of a more detailed study of spacetime (\ref{eq:reg}). We refer the interested reader to the instructive review~\cite{Lan:2023cvz}. Notably, the definition of a regular BH as having finite curvature invariants~\cite{1982ZhPmR..36..214M,Frolov:1988vj,Frolov:2016pav,Chamseddine:2016ktu} is not the only one. One might rather ask for the spacetime to be \textbf{geodesically complete}~\cite{Hawking:1973uf,Wald:1984rg} (that is, its causal geodesics have an affine parameter extending to infinity). Both definitions are not equivalent~\cite{Geroch:1968ut,Olmo:2015bya}, and geodesic completeness remains to be studied as regards the BH (\ref{eq:reg}).
\\

Still, 'regular BHs' (with finite curvature invariants) are in general constructed in the framework of GR, by postulating a metric which has no curvature singularity, and plugging it into Einstein's equations. Such a metric is not Ricci flat, thus is supported by a corresponding \textit{ad hoc} energy-momentum tensor $T_{\mu\nu}$. The first example of such a construction is the Bardeen BH~\cite{1968qtr..conf...87B}. More than thirty years after this proposal, the corresponding $T_{\mu\nu}$ was interpreted as a magnetic monopole in the context of nonlinear electrodynamics~\cite{Ayon-Beato:2000mjt}. Such an interpretation has been generalized to other regular BHs, see e.g.~\cite{Bronnikov:2000vy}. On the contrary, the regular BHs we constructed are not solution of GR with an \textit{ad hoc} energy-momentum tensor, but vacuum solutions of a scalar-tensor theory. Note however that regular BHs were already constructed in DHOST theory~\cite{Babichev:2020qpr,Baake:2021jzv}. The real novelty of our regular BHs is their interpretation as a BH with primary scalar hair $q$, for which $q$ and the mass $M$ are related so as to evade the curvature singularity.
\subsection{Simple cases $n=5/2$ and $n=2$}
\begin{figure}[t]
    \centering
    \begin{subfigure}[b]{0.485\textwidth}
    \includegraphics[width=1\textwidth]{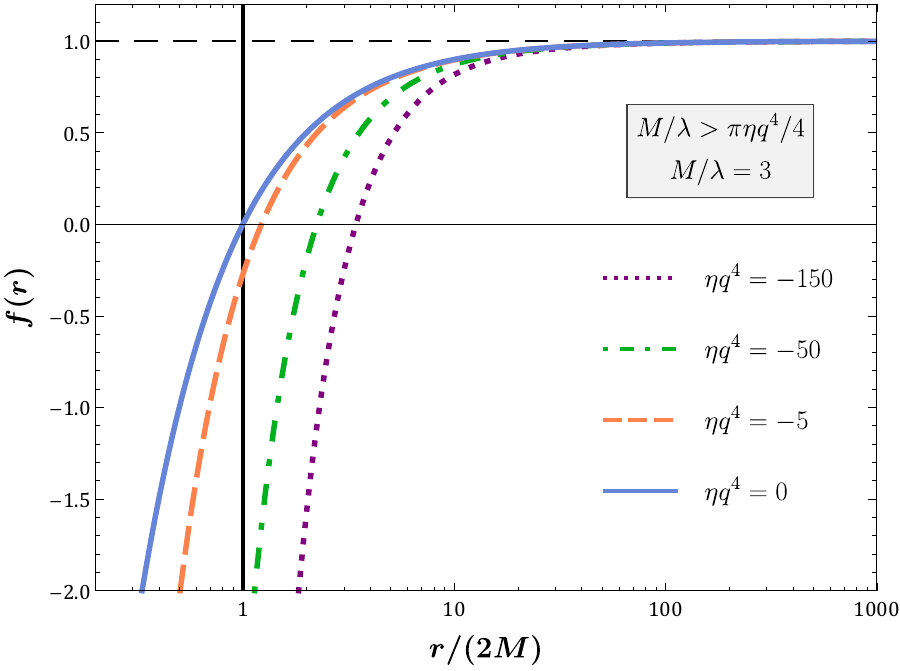}
    \caption{\hspace*{-1.5em}}
    \label{subf:sparse}
    \end{subfigure}
    \hfill
    \begin{subfigure}[b]{0.48\textwidth}
    \includegraphics[width=1\textwidth]{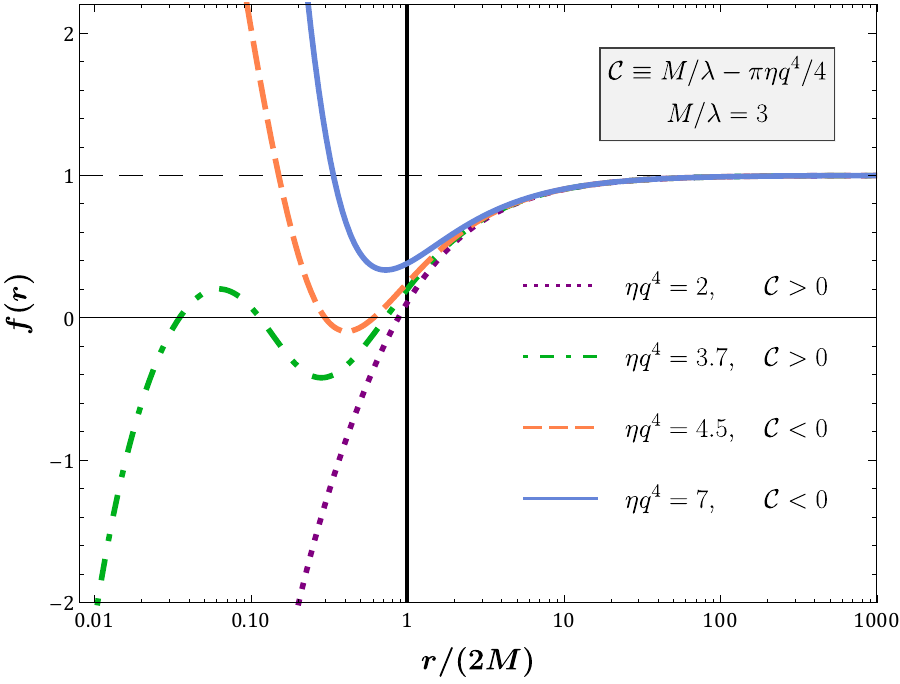}
    \caption{\hspace*{-1.5em}}
    \label{subf:ult-com}
    \end{subfigure}
    \caption{Spacetime~(\ref{eq:negal2}) with primary hair. (a) $\eta<0$, BH solutions with a single horizon larger than the Schwarzschild radius. (b) $\eta>0$, BH solutions with one, two or three horizons smaller than the Schwarzschild radius, and a naked singularity (no horizon). The horizontal axis is logarithmic in both figures.}
    \label{fig:bh-plots}
\end{figure}

\begin{figure}[h]
    \centering
    \begin{subfigure}[b]{0.49\textwidth}
    \includegraphics[width=1\textwidth]{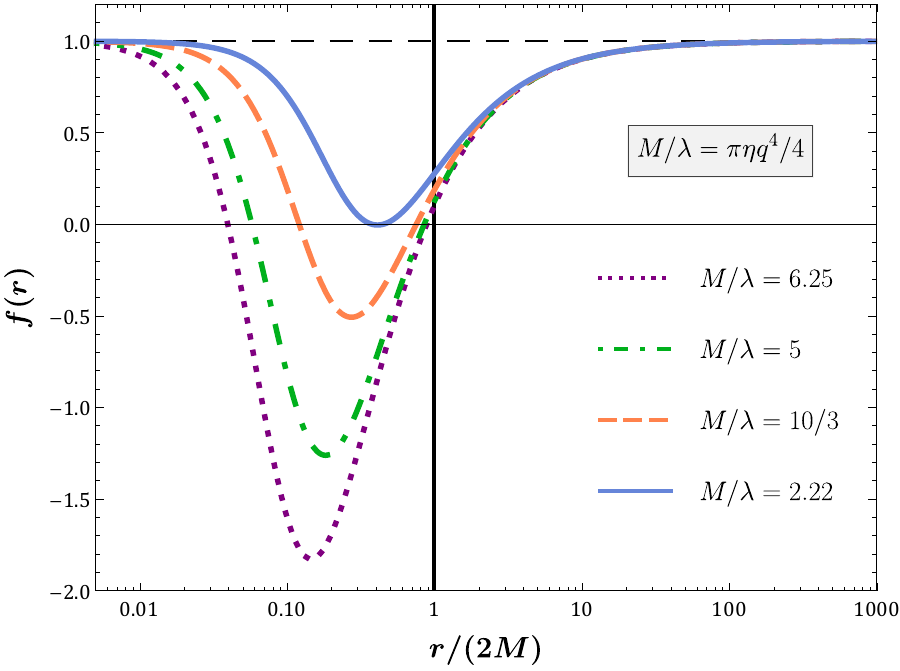}
    \caption{\hspace*{-1.5em}}
    \label{subf: rbh}
    \end{subfigure}
    \hfill
    \begin{subfigure}[b]{0.485\textwidth}
    \includegraphics[width=1\textwidth]{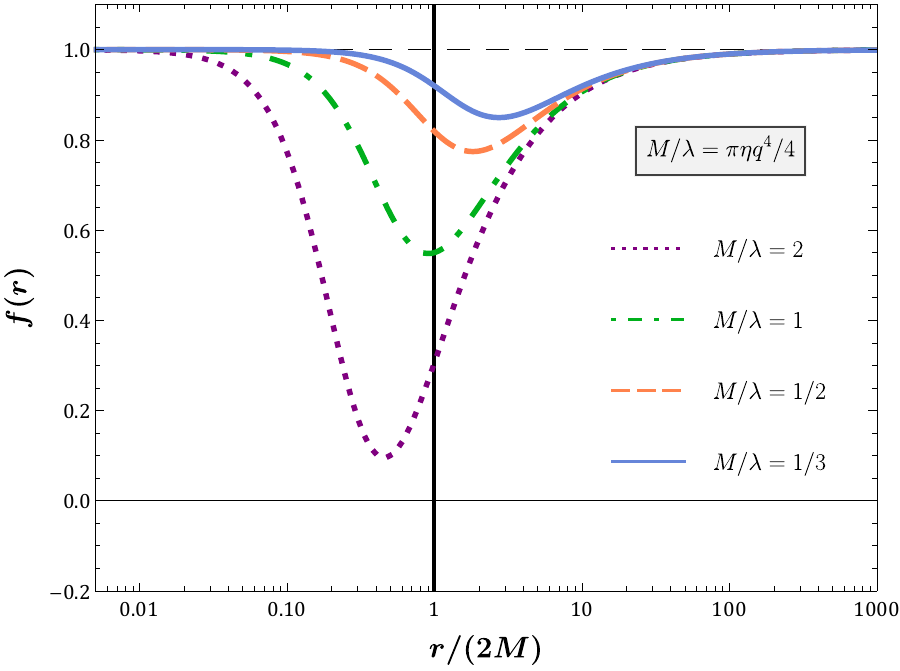}
    \caption{\hspace*{-1.5em}}
    \label{subf: sol}
    \end{subfigure}
    \caption{Regular spacetime~(\ref{eq:negal2reg}). (a) Regular BH solutions. (b) Regular solitonic solutions. The horizontal axis is logarithmic in both figures.}
    \label{fig:rbh-sol-plot}
\end{figure}

The hypergeometric function appearing in (\ref{eq:gamma}) simplifies to explicit expressions, either algebraic in the radial coordinate $r$ if $n=p+1/2$ with $p\geq 2$ integer, or involving $\arctan\left(r/\lambda\right)$ for $n=2$ and $n=3$. The simplest examples are thus for $n=5/2$,
\begin{equation}
h\left(r\right)=1-\frac{2M}{r}-r^2\frac{\Lambda}{3}+\frac{2\eta q^5}{3r}\left(1-\frac{r^3}{\left(r^2+\lambda^2\right)^{3/2}}\right),\label{eq:negal52}
\end{equation}
and for $n=2$,
\begin{equation}
h\left(r\right)=1-\frac{2M}{r}-r^2\frac{\Lambda}{3}+\eta q^4\left(\frac{\pi/2-\arctan\left(r/\lambda\right)}{r/\lambda}+\frac{1}{1+\left(r/\lambda\right)^2}\right).\label{eq:negal2}
\end{equation}
When $q$ is related to the mass as in (\ref{eq:qm}), the corresponding regular BHs of (\ref{eq:reg}) are, for $n=5/2$,
\begin{equation}
h\left(r\right)=1-r^2\frac{\Lambda}{3}-\frac{2Mr^2}{\left(r^2+\lambda^2\right)^{3/2}},\label{eq:negal52reg}
\end{equation}
and for $n=2$,
\begin{equation}
h\left(r\right)=1-r^2\frac{\Lambda}{3}-\frac{4M}{\pi\lambda}\left(\frac{\arctan\left(r/\lambda\right)}{r/\lambda}-\frac{1}{1+\left(r/\lambda\right)^2}\right).\label{eq:negal2reg}
\end{equation}
As an illustration, the metric function $h(r)$, for the case $n=2$ with $\Lambda=0$, is presented on Figs.~\ref{fig:bh-plots} and~\ref{fig:rbh-sol-plot}. Fig.~\ref{fig:bh-plots} corresponds to Eq.~(\ref{eq:negal2}), with $h(r)\to\pm\infty$ as $r\to 0$, while Fig.~\ref{fig:rbh-sol-plot} shows the case of the regular spacetime, Eq.~(\ref{eq:negal2reg}). \\

Let us start with the case of~(\ref{eq:negal2}), i.e. Fig.~\ref{fig:bh-plots}. When $\eta<0$, $h(r)\to -\infty$ as $r\to 0$, so there is an event horizon, with a size greater than the Schwarzschild radius $r_\text{Sch}=2M$. As the scalar charge $q$ increases, the event horizon increases accordingly away from its GR size. \\

For $\eta>0$, the situation is the following. For small scalar hair, i.e. for $M/\lambda>\pi\eta q^4/4$, again $h(r)\to -\infty$ as $r\to 0$, so there is at least one horizon. However, in this case, the horizon size is always smaller than the Schwarzschild radius. As the hair $q$ increases, the event horizon shrinks in size, and three horizons emerge when $M/\lambda > \left(4+\pi\right)/4$. As the ratio $M/\lambda$ exceeds more and more this threshold\footnote{The threshold value $M/\lambda=\left(4+\pi\right)/4$ is found by solving the system of equations for a triple horizon, which is then found to be located at $r=\lambda$ for the unique value $\eta q^4=2$.}, the range of values of $\eta q^4$ allowing for three horizons increases. For large scalar hair, on the other hand, i.e. for $M/\lambda<\pi\eta q^4/4$, $h(r)\rightarrow +\infty$ as $r\rightarrow 0$. As the scalar hair gradually increases, one first obtains a BH with two horizons, then a naked singularity. Note also that the metric reduces to flat spacetime only when both $M$ and $q$ vanish. If $M=0$ but $q\neq 0$, the spacetime has zero mass but is not trivial: it is a BH if $\eta<0$, and a naked singularity if $\eta>0$.\\

On the other hand, for the regular spacetime (\ref{eq:negal2reg}) corresponding to Fig.~\ref{fig:rbh-sol-plot}, there exists a threshold value $a\approx 2.2116$ of the ratio $M/\lambda$ that discriminates between different types of regular solutions. The spacetime describes a regular BH with two horizons if $M/\lambda>a$, and a regular soliton with no horizon if $M/\lambda<a$.
\\

To cut a long story short, in this Chap.~\ref{chap:4}, many closed-form BH solutions were obtained: a stealth Schwarzschild solution with non-constant kinetic term (\ref{eq:g2g4newst}-\ref{eq:phinewst}); new solutions with secondary hair, both homogeneous (\ref{eq:hrcanon}) and non-homogeneous (\ref{eq:nonhombh}); BHs with primary scalar hair (\ref{eq:gamma}); and regular BHs (\ref{eq:reg}). Also, as regards theories without parity symmetry, compatibility conditions generalizing the 4DEGB case were found, and lead to BH solutions defined by roots of algebraic equations, see Eqs. (\ref{eq:g5comp}-\ref{eq:phiprimenonpar}).
\\

The decisive assumption of this chapter is shift symmetry, which enables from a start the compact rewriting of the field equations, in particular using the vanishing of the radial component of the Noether current, $\mathcal{J}^r_\text{shift}=0$. Also, shift symmetry is one of the assumptions of the Hui-Nicolis no-hair theorem of paragraph~\ref{subsec:huinic}. As a consequence, keeping shift symmetry implies the violation of one of the other assumptions of the theorem, by for example having a linear time dependence in the scalar field, or displaying non-analytic Horndeski functions. \\

On the other hand, removing the assumption of shift symmetry, one gains access to theories which are much more difficult to analyze, but completely free from no-scalar-hair theorems. Turning on to the study of such theories in all generality would be too ambitious. Rather, one can remember paragraph~\ref{subsec:bhegb}: the 4DEGB theory allows for closed-form BH solutions even in the absence of shift symmetry. This theory remains remarkable by the couplings it involves between the scalar field and a Lovelock invariant, namely the GB scalar; and by its generalized conformal invariance. The next chapter thus elaborates on both these considerations in order to build new BH solutions.
\chapter[Lovelock invariants and conformally-coupled scalar field]{Lovelock invariants and conformally-coupled scalar field}\label{chap:5}
\chaptermark{Chap.~5~~Lovelock invariants and conformally-coupled scalar field}
In Sec.~\ref{sec:lovelock_th}, we presented Lovelock's theorem, leading to the following action functional for a metric theory of gravity in $D$ dimensions:
\begin{equation}
S\left[g_{\mu\nu}\right] = \int\mathrm{d}^Dx\sqrt{-g}\sum_{k=0}^{\lfloor\frac{D-1}{2}\rfloor}\alpha_k\mathcal{R}^{\left(k\right)}.
\end{equation} 
The $\mathcal{R}^{\left(k\right)}$ are the Lovelock invariants,
\begin{equation}
\mathcal{R}^{\left(k\right)} \equiv \frac{1}{2^k}\delta^{\mu_1\nu_1\cdots\mu_k\nu_k}_{\alpha_1\beta_1\cdots\alpha_k\beta_k}\prod_{i=1}^{k} R^{\alpha_i\beta_i}_{\quad\,\mu_i\nu_i}.\label{eq:lov_lov}
\end{equation}
Restricting to $k\leq 2$ leads to the EGB action,
\begin{equation}
S\left[g_{\mu\nu}\right]=\int\mathrm{d}^Dx\sqrt{-g}\left(R-2\Lambda+\alpha_2\mathcal{G}\right),\label{eq:egbDdim}
\end{equation}
which is the first correction to the Einstein-Hilbert action. The GB invariant $\mathcal{G}$ is a boundary term in $D=4$. However, Secs.~\ref{sec:compact} and~\ref{sec:fourd} detail the obtention of the 4DEGB theory, by performing a diagonal KK compactification followed by a singular limit where the dimension of the internal space of the compactification is sent to zero. As a reminder, the action reads
\begin{align}
S\left[g_{\mu\nu},\phi\right] =\int\mathrm{d}^4x\sqrt{-g}{}&{}\Bigl\{R-2\Lambda-2\lambda\mathrm{e}^{4\phi}-\beta\mathrm{e}^{2\phi}\Bigl[R+6\left(\partial\phi\right)^2\Bigr]\nonumber\\ {}&{}+\alpha\Bigl[-\phi\mathcal{G}+4G^{\mu\nu}\phi_\mu\phi_\nu+4\Box\phi\left(\partial\phi\right)^2+2\left(\partial\phi\right)^4\Bigr]\Bigr\}.\label{eq:fern}
\end{align} 
The 4DEGB theory is a four-dimensional scalar-tensor theory which reproduces features of the higher-dimensional, pure metric EGB theory, and allows for closed-form BH solutions, see paragraph~\ref{subsec:bhegb}. Notably, there exists a pure geometric equation $\mathfrak{R}=0$, where
\begin{equation}
\mathfrak{R}=g^{\mu\nu}\mathcal{E}_{\mu\nu}+\mathcal{E}_\phi=8\Lambda-2R-\alpha\mathcal{G}.\label{eq:geom_chap_5}
\end{equation}
The existence of this geometric equation relates to the fact that the 4DEGB theory displays a generalized conformal invariance, see paragraph~\ref{subsec:local},
\begin{equation}
\frac{\delta S}{\delta \phi}\left[g^\epsilon_{\mu\nu},\phi^\epsilon\right]=\frac{\delta S}{\delta \phi}\left[g_{\mu\nu},\phi\right],\quad g^\epsilon_{\mu\nu}=(1-2\epsilon)g_{\mu\nu},\quad \phi^\epsilon= \phi+\epsilon,\label{eq:gen_conf_sym}
\end{equation}
for $\epsilon(x)$ an arbitrary infinitesimal function on spacetime. \\

This Chap.~\ref{chap:5} elaborates around these results. In a first time, Sec.~\ref{sec:selecting}, we look for the most general action with a similar form to (\ref{eq:fern}), and allowing for closed-form solutions. In a second time, Sec.~\ref{sec:higherdim} studies the context of \textbf{higher-dimensional} scalar-tensor theories with generalized conformal invariance. Finally, Sec.~\ref{sec:diagkk} performs a \textbf{KK dimensional reduction of generic Lovelock invariants}.
\section{Generalized couplings to the Gauss-Bonnet invariant}\label{sec:selecting}
\sectionmark{5.1~~Generalized couplings to the Gauss-Bonnet invariant}
This Sec.~\ref{sec:selecting} is based on the article~\cite{Babichev:2023dhs}, written among others by the present author. In order to generalize action (\ref{eq:fern}), let us consider
\begin{align}
S = \int \mathrm{d}^4x\sqrt{-g}\Bigl\{{}&{}\left(1+W\left(\phi\right)\right)R-
\frac{1}{2}V_k\left(\phi\right)\left(\partial\phi\right)^2+Z\left(\phi\right)+
V\left(\phi\right)\mathcal{G}\nonumber\\
 {}&{}+V_2\left(\phi\right)G^{\mu\nu}\partial_\mu\phi\,\partial_\nu\phi +V_3\left(\phi\right)\left(\partial\phi\right)^4+V_4\left(\phi\right)\Box\phi\left(\partial\phi\right)^2\Bigr\}. \label{eq:action}
\end{align}
There is a kinetic term $\frac{1}{2}V_k\left(\phi\right)\left(\partial\phi\right)^2$ in order to evade possible strong coupling
issues. The potential $W(\phi)$ determines if the
scalar is minimally coupled to the Ricci scalar (to lowest order) or
not. The potential $V\left(\phi\right)$ multiplies the GB
invariant, so is defined up to an additive constant which would only
yield a boundary term. The potential term $Z(\phi)$ may include the
cosmological constant, as well as Liouville exponential terms, most commonly
present in non-critical string theories~\cite{Dixon:1986iz}, in self-tuning scenarios~\cite{Charmousis:2017rof},
and holographic gravitational backgrounds, see e.g.~\cite{Charmousis:2012dw}. The Horndeski functions $G_k\left(\phi,X\right)$ are
\begin{align}
G_2={}&{} Z+XV_k+4X^2V_3+8X^2\left(3-\ln\left\lvert X\right\rvert\right)V_{\phi\phi\phi\phi},\\ G_3={}&{} 2XV_4+4X\left(7-3\ln\left\lvert X\right\rvert\right)V_{\phi\phi\phi},\\
G_4={}&{} 1+W+4X\left(2-\ln\left\lvert X\right\rvert\right)V_{\phi\phi},\\
G_5={}&{} -4V_\phi \ln\left\lvert X\right\rvert-\int V_2\,\mathrm{d}\phi,
\end{align}
see paragraph~\ref{subsec:common}. The 4DEGB action (\ref{eq:fern}) fits into the framework of
action (\ref{eq:action}) with the following potentials,
\begin{align}
W ={}&{} -\beta\mathrm{e}^{2\phi},\quad V_k = 12\beta\mathrm{e}^{2\phi},\quad Z = -2\lambda \mathrm{e}^{4\phi}-2\Lambda,
\\ V ={}&{} -\alpha\phi,\quad V_2 = 4\alpha = V_4 ,\quad V_3 = 2\alpha,\label{eq:pot_fern}
\end{align}
with three coupling constants $\alpha$, $\beta$ and $\lambda$. In
this case, the action acquires generalized conformal invariance, and as
such leads to the geometric constraint (\ref{eq:geom_chap_5}). This is
no longer true for the general action (\ref{eq:action}).
\\

As in the previous chapter, the interest lies in static and spherically symmetric spacetimes. The following metric ansatz is considered,
\begin{equation}
\label{metric}
\mathrm{d}s^2 = -f\left(r\right)\mathrm{d}t^2+\mathrm{d}r^2/f\left(r\right)+r^2\mathrm{d}\Omega^2,\quad \phi=\phi\left(r\right).
\end{equation}
What follows presents a way to filter out
theories, reducing the above general action to a more tractable, yet
quite general theory. We assume from now on the most interesting case of a non-constant scalar field $\phi(r)$.
\\

This section is organized as follows: in paragraph~\ref{subsec:int_comp}, we present conditions
on the potentials which enable to rewrite the field equations as
three simple compatibility conditions, parameterized by a unique
real function $\mu(r)$. Then, paragraph~\ref{subsec:mu1} studies
two novel BH solutions obtained for the case $\mu(r)\equiv
1$. We also briefly comment on the case of constant $\mu(r)\neq
1$, which leads to BHs with unsatisfactory asymptotics.

\subsection{Integrability and compatibility}\label{subsec:int_comp}
Given the ansatz~(\ref{metric}), it
is quite common to consider the combination of metric field equations
$\mathcal{E}^t_t-\mathcal{E}^r_r=0$ in order to determine the
expression of the scalar field. This is for example what is done for the 4DEGB theory, see~\cite{Fernandes:2021dsb}. One obtains
\begin{align}
\frac{\phi''}{\left(\phi'\right)^2}\Bigl[r^2W_\phi +{}&{}
4\left(1-f\right) V_\phi + 2frV_2 \phi' + fr^2
V_4\left(\phi'\right)^2\Bigr] \nonumber\\ ={}&{}
-\frac{r^2}{2}\left(V_k+2W_{\phi\phi}\right) -
\left(V_2+4V_{\phi\phi}\right)\left(1-f\right)\nonumber\\ {}&{}-
fr\left(V_{2\phi}-2V_4\right)\phi' - fr^2
\left(V_{4\phi}-2V_3\right)\left(\phi'\right)^2, \label{eqwfactor}
\end{align}
where prime stands for radial derivative, while a subscript
$\phi$ denotes derivation with respect to $\phi$. Upon close inspection, one sees that,
choosing the potentials of the theory as follows,
\begin{align}
V_k+2W_{\phi\phi} ={}&{} \frac{2}{d(\phi)}W_\phi,\quad V_2+4V_{\phi\phi}
= \frac{4}{d(\phi)}V_\phi,\nonumber\\ V_{2\phi}-2V_4 ={}&{}
\frac{2}{d(\phi)}V_2,\quad V_{4\phi}-2V_3 = \frac{1}{d(\phi)}V_4,
\label{conditions}
\end{align}
where $d(\phi)$ is an arbitrary function, Eq.~(\ref{eqwfactor}) is factorized in a simple and elegant
way,
\begin{equation}
\Biggl[\frac{\phi''}{\left(\phi'\right)^2}+\frac{1}{d(\phi)}\Biggr]\Biggl[r^2W_\phi
+ 4\left(1-f\right) V_\phi + 2frV_2 \phi' + fr^2
V_4\left(\phi'\right)^2\Biggr]=0. \label{eq:simple_equation}
\end{equation}
Under the conditions~(\ref{conditions}), the potentials $V_k$
and $V_{2,3,4}$ can be parameterized in terms of the
Einstein-Hilbert and GB potentials $W$ and $V$ as
\begin{align}
V_k = {}& \frac{2}{d}W_\phi-2W_{\phi\phi},\label{eq:vk}\\
V_2 = {}& \frac{4}{d}V_\phi-4V_{\phi\phi},\\
V_4 ={}& -\frac{2}{d^2}\left(2+d_\phi\right)V_\phi+\frac{6}{d}V_{\phi\phi}-2V_{\phi\phi\phi},\\
V_3 ={}&
\frac{1}{d^2}\left[\frac{1}{d}\left(1+2d_\phi\right)\left(2+d_\phi\right)-d_{\phi\phi}\right]
V_\phi-\frac{1}{d^2}\left(5+4d_\phi\right)V_{\phi\phi}+\frac{4}{d}V_{\phi\phi\phi}-V_{\phi\phi\phi\phi}.
\label{eq:v3}
\end{align}
In other words, the factorization (\ref{eq:simple_equation}) is made
possible with any action (\ref{eq:action}) parameterized by three
independent potentials, namely $W$, $Z$ and $V$, provided that the
remaining potentials are fixed by the above equations.
\\

Take now a closer look at the factorization
(\ref{eq:simple_equation}) and at its possible consequences for our
purpose. First of all, the potentials
(\ref{eq:pot_fern}) of the 4DEGB theory (\ref{eq:fern}) correspond to a
constant function $d$ given by $d\left(\phi\right)=-1$. In the 4DEGB case, if we forget about the shift-symmetric case, there were two branches of solutions for the scalar field, namely (\ref{eq:phi_lu_pang}) and (\ref{eq:log_scal}). Analogously, Eq. (\ref{eq:simple_equation}) offers the possibility of two branches of
solutions for the scalar field, depending on which bracket vanishes.\\

The first branch,
corresponding to the vanishing of the first bracket in (\ref{eq:simple_equation}),
does not involve the coupling functions of the theory nor the
metric function. It is a simple differential equation giving the scalar field independently of the geometry. The second branch is much more involved because
the equation involves explicitly the coupling potentials of the
theory and the metric function. The study of this latter case is performed in our original article~\cite{Babichev:2023dhs}, but we omit it here, since it does not lead to new solutions. We hence focus on the first branch, for which the scalar
field satisfies the equation
\begin{equation}
\phi'' =
-\frac{\left(\phi'\right)^2}{d\left(\phi\right)}.\label{eq:first_branch}
\end{equation}
To go further, one must show the compatibility of the remaining
equations with (\ref{eq:first_branch}), and with the ansatz for the
metric~(\ref{metric}). This requires fixing the potentials $W$, $Z$
and $V$ in such a way that the two remaining independent equations
admit the same metric function $f$ as solution. It is quite remarkable that, taking into account the
expression of $\phi''$ from (\ref{eq:first_branch}), the two
remaining equations\footnote{For spherical symmetry, the non-vanishing metric field equations are $\mathcal{E}_{tt}$, $\mathcal{E}_{rr}$, $\mathcal{E}_{\theta\theta}$ and $\mathcal{E}_{\varphi\varphi}=\sin^2\theta\,\mathcal{E}_{\theta\theta}$. On the other hand, the scalar field equation is implied by the metric field equations by virtue of the Bianchi identity (\ref{eq:bianchi}). Therefore the remaining field equations are indeed just $\mathcal{E}_{rr}$ and $\mathcal{E}_{\theta\theta}$.}, $\mathcal{E}_{rr}=0$ and
$\mathcal{E}_{\theta\theta}=0$, can be integrated once and twice
respectively, giving
\begin{equation}
\mathcal{E}_{rr} \propto I_1'\left(r\right),\quad
\mathcal{E}_{\theta\theta}\propto I_2''\left(r\right).
\label{eq:errethth}
\end{equation}
The following notations are used,
\begin{align}
I_1\left(r\right)\equiv{}& f^2\left(r^2V\right)'''-f\left(2r\left(1+\mathcal{W}'\right)+4V'+r^2\mathcal{W}''\right)+2r+2\mathcal{W}+r\mathcal{Z}'-\mathcal{Z},\\
I_2\left(r\right)\equiv{}& f^2\left(r V\right)'' - f
r\left(1+\mathcal{W}'\right) + \mathcal{Z}.
\end{align}
and are introduced for clarity two auxiliary functions $\mathcal{W}$ and
$\mathcal{Z}$, determined by,
\begin{equation}
W = \mathcal{W}\,',\quad rZ = \mathcal{Z}''. \label{eq:wz}
\end{equation}
The integration of the equations (\ref{eq:errethth}) implies the
existence of three integration constants, $d_1$, $c_1$ and $c_2$, such that
\begin{equation}
I_1 - d_1 = 0,\qquad I_2 -c_2+c_1 r =0. \label{eq:eqforf}
\end{equation}
As the following calculations show, the integration constants $c_1$, $c_2$, $d_1$ are
not independent and are either gauged away or related to the mass of the
BH. Compatibility of the field equations is ensured, once the two quadratic
equations (\ref{eq:eqforf}), defining the metric function $f$, are
proportional. Denoting by $2\mu\left(r\right)$ this proportionality
factor, which is \textit{a priori} an arbitrary non-vanishing function of $r$, one obtains the following system of
equations,
\begin{align}
\left(r^2 V\right)'''  ={}& 2\mu \left(r V\right)'',\label{eq:comp1}\\
4 V' = {}& 2 \left(\mu-1\right)r\left(\mathcal{W}'+1\right)-r^2\mathcal{W}'',\label{eq:comp2}\\
2r+2\mathcal{W} ={}& d_1-2\mu c_2+2\mu
c_1r+\left(2\mu+1\right)\mathcal{Z}-r\mathcal{Z}',\label{eq:comp3}
\end{align}
where it is assumed that the factors in front of different powers of
$f$ are proportional independently. For a given
proportionality factor $\mu\left(r\right)$, these equations will
determine the unfixed potentials $W$, $Z$ and $V$ as functions of
$r$ (or equivalently of $\phi$), while the quadratic equations (\ref{eq:eqforf}) will give the metric function $f(r)$.
\\

As one may notice, the above conditions for compatibility of the
equations are independent of the choice of  $d(\phi)$, indicating
that changing $d(\phi)$ does not change the physical results.
Indeed,  for any scalar field satisfying (\ref{eq:first_branch}),
the redefined scalar $\phi\to \int H(\phi)\phi' dr$
satisfies (\ref{eq:first_branch}) with $d(\phi)=-1$, provided that
$H$ solves the ordinary differential equation
$H_{\phi}-H^2-\frac{H}{d}=0$. One can therefore, without any loss of
generality, fix $d\left(\phi\right)=-1$. Then, the general solution
of Eq. (\ref{eq:first_branch}) is
\begin{equation}
\phi\left(r\right) = \ln \left(\frac{c}{r+\tilde{c}}\right), \label{eq:ccprime}
\end{equation}
where $c$ and $\tilde{c}$ are two integration constants. Note that the
constant $\tilde{c}$ can be further fixed to have a specific value for
convenience\footnote{Clearly, different choice of $\tilde{c}$
in (\ref{eq:ccprime}) amounts to a redefinition of the scalar field.
We can use the residual freedom to redefine the scalar by choosing
the constant of integration $\tilde{c}$. Indeed, the function $H(\phi)$ that
provides $d(\phi)=-1$ is defined up to an integration constant,
since it satisfies $H_{\phi}-H^2-\frac{H}{d}=0$. One can show that,
by adjusting this integration constant, one can change $\tilde{c}$.}. In
particular, one can choose $\tilde{c}\propto c$, as in the examples just
below, or $\tilde{c}=0$ for the solutions presented in the following paragraph~\ref{subsec:mu1}. Also, note the analogy with the logarithmic scalar field of Eq. (\ref{eq:log_scal}), which corresponds to one of the solutions of the 4DEGB theory. \textbf{It is possible to summarize the procedure described above by stating that, given an action (\ref{eq:action}) which generalizes the 4DEGB action, we are looking for compatibility conditions which enable the theory to admit a solution with a logarithmic scalar field similar to the one (\ref{eq:log_scal}) of the 4DEGB theory}.
\\

Let us first demonstrate how one can reproduce certain known
solutions by using our formalism. If the GB potential
$V=0$, then (\ref{eq:comp1}) is satisfied automatically, and
Eqs. (\ref{eq:vk})--(\ref{eq:v3}) show that the action only has the
Einstein-Hilbert potential $W$, the kinetic potential $V_k$ and the
self-interaction and cosmological constant in $Z$. In this case
therefore, all higher-order terms in action (\ref{eq:action}) are
missing, and we are left with an action with at most two derivatives. This  encompasses the BBMB and MTZ BHs, that we have recalled in paragraph~\ref{subsec:bbmbbh}. Indeed,
consider the following potentials,
\begin{equation}
V=0,\quad W=-\beta\mathrm{e}^{2\phi},\quad Z = -2\lambda\mathrm{e}^{4\phi}-2\Lambda,\quad V_k = 12\beta\mathrm{e}^{2\phi},
\label{chile}
\end{equation}
where $V_k$ is determined by $W$ according to (\ref{eq:vk}). Then, take into account the scalar field profile (\ref{eq:ccprime}),
and solve the compatibility
conditions (\ref{eq:comp2}-\ref{eq:comp3}). Finally finding the
metric function $f$ from (\ref{eq:eqforf}), one gets the solution
\begin{equation}
\phi = \ln\left(\frac{M}{\sqrt{\beta}\left(r-M\right)}\right),\quad f = \left(1-\frac{M}{r}\right)^2-\frac{\Lambda r^2}{3}, \label{eq:bbmbmtznotyet}
\end{equation}
provided the relation $\lambda=-\Lambda\beta^2$ holds.
Eq. (\ref{eq:bbmbmtz}) is either the BBMB BH (for
$\Lambda=0$) or the MTZ BH (for $\Lambda\neq 0$), with a
unique integration constant $M$ playing the role of the BH
mass. The value of the function $\mu(r)$ in both these cases is
$\mu(r)=1+M^2/\left(2M^2-3Mr+r^2\right)$.
\\

Another solution, for non-zero GB potential $V$, is the one of the 4DEGB theory (\ref{eq:fern}), i.e. potentials (\ref{eq:pot_fern}). Indeed, for $\lambda=\beta^2/(4\alpha)$, the scalar field reads
\begin{equation}
\phi\left(r\right)=\ln\left(\frac{\sqrt{-2\alpha/\beta}}{r}\right),
\end{equation}
see Eq. (\ref{eq:log_scal}). In the current formalism, this corresponds to the choice
$c=\sqrt{-2\alpha/\beta}$ and $\tilde{c}=0$ in the solution (\ref{eq:ccprime}) for $\phi$. The corresponding value of the function $\mu(r)$ is a constant, $\mu(r)=1$.
\\

As mentioned above, there is a possibility to fix the constant of
integration $\tilde{c}$ in (\ref{eq:ccprime}). For the BBMB and MTZ
solutions, we took $\tilde{c}\neq 0$ in order to stick with the standard
form of these solutions. On the other hand, to retrieve the solution
of the 4DEGB theory, we chose $\tilde{c}=0$, again to be in agreement with the original solution (\ref{eq:log_scal}). From now on, we will
set $\tilde{c}=0$ and thus consider
\begin{equation}
\phi\left(r\right) = \ln \left(\frac{c}{r}\right),
\label{eq:choice_of_phi}
\end{equation}
where $c>0$ is a constant with dimension $1$. As will be made clear below, the
constant $c$ of the scalar field solution (\ref{eq:choice_of_phi})
is related to the coupling constants of the theory, once the compatibility conditions (\ref{eq:comp1}-\ref{eq:comp3}) are solved.
\\

It turns out that compatibility conditions
(\ref{eq:comp1}-\ref{eq:comp3}) can be solved for any constant
$\mu$, the most interesting case being $\mu=1$. Indeed, it was shown above that the solution of 4DEGB corresponds to $\mu=1$. Conversely, it appears that solving the
compatibility conditions with $\mu=1$ leads to a more general action than the 4DEGB action (\ref{eq:fern}). This action admits BH solutions with far away Schwarzschild asymptotics. The following paragraph is thus dedicated to this $\mu=1$ case. The more general case of constant $\mu\neq 1$ is explained in our original article~\cite{Babichev:2023dhs}. In this case, there also exist closed-form BH solutions, but they do not have the correct asymptotic behaviour $f(r)=1-2M/r+\cdots$. Therefore, we will not study this case here and rather refer the interested reader to the original article.
\newpage
\subsection{New black hole solutions}\label{subsec:mu1}

For a constant
proportionality factor $\mu\left(r\right) = \mu =
\text{cst.}$, the compatibility conditions
(\ref{eq:comp1}-\ref{eq:comp3}) are integrable, and new explicit
solutions can be found. Different choices of constant
$\mu$ yield solutions of differing far away asymptotics, and only
for $\mu=1$ do the metric solutions have a standard four-dimensional Newtonian
behaviour at infinity, i.e. $f\sim 1-2M/r-(\Lambda r^2/3)$. Here, the optional
$\Lambda$-term, in parenthesis, stands for the cosmological constant if present in the action, while $M$ is the mass of the solution.
\\

Let us thus focus on $\mu=1$, and present the results in a way which enables to interpret them easily. We consider the following potentials $W, Z$ and $V$,
\begin{equation}
W = -\beta_4\mathrm{e}^{2\phi}-\beta_5\mathrm{e}^{3\phi},\quad Z =
-2\lambda_4\mathrm{e}^{4\phi}-2\lambda_5\mathrm{e}^{5\phi}-2\Lambda,\quad
V=-\alpha_4\phi-\alpha_5\mathrm{e}^{\phi},\label{eq:pot_mu_egal_1}
\end{equation}
where $\beta_4$, $\beta_5$, $\lambda_4$, $\lambda_5$, $\alpha_4$ and
$\alpha_5$ are six coupling constants, and $\Lambda$ is the usual cosmological constant. The choice of subscripts $4$ and
$5$ will become clear momentarily. The
remaining potentials are given by the compatibility conditions
(\ref{eq:vk}-\ref{eq:v3}) with $d=-1$, giving the following action,
\begin{align}
S = {}&\int\mathrm{d}^4x\sqrt{-g}\Biggl\{
R-2\Lambda-2\lambda_4\mathrm{e}^{4\phi}-2\lambda_5\mathrm{e}^{5\phi}-
\beta_4\mathrm{e}^{2\phi}\Bigl[R+6\left(\partial\phi\right)^2\Bigr]\nonumber\\{}&{}-
\beta_5\mathrm{e}^{3\phi}\Bigl[R+12\left(\partial\phi\right)^2\Bigr]
-\alpha_4\Bigl[\phi\mathcal{G}-4G^{\mu\nu}\phi_\mu\phi_\nu-
4\Box\phi\left(\partial\phi\right)^2-2\left(\partial\phi\right)^4\Bigr]\nonumber\\{}&{}-\alpha_5
\mathrm{e}^{\phi}\Bigl[\mathcal{G}-8G^{\mu\nu}\phi_\mu\phi_\nu-12\Box\phi
\left(\partial\phi\right)^2-12\left(\partial\phi\right)^4\Bigr]\Biggr\}.
\label{eq:complete}
\end{align}
The resulting action for
$\lambda_5=\beta_5=\alpha_5=0$ coincides with the 4DEGB theory (\ref{eq:fern}). We recall
that for this theory, $\delta S/\delta\phi$ is conformally
invariant, although the $\alpha_4$ term of
the action is not.
\\

As regards the parts of the action
depending on $\lambda_5$, $\beta_5$, $\alpha_5$, their interpretation is the following. They correspond to the most general action with \textbf{local conformal invariance in
five dimensions}, see for
example~\cite{Oliva:2011np,Giribet:2014bva}. To be perfectly clear, this means that the five-dimensional scalar-tensor action
\begin{align}
S = {}&\int\mathrm{d}^5x\sqrt{-g}\Biggl\{-2\lambda_5\mathrm{e}^{5\phi}-
\beta_5\mathrm{e}^{3\phi}\left(R+12\left(\partial\phi\right)^2\right)
\nonumber\\{}&{}-\alpha_5
\mathrm{e}^{\phi}\left(\mathcal{G}-8G^{\mu\nu}\phi_\mu\phi_\nu-12\Box\phi
\left(\partial\phi\right)^2-12\left(\partial\phi\right)^4\right)\Biggr\}
\end{align}
(note the five-dimensional element $\mathrm{d}^5x$) has local conformal invariance under $g_{\mu\nu}\to\mathrm{e}^{2\sigma}g_{\mu\nu}$, $\phi\to\phi-\sigma$. This motivates the use
of subscripts $4$ and $5$ for the parameterization of the full
action under consideration, Eq. (\ref{eq:complete}). Very importantly, the full action does not have conformal invariance, not even for $\delta S/\delta\phi$. The parts in $\lambda_5$, $\beta_5$, $\alpha_5$ have absolutely no symmetry in four dimensions. In the Horndeski vocabulary, this action corresponds to
\begin{align}
G_2 = {}&{}-2\Lambda+8X^2\left(\alpha_4+3\alpha_5\mathrm{e}^\phi\right)+12X\left(\beta_4\mathrm{e}^{2\phi}+2\beta_5\mathrm{e}^{3\phi}\right)-2\left(\lambda_4\mathrm{e}^{4\phi}+\lambda_5\mathrm{e}^{5\phi}\right)\nonumber\\{}&{}+8\alpha_5\mathrm{e}^\phi X^2\ln\left\lvert X\right\rvert,\\
G_3 = {}&{}8\alpha_4X+4\alpha_5\mathrm{e}^\phi X\left(3\ln\left\lvert X\right\rvert-1\right),\\ G_4 ={}&{} 1-\beta_4\mathrm{e}^{2\phi}-\beta_5\mathrm{e}^{3\phi}+4\alpha_4X+4\alpha_5\mathrm{e}^\phi X\left(\ln\left\lvert X\right\rvert-2\right),\\
G_5 ={}&{} -8\alpha_5\mathrm{e}^\phi +4\left(\alpha_4+\alpha_5\mathrm{e}^\phi\right)\ln\left\lvert X\right\rvert.
\end{align}
The link between the potentials (\ref{eq:pot_mu_egal_1}) and the framework we developed above, with the function $\mu(r)$, is as follows. \textbf{The potentials (\ref{eq:pot_mu_egal_1}), along with
the scalar field (\ref{eq:choice_of_phi}), solve the compatibility
conditions (\ref{eq:comp1}-\ref{eq:comp3}) for $\mu(r)\equiv 1$, for two
distinct sets of relations between the coupling constants $\alpha_{4,5}$, $\beta_{4,5}$, $\lambda_{4,5}$, and the constant $c$ appearing in the scalar field $\phi=\ln(c/r)$. This therefore yields two
distinct metric solutions of the form~(\ref{metric}) that we are going to detail now}.

\subsubsection{First solution}
The first
solution exists with all coupling constants switched on, namely,
\begin{equation}
\lambda_4 = \frac{\beta_4^2}{4\alpha_4},\quad \lambda_5 =
\frac{9\beta_5^2}{20\alpha_5},\quad\frac{\beta_5}{\beta_4} =
\frac{2\alpha_5}{3\alpha_4}.\label{eq:coupling_non_vanishing}
\end{equation}
The solution reads,
\begin{align}
\phi={}&\ln\left(\frac{\eta}{r}\right),\quad \eta \equiv \sqrt{\frac{-2\alpha_4}{\beta_4}},\nonumber\\
f\left(r\right) = {} & 1 + \frac{2\alpha_5\eta}{3\alpha_4r}
+\frac{r^2}{2\alpha_4}\left[1\pm
\sqrt{\left(1+\frac{4\alpha_5\eta}{3r^3}\right)^2+4\alpha_4\left(\frac{\Lambda}{3}+\frac{2
M}{r^3}+\frac{2\alpha_4}{r^4}+\frac{8\alpha_5\eta}{5r^5}\right)}\right],\label{eq:sol_1}
\end{align}
where $M$ is a free integration constant. A number of comments can
be made concerning this solution, which is of the Boulware-Deser
type~\cite{Boulware:1985wk}, typical of higher-order metric theories,
admitting two branches. The $+$ branch is asymptotically of
dS or adS type much like~\cite{Boulware:1985wk}, even for a vanishing $\Lambda$.
In Lovelock theory, this upper branch is perturbatively
unstable~\cite{Charmousis:2008ce}, although no direct analogy can be
made with the case here. We will not consider this branch any
further, as we are mostly interested in asymptotically flat
spacetimes. Therefore, for simplicity, let us set $\Lambda=0$, and consider
the $-$ branch, in order to discuss some properties of the
solution. 
\\

This branch displays a Schwarzschild limit as
the coupling constants $\alpha_4$, $\alpha_5$ tend to zero. Also,
for $\alpha_5=0$, which automatically implies $\beta_5=\lambda_5=0$,
the solution of the 4DEGB theory with logarithmic scalar field, (\ref{eq:prod_sphere_sol},\ref{eq:log_scal}), is
recovered. The asymptotics $r\to\infty$ of the full solution are
given by,
\begin{equation}
f(r)=1-\frac{2M}{r}-\frac{2\alpha_4}{r^2}-\frac{8\alpha_5\eta}{5r^3}+\mathcal{O}\left(\frac{1}{r^4}\right),
\end{equation}
and $M$ is therefore the ADM mass. The function $f$ has the same
behavior as the Schwarzschild metric up to first order, while the
next order is controlled by the coupling $\alpha_4$, and the other
couplings $\alpha_5$ and $\beta_4$ appear via $\eta$ in the higher
corrections. When we have a horizon, the BH has therefore
secondary hair, as the only integration constant appearing in the
metric is $M$, while all other constants are fixed by the theory.
Note that even if $M=0$, the spacetime is not trivial, and is in
fact a BH or naked singularity. On the other hand, as $r\to
0$, the metric function (\ref{eq:sol_1}) behaves as
\begin{equation}
f(r)=\left\lbrace\begin{split}& -\frac{1}{5}-\frac{21\alpha_4 r}{50\alpha_5\eta}+\mathcal{O}\left(r^2\right) {}&{} \text{if }\alpha_5>0, \\ & \frac{4\alpha_5\eta}{3\alpha_4 r}+\frac{11}{5}+\mathcal{O}\left(r\right) {}&{} \text{if }\alpha_5<0, \end{split}\right. 
\end{equation}
and while $f(0)$ is finite for $\alpha_5>0$, spacetime curvature is
infinite at $r=0$ since $f(r)$ does not possess a regular
core $f(r)=1+\mathcal{O}\left(r^2\right)$~\cite{Burzilla:2020utr}. In fact, the spacetime might not be defined in the whole
$r\in\left(0,\infty\right)$ but only on $\left(r_S,\infty\right)$
where $r_S>0$ is such that the square root ceases to be well-defined
below $r_S$. This branch singularity is typical for Lovelock
spacetimes~\cite{Charmousis:2008kc}. Before discussing the horizon
structure of the solution (\ref{eq:sol_1}), let us present the
second $\mu(r)\equiv 1$ solution arising from action (\ref{eq:complete}).
\subsubsection{Second solution}
Indeed a second, quite distinct class of solutions exists,
provided the couplings of the 4DEGB action are switched off,
\begin{equation}
\lambda_4=\beta_4=\alpha_4=0,\quad
\lambda_5=\frac{9\beta_5^2}{20\alpha_5},\label{eq:coupling_vanishing}
\end{equation}
with a scalar field and a metric function given by
\begin{equation}
\phi=\ln\left(\frac{\eta}{r}\right),\quad \eta =
2\sqrt{\frac{-\alpha_5}{3\beta_5}},\quad
f\left(r\right)=\frac{1}{1+\frac{4\alpha_5\eta}{3r^3}}\left[1-\frac{\Lambda
r^2}{3}-\frac{2M}{r}-\frac{4\alpha_5\eta}{15
r^3}\right],\label{eq:sol_2}
\end{equation}
where $M$ is a mass integration constant and $\Lambda$ the
cosmological constant which we again set to zero for simplicity.
Surprisingly, although the action includes higher-order terms, these do
not yield branching solutions with square roots. Asymptotically as $r\to\infty$, the metric function behaves
as
\begin{equation}
f(r) = 1-\frac{2M}{r}-\frac{8\alpha_5\eta}{5r^3}+\mathcal{O}\left(\frac{1}{r^4}\right).
\end{equation}
On the other hand, close to the origin the metric function does not
blow up and behaves as
\begin{equation}
f(r) =-\frac{1}{5}-\frac{3Mr^2}{2\alpha_5\eta}+\mathcal{O}\left(r^3\right).
\end{equation}
Note that although the $r^2$ term is what is needed for a regular BH, regularity is
spoiled by $f(0)=-1/5$. Hence, at $r=0$, there is a curvature
singularity. If $\alpha_5<0$, the spacetime becomes singular at
$0<r_S = \left(-4\alpha_5\eta/3\right)^{1/3}$, unless the numerator
of (\ref{eq:sol_2}) also vanishes at $r_S$. This occurs for a mass
$M=M_S$ with
\begin{equation}
M_S =\frac{6^{2/3}\left(-\alpha_5\eta\right)^{1/3}}{5}.\label{eq:ms}
\end{equation}
If $\alpha_5>0$ on the other hand, the spacetime is a BH, even for the case $M=0$.
\subsubsection{Horizon structure of both solutions}
More generally, as regards the horizon of both spacetimes (\ref{eq:sol_1}) and (\ref{eq:sol_2}), they turn out to be given by a cubic polynomial equation,
\begin{equation}
15r_h^3-30Mr_h^2-15\alpha_4r_h-4\alpha_5\eta=0,\label{eq:equation_horizon}
\end{equation}
where $r=r_h$ is the location of the event horizon, and with of course $\alpha_4=0$ in the case of (\ref{eq:sol_2}). This condition is necessary, but not sufficient. It is sufficient in the case of (\ref{eq:sol_1}) if $\alpha_4>0$ and $\alpha_5>0$, and in the case of (\ref{eq:sol_2}) if $\alpha_5>0$, or if $\alpha_5<0$ and $M> M_S$.
\\

\begin{figure}[!htb]
\begin{subfigure}{5.2cm}
\includegraphics[width=\linewidth]{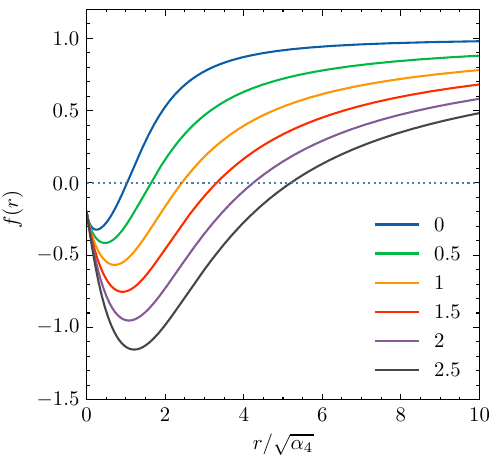}
\end{subfigure}
\begin{subfigure}{4.9cm}
\includegraphics[width=\linewidth]{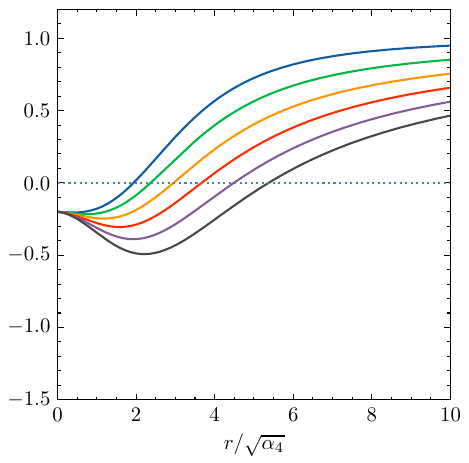}
\end{subfigure}
\begin{subfigure}{4.9cm}
\includegraphics[width=\linewidth]{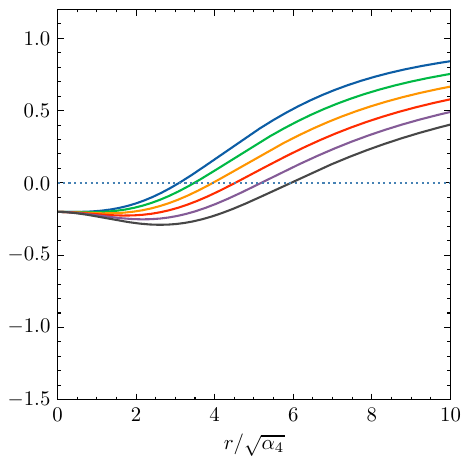}
\end{subfigure}
\caption{Metric profile $f(r)$ of Eq. (\ref{eq:sol_1}) for
$\alpha_4>0$ and different values of the mass (in units of
$\sqrt{\alpha_4}$, indicated by the colors) and different positive
values of the product $\alpha_5\eta$ (in units of $\alpha_4^{3/2}$),
namely: on the left $\alpha_5\eta=0.25$, in the middle
$\alpha_5\eta=20$, on the right $\alpha_5\eta=100$. The spacetime is
a BH for any mass, with hidden singularity at $r=0$.}
\label{fig:1Sel}
\end{figure}

\begin{figure}[!htb]
\begin{subfigure}{5.3cm}
\includegraphics[width=\linewidth]{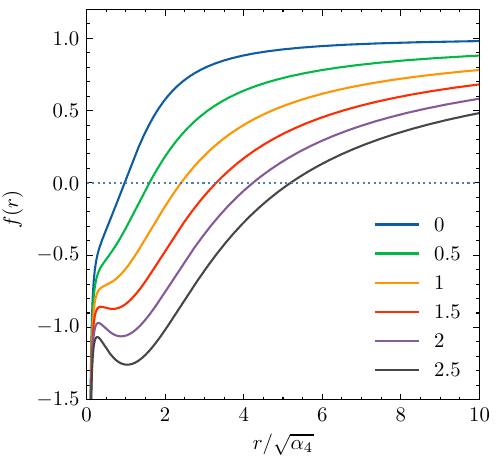}
\end{subfigure}
\begin{subfigure}{4.9cm}
\includegraphics[width=\linewidth]{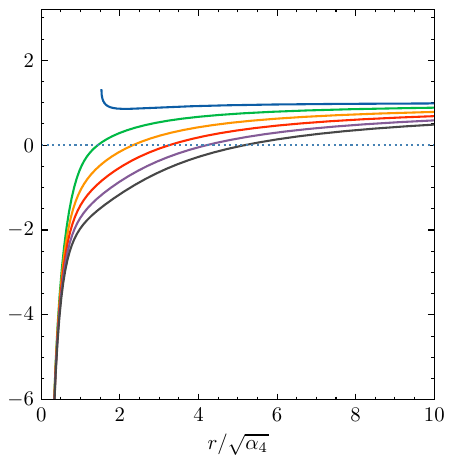}
\end{subfigure}
\begin{subfigure}{4.9cm}
\includegraphics[width=\linewidth]{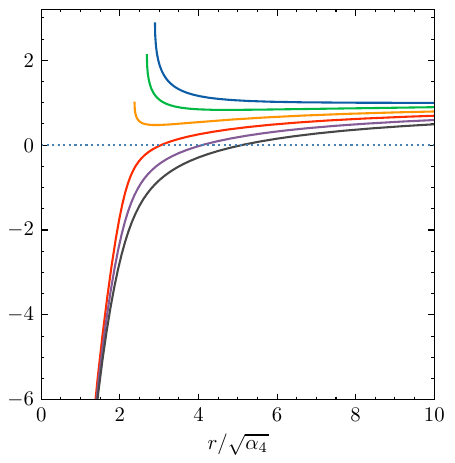}
\end{subfigure}
\caption{Metric profile $f(r)$ of Eq. (\ref{eq:sol_1}) for
$\alpha_4>0$ and different values of the mass (in units of
$\sqrt{\alpha_4}$, indicated by the colors) and different negative
values of the product $\alpha_5\eta$ (in units of $\alpha_4^{3/2}$),
namely: on the left $\alpha_5\eta=-0.25$, in the middle
$\alpha_5\eta=-2$, on the right $\alpha_5\eta=-10$. On the left, the
spacetime is a BH for any mass. When
$\left\lvert\alpha_5\eta\right\rvert$ increases (middle and right
plots), the light spacetimes acquire a naked singularity at a radius
$r_S>0$, while the heavier spacetimes remain BHs with hidden
singularity at $r=0$.} \label{fig:2Sel}
\end{figure}

\begin{figure}[!htb]
\begin{subfigure}{5.2cm}
\includegraphics[width=\linewidth]{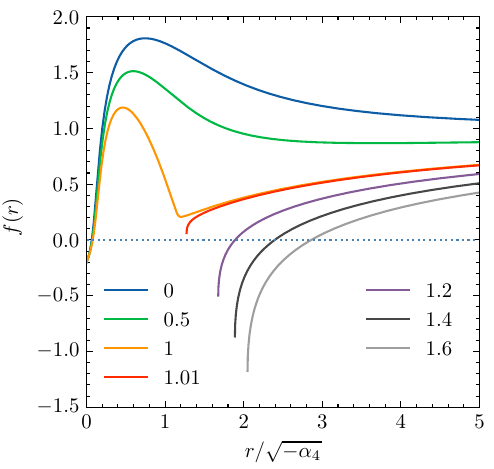}
\end{subfigure}
\begin{subfigure}{4.9cm}
\includegraphics[width=\linewidth]{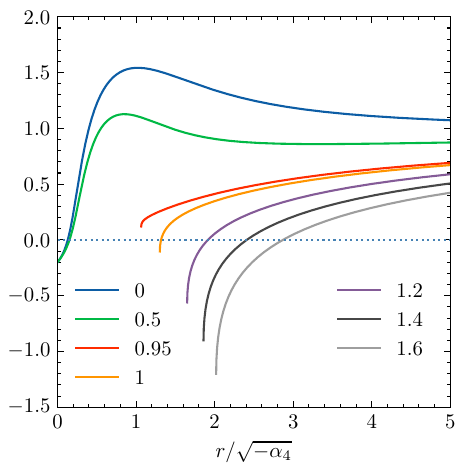}
\end{subfigure}
\begin{subfigure}{4.9cm}
\includegraphics[width=\linewidth]{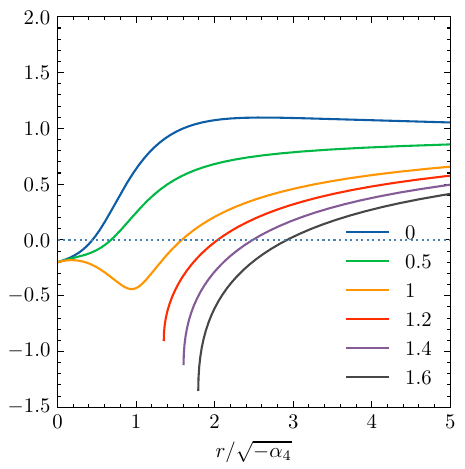}
\end{subfigure}
\caption{Metric profile $f(r)$ of Eq. (\ref{eq:sol_1}) for
$\alpha_4<0$ and different values of the mass (in units of
$\sqrt{-\alpha_4}$, indicated by the colors) and different positive
values of the product $\alpha_5\eta$ (in units of
$\left(-\alpha_4\right)^{3/2}$), namely: on the left
$\alpha_5\eta=0.25$, in the middle $\alpha_5\eta=0.5$, on the right
$\alpha_5\eta=2$. For small $\alpha_5\eta$ (left and middle plots),
light and heavy masses give BHs, with hidden singularity at
$r=0$ or at $r_S>0$ respectively, while intermediate masses give a
naked singularity at $r_S>0$ (see the red curves on the left and
middle plots). For sufficiently large $\alpha_5\eta$ (right plot),
all spacetimes are BHs, with hidden singularity at $r=0$ for
light masses or at $r_S>0$ for large masses.} \label{fig:3Sel}
\end{figure}

\begin{figure}[!htb]
\begin{subfigure}{5.2cm}
\includegraphics[width=\linewidth]{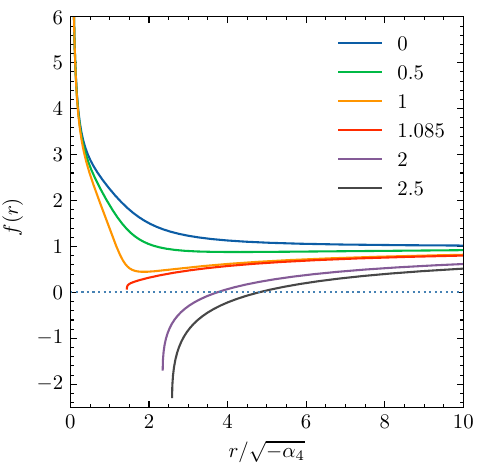}
\end{subfigure}
\begin{subfigure}{4.9cm}
\includegraphics[width=\linewidth]{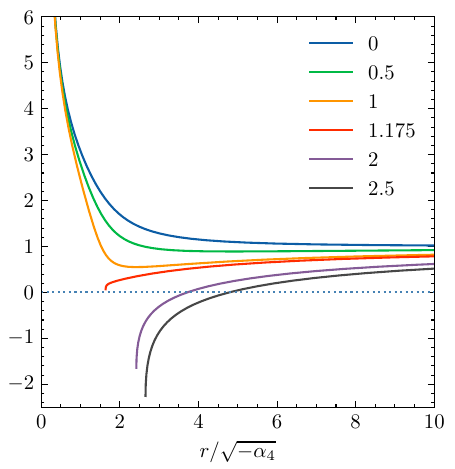}
\end{subfigure}
\begin{subfigure}{4.9cm}
\includegraphics[width=\linewidth]{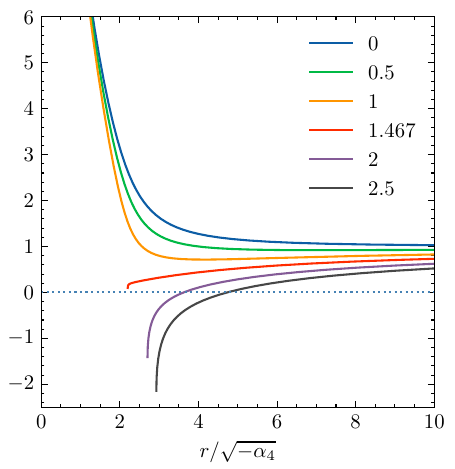}
\end{subfigure}
\caption{Metric profile $f(r)$ of Eq. (\ref{eq:sol_1}) for
$\alpha_4<0$ and different values of the mass (in units of
$\sqrt{-\alpha_4}$, indicated by the colors) and different negative
values of the product $\alpha_5\eta$ (in units of
$\left(-\alpha_4\right)^{3/2}$), namely: on the left
$\alpha_5\eta=-0.25$, in the middle $\alpha_5\eta=-1$, on the right
$\alpha_5\eta=-5$. The spacetime is a naked singularity at $r=0$ for
light masses, a naked singularity at $r_S>0$ for intermediate masses
(see the red curves), and a BH with hidden singularity at
$r_S>0$ for large masses.} \label{fig:4Sel}
\end{figure}

In order to sketch the general aspect of the spacetimes and their horizons, we present various plots of the functions $f(r)$ of Eqs. (\ref{eq:sol_1}) and (\ref{eq:sol_2}). The plots in
Figs.~\ref{fig:1Sel}, \ref{fig:2Sel}, \ref{fig:3Sel} and \ref{fig:4Sel} present respectively the cases ($\alpha_4>0$, $\alpha_5>0$),
($\alpha_4>0$, $\alpha_5<0$), ($\alpha_4<0$, $\alpha_5>0$) and
($\alpha_4<0$, $\alpha_5<0$) of Eq.~(\ref{eq:sol_1}). The obtained spacetimes have at most
one horizon. It is only when $\alpha_4$ and $\alpha_5$ are positive that there is always a horizon (even for $M=0$). This is due to the fact that the square root is never zero and no branch singularity is possible. For all the
other cases however, there may be naked singularities for certain values
of the coupling constants. An exotic result, in the case of Fig.
\ref{fig:3Sel}, left and middle plots, is a mass gap and horizon gap
between light BHs and heavy BHs: there exist masses $M_1$ and $M_2$ such that masses
$M\in\left(M_1,M_2\right)$ give rise to naked singularities, while
$M< M_1$ or $M> M_2$ give BHs, the BHs with
$M< M_1$ having very tiny horizons.\\

\begin{figure}[!htb]
\begin{subfigure}{8cm}
\includegraphics[width=\linewidth]{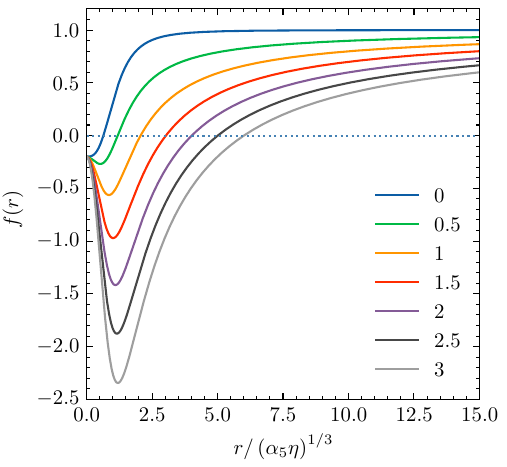}
\end{subfigure}
\begin{subfigure}{7.3cm}
\includegraphics[width=\linewidth]{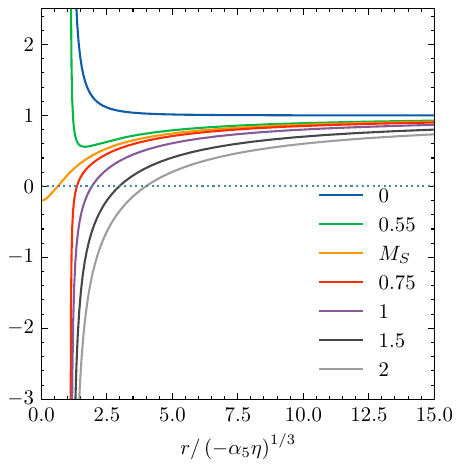}
\end{subfigure}
\caption{Metric profile $f(r)$ of Eq. (\ref{eq:sol_2}) for
$\alpha_5>0$ (left plot) or $\alpha_5<0$ (right plot) and different
values of the mass (in units of
$\left\lvert\alpha_5\eta\right\rvert^{1/3}$, indicated by the
colors). For $\alpha_5>0$, the spacetime is a BH for any
mass. For $\alpha_5<0$, the only mass giving a singularity at $r=0$
is $M_S$, see Eq. (\ref{eq:ms}), and this singularity is hidden by a
horizon. For other masses, there is a singularity at
$r_S=\left(-4\alpha_5\eta/3\right)^{1/3}$, either naked for $M<
M_S$ or hidden by a horizon if $M> M_S$.} \label{fig:5}
\end{figure}

Concerning (\ref{eq:sol_2}),
its profile is presented in Fig.~\ref{fig:5}: if $\alpha_5>0$, it is
a BH for any mass, while for $\alpha_5<0$, it is a BH only for $M\geq M_S$ where $M_S$ is given by (\ref{eq:ms}).
\\

The two cases where a horizon exists for any mass, that is to say, Fig.~\ref{fig:1Sel} [solution (\ref{eq:sol_1}) with $\alpha_4>0$, $\alpha_5>0$] and Fig.~\ref{fig:5}, left plot [solution (\ref{eq:sol_2}) with $\alpha_5>0$], are in fact strongly constrained by the following argument, which was formerly developed in~\cite{Charmousis:2021npl}. One assumes that the considered solutions verify a Birkhoff type argument~\cite{Israel:1967wq}, more precisely, that they are the unique static, spherically-symmetric solutions of their respective theories. If this assumption is true, these solutions must in particular represent the gravitational field created by an atomic nucleus, of radius $R\sim 10^{-15}\,$m and mass $M\sim 10^{-54}\,$m. Since these nuclei can be experimentally probed, they are not covered by a horizon. Therefore $r_h<R$, where $r_h$ is root of Eq. (\ref{eq:equation_horizon}). It is easy to show that, for the considered two cases, this leads to the constraints
\begin{equation}
0<\alpha_4<R\left(R-2M\right)\sim 10^{-30}\,\text{m}^2,\quad 0<\alpha_5\eta<\frac{15}{4}R^2\left(R-2M\right)\sim 10^{-45}\,\text{m}^3.
\end{equation}
Of course, the first inequality does not concern Fig.~\ref{fig:5}, left plot, which already has $\alpha_4=0$. Such stringent bounds make the associated gravitational effects probably undetectable.
\\

In the case of solution (\ref{eq:sol_2}), thus remains most likely a unique case, $\alpha_5<0$ (Fig.~\ref{fig:5}, right plot), where BHs have a minimal mass $M_S$ given by (\ref{eq:ms}). Another argument from~\cite{Charmousis:2021npl} can then be used to constrain the value of $\left\lvert\alpha_5\eta\right\rvert$ for this theory. Indeed, the minimal mass $M_S$ must be lower than the mass of experimentally detected BHs. In GW200115~\cite{LIGOScientific:2021qlt}, the second object is a BH of mass $M=5.7^{+1.8}_{-2.1}M_\odot$ at $90\%$ credible interval, giving,
\begin{equation}
\left\lvert\alpha_5\eta\right\rvert \lesssim 2070^{+565}_{-659}\,\text{km}^3.
\end{equation}
If one takes into account other events, for which the second object is lighter but it is not sure that it is a BH, namely GW170817~\cite{LIGOScientific:2017ync,LIGOScientific:2017vwq} and GW190814~\cite{LIGOScientific:2020zkf,Fattoyev:2020cws}, one rather gets $\left\lvert\alpha_5\eta\right\rvert \lesssim 230\,\text{km}^3$ and $\left\lvert\alpha_5\eta\right\rvert \lesssim 194\,\text{km}^3$. Finding such simple constraint would not be possible in the other cases, of Figs.~\ref{fig:2Sel},~\ref{fig:3Sel} and~\ref{fig:4Sel}. Indeed, in these cases, the minimal mass depends non trivially on both $\alpha_4$ and $\alpha_5\eta$.
\\

This completes the study of solutions of the form~(\ref{metric}) to
action (\ref{eq:complete}) with a non-trivial scalar field having a logarithmic running. One can also question the existence of solutions with a constant
scalar field, $\phi=\phi_0$. It is easy to see that, for
the considered action (\ref{eq:complete}), a solution exists, provided the couplings satisfy
\begin{align}
0={}&\alpha_4^2\left[\alpha_4^3 \alpha_5 (\beta_4 \lambda_5-2 \beta_5 \lambda_4)+
\alpha_4^4 \beta_5 \lambda_5-5 \alpha_4 \alpha_5^3 \lambda_5+4 \alpha_5^4 \lambda_4\right]+2\alpha_5^5\left(2\alpha_5\beta_4-3\alpha_4\beta_5\right)\Lambda , \label{eq:cond_cst}\\
0\neq{}&\alpha_5^3-\alpha_4^2\alpha_5\beta_4+\alpha_4^3\beta_5.\label{eq:cond_cst_2}
\end{align}
The solution is a Schwarzschild-(a)dS BH,
\begin{equation}
\phi_0 = \ln\left(-\frac{\alpha_4}{\alpha_5}\right),\quad
f(r)=1-\frac{2M}{r}+\frac{\alpha_4^4\left(\alpha_4\lambda_5-\alpha_5\lambda_4\right)-
\Lambda\alpha_5^5}{3\alpha_5^2\left(\alpha_5^3-\alpha_4^2\alpha_5\beta_4+\alpha_4^3\beta_5\right)}r^2.\label{eq:sads}
\end{equation}
Interestingly enough, this solution is valid for the
theory with coupling constants given by
(\ref{eq:coupling_non_vanishing}), which also admits the BH solution (\ref{eq:sol_1}). However, while (\ref{eq:sol_1}) is asymptotically flat if $\Lambda=0$, this is not the case of (\ref{eq:sads}), which
has an effective cosmological constant.
\\

In a word, starting from a generic action
\begin{align}
S = \int \mathrm{d}^4x\sqrt{-g}\Bigl\{{}&{}\left(1+W\left(\phi\right)\right)R-
\frac{1}{2}V_k\left(\phi\right)\left(\partial\phi\right)^2+Z\left(\phi\right)+
V\left(\phi\right)\mathcal{G}\nonumber\\
 {}&{}+V_2\left(\phi\right)G^{\mu\nu}\partial_\mu\phi\,\partial_\nu\phi +V_3\left(\phi\right)\left(\partial\phi\right)^4+V_4\left(\phi\right)\Box\phi\left(\partial\phi\right)^2\Bigr\} \label{eq:action_no_sym}
\end{align}
and imposing the scalar field to satisfy the equation $\mathcal{E}^t_t-\mathcal{E}^r_r=0$, enables to find new BH solutions with a relevant GR limit.\\

An important aspect of this work might be the following. All closed-form BH solutions found before this work, and which have been recalled in the previous chapters, were found in scalar-tensor theories displaying some kind of symmetry. Indeed, the theories either had shift symmetry (which is an exact, continuous symmetry at the level of the action), or generalized conformal symmetry (which is a local symmetry bearing only on $\delta S/\delta\phi$). On the contrary, the newly presented solutions (\ref{eq:sol_1}) and (\ref{eq:sol_2}) exist for the action (\ref{eq:complete}). \textbf{This latter action does not display any of these symmetries}. Rather, this action presents a - yet unexplained - link with a five-dimensional local conformal symmetry, which of course does not lead to any symmetry in four dimensions. 
\\

The starting action (\ref{eq:action_no_sym}) displays couplings up to the GB invariant $\mathcal{G}$, as inspired by the 4DEGB action (\ref{eq:fern}). Let us now consider couplings with higher-order Lovelock invariants. Lovelock invariants of order higher than $\mathcal{G}$ are non-vanishing only in spacetime dimension $D\geq 6$. Therefore, one must start with such higher-dimensional considerations (this is the focus of the next section), before finding consequences in four dimensions (this will be dealt with in Sec.~\ref{sec:diagkk}).
\section[Scalar field conformally coupled to Lovelock invariants in arbitrary dimension]{Scalar field conformally coupled to Lovelock invariants in arbitrary dimension}\label{sec:higherdim}
\sectionmark{5.2~~Conformal couplings with Lovelock invariants in arbitrary dimension}
\subsection{Conformal couplings}
As explained in paragraph~\ref{subsec:local} and Sec.~\ref{sec:fourd}, the 4DEGB theory coincides with \textbf{the most general scalar-tensor action with generalized conformal invariance}, under
\begin{equation}
g_{\mu\nu}\to\mathrm{e}^{2\sigma(x)} g_{\mu\nu},\quad\phi\to\phi-\sigma(x),\label{eq:loc_inv_conf_lov}
\end{equation}
where $\sigma(x)$ is an arbitrary function on spacetime. One of the building blocks of such an action is \textbf{the most general action with local conformal invariance}, $S_{\text{loc}}$. $S_{\text{loc}}$ was constructed in paragraph~\ref{subsec:local}, see Eq. (\ref{eq:slocpara}), by noticing that its corresponding Lagrangian density had to be a Lorentz scalar density of unit weight, built out of the metric $\hat{g}_{\mu\nu}$ invariant under the transformation (\ref{eq:loc_inv_conf_lov}),
\begin{equation}
\hat{g}_{\mu\nu} = \mathrm{e}^{2\phi} g_{\mu\nu}.
\end{equation}
By restricting to second-order field equations, this unambiguously led to
\begin{equation}
S_{\text{loc}}\left[g_{\mu\nu},\phi\right]=\int\mathrm{d}^4x\sqrt{-\hat{g}}\left(-\beta\hat{R}-2\lambda\right)=\int\mathrm{d}^4x\sqrt{-g}\Bigl[-\beta\mathrm{e}^{2\phi}\left(R+6\left(\partial\phi\right)^2\right)-2\lambda\mathrm{e}^{4\phi}\Bigr].\label{eq:sloc4}
\end{equation}
Indeed, the intermediate expression is just the Einstein-Hilbert (with cosmological constant) action for the metric $\hat{g}_{\mu\nu}$, with $\beta$ and $\lambda$ two coupling constants. Having this in mind, it becomes straightforward to generalize this construction in arbitrary spacetime dimension $D$. Let us thus note $S_{(D),\,\text{loc}}\left[g_{\mu\nu},\phi\right]$ the most general scalar-tensor action in $D$ dimensions, invariant under the local conformal transformation (\ref{eq:loc_inv_conf_lov}), and yielding second-order field equations. This action was constructed by Oliva and Ray~\cite{Oliva:2011np} in the following manner. This time, $S_{(D),\,\text{loc}}$ will contain not only the Einstein-Hilbert and cosmological constant terms for the metric $\hat{g}_{\mu\nu}$, but also all the remaining Lovelock invariants of the metric $\hat{g}_{\mu\nu}$. Consequently,
\begin{equation}
S_{(D),\,\text{loc}}\left[g_{\mu\nu},\phi\right]=\int\mathrm{d}^Dx\sqrt{-\hat{g}}\sum_{k=0}^{\lfloor \frac{D-1}{2}\rfloor}\beta_k\hat{\mathcal{R}}^{(k)},
\end{equation}
where the $\beta_k$'s are couplings constants, and $\hat{\mathcal{R}}^{(k)}$ is defined by (\ref{eq:lov_lov}), but for the metric $\hat{g}_{\mu\nu}$, i.e., on the right hand side of (\ref{eq:lov_lov}) appears the Riemann tensor of $\hat{g}_{\mu\nu}$~\cite{Carneiro:2004rt,Dabrowski:2008kx},
\begin{equation}
\hat{R}^{\mu\nu}_{\,\,\,\,\,\,\rho\sigma}=\mathrm{e}^{-2\phi}S^{\mu\nu}_{\,\,\,\,\,\,\rho\sigma},\quad S^{\mu\nu}_{\,\,\,\,\,\,\rho\sigma}\equiv R^{\mu\nu}_{\,\,\,\,\,\,\rho\sigma}-4\delta^{[\mu}_{[\rho}\phi^{\nu]}_{\sigma]}+4\delta^{[\mu}_{[\rho}\phi^{\nu]}\phi_{\sigma]}-\left(\partial\phi\right)^2\delta^{\mu\nu}_{\rho\sigma}.\label{eq:s1}
\end{equation}
In fact, because of the Kronecker symbols which appear in $S^{\mu\nu}_{\,\,\,\,\,\,\rho\sigma}$, the expression for $\hat{\mathcal{R}}^{(k)}$ depends on the dimension $D$ of spacetime, which is not the case for $\mathcal{R}^{(k)}$.\footnote{We are here talking of the covariant expression. Of course, once specifying to a basis, the value of $\mathcal{R}^{(k)}$ does depend on the dimension. For instance, one has $\mathcal{R}^{(1)}=R$ (Ricci scalar, same expression as a Lorentz scalar for any dimension), but for a sphere of dimension $D$ and unit radius, $R=D$ depends on the dimension.} We thus indicate this dependence on the dimension by a subscript, $\hat{\mathcal{R}}^{(k)}_{(D)}$, and define
\begin{equation}
\mathcal{S}^{(k)}_{(D)}\equiv \mathrm{e}^{2k\phi}\,\hat{\mathcal{R}}^{(k)}_{(D)}= \frac{1}{2^k}\delta^{\mu_1\nu_1\cdots\mu_k\nu_k}_{\alpha_1\beta_1\cdots\alpha_k\beta_k}\prod_{i=1}^{k}S^{\alpha_i\beta_i}_{\quad\,\mu_i\nu_i},\label{eq:s2}
\end{equation}
where it is understood that the Kronecker symbols appearing both in $\delta^{\mu_1\nu_1\cdots\mu_k\nu_k}_{\alpha_1\beta_1\cdots\alpha_k\beta_k}$ and in $S^{\alpha_i\beta_i}_{\quad\,\mu_i\nu_i}$ are accordingly the $D$-dimensional ones. Then, the invariant action is a linear combination of Lagrangians $\mathcal{L}^{(k)}_{(D)}$,
\begin{equation}
S_{(D),\,\text{loc}}\left[g_{\mu\nu},\phi\right]=\int\mathrm{d}^Dx\sqrt{-g}\sum_{k=0}^{\lfloor \frac{D-1}{2}\rfloor}\beta_k\,\mathcal{L}^{(k)}_{(D)},\quad \mathcal{L}^{(k)}_{(D)} \equiv\mathrm{e}^{(D-2k)\phi}\,\mathcal{S}^{(k)}_{(D)}. \label{eq:slocD}
\end{equation}
Remember that $S_{(D),\,\text{loc}}$ is not interesting in itself: the local conformal invariance removes the additional degree of freedom that one is precisely trying to introduce with the scalar field. One has to add terms in the action which break conformal invariance. In four dimensions, adding the Einstein-Hilbert terms, $R-2\Lambda$, led to the action $S_\text{BBMB/MTZ}$, see (\ref{eq:bbmbmtz}), which admits the BBMB and MTZ solutions. In $D$ dimensions, one can naturally add not only the Einstein-Hilbert term but also the whole set of Lovelock invariants,
\begin{equation}
S_{(D),\,\text{Lov}}\left[g_{\mu\nu}\right]=\int\mathrm{d}^Dx\sqrt{-g}\sum_{k=0}^{\lfloor \frac{D-1}{2}\rfloor}\alpha_k\mathcal{R}^{(k)},
\end{equation}
thus getting a total action where the local conformal invariance is broken,
\begin{align}
S_{(D),\,\text{brok}}\left[g_{\mu\nu},\phi\right] {}&{}\equiv S_{(D),\,\text{Lov}}\left[g_{\mu\nu}\right]+S_{(D),\,\text{loc}}\left[g_{\mu\nu},\phi\right]\nonumber\\{}&{}=\int\mathrm{d}^Dx\sqrt{-g}\sum_{k=0}^{\lfloor \frac{D-1}{2}\rfloor}\left(\alpha_k\mathcal{R}^{(k)}+\beta_k\,\mathcal{L}^{(k)}_{(D)}\right).\label{eq:action_brok_D}
\end{align}
We note $\mathcal{E}_{\mu\nu}$ and $\mathcal{E}_\phi$ the field equations of $S_{(D),\,\text{brok}}$. This action, which differs from $S_{(D),\,\text{loc}}$ only by a pure metric action, has generalized conformal invariance, i.e. conformal invariance of $\delta S/\delta\phi$. In paragraph~\ref{subsec:local}, we recalled that generalized conformal invariance is equivalent to the existence of a pure geometric equation, see Eq. (\ref{eq:puregeomfisrtoc}). The proof remains of course unaffected by the mere change of dimension. Therefore, the field equations of $S_{(D),\,\text{brok}}$ combine to yield a purely geometric equation,
\begin{equation}
\mathfrak{R}=\mathcal{E}^\mu_\mu+\mathcal{E}_\phi=\mathcal{E}^\mu_{\text{Lov}\,\mu}+\mathcal{E}^\mu_{\text{loc}\,\mu}+\mathcal{E}_{\text{loc}\,\phi}=\mathcal{E}^\mu_{\text{Lov}\,\mu}.
\end{equation}
Indeed, $\mathcal{E}^\mu_{\text{loc}\,\mu}+\mathcal{E}_{\text{loc}\,\phi}=0$ because of the local conformal invariance of $S_{(D),\,\text{loc}}$, see Eq. (\ref{eq:geomegal0}). Using (\ref{eq:lov_eq}) and (\ref{eq:lov_trace}) gives
\begin{equation}
\mathfrak{R}=\sum_{k=0}^{\lfloor \frac{D-1}{2}\rfloor}\alpha_k\left(2k-D\right)\mathcal{R}^{(k)}.\label{eq:geom_D}
\end{equation}
Of course, this can be verified directly from the full field equations of the theory, which are
\begin{equation}
\mathcal{E}_{\mu\nu}=2\sum_{k=0}^{\lfloor \frac{D-1}{2}\rfloor}\left(\alpha_k\mathcal{H}^{(k)}_{\mu\nu}+\beta_k\mathcal{T}^{(k)}_{(D)\mu\nu}\right),\quad\mathcal{E}_\phi = \sum_{k=0}^{\lfloor \frac{D-1}{2}\rfloor}\beta_k\left(D-2k\right)\mathrm{e}^{(D-2k)\phi}\,\mathcal{S}^{(k)}_{(D)}.\label{eq:field_eq_D}
\end{equation}
$\mathcal{H}^{(k)}_{\mu\nu}$ is defined in (\ref{eq:lov_trace}) and $\mathcal{T}^{(k)}_{(D)\mu\nu}$ is analogous,
\begin{equation}
\mathcal{H}_{\quad\nu}^{(k)\mu}=\frac{-1}{2^{k+1}}\delta^{\mu\,\mu_1\nu_1\cdots\mu_k\nu_k}_{\nu\,\alpha_1\beta_1\cdots\alpha_k\beta_k}\prod_{i=1}^{k} R^{\alpha_i\beta_i}_{\quad\,\mu_i\nu_i},\quad\mathcal{T}^{(k)\mu}_{(D)\nu}\equiv \frac{-\mathrm{e}^{(D-2k)\phi}}{2^{k+1}}\delta^{\mu\,\mu_1\nu_1\cdots\mu_k\nu_k}_{\nu\,\alpha_1\beta_1\cdots\alpha_k\beta_k}\prod_{i=1}^{k} S^{\alpha_i\beta_i}_{\quad\,\mu_i\nu_i},\label{eq:handt}
\end{equation}
with traces
\begin{equation}
\mathcal{H}_{\quad\mu}^{(k)\mu}=\frac{2k-D}{2}\mathcal{R}^{(k)},\quad \mathcal{T}^{(k)\mu}_{(D)\mu}=\frac{2k-D}{2}\mathrm{e}^{(D-2k)\phi}\mathcal{S}^{(k)}_{(D)}.\label{eq:traceht}
\end{equation}
In a word, we are considering the action (\ref{eq:action_brok_D}), with field equations (\ref{eq:field_eq_D}), which combine to give the geometric equation (\ref{eq:geom_D}). Let us now make the somehow obscure conformal Lagrangians $\mathcal{L}^{(k)}_{(D)}$ appearing in $S_{(D),\,\text{loc}}$, Eq. (\ref{eq:slocD}), more explicit by studying the simplest cases, $k\leq 3$. The following considerations are based on the article~\cite{Babichev:2023rhn}, of which the present author is one of the authors.\newpage
\subsection{Simple examples} \label{subsec:simple_ex}
\subsubsection{$k=0$: self-interacting conformal potential}
$\mathcal{L}^{(0)}_{(D)}$ is of course the self-interacting potential with local conformal invariance in $D$ dimensions,
\begin{equation}
\int\mathrm{d}^Dx\sqrt{-g}\,\mathcal{L}^{(0)}_{(D)} = \int\mathrm{d}^Dx\sqrt{-g}\,\mathrm{e}^{D\phi}.\label{eq:l0}
\end{equation}
\subsubsection{$k=1$: conformal coupling to the Ricci scalar}
$\mathcal{L}^{(1)}_{(D)}$ is the conformal coupling to the Ricci scalar,
\begin{align}
\int\mathrm{d}^Dx\sqrt{-g}\,\mathcal{L}^{(1)}_{(D)}={}&{}\int\mathrm{d}^Dx\sqrt{-g}\,\mathrm{e}^{(D-2)\phi}\Bigl(R-2\left(D-1\right)\Box\phi-\left(D-1\right)\left(D-2\right)\left(\partial\phi\right)^2\Bigr)\label{eq:l1_first}\\
\underset{\text{IBP}}{=}{}&{}\int\mathrm{d}^Dx\sqrt{-g}\,\mathrm{e}^{(D-2)\phi}\Bigl(R+\left(D-1\right)\left(D-2\right)\left(\partial\phi\right)^2\Bigr),\label{eq:l1}
\end{align}
where $\underset{\text{IBP}}{=}$ means that an integration by parts is performed and associated boundary terms are dropped. Taking $D=4$, both the self-interaction and the conformal coupling to $R$ are of course consistent with the terms of the four-dimensional invariant action (\ref{eq:sloc4}).
\subsubsection{$k=2$: conformal coupling to the Gauss-Bonnet invariant}
$\mathcal{L}^{(2)}_{(D)}$ corresponds to the conformal coupling with the GB invariant,
\begin{align}
\int\mathrm{d}^Dx\sqrt{-g}\,\mathcal{L}^{(2)}_{(D)}={}&{}\int\mathrm{d}^Dx\sqrt{-g}\,\mathrm{e}^{(D-4)\phi}\Bigl\{\mathcal{G}+8\left(D-3\right)G^{\mu\nu}\phi_{\mu\nu}\nonumber\\{}&{}+4\left(D-2\right)\left(D-3\right)\left[\left(\Box\phi\right)^2-\phi_{\mu\nu}\phi^{\mu\nu}\right]\nonumber\\{}&{}-2\left(D-3\right)\left(D-4\right)R\left(\partial\phi\right)^2-8\left(D-3\right)R_{\mu\nu}\phi^\mu\phi^\nu\nonumber\\{}&{}+8\left(D-2\right)\left(D-3\right)\phi^\mu\phi_{\mu\nu}\phi^\nu+4\left(D-2\right)\left(D-3\right)^2\left(\partial\phi\right)^2\Box\phi\nonumber\\{}&{}+\left(D-1\right)\left(D-2\right)\left(D-3\right)\left(D-4\right)\left(\partial\phi\right)^4\Bigr\}\label{eq:l2_first}\\
\underset{\text{IBP}}{=}{}&{}\int\mathrm{d}^Dx\sqrt{-g}\,\mathrm{e}^{(D-4)\phi}\Bigl[\mathcal{G}-4\left(D-3\right)\left(D-4\right)G^{\mu\nu}\phi_\mu\phi_\nu\nonumber\\{}&{}-2\left(D-2\right)\left(D-3\right)\left(D-4\right)\left(\partial\phi\right)^2\Box\phi\nonumber\\{}&{}-\left(D-2\right)\left(D-3\right)^2\left(D-4\right)\left(\partial\phi\right)^4\Bigr].\label{eq:l2}
\end{align}
Remember that this Lagrangian emerges as a term $\int\mathrm{d}^Dx\sqrt{-\hat{g}}\,\hat{\mathcal{G}}_{(D)}$ for the metric $\hat{g}_{\mu\nu}$. In four dimensions, this integral is proportional to the Euler characteristic of the spacetime manifold by the Chern theorem, so $\mathcal{L}^{(2)}_{(D)}$ should be a boundary term. One can verify it immediately by setting $D=4$ in the last expression: all terms vanish but $\mathcal{G}$, which is indeed a boundary term. However, we have previously encountered, in the 4DEGB theory (\ref{eq:fern}), an action in $D=4$ where $\phi$ is conformally coupled to $\mathcal{G}$, namely
\begin{equation}
S=\int\mathrm{d}^4x\sqrt{-g}\,\mathcal{L},\quad \mathcal{L}=\phi\mathcal{G}-4G^{\mu\nu}\phi_\mu\phi_\nu-4\left(\partial\phi\right)^2\Box\phi-2\left(\partial\phi\right)^4.\label{eq:4dDd}
\end{equation}
This action satisfies the generalized conformal invariance, i.e., $\delta S/\delta \phi$ is locally conformally invariant, although $S$ itself is not. In fact, this action was precisely constructed by Fernandes in~\cite{Fernandes:2021dsb} by imposing this condition on $\delta S/\delta \phi$, and then 'integrating' this condition to find $S$. Alternatively, we presented in Sec.~\ref{sec:fourd} how $S$ can be obtained by a KK dimensional reduction followed by a singular limit procedure, where the dimension of the internal space is sent to zero. \textbf{Here, thanks to the construction of $\mathcal{L}^{(2)}_{(D)}$ for arbitrary $D$, one obtains a third way to reach the four-dimensional action $S$ of Eq. (\ref{eq:4dDd}), or equivalently its corresponding Lagrangian $\mathcal{L}$}. One first writes
\begin{equation}
\mathrm{e}^{\left(D-4\right)\phi}=1+\left(D-4\right)\phi+\mathcal{O}\left(\left(D-4\right)^2\right),
\end{equation}
and multiplies $\mathcal{L}_2$ by its coupling constant $\beta_2$, getting
\begin{align}
\beta_2\mathcal{L}_2={}&{}\beta_2\mathcal{G}+\beta_2\left(D-4\right)\Bigl[\phi\mathcal{G}-4\left(D-3\right)G_{\mu\nu}\phi^\mu\phi^\nu-2\left(D-2\right)\left(D-3\right)\left(\partial\phi\right)^2\Box\phi\nonumber\\{}&{}\hspace{3cm}-\left(D-2\right)\left(D-3\right)^2\left(\partial\phi\right)^4\Bigr]+\beta_2\,\mathcal{O}\left(\left(D-4\right)^2\right).
\end{align}
We then define the limit $D\to 4$ in the following way. First, $\beta_2\mathcal{G}$ is a boundary term in $D=4$ and is thus removed. Second, one introduces a new coupling $\tilde{\beta}_2\equiv \beta_2\left(D-4\right)$ and imposes $\tilde{\beta}_2$ to remain constant when taking the limit $D\to 4$ (so the original coupling $\beta_2$ is sent to infinity). So the last term, $\beta_2\,\mathcal{O}\left(\left(D-4\right)^2\right)=\tilde{\beta}_2\,\mathcal{O}(D-4)$, vanishes in the limit $D\to 4$. Therefore, once the limit is taken, only remain the terms between brackets, evaluated at $D=4$, which correspond as announced to the four-dimensional Lagrangian $\mathcal{L}$ of Eq. (\ref{eq:4dDd}). \\

We have done this procedure to recover the four-dimensional generalized conformal coupling to GB, but we could of course do the same with the conformal coupling to the Ricci curvature, with $\mathcal{L}^{(1)}_{(D)}$ defined by (\ref{eq:l1}). This time, the factor in $\mathcal{L}^{(1)}_{(D)}$ is $\mathrm{e}^{(D-2)\phi}$, and $R$, which is the Lovelock invariant of order $k=1$, is a boundary term in spacetime dimension $D=2k=2$, so the singular limit procedure is exactly the same but with an expansion around $D=2$ this time. The obtained two-dimensional scalar-tensor action is
\begin{equation}
S=\int\mathrm{d}^2x\sqrt{-g}\left[\phi R+\left(\partial\phi\right)^2\right].\label{eq:liouville}
\end{equation}
This is a usual tentative to define gravity in two dimensions, the naive Einstein-Hilbert action being trivial because $R$ is a boundary term. This Jackiw-Teitelboim action\footnote{Also called $R=T$ model, because of its field equations when taking into account an energy-momentum tensor $T_{\mu\nu}$ with trace $T$.}, introduced by Jackiw~\cite{Jackiw:1984je}, has been largely studied in the literature, e.g.~\cite{Sikkema:1989ib,Mann:1989gh,Mann:1991md,Mann:1992np,Mann:1992ar}, see also~\cite{Strobl:1999wv} for a review on two-dimensional gravity. Very importantly, the original form $\mathcal{L}^{(k)}_{(D)}\equiv\mathrm{e}^{(D-2k)\phi}\mathcal{S}^{(k)}_{(D)}$, which yields (\ref{eq:l1_first}) and (\ref{eq:l2_first}) for $\mathcal{L}^{(1)}_{(D)}$ and $\mathcal{L}^{(2)}_{(D)}$ respectively, does not enable to perform the singular limit. One must first carry out integration by parts so as to put the Lagrangians under the form $\mathrm{e}^{(D-2k)\phi}\left[\mathcal{R}^{(k)}+\mathcal{O}(D-2k)\right]$, see Eqs. (\ref{eq:l1}) and (\ref{eq:l2}), which then permits the singular limit. 
\subsubsection{$k=3$: conformal coupling to the cubic Lovelock invariant}
The conformal Lagrangians $\mathcal{L}^{(1)}_{(D)}$ and $\mathcal{L}^{(2)}_{(D)}$ just enabled, through a singular limit, to recover already known Lagrangians in two and four-dimensional gravity. Let us now obtain \textbf{a new Lagrangian} by applying the same reasoning to $\mathcal{L}^{(3)}_{(D)}$, the conformal coupling to the cubic Lovelock invariant $\mathcal{R}^{(3)}$. Again, the first form $\mathcal{L}^{(3)}_{(D)}\equiv\mathrm{e}^{(D-6)\phi}\mathcal{S}^{(3)}_{(D)}$ is quite intricate and there is no point in writing it down, but integrations by parts yield the following form,
\begin{align}
\int\mathrm{d}^Dx\sqrt{-g}\,{}&{}\mathcal{L}^{(3)}_{(D)}\underset{\text{IBP}}{=} \int\mathrm{d}^Dx\sqrt{-g}\,\mathrm{e}^{(D-6)\phi}\Bigl\{\mathcal{R}^{(3)}-6(D-5)(D-6)\mathcal{H}^{(2)}_{\mu\nu}\phi^\mu\phi^\nu\nonumber\\{}&{}-8(D-4)(D-5)(D-6)P^{\mu\rho\nu\sigma}\phi_\mu\phi_\nu\phi_{\rho\sigma}\nonumber\\{}&{}-(D-3)(D-4)(D-5)(D-6)\Bigl[R\left(\partial\phi\right)^4-4\left(\partial\phi\right)^2\left[\left(\Box\phi\right)^2-\phi_{\mu\nu}\phi^{\mu\nu}\right]\Bigr]\nonumber\\{}&{}+2(D-3)(D-4)(D-5)(D-6)(2D-9)\left(\partial\phi\right)^4\Box\phi\nonumber\\{}&{}+(D-3)(D-4)(D-5)(D-6)(22+D(D-9))\left(\partial\phi\right)^6\Bigr\}\label{eq:l3},
\end{align}
where, on the first line, appears
\begin{equation}
P_{\mu\rho\nu\sigma} \equiv R_{\mu\rho\nu\sigma}+g_{\mu\sigma}R_{\rho\nu}+g_{\rho\nu}R_{\mu\sigma}-g_{\mu\nu}R_{\rho\sigma}-g_{\rho\sigma}R_{\mu\nu}+\frac{1}{2}\left(g_{\mu\nu}g_{\rho\sigma}-g_{\mu\sigma}g_{\rho\nu}\right)R.\label{eq:doubledula}
\end{equation}
We already encountered this tensor, which would coincide with the double dual of the Riemann tensor in four dimensions, when discussing the Paul term of the Fab Four, see Eq. (\ref{eq:paulterm}). The Paul term is precisely the term $P^{\mu\rho\nu\sigma}\phi_\mu\phi_\nu\phi_{\rho\sigma}$ which appears here in the action.\\

$\mathcal{L}^{(3)}_{(D)}$ has been put under a form which enables to consider again the singular limit, this time $D\to 6$. Having by now got used to this singular limit, note that it is most easily performed in the following way. First, remove the factor $\mathrm{e}^{(D-2k)\phi}$. Second, multiply $\mathcal{R}^{(k)}$ by $\phi$. Third, evaluate all remaining terms at $D=2k$ without taking into account the factors $(D-2k)$, i.e., setting them equal to one. This yields a six-dimensional scalar-tensor action,
\begin{align}
S=\int \mathrm{d}^6x\sqrt{-g}{}&{}\Bigl\{\phi\mathcal{R}^{(3)}-6 \mathcal{H}^{(2)}_{\mu\nu}\phi^\mu\phi^\nu-16 P^{\mu\rho\nu\sigma}\phi_\mu\phi_\nu\phi_{\rho\sigma}+36\left(\partial\phi\right)^4\Box\phi\nonumber\\{}&{}- 6\Bigl[R\left(\partial\phi\right)^4-4\left(\partial\phi\right)^2\left[\left(\Box\phi\right)^2-\phi_{\mu\nu}\phi^{\mu\nu}\right]\Bigr]+24\left(\partial\phi\right)^6\Bigr\}. \label{eq:l3_reg}
\end{align}
\subsection{Regularization and generalized conformal invariance}\label{subsec:reg}
It is not hard to generalize the previous results, and to prove that \textbf{the Lagrangian $\beta_k\,\mathcal{L}^{(k)}_{(D)}$ (which is non-trivial only in dimension $D\geq 2k+1$, and is a boundary term in $D=2k$) gives rise to a Lagrangian} $\tilde{\beta}_k\mathcal{L}^{(k)}_{(2k),\,\text{reg}}$ \textbf{in dimension $D=2k$, following a singular limit procedure described in the following proof}. First of all, 
\begin{equation}
\int\mathrm{d}^Dx\sqrt{-g}\mathcal{L}^{(k)}_{(D)}=\int\mathrm{d}^Dx\sqrt{-\hat{g}}\hat{\mathcal{R}}^{(k)}_{(D)}
\end{equation}
so $\mathcal{L}^{(k)}_{(D)}$ is indeed a boundary term in $D=2k$ by the Chern theorem. Therefore, integrating by parts,
\begin{equation}
\int\mathrm{d}^Dx\sqrt{-g}\mathcal{L}^{(k)}_{(D)}\underset{\text{IBP}}{=}\mathcal{O}\left(D-2k\right).\label{eq:lk}
\end{equation}
One can define
\begin{equation}
\mathcal{U}^{(k)}_{(D)}\equiv \mathcal{S}^{(k)}_{(D)}-\mathcal{R}^{(k)},\label{eq:ukskrk}
\end{equation}
then, expanding the exponential,
\begin{equation}
\mathcal{L}^{(k)}_{(D)}=\mathrm{e}^{(D-2k)\phi}\mathcal{S}^{(k)}_{(D)} = \mathrm{e}^{(D-2k)\phi}\mathcal{U}^{(k)}_{(D)}+\mathcal{R}^{(k)}+(D-2k)\phi\mathcal{R}^{(k)}+\mathcal{O}\left((D-2k)^2\right).\label{eq:uk}
\end{equation}
However, the Lovelock invariant $\mathcal{R}^{(k)}$ is a boundary term in $2k$ dimensions, so 
\begin{equation}
\int\mathrm{d}^Dx\sqrt{-g}\mathcal{R}^{(k)}\underset{\text{IBP}}{=}\mathcal{O}(D-2k).
\end{equation} 
This, together with (\ref{eq:lk}) and (\ref{eq:uk}), imply that there exists a Lorentz scalar $\mathcal{W}^{(k)}_{(D)}$, regular when $D\to 2k$, such that
\begin{equation}
\int\mathrm{d}^Dx\sqrt{-g}\,\mathrm{e}^{(D-2k)\phi}\mathcal{U}^{(k)}_{(D)}\underset{\text{IBP}}{=}\int\mathrm{d}^Dx\sqrt{-g}\left(D-2k\right)\mathcal{W}^{(k)}_{(D)}+\mathcal{O}\left((D-2k)^2\right).\label{eq:ukwk}
\end{equation}
So
\begin{align}
\int\mathrm{d}^Dx\sqrt{-g}\beta_k\mathcal{L}^{(k)}_{(D)}\underset{\text{IBP}}{=}\int\mathrm{d}^Dx\sqrt{-g}\Bigl[\beta_k\mathcal{R}^{(k)}+\beta_k\left(D-2k\right){}&{}\mathcal{W}^{(k)}_{(D)}+\beta_k(D-2k)\phi\mathcal{R}^{(k)}\Bigr]\nonumber\\{}&{}+\beta_k\,\mathcal{O}\left((D-2k)^2\right).\label{eq:betaklklast}
\end{align}
Then, one defines the singular limit procedure $D\to 2k$ exactly as for the particular cases. First, $\beta_k\mathcal{R}^{(k)}$ is a boundary term in $2k$ dimensions and is thus removed. Then, introduce a new coupling $\tilde{\beta}_k\equiv\beta_k(D-2k)$, and impose $\tilde{\beta}_k$ to remain constant when taking $D\to 2k$ (so the original coupling $\beta_k$ becomes infinite). As a result, the last term in (\ref{eq:betaklklast}) is $\tilde{\beta}_k\,\mathcal{O}(D-2k)$ and vanishes when $D\to 2k$, leaving as announced a 'regularized action' in dimension $D=2k$, 
\begin{equation}
S_{(2k),\,\text{reg}}=\int\mathrm{d}^{2k}x\sqrt{-g}\,\tilde{\beta}_k\mathcal{L}^{(k)}_{(2k),\,\text{reg}},\quad \mathcal{L}^{(k)}_{(2k),\,\text{reg}}\equiv\phi\mathcal{R}^{(k)}+\mathcal{W}^{(k)}_{(2k)}.\label{eq:sreglktilde}
\end{equation}
This proof is not particularly illuminating, because only the existence of $\mathcal{W}^{(k)}_{(2k)}$ is proved, but not its precise form. Just as in the particular cases, there is a factorially increasing amount of work to perform the integration by parts involved in Eq. (\ref{eq:ukwk}). This laborious task enables to find $\mathcal{W}^{(k)}_{(D)}$ and then to evaluate it at $D=2k$ to get $\mathcal{W}^{(k)}_{(2k)}$.
\subsubsection{Field equations and generalized conformal invariance}
It is fairly easy to deduce the field equations associated to this regularized action. The scalar field equation corresponding to the initial Lagrangian $\beta_k\mathcal{L}^{(k)}_{(D)}$, in dimension $D>2k$, is given by the corresponding term of the sum (\ref{eq:field_eq_D}),
\begin{equation}
\mathcal{E}_\phi =\beta_k(D-2k)\mathrm{e}^{(D-2k)\phi}\mathcal{S}^{(k)}_{(D)}.
\end{equation}
But we have rewritten $\beta_k\mathcal{L}^{(k)}_{(D)}$ as in (\ref{eq:betaklklast}), from which the singular limit procedure is just two steps: 1) dropping the first term $\beta_k\mathcal{R}^{(k)}$, 2) defining $\tilde{\beta}_k=\beta_k(D-2k)$ and taking $D\to 2k$. The first step regards a term which does not involve the scalar field, so does not affect the scalar field equation. Thus the scalar field equation obtained from $S_{(2k),\,\text{reg}}$ comes from the second step applied to $\mathcal{E}_\phi$, yielding
\begin{equation}
\mathcal{E}_{\text{reg}\,\phi}=\tilde{\beta}_k\mathcal{S}^{(k)}_{(2k)}.\label{eq:eregphi}
\end{equation} 
As regards the metric field equations, see (\ref{eq:field_eq_D}),
\begin{equation}
\mathcal{E}_{\mu\nu}=2\beta_k\mathcal{T}^{(k)}_{(D)\mu\nu}=2\beta_k\mathcal{H}^{(k)}_{\mu\nu}+2\beta_k\left(\mathcal{T}^{(k)}_{(D)\mu\nu}-\mathcal{H}^{(k)}_{\mu\nu}\right)
\end{equation}
In the last rewriting, we have made explicit $2\beta_k\mathcal{H}^{(k)}_{\mu\nu}$, which is nothing but the equation associated to the variation of $\beta_k\mathcal{R}^{(k)}$. So, dropping $\beta_k\mathcal{R}^{(k)}$ out of the Lagrangian (step 1 above) amounts to dropping $2\beta_k\mathcal{H}^{(k)}_{\mu\nu}$ from $\mathcal{E}_{\mu\nu}$. Thus remains $2\beta_k(\mathcal{T}^{(k)}_{(D)\mu\nu}-\mathcal{H}^{(k)}_{\mu\nu})$. But $\mathcal{T}^{(k)}_{(D)\mu\nu}$ and $\mathcal{H}^{(k)}_{\mu\nu}$ vanish when $D=2k$, since they are defined by an antisymmetrization over $2k+1$ indices, see (\ref{eq:handt}). So there exists Lorentz scalars $\mathcal{T}^{(k)}_{(D)\text{reg}\,\mu\nu}$ and $\mathcal{H}^{(k)}_{\text{reg}\,\mu\nu}$ such that
\begin{align}
\mathcal{T}^{(k)}_{(D)\mu\nu} {}&{}= \left(D-2k\right)\mathcal{T}^{(k)}_{(D)\text{reg}\,\mu\nu}+\mathcal{O}\left((D-2k)^2\right),\nonumber\\ \mathcal{H}^{(k)}_{\mu\nu} {}&{}= \left(D-2k\right)\mathcal{H}^{(k)}_{\text{reg}\,\mu\nu}+\mathcal{O}\left((D-2k)^2\right),\label{eq:ttilde}
\end{align}
hence applying the second step of the singular limit $D\to 2k$ gives
\begin{equation}
\mathcal{E}_{\text{reg}\,\mu\nu}=2\tilde{\beta}_k\left(\mathcal{T}^{(k)}_{(2k)\text{reg}\,\mu\nu}-\mathcal{H}^{(k)}_{\text{reg}\,\mu\nu}\right).
\end{equation}
This is not very useful because $\mathcal{T}^{(k)}_{(2k)\text{reg}\,\mu\nu}$ and $\mathcal{H}^{(k)}_{\text{reg}\,\mu\nu}$ are not constructed explicitly. However, the trace of (\ref{eq:ttilde}) directly yields
\begin{equation}
\mathcal{T}_{(D)\text{reg}\,\mu}^{(k)\quad\mu}=-\frac{\mathcal{S}^{(k)}}{2},\quad \mathcal{H}_{\text{reg}\,\mu}^{(k)\mu}=-\frac{\mathcal{R}^{(k)}}{2},
\end{equation}
where we have used (\ref{eq:traceht}). As a result,
\begin{equation}
\mathcal{E}_{\text{reg}\,\mu}^\mu+\mathcal{E}_{\text{reg}\,\phi} = -\tilde{\beta}_k\mathcal{R}^{(k)},\label{eq:geom_reg}
\end{equation}
Hence, the combination $\mathcal{E}_{\text{reg}\,\mu}^\mu+\mathcal{E}_{\text{reg}\,\phi}$ is a purely geometric quantity. We have shown in paragraph~\ref{subsec:local} that this is equivalent to the generalized conformal invariance of the corresponding action, here $S_{(2k),\,\text{reg}}$. In a word, \textbf{just as in four dimensions, where there exists (in the 4DEGB action) a coupling between the scalar field and $\mathcal{R}^{(2)}=\mathcal{G}$ which is not conformally invariant, but is such that $\delta S/\delta\phi$ is conformally invariant, in any even dimension $D=2k$, there exists such a similar coupling between $\phi$ and $\mathcal{R}^{(k)}$}. We have in fact directly proved the conformal invariance of $\delta S_{(2k),\,\text{reg}}/\delta\phi$, since
\begin{equation}
\frac{\delta S_{(2k),\,\text{reg}}}{\delta\phi}=\sqrt{-g}\mathcal{E}_{\text{reg}\,\phi} = \tilde{\beta}_k\sqrt{-g}\mathcal{S}^{(k)}_{(2k)}
\end{equation}
and, under the local conformal transformation (\ref{eq:loc_inv_conf_lov}), $\mathcal{S}^{(k)}_{(2k)}\to\mathrm{e}^{-2k\sigma}\mathcal{S}^{(k)}_{(2k)}$ by virtue of (\ref{eq:s2}), while $\sqrt{-g}\to\sqrt{-g}\,\mathrm{e}^{2k\sigma}$ in dimension $D=2k$. \textbf{As a conclusion, for arbitrary dimension $D$, we have constructed the most general scalar-tensor action with generalized conformal invariance and with second-order field equations. If $D=2p+1$ is odd,
\begin{equation}
S=\int\mathrm{d}^{2p+1}x\sqrt{-g}\sum_{k=0}^{p}\left(\alpha_k\mathcal{R}^{(k)}+\beta_k\,\mathrm{e}^{(2p+1-2k)\phi}\,\mathcal{S}^{(k)}_{(2p+1)}\right),\label{eq:action_odd}
\end{equation}
while if $D=2p$ is even,
\begin{equation}
S=\int\mathrm{d}^{2p}x\sqrt{-g}\left[\sum_{k=0}^{p-1}\left(\alpha_k\mathcal{R}^{(k)}+\beta_k\,\mathrm{e}^{(2p-2k)\phi}\,\mathcal{S}^{(k)}_{(2p)}\right)+\tilde{\beta}_p\mathcal{L}^{(p)}_{(2p),\,\text{reg}}\right].\label{eq:action_even}
\end{equation}}The terms in $\alpha_k$ are the Lovelock terms and are not conformally invariant, the terms in $\beta_k$ have local conformal invariance, and the term in $\tilde{\beta}_p$ is not conformally invariant but its derivative $\delta/\delta\phi$ has local conformal invariance. $\mathcal{L}^{(p)}_{(2p),\,\text{reg}}$ is not known explicitly but must be computed by the procedure described by Eqs. (\ref{eq:ukskrk}), (\ref{eq:ukwk}) and (\ref{eq:sreglktilde}). We have given its expression for $p=1,2,3$, respectively (\ref{eq:liouville}), (\ref{eq:4dDd}), (\ref{eq:l3_reg}). We emphasize that the new contribution of our article~\cite{Babichev:2023rhn} concerns all what regards the regularized term $\tilde{\beta}_p\mathcal{L}^{(p)}_{(2p),\,\text{reg}}$. The other terms were already constructed by~\cite{Oliva:2011np}.
\subsection{Black hole solutions}
We now proceed with the BH solutions of the obtained theories, following again our article~\cite{Babichev:2023rhn}. First of all, we do not consider the regularized Lagrangians $\mathcal{L}^{(p)}_{(2p)\text{, reg}}$, and focus on the action
\begin{equation}
S=\int\mathrm{d}^{D}x\sqrt{-g}\sum_{k=0}^{\left\lfloor\frac{D-1}{2}\right\rfloor}\left(\alpha_k\mathcal{R}^{(k)}+\beta_k\,\mathrm{e}^{(D-2k)\phi}\,\mathcal{S}^{(k)}_{(D)}\right).\label{eq:action_rdy_to_study}
\end{equation}
We recall that $\alpha_k$ and $\beta_k$ are couplings, $\mathcal{R}^{(k)}$ is the Lovelock invariant of order $k$ and $\mathrm{e}^{(D-2k)\phi}\,\mathcal{S}^{(k)}_{(D)}$ is the conformally-invariant Lagrangian, coupling the scalar field to the Lovelock invariant of order $k$, as defined in Eqs. (\ref{eq:s1}-\ref{eq:s2}). We recall the field equations,
\begin{equation}
\mathcal{E}_{\mu\nu}=2\sum_{k=0}^{\lfloor \frac{D-1}{2}\rfloor}\left(\alpha_k\mathcal{H}^{(k)}_{\mu\nu}+\beta_k\mathcal{T}^{(k)}_{(D)\mu\nu}\right),\quad\mathcal{E}_\phi = \sum_{k=0}^{\lfloor \frac{D-1}{2}\rfloor}\beta_k\left(D-2k\right)\mathrm{e}^{(D-2k)\phi}\,\mathcal{S}^{(k)}_{(D)},\label{eq:field_eq_D_bis}
\end{equation}
with
\begin{equation}
\mathcal{H}_{\quad\nu}^{(k)\mu}=\frac{-1}{2^{k+1}}\delta^{\mu\,\mu_1\nu_1\cdots\mu_k\nu_k}_{\nu\,\alpha_1\beta_1\cdots\alpha_k\beta_k}\prod_{i=1}^{k} R^{\alpha_i\beta_i}_{\quad\,\mu_i\nu_i},\quad\mathcal{T}^{(k)\mu}_{(D)\nu}\equiv \frac{-\mathrm{e}^{(D-2k)\phi}}{2^{k+1}}\delta^{\mu\,\mu_1\nu_1\cdots\mu_k\nu_k}_{\nu\,\alpha_1\beta_1\cdots\alpha_k\beta_k}\prod_{i=1}^{k} S^{\alpha_i\beta_i}_{\quad\,\mu_i\nu_i}.\label{eq:handt_bis}
\end{equation} 
\subsubsection{Static scalar field}
As is by now usual, we restrict to static and spherically-symmetric spacetimes, with also a static scalar field,
\begin{equation}
\mathrm{d}s^{2}=-f(r)\mathrm{d}t^{2}+\mathrm{d}r^{2}/f(r)+r^{2}\mathrm{d}\Sigma_{D-2,\gamma}^{2},\quad
\phi=\phi(r) \label{ansatz}
\end{equation}
where $d\Sigma_{D-2,\gamma}^{2}$ is the metric of a
($D-2$)-dimensional Euclidean space of  constant curvature
$\gamma\left(D-2\right)\left(D-3\right)$ with $\gamma=0,\pm 1$.
\\

When the $\alpha_k$-part of the action only contains the Einstein-Hilbert
term with (potentially) a cosmological constant, that is, $\alpha_k=0$ for
$k>1$, two different analytic classes of solutions can be found for
the ansatz~(\ref{ansatz}). These two classes correspond to two
different relations between the coupling constants of the action. Setting $\alpha_0=-2\Lambda$ and $\alpha_1=1$, the solutions can be generically given in terms of the metric
functions
\begin{align}
f^{(i)}(r)  &  =\gamma-\frac{M}{r^{D-3}}-
\frac{2\Lambda}{(D-1)(D-2)}r^{2}+\frac{q^{(i)}}{r^{D-2}},
\label{stealthmetricEinsteincase}
\end{align}
dressed with a scalar field given by
\begin{align}
\label{second}
&\phi^{(1)}(r)=\ln\left(\frac{N}{r}\right),\\
&\phi^{(2)}(r) =\ln\left(\frac{N}{r\sigma_{\gamma}\left(c\pm\int
\frac{\mathrm{d}r}{r\sqrt{f^{(2)}(r)}}\right)}\right). \label{sfstealth}
\end{align}
The superscript $(i)$ denotes the first and the second class of solution, and the function $\sigma_{\gamma}$ depends on the topology
of the base manifold:
\begin{equation}
\sigma_{1}(X)=\cosh(X),\quad \sigma_{-1}(X)=\cos(X),\quad
\sigma_{0}(X)=X.
\end{equation}
In the above expressions, $M$ appearing in the metric is an integration constant
proportional to the mass, while the constant $c$ appearing in the
scalar field, for the second  class of solutions (\ref{sfstealth}),
is arbitrary. 
The constant $N$ of both scalar fields
(\ref{second})--(\ref{sfstealth}) is fixed in terms of the coupling
constants of the theory through the relation
\begin{align}
\sum\limits_{k=1}^{\left\lfloor\frac{D-1}{2}\right\rfloor  }\frac{k\beta_{k}}{N^{2k}(D-2k-1)!} \tilde{\gamma}_{(i)}^{k-1}  &
=0\text{,}\label{ConstN}
\end{align}
while the coupling of the conformal potential $\beta_0^{(i)}$ is fixed
in terms of the other couplings as
\begin{align}
 \frac{D(D-1)}{(D-1)!}\beta_0^{(i)}+\sum\limits_{k=1}^{\left\lfloor\frac{D-1}{2}\right\rfloor  }\frac{\left(  D\left( D-1\right)
+4\epsilon_k^{(i)}\right)\beta_{k}\tilde{\gamma}_{(i)}^{k}}{N^{2k}(D-2k-1)!}
& =0. \label{couplingpotential}
\end{align}
We introduced $\epsilon_k^{(1)}=k^2$, $\epsilon_k^{(2)}=0$,
$\tilde{\gamma}_{(1)}=\gamma$ and
$\tilde{\gamma}_{(2)}=\gamma-\delta_{\gamma,0}$ (where $\delta_{\gamma,0}$ is the Kronecker symbol). Finally, for both solutions, the constant $q^{(i)}$ appearing in the metric function
(\ref{stealthmetricEinsteincase}) is fixed in terms of the coupling
constants as
\begin{eqnarray}
\frac{q^{(i)}}{N^D}=-\frac{\beta_0^{(i)}}{(D-2)}-\sum\limits_{k=1}^{\left\lfloor\frac{D-1}{2}\right\rfloor
}\frac{\beta_k(D-3)!\tilde{\gamma}_{(i)}^k}{N^{2k}(D-2k-2)!}.
\label{qi}
\end{eqnarray}
The first class of solutions  with $i=1$ has $q^{(1)}\not=0$ for
$\gamma\neq 0$ and was already found in~\cite{Giribet:2014bva}. For the second class of solutions, $i=2$, one can verify easily from the above that $q^{(2)}=0$. Consequently, for $i=2$, the solution~(\ref{stealthmetricEinsteincase})
can be interpreted as a stealth solution on the $D$-dimensional Schwarzschild-(a)dS
spacetime, also known as Schwarzschild-Tangherlini spacetime~\cite{Tangherlini:1963bw}. Importantly, the two classes of spacetimes $i=1,2$ are solutions of distinct theories since $\beta_0^{(1)}\neq \beta_0^{(2)}$ as shown by~(\ref{couplingpotential}). 
\\

In the general Lovelock case, where $\alpha_k\not=0$ for at least one
$k>1$, similar classes of solutions exist. The scalar field profiles
keep the same form~(\ref{second}-\ref{sfstealth}) and must obey the same conditions~(\ref{ConstN}) and~(\ref{couplingpotential}). The metric functions $f^{(i)}$
have a different form, which can be understood from the geometric equation (\ref{eq:geom_D}). They are now given by a polynomial equation of
order $\left\lfloor\frac{D-1}{2}\right\rfloor$,
\begin{align}
\sum\limits_{k=0}^{\left\lfloor\frac{D-1}{2}\right\rfloor }\frac{\alpha_k(D-1)!}
{(D-2k-1)!}\left(  \frac{\gamma-f^{(i)}\left(  r\right)
}{r^{2}}\right)^{k} =
\frac{M(D-1)(D-2)}{ r^{D-1}} -\frac{q^{(i)}(D-1)(D-2)}{r^D}\ ,
\label{poly}
\end{align}
where $M$ is an arbitrary constant related to the mass, and
$q^{(i)}$ are given again by~(\ref{qi}), meaning in particular that
$q^{(2)}=0$. It follows then that the second class of solutions can
be interpreted as stealth BHs of Lovelock theory, see~\cite{Wheeler:1985nh,Myers:1988ze}. In the quadratic
case $\alpha_k=0$ for $k>2$, the real roots of this polynomial can be
easily written down and correspond to a Boulware-Deser BH~\cite{Boulware:1985wk}. For the other cases, the expression for $f$ 
is quite cumbersome and does not exist explicitly for $k\geq 5$, except the case when the polynomial equation
(\ref{poly}) has a single root.
This occurs for the particular choice 
of the coupling constants
\begin{equation}
\alpha_k=\binom{\left\lfloor\frac{D-1}{2}\right\rfloor}{k}\frac{(D-2k-1)!}{(D-1)!},
\end{equation}
where the first factor is a binomial factor. For this particular choice, one can easily express the solution in odd dimension as
\begin{eqnarray}
f^{(i)}(r)=\gamma+r^2-\left(\tilde{M}-\frac{\tilde{q}^{(i)}}{r}\right)^{\frac{2}{D-1}},
\label{polyCSodd}
\end{eqnarray}
while in even dimension,
\begin{eqnarray}
f^{(i)}(r)=
\gamma+r^2-\left(\frac{\tilde{M}}{r}-\frac{\tilde{q}^{(i)}}{r^2}\right)^{\frac{2}{D-2}},
\label{polyCSeven}
\end{eqnarray}
where we have defined $\tilde{M}=M(D-1)(D-2)$ and
$\tilde{q}^{(i)}=q^{(i)}(D-1)(D-2)$. For the second solution,
$\tilde{q}^{(2)}=0$, the spacetime corresponds to the BH solutions obtained in~\cite{Banados:1993ur}.
\subsubsection{Time dependent scalar field in theories with no shift symmetry}
As we already emphasized many times in this manuscript, scalar-tensor
theories with shift symmetry $\phi\to\phi+\mbox{constant}$ may
accommodate BH solutions with a scalar field depending
linearily on time. The underlying idea of this feature is that the
field equations only involve derivatives of the scalar field, and
hence its explicit time dependence does not appear at the level of
the field equations.
Here, the action (\ref{eq:action_rdy_to_study}) is not shift-symmetric. Nevertheless, if
$\beta_0=\beta_1=0$ in action (\ref{eq:action_rdy_to_study}), the stealth metric
function $f^{(2)}\left(r\right)$ with $q^{(2)}=0$, see Eq.~(\ref{stealthmetricEinsteincase}), can be dressed
with a time-dependent scalar field. It is given by
\begin{equation}
\phi\left(t,r\right) =c+\zeta\,
t+\hspace{-0.1cm}\int\frac{\pm\sqrt{\gamma
f^{(2)}(r)+\zeta^2r^2}/f^{(2)}(r)-1}{r}\mathrm{d}r, \label{timedep}
\end{equation}
where $c$ and $\zeta$ are arbitrary constants. The emergence of such stealth
solutions in spite of the absence of shift symmetry in the theory
under consideration can be understood as follows. The solution is stealth, so the effective energy-momentum tensor of the scalar field must vanish. According to the field equations (\ref{eq:field_eq_D_bis}-\ref{eq:handt_bis}), the vanishing of this effective energy-momentum tensor corresponds to the following equation,
\begin{equation}
0=\sum_{k=2}^{\lfloor \frac{D-1}{2}\rfloor}\beta_k\frac{\mathrm{e}^{(D-2k)\phi}}{2^{k}}\delta^{\mu\,\mu_1\nu_1\cdots\mu_k\nu_k}_{\nu\,\alpha_1\beta_1\cdots\alpha_k\beta_k}\prod_{i=1}^{k} S^{\alpha_i\beta_i}_{\quad\,\mu_i\nu_i},
\end{equation}
where the sum starts at $k=2$ because of the present assumption $\beta_0=\beta_1=0$. The above expression is clearly not shift-symmetric, since it depends on the non-derivated scalar field via $\mathrm{e}^{(D-2k)\phi}$, in agreement with the fact that the
action is not shift-symmetric. However, the tensor $S^{\mu\nu}_{\,\,\,\,\,\,\rho\sigma}$, see Eq. (\ref{eq:s1}), only depends on the
derivatives of $\phi$. One can verify that for the
stealth configuration described by the metric function $f^{(2)}(r)$
and the time-dependent scalar field~(\ref{timedep}),
each term 
\begin{equation}
\delta^{\mu\,\mu_1\nu_1\cdots\mu_k\nu_k}_{\nu\,\alpha_1\beta_1\cdots\alpha_k\beta_k}\prod_{i=1}^{k} S^{\alpha_i\beta_i}_{\quad\,\mu_i\nu_i}
\end{equation}
vanishes independently. Consequently, one gets a solution which is effectively shift-symmetric in
$\phi$, as highlighted by the form of~(\ref{timedep}).
\subsubsection{Solutions to the regularized action}
In the case of even dimension $D=2p$, we now include the regularized Lagrangian $\mathcal{L}^{(p)}_{(2p),\,\text{reg}}$,
\begin{equation}
S=\int\mathrm{d}^{2p}x\sqrt{-g}\left[\sum_{k=0}^{p-1}\left(\alpha_k\mathcal{R}^{(k)}+\beta_k\,\mathrm{e}^{(2p-2k)\phi}\,\mathcal{S}^{(k)}_{(2p)}\right)+\tilde{\beta}_p\mathcal{L}^{(p)}_{(2p),\,\text{reg}}\right].\label{eq:act_reg_par}
\end{equation}
Remember that, as opposed to the Lagrangians $\mathrm{e}^{(2p-2k)\phi}\,\mathcal{S}^{(k)}_{(2p)}$, the regularized Lagrangian $\mathcal{L}^{(p)}_{(2p),\,\text{reg}}$ does not have conformal invariance, however, it yields a conformally-invariant contribution to $\delta S/\delta \phi$. 
\\

This action comes from a regularization procedure, described in paragraph~\ref{subsec:reg}. In particular, the coupling $\tilde{\beta}_p$ comes from the redefinition of the original coupling $\beta_p$ as $\tilde{\beta}_p=(D-2p)\beta_p$ in the dimensional limit $D\to 2p$. From this, one can immediately deduce the spherically-symmetric BH solutions of the regularized action. Indeed, before regularization, one has two solutions with static scalar field~(\ref{second}) or~(\ref{sfstealth}) and respective metric function $f^{(i)}(r)$ given by~(\ref{poly}), with $i=1,2$. These solutions exist provided the integration constant $N$ appearing in the scalar field, the coupling constants $\beta_k$, and the constant $q^{(i)}$ of the metric, are related by the three equations~(\ref{ConstN}-\ref{qi}). Let us write down these relations in dimension $D>2p$ \textbf{before regularization},
\begin{align}
{}&{}0=\sum\limits_{k=1}^{p-1}\frac{k\beta_{k}}{N^{2k}(D-2k-1)!} \tilde{\gamma}_{(i)}^{k-1}+\frac{p\beta_{p}(D-2p)}{N^{2p}(D-2p)!} \tilde{\gamma}_{(i)}^{p-1},\\
{}&{}0=\frac{D(D-1)}{(D-1)!}\beta_0^{(i)}+\sum\limits_{k=1}^{p-1}\frac{\left(  D\left( D-1\right)
+4\epsilon_k^{(i)}\right)\beta_{k}\tilde{\gamma}_{(i)}^{k}}{N^{2k}(D-2k-1)!}\nonumber\\{}&{}\qquad\qquad\qquad\qquad+\frac{\left(  D\left( D-1\right)
+4\epsilon_p^{(i)}\right)\beta_{p}(D-2p)\tilde{\gamma}_{(i)}^{p}}{N^{2p}(D-2p)!},\\
{}&{}\frac{q^{(i)}}{N^D}=-\frac{\beta_0^{(i)}}{(D-2)}-\sum\limits_{k=1}^{p-1
}\frac{\beta_k(D-3)!\tilde{\gamma}_{(i)}^k}{N^{2k}(D-2k-2)!}-\frac{\beta_p(D-2p)(D-2p-1)(D-3)!\tilde{\gamma}_{(i)}^p}{N^{2p}(D-2p)!}.
\end{align}
We recall that $\epsilon_k^{(1)}=k^2$, $\epsilon_k^{(2)}=0$,
$\tilde{\gamma}_{(1)}=\gamma$ and
$\tilde{\gamma}_{(2)}=\gamma-\delta_{\gamma,0}$, where $\gamma=0,\pm 1$ corresponds to the topology of the horizon.
\\

The BH solutions with static scalar field for the regularized action (\ref{eq:act_reg_par}) can thus be read from the above relations, by defining $\tilde{\beta}_p=(D-2p)\beta_p$ and making $D\to 2p$. This leads to the three relations
\begin{align}
0={}&{}\sum\limits_{k=1}^{p-1}\frac{k\beta_{k}}{N^{2k}(2(p-k)-1)!} \tilde{\gamma}_{(i)}^{k-1}+\frac{p\tilde{\beta}_{p}}{N^{2p}} \tilde{\gamma}_{(i)}^{p-1},\label{eq:rel1}\\
0={}&{}\frac{2p(2p-1)}{(2p-1)!}\beta_0^{(i)}+\sum\limits_{k=1}^{p-1}\frac{\left(  2p(2p-1)
+4\epsilon_k^{(i)}\right)\beta_{k}\tilde{\gamma}_{(i)}^{k}}{N^{2k}(2(p-k)-1)!}+\frac{\left(  2p(2p-1)
+4\epsilon_p^{(i)}\right)\tilde{\beta}_{p}\tilde{\gamma}_{(i)}^{p}}{N^{2p}},\\
\frac{q^{(i)}}{N^{2p}}={}&{}-\frac{\beta_0^{(i)}}{(2p-2)}-\sum\limits_{k=1}^{p-1
}\frac{\beta_k(2p-3)!\tilde{\gamma}_{(i)}^k}{N^{2k}(2(p-k)-2)!}+\frac{\tilde{\beta}_{p}(2p-3)!\tilde{\gamma}_{(i)}^p}{N^{2p}}.\label{eq:rel3}
\end{align}
If these three relations are verified, the regularized action (\ref{eq:act_reg_par}) admits the static BH~(\ref{ansatz}) with static scalar field~(\ref{second}) or~(\ref{sfstealth}) for respectively $i=1,2$. We recall that, before regularization, the polynomial equation verified by the metric function $f^{(i)}(r)$ is,
\begin{equation}
\sum\limits_{k=0}^{p-1}\frac{\alpha_k(D-1)!}{(D-2k-1)!}\left(  \frac{\gamma-f^{(i)}\left(  r\right)}{r^{2}}\right)^{k} =\frac{M(D-1)(D-2)}{ r^{D-1}} -\frac{q^{(i)}(D-1)(D-2)}{r^D}.
\end{equation}
This polynomial equation came from the geometric equation (\ref{eq:geom_D}). However, this equation now gets modified by the regularization procedure, see Eq. (\ref{eq:geom_reg}). Therefore, after the regularization, the polynomial equation verified by the metric function $f^{(i)}(r)$ becomes
\begin{align}
\sum\limits_{k=0}^{p-1}\frac{\alpha_k(2p-1)!}{(2(p-k)-1)!}{}&{}\left(  \frac{\gamma-f^{(i)}\left(  r\right)}{r^{2}}\right)^{k}+\tilde{\beta}_p(2p-1)!\left(  \frac{\gamma-f^{(i)}\left(  r\right)}{r^{2}}\right)^{p}\nonumber\\{}&{} =\frac{M(2p-1)(2p-2)}{ r^{D-1}} -\frac{q^{(i)}(2p-1)(2p-2)}{r^D}.\label{poly-reg}
\end{align}
In a word, the regularized action (\ref{eq:act_reg_par}) admits the static BH~(\ref{ansatz}), with metric function~(\ref{poly-reg}), dressed with a static scalar field~(\ref{second}) or~(\ref{sfstealth}) for respectively $i=1,2$. As before, it is easy to verify that $q^{(1)}\neq 0$ for $\gamma\neq 0$, while $q^{(2)}=0$.
\\

Importantly, for the case without regularized Lagrangian $\tilde{\beta}_p\mathcal{L}^{(p)}_{(2p),\,\text{reg}}$, the solution $i=1$ (with $\gamma\neq 0$) was not a stealth solution because $q^{(1)}\neq 0$, while the solution $i=2$ was stealth. In other words, the polynomial before regularization, Eq.~(\ref{poly}), was the same as for pure Lovelock gravity, because it had $q^{(2)}=0$. But now, for the regularized action, both $i=1$ and $i=2$ solutions are not stealth, even though $q^{(2)}=0$. This is because the term in $\tilde{\beta}_p$ in Eq.~(\ref{poly-reg}) comes from the regularized scalar-tensor Lagrangian $\tilde{\beta}_p\mathcal{L}^{(p)}_{(2p),\,\text{reg}}$, but would not appear if one were only considering pure Lovelock gravity in dimension $D=2p$. The metric function has thus no equivalent in pure metric gravity and the solution is not a stealth one.
\\

Last but not least, before regularization, there also exists a static BH solution with time-dependent scalar field~(\ref{timedep}), if $\beta_0=\beta_1=0$. As regards the metric function, this BH is identical to the solution $i=2$, only the scalar field is different. This remains true after the regularization procedure: if $\beta_0=\beta_1=0$, the regularized action (\ref{eq:act_reg_par}) admits the static BH given by (\ref{poly-reg}), with $q=q^{(2)}=0$, as a solution, dressed with a time-dependent scalar field (\ref{timedep}). \\

Of course, the attentive reader has already remarked the great similarity between the three scalar field profiles~(\ref{second},\ref{sfstealth},\ref{timedep}) and the scalar fields for the 4DEGB action, Eqs. (\ref{eq:phi_lu_pang}), (\ref{eq:cured}) and (\ref{eq:log_scal}). This is not surprising, since the conformally-invariant Lagrangians and their regularization procedure is just a generalization of a procedure which, as we demonstrated in paragraph~\ref{subsec:simple_ex}, enables to obtain the 4DEGB action (although it was not the procedure followed in the original articles~\cite{Lu:2020iav,Fernandes:2020nbq,Hennigar:2020lsl,Fernandes:2021dsb}).
\\

At this point, \textbf{we have fully described the static, spherically-symmetric BH solutions for a conformally-coupled scalar field in Lovelock gravity}. The starting point was the construction of the most general action with local conformal invariance in spacetime dimension $D$~\cite{Oliva:2011np},
\begin{equation}
S_{(D),\,\text{loc}}\left[g_{\mu\nu},\phi\right]=\int\mathrm{d}^Dx\sqrt{-g}\sum_{k=0}^{\lfloor \frac{D-1}{2}\rfloor}\beta_k\,\mathcal{L}^{(k)}_{(D)},\quad \mathcal{L}^{(k)}_{(D)} =\mathrm{e}^{(D-2k)\phi}\,\mathcal{S}^{(k)}_{(D)}.
\end{equation}
In this action, we recall that $\mathcal{S}^{(k)}_{(D)}$ is (up to an exponential factor) the Lovelock invariant of the metric $\hat{g}_{\mu\nu}=\mathrm{e}^{2\phi}g_{\mu\nu}$,
\begin{equation}
\mathcal{S}^{(k)}_{(D)}= \mathrm{e}^{2k\phi}\,\hat{\mathcal{R}}^{(k)}_{(D)}= \frac{1}{2^k}\delta^{\mu_1\nu_1\cdots\mu_k\nu_k}_{\alpha_1\beta_1\cdots\alpha_k\beta_k}\prod_{i=1}^{k}S^{\alpha_i\beta_i}_{\quad\,\mu_i\nu_i}.
\end{equation}
Then, the conformally-coupled scalar field that we have considered is obtained by breaking the conformal invariance, through the addition of the usual pure geometric Lovelock Lagrangians. Also, a regularization procedure enables, for every even dimension $D=2p$, to construct a Lagrangian coupling the scalar field with the Lovelock invariant of order $p$. This Lagrangian (or rather its associated action integral) does not display local conformal invariance, but its functional derivative $\delta/\delta\phi$ does. The BH solutions to this regularized action are easily obtained from the BH solutions before regularization.
\\

It turns out that the conformally-invariant Lagrangians $\mathcal{L}^{(k)}_{(D)} =\mathrm{e}^{(D-2k)\phi}\,\mathcal{S}^{(k)}_{(D)}$ play an interesting role when it comes to considering the KK reduction of Lovelock theories down to lower dimensions. The next section describes this role in detail, and performs the \textbf{diagonal KK reduction of Lovelock invariants of arbitrary order}. These results have not yet been presented in any (pre-)publication. We however refer the interested reader to~\cite{Casalino:2020kbt,Colleaux:2020wfv}, where many considerations regarding the regularization of Lovelock gravity into four dimensions, and its possible consequences as concerns the obtention of regular spacetimes, are developed.

\section{Diagonal Kaluza-Klein reduction of Lovelock theory}\label{sec:diagkk}
\sectionmark{5.3~~Diagonal Kaluza-Klein reduction of Lovelock theory}
\subsection{Diagonal decomposition}
Let us thus study the link between KK dimensional reduction and the Lagrangians $\mathcal{L}^{(k)}_{(D)}$. Consider a $\mathfrak{D}$-dimensional spacetime, with metric $g_{(\mathfrak{D})\,AB}$ (capital Latin indices), split, in a diagonal way, between a $D$-dimensional \textbf{target} spacetime $g_{(D)\,\mu\nu}$ (Greek indices) and an $n$-dimensional space $g_{(n)\,ab}$ (small Latin indices),
\begin{equation}
g_{(\mathfrak{D})AB}\mathrm{d}x^A\mathrm{d}x^B=g_{(D)\mu\nu}\mathrm{d}x^\mu\mathrm{d}x^\nu+g_{(n)ab}\mathrm{d}x^a\mathrm{d}x^b.\label{eq:kkdiagdec}
\end{equation}
The $n$-dimensional metric is then written as the product of a conformal factor $\mathrm{e}^{-2\phi}$, where $\phi$ is a scalar field depending only on the target spacetime coordinates $x^\mu$, and a metric $\widetilde{g}_{(n)ab}$ of a so-called \textbf{internal} space,
\begin{equation}
g_{(n)ab}=\mathrm{e}^{-2\phi}\widetilde{g}_{(n)ab}.\label{eq:internal}
\end{equation}
Under this decomposition, the Riemann tensor $R^{AB}_{(\mathfrak{D})\,CD}$ of the $\mathfrak{D}$-dimensional metric has the following non-vanishing components,
\begin{align}
R^{\mu\nu}_{(\mathfrak{D})\,\rho\sigma}={}&{}R^{\mu\nu}_{(D)\,\rho\sigma},\label{eq:r1}\\
R^{ab}_{(\mathfrak{D})\,cd}={}&{}\mathrm{e}^{2\phi}\widetilde{R}^{ab}_{(n)\,\rho\sigma}-\left(\partial\phi\right)^2\delta^{ab}_{cd},\label{eq:r2}\\
R^{\mu a}_{(\mathfrak{D})\,\nu b}={}&{} \delta^a_b\left(\phi^{\mu}_\nu -\phi^\mu\phi_\nu\right),\label{eq:r3}
\end{align}
where $R^{\mu\nu}_{(D)\,\rho\sigma}$ is the Riemann tensor of the target spacetime and $\widetilde{R}^{ab}_{(n)\,\rho\sigma}$ is the Riemann tensor of the metric $\widetilde{g}_{(n)\,ab}$ of the internal space. Of course, for the last line with mixed indices (Greek and Latin), one must take into account the other three components related by the symmetries of the Riemann tensor, namely $R^{a\mu}_{(\mathfrak{D})\,\nu b}$, $R^{\mu a}_{(\mathfrak{D})\,b\nu}$ and $R^{a\mu}_{(\mathfrak{D})\,b\nu}$.
\\

One easily computes the non-vanishing components of the Ricci tensor,
\begin{align}
R^\alpha_{(\mathfrak{D})\beta}={}&{}R^{\mu\alpha}_{(\mathfrak{D})\,\mu\beta}+R^{a\alpha}_{(\mathfrak{D})\,a\beta}=R^{\alpha}_{(D)\beta}+n\left(\phi^{\alpha}_\beta-\phi^\alpha\phi_\beta\right),\\
R^a_{(\mathfrak{D})b}={}&{}R^{\mu a}_{(\mathfrak{D})\,\mu b}+R^{ca}_{(\mathfrak{D})\,cb}=\delta^a_b\Box\phi-n\delta^a_b\left(\partial\phi\right)^2+\mathrm{e}^{2\phi}\widetilde{R}^a_{(n)b}.
\end{align}
The Ricci scalar is then
\begin{equation}
R_{(\mathfrak{D})}=R_{(D)}+2n\Box\phi-n(n+1)\left(\partial\phi\right)^2+\mathrm{e}^{2\phi}\widetilde{R}_{(n)}.\label{eq:Riccikk}
\end{equation}
Of course, in the above, $\widetilde{R}^a_{(n)b}$ and $\widetilde{R}_{(n)}$ are respectively the Ricci tensor and Ricci scalar of the internal space, with metric $\widetilde{g}_{(n)ab}$.
\subsection{The simple cases of Ricci and Gauss-Bonnet scalars: conjecture for the Lagrangians}\label{subsec:simple_conj_lag}
The Ricci scalar in the KK picture is thus computed, Eq. (\ref{eq:Riccikk}). One can also compute the GB scalar:
\begin{equation}
\mathcal{G}_{(\mathfrak{D})}=R_{(\mathfrak{D})}^2-4R^A_{(\mathfrak{D})B}R^B_{(\mathfrak{D})A}+R^{AB}_{(\mathfrak{D})CD}R^{CD}_{(\mathfrak{D})AB}.
\end{equation}
Using (\ref{eq:r1}-\ref{eq:Riccikk}), a short calculation leads to
\begin{align}
\mathcal{G}_{(\mathfrak{D})}={}&{}\mathcal{G}_{(D)}+(n+1)n(n-1)(n-2)\left(\partial\phi\right)^4-4n^2(n-1)\Box\phi\left(\partial\phi\right)^2\nonumber\\{}&{}+4n(n-1)\left[\left(\Box\phi\right)^2-\phi_{\mu\nu}\phi^{\mu\nu}\right]-8nG^{\mu\nu}_{(D)}\phi_{\mu\nu}+8nR^{\mu\nu}_{(D)}\phi_\mu\phi_\nu\nonumber\\{}&{}-2n(n+1)R_{(D)}\left(\partial\phi\right)^2+8n(n-1)\phi^\mu\phi_{\mu\nu}\phi^\nu\nonumber\\{}&{}+2\widetilde{R}_{(n)}\mathrm{e}^{2\phi}\left[R+2(n-2)\Box\phi-(n-1)(n-2)\left(\partial\phi\right)^2\right]+\mathrm{e}^{4\phi}\widetilde{\mathcal{G}}_{(n)}.\label{eq:gbkk}
\end{align}
The expressions for the conformal Lagrangians $\mathcal{L}^{(0)}_{(d)}$, $\mathcal{L}^{(1)}_{(d)}$ and $\mathcal{L}^{(2)}_{(d)}$ were given in paragraph~\ref{subsec:simple_ex}. \textbf{Importantly, what matters for the moment are the expressions before any integration by parts is performed}. This corresponds to Eqs. (\ref{eq:l0}), (\ref{eq:l1_first}) and (\ref{eq:l2_first}). To be clear, let us recall these expressions here,
\begin{align}
\mathcal{L}^{(0)}_{(d)}={}&{}\mathrm{e}^{d\phi},\label{eq:l0d}\\
\mathcal{L}^{(1)}_{(d)}={}&{}\mathrm{e}^{(d-2)\phi}\Bigl(R-2\left(d-1\right)\Box\phi-\left(d-1\right)\left(d-2\right)\left(\partial\phi\right)^2\Bigr),\label{eq:l1d}\\
\mathcal{L}^{(2)}_{(d)}={}&{}\mathrm{e}^{(d-4)\phi}\Bigl\{\mathcal{G}+8\left(d-3\right)G^{\mu\nu}\phi_{\mu\nu}+4\left(d-2\right)\left(d-3\right)\left[\left(\Box\phi\right)^2-\phi_{\mu\nu}\phi^{\mu\nu}\right]\nonumber\\{}&{}-2\left(d-3\right)\left(d-4\right)R\left(\partial\phi\right)^2-8\left(d-3\right)R_{\mu\nu}\phi^\mu\phi^\nu+8\left(d-2\right)\left(d-3\right)\phi^\mu\phi_{\mu\nu}\phi^\nu\nonumber\\{}&{}+4\left(d-2\right)\left(d-3\right)^2\left(\partial\phi\right)^2\Box\phi+\left(d-1\right)\left(d-2\right)\left(d-3\right)\left(d-4\right)\left(\partial\phi\right)^4\Bigr\}.\label{eq:l2d}
\end{align}
This enables to insist on the following fact: \textbf{the subscript $(d)$ must here be considered as a parameter of the Lagrangian, and not as a spacetime dimension}. The Lagrangians $\mathcal{L}^{(k)}_{(d)}$ were first introduced as the Lagrangians such that the $d$-dimensional action integral $\int\mathrm{d}^dx\sqrt{-g}\mathcal{L}^{(k)}_{(d)}$ has local conformal invariance. However, once the explicit expression of $\mathcal{L}^{(k)}_{(d)}$ is computed, nothing prevents from considering these Lagrangians in a dimension different from $d$. We will come back on this point later, in paragraph~\ref{subsec:specific}, showing explicitly how to consider not only the first Lagrangians $\mathcal{L}^{(0)}_{(d)}$, $\mathcal{L}^{(1)}_{(d)}$ and $\mathcal{L}^{(2)}_{(d)}$, but also the Lagrangian of arbitrary order $k$, $\mathcal{L}^{(k)}_{(d)}$, in other dimension than the one (namely $d$) where it has conformal invariance. 
\\

Looking at the expressions for the Ricci and GB invariants in the KK picture, Eqs. (\ref{eq:Riccikk}) and (\ref{eq:gbkk}), one sees that
\begin{align}
R_{(\mathfrak{D})}={}&{}\mathrm{e}^{(n+1)\phi}\left(\mathcal{L}^{(1)}_{(1-n)}+\widetilde{R}_{(n)}\mathcal{L}^{(0)}_{(1-n)}\right),\label{rmfd}\\
\mathcal{G}_{(\mathfrak{D})}={}&{}\mathrm{e}^{(n+1)\phi}\left(\mathcal{L}^{(2)}_{(3-n)}+2\widetilde{R}_{(n)}\mathcal{L}^{(1)}_{(3-n)}+\widetilde{\mathcal{G}}_{(n)}\mathcal{L}^{(0)}_{(3-n)}\right).\label{gmfd}
\end{align}
What misses in the notation $\mathcal{L}^{(k)}_{(d)}$ is \textbf{the dimension of spacetime metric on which it is evaluated}. We decide to make it implicit: it must be understood that the $\mathcal{L}^{(k)}_{(d)}$'s above are evaluated for the target spacetime metric $g_{(D)\mu\nu}$. This means that, for instance, the Ricci scalar and covariant derivatives appearing in the expression (\ref{eq:l1d}) of $\mathcal{L}^{(1)}_{(d)}$ are the ones of the metric $g_{(D)\mu\nu}$.\\

To be clear, for the KK decomposition (\ref{eq:kkdiagdec}-\ref{eq:internal}), Eqs.~(\ref{rmfd}) and~(\ref{gmfd}) relate
\begin{enumerate}
\item on the left hand side, the Ricci and GB scalars of the total $\mathfrak{D}$-dimensional spacetime,
\item on the right hand side, Lagrangians $\mathcal{L}^{(k)}_{(d)}$, evaluated for the target spacetime metric $g_{(D)\mu\nu}$, and Lovelock invariants of the internal space metric $\widetilde{g}_{(n)ab}$.
\end{enumerate}
Let us insist again on the fact that the Lagrangians $\mathcal{L}^{(k)}_{(d)}$ (with $d=1-n$ and $d=3-n$) on the right hand side are \textbf{not conformally-invariant Lagrangians} in general for the metric $g_{(D)\mu\nu}$. They would be conformally invariant only if the target spacetime dimension $D$ were coinciding with $d$.
\\

Eqs.~(\ref{rmfd}) and~(\ref{gmfd}) are similar in form and \textbf{lead to the following conjecture}: the Lovelock invariant of order $k$ of the total $\mathfrak{D}$-dimensional spacetime, $\mathcal{R}^{(k)}_{(\mathfrak{D})}$, can be expressed as a linear combination of Lagrangians $\mathcal{L}^{(p)}_{(2k-1-n)}$ evaluated on the target spacetime metric $g_{(D)\mu\nu}$, with coefficients being the Lovelock invariants $\widetilde{\mathcal{R}}^{(k)}_{(n)}$ of the internal space. Explicitly,
\begin{equation}
\mathcal{R}^{(k)}_{(\mathfrak{D})}\stackrel{?}{=}\mathrm{e}^{(n+1)\phi}\sum_{p=0}^k\binom{k}{p}\widetilde{\mathcal{R}}_{(n)}^{(p)}\mathcal{L}^{(k-p)}_{(2k-n-1)}.\label{eq:conjecture}
\end{equation}
The question mark above the equal sign puts the stress on the fact that this result has only been shown above for $k=1$ and $k=2$ (and is trivial for $k=0$), but remains a conjecture for more general $k$. It will be proved below, in paragraph~\ref{subsec:proof_lag}. Before moving on to the proof, we may first develop a similar conjecture regarding not the Lagrangians themselves, but the action integrals.
\subsection{The simple cases of Ricci and Gauss-Bonnet scalars: conjecture for the action integrals}
Indeed, Eqs.~(\ref{rmfd}) and~(\ref{gmfd}) are results bearing on the Lagrangians, they do not require any integration over spacetime nor integration by parts. Let us now investigate their consequences when, precisely, one integrates over spacetime. If the internal space is such that its Ricci and GB scalars $\widetilde{R}_{(n)}$ and $\widetilde{\mathcal{G}}_{(n)}$ are constant numbers, then the integral over the total spacetime factorizes into an integral over the internal space and an integral over the $D$-dimensional target spacetime. This reads
\begin{align}
\int\mathrm{d}^\mathfrak{D}x\sqrt{-g_{(\mathfrak{D})}} R_{(\mathfrak{D})}={}&{}\mathcal{V}_{(n)}\int\mathrm{d}^Dx\sqrt{-g}\,\mathrm{e}^{\phi}\left(\mathcal{L}^{(1)}_{(1-n)}+\widetilde{R}_{(n)}\mathcal{L}^{(0)}_{(1-n)}\right),\label{eq:kkact1}\\
\int\mathrm{d}^\mathfrak{D}x\sqrt{-g_{(\mathfrak{D})}}\mathcal{G}_{(\mathfrak{D})}={}&{}\mathcal{V}_{(n)}\int\mathrm{d}^Dx\sqrt{-g}\,\mathrm{e}^{\phi}\left(\mathcal{L}^{(2)}_{(3-n)}+2\widetilde{R}_{(n)}\mathcal{L}^{(1)}_{(3-n)}+\widetilde{\mathcal{G}}_{(n)}\mathcal{L}^{(0)}_{(3-n)}\right).\label{eq:kkact2}
\end{align}
$\mathcal{V}_{(n)}=\int\mathrm{d}^nx\sqrt{-\widetilde{g}_{(n)}}$ is the volume of the internal space. For brevity, in the integral $\int\mathrm{d}^Dx\sqrt{-g}$, we only write $\sqrt{-g}$, making implicit that this is the determinant of the target spacetime metric, $\sqrt{-g_{(D)}}$. This is consistent with the convention described above, that the Lagrangians $\mathcal{L}^{(k)}_{(d)}$ are understood to be evaluated for the metric $g_{(D)\mu\nu}$. In other words, the integrals $\int\mathrm{d}^Dx\sqrt{-g}\cdots$ really are scalar-tensor actions for the metric $g_{(D)\mu\nu}$ and the scalar field $\phi$. Thus, the original pure metric Lovelock gravity in dimension $\mathfrak{D}$ leads to a scalar-tensor action in dimension $D$. 
\\

It is possible to simplify these expressions further. Indeed, take $d$ an arbitrary number. Then,
\begin{align}
\int\mathrm{d}^Dx\sqrt{-g}\,\mathrm{e}^\phi\mathcal{L}^{(0)}_{(d)}={}&{}\int\mathrm{d}^Dx\sqrt{-g}\,\mathrm{e}^{(d+1)\phi}=\int\mathrm{d}^Dx\sqrt{-g}\,\mathcal{L}^{(0)}_{(d+1)}.\label{eq:ipp0}
\end{align}
For $\mathcal{L}^{(1)}_{(d)}$ and $\mathcal{L}^{(2)}_{(d)}$, one can proceed with integration by parts (denoted again by 'IBP'). For $\mathcal{L}^{(1)}_{(d)}$ first, take its expression (\ref{eq:l1d}). On the one hand,
\begin{align}
\int\mathrm{d}^Dx\sqrt{-g}\,\mathrm{e}^\phi\mathcal{L}^{(1)}_{(d)}={}&{}\int\mathrm{d}^Dx\sqrt{-g}\,\mathrm{e}^{(d-1)\phi}\Bigl(R-2\left(d-1\right)\Box\phi-\left(d-1\right)\left(d-2\right)\left(\partial\phi\right)^2\Bigr)\nonumber\\
{}\underset{\text{IBP}}{=}{}&{}\int\mathrm{d}^Dx\sqrt{-g}\,\mathrm{e}^{(d-1)\phi}\Bigl(R+d\left(d-1\right)\left(\partial\phi\right)^2\Bigr),
\end{align}
and on the other hand,
\begin{align}
\int\mathrm{d}^Dx\sqrt{-g}\,\mathcal{L}^{(1)}_{(d+1)}={}&{}\int\mathrm{d}^Dx\sqrt{-g}\,\mathrm{e}^{(d-1)\phi}\Bigl(R-2d\Box\phi-d\left(d-1\right)\left(\partial\phi\right)^2\Bigr)\nonumber\\
{}\underset{\text{IBP}}{=}{}&{}\int\mathrm{d}^Dx\sqrt{-g}\,\mathrm{e}^{(d-1)\phi}\Bigl(R+d\left(d-1\right)\left(\partial\phi\right)^2\Bigr).
\end{align}
Therefore,
\begin{equation}
\int\mathrm{d}^Dx\sqrt{-g}\,\mathrm{e}^\phi\mathcal{L}^{(1)}_{(d)}\underset{\text{IBP}}{=}\int\mathrm{d}^Dx\sqrt{-g}\mathcal{L}^{(1)}_{(d+1)}.\label{eq:ipp1}
\end{equation}
Regarding now $\mathcal{L}^{(2)}_{(d)}$, start from its expression (\ref{eq:l2d}). On the one hand,
\begin{align}
\int\mathrm{d}^Dx\sqrt{-g}\,\mathrm{e}^\phi\mathcal{L}^{(2)}_{(d)}={}&{}\int\mathrm{d}^Dx\sqrt{-g}\,\mathrm{e}^{(d-3)\phi}\Bigl\{\mathcal{G}+8\left(d-3\right)G^{\mu\nu}\phi_{\mu\nu}-8\left(d-3\right)R_{\mu\nu}\phi^\mu\phi^\nu\nonumber\\{}&{}+4\left(d-2\right)\left(d-3\right)\left[\left(\Box\phi\right)^2-\phi_{\mu\nu}\phi^{\mu\nu}\right]-2\left(d-3\right)\left(d-4\right)R\left(\partial\phi\right)^2\nonumber\\{}&{}+8\left(d-2\right)\left(d-3\right)\phi^\mu\phi_{\mu\nu}\phi^\nu+4\left(d-2\right)\left(d-3\right)^2\left(\partial\phi\right)^2\Box\phi\nonumber\\{}&{}+\left(d-1\right)\left(d-2\right)\left(d-3\right)\left(d-4\right)\left(\partial\phi\right)^4\Bigr\}\nonumber\\
{}\underset{\text{IBP}}{=}{}&{}\int\mathrm{d}^Dx\sqrt{-g}\,\mathrm{e}^{(d-3)\phi}\Bigl[\mathcal{G}-4\left(d-2\right)\left(d-3\right)G^{\mu\nu}\phi_\mu\phi_\nu\nonumber\\{}&{}\hspace{-2.4cm}-2\left(d-1\right)\left(d-2\right)\left(d-3\right)\left(\partial\phi\right)^2\Box\phi-\left(d-1\right)\left(d-2\right)^2\left(d-3\right)\left(\partial\phi\right)^4\Bigr].
\end{align}
But, on the other hand,
\begin{align}
\int\mathrm{d}^Dx\sqrt{-g}\,\mathcal{L}^{(2)}_{(d+1)}={}&{}\int\mathrm{d}^Dx\sqrt{-g}\,\mathrm{e}^{(d-3)\phi}\Bigl\{\mathcal{G}+8\left(d-2\right)G^{\mu\nu}\phi_{\mu\nu}-8\left(d-2\right)R_{\mu\nu}\phi^\mu\phi^\nu\nonumber\\{}&{}+4\left(d-1\right)\left(d-2\right)\left[\left(\Box\phi\right)^2-\phi_{\mu\nu}\phi^{\mu\nu}\right]-2\left(d-2\right)\left(d-3\right)R\left(\partial\phi\right)^2\nonumber\\{}&{}+8\left(d-1\right)\left(d-2\right)\phi^\mu\phi_{\mu\nu}\phi^\nu+4\left(d-1\right)\left(d-2\right)^2\left(\partial\phi\right)^2\Box\phi\nonumber\\{}&{}+d\left(d-1\right)\left(d-2\right)\left(d-3\right)\left(\partial\phi\right)^4\Bigr\}\nonumber\\
{}\underset{\text{IBP}}{=}{}&{}\int\mathrm{d}^Dx\sqrt{-g}\,\mathrm{e}^{(d-3)\phi}\Bigl[\mathcal{G}-4\left(d-2\right)\left(d-3\right)G^{\mu\nu}\phi_\mu\phi_\nu\nonumber\\{}&{}\hspace{-2.4cm}-2\left(d-1\right)\left(d-2\right)\left(d-3\right)\left(\partial\phi\right)^2\Box\phi-\left(d-1\right)\left(d-2\right)^2\left(d-3\right)\left(\partial\phi\right)^4\Bigr].
\end{align}
As a consequence,
\begin{equation}
\int\mathrm{d}^Dx\sqrt{-g}\,\mathrm{e}^\phi\mathcal{L}^{(2)}_{(d)}\underset{\text{IBP}}{=}\int\mathrm{d}^Dx\sqrt{-g}\mathcal{L}^{(2)}_{(d+1)}.\label{eq:ipp2}
\end{equation}
Using Eqs. (\ref{eq:ipp0}), (\ref{eq:ipp1}) and (\ref{eq:ipp2}) into (\ref{eq:kkact1}) and (\ref{eq:kkact2}) leads to
\begin{align}
\int\mathrm{d}^\mathfrak{D}x\sqrt{-g_{(\mathfrak{D})}} R_{(\mathfrak{D})}\underset{\text{IBP}}{=}{}&{}\mathcal{V}_{(n)}\int\mathrm{d}^Dx\sqrt{-g}\left(\mathcal{L}^{(1)}_{(2-n)}+\widetilde{R}_{(n)}\mathcal{L}^{(0)}_{(2-n)}\right),\\
\int\mathrm{d}^\mathfrak{D}x\sqrt{-g_{(\mathfrak{D})}}\mathcal{G}_{(\mathfrak{D})}\underset{\text{IBP}}{=}{}&{}\mathcal{V}_{(n)}\int\mathrm{d}^Dx\sqrt{-g}\left(\mathcal{L}^{(2)}_{(4-n)}+2\widetilde{R}_{(n)}\mathcal{L}^{(1)}_{(4-n)}+\widetilde{\mathcal{G}}_{(n)}\mathcal{L}^{(0)}_{(4-n)}\right).
\end{align}
From the particular cases $k=0$ (which is trivial), $k=1$ and $k=2$, we had formulated a conjecture (\ref{eq:conjecture}) for all $k$, bearing on the KK decomposition of the Lovelock \textbf{Lagrangians}. In a similar way, we can now \textbf{propose a conjecture} for the KK decomposition of the Lovelock \textbf{action integrals},
\begin{equation}
\int\mathrm{d}^\mathfrak{D}x\sqrt{-g_{(\mathfrak{D})}}\mathcal{R}^{(k)}_{(\mathfrak{D})}\underset{\text{IBP}}{\stackrel{?}{=}}\mathcal{V}_{(n)}\int\mathrm{d}^Dx\sqrt{-g}\sum_{p=0}^k\binom{k}{p}\widetilde{\mathcal{R}}_{(n)}^{(p)}\mathcal{L}^{(k-p)}_{(2k-n)}.\label{eq:conjecture_act}
\end{equation}
Of course, this requires the additional assumption that the Lovelock invariants $\widetilde{\mathcal{R}}_{(n)}^{(p)}$ of the internal space are constant numbers. For the simple and explicitly demonstrated cases $k=0,1,2$, going from the Lagrangian result to the action result required the use of $\int\mathrm{d}^Dx\sqrt{-g}\,\mathrm{e}^\phi\mathcal{L}^{(k)}_{(d)}\underset{\text{IBP}}{=}\int\mathrm{d}^Dx\sqrt{-g}\mathcal{L}^{(k)}_{(d+1)}$, which we proved for $k=0,1,2$, see Eqs. (\ref{eq:ipp0}), (\ref{eq:ipp1}) and (\ref{eq:ipp2}). Similarly, the Lagrangian conjecture (\ref{eq:conjecture}) immediately implies the action conjecture (\ref{eq:conjecture_act}) provided the same lemma holds for all order $k$,
\begin{equation}
\int\mathrm{d}^Dx\sqrt{-g}\,\mathrm{e}^\phi\mathcal{L}^{(k)}_{(d)}\underset{\text{IBP}}{\stackrel{?}{=}}\int\mathrm{d}^Dx\sqrt{-g}\mathcal{L}^{(k)}_{(d+1)}.\label{eq:conjecture_lemma}
\end{equation}
\subsection{Lagrangians with local conformal invariance in specific conformal dimension}\label{subsec:specific}
Let us summarize: \textbf{in the context of the KK diagonal decomposition (\ref{eq:kkdiagdec}-\ref{eq:internal}), there exist two conjectures regarding the decomposition of the Lovelock invariant of order $k$ of the total $\mathfrak{D}$-dimensional space, $\mathcal{R}^{(k)}_{(\mathfrak{D})}$:
\begin{align}
\mathcal{R}^{(k)}_{(\mathfrak{D})}\stackrel{?}{=}{}&{}\mathrm{e}^{(n+1)\phi}\sum_{p=0}^k\binom{k}{p}\widetilde{\mathcal{R}}_{(n)}^{(p)}\mathcal{L}^{(k-p)}_{(2k-n-1)},\label{eq:conjecture_lag_align}\\
\int\mathrm{d}^\mathfrak{D}x\sqrt{-g_{(\mathfrak{D})}}\mathcal{R}^{(k)}_{(\mathfrak{D})}\underset{\text{IBP}}{\stackrel{?}{=}}{}&{}\mathcal{V}_{(n)}\int\mathrm{d}^Dx\sqrt{-g}\sum_{p=0}^k\binom{k}{p}\widetilde{\mathcal{R}}_{(n)}^{(p)}\mathcal{L}^{(k-p)}_{(2k-n)},\label{eq:conjecture_act_align}
\end{align}
and the first conjecture implies the second one provided the following lemma holds,
\begin{equation}
\int\mathrm{d}^Dx\sqrt{-g}\,\mathrm{e}^\phi\mathcal{L}^{(k)}_{(d)}\underset{\text{IBP}}{\stackrel{?}{=}}\int\mathrm{d}^Dx\sqrt{-g}\mathcal{L}^{(k)}_{(d+1)}.\label{eq:conjecture_lemma_align}
\end{equation}
All these have been demonstrated for $k=0,1,2$, but remains to be shown for arbitrary $k$}. However, for the moment, the explicit meaning of $\mathcal{L}^{(k)}_{(d)}$ remains unclear in general. Indeed, let us forget for a moment about the KK context, and simply remember that $\mathcal{L}^{(k)}_{(d)}$ was first defined by Eqs. (\ref{eq:s1}-\ref{eq:slocD}), in relation to the $k$-th Lovelock invariant $\hat{\mathcal{R}}^{(k)}_{(d)}$ of a $d$-dimensional metric $\hat{g}_{\mu\nu}= \mathrm{e}^{2\phi}g_{\mu\nu}$:
\begin{equation}
\mathcal{L}^{(k)}_{(d)}=\mathrm{e}^{(d-2k)\phi}\mathcal{S}^{(k)}_{(d)},\quad \mathcal{S}^{(k)}_{(d)}= \mathrm{e}^{2k\phi}\,\hat{\mathcal{R}}^{(k)}_{(d)}= \frac{1}{2^k}\delta^{\mu_1\nu_1\cdots\mu_k\nu_k}_{\alpha_1\beta_1\cdots\alpha_k\beta_k}\prod_{i=1}^{k}S^{\alpha_i\beta_i}_{\quad\,\mu_i\nu_i}.\label{eq:lkd_reminder_1}
\end{equation}
The tensor $S^{\mu\nu}_{\,\,\,\,\,\,\rho\sigma}$ is linked to the Riemann tensor $\hat{R}^{\mu\nu}_{\,\,\,\,\,\,\rho\sigma}$ of $\hat{g}_{\mu\nu}$,
\begin{equation}
\hat{R}^{\mu\nu}_{\,\,\,\,\,\,\rho\sigma}=\mathrm{e}^{-2\phi}S^{\mu\nu}_{\,\,\,\,\,\,\rho\sigma},\quad S^{\mu\nu}_{\,\,\,\,\,\,\rho\sigma}\equiv R^{\mu\nu}_{\,\,\,\,\,\,\rho\sigma}-4\delta^{[\mu}_{[\rho}\phi^{\nu]}_{\sigma]}+4\delta^{[\mu}_{[\rho}\phi^{\nu]}\phi_{\sigma]}-\left(\partial\phi\right)^2\delta^{\mu\nu}_{\rho\sigma}.\label{eq:lkd_reminder_2}
\end{equation}
It thus seems for the moment that $\mathcal{L}^{(k)}_{(d)}$ is clearly defined only in a $d$-dimensional spacetime, whereas the conjectures (\ref{eq:conjecture_lag_align}-\ref{eq:conjecture_lemma_align}) are writing down $\mathcal{L}^{(k)}_{(d)}$ in a $D$-dimensional spacetime (namely the target spacetime with metric $g_{(D)\mu\nu}$ in the KK decomposition). These conjectures come from the demonstrated results for $k=0,1,2$. In these simple cases, $\mathcal{L}^{(0)}_{(d)}$, $\mathcal{L}^{(1)}_{(d)}$ and $\mathcal{L}^{(2)}_{(d)}$ were defined in the following manner (see the discussion at the beginning of paragraph~\ref{subsec:simple_conj_lag}):
\begin{enumerate}
\item First, compute explicitly $\mathcal{L}^{(k)}_{(d)}$ from its natural definition in a $d$-dimensional spacetime, making explicit all factors involving $d$, which arise from the contractions of the Kronecker symbols. This yields a Lorentz scalar where there are numerical factors involving $d$, and expressions involving only the metric, the scalar field and its covariant derivatives, for instance $R$, $\Box\phi$, $G^{\mu\nu}\phi_\mu\phi_\nu$, etc.
\item Simply take the previous Lorentz scalar and consider it no more for a $d$-dimensional metric, but for a $D$-dimensional metric where \textit{a priori} $D\neq d$. This means that $R$, $\Box\phi$, $G^{\mu\nu}\phi_\mu\phi_\nu$, etc., are those of a $D$-dimensional metric and of a scalar field living in this $D$-dimensional spacetime. This defines $\mathcal{L}^{(k)}_{(d)}$ in dimension $D\neq d$, in which case the subscript $d$ shall be seen as a mere numerical parameter of a $D$-dimensional Lorentz scalar. 
\end{enumerate}
Importantly, \textbf{this procedure is very different from taking (\ref{eq:lkd_reminder_1}-\ref{eq:lkd_reminder_2}) and directly computing it for a $D$-dimensional spacetime, as the Kronecker contractions would yield different factors. With the above prescription, the Kronecker contractions are the same as would be computed in $d$ dimensions, but once these contractions have been performed, the resulting Lorentz scalar is considered in $D$ dimensions}. 
\\

This prescription was easy to perform for the first few Lagrangians, of order $k=0,1,2$, see Eqs. (\ref{eq:l0d}-\ref{eq:l2d}), and even for order $k=3$, see (\ref{eq:l3}). One must now find a way of applying the above prescription for arbitrary order $k$. We will need the rule for contractions of Kronecker symbols,
\begin{equation}
\text{In a space of dimension }d,\quad \delta^{y_1\cdots y_n}_{x_1\cdots x_n}\delta^{x_1\cdots x_nx_{n+1}\cdots x_p}_{y_1\cdots y_n\,y_{n+1}\cdots y_p}=n!\frac{(d+n-p)!}{(d-p)!}\delta^{x_{n+1}\cdots x_p}_{y_{n+1}\cdots\, y_p}.\label{eq:delta_contr}
\end{equation} 
With this rule at hand, let us simplify $\mathcal{L}^{(k)}_{(d)}$, or equivalently $\mathcal{S}^{(k)}_{(d)}$, see Eq. (\ref{eq:lkd_reminder_1}), keeping in mind that the subscript $(d)$ means that all Kronecker contractions yield factors corresponding to $d$ dimensions. The following computation lasts about one page and the final result is Eq. (\ref{eq:skdfinal}). First,
\begin{align}
\mathcal{S}^{(k)}_{(d)}{}&{}=\frac{1}{2^k}\delta^{\mu_1\nu_1\cdots\mu_k\nu_k}_{\alpha_1\beta_1\cdots\alpha_k\beta_k}\prod_{i=1}^k\left(R^{\alpha_i\beta_i}_{\,\,\,\,\,\,\,\,\mu_i\nu_i}-4\delta^{[\alpha_i}_{[\mu_i}\phi^{\beta_i]}_{\nu_i]}+4\delta^{[\alpha_i}_{[\mu_i}\phi^{\beta_i]}\phi_{\nu_i]}-\left(\partial\phi\right)^2\delta^{\alpha_i\beta_i}_{\mu_i\nu_i}\right)\nonumber\\{}&{}=\frac{1}{2^k}\delta^{\mu_1\nu_1\cdots\mu_k\nu_k}_{\alpha_1\beta_1\cdots\alpha_k\beta_k}\sum_{q=0}^k\,\sum_{\substack{I\subset\left[1,k\right]\\ \left\lvert I\right\rvert=q}}\,\prod_{i\in I}\left(R^{\alpha_i\beta_i}_{\quad\mu_i\nu_i}-\left(\partial\phi\right)^2\delta^{\alpha_i\beta_i}_{\mu_i\nu_i}\right)\prod_{i\not\in I}\left(4\delta^{[\alpha_i}_{[\mu_i}\phi^{\beta_i]}\phi_{\nu_i]}-4\delta^{[\alpha_i}_{[\mu_i}\phi^{\beta_i]}_{\nu_i]}\right)\nonumber\\{}&{}=\frac{1}{2^k}\delta^{\mu_1\nu_1\cdots\mu_k\nu_k}_{\alpha_1\beta_1\cdots\alpha_k\beta_k}\sum_{q=0}^k\binom{k}{q}\prod_{i=1}^q\left(R^{\alpha_i\beta_i}_{\quad\mu_i\nu_i}-\left(\partial\phi\right)^2\delta^{\alpha_i\beta_i}_{\mu_i\nu_i}\right)\prod_{i=q+1}^k\left(4\delta^{[\alpha_i}_{[\mu_i}\phi^{\beta_i]}\phi_{\nu_i]}-4\delta^{[\alpha_i}_{[\mu_i}\phi^{\beta_i]}_{\nu_i]}\right).
\end{align}
The notation $\left\lvert I\right\rvert =q$ means the cardinality of the set $I$. Due to the (anti)symmetries of the Kronecker symbol, the Riemann tensor, and the antisymmetrized bracketed indices, all different subsets $I\subset [1,k]$ with given cardinal $\lvert I\rvert=q$ give the same contribution once contracted with $\delta^{\mu_1\nu_1\cdots\mu_k\nu_k}_{\alpha_1\beta_1\cdots\alpha_k\beta_k}$, yielding the usual binomial factor $\binom{k}{q}$. This will often happen, for instance in the following line, and we do not detail it anymore. Using the contraction of Kronecker symbols (\ref{eq:delta_contr}),
\begin{align}
{}&{}\delta^{\mu_1\nu_1\cdots\mu_k\nu_k}_{\alpha_1\beta_1\cdots\alpha_k\beta_k}\prod_{i=1}^q\left(R^{\alpha_i\beta_i}_{\quad\mu_i\nu_i}-\left(\partial\phi\right)^2\delta^{\alpha_i\beta_i}_{\mu_i\nu_i}\right)\nonumber\\{}&{}=\delta^{\mu_1\nu_1\cdots\mu_k\nu_k}_{\alpha_1\beta_1\cdots\alpha_k\beta_k}\sum_{j=0}^q\binom{q}{j}\left(-\left(\partial\phi\right)^2\right)^{q-j}\left(\prod_{i=1}^jR^{\alpha_i\beta_i}_{\quad\mu_i\nu_i}\right)\left(\prod_{i=j+1}^q\delta^{\alpha_i\beta_i}_{\mu_i\nu_i}\right)\nonumber\\{}&{}=\sum_{j=0}^q\frac{2^{q-j}\binom{q}{j}(d-2(k+j-q))!}{(d-2k)!}\left(-\left(\partial\phi\right)^2\right)^{q-j}\delta^{\mu_1\nu_1\cdots\mu_j\,\nu_j\mu_{q+1}\nu_{q+1}\cdots\,\mu_k\nu_k}_{\alpha_1\beta_1\cdots\alpha_j\beta_j\alpha_{q+1}\beta_{q+1}\cdots\alpha_k\beta_k}\left(\prod_{i=1}^jR^{\alpha_i\beta_i}_{\quad\mu_i\nu_i}\right).
\end{align}
Then,
\begin{align}
{}&{}\delta^{\mu_1\nu_1\cdots\mu_j\,\nu_j\mu_{q+1}\nu_{q+1}\cdots\,\mu_k\nu_k}_{\alpha_1\beta_1\cdots\alpha_j\beta_j\alpha_{q+1}\beta_{q+1}\cdots\alpha_k\beta_k}\prod_{i=q+1}^k\left(4\delta^{[\alpha_i}_{[\mu_i}\phi^{\beta_i]}\phi_{\nu_i]}-4\delta^{[\alpha_i}_{[\mu_i}\phi^{\beta_i]}_{\nu_i]}\right)\nonumber\\{}&{}=4^{k-q}\frac{(d-(2j+k-q))!}{(d-2(k-q+j))!}\delta^{\mu_1\nu_1\cdots\mu_j\,\nu_j\nu_{q+1}\cdots\nu_k}_{\alpha_1\beta_1\cdots\alpha_j\beta_j\beta_{q+1}\cdots\beta_k}\prod_{i=q+1}^k\left(\phi^{\beta_i}\phi_{\nu_i}-\phi^{\beta_i}_{\nu_i}\right),
\end{align}
where the $4^{k-q}$ is because, for each $i\in\left[q+1,k\right]$, each of the four terms implied by the antisymmetrization over $(\alpha_i,\beta_i)$ and $(\mu_i,\nu_i)$ contributes the same because of the symmetries of the Kronecker symbol. All in all,
\begin{align}
\mathcal{S}^{(k)}_{(d)}=\sum_{q=0}^k\sum_{j=0}^q\binom{k}{q}\binom{q}{j}{}&{}2^{k-q-j}\frac{(d-(2j+k-q))!}{(d-2k)!}\left(-\left(\partial\phi\right)^2\right)^{q-j}\nonumber\\{}&{}\delta^{\mu_1\nu_1\cdots\mu_j\nu_j\nu_{q+1}\cdots\nu_k}_{\alpha_1\beta_1\cdots\alpha_j\beta_j\beta_{q+1}\cdots\beta_k}\left(\prod_{i=1}^jR^{\alpha_i\beta_i}_{\quad\mu_i\nu_i}\right)\prod_{i=q+1}^k\left(\phi^{\beta_i}\phi_{\nu_i}-\phi^{\beta_i}_{\nu_i}\right).
\end{align}
The computation is over, but for further convenience, we perform the following manipulations on the sums: 
\begin{enumerate}
\item In the sum over $j$, make $q-j\to j$,
\item Interchange the sums over $q$ and $j$, yielding $\sum_{j=0}^{k}\sum_{q=j}^{k}\cdots$,
\item In the sum over $q$, make $k-q\to q$,
\item Use the identity $\binom{k-q}{j}\binom{k}{q}=\binom{k-j}{q}\binom{k}{j}$,
\item Rename $j$ as $i$ and $q$ as $j$.
\end{enumerate}
One finally obtains, for $\mathcal{L}^{(k)}_{(d)}=\mathrm{e}^{(d-2k)\phi}\mathcal{S}^{(k)}_{(d)}$,
\begin{align}
\mathcal{L}^{(k)}_{(d)}=\mathrm{e}^{(d-2k)\phi}\,2^{-k}{}&{}\sum_{i=0}^{k}2^i\binom{k}{i}\left(-\left(\partial\phi\right)^2\right)^i\sum_{j=0}^{k-i}4^j\binom{k-i}{j}\left(-1\right)^j\frac{(d-2k+2i+j)!}{(d-2k)!}\nonumber\\{}&{}\delta^{\tau_1\cdots\,\,\tau_j\rho_1\sigma_1\cdots\rho_{k-i-j}\,\sigma_{k-i-j}}_{\lambda_1\cdots\lambda_j\alpha_1\beta_1\cdots\alpha_{k-i-j}\beta_{k-i-j}}\left(\prod_{q=1}^{k-i-j}R^{\alpha_q\beta_q}_{\quad\,\,\rho_q\sigma_q}\right)\prod_{q=1}^j\left(\phi^{\lambda_q}_{\tau_q}-\phi^{\lambda_q}\phi_{\tau_q}\right).\label{eq:skdfinal}
\end{align}
Note that, in the second line, we have renamed the contracted indices. Also, there is a factor $(-1)^j$ because in the final product, we have taken a $-1$ out of each of the $j$ factors.\\

In accordance with our aim, the dependence on the dimension $d$ is completely explicit in (\ref{eq:skdfinal}). Indeed, the covariant expression of $R^{\alpha_q\beta_q}_{\quad\,\,\rho_q\sigma_q}$ and $\phi^{\lambda_q}_{\tau_q}$ involves the metric, the scalar field and their derivatives, but not the dimension nor Kronecker symbols, so the remaining contractions of spacetime indices on the second line do not yield any additional occurence of $d$. As a consequence, the expression (\ref{eq:skdfinal}) can be evaluated for the metric of a spacetime of any dimension $D$, meaning that the Riemann tensor and covariant derivatives are those associated with this latter metric, and that the remaining Kronecker symbol $\delta^{\tau_1\cdots\,\,\tau_j\rho_1\sigma_1\cdots\rho_{k-i-j}\,\sigma_{k-i-j}}_{\lambda_1\cdots\lambda_j\alpha_1\beta_1\cdots\alpha_{k-i-j}\beta_{k-i-j}}$ also lives in this $D$-dimensional spacetime. To put it simply, this amounts to taking all spacetime indices of (\ref{eq:skdfinal}) to be $D$-dimensional.
\\

In a nutshell, $\mathcal{L}^{(k)}_{(d)}$, as given by (\ref{eq:skdfinal}), is a Lorentz scalar whose expression depends on both its \textbf{order} $k$ and its \textbf{specific conformal dimension} $d$. In any spacetime dimension $D$, it gives rise to a scalar-tensor action
\begin{equation}
S\left[g_{\mu\nu},\phi\right]=\int\mathrm{d}^Dx\sqrt{-g}\,\mathcal{L}^{(k)}_{(d)}.\label{eq:nutshell}
\end{equation} 
This scalar-tensor action is locally conformally-invariant if the \textbf{specific conformal dimension} $d$ of the Lagrangian $\mathcal{L}^{(k)}_{(d)}$ matches the spacetime dimension $D$, that is, $d=D$. \textbf{In the following, the Lagrangians $\mathcal{L}^{(k)}_{(d)}$ may be generically called conformal Lagrangians, while it must be understood that their corresponding action enjoys local conformal invariance only if the specific conformal dimension $d$ matches the spacetime dimension $D$}.
\subsection{Proof of the first conjecture (conjecture for the Lagrangians)}\label{subsec:proof_lag}
Now that the meaning of $\mathcal{L}^{(k)}_{(d)}$ is clear, it becomes possible to prove the first of the above conjectures, regarding the exact (without integration by parts) writing of $\mathcal{R}^{(k)}_{(\mathfrak{D})}$, the $k$-th Lovelock invariant of the total spacetime,
\begin{equation*}
\mathcal{R}^{(k)}_{(\mathfrak{D})}\stackrel{?}{=}\mathrm{e}^{(n+1)\phi}\sum_{p=0}^k\binom{k}{p}\widetilde{\mathcal{R}}_{(n)}^{(p)}\mathcal{L}^{(k-p)}_{(2k-n-1)}.
\end{equation*}
On the right hand side, appears a linear combination of the 'conformal' Lagrangians, with coefficients being the Lovelock invariants $\widetilde{\mathcal{R}}_{(n)}^{(p)}$ of the internal space. Note that Ref.~\cite{VanAcoleyen:2011mj} performs calculations similar to the following ones, but with different simplifications which do not make the 'conformal' Lagrangians apparent.\\

The reader having no interest in the following proof may directly move on to its conclusions, around Eq. (\ref{eq:diag_fin_proof}). The $k$-th Lovelock invariant of the $\mathfrak{D}$-dimensional spacetime is
\begin{equation}
\mathcal{R}^{(k)}_{(\mathfrak{D})}=\frac{1}{2^k}\delta^{R_1S_1\cdots R_kS_k}_{A_1B_1\cdots A_kB_k}\prod_{i=1}^kR^{A_iB_i}_{(\mathfrak{D})\,R_iS_i},
\end{equation}
and each of the $R^{A_iB_i}_{(\mathfrak{D})\,R_iS_i}$ can be of one of the three types (\ref{eq:r1}-\ref{eq:r3}), which we recall,
\begin{align}
R^{\mu\nu}_{(\mathfrak{D})\,\rho\sigma}={}&{}R^{\mu\nu}_{(D)\,\rho\sigma},\label{eq:r1bis}\\
R^{ab}_{(\mathfrak{D})\,cd}={}&{}\mathrm{e}^{2\phi}\widetilde{R}^{ab}_{(n)\,\rho\sigma}-\left(\partial\phi\right)^2\delta^{ab}_{cd},\label{eq:r2bis}\\
R^{\mu a}_{(\mathfrak{D})\,\nu b}={}&{} \delta^a_b\left(\phi^{\mu}_\nu -\phi^\mu\phi_\nu\right).\label{eq:r3bis}
\end{align}
Therefore,
\begin{align}
\mathcal{R}^{(k)}_{(\mathfrak{D})}=\frac{1}{2^k}\sum_{i+j\leq k}4^j\binom{k}{i}\binom{k-i}{j}\,{}&{}\delta^{r_1s_1\cdots r_is_it_1\cdots t_j}_{a_1b_1\cdots a_ib_il_1\cdots l_j}\,\delta^{\tau_1\cdots\,\,\tau_j\rho_1\sigma_1\cdots\rho_{k-i-j}\sigma_{k-i-j}}_{\lambda_1\cdots\lambda_j\alpha_1\beta_1\cdots\alpha_{k-i-j}\beta_{k-i-j}}\nonumber\\{}&{}\left(\prod_{q=1}^iR^{a_qb_q}_{(\mathfrak{D})\,r_qs_q}\right)\left(\prod_{q=1}^{k-i-j}R^{\alpha_q\beta_q}_{(\mathfrak{D})\,\rho_q\sigma_q}\right)\prod_{q=1}^jR^{\lambda_ql_q}_{(\mathfrak{D})\,\tau_qt_q}.
\end{align}
Indeed, the binomial coefficient $\binom{k}{i}$ is for the choice, among the $k$ pairs of down indices $(R_i,S_i)$, of $i$ pairs of down Latin indices $(r_q,s_q)$. Then, among the remaining $k-i$ pairs of down $(R_i,S_i)$, one chooses $k-i-j$ pairs of down Greek indices $(\rho_q,\sigma_q)$, hence the factor $\binom{k-i}{j}$. The remaining pairs $(R_i,S_i)$ are automatically $j$ mixed ones and do not give rise to binomial factors, however, one must account for the four distinct contributions $R^{\mu a}_{(\mathfrak{D})\,\nu b}$, $R^{a\mu}_{(\mathfrak{D})\,\nu b}$, $R^{\mu a}_{(\mathfrak{D})\,b\nu}$ and $R^{a\mu}_{(\mathfrak{D})\,b\nu}$, hence the factor $4^{j}$ (these contributions all come with the same sign since the antisymmetry of the Riemann tensor is compensated by the antisymmetry of the generalized Kronecker symbol). In the following, extensive use is made of the contraction rule for Kronecker symbols (\ref{eq:delta_contr}). Using (\ref{eq:r3bis}), one has
\begin{align}
\delta^{r_1s_1\cdots r_is_it_1\cdots t_j}_{a_1b_1\cdots a_ib_il_1\cdots l_j}\prod_{q=1}^jR^{\lambda_ql_q}_{(\mathfrak{D})\,\tau_qt_q}={}&{}\delta^{r_1s_1\cdots r_is_it_1\cdots t_j}_{a_1b_1\cdots a_ib_il_1\cdots l_j}\prod_{q=1}^j\delta^{l_q}_{t_q}\left(\phi^{\lambda_q}_{\tau_q}-\phi^{\lambda_q}\phi_{\tau_q}\right)\nonumber\\{}={}&{}\frac{(n-2i)!}{(n-j-2i)!}\delta^{r_1s_1\cdots r_is_i}_{a_1b_1\cdots a_ib_i}\prod_{q=1}^j\left(\phi^{\lambda_q}_{\tau_q}-\phi^{\lambda_q}\phi_{\tau_q}\right),
\end{align}
On the other hand, (\ref{eq:r2bis}) leads to
\begin{align}
\delta^{r_1s_1\cdots r_is_i}_{a_1b_1\cdots a_ib_i}{}&{}\prod_{q=1}^iR^{a_qb_q}_{(\mathfrak{D})\,r_qs_q}\nonumber\\{}&{}=\delta^{r_1s_1\cdots r_is_i}_{a_1b_1\cdots a_ib_i}\sum_{p=0}^i\mathrm{e}^{2p\phi}\left(-\left(\partial\phi\right)^2\right)^{i-p}\sum_{I\subset [1,i],\lvert I\rvert=p}\left(\prod_{q\in I} \widetilde{R}^{a_qb_q}_{(n)\,r_qs_q}\right)\left(\prod_{q\not\in I} \delta^{a_qb_q}_{r_qs_q}\right)\nonumber\\{}&{}=2^i\sum_{p=0}^i\binom{i}{p}\frac{(n-2p)!}{(n-2i)!}\mathrm{e}^{2p\phi}\widetilde{\mathcal{R}}_{(n)}^{(p)}\left(-\left(\partial\phi\right)^2\right)^{i-p},
\end{align}
where $\widetilde{\mathcal{R}}_{(n)}^{(p)}$ is the $p$-th Lovelock invariant of the metric $\widetilde{g}_{(n)\,ab}$ of the internal space. Taking all the previous into account, as well as (\ref{eq:r1bis}), gives
\begin{align}
\mathcal{R}^{(k)}_{(\mathfrak{D})}=\sum_{i+j\leq k}{}&{}\frac{2^{i-k}4^j\binom{k}{i}\binom{k-i}{j}}{(n-j-2i)!}\delta^{\tau_1\cdots\,\,\tau_j\rho_1\sigma_1\cdots\rho_{k-i-j}\sigma_{k-i-j}}_{\lambda_1\cdots\lambda_j\alpha_1\beta_1\cdots\alpha_{k-i-j}\beta_{k-i-j}}\left(\prod_{q=1}^{k-i-j}R^{\alpha_q\beta_q}_{(D)\,\rho_q\sigma_q}\right)\nonumber\\{}&{}\left(\prod_{q=1}^j\left(\phi^{\lambda_q}_{\tau_q}-\phi^{\lambda_q}\phi_{\tau_q}\right)\right)\sum_{p=0}^i\binom{i}{p}(n-2p)!\,\mathrm{e}^{2p\phi}\widetilde{\mathcal{R}}_{(n)}^{(p)}\left(-\left(\partial\phi\right)^2\right)^{i-p}.
\end{align}
By direct computation, $\binom{k}{i}\binom{i}{p}=\binom{k}{p}\binom{k-p}{k-i}$. Putting in addition the sum over $p$ in the first place, one gets
\begin{align}
\mathcal{R}^{(k)}_{(\mathfrak{D})}=\sum_{p=0}^k{}&{}\binom{k}{p}\mathrm{e}^{2p\phi}\widetilde{\mathcal{R}}_{(n)}^{(p)}\sum_{i=p}^k2^{i-k}\left(-\left(\partial\phi\right)^2\right)^{i-p}\sum_{j=0}^{k-i}\frac{4^j\binom{k-i}{j}\binom{k-p}{k-i}(n-2p)!}{(n-j-2i)!}\nonumber\\{}&{}\delta^{\tau_1\cdots\,\,\tau_j\rho_1\sigma_1\cdots\rho_{k-i-j}\sigma_{k-i-j}}_{\lambda_1\cdots\lambda_j\alpha_1\beta_1\cdots\alpha_{k-i-j}\beta_{k-i-j}}\left(\prod_{q=1}^{k-i-j}R^{\alpha_q\beta_q}_{(D)\,\rho_q\sigma_q}\right)\prod_{q=1}^j\left(\phi^{\lambda_q}_{\tau_q}-\phi^{\lambda_q}\phi_{\tau_q}\right).
\end{align}
Make $i-p\to i$ in the sum over $i$, and obtain finally,
\begin{align}
\mathcal{R}^{(k)}_{(\mathfrak{D})}=\sum_{p=0}^k{}&{}\binom{k}{p}\mathrm{e}^{2p\phi}\widetilde{\mathcal{R}}_{(n)}^{(p)}2^{p-k}\sum_{i=0}^{k-p}2^{i}\binom{k-p}{i}\left(-\left(\partial\phi\right)^2\right)^{i}\sum_{j=0}^{k-p-i}\frac{4^j\binom{k-p-i}{j}(n-2p)!}{(n-2p-2i-j)!}\nonumber\\{}&{}\delta^{\tau_1\cdots\,\,\tau_j\rho_1\sigma_1\cdots\rho_{k-p-i-j}\sigma_{k-p-i-j}}_{\lambda_1\cdots\lambda_j\alpha_1\beta_1\cdots\alpha_{k-p-i-j}\beta_{k-p-i-j}}\left(\prod_{q=1}^{k-p-i-j}R^{\alpha_q\beta_q}_{(D)\,\rho_q\sigma_q}\right)\prod_{q=1}^j\left(\phi^{\lambda_q}_{\tau_q}-\phi^{\lambda_q}\phi_{\tau_q}\right).\label{eq:rkdfinal}
\end{align}
Comparing this last expression with the expression (\ref{eq:skdfinal}) for $\mathcal{L}^{(k)}_{(d)}$, one immediately sees that
\begin{equation}
\mathcal{R}^{(k)}_{(\mathfrak{D})}=\mathrm{e}^{(2k-d)\phi}\sum_{p=0}^k\binom{k}{p}\widetilde{\mathcal{R}}_{(n)}^{(p)}\mathcal{L}^{(k-p)}_{(d)},\label{eq:relskpdfirst}
\end{equation}
provided the dimension $n$ of the internal space in the KK decomposition, and the specific conformal dimension $d$ of $\mathcal{L}^{(k-p)}_{(d)}$, are related such that
\begin{equation}
\frac{(n-2p)!}{(n-2p-2i-j)!}\stackrel{!}{=}\left(-1\right)^j\frac{(d-2k+2p+2i+j)!}{(d-2k+2p)!},\label{eq:point_exc}
\end{equation}
where $\stackrel{!}{=}$ means that this equality is not yet verified but is a requirement in order to get (\ref{eq:relskpdfirst}). Yet, the left hand side is
\begin{equation}
\frac{(n-2p)!}{(n-2p-2i-j)!} = \prod_{m=1}^{2i+j}(n+1-2p-m),
\end{equation}
while the right hand side is
\begin{align*}
\left(-1\right)^j\frac{(d-2k+2p+2i+j)!}{(d-2k+2p)!}={}&{}\left(-1\right)^j\prod_{m=1}^{2i+j}(d-2k+2p+m)\\ {}={}&{}\prod_{m=1}^{2i+j}(2k-d-2p-m).\label{eq:moins1j}
\end{align*}
Consequently, $2k-d=n+1$ in order for (\ref{eq:point_exc}), and thus (\ref{eq:relskpdfirst}), to be verified. Therefore, 
\begin{equation}
\mathcal{R}^{(k)}_{(\mathfrak{D})}=\mathrm{e}^{(n+1)\phi}\sum_{p=0}^k\binom{k}{p}\widetilde{\mathcal{R}}_{(n)}^{(p)}\mathcal{L}^{(k-p)}_{(2k-n-1)},\label{eq:final_result}
\end{equation}
which is exactly the first conjecture. \textbf{To cut a long story short, we have thus proved that, for a KK diagonal decomposition
\begin{equation}
g_{(\mathfrak{D})AB}\mathrm{d}x^A\mathrm{d}x^B=g_{(D)\mu\nu}\mathrm{d}x^\mu\mathrm{d}x^\nu+\mathrm{e}^{-2\phi\left(x^\mu\right)}\widetilde{g}_{(n)ab}\mathrm{d}x^a\mathrm{d}x^b,\label{eq:diag_fin_proof}
\end{equation}
the Lovelock invariant $\mathcal{R}^{(k)}_{(\mathfrak{D})}$ writes as in (\ref{eq:final_result}). The Lagrangians $\mathcal{L}^{(k-p)}_{(2k-n-1)}$ are scalar-tensor Lagrangians for the metric $g_{(D)\mu\nu}$ and the scalar field $\phi$. They are Lagrangians with conformal invariance in another dimension: the $(2k-n-1)$-dimensional scalar-tensor action $\int\mathrm{d}^{2k-n-1}x\sqrt{-g}\mathcal{L}^{(k-p)}_{(2k-n-1)}$ has local conformal invariance under $g_{\mu\nu}\to\mathrm{e}^{2\sigma}g_{\mu\nu}$, $\phi\to\phi-\sigma$, with $\sigma$ an arbitrary function on spacetime.}
\subsection{Considerations on the second conjecture (conjecture for the action integrals)}
The first conjecture has been demonstrated. Therefore, if the Lovelock invariants of the internal space are constant numbers, one has the following result regarding the KK reduction of the action functional,
\begin{equation}
\int\mathrm{d}^\mathfrak{D}x\sqrt{-g_{(\mathfrak{D})}}\mathcal{R}^{(k)}_{(\mathfrak{D})}=\mathcal{V}_{(n)}\int\mathrm{d}^Dx\sqrt{-g}\,\mathrm{e}^{\phi}\sum_{p=0}^k\binom{k}{p}\widetilde{\mathcal{R}}_{(n)}^{(p)}\mathcal{L}^{(k-p)}_{(2k-n-1)}.\label{eq:kkaction1}
\end{equation}
As a consequence, demonstrating the second conjecture (\ref{eq:conjecture_act}), namely
\begin{align}
\int\mathrm{d}^\mathfrak{D}x\sqrt{-g_{(\mathfrak{D})}}\mathcal{R}^{(k)}_{(\mathfrak{D})}\underset{\text{IBP}}{\stackrel{?}{=}}{}&{}\mathcal{V}_{(n)}\int\mathrm{d}^Dx\sqrt{-g}\sum_{p=0}^k\binom{k}{p}\widetilde{\mathcal{R}}_{(n)}^{(p)}\mathcal{L}^{(k-p)}_{(2k-n)},\label{eq:kkaction2}
\end{align}
amounts to demonstrating the lemma (\ref{eq:conjecture_lemma}) mentioned above,
\begin{equation}
\int\mathrm{d}^Dx\sqrt{-g}\,\mathrm{e}^\phi\mathcal{L}^{(k)}_{(d)}\underset{\text{IBP}}{\stackrel{?}{=}}\int\mathrm{d}^Dx\sqrt{-g}\mathcal{L}^{(k)}_{(d+1)}.\label{eq:kkaction3}
\end{equation}
\textbf{We were not able to prove this lemma, therefore, the form (\ref{eq:kkaction2}) of the KK reduced action remains a conjecture and has not been demonstrated}. One may wonder why we are interested in getting the form (\ref{eq:kkaction2}), whereas we already have the first form (\ref{eq:kkaction1}), which is truly demonstrated. The reason is the following. In the particular case of GB gravity, $k=2$, the form (\ref{eq:kkaction2}) is not a conjecture but a demonstrated result. As explained in Sec.~\ref{sec:fourd}, if $D=4$, the resulting four-dimensional scalar-tensor action (coupled to the pure Einstein-Hilbert term) leads to static BH solutions which do not have the usual asymptotic behaviour: the metric function behaves as $f\approx 1/(n+1)(1-2M/r^{n+1})$. BH solutions with proper Schwarzschild asymptotics are rather obtained by performing a singular limit where the dimension $n$ of the internal space is sent to zero, leading to the 4DEGB action that we have discussed extensively.\\

One is thus tempted to perform such a singular limit $n\to 0$ for arbitrary $k$. \textbf{Precisely, the form (\ref{eq:kkaction2}) would enable to perform easily such a singular limit. This is the reason why we are not satisfied with the properly demonstrated form (\ref{eq:kkaction1}), and we would prefer the not yet demonstrated form (\ref{eq:kkaction2})}. We recall that the result (\ref{eq:kkaction2}) has been demonstrated for $k=0,1,2$. To convince ourselves of the apparent relevance of this result, we have also demonstrated it for $k=3$, i.e. for the KK reduction of the cubic Lovelock invariant. Indeed, using (\ref{eq:r1}-\ref{eq:r3}), lengthy calculations lead to
\begin{align}
{}&{}\int\mathrm{d}^\mathfrak{D} x\sqrt{-g_{(\mathfrak{D})}}\mathcal{R}^{(3)}_{(\mathfrak{D})} \nonumber\\{}&{}\underset{\text{IBP}}{=}\mathcal{V}_{(n)}\int\mathrm{d}^Dx\sqrt{-g}\,\mathrm{e}^{-n\phi}\Biggl\{\widetilde{\mathcal{R}}^{(3)}_{(n)}\mathrm{e}^{6\phi}+3\widetilde{\mathcal{G}}_{(n)}\mathrm{e}^{4\phi}\Bigl[R+\left(n-4\right)\left(n-5\right)\left(\partial\phi\right)^2\Bigr]\nonumber\\{}&{}+3\widetilde{R}_{(n)}\mathrm{e}^{2\phi}\Bigl[\mathcal{G}-4\left(n-2\right)\left(n-3\right)G^{\mu\nu}\phi_\mu\phi_\nu+2\left(n-2\right)\left(n-3\right)\left(n-4\right)\Box\phi\left(\partial\phi\right)^2\nonumber\\{}&{}-\left(n-2\right)\left(n-3\right)^2\left(n-4\right)\left(\partial\phi\right)^4\Bigr]+\mathcal{R}^{(3)}-6n\left(n-1\right)\mathcal{H}^{(2)}_{\mu\nu}\phi^\mu\phi^\nu\nonumber\\{}&{}+8n\left(n-1\right)\left(n-2\right)\Bigl[3\left(\partial\phi\right)^2G^{\mu\nu}\phi_{\mu\nu}+\Bigl(\left(\Box\phi\right)^3-3\Box\phi\phi^{\mu\nu}\phi_{\mu\nu}+2\phi_{\mu\nu}\phi^{\nu\rho}\phi^\mu_\rho\Bigl)\Bigr]\nonumber\\{}&{}+3\left(n+1\right)n\left(n-1\right)\left(n-2\right)\Bigl[\left(\partial\phi\right)^4R-4\left(\partial\phi\right)^2\Bigl(\left(\Box\phi\right)^2-\phi_{\mu\nu}\phi^{\mu\nu}\Bigr)\Bigr]\nonumber\\{}&{}+6n\left(n\left(15-10n+n^3\right)-6\right)\Box\phi\left(\partial\phi\right)^4+24n\left(n-1\right)\left(n-2\right)P^{\mu\nu\rho\sigma}\phi_{\mu\rho}\phi_\nu\phi_\sigma\nonumber\\{}&{}-n\left(n-1\right)\left(n-2\right)\left(12+n\left(n\left(6+n\right)-13\right)\right)\left(\partial\phi\right)^6\Biggr\}.
\end{align}
One immediately sees the identity between the terms in $\mathrm{e}^{6\phi}$, $\mathrm{e}^{4\phi}$ and $\mathrm{e}^{2\phi}$, and respectively Eqs. (\ref{eq:l0}), (\ref{eq:l1}) and (\ref{eq:l2}), i.e. $\mathcal{L}^{(0)}_{(d)}$, $\mathcal{L}^{(1)}_{(d)}$ and $\mathcal{L}^{(2)}_{(d)}$, provided $d=6-n$. Regarding the remaining terms (starting with $\mathcal{R}^{(3)}$), one can check that they correspond up to integration by parts to Eq. (\ref{eq:l3}), i.e. $\mathcal{L}^{(3)}_{(d)}$, provided again $d=6-n$. The easiest way of verifying this equality up to integration by parts is to compute the field equations and remark that they are the same. Consequently, 
\begin{equation}
\int\mathrm{d}^\mathfrak{D} x\sqrt{-g_{(\mathfrak{D})}}\mathcal{R}^{(3)}_{(\mathfrak{D})}\underset{\text{IBP}}{=}\mathcal{V}_{(n)}\int\mathrm{d}^Dx\sqrt{-g}\sum_{p=0}^3\binom{3}{p}\widetilde{\mathcal{R}}_{(n)}^{(p)}\mathcal{L}^{(3-p)}_{(6-n)},
\end{equation}
or in other words, the second conjecture (\ref{eq:kkaction2}) is also true for $k=3$. 
\\

\textbf{Nothing proves formally that the second conjecture (\ref{eq:kkaction2}) remains true for $k\geq 4$. Again, the only remaining step is to prove the lemma (\ref{eq:kkaction3}) for arbitrary $k$. Nevertheless, the author thinks it probable for (\ref{eq:kkaction2}) to be indeed true, and thinks that, even in the current absence of a proof, it remains worth examining the corollaries of this conjectured result}. As mentioned previously, if (\ref{eq:kkaction2}) were true, one could easily perform the singular limit $n\to 0$, as was done for the 4DEGB gravity. The proper way to do it is as we have always done the previous regularizations: multiply the action by a coupling constant $\alpha$, set a new coupling $\bar{\alpha}=\alpha\,n$ and make $n\to 0$ while maintaining $\bar{\alpha}$ constant. Since this procedure is by now familiar, let us briefly describe the result. For $p>0$, the Lovelock invariant of order $p$ of a space of zero dimension is clearly vanishing, so there exists a Lorentz scalar $\widetilde{\mathcal{R}}_{(0),\,\text{reg}}^{(p)}$ such that 
\begin{equation}
\widetilde{\mathcal{R}}_{(n)}^{(p)}=n\widetilde{\mathcal{R}}_{(0),\,\text{reg}}^{(p)}+\mathcal{O}\left(n^2\right).
\end{equation}  
So for $p>0$, the regularization $n\to 0$ of the term $\widetilde{\mathcal{R}}_{(n)}^{(p)}\mathcal{L}^{(k-p)}_{(2k-n)}$ appearing in (\ref{eq:kkaction2}) simply gives 
\begin{equation}
\widetilde{\mathcal{R}}_{(0),\,\text{reg}}^{(p)}\mathcal{L}^{(k-p)}_{(2k)}.
\end{equation}
On the other hand, for $p=0$, $\widetilde{\mathcal{R}}_{(n)}^{(0)}=1$, and the term to be regularized is $\mathcal{L}^{(k)}_{(2k-n)}$. Fortunately, we have already studied in paragraph~\ref{subsec:reg} the regularization of the Lagrangian $\mathcal{L}^{(k)}_{(d)}$ to a Lagrangian $\mathcal{L}^{(k)}_{(2k),\,\text{reg}}$ for $d\to 2k$. This is equivalent to sending $n\to 0$ in $\mathcal{L}^{(k)}_{(2k-n)}$. As a result, the $n\to 0$ regularization of the KK conjectured action (\ref{eq:kkaction2}) of order $k$ (i.e. coming from the Lovelock invariant of order $k$) gives the scalar-tensor action
\begin{equation}
S^{(k)}_{\text{KK},\,\text{reg}}\left[g_{\mu\nu},\phi\right]=\int\mathrm{d}^Dx \sqrt{-g}\Bigl[\mathcal{L}^{(k)}_{(2k),\,\text{reg}}+\sum_{p=1}^k\binom{k}{p}\widetilde{\mathcal{R}}_{(0),\,\text{reg}}^{(p)}\mathcal{L}^{(k-p)}_{(2k)}\Bigr],\label{eq:kkregultimate}
\end{equation}
where $\mathcal{L}^{(k)}_{(2k),\,\text{reg}}$ is defined in (\ref{eq:sreglktilde}). To be rigorous, the proof of existence of $\mathcal{L}^{(k)}_{(2k),\,\text{reg}}$ consisted in a regularization of $\int\mathrm{d}^Dx\sqrt{-g}\beta_k\mathcal{L}^{(k)}_{(D)}$ in the limit $D\to 2k$, and this proof can indeed be repeated identically but for the step where the term $\beta_k\mathcal{R}^{(k)}$ in Eq. (\ref{eq:betaklklast}) is dropped from the action with the reason that '$\mathcal{R}^{(k)}$ is a boundary term in $D=2k$ dimensions'. Here, in the KK action before regularization (\ref{eq:kkaction2}), this step remains valid only if $\mathcal{R}^{(k)}$ is a boundary term or vanishes in dimension $D$, that is, only if $D\leq 2k$, so the regularized action (\ref{eq:kkregultimate}) indeed corresponds to a KK regularization procedure only for $D\leq 2k$. This does not put any serious restriction: the aim of the KK procedure is to make sense in lower dimensions of an action which seems \textit{a priori} trivial in such lower dimensions, but for $D>2k$, the Lovelock action of order $k$ is not trivial, so there is no reason of making a KK of the $k$-th Lovelock invariant if the dimension is not $\leq 2k$. 
\\

\textbf{The interpretation of (\ref{eq:kkregultimate}) as the KK regularized action of the Lovelock invariant of order $k$ will of course become true only when a proof of the conjectured lemma (\ref{eq:kkaction3}) is performed. For the moment, only the cases up to $k= 3$ are fully treated. Still, this is enough to get interesting results in four dimensions, as we now describe}.
\subsection{Kaluza-Klein reduction of cubic Lovelock as a Horndeski theory}\label{subsec:lastconj}
Indeed, the case of the cubic Lovelock invariant, $k=3$, that we have explicitly demonstrated above in order to convince us of the relevance of the conjecture, can now be explored in detail. For simplicity, let us consider the KK reduction along a flat internal space. Its Lovelock invariants are all identically vanishing. Therefore, in (\ref{eq:kkregultimate}), all terms in the sum $\sum_{p=1}^3$ vanish, and the corresponding regularized KK action reduces to
\begin{align}
S^{(3)}_{\text{KK},\,\text{reg}} ={}&{} \int\mathrm{d}^Dx\sqrt{-g}\mathcal{L}^{(3)}_{(6),\,\text{reg}}\nonumber\\
{}={}&{}\int \mathrm{d}^Dx\sqrt{-g}\Bigl\{\phi\mathcal{R}^{(3)}-6 \mathcal{H}^{(2)}_{\mu\nu}\phi^\mu\phi^\nu-16 P^{\mu\rho\nu\sigma}\phi_\mu\phi_\nu\phi_{\rho\sigma}+36\left(\partial\phi\right)^4\Box\phi\nonumber\\{}&{}\qquad\qquad- 6\Bigl[R\left(\partial\phi\right)^4-4\left(\partial\phi\right)^2\left[\left(\Box\phi\right)^2-\phi_{\mu\nu}\phi^{\mu\nu}\right]\Bigr]+24\left(\partial\phi\right)^6\Bigr\}.
\end{align}
The second line comes from the previously computed expression for $\mathcal{L}^{(3)}_{(6),\,\text{reg}}$, see (\ref{eq:l3_reg}). Importantly, at that time, this was computed only as a formal regularization in dimension $D=6$, but we have just shown that this is precisely the same Lagrangian which arises when performing a KK reduction of the cubic Lovelock invariant down to any dimension $D\leq 6$. In particular, in the most usual case of a four-dimensional spacetime, $D=4$, the cubic Lovelock invariant, $\mathcal{R}^{(3)}$, and the variation of the GB scalar, $\mathcal{H}^{(2)}_{\mu\nu}$, both vanish identically. Moreover, one may add a usual Einstein-Hilbert action $S_{\text{EH}}$ to the KK action, in order to get a well-defined GR limit. The full four-dimensional action under consideration then reads
\begin{align}
S={}&{}S_{\text{EH}}-\frac{\alpha_1}{24} S^{(3)}_{\text{KK},\,\text{reg}}\nonumber\\
{}={}&{}\int \mathrm{d}^4x\sqrt{-g}\Bigl\{R-\frac{\alpha_1}{24}\Bigl(-16 P^{\mu\rho\nu\sigma}\phi_\mu\phi_\nu\phi_{\rho\sigma}+36\left(\partial\phi\right)^4\Box\phi\nonumber\\{}&{}- 6\Bigl[R\left(\partial\phi\right)^4-4\left(\partial\phi\right)^2\left[\left(\Box\phi\right)^2-\phi_{\mu\nu}\phi^{\mu\nu}\right]\Bigr]+24\left(\partial\phi\right)^6\Bigr)\Bigr\},\label{eq:actionpresent}
\end{align}
where $\alpha_1$ is a coupling constant, and the factor $-1/24$ has been chosen so as to get simple expressions below, Eq. (\ref{eq:cubic_lov_horn}). The Paul term $P^{\mu\rho\nu\sigma}\phi_\mu\phi_\nu\phi_{\rho\sigma}$ is a Horndeski term $G_5= 3X$, see Eq. (\ref{eq:paultermnari}), and it has been known since~\cite{VanAcoleyen:2011mj} that it emerges from KK reduction of the cubic Lovelock invariant, see also~\cite{Charmousis:2011bf,Charmousis:2011ea}. Action (\ref{eq:actionpresent}) is a scalar-tensor action belonging to the shift-symmetric Horndeski class, with Horndeski functionals
\begin{equation}
G_2 = 8\alpha_1 X^3,\quad G_3 =6\alpha_1 X^2,\quad G_4 = 1+\alpha_1 X^2,\quad G_5=2\alpha_1 X,\label{eq:cubic_lov_horn}
\end{equation}
where $X=-\phi^\mu\,\phi_\mu/2$ is the kinetic term. Remarkably, this theory has already been encountered in this report, from a very different approach. Indeed, section~\ref{sec:nonpar} presented sufficient compatibility conditions ensuring the solving of (beyond) Horndeski field equations with shift symmetry but no parity symmetry [i.e. including functions $G_3(X)$ and $G_5(X)$], for a spherically-symmetric and static BH ansatz. For clarity, let us briefly recall the results, Eq (\ref{eq:g5comp}) and below. The ansatz reads
\begin{equation}
\mathrm{d}s^2=-f\mathrm{d}t^2+\mathrm{d}r^2/f+r^2\mathrm{d}\Omega^2,\quad \phi=\phi(r).\label{eq:ansatz_cub_lov}
\end{equation}
If the Horndeski functionals read
\begin{align}
G_{2}={}&{}4\alpha n(2n-1)\frac{(-2X)^{n+1}}{n+1},\quad G_{3}=-4\alpha (2n-1)(-2X)^{n},\nonumber\\ G_4 ={}&{} 1-2\alpha(-2X)^n,\quad G_{5}=4\alpha n\frac{(-2X)^{n-1}}{n-1},\label{eq:g5compbis}
\end{align}
then there exists a BH solution, given as the root of an algebraic equation (with $M$ the ADM mass),
\begin{align}
    {}&{} 0=(n+1)\left(2n-1\right)^n r^{2n-1}\left[(2n-1)(2M-r)+rF^2\right]-2\alpha\left(1-F\right)^{2n}\left(1+2nF+F^2\right),\nonumber\\ {}&{}F(r)^2\equiv (2n-1)f(r),\label{eq:alg_not_good}
\end{align}
supported by a scalar field
\begin{equation}
\phi(r) = \int\frac{1-\sqrt{(2n-1)f(r)}}{r\sqrt{(2n-1)f(r)}}\mathrm{d}r.\label{eq:scalar_cub_lov}
\end{equation}
The KK of cubic Lovelock, Eq. (\ref{eq:cubic_lov_horn}), is seen to match the case $n=2$ of (\ref{eq:g5compbis}), with $\alpha=-\alpha_1/8$. Note that, in section~\ref{sec:nonpar}, the general case (for arbitrary $n$) was studied, and the best we could obtain was the algebraic equation (\ref{eq:alg_not_good}), which bears on $F\propto \sqrt{f}$ and hence is meaningful only for $f(r)\geq 0$. However, if we focus specifically on the present case corresponding to the KK of cubic Lovelock, $n=2$, it becomes possible to find a polynomial equation for $f$ itself. Indeed, the integration of the field equations leads to
\begin{align}
0 ={}&{} 729 \alpha_1 ^2f^6 -4374 \alpha_1 ^2f^5 +\left(5103 \alpha_1 ^2+17496 \alpha_1  r^4\right)f^4 \nonumber\\{}&{}+\left(-2484 \alpha_1 ^2-69984 \alpha_1  r^4+34992 \alpha_1  M  r^3\right)f^3 \nonumber\\{}&{}+ \left(567 \alpha_1 ^2+104976 r^8+34992 \alpha_1  r^4-104976 \alpha_1  M  r^3\right)f^2\nonumber\\{}&{}+ \left(-54 \alpha_1 ^2-209952 r^8+419904 M  r^7+18144 \alpha_1  r^4-34992 \alpha_1  M  r^3\right)f\nonumber\\{}&{}+\alpha_1 ^2+104976 r^8-419904 M  r^7+419904 M ^2 r^6-648 \alpha_1  r^4+1296 \alpha_1  M  r^3.\label{eq:polykklov3}
\end{align}
This polynomial equation of order 6 is this time valid for any sign of $f(r)$. There is no contradiction: it is equivalent to (\ref{eq:alg_not_good}), with $n=2$, when $f(r)\geq 0$. The only difference is that (\ref{eq:alg_not_good}) was obtained when integrating the field equations of (\ref{eq:g5compbis}) for arbitrary $n$, whereas the better result (\ref{eq:polykklov3}) comes from the easier task of integrating the field equations of only the particular case $n=2$. 
\\

\textbf{In a word, the KK of cubic Lovelock along a flat internal space corresponds to a shift-symmetric Horndeski theory (\ref{eq:cubic_lov_horn}), which admits a static, spherically-symmetric BH solution (\ref{eq:ansatz_cub_lov}), with metric function $f(r)$ given implicitly by the polynomial equation (\ref{eq:polykklov3}), and scalar field (\ref{eq:scalar_cub_lov}) with $n=2$, i.e.
\begin{equation}
\phi(r)=\int\frac{1-\sqrt{3f(r)}}{r\sqrt{3f(r)}}\mathrm{d}r.\label{eq:phikklov3finalfinal}
\end{equation}}The scalar field $\phi$ is seen to be well-defined outside and at the horizon, while it becomes imaginary inside. The kinetic term itself becomes imaginary below the horizon,
\begin{equation}
X = -\frac{\left(1-\sqrt{3f}\right)^2}{6r^2}.
\end{equation}
On the other hand, the polynomial equation (\ref{eq:polykklov3}) implies the following asymptotic behaviour when $r\to\infty$,
\begin{equation}
f(r)=1-\frac{2 M }{r}+\mathcal{O}\left(\frac{1}{r^4}\right),
\end{equation}
which is like Schwarzschild at leading order.\\

The KK reduction of cubic Lovelock corresponds to the case $n=2$ of the compatible shift-symmetric Horndeski theories found in Sec.~\ref{sec:nonpar}. We have also seen in Sec.~\ref{sec:nonpar} that the shift-symmetric 4DEGB theory, that is, the KK reduction of quadratic Lovelock along a flat internal space, corresponds to the case $n=1$. It is therefore tempting to assume the following: \textbf{the KK reduction, along a flat internal space, of the Lovelock action of order $k$, corresponds to the Horndeski theory (\ref{eq:g5compbis}) of section~\ref{sec:nonpar} with $n=k-1$.} However, since the form of the regularized KK action (\ref{eq:kkregultimate}) itself is already a conjecture for $k\geq 4$, we will not elaborate further on this point. Before moving on to the conclusions of this Chap.~\ref{chap:5}, we refer the interested reader to the article~\cite{Alkac:2022fuc}, where an attempt to regularize cubic Lovelock gravity in four dimensions from a different approach is performed. However, the BH solution (\ref{eq:polykklov3}-\ref{eq:phikklov3finalfinal}) described above is not obtained in this latter article.
\\

The results gathered in this Chap.~\ref{chap:5} are all related by their common link to Lovelock invariants and conformal invariance, but remain quite diverse. Sec.~\ref{sec:selecting} has somehow generalized the 4DEGB action (\ref{eq:fern}): starting from an action (\ref{eq:action}) with a similar form and imposing the existence of a logarithmic scalar field, one can find compatibility conditions bearing on the generalized potentials appearing in the action. This selects an action (\ref{eq:complete}) which admits closed-form BH solutions (\ref{eq:sol_1}) and (\ref{eq:sol_2}) . This new action displays terms of the 4DEGB action, which has generalized conformal invariance, but also terms which would have conformal invariance in five dimensions but do not seem to have any symmetry in four dimensions. The fact that such scalar-tensor terms with no symmetry allow for closed-form solutions remains unprecedented. Beyond the purely mathematical arguments, the appearance of such terms remains mysterious from a physical point of view. 
\\

Then, Sec.~\ref{sec:higherdim} studies conformal couplings with more general Lovelock invariants than the GB scalar. This requires to study higher-dimensional scalar-tensor actions. A novel way of understanding the 4DEGB action, from a $D\to 4$ regularization of higher-dimensional conformally-invariant Lagrangians, is presented. This procedure is extended to arbitrary dimensions, and the static BH solutions of the corresponding theories are found. 
\\

Our interest however lies in four-dimensional results, and it turns out that the higher-dimensional considerations of Sec.~\ref{sec:higherdim} greatly simplify the understanding of Sec.~\ref{sec:diagkk}, which performs a diagonal KK reduction of the Lovelock invariant of order $k$, with arbitrary $k$. This computation is motivated by the fact that, for $k=2$ (GB invariant), the KK reduction led to the 4DEGB theory extensively described in this thesis. Remarkably, in the course of the KK compactification down to $D$ dimensions, appear the Lagrangians $\mathcal{L}^{(k)}_{(d)}$ which are the Lagrangians of Sec.~\ref{sec:higherdim} with conformal invariance in dimension $d$, but without such invariance in dimension $D$. The KK decomposition of the Lovelock \textbf{Lagrangian} of order $k$ is conjectured then demonstrated, leading to the result (\ref{eq:final_result}). As regards the KK of the Lovelock \textbf{action}, only a conjecture (\ref{eq:kkaction2}) can be formulated, but not demonstrated except for low orders $k=0,1,2,3$. 
\\

The case $k=3$ (KK of cubic Lovelock) still shows the interesting consequences of such compactification down to four dimensions, leading to exact BH solutions. Surprisingly, the KK reduction of both GB and cubic Lovelock are seen to fit into the integration of Horndeski theories with no parity symmetry presented from a completely different point of view in Sec.~\ref{sec:nonpar}. 
\\

In a word, Chap.~\ref{chap:5} was able to obtain BH solutions in theories which possess a direct link with Lovelock invariants and with conformal invariance in other spacetime dimensions. The approach used in this chapter thus focused on finding relevant theories, potentially motivated from higher-dimensional considerations. This appears to be quite different from Chap.~\ref{chap:4}, where we undertook a systematic study of shift-symmetric field equations, enabling to find theories allowing closed-form solutions, but paying little attention to the meaning or interpretation of the obtained scalar-tensor theories.
\\

This completes the presentation of the new static and spherically-symmetric BH solutions obtained during this thesis. The next and last chapter will indeed construct distinct types of solutions in scalar-tensor theories, namely a wormhole, Sec.~\ref{sec:wormhole} and a BH which is not stationary since it is embedded in an FLRW spacetime, Sec.~\ref{sec:kerrflrw}. Both these new solutions are obtained by using the method of generation of solutions through conformal-disformal transformations described in paragraph~\ref{subsec:intconfdisf}.

\chapter[Generation of solutions with conformal and disformal transformations]{Generation of solutions with conformal and disformal transformations}\label{chap:6}
\chaptermark{Chap.~6~~Generation of solutions with conformal-disformal transformations}
Paragraph~\ref{subsec:intconfdisf} describes the conformal-disformal transformations of a scalar-tensor theory,
\begin{equation}
g_{\mu\nu}\to \tilde{g}_{\mu\nu}=C(\phi,X)g_{\mu\nu}+D(\phi,X)\phi_\mu\phi_\nu,\quad \phi\to\tilde{\phi}=\phi.\label{eq:disf_final}
\end{equation}
Under such a transformation, an initial DHOST action $S\left[g_{\mu\nu},\phi\right]$ is mapped to another DHOST action $\tilde{S}\left[\tilde{g}_{\mu\nu},\phi\right]$,
\begin{equation}
\tilde{S}\left[\tilde{g}_{\mu\nu},\phi\right]\equiv S\left[g_{\mu\nu},\phi\right].
\end{equation}
If $\left(g_{\mu\nu},\phi\right)$ is a solution to the variational principle of action $S$, and the conformal-disformal transformation is invertible, then $(\tilde{g}_{\mu\nu},\phi)$ is a solution to the variational principle of action $\tilde{S}$. This \textbf{generation of solutions}~\cite{BenAchour:2020wiw} enables to get new solutions almost for free. For instance, in paragraph~\ref{subsec:disf_kerr}, we recalled how a disformal transformation of a seed Kerr metric leads to the so-called disformal Kerr metric, which is a rotating BH with very distinct features as compared to the usual Kerr BH.
\\

The present chapter uses the generation of solutions in two different contexts, in order to obtain new exact solutions in scalar-tensor theories. The first section describes the pure disformal transformation, i.e. $C=0$ in (\ref{eq:disf_final}), of a seed static, spherically-symmetric BH. The obtained transformed solution is a \textbf{wormhole}. The second section studies the pure conformal  transformation, i.e. $D=0$ in (\ref{eq:disf_final}), of a seed stealth Kerr spacetime. The obtained transformed solution is a \textbf{non-stationary spacetime which combines the features of both a Kerr BH and a cosmological FLRW spacetime}.
\section[Generation of a wormhole spacetime by disformal transformation]{Generation of a wormhole spacetime by disformal transformation}\label{sec:wormhole}
\sectionmark{6.1~~Generation of a wormhole spacetime by disformal transformation}
The existence of wormholes in the literature dates back to the works of Flamm in 1916~\cite{cite-key}, made more popular by Einstein and Rosen with their Einstein-Rosen bridge~\cite{Einstein:1935tc}. However, it was only in 1988 that Morris and Thorne proposed 'a new class of solutions of the Einstein field equations [...] which describe wormholes that, in principle, could be traversed by human beings'~\cite{Morris:1988cz}. However, wormholes are not vacuum solutions in GR: they must be supported by matter fields whose energy-momentum tensor violates the standard energy conditions~\cite{Visser:1995cc} (for a review on energy conditions, see e.g.~\cite{Kontou:2020bta}). 
\\

However, in modified theories of gravity, it becomes possible to construct wormholes supported by matter satisfying the energy conditions, or even as vacuum solutions of the modified field equations. This task was for instance performed in $f(R)$ gravity~\cite{Lobo:2009ip}, in Brans-Dicke theory~\cite{Agnese:1995kd,Anchordoqui:1996jh} or in higher-order scalar-tensor theories~\cite{Antoniou:2019awm,Bakopoulos:2021liw}. We refer the interested reader to the review~\cite{Alcubierre:2017pqm} for more examples, or to the recent article~\cite{Bakopoulos:2023tso} which in addition constructs a stealth Ellis wormhole~\cite{Ellis:1973yv} in Horndeski theory. 
\\

In the present section, we start from a seed scalar-tensor BH solution and perform a disformal transformation, which maps it to a wormhole spacetime. Importantly, the \textbf{throat} of the obtained wormhole is not a fixed parameter of the theory: the obtained solutions are as usual parameterized by their mass $M$, which is a free integration constant, and the throat of the wormhole is a function of the mass. The results of this section are coming from the article~\cite{Babichev:2022awg} written by the present author, among others. 
\subsection{Disformal transformation for static and spherical symmetry}
Consider a seed scalar-tensor BH solution, assumed to be static, spherically-symmetric, and dressed with a static scalar field:
\begin{equation}
\mathrm{d}s^2=-f(r)\mathrm{d}t^2+\mathrm{d}r^2/f(r)+r^2\mathrm{d}\Omega^2,\quad \phi=\phi(r).\label{eq:ansatz_last_chapter}
\end{equation}
The equality $g_{tt}=-g^{rr}$ is also assumed. Then, by direct computation, the disformed metric
\begin{equation}
\tilde{g}_{\mu\nu}=g_{\mu\nu}+D(\phi,X)\phi_\mu\phi_\nu
\end{equation}
has the following line element~\cite{Bakopoulos:2021liw},
\begin{equation}
\mathrm{d}\tilde{s}^2=-f(r)\,\mathrm{d}t^2+\frac{\mathrm{d}r^2}{f(r)\,
W^{-1}\left(\phi,X\right)}+r^2\mathrm{d}\Omega^2,\label{eq:dstilde}
\end{equation}
where
\begin{equation}
W\left(\phi,X\right)\equiv 1-2D\left(\phi,X\right)X.
\end{equation}
\subsection{The seed black hole solution}\label{sec:seedbhsol}
\begin{figure}
\begin{subfigure}{7.5cm}
\includegraphics[width=\linewidth]{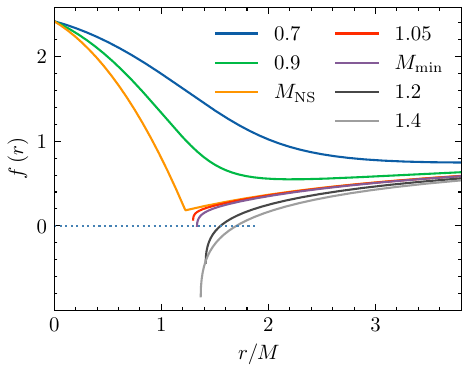}
\end{subfigure}
\begin{subfigure}{7.9cm}
\includegraphics[width=\linewidth]{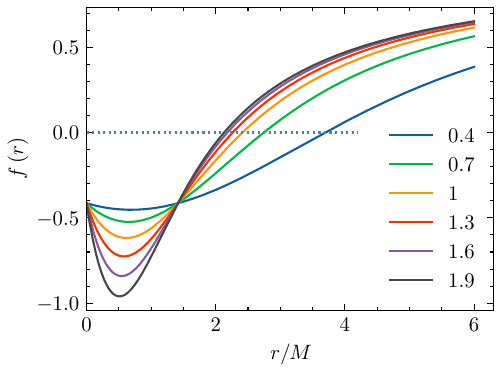}
\end{subfigure}
\caption{Metric function $f(r)$ for different values of $M/\sqrt{\left\lvert\alpha\right\rvert}$ for negative $\alpha$ (left plot) and positive $\alpha$ (right plot).
On the left panel, for $M\leq M_\text{NS}$, the upper curves correspond to the spacetime with a naked singularity at $r=0$.
For $M_\text{NS}<M< M_\text{min}$, the spacetime has a naked singularity at $r=r_P$, while for $M\geq M_{\text{min}}$ the metric describes a BH.
On the right panel, the spacetime admits a singularity at $r=0$, always covered by the horizon.}
\label{fig:f(r)fern}
\end{figure}

The seed solution is taken to be one of the solutions of the 4DEGB theory, namely (\ref{eq:prod_sphere_sol},\ref{eq:log_scal}), which corresponds to a regularized KK reduction along an internal space which is a product of two-spheres. For clarity, the initial action functional is 
\begin{align}
S =\int\mathrm{d}^4x\sqrt{-g}{}&{}\Bigl\{R-\frac{\beta^2}{2\alpha}\mathrm{e}^{4\phi}-\beta\mathrm{e}^{2\phi}\Bigl[R+6\left(\partial\phi\right)^2\Bigr]\nonumber\\ {}&{}+\alpha\Bigl[-\phi\mathcal{G}+4G^{\mu\nu}\phi_\mu\phi_\nu+4\Box\phi\left(\partial\phi\right)^2+2\left(\partial\phi\right)^4\Bigr]\Bigr\},\label{eq:action_4degb_final}
\end{align}
and the scalar-tensor BH solution is of the form (\ref{eq:ansatz_last_chapter}) with
\begin{equation}
f\left(r\right) = 1+\frac{r^2}{2\alpha}\left(1-\sqrt{1+\frac{8\alpha M}{r^3}+\frac{8\alpha^2}{r^4}}\right),\quad \phi\left(r\right) = \ln\left(\frac{\sqrt{-2\alpha/\beta}}{r}\right).\label{eq:solutionA}
\end{equation}
The coupling constants $\alpha$ and $\beta$ must have different signs for $\phi$ to be well-defined. This BH displays very distinct properties depending on the sign of the coupling constant $\alpha$. For $\alpha<0$ (and hence $\beta>0$),
the standard kinetic term has the usual sign in the
action\footnote{This can be seen from the scalar field redefinition
$\Phi=\mathrm{exp}\left(\phi\right)$.}.
For convenience, we rewrite the spacetime (\ref{eq:solutionA}) for the choice $\alpha<0$ as follows,
\begin{equation}
f\left(r\right) =
1-\frac{r^2}{2\left\vert\alpha\right\vert}+\frac{\sqrt{P\left(r\right)}}{2\left\vert\alpha\right\vert},\quad
P\left(r\right)\equiv r^4-8\left\vert\alpha\right\vert Mr + 8
\left\vert\alpha\right\vert^2,\label{eq:fF}
\end{equation}
and we define the radius $r=r_P$ and the values $M_\text{NS}$ and $M_\text{min}$ by
\begin{equation}
P\left(r_P\right)\equiv 0,\qquad
\frac{\left\vert\alpha\right\vert}{M_\text{NS}^2}\equiv
\frac{3}{4}\sqrt{\frac{3}{2}},\qquad
\frac{\left\vert\alpha\right\vert}{M_{\text{min}}^2}\equiv
\frac{8}{9}.
\label{M0}
\end{equation}
It is easy to see that, for $0\leq M\leq M_\text{NS}$, the spacetime
admits a naked singularity at $r=0$, while if $M_\text{NS}<M<
M_\text{min}$, the naked singularity is brought forward to $r=r_P$.
Only for larger masses $M\geq M_\text{min}$ does the spacetime describe a BH, with a single event horizon at
$r_+=M+\sqrt{M^2-\left\vert\alpha\right\vert}$, covering the
singularity at $r=r_P$. Note that, for
$\alpha<0$, the event horizon has a smaller
size as compared to the standard Schwarzschild radius $r_\text{Sch}=2M$. In particular the minimal horizon size is $r_{+\text{, min}}=\sqrt{2\vert\alpha\vert}=4M_\text{min}/3$. The behavior of the metric function is illustrated in
Fig.~\ref{fig:f(r)fern} (left panel), where $f\left(r\right)$ is
shown for different $M/\sqrt{\left\lvert\alpha\right\rvert}$.\\

The case $\alpha>0$ is more straightforward to analyze:
independently of the value of $\alpha$, the solution (\ref{eq:solutionA}) describes a BH for any mass $M$, and with a unique horizon $r_+ = M+\sqrt{M^2+\alpha}$, covering the
singularity $r=0$. The horizon is now at $r_+>r_\text{Sch}=2M$. The behavior of the function $f(r)$ is illustrated
in Fig.~\ref{fig:f(r)fern}, right panel.
\\

To conclude the discussion, we would like to mention, in the spirit
of~\cite{Charmousis:2021npl}, that if a Birkhoff-like uniqueness
theorem~\cite{Israel:1967wq} were valid for the
solution (\ref{eq:solutionA}), it would
inevitably lead to the constraint $\alpha<0$. Indeed, if the solution (\ref{eq:solutionA}) were unique, any static and
spherically symmetric object of mass $M$ would create an exterior
gravitational field given by (\ref{eq:solutionA}). If $\alpha>0$, this object would therefore be a BH with horizon $r_+=M+\sqrt{M^2+\alpha}$, unless this event horizon is hidden below the surface of the object. An atomic nucleus has radius $R\sim
10^{-15}\,\mbox{m}$, and is not a BH since it can be experimentally probed, therefore $r_+<R$. This in turn implies
\begin{equation}
0<\alpha < R\left(R-2M\right) \sim 10^{-30}\,\mbox{m}^2,
\label{eq:constraint}
\end{equation}
essentially rendering $\alpha>0$ irrelevant. 
\subsection{Wormhole construction}
\begin{figure}
\begin{subfigure}{0.51\textwidth}
\includegraphics[width=\linewidth]{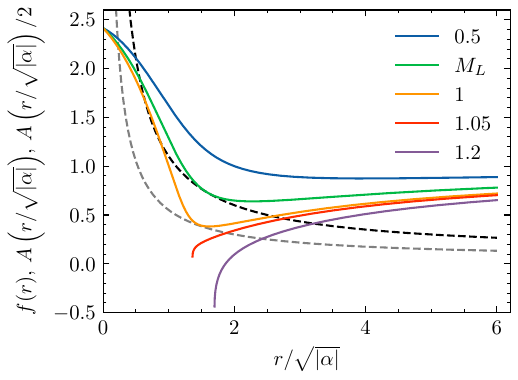}
\end{subfigure}
\begin{subfigure}{0.48\textwidth}
\includegraphics[width=\linewidth]{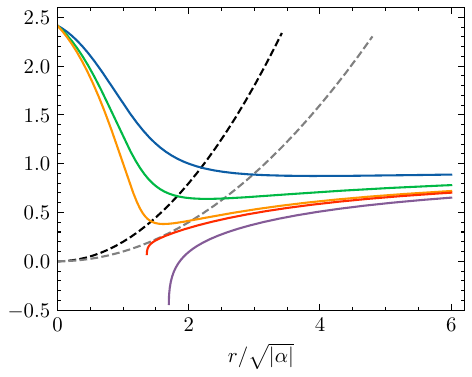}
\end{subfigure}
\caption{The functions $A$ (black curve) and $A/2$ (grey curve) are shown as functions of $r/\sqrt{|\alpha|}$ for two different cases: (\ref{eq:ourA}) with $a=0.1$ (left plot), and
$A\left(r/\sqrt{\left\lvert\alpha\right\rvert}\right)=r^2/\left(5\left\lvert\alpha\right\rvert\right)$ (right plot); while the metric function $f$ is shown for several values of $M/\sqrt{\left\lvert\alpha\right\rvert}$, in color. The throat radius $r_0$ (the singular radius $r_*$) is the largest intersection of $f$ with the black (grey) curve. On the left plot, $r_*$ is covered by the wormhole throat and the conditions for the disformal transformation formulated in the text are satisfied. This is not the case for the right plot. The meaning of $M_L/\sqrt{\left\lvert\alpha\right\rvert}\approx 0.8213$ will be made clear later in the text.}
\label{fig:Aperfect}
\end{figure}

From now on, we thus restrict to the case $\alpha<0$, which, as seen above, presents naked singularities, at a radius $r=r_S\in\lbrace 0,r_P\rbrace$, for small masses, and a horizon, at a radius $r=r_+$, for sufficiently large masses. The present paragraph shows \textbf{how the singularities can be removed by a disformal transformation, which maps the seed BH into a regular wormhole metric}.
Given the solution (\ref{eq:solutionA}) for the scalar field and to simplify expressions, we redefine the scalar field as
\begin{eqnarray}
\psi=\sqrt{-\frac{2\alpha}{\beta}} \mathrm{e}^{-\phi}\Longrightarrow
\psi\underset{\text{on shell}}{=}r, \label{defpsi}
\end{eqnarray}
with $\psi$ of  dimension 1.
We look for such $W\left(\psi,X\right)=1-2D(\psi,X)X$ that the disformed metric (\ref{eq:dstilde}) describes a wormhole geometry. We have to impose three requirements on $W\left(\psi,X\right)$:
\begin{enumerate}
\item We require that $W^{-1}$ vanishes at a point $r=r_0$ such that $r_0>r_S$ and $r_0>r_+$, if the spacetime admits a naked singularity $r=r_S$ or an event horizon $r=r_+$. This ensures that $r=r_0$ corresponds to the wormhole throat, since
$\tilde{g}^{rr}\left(r_0\right)=0$ while
$\tilde{g}_{tt}\left(r\right)>0$ for any $r\geq r_0$, see e.g.~\cite{Morris:1988cz}.
\item The asymptotic flatness and the absence of solid deficit angle of the disformed metric is obtained by imposing that $W\to 1$ as $r\to\infty$.
\item The disformal transformation should be invertible, which implies that the determinant
of the Jacobian of the metric transformation (\ref{eq:disf_final}) is not zero or infinity. As explained in paragraph~\ref{subsec:ineq}, the transformation is invertible if
\begin{equation}
1-2DX\not \in\left\lbrace 0,\infty\right\rbrace,\quad 1+2X^2D_X\text{ keeps a constant sign, }>0 \text{ or }<0.
\end{equation}
More precisely, $1-2DX$ can vanish or become infinite, but only on a subset of spacetime of vanishing measure. Typically, the wormhole throat $r=r_0$ corresponds to $\infty=W=1-2DX$ according to the first requirement, but this happens at a unique point of spacetime, so is harmless. The invertibility requirement is not
manifest in the solution itself, but is essential for the wormhole to be solution of a well-defined variational principle, as will be made clear later.
\end{enumerate}
To satisfy these requirements, we choose $W\left(\psi,X\right)$ to have the relatively simple form,
\begin{eqnarray}
W^{-1}\left(\psi,X\right)=\left(1-1/a\right)^{-1}\left(1+\frac{2\psi^2
X}{A\left(\psi/\sqrt{\left\lvert\alpha\right\rvert}\right)}\right),
\label{eq:condjacx}
\end{eqnarray}
where $a\in(0,1)$, while $A$ is a non-negative function. For our seed BH solution (\ref{eq:solutionA}), the kinetic term of the scalar field is $X=-f(r)/\left(2r^2\right)$, so
\begin{equation}
W^{-1}\left(\psi,X\right)\underset{\text{on shell}}{=}\left(1-1/a\right)^{-1}\left(1-\frac{f(r)}{A\left(r/\sqrt{\left\lvert\alpha\right\rvert}\right)}\right).
\end{equation}
Consequently, and taking into account that $f(r)\to 1$ when $r\to\infty$, one must impose $A\left(r\to\infty\right)= a$  in order for condition 2 (asymptotic flatness) to be fulfilled. On the other hand, the throat $r=r_0$ of the wormhole (condition 1) is obtained by solving $W^{-1}=0$. This corresponds to the intersection of $f\left(r\right)$ with $A\left(r/\sqrt{\left\lvert\alpha\right\rvert}\right)$:
\begin{eqnarray}
f(r_0)=A\left(\frac{r_0}{\sqrt{\left\lvert\alpha\right\rvert}}\right). \label{eq:vanishing}
\end{eqnarray}
As regards now condition 3, the disformal transformation becomes
non-invertible at two points. First at the throat $r=r_0$, due
to the infinite determinant of the transformed metric, the
disformed spacetime cannot be mapped to the original spacetime. This
is however a mere coordinate singularity for the resulting wormhole spacetime, as will be seen below, in
Eqs.~(\ref{goodcoords}-\ref{goodcoordsbis}). The second
singular point is given by the equation $1+2X^2 D_X=0$, where $1+2X^2 D_X$ changes sign in the considered case. For our choice of $W$ as in
(\ref{eq:condjacx}), this point is located at radius
$r=r_*$ such that
\begin{equation}
f\left(r_*\right) = \frac{1}{2} A\left(\frac{r_*}{\sqrt{\left\lvert\alpha\right\rvert}}\right). \label{eq:condjac}
\end{equation}
At $r=r_*$, the transformation (\ref{eq:disf_final}) becomes
non-invertible since the determinant of the Jacobian becomes
infinite\footnote{As shown in the following paragraph~\ref{subsec:disfbh}, the presence of
$r=r_*$ prevents the disformed metric from solving a well-defined
variational principle for the beyond Horndeski action obtained via
the transformation (\ref{eq:disf_final}).}, i.e. condition 3
above is not satisfied. In order for the wormhole solution to
originate from a unique well-defined action, the function $A$ should be chosen
such that the location $r=r_*$ does not appear in the wormhole spacetime. But the wormhole spacetime corresponds to $r\geq r_0$. Therefore, $r_\star$ must be smaller than the throat location $r_0$, that is $r_*<r_0$. This allows infinitely many possibilities for $A$, but for our purposes, one can easily prove that the simple choice
\begin{equation}
A\left(\frac{\psi}{\sqrt{\left\lvert\alpha\right\rvert}}\right) = a+\frac{\sqrt{\left\lvert\alpha\right\rvert}}{\psi} \label{eq:ourA}
\end{equation}
satisfies these requirements for any $0<a<1$. This is illustrated on the left plot of Fig.~\ref{fig:Aperfect}. Conversely, on the right plot\footnote{Of course, the choice of $A$ on the right plot of Fig.~\ref{fig:Aperfect} does not satisfy anyway the requirement for asymptotic flatness, $A(r\to\infty)=a$. The purpose is illustrative: one could not choose a function $A$ which converges to $a$ as $r\to\infty$ but behaving for small $r$ like the function $A$ on the right plot.}, the disformal mapping $D(\psi,X)$ depends only on $X$ but not on the scalar field $\psi$, that is to say $A\propto\psi^2$ [see (\ref{eq:condjacx})]. As a result, condition 3 is not satisfied, because the singularity of the disformal transformation at $r=r_*$ is hit before the throat, $r_0<r_*$. Note that the crossing point $r=r_*$ is not a singular point of the disformed metric, but the disformed metric ceases to solve well-defined field equations below $r=r_*$. 
\\

\textbf{At the end of the day, with the convenient choice (\ref{eq:ourA}), the wormhole solution satisfying all three requirements reads (reinstating the original
scalar $\phi$),
\begin{eqnarray}
\mathrm{d}\tilde{s}^2 &=&-f\left(r\right)\mathrm{d}t^2+\mathrm{d}r^2/h(r)+r^2\mathrm{d}\Omega^2,\label{eq:wormholesolution}\\
\phi(r) &=&\ln\left(\frac{\sqrt{-2\alpha/\beta}}{r}\right),\label{eq:wormholesolution_bis}
\end{eqnarray}
where
\begin{equation}
\label{eq:wormholesolution_bisbis}
h(r)=\frac{f\left(r\right)}{1-1/a}\left(1-\frac{f\left(r\right)}{a+\frac{\sqrt{\left\lvert\alpha\right\rvert}}{r}}\right),
\end{equation}
and $f(r)$ is given in (\ref{eq:solutionA})}. The wormhole
configuration (\ref{eq:wormholesolution}-\ref{eq:wormholesolution_bisbis})
is a solution of a beyond Horndeski theory, which is given in the next paragraph~\ref{subsec:disfbh} as well as in appendix~\ref{chap:g2g3}. $M$ is an integration constant of the solution and can be interpreted as its mass by looking at the asymptotic behaviour. In addition to the parameters $\alpha$ and $\beta$ of
the original theory (\ref{eq:action_4degb_final}), the new theory is also
parameterized by the dimensionless
parameter $a\in(0,1)$. 
\\

The throat radius $r_0$ can be computed as a function of the mass $M$ of the wormhole, provided the function $A$ is invertible [which is of course the case for (\ref{eq:ourA})]. Let $f_0$ be the value of the metric function at the throat,
\begin{equation}
f_0\equiv f\left(r_0\right)=A\left(\frac{r_0}{\sqrt{\left\lvert\alpha\right\rvert}}\right)=a +\frac{\sqrt{\left\lvert\alpha\right\rvert}}{r_0}. \label{eq:f0}
\end{equation}
$f_0$ quantifies the
compactness of the wormhole. Indeed, if $f_0\ll 1$, then\footnote{We will see that $r_0\to \infty$ for large $M$, so $f_0\sim a$, therefore $f_0\ll 1$ happens if $a\ll 1$.} the redshift at the throat is large, and the wormhole behaves very much like a BH horizon for far away observers, see for example~\cite{Damour:2007ap}. \\

\begin{figure}
\begin{subfigure}{0.51\textwidth}
\includegraphics[width=\linewidth]{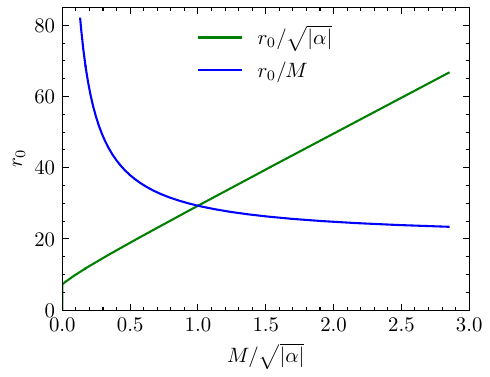}
\end{subfigure}
\begin{subfigure}{0.48\textwidth}
\includegraphics[width=\linewidth]{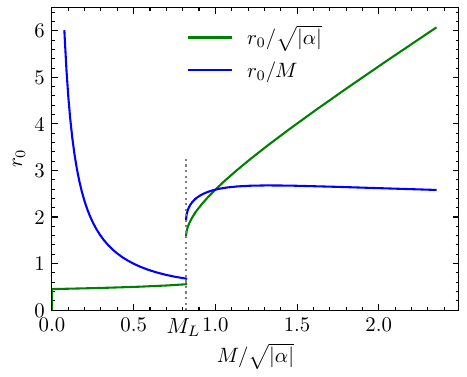}
\end{subfigure}
\caption{The plot shows the throat radius $r_0$ as a function of
$M/\sqrt{|\alpha|}$, for $a=0.9$ (left plot, no discontinuity in $r_0$) and $a=0.1$ (right plot, discontinuity at $M_L$). The discontinuity corresponds to a change of branch in
the solution of Eq. (\ref{eq:vanishing}).}
\label{fig:rzero}
\end{figure}

Eq. (\ref{eq:f0}) enables to get $r_0$ and $M$ as functions of $f_0$. Inverting the latter relation yields $f_0$ as a function of $M$, and a last step then yields $r_0$ as a function of $M$. This procedure enables to show that there exists a value\footnote{More precisely, $a_0$ is the unique root in $\left]0,1\right[$ of the equation $-1127+2956a-2948a^2+1532a^3-120a^4-480a^5+224a^6-32a^7=0$.} $a_0\approx 0.87396$ of the parameter $a$, such that for $a\geq a_0$, $r_0$ is a smooth function of $M$, while for $a<a_0$, there is a discontinuity in $r_0$ at a mass $M_L$ (which depends of course on $a$). Fig.~\ref{fig:rzero} illustrates these different behaviours for the values $a=0.9$ (left plot) and $a=0.1$ (right plot). One can easily understand this behaviour by taking a look at the left plot of Fig.~\ref{fig:Aperfect}, which corresponds to $a=0.1$: for $M<M_L$ (blue curve), the throat is close to the origin and blueshifted, while for $M>M_L$ (yellow curve), the throat is at a bigger radius and redshifted. 
\\

The size of the throat increases with the parameter $a$. For example, it is easy to show that the throat radius quickly converges towards $r_0\approx 2M/\left(1-a\right)$ as soon as $M>\sqrt{\left\lvert\alpha\right\rvert}$. Therefore, for sufficiently large masses, the throat radius is enhanced by a factor $\left(1-a\right)^{-1}$ with respect to the Schwarzschild radius for the corresponding mass.
\\

\begin{figure}
\begin{subfigure}{0.495\textwidth}
\includegraphics[width=\linewidth]{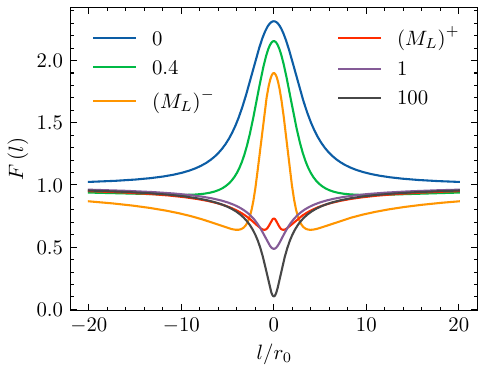}
\end{subfigure}
\begin{subfigure}{0.495\textwidth}
\includegraphics[width=\linewidth]{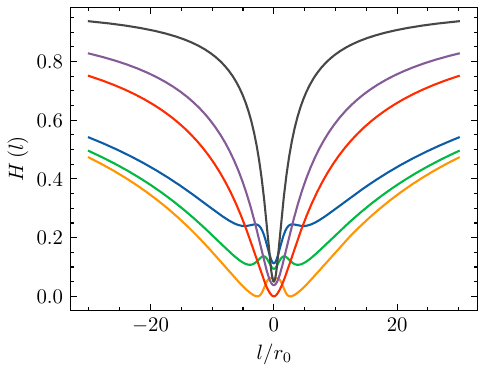}
\end{subfigure}
\caption{Functions $F\left(l\right)$ and $H\left(l\right)$ of
metric (\ref{goodcoords}) (with parameter $a=0.1$), for different values of
$M/\sqrt{\left\lvert\alpha\right\rvert}$ given by the legend. The values $\left(M_L\right)^-$ and $\left(M_L\right)^+$ are as close as possible to the limit mass $M_L$ with our numerical precision, namely $\left(M_L\right)^\pm=M_L\left(1\pm 10^{-15}\right)$, illustrating the discontinuity occurring at this mass. For huge masses, the redshift function $F(l)$ converges to the value $a$($=0.1$ here) at the throat.}\label{fig:f2f3}
\end{figure}

We conclude our discussion by presenting the wormhole solution using coordinates which are everywhere non-singular, including at the throat. The radial coordinate $r$ is changed to $l$, with range $l\in(-\infty,\infty)$, defined by
\begin{equation}
r^2 \equiv l^2 + r_0^2.
\end{equation}
In this coordinate system, any wormhole metric of the form (\ref{eq:wormholesolution}), is given by
\begin{equation}
\label{goodcoords}
\mathrm{d}\tilde{s}^2 = -F\left(l\right)\mathrm{d}t^2+\mathrm{d}l^2/H\left(l\right)+\left(l^2+r_0^2\right)\mathrm{d}\Omega^2,
\end{equation}
where
\begin{equation}
\label{goodcoordsbis}
F\left(l\right) \equiv f\left(\sqrt{l^2+r_0^2}\right),\quad
H\left(l\right) \equiv h\left(\sqrt{l^2+r_0^2}\right)\frac{l^2+r_0^2}{l^2}.
\end{equation}
The function $H\left(l\right)$ is regular everywhere. In particular, at the throat $l\to 0$,
\begin{equation}
H\left(l\right)=\frac{r_0}{2}h'\left(r_0\right)+\mathcal{O}\left(l^2\right).
\label{eq:f3lf1prime}
\end{equation}
Since $h\left(r>r_0\right)> 0$, $H\left(l\right)\geq0$
everywhere\footnote{$H\left(l\right)= 0$ occurs for $l=0$ and $h'\left(r_0\right)=0$. This corresponds to the particular value of $M$ where a discontinuity in $r_0$ occurs, see Fig.~\ref{fig:rzero}, right plot.}.
The other metric function, $F\left(l\right)$, is regular and non-negative everywhere. In Fig.~\ref{fig:f2f3}, the
functions $F\left(l\right)$ and $H\left(l\right)$ are plotted for different
masses $M$, when $a=0.1$. The masses of the yellow and red plots are chosen very
close to the mass $M_L$ where occurs the discontinuity in $r_0$: for
the yellow plot, the mass is still sufficiently low so that the
throat $r_0$ is close to $r=0$ and blueshifted (i.e. $f>1$),
while for the red plot, the throat $r_0$ is much larger and the
spacetime is redshifted ($f<1$) there. This is not just a sharp evolution of
the behaviour of $F\left(l\right)$ as a function of the mass, but a
true discontinuity at $M=M_L$. \\

The corresponding beyond Horndeski theory which admits this wormhole as a solution is quite intricate, and we give it in the below paragraph and in appendix~\ref{chap:g2g3} only for completeness. One should not wonder too much about the physical meaning of such a complicated theory. The aim of the present section was rather to show how, in principle, a disformal transformation can map a seed solution with curvature singularities (either naked or hidden inside a horizon) to a new, perfectly regular, wormhole solution.
\subsection{Associated variational principle}\label{subsec:disfbh}
The generic transformation of a seed Horndeski action under the disformal transformation (\ref{eq:disf_final}) is given in appendix~\ref{chap:appdisf}. One thus only needs to apply these formulas to the seed action (\ref{eq:action_4degb_final}), which has Horndeski functions
\begin{align}
G_2 ={}&{} -\frac{\beta^2}{2\alpha}\mathrm{e}^{4\phi}+12\beta\mathrm{e}^{2\phi}X+8\alpha X^2,\quad G_3 = 8\alpha X,\nonumber\\ G_4 ={}&{} 1-\beta\mathrm{e}^{2\phi}+4\alpha X,\quad G_5 = 4\alpha\ln\left\lvert X\right\rvert.
\end{align}
The choice of disformal factor $D(\phi,X)$ is such that
\begin{equation}
\left(1-2D\left(\phi,X\right)X\right)^{-1}=W^{-1}\left(\phi,X\right)=
\left(1-1/a\right)^{-1}\left(1+2B\left(\phi\right)
X\right),\quad 0<a<1,
\end{equation}
see eq. (\ref{eq:condjacx}), with
\begin{equation}
B(\phi)\equiv\frac{\psi^2}{A\left(\psi/\sqrt{\left\lvert\alpha\right\rvert}\right)},\quad \psi=\sqrt{\frac{-2\alpha}{\beta}}\mathrm{e}^{-\phi}.
\end{equation}
To reformulate the initial Horndeski functions, which depend on $X$, into new functions depending on the disformed kinetic term $\tilde{X}$, one must compute $X$ as a function of $\tilde{X}$. The disformed kinetic term is $\tilde{X}=XW^{-1}\left(\phi,X\right)$, see Eq. (\ref{eq:xtildeappendixfinal}). This yields a second-order polynomial in $X$, and one gets two possible solutions for $X$, given by
\begin{equation}
X = \frac{-1}{4B\left(\phi\right)}\left(1\pm
\Xi\left(\phi,\tilde{X}\right)\right),\quad
\Xi\left(\phi,\tilde{X}\right)\equiv
\sqrt{1+8B\left(\phi\right)\left(1-\frac{1}{a}\right)\tilde{X}}.
\label{eq:xxtilde}
\end{equation}
Depending on which sign is chosen ($+$ or $-$), one is led to two
distinct disformed actions, $\tilde{S}_+$ and $\tilde{S}_-$ respectively. One must therefore identify which variational principle is
solved by the disformed metric (\ref{eq:wormholesolution}-\ref{eq:wormholesolution_bisbis}). To this aim, one has to analyze the situation on shell, where
\begin{equation}
\Xi\left(\phi,\tilde{X}\right) = \left\lvert
\xi\left(r\right)\right\rvert,\quad \xi\left(r\right)\equiv 
1-2B\left(\phi\right)\frac{f\left(r\right)}{r^2},\quad
\phi=\ln\left(\frac{\sqrt{-2\alpha/\beta}}{r}\right). \label{eq:Ss}
\end{equation}
Eqs. (\ref{eq:xxtilde}) and (\ref{eq:Ss}) in turn imply that
\begin{equation}
\frac{-f\left(r\right)}{2r^2} =
\frac{-1}{4B\left(\phi\right)}\left(1\pm\left\lvert
\xi\left(r\right)\right\rvert\right).
\end{equation}
This is seen to be consistent only by choosing the $+$ sign when
$\xi\left(r\right)\leq 0$, and the $-$ sign when $\xi\left(r\right)\geq 0$. As a consequence, the disformed metric solves the
equations of motion of $\tilde{S}_+$ (respectively $\tilde{S}_-$) if and only if
$\xi\left(r\right)\leq 0$ (respectively $\xi\left(r\right)\geq 0$). In
particular, it is not possible to define an action principle for
the disformed theory if the function $\xi\left(r\right)$ has a non-constant sign. Note that $\xi\left(r\right)$ changes sign precisely at the singular radius $r_*$ identified in (\ref{eq:condjac}). Thus, one retrieves the necessity of removing $r_*$ from the spacetime, by hiding it below the wormhole throat. This is for instance ensured by our choice (\ref{eq:ourA}), for which $\xi\left(r\right)<0$ in the whole
physical spacetime, and hence a well-defined action principle is
shown to exist.\\

The corresponding beyond Horndeski theory is computed by using the formulas of appendix~\ref{chap:appdisf}. For readability, we write coefficients as functions of variables
$\left(y,x\right)$, where $y$ stands for $\phi$ and $x$ for
$\tilde{X}$. For instance, one must understand $\Xi=\Xi(y,x)$ and $B=B(y)$. Subscripts $y$ and $x$ mean derivation with respect to $y$ and $x$. The beyond Horndeski functions are:
\begin{flalign}
\widetilde{F_5}\left(y,x\right) = \frac{2 (a-1) \alpha
\sqrt{-\frac{x B} {\Xi+1}} \left(a
\Xi+4 (a-1) x B-2 \Xi+a\right)}
{3 a x^2 \Xi \left(a \left(\Xi-1\right)-4 (a-1) x B\right)},&&
\end{flalign}
\begin{flalign}
\widetilde{G_5}\left(y,x\right) = \frac{2 \alpha  \ln
\left(\frac{\Xi+1}{4 B}\right)} {\sqrt{-\frac{x
B}{\Xi+1}}}+\frac{8 \alpha
\sqrt{\Xi-1}
\arctan\left(\frac{\sqrt{\Xi-1}}{\sqrt{2}}\right)-4
\sqrt{2}
 \alpha  \ln \left(\frac{\Xi+1}{4 B}\right)}{\sqrt{\frac{a-a \Xi}{a-1}}},&&
\end{flalign}
\begin{flalign}
\widetilde{G_4}\left(y,x\right) ={}&\frac{1}{B \sqrt{-\frac{x
B}{\Xi+1}} \sqrt{\frac{a-a \Xi}{a-1}}
\left(8 (a-1) x
B+a-a\Xi\right)}\Biggl\{ 4 \alpha  x B_y \Biggl[(a-1) x
B\Biggl[8 \sqrt{2} \nonumber \\{}&{}\sqrt{-\frac{x
B}{\Xi+1}}-\Biggl(\sqrt{\frac{a-a
\Xi}{a-1}}-2 \sqrt{2} \sqrt{-\frac{x
B}{\Xi+1}}\Biggr) \ln
\left(\frac{\Xi+1}{4
B}\right)\Biggr]-\sqrt{2} a
\left(\Xi-1\right)\nonumber\\{}&{}\sqrt{-\frac{x
B}{\Xi+1}}\Biggr] \Biggr\}+2 \sqrt{-\frac{x
B}{\Xi+1}} \left(1-\frac{\alpha
\left(\Xi+1\right)}{B}-\beta  \mathrm{e}^{2 y}\right).&&
\end{flalign}
As regards $\widetilde{G_2}$, $\widetilde{G_3}$ and $\widetilde{F_4}$, their lengthy expressions are reported in appendix~\ref{chap:g2g3}.
\\

We have seen how a pure disformal transformation enables to transform a non-stealth BH into a wormhole. Let us now peform a pure conformal transformation on a seed stealth Kerr BH, and obtain a non-stationary BH. The following section is based on the article~\cite{Babichev:2023mgk}, of which the present author is one of the authors.
\section[Generation of a Kerr-FLRW spacetime by conformal transformation]{Generation of a Kerr-FLRW spacetime by conformal transformation}\label{sec:kerrflrw}
\sectionmark{6.2~~Generation of a Kerr-FLRW spacetime by conformal transformation}
\subsection{Motivations and outline of the construction}
While BHs, and wormholes, have been the focus of attention up to now, as regards cosmology, the FLRW metric is the GR solution which describes a homogeneous isotropic universe,
\begin{equation}
\mathrm{d}s^2=-\mathrm{d}t^2+A(t)^2\left(\frac{\mathrm{d}r^2}{1-\kappa r^2}+r^2\mathrm{d}\Omega^2\right).
\end{equation}
Under this form, the metric is said to be written using \textbf{cosmological time} $t$. The parameter $\kappa$ has value $1$, $0$ or $-1$, corresponding respectively to a universe with spherical, flat or elliptic spatial sections, i.e. sections of constant $t$. The \textbf{scale factor} is the function\footnote{It has obviously nothing to do with the function $A$ of the previous Sec.~\ref{sec:wormhole}.} $A(t)$. It is usually written $a(t)$ in the literature, as was done in Eq. (\ref{eq:firstflrw}), but we use a capital letter to distinguish it from the rotation parameter $a$ of the Kerr metric. Indeed, \textbf{the aim of the present section is to construct a metric combining the features of an FLRW spacetime with those of a rotating Kerr BH}. Before moving on to to this construction, let us recall~\cite{Ibison:2007dv} that the FLRW metric can be written in \textbf{conformal time} $\tau$, defined by the relation
\begin{equation}
A(\tau)\mathrm{d}\tau=\mathrm{d}t.
\end{equation}
The FLRW metric in conformal time reads
\begin{equation}
\mathrm{d}s^2=A(\tau)^2\left(-\mathrm{d}\tau^2+\frac{\mathrm{d}r^2}{1-\kappa r^2}+r^2\mathrm{d}\Omega^2\right).\label{eq:flrw_conf}
\end{equation}
The task of combining the features of a BH and of a Friedmann universe, apart from being an exciting theoretical challenge, may also present practical interest, to understand the formation and evolution of primordial BHs~\cite{Carr:2020xqk,Green:2020jor,Escriva:2022duf} in the early universe, for instance.
\\

In spherical symmetry and for an exponentially expanding universe, such a solution is well-known and given by the Schwarzschild-dS metric (\ref{eq:sdsfirst}). 
The relative simplicity of this solution, and in particular the existence of coordinates where the metric is static, can be accounted for by the form of the 'matter', which is a mere cosmological constant. Indeed, the cosmological constant has a constant density during the evolution, and hence does not accrete onto a BH, thus keeping the mass of the BH constant. 
\\

For a more general FLRW behaviour, McVittie proposed a metric, which was believed to describe a point-like object embedded in an FLRW universe~\cite{McVittie:1933zz}. 
It was however later understood that it cannot describe a point-like object, because of a coordinate singularity inherited in the solution. Likewise, the McVittie metric cannot describe a BH in the universe, due to presence of the same singularity at the would-be event horizon{\footnote{
This particular problem is alleviated for McVittie spacetimes if late time cosmology is dominated by a positive cosmological constant~\cite{Kaloper:2010ec}.}}. Besides, the density and pressure of the McVittie fluid source are not related by an equation of state: for this solution, the pressure is a function of not only the density, but also the coordinates.
\\

Physically, the problems of the McVittie solution are related to postulating rather than finding the metric via the Einstein equations. In particular, a zero flow of fluid onto the BH is assumed, which is only justified in the case of the cosmological constant: at least in the test fluid approximation, BHs do accrete surrounding fluid~\cite{Bondi:1952ni,Babichev:2004yx,Babichev:2005py,Babichev:2013vji}. 
\\

The reader may be interested by Ref.~\cite{Faraoni:2015ula}, where other time-dependent metrics aiming at describing BHs in a cosmological background are considered in detail; and by article \textit{Conformally Schwarzschild cosmological black holes}, Ref.~\cite{Sato:2022yto}. This latter article summarizes the more or less relevant attempts of merely starting with a Schwarzschild metric, and mutliplying it by a conformal scale factor. Importantly, \textbf{the set of coordinates of the initial Schwarzschild metric is paramount in these constructions}. For instance, the Thakurta spacetime is
\begin{equation}
\mathrm{d}s^2=A(\tau)^2\left[-f(r)\mathrm{d}\tau^2+\mathrm{d}r^2/f(r)+r^2\mathrm{d}\Omega^2\right],\quad f(r)=1-2M/r.\label{eq:thakurta}
\end{equation}
It obviously behaves as an FLRW spacetime (with flat spatial sections) when $r\to\infty$, but there is a singularity at $r=2M$. In the Schwarzschild case, this singularity is illusory, and one can remove it by redefining the time coordinate. But for the Thakurta metric, such a redefinition does not work, because the metric depends on the time coordinate \textit{via} the conformal scale factor $A(\tau)$. Other attempts are more promising. We only present the one which will be useful for the rest of this section, namely, the \textbf{Culetu spacetime}~\cite{Culetu:2012ih}. It reads
\begin{align}
\mathrm{d}s^2 ={}& A(\tau)^2\Biggl[-\left(1-\frac{2M}{r}\right)\mathrm{d}\tau^2+2\sqrt{\frac{2M}{r}}\mathrm{d}\tau \mathrm{d}r+\mathrm{d}r^2+r^2\mathrm{d}\Omega^2\Biggr].\label{eq:culetu_first}
\end{align}
One sees that the metric between brackets is Schwarzschild, but now written in the \textbf{Painlev\'e-G\"ullstrand} coordinates~\cite{1921CR....173..677P,Gullstrand:1922tfa}, as opposed to the Thakurta class (\ref{eq:thakurta}), where the 'usual' coordinates of Schwarzschild were used. In the absence of conformal scale factor, both choice of coordinates describe the same Schwarzschild metric in different coordinates. However, with the  $A(\tau)$, this is no more the case. In particular, the Culetu metric has no curvature singularity at $r=2M$: there are only the usual BH singularity at $r=0$, and the Big Bang singularity at $A=0$. As regards field equations, the Culetu metric is postulated \textit{ad hoc} as a solution of the Einstein field equations of GR with matter,
\begin{equation}
G_{\mu\nu}=T_{\mu\nu}.
\end{equation}
The Einstein tensor $G_{\mu\nu}$ of metric (\ref{eq:culetu_first}) is then computed, leading to the energy-momentum tensor $T_{\mu\nu}$ of the associated matter fields. Ref.~\cite{Sato:2022yto} proved that this matter violates the usual energy conditions.
\\

All the approaches mentioned above only attempt to find non-rotating BHs in an FLRW universe. For rotating BHs, the only relevant solution is the Kerr-dS solution (\ref{eq:kerr-ds}), discovered by Carter~\cite{CARTER1968399}. It describes a rotating BH in an exponentially expanding universe, with expansion driven by the cosmological constant. Given the issues related to the construction of spherically symmetric BHs in FLRW, the task of finding rotating analogs embedded in general Friedmann universe in GR seems extremely difficult.
\\

The present section, as all this thesis, however focuses on scalar-tensor gravity. The stability of DHOST theory under conformal transformations will be used to construct \textbf{non-rotating and rotating BHs embedded in generic FLRW backgrounds as exact solutions of DHOST theory in vacuum}. 'In vacuum' of course means that there are no matter fields, so the associated FLRW expansion is uniquely driven by the scalar field $\phi$.\\

As a positive outcome, we will for instance retrieve the Culetu spacetime (\ref{eq:culetu_first}), but without violations of the energy conditions, which occur when this metric is seen as an \textit{ad hoc} solution of GR field equations with matter. Of course, the construction remains not completely satisfactory, since there are matter fields in the universe, and they should obviously contribute to the FLRW expansion. The construction of the present section should therefore be seen as a starting point, and a proof of principle that a scalar field can remove violations of the energy conditions in the case of expanding BHs. More realistic models, including both the scalar field $\phi$ and matter fields, may be thought about in future works.
\\

The construction is based on the stealth Kerr solution, described in paragraph~\ref{subsec:stealthkerr}. As a reminder, this solution reads
\begin{align}
\mathrm{d}s^2={}&-\left(1-\frac{2Mr}{\Sigma}\right)\mathrm{d}t^2+\frac{\Sigma}{\Delta}\mathrm{d}r^2+\Sigma\mathrm{d}\theta^2+\frac{\sin^2\theta}{\Sigma}\Upsilon\mathrm{d}\varphi^2-\frac{4Mar\sin^2\theta}{\Sigma}\mathrm{d}t\mathrm{d}\varphi,\label{eq:kerr_before_conf}\\
\phi ={}& q\left(t+\int\psi\left(r\right)\mathrm{d}r\right),\quad \psi(r)\equiv\frac{\sqrt{2Mr\left(r^2+a^2\right)}}{\Delta}.\label{eq:HJ_before_conf}
\end{align}
The following notations are used,
\begin{equation}
\Sigma=r^2+a^2\cos^2\theta, \quad \Delta = r^2+a^2-2Mr,\quad \Upsilon = \left(r^2+a^2\right)\Sigma+2Mra^2\sin^2\theta.\label{eq:notations_new}
\end{equation}
The parameter $q$ appearing in the scalar field $\phi$ is linked to the kinetic term $X=\partial_\mu\phi\,\partial^\mu\phi$, which is constant: $X=X_0\equiv -q^2$. The stealth Kerr is a solution of the following quadratic DHOST theory with shift and parity symmetry, and speed of GWs equal to speed of light,
\begin{align}
S=\int\mathrm{d}^4x\sqrt{-g}\Bigl\{F\left(X\right)R+{}&{}P\left(X\right)+A_3\left(X\right)\Box\phi\phi^\mu\phi_{\mu\nu}\phi^\nu\nonumber \\ {}&{} +A_4\left(X\right)\phi^\mu\phi_{\mu\nu}\phi^{\nu\rho}\phi_\rho+A_5\left(X\right)\left(\phi^\mu\phi_{\mu\nu}\phi^\nu\right)^2\Bigr\},\label{eq:dhost_shift_par_before_conf}
\end{align}
provided the following conditions are satisfied,
\begin{equation}
A_3\left(X_0\right) = 0,\quad P_X\left(X_0\right) = 0,\quad P\left(X_0\right)=0. 
\end{equation}
Also, as always in DHOST, the theory functions $A_4$ and $A_5$ are determined by the remaining theory functions in order to ensure degeneracy of the kinetic matrix, see Eqs. (\ref{eq:a4dhost}-\ref{eq:a5dhost}).
\\

Paragraph~\ref{subsec:conf_kerr} constructs the conformal Kerr solution and the DHOST theory of which it is an exact solution. The link between this solution and FLRW spacetime is quite clear when the usual radial coordinate $r$ is large. However, since the conformal Kerr spacetime is not stationary, highlighting its BH features requires the formalism of \textbf{double-null foliations and trapping surfaces}. This formalism is presented in paragraph~\ref{subsec:foliations}. The last two paragraphs apply this formalism to the conformal Kerr spacetime, first in the easier case of spherical symmetry (paragraph~\ref{sub:culetu}), then in the general case with rotation (paragraph~\ref{subsec:conf_kerr_study}).

\subsection{Cosmological Kerr black holes as vacuum solutions of DHOST theory}\label{subsec:conf_kerr}
We start with the stealth Kerr solution (\ref{eq:kerr_before_conf}-\ref{eq:HJ_before_conf}). In particular, $g_{\mu\nu}$ refers to the Kerr metric. We then perform the following conformal transformation,
\begin{equation}
g_{\mu\nu}\mapsto\tilde{g}_{\mu\nu}=C\left(\phi\right)g_{\mu\nu}.\label{eq:conformal_map}
\end{equation}
The new metric $\tilde{g}_{\mu\nu}$, with the unchanged scalar field (\ref{eq:HJ_before_conf}), is a solution of a new DHOST action,
\begin{align}
\tilde{S}\left[\tilde{g}_{\mu\nu},\phi\right] = \int\mathrm{d}^4x\sqrt{-\tilde{g}}\Biggl[\tilde{F}\tilde{R} +\tilde{P}+\tilde{Q}\tilde{\Box}\phi+ \tilde{A_3}{}&{} \tilde{\phi^\mu}\widetilde{\phi_{\mu\nu}}\tilde{\phi^\nu}\tilde{\Box}\phi + \tilde{A_4} \tilde{\phi^\mu}\widetilde{\phi_{\mu\nu}}\tilde{\phi^{\nu\rho}}\tilde{\phi_\rho} \nonumber\\{}&{}+ \tilde{A_5}\left(\tilde{\phi^\mu}\widetilde{\phi_{\mu\nu}}\tilde{\phi^\nu}\right)^2\Biggr].\label{eq:resulting_dhost}
\end{align}
The new action has the same form as the initial action (\ref{eq:dhost_shift_par_before_conf}), apart from the $\tilde{Q}\tilde{\Box}\phi$ term. In particular, the speed of GWs remains equal to the speed of light (this feature is not affected by the new term). It is understood that indices are contracted with the metric $\tilde{g}_{\mu\nu}$, and $\widetilde{\phi_{\mu\nu}} = \tilde{\nabla}_\mu\tilde{\nabla}_\nu\phi$, etc. The coefficients of the new theory are functions of the new kinetic term $\tilde{X} = \tilde{\phi^\mu}\tilde{\phi_\mu}$, and also of $\phi$, so the new theory is no more shift symmetric. They read
\begin{align}
\tilde{P} ={}& \frac{P}{C^2} + \frac{3\tilde{X}}{C^2}\left(C_{\phi\phi}-\frac{3C_\phi^2}{2C}\right)F - \frac{\tilde{X}^3C_\phi^2}{2C}A_3+\frac{\tilde{X}^3C_\phi^2}{4C}A_4 + \frac{\tilde{X}^4C_\phi^2}{4}A_5\nonumber\\{}&{}-\tilde{X}\left(K_3+K_4+K_5\right)_\phi,\\
\tilde{Q} ={}& \frac{3C_\phi}{C^2}G+\frac{\tilde{X}^2C_\phi}{2}A_3 - \left(K_3+K_4+K_5\right),\\
\tilde{F} ={}& \frac{F}{C},\\
\tilde{A_3} ={}& CA_3,\\
\tilde{A_4} ={}& CA_4,\\
\tilde{A_5} ={}& C^2A_5,\label{eq:resulting_dhost_2}
\end{align}
where the subscript $\phi$ means derivation with respect to $\phi$, and the three following functions are introduced for convenience,
\begin{equation}
K_3 = -\frac{C_\phi}{2}\int\mathrm{d}\tilde{X} \tilde{X}A_3,\quad K_4 = \frac{C_\phi}{2}\int\mathrm{d}\tilde{X} \tilde{X}A_4,\quad K_5 = \frac{CC_\phi}{2}\int\mathrm{d}\tilde{X} \tilde{X}^2A_5.
\end{equation}
In (\ref{eq:conformal_map}), $C(\phi)$ is the conformal factor. Moreover, the seed metric $g_{\mu\nu}$ is asymptotically flat, i.e. $g_{\mu\nu}\to \eta_{\mu\nu}$ when $r\to\infty$. Therefore, when $r\to\infty$, the resulting metric $\tilde{g}_{\mu\nu}$ is seen to have the form (\ref{eq:flrw_conf}) of an FLRW spacetime in conformal time (with $\kappa=0$), \textbf{provided the scalar field $\phi$ coincides with the conformal time $\tau$ (up to a factor), 
\begin{equation}
\tau \equiv \frac{\phi}{q},\label{eq:tau}
\end{equation}
and the conformal factor $C(\phi)$ plays the role of the conformal scale factor $A(\tau)$,
\begin{equation}
C\left(\phi\right)\equiv A\left(\phi/q\right)^2=A\left(\tau\right)^2. \label{eq:conformal_scale}
\end{equation}}Using (\ref{eq:HJ_before_conf}) and (\ref{eq:tau}), one can rewrite the metric (\ref{eq:kerr_before_conf}) in terms of the new coordinates $\left(\tau,r,\theta,\varphi\right)$, which are an extension of Painlev\'e-Gullstrand coordinates for the Kerr metric. Indeed, $\mathrm{d}\tau$ is the proper time interval of a freely falling particle with vanishing speed at infinity, and vanishing angular momentum. 
Multiplying the resulting metric by the conformal factor $A\left(\tau\right)^2$ finally leads to the conformally-related configuration $\left(\tilde{g}_{\mu\nu},\phi\right)$,
\begin{align}
\mathrm{d}\tilde{s}^2={}&{}A\left(\tau\right)^2\Biggl\{ {}&-\left(1-\frac{2Mr}{\Sigma}\right)\mathrm{d}\tau^2+\left[\frac{\Sigma}{\Delta}-\psi\left(r\right)^2\left(1-\frac{2Mr}{\Sigma}\right)\right]\mathrm{d}r^2+\Sigma\mathrm{d}\theta^2\nonumber \\ &{}&{}+\frac{\sin^2\theta}{\Sigma}\Upsilon\mathrm{d}\varphi^2 -\frac{4Mar\sin^2\theta}{\Sigma}\mathrm{d}\tau\mathrm{d}\varphi+\frac{4Mar\psi\left(r\right)\sin^2\theta}{\Sigma}\mathrm{d}r\mathrm{d}\varphi\nonumber \\ &{}&{}+2\left(1-\frac{2Mr}{\Sigma}\right)\psi\left(r\right)\mathrm{d}\tau\mathrm{d}r\Biggr\}, \label{eq:conf_kerr}\\
\phi={}& q\tau. &{} \label{eq:phi_conf_kerr}
\end{align}
The obtained spacetime (\ref{eq:conf_kerr}) is axisymmetric but no longer stationary. Expanding the metric as $r\to\infty$ gives
\begin{align}
\mathrm{d}\tilde{s}^2= A\left(\tau\right)^2\Bigl\{{}&{} -\left[1+\mathcal{O}\left(r^{-1}\right)\right]\mathrm{d}\tau^2 +\left[1+\mathcal{O}\left(r^{-1}\right)\right]\mathrm{d}r^2 + r^2\left[1+\mathcal{O}\left(r^{-2}\right)\right] \mathrm{d}\theta^2\nonumber\\ {}&{}+r^2\sin^2\theta\left[1+\mathcal{O}\left(r^{-2}\right)\right] \mathrm{d}\varphi^2  + \mathcal{O}\left(r^{-1}\right)\mathrm{d}\tau \mathrm{d}\varphi +\mathcal{O}\left(r^{-3/2}\right)\mathrm{d}r \mathrm{d}\varphi\nonumber\\ {}&{}+\mathcal{O}\left(r^{-1/2}\right)\mathrm{d}\tau \mathrm{d}r\Bigr\},
\end{align}
which shows that in this limit, the spacetime is as announced a spatially flat FLRW spacetime written in conformal time $\tau$. 
Note also that for $M=0$, the metric (\ref{eq:conf_kerr}) reduces exactly to a flat FLRW universe in conformal time $\tau$, although the spatial part is written in ellipsoidal coordinates.
The apparent singularity of the metric (\ref{eq:conf_kerr}) at $\Delta=0$ is due to a bad behaviour of coordinates there. It can be removed by a change of the coordinate $\varphi$, by defining for instance
\begin{equation}
\mathrm{d}\varphi_+ = \mathrm{d}\varphi + \frac{a}{\Delta}\mathrm{d}r. \label{eq:phi+}
\end{equation}
This brings the metric to the following form, explicitly regular at $\Delta=0$,
\begin{align}
\mathrm{d}\tilde{s}^2={}&A\left(\tau\right)^2\Biggl\{ -\left(1-\frac{2Mr}{\Sigma}\right)\mathrm{d}\tau^2+\left(2-\frac{1-2Mr/\Sigma}{1+\sqrt{\frac{2Mr}{r^2+a^2}}}\right)\frac{\mathrm{d}r^2}{1+\sqrt{\frac{2Mr}{r^2+a^2}}}+\Sigma\mathrm{d}\theta^2\nonumber\\{}&{}+\frac{\sin^2\theta}{\Sigma}\Upsilon\mathrm{d}\varphi_+^2 -\frac{4Mar\sin^2\theta}{\Sigma}\mathrm{d}\tau\mathrm{d}\varphi_+-2a\sin^2\theta\left(1+\frac{2Mr/\Sigma}{1+\sqrt{\frac{2Mr}{r^2+a^2}}}\right)\mathrm{d}r\mathrm{d}\varphi_+\nonumber\\{}&{}+2\left(1-\frac{1-2Mr/\Sigma}{1+\sqrt{\frac{2Mr}{r^2+a^2}}}\right)\mathrm{d}\tau\mathrm{d}r\Biggr\}.\label{eq:good_coord}
\end{align}
The above metric has a Big Bang singularity at $A(\tau)=0$, and the usual Kerr singularity at $\Sigma=0$, which is a ring singularity. However, the various square roots in the above expression are not twice differentiable at $r=0$, therefore the spacetime has in fact a curvature singularity at $r=0$, which is a disk singularity comprising the ring singularity $\Sigma=0$. This is seen on the Ricci scalar,
\begin{equation}
R = \frac{3}{A^2}\left[2\frac{\ddot{A}}{A}-\frac{3r^2+a^2}{\Sigma\sqrt{r^2+a^2}}\sqrt{\frac{2M}{r}}\frac{\dot{A}}{A}\right],\label{eq:ric_conf_kerr}
\end{equation} 
which diverges only at $r=0$ and at $A\left(\tau\right)=0$. A dot means derivation with respect to $\tau$. The absence of singularity at $\Delta=0$ is not trivial \textit{a priori} and is made possible because $\tau$ is a 'good' time coordinate. For instance, if one were taking the Kerr metric in usual coordinates $(t,r,\theta,\varphi)$, Eq. (\ref{eq:kerr_before_conf}), and multiplying it by a scale factor $A(t)^2$, there would be a true curvature singularity at $\Delta=0$. 
\\

In the following, the calculations will be illustrated with a power-law scale factor~\cite{Misner:1973prb,Wald:1984rg,Carroll:2004st}, 
\begin{equation}
\label{alpha}
A(\tau)=A_0|\tau |^\alpha,
\end{equation}
where the range of the time coordinate is $\tau>0$ for $\alpha>0$, and $\tau<0$ for $\alpha<0$. The solution mimicks asymptotically (as $r\to\infty$) an FLRW universe sourced by a perfect fluid with equation of state $w=(2-\alpha)/(3\alpha)$. The term 'mimicking' is important, since there is no matter for the conformal Kerr solution, which is a solution of a DHOST theory in vacuum. The expansion is fully driven by the scalar field. In particular, the power $\alpha$ appears in the conformal factor, see Eq. (\ref{eq:conformal_scale}),
\begin{equation}
C(\phi)=A_0^2\left\lvert\phi/q\right\rvert^{2\alpha}.
\end{equation}
As a consequence, $\alpha$ is a parameter of the resulting DHOST theory (\ref{eq:resulting_dhost}-\ref{eq:resulting_dhost_2}). Each $\alpha$ labelling a particular equation of state stands for a different conformal Kerr solution of a different DHOST theory. 
\\

Note that the standard cosmological time $t$ is related to the conformal time via the relation $A(\tau)\mathrm{d}\tau=\mathrm{d}t$. 
Depending on $\alpha$, the latter can be integrated to give
\begin{equation}
\label{eq:T}
t=
\begin{cases}
	\frac{A_0}{|\alpha+1|}|\tau|^{\alpha+1} & \text{for $\alpha>0$ and $\alpha<-1$}, \\
	-A_0\ln\left(-\tau/A_0\right) & \text{for $\alpha=-1$},\\
	t_0-\frac{A_0}{\alpha+1}|\tau|^{\alpha+1}, & \text{for $-1<\alpha<0$}.
		 \end{cases}
\end{equation}
For $\alpha=-1$, the constant $1/A_0$ corresponds to the constant Hubble parameter of dS, $H_0=1/A_0$. For $-1<\alpha<0$, the constant $t_0$ is the time of the so-called Big Rip~\cite{Caldwell:1999ew,Caldwell:2003vq}, where the scale factor of the universe diverges at a finite cosmological time. The scale factor in terms of the cosmological time $t$ is given by
\begin{equation}
\label{eq:AT}
A(t) \propto
\begin{cases}
	t^\frac{\alpha}{\alpha+1} & \text{for $\alpha>0$ and $\alpha<-1$},\\
	e^{t/A_0} & \text{for $\alpha=-1$},\\
	\left(t_0-t\right)^\frac{\alpha}{\alpha+1}, & \text{for $-1<\alpha<0$}.
		 \end{cases}
\end{equation}
In particular, $\alpha=2,1,-1$ corresponds to matter, radiation and cosmological constant respectively.
\\

In a nutshell, \textbf{the metric (\ref{eq:conf_kerr}) or (\ref{eq:good_coord}), dressed with the scalar field (\ref{eq:phi_conf_kerr}), is a vacuum solution of the DHOST theory (\ref{eq:resulting_dhost}-\ref{eq:resulting_dhost_2}). It is seen to behave as an FLRW spacetime when $r\to\infty$}. It thus remains to characterize this spacetime as a BH. For such a non-stationary spacetime, this requires the formalism of double-null foliations and trapping surfaces, which is introduced in the following paragraph.

\subsection{Double-null foliations and trapping surfaces}
\label{subsec:foliations}
\subsubsection{Foliations, expansions and trapping horizons}
We will mostly follow the \textbf{2+2 formalism} initiated by Hayward~\cite{Hayward:1993wb}, adapting it to our current purposes when needed\footnote{Ref.~\cite{Hayward:1993wb} does not require null congruences to be geodesics. However most works, including the pioneering Ref.~\cite{Penrose:1964wq} or more recent, e.g.~\cite{Faraoni:2015ula} or~\cite{Faraoni:2016xgy}, consider only congruences of null geodesics. We adopt the latter, more physical notion.\label{fn:geod}}. 
Note that the usual formalism leads to ambiguities in the identification of trapping horizons, see~\cite{Faraoni:2016xgy} and references therein. In recalling the formalism, we point out sources of ambiguities and propose some prescriptions to get rid of some of them.
\\

In the approach of~\cite{Hayward:1993wb}, the spacetime manifold $\mathcal{M}$ with metric $g_{\mu\nu}$ is foliated by space-like 2-surfaces $\mathcal{S}$ which are the intersection of two families of null 3-surfaces $\Sigma_u$ and $\Sigma_v$. 
The surfaces $\Sigma_u$ and $\Sigma_v$ are defined as the surfaces of constant $u$ and $v$, where $u$ and $v$ are functions on spacetime. 
The normal one-forms $-\mathrm{d}u\equiv L=L_\mu \mathrm{d}x^\mu$ and $-\mathrm{d}v\equiv N=N_\mu \mathrm{d}x^\mu$ are null, and the dual vectors $L^\mu\partial_\mu$ and $N^\mu\partial_\mu$ are future-directed. As explained in footnote~\ref{fn:geod}, we impose these vectors to be geodesic, that is to say, $g^{\mu\nu}L_\mu\nabla_\nu L_\rho\propto L_\rho$ and $g^{\mu\nu}N_\mu\nabla_\nu N_\rho\propto N_\rho$. In a word, \textbf{we assume that spacetime has a pair of null coordinates $u$ and $v$, associated to the null geodesic one-forms $L=-\mathrm{d}u$ and $N=-\mathrm{d}v$}.\\

Nevertheless, it will turn out that the normalization of $L$ and $N$ can be crucial for identifying the trapping horizons, see the discussion below, after Eq. (\ref{eq:rescalings}). For later purposes, we thus introduce $l=\epsilon^2 L$ and $n=\delta^2 N$ which are just rescalings of $L$ and $N$, with $\epsilon$ and $\delta$ functions on spacetime. The scalar product between two future-directed vectors is negative, thus we write it as
\begin{equation}
g^{\mu\nu}l_\mu n_\nu = -\frac{1}{F^2},\label{eq:scalar_prod}
\end{equation} 
where $F$ is also a function of spacetime.
The first fundamental form is defined as usual,
\begin{equation}
h_{\mu\nu}=g_{\mu\nu}+F^2\left(l_\mu n_\nu+n_\mu l_\nu\right).\label{eq:hmunu}
\end{equation}
The tensor $h_{\mu\nu}$ is the induced metric on $\mathcal{S}$ (the spacelike 2-surfaces orthogonal to both $L$ and $N$); $h_{\mu\nu}$ projects every vector field of $\mathcal{M}$ onto $\mathcal{S}$. The seeked for relevant quantities are then the \textbf{expansions} $\theta_\pm$, defined by
\begin{equation}
\theta_{\pm}=\frac{1}{2}h^{\mu\nu}\mathcal{L}_{\pm}h_{\mu\nu}.\label{eq:expansion}
\end{equation}
$\mathcal{L}_{\pm}$ denotes the Lie derivative with respect to the vector fields $F^2 l^\mu\partial_\mu$ and $F^2 n^\mu\partial_\mu$ respectively. \textbf{For the expansions, we use the usual notations of the literature, and insist that these $\theta_\pm$ have nothing to do with the angular coordinate $\theta$}. A more explicit form of (\ref{eq:expansion}) is easily computed:
\begin{equation}
\theta_+ = F^2\nabla^\mu l_\mu+F^4n^\nu l^\mu\nabla_\mu l_\nu,\quad \theta_- = F^2\nabla^\mu n_\mu+F^4l^\nu n^\mu\nabla_\mu n_\nu.\label{eq:expansions}
\end{equation}
Apart from the expansions (\ref{eq:expansions}), the other important quantity is the evolution of one expansion along the other geodesic, that is, $\mathcal{L}_-\theta_+$ and $\mathcal{L}_+\theta_-$. 
Indeed one defines a \textbf{future outer trapping horizon} as a 3-surface $\mathcal{H}$ on which three properties hold~\cite{Hayward:1993wb}:\begin{enumerate}
\item ingoing light rays converge, $\theta_-\rvert_\mathcal{H}<0$,
\item outgoing light rays are parallel on the surface, $\theta_+\rvert_\mathcal{H}=0$,
\item in addition, $\mathcal{L}_-\theta_+\rvert_\mathcal{H}<0$, which implies that outgoing light rays are diverging outside and converging inside the surface.
\end{enumerate}
\textbf{The existence of a future outer trapping horizon therefore defines a BH in the non-stationary context}. 
\\

More generally, a \textbf{trapping horizon} $\mathcal{H}$, defined as a 3-surface on which $\theta_+\rvert_\mathcal{H}=0$, is characterized by two properties: \textbf{future or past; and outer or inner}. It is said to be future, respectively past, if $\theta_-\rvert_H<0$, respectively $\theta_-\rvert_\mathcal{H}>0$. It is outer, respectively inner, if $\mathcal{L}_-\theta_+\rvert_\mathcal{H}<0$, respectively $\mathcal{L}_-\theta_+\rvert_\mathcal{H}>0$. For example, in a Reissner-Nordstr\"om BH, the outer (event) horizon is a future outer trapping horizon, while the inner (Cauchy) horizon is a future inner trapping horizon. For a maximally-extended Schwarzschild BH, the BH horizon is a future outer trapping horizon, while the white hole horizon is a past outer trapping horizon\footnote{These results apply when the normalizations of $l$ and $n$ are chosen appropriately. The discussion which follows precisely gives counter-examples, which in turn enable to give rules for the proper normalization.}. In the above definitions, we have fixed the vanishing expansion to be $\theta_+$, but one could obviously rewrite these definitions symmetrically with this time $\theta_-\rvert_\mathcal{H}=0$.
\\

The existence and nature of trapping horizons do not depend on the normalization of $l$ and $n$, unless the normalization is singular. Indeed, if one rescales $l\to \gamma^2 l$, $n\to \beta^2 n$  with $\gamma$ and $\beta$ functions on spacetime, one can compute from (\ref{eq:scalar_prod}) and (\ref{eq:expansions}),
\begin{equation}
F^2\to\frac{F^2}{\gamma^2\beta^2},\quad \theta_+\to\frac{\theta_+}{\beta^2},\quad \theta_-\to \frac{\theta_-}{\gamma^2}\quad \mathcal{L}_-\theta_+\rvert_\mathcal{H}\to \frac{\mathcal{L}_-\theta_+}{\gamma^2\beta^2}\Big\rvert_H,\quad \mathcal{L}_+\theta_-\rvert_H\to\frac{\mathcal{L}_+\theta_-}{\gamma^2\beta^2}\Big\rvert_\mathcal{H}.\label{eq:rescalings}
\end{equation}
Therefore, the trapping horizons are identified unambiguously under such rescaling, unless one of the functions $\gamma$ or $\beta$ diverges or vanishes. 
\subsubsection{An illustrative example: Schwarzschild spacetime in coordinates of Painlev\'e-G\"ullstrand}
Let us illustrate this abstract discussion by a concrete example, namely Schwarzschild spacetime in Painlev\'e-G\"ullstrand coordinates,
\begin{equation}
\mathrm{d}s^2 = -\left(1-\frac{2M}{r}\right)\mathrm{d}\tau^2+2\sqrt{\frac{2M}{r}}\mathrm{d}\tau \mathrm{d}r+\mathrm{d}r^2+r^2\mathrm{d}\Omega^2.
\end{equation}
The spacetime has spherical symmetry, and it is therefore natural to look for a double-null foliation which respects this symmetry, that is to say, with $L=-\mathrm{d}u$ and $N=-\mathrm{d}v$ orthogonal to the coordinate vectors $\partial_\theta$ and $\partial_\varphi$. Up to global rescalings, a unique such pair exists,
\begin{equation}
L_\mu \mathrm{d}x^\mu = -\mathrm{d}\tau+\frac{\mathrm{d}r}{1-\sqrt{2M/r}},\quad N_\mu \mathrm{d}x^\mu = -\mathrm{d}\tau-\frac{\mathrm{d}r}{1+\sqrt{2M/r}},\label{eq:LN_Sch}
\end{equation}
with associated null coordinates
\begin{align}
u ={}&{} \tau -r-2\sqrt{2Mr}-4M\ln\left\lvert\sqrt{\frac{r}{2M}}-1\right\rvert,\\ v ={}&{} \tau + r+2\sqrt{2Mr}+4M\ln\left\lvert\sqrt{\frac{r}{2M}}+1\right\rvert.
\end{align}
First, if one does not rescale $L$ and $N$, that is, takes $l=L$ and $n=N$, one readily computes $F^2 = \left(r-2M\right)/\left(2r\right)$, $\theta_+ = \left(r-2M\right)/r^2$ and $\theta_- = -\theta_+$. Both expansions vanish at $r=2M$, while one expects an expansion to vanish and the other to be negative. \\

In view of (\ref{eq:rescalings}), one can even choose worse normalisations, like $l=\left(r-2M\right)L$ and $n=\left(r-2M\right)N$, leading to $F^2=1/\left(2r\left(r-2M\right)\right)$ and $\theta_+=1/r^2=-\theta_-$ which do not vanish at all. One might argue that in both these cases, $F^2$ either diverges or vanishes at the horizon, which may be the source of the problems. 
\\

However, problematic cases can arise even if $F^2$ is regular, say $F^2=1$, by setting for example $l = \left(r-2M\right)^2/\left(2r\right)L$ and $n=N/\left(r-2M\right)$. Then $\theta_+=\left(r-2M\right)^2/r^2$ and $\theta_-=-2/\left(r\left(r-2M\right)\right)$: $\theta_+$ vanishes appropriately at $r=2M$, but $\theta_-$ diverges towards $+\infty$ if $r=\left(2M\right)^-$ and towards $-\infty$ if $r=\left(2M\right)^+$. In this latest case, although $F^2$ and $l$ are regular, $n$ is diverging at $r=2M$.
\\

If, finally, one requires $F^2=1$, and $l$ and $n$ are both regular (apart of course at the spacetime singularity $r=0$), the Schwarzschild horizon is correctly identified as a future outer trapping horizon. If indeed
\begin{equation}
l=\frac{1}{\sqrt{2}}\left(1-\sqrt{\frac{2M}{r}}\right)L,\quad n = \frac{1}{\sqrt{2}}\left(1+\sqrt{\frac{2M}{r}}\right)N,\label{eq:ln_sch}
\end{equation}  
then $F^2=1$, both $l$ and $n$ are well-defined if $r\neq 0$, and one gets
\begin{equation}
\theta_\pm = \pm\frac{\sqrt{2}}{r}\left(1\mp\sqrt{\frac{2M}{r}}\right).\label{eq:exp_sch}
\end{equation}
This gives $\theta_-<0$ everywhere, so in particular at $r=2M$, which is the unique vanishing point of $\theta_+$. In addition, $\mathcal{L}_-\theta_+=-\left(2M\right)^{-2}<0$ at $r=2M$, so one properly identifies a future outer trapping horizon at $r=2M$.
\subsubsection{Rules for normalization}
Having in mind this example, as well as the expression (\ref{eq:expansions}) for the expansions and (\ref{eq:rescalings}) for their behaviour under rescaling, we give two requirements for the proper normalization of $l$ and $n$.\\
 
\textbf{First, we fix their scalar product (\ref{eq:scalar_prod}) to be finite, in particular, without loss of generality, we impose $F^2=1$}. 
This normalization ensures that $F^2$ does not vanish nor diverge, avoiding thus possible unphysical zeros of the expansions according to (\ref{eq:expansions}). 
\\

\textbf{Second, we impose $l$ and $n$ to be well-defined in the whole spacetime (apart of course from true curvature singularities)}. Indeed, a rescaling of the form $l\to\gamma^2 l$, $n\to n/\gamma^2$, although preserving $F^2=1$, could give unphysical vanishing or divergence of the expansions at the roots or poles of $\gamma^2$ according to (\ref{eq:rescalings}).\\

As a summary, \textbf{to study the trapping horizons of a spacetime, one first finds double null coordinates $u$ and $v$ associated to null, geodesic, future-directed one-forms $L=-\mathrm{d}u$ and $N=-\mathrm{d}v$, and then defines $l$ and $n$, proportional respectively to $L$ and $N$, such that $l$ and $n$:
\begin{enumerate}
\item have unit scalar product $g^{\mu\nu}l_\mu n_\nu=-1$,
\item are well-defined everywhere (apart from curvature singularities).
\end{enumerate}
With this convention, the expansions (\ref{eq:expansions}) are then given by
\begin{equation}
\theta_+ = \nabla^\mu l_\mu+n^\nu l^\mu\nabla_\mu l_\nu,\quad \theta_- = \nabla^\mu n_\mu+l^\nu n^\mu\nabla_\mu n_\nu.\label{eq:expansions_bis}
\end{equation}}Using (\ref{eq:expansions_bis}) along with $\mathcal{L}_-\theta_+ = n^\mu\partial_\mu\theta_+$ and $\mathcal{L}_+\theta_- = l^\mu\partial_\mu\theta_-$, one can identify appropriately the future/past, outer/inner trapping horizons $\mathcal{H}$ as defined by~\cite{Hayward:1993wb} and recalled above. These trapping horizons do not depend on rescalings of $l$ and $n$ preserving conditions 1 and 2 above. 
Indeed, the only possible rescaling preserving the scalar product is $l\to\gamma^2 l$, $n\to n/\gamma^2$, and (\ref{eq:rescalings}) shows that $\mathcal{L}_\pm\theta_\mp\rvert_\mathcal{H}$ are invariant, while $\theta_+\to \gamma^2\theta_+$ and $\theta_-\to \theta_-/\gamma^2$. 
Therefore, given a choice of $L=-\mathrm{d}u$ and $N=-\mathrm{d}v$, the requirements 1 and 2 enable to identify trapping horizons without ambiguity. \\

\textbf{The only ambiguity in the identification of trapping horizons of the spacetime thus regards the initial choice of double null geodesic coordinates $u$ and $v$ (that is, of $L$ and $N$), but we will see that, for the cases of interest regarding the conformal Kerr spacetime, such an ambiguity does not arise}. 
\subsubsection{Conformally-related metrics}
The formalism above is applicable to the case of conformal Kerr, where there are two conformally-related metrics on the same manifold, $\tilde{g}_{\mu\nu} = A^2g_{\mu\nu}$, with $A$ a function on spacetime. In this case, it is easy to relate the 2+2 foliation for $g_{\mu\nu}$ and $\tilde{g}_{\mu\nu}$ metrics. Indeed, the exact one-forms $L=-\mathrm{d}u$ and $N=-\mathrm{d}v$ are defined independently of the metric, while their null norm is preserved by the conformal change.
In addition, the one-forms $L=-\mathrm{d}u$ and $N=-\mathrm{d}v$ are also geodesic in the new metric: since the Christoffel coefficients are modified as~\cite{Carneiro:2004rt,Dabrowski:2008kx}
\begin{equation}
\tilde{\Gamma}^\lambda_{\nu\rho}=\Gamma^\lambda_{\nu\rho}+\frac{\partial_\nu A}{A}\delta^\lambda_\rho+\frac{\partial_\rho A}{A}\delta^\lambda_\nu-\frac{\partial_\sigma A}{A}g^{\sigma\lambda}g_{\nu\rho},
\end{equation} 
one computes
\begin{equation}
\tilde{g}^{\mu\nu}L_\mu\tilde{\nabla}_\nu L_\rho = A^{-2}g^{\mu\nu}L_\mu\left\lbrace \nabla_\nu L_\rho -\frac{\partial_\nu A}{A}L_\rho-\frac{\partial_\rho A}{A}L_\nu+\frac{\partial_\sigma A}{A}g^{\sigma\lambda}L_\lambda g_{\nu\rho}\right\rbrace.
\end{equation}
The second and fourth term compensate, while the third vanishes because $L$ is null. If $L$ is geodesic for $g_{\mu\nu}$, i.e. $g^{\mu\nu}L_\mu\nabla_\nu L_\rho\propto L_\rho$, one then has $\tilde{g}^{\mu\nu}L_\mu\widetilde{\nabla}_\nu L_\rho\propto L_\rho$. Therefore, $L$ is geodesic for $\tilde{g}_{\mu\nu}$, and the same holds for $N$. Thus, one shall use the same $L$ and $N$ for both spacetimes. \\

Then, if $l\propto L$ and $n\propto N$ are the associated normalized one-forms for $g_{\mu\nu}$, with $g^{\mu\nu}l_\mu n_\nu=-1$, then the one-forms $\tilde{l}$ and $\tilde{n}$ for $\tilde{g}_{\mu\nu}$ are $\tilde{l}=Al$, $\tilde{n}=An$, so that they are normalized with respect to the metric $\tilde{g}_{\mu\nu}$, $\tilde{g}^{\mu\nu}\widetilde{l}_\mu \widetilde{n}_\nu=-1$. \\

If the expansions $\theta_\pm$ for the metric $g_{\mu\nu}$ are known, see (\ref{eq:expansions_bis}), then the expansions $\tilde{\theta}_\pm$ for the conformally related metric read
\begin{equation}
\tilde{\theta}_+ = \frac{1}{A}\left(\theta_+ + \frac{2}{A}g^{\mu\nu}l_\mu\partial_\nu A\right),\quad \tilde{\theta}_- = \frac{1}{A}\left(\theta_- + \frac{2}{A}g^{\mu\nu}n_\mu\partial_\nu A\right).\label{eq:thetatilde}
\end{equation}
In the case of present interest, $g_{\mu\nu}$ is the stationary, axisymmetric Kerr metric, and $A=A(\tau)$ where $\tau$ is a conformal time. The equations $\tilde{\theta}_\pm\rvert_{\tilde{\mathcal{H}}}=0$, defining trapping horizons $\tilde{\mathcal{H}}$ of the conformal metric $\tilde{g}_{\mu\nu}$, yield 
\begin{equation}
\frac{\dot{A}}{A}\Big\rvert_{\tilde{\theta}_+=0} = -\frac{\theta_+}{2l^\tau},\quad \frac{\dot{A}}{A}\Big\rvert_{\tilde{\theta}_-=0} = -\frac{\theta_-}{2n^\tau},
\end{equation}
where of course $l^\tau=g^{\tau\mu}l_\mu$ and $n^\tau=g^{\tau\mu}n_\mu$. \textbf{The signs of $\tilde{\theta}_-$ when $\tilde{\theta}_+$ vanishes, and vice-versa, determine if the trapping horizons are future or past. They are given by
\begin{equation}
\tilde{\theta}_-\rvert_{\tilde{\theta}_+=0}=\frac{1}{A}\left(\theta_--\frac{n^\tau}{l^\tau}\theta_+\right),\quad \tilde{\theta}_+\rvert_{\tilde{\theta}_-=0}=\frac{1}{A}\left(\theta_+-\frac{l^\tau}{n^\tau}\theta_-\right).\label{eq:signs_theta}
\end{equation}}Assuming a power-law behaviour of the scale factor, $A(\tau)=A_0\left\lvert\tau\right\rvert^\alpha$, Eq. (\ref{eq:thetatilde}) implies that \textbf{the trapping horizons are given by the equations $\tau=\tau_\pm\left(r,\theta\right)$ where
\begin{equation}
\tau_+ = -\alpha\frac{2l^\tau}{\theta_+},\quad \tau_- = -\alpha\frac{2n^\tau}{\theta_-}.\label{eq:tau_plus_minus}
\end{equation}}Finally, for such a scale factor, the Lie derivatives for the conformal spacetime, $\widetilde{\mathcal{L}_\mp\theta_\pm}$, whose signs define whether the trapping horizons are inner or outer, are computed to be
\begin{align}
\widetilde{\mathcal{L}_-\theta_+}\rvert_{\widetilde{\theta}_+=0} ={}&{} \frac{1}{2\alpha l^\tau A^2}\left\lbrace 2\alpha \left(l^\tau n^i\partial_i \theta_+ - \theta_+ n^i\partial_i l^\tau\right)-n^\tau \theta_+^2\right\rbrace,\label{eq:signs_lie_1} \\
\widetilde{\mathcal{L}_+\theta_-}\rvert_{\widetilde{\theta}_-=0} ={}&{} \frac{1}{2\alpha n^\tau A^2}\left\lbrace 2\alpha \left(n^\tau l^i\partial_i \theta_- - \theta_- l^i\partial_i n^\tau\right)-l^\tau \theta_-^2\right\rbrace,\label{eq:signs_lie}
\end{align}
where $l^i = g^{i\mu}l_\mu$ and $n^i = g^{i\mu}n_\mu$ with index $i$ running over spatial coordinates. 
Eqs. (\ref{eq:signs_theta}-\ref{eq:signs_lie}) show that, \textbf{in order to find the trapping horizons and their nature for the conformal spacetime $\tilde{g}_{\mu\nu} = A\left(\tau\right)^2g_{\mu\nu}$, one only needs to compute the two expansions $\theta_+$ and $\theta_-$, and the two contravariant vectors $l^\mu$ and $n^\mu$, for the seed metric $g_{\mu\nu}$}.
\\

In the following paragraphs, we apply these results to the case of the conformal Kerr spacetime, starting with the easier case of vanishing rotation, $a=0$.

\subsection{Spherically symmetric case: Culetu spacetime}
\label{sub:culetu}
In this case, the seed metric (\ref{eq:kerr_before_conf}) is the Schwarzschild metric. From (\ref{eq:conf_kerr}-\ref{eq:phi_conf_kerr}), the solution for the conformal metric and the scalar field follows,
\begin{align}
\mathrm{d}\tilde{s}^2 ={}& A(\tau)^2\Biggl\{-\left(1-\frac{2M}{r}\right)\mathrm{d}\tau^2+2\sqrt{\frac{2M}{r}}\mathrm{d}\tau \mathrm{d}r+\mathrm{d}r^2+r^2\mathrm{d}\Omega^2\Biggr\},\label{eq:sch}\\
\phi ={}& q\tau.\label{eq:HJ_sch}
\end{align}
This is nothing but the Culetu metric (\ref{eq:culetu_first}). It was recently studied in detail in~\cite{Sato:2022yto}, for positive exponents of the scale factor, $\alpha>0$, in the context of GR. The spacetime solves the Einstein equations, sourced by an energy-momentum tensor violating standard energy conditions. This is in contrast with the context of the present paper, in which the Culetu spacetime is a vacuum solution of a DHOST theory. In order to be self-contained, we will re-derive the results for the trapping horizons in the case of Culetu spacetime studied in~\cite{Sato:2022yto} for $\alpha>0$. In the case of positive $\alpha$ our results fully agree with the findings of~\cite{Sato:2022yto}, while the results for negative $\alpha$ are new.
\\

\begin{figure}
\begin{subfigure}{8cm}
\includegraphics[width=\linewidth]{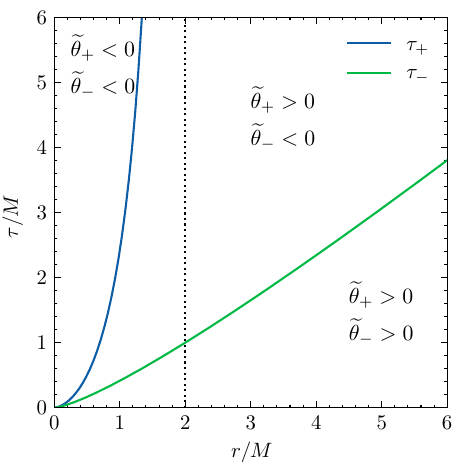}
\end{subfigure}
\begin{subfigure}{7.8cm}
\includegraphics[width=\linewidth]{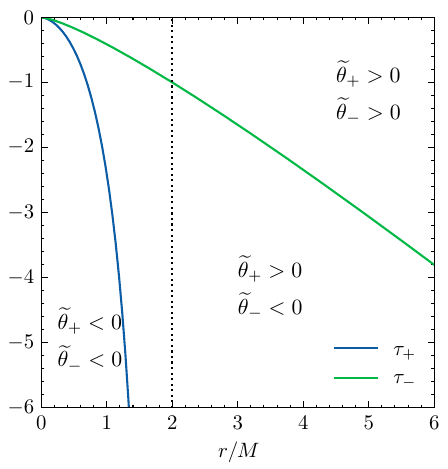}
\end{subfigure}
\caption{Trapping horizons $\tau=\tau_+(r)$ and $\tau=\tau_-(r)$ for the Culetu spacetime, where $r$ is the coordinate radius, for $\alpha=1$ (radiation, left) and $\alpha=-1$ (cosmological constant, right). $\tau_+$ diverges at $r=2M$ which is indicated by the dotted line.}
\label{fig:1}
\end{figure}

\begin{figure}
\begin{subfigure}{8cm}
\includegraphics[width=\linewidth]{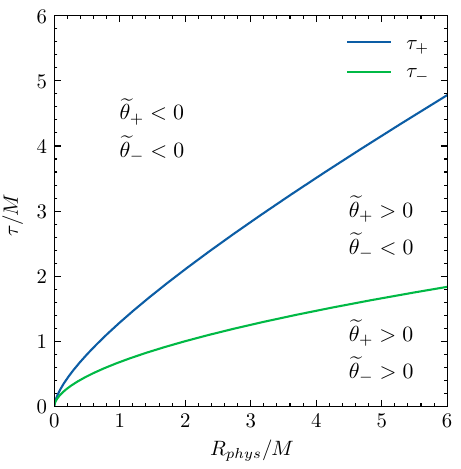}
\end{subfigure}
\begin{subfigure}{7.8cm}
\includegraphics[width=\linewidth]{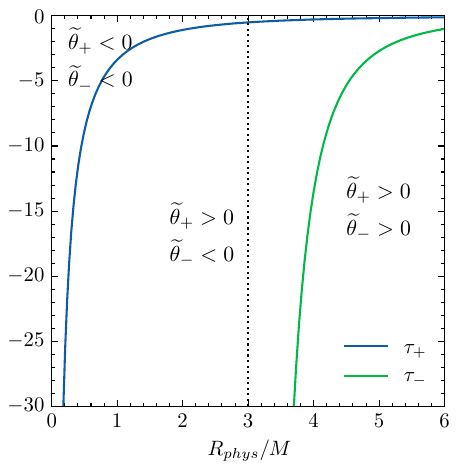}
\end{subfigure}
\caption{Trapping horizons $\tau=\tau_+\left(R_{\text{phys}}\right)$ and $\tau=\tau_-\left(R_{\text{phys}}\right)$ for the Culetu spacetime, where $R_{\text{phys}}=A\left(\tau\right)r$ is the physical radius, for $\alpha=1$ (radiation, left) and $\alpha=-1$ (cosmological constant, right). On this last plot, $\tau_-$ diverges at $R_{\text{phys}}=1/H_0$ which is indicated by the dotted line, where $H_0$ is the constant cosmological Hubble rate, taken for the plot to be $H_0=1/\left(3M\right)$. For the left plot, the factor $A_0$ appearing in $A=A_0\tau^\alpha$ is set to unity.}
\label{fig:2}
\end{figure}

As presented in the previous paragraph~\ref{subsec:foliations}, there is a unique (normalized and well-defined) double-null foliation preserving the spherical symmetry of the Schwarzschild spacetime, see Eqs. (\ref{eq:LN_Sch}) and (\ref{eq:ln_sch}). The expansions $\theta_\pm$ for the Schwarzschild seed spacetime are given by (\ref{eq:exp_sch}), while the seed contravariant vectors are readily computed to be
\begin{equation}
l^\mu \partial_\mu = \frac{1}{\sqrt{2}}\left[\partial_\tau + \left(1-\sqrt{\frac{2M}{r}}\right)\partial_r\right],\quad n^\mu \partial_\mu = \frac{1}{\sqrt{2}}\left[\partial_\tau - \left(1+\sqrt{\frac{2M}{r}}\right)\partial_r\right].
\end{equation}
Using the above expressions in (\ref{eq:signs_theta}) and (\ref{eq:tau_plus_minus}), one immediately gets that for $A=A_0\left\lvert\tau\right\rvert^\alpha$, the expansions $\tilde{\theta}_\pm$ vanish at $\tau=\tau_\pm$ with 
\begin{equation}
\tau_+ = -\alpha r\left(1-\sqrt{\frac{2M}{r}}\right)^{-1},\quad \tau_- = \alpha r \left(1+\sqrt{\frac{2M}{r}}\right)^{-1}.\label{eq:taupm}
\end{equation}
Since the coordinate $\tau$ has by definition the same sign as the exponent $\alpha$, the trapping horizon $\tau=\tau_+$ must have $r<2M$, while the trapping horizon $\tau=\tau_-$ extends for all $r$. We plot these trapping horizons in Fig.~\ref{fig:1} for positive and negative $\alpha$. 
The graph for $\alpha=-1$ shows that the radial coordinate $r$ for both trapping horizons is shrinking to zero at  $\tau=0$ (future infinity). 
This happens because $r$ is not a physical distance but a comoving coordinate, while the physical radius (as measured by a far away observer) is 
\begin{equation}
R_{\text{phys}}=A\left(\tau\right)r. \label{eq:Rphys}
\end{equation}
We thus present in Fig.~\ref{fig:2} the trapping horizons in terms of the physical radius $R_{\text{phys}}$. 
Both physical horizons expand for all presented cases. 
Now, let us examine the nature of these trapping horizons. Using (\ref{eq:signs_theta}),
\begin{equation}
\tilde{\theta}_-\rvert_{\tau_+} = -\frac{2\sqrt{2}}{Ar}<0,\quad \tilde{\theta}_+\rvert_{\tau_-}= \frac{2\sqrt{2}}{Ar}>0,
\end{equation}
showing that $\tau=\tau_+$ is a future trapping horizon, while $\tau=\tau_-$ is a past trapping horizon. Finally, (\ref{eq:signs_lie_1}-\ref{eq:signs_lie}) give
\begin{align}
\widetilde{\mathcal{L}_-\theta_+}\rvert_{\tau_+} ={}&{} \frac{1}{\alpha A^2r^2}\left[\alpha\left(1+\sqrt{\frac{2M}{r}}\right)\left(1-\frac{3}{2}\sqrt{\frac{2M}{r}}\right)-\left(1-\sqrt{\frac{2M}{r}}\right)^2\right],\label{eq:lmtp_culetu}\\ 
\widetilde{\mathcal{L}_+\theta_-}\rvert_{\tau_-} ={}&{} \frac{1}{\alpha A^2r^2}\left[\alpha\left(1-\sqrt{\frac{2M}{r}}\right)\left(1+\frac{3}{2}\sqrt{\frac{2M}{r}}\right)-\left(1+\sqrt{\frac{2M}{r}}\right)^2\right].\label{eq:lptm_culetu}
\end{align}
Let us for the moment focus of the case of decelerating universe, $\alpha>0$. In this case, $\tau=\tau_+$ exists for $r<2M$. Eq (\ref{eq:lmtp_culetu}) thus implies that $\widetilde{\mathcal{L}_-\theta_+}\rvert_{\tau_+}<0$, \textbf{therefore $\tau=\tau_+$ is a future outer trapping horizon, and the spacetime is indeed a cosmological BH}. 
As regards the past trapping horizon $\tau=\tau_-$, there are two separate cases depending on whether $\alpha>1$ or $\alpha<1$. It is easy to show from (\ref{eq:lptm_culetu}) that $\widetilde{\mathcal{L}_+\theta_-}\rvert_{\tau_-}$ has a zero at 
\begin{equation}
r_{1} = \frac{8+\alpha\left(13\alpha-12\right)+\left(4-\alpha\right)\sqrt{\alpha\left(25\alpha-16\right)}}{4\left(\alpha-1\right)^2}M,
\end{equation}
if and only if $\alpha>1$. Therefore, for $0<\alpha\leq 1$, the past trapping horizon is outer; while for $\alpha>1$, it is outer for $r<r_{1}$ and inner for $r>r_{1}$. 
\\

Let us now describe the case of accelerating universe, $\alpha<0$. On the one hand, for $\alpha<-2/3$ (including asymptotically dS case $\alpha=-1$), $\tau=\tau_+$ is a future outer trapping horizon (BH horizon), while the past trapping horizon $\tau=\tau_-$ (cosmological horizon) is outer for $r<r_{2}$ and inner for $r>r_{2}$ where
\begin{equation}
r_{2} = \frac{8+\alpha\left(13\alpha-12\right)-\left(4-\alpha\right)\sqrt{\alpha\left(25\alpha-16\right)}}{4\left(\alpha-1\right)^2}M.
\end{equation}
On the other hand, for $0>\alpha>-2/3$, $\tau=\tau_-$ is a past inner trapping horizon, while the future trapping horizon $\tau=\tau_+$ is inner for $r<r_{2}$ and outer for $r>r_{2}$. In this case, $r_{2}<2M$, so there is indeed a region $2M>r>r_{2}$ where $\tau=\tau_+$ is future outer trapping horizon. 
Note that in this case the nature of the future trapping horizon changes with time, being inner at early times and outer later. It is the only range of $\alpha$ for which this happens. Finally, for $\alpha=-2/3$, $\tau=\tau_+$ is future outer and $\tau=\tau_-$ is past inner. 
\\

\textbf{In a word, for all values of $\alpha$, the Culetu spacetime possesses a future outer trapping horizon and is therefore a cosmological BH}. We now turn to the general case of conformal Kerr spacetime (\ref{eq:good_coord}) with non-zero rotation. 
\subsection{Conformal Kerr spacetime}\label{subsec:conf_kerr_study}
First and foremost, we identify 2+2 foliations of Kerr spacetime in Boyer-Lindquist coordinates $(t,r,\theta,\varphi)$, and then write them in coordinates $(\tau,r,\theta,\varphi_+)$ of (\ref{eq:good_coord}). As seen just above, in the spherically-symmetric case, there is a unique such foliation, provided one requires it to respect the spherical symmetry of the spacetime. When rotation is present, the situation becomes \textit{a priori} more intricate, but will turn out to lead again to a unique foliation with natural requirements.
\\

Following paragraph~\ref{subsec:foliations}, we start by looking for an exact, null geodesic one-form $p=p_\mu \mathrm{d}x^\mu=-\mathrm{d}u$. 
Carter showed~\cite{Carter:1968rr} that $p$ is given by the following expression\footnote{Strictly speaking, this result holds true if the integral geodesic curves with tangent vector $p^\mu=g^{\mu\nu}p_\nu$ have affine parameterization, i.e. $p^\mu\nabla_\mu p^\nu=0$. This is always the case up to rescaling of $p$. Note also that the two $\pm$ of Eq. (\ref{eq:p}) are independent.}
\begin{equation}
p_\mu \mathrm{d}x^\mu = -E \mathrm{d}t+L_z\mathrm{d}\varphi\pm\frac{\sqrt{\mathcal{R}}}{\Delta}\mathrm{d}r\pm\sqrt{\Theta}\,\mathrm{d}\theta, \label{eq:p}
\end{equation}
where
\begin{equation}
\mathcal{R} \equiv \left[E\left(r^2+a^2\right)-aL_z\right]^2-\Delta K,\quad\Theta \equiv K - \sin^2\theta\left(aE-\frac{L_z}{\sin^2\theta}\right)^2.\label{eq:RK}
\end{equation}
There are three constants of motion: the energy $E$, the angular momentum $L_z$, and the \textbf{Carter's constant} $K$, which guarantees integrability of the geodesic equations.  
Note that $p$ respects the symmetries of the Kerr spacetime, that is, $p_\mu=p_\mu(r,\theta)$. Very importantly, \textbf{along each individual geodesic having $p^\mu=g^{\mu\nu}p_\nu$ as tangent vector, $E$, $L_z$ and $K$ are constants. Nevertheless, the geodesic congruence as a whole may \textit{a priori} have $E$, $L_z$ and $K$ which depend on $r$ and $\theta$ (not however on $t$ or $\varphi$ due to the axial symmetry)}. For example, for the well-known 'principal null congruence' of Kerr, see e.g.~\cite{Wald:1984rg}, $E$ is constant throughout spacetime, but $L_z = a E \sin^2\theta$ depends on the angle $\theta$. $L_z$ is still a constant along each geodesic of the principal null congruence, since each geodesic lies in a plane of constant $\theta$. 
\\

Having the double-null construction of paragraph~\ref{subsec:foliations} in mind, $p$ must be an exact form, $\mathrm{d}p=0$. Writing this requirement explicitly for each component leads to:
\begin{equation}
p_t=\text{constant},\quad p_\varphi=\text{constant},\quad\partial_r p_\theta = \partial_\theta p_r. \label{eq:dpegal0}
\end{equation} 
Since $p_t$ and $p_\varphi$ are constants, $E$ and $L_z$ must be constants throughout the spacetime (note the difference with respect to the principal null congruence).
Given the expression of $\Theta$, regularity at the poles $\theta=0,\pi$ then implies $L_z=0$. One can also normalize the affine parameter to have $E=1$, without loss of generality. It will be useful for the following to introduce an auxiliary function $k$, related to Carter's 'constant' $K$ as
\begin{equation}
K(r,\theta) \equiv k^2(r,\theta)+a^2\sin^2\theta. \label{eq:Kk}
\end{equation}
Taking into account the conditions above, from (\ref{eq:p}) and (\ref{eq:RK}), we identify the pair $(L,N)$, with $L$ outgoing and $N$ ingoing, of exact, null, geodesic congruence in Kerr spacetime\footnote{One could as well have $-k\,\mathrm{d}\theta$ in $L$ and $+k\,\mathrm{d}\theta$ in $N$. This amounts to a change $k\to -k$ coming from the fact that $k$ is defined through its square, Eq. (\ref{eq:Kk}), and from the fact that $\pm\sqrt{\Theta}=\pm\left\lvert k\right\rvert$. This choice is of course irrelevant.},
\begin{equation}
L = - \mathrm{d}t+\frac{\sqrt{\mathcal{R}}}{\Delta}\mathrm{d}r+k\,\mathrm{d}\theta,\quad N = - \mathrm{d}t-\frac{\sqrt{\mathcal{R}}}{\Delta}\mathrm{d}r-k\,\mathrm{d}\theta,\label{eq:nullpairkerr}
\end{equation}
where
\begin{equation}
\mathcal{R} = \left(r^2+a^2\right)^2-\Delta \left(k^2+a^2\sin^2\theta\right).\label{eq:beforepde}
\end{equation}
Moreover, \textbf{the third condition in Eq. (\ref{eq:dpegal0}) amounts to imposing the following partial differential equation on $k(r,\theta)$,
\begin{equation}
\sqrt{\mathcal{R}}\partial_r k + k\partial_\theta k = -a^2\sin\theta\cos\theta. \label{eq:pde}
\end{equation}}The fact that $L$ and $N$ are exact forms, $L = -\mathrm{d}u$ and $N = -\mathrm{d}v$, can be explicitly checked. The associated null coordinates $u$ and $v$ are~\cite{Arganaraz:2021fpm}
\begin{equation}
u=t-r_s,\quad v=t+r_s,\quad r_s\left(r,\theta\right) \equiv \int^r \frac{\sqrt{\mathcal{R}\left(r',\theta=0\right)}}{\Delta\left(r'\right)}\mathrm{d}r'+\int_0^\theta k\left(r,\theta'\right)\mathrm{d}\theta'.
\end{equation}
The PDE (\ref{eq:pde}) does not define $k=k(r,\theta)$ uniquely, since one needs to supply a boundary condition. 
An obvious solution of the PDE is $k=a\cos\theta$, giving a Carter's function $K(r,\theta)$ which is constant throughout spacetime, $K=a^2$, see (\ref{eq:Kk}). This corresponds in fact to a choice of foliation made by Hayward in 2004~\cite{Hayward:2004ih} (hence the following subscript 'H'),
\begin{equation}
L_{\text{H}} = - \mathrm{d}t+\frac{\sqrt{\mathcal{R}_{\text{H}}}}{\Delta}\mathrm{d}r+a\cos\theta\,\mathrm{d}\theta,\quad N_{\text{H}} = - \mathrm{d}t-\frac{\sqrt{\mathcal{R}_{\text{H}}}}{\Delta}\mathrm{d}r-a\cos\theta\,\mathrm{d}\theta,\label{eq:hay_choice}
\end{equation}
where 
\begin{equation}
\mathcal{R}_{\text{H}} = \left(r^2+a^2\right)^2-\Delta a^2.
\end{equation}
This foliation of Kerr by Hayward is however singular at the poles, as was pointed out recently by two authors, Arga\~naraz and Moreschi~\cite{Arganaraz:2021fpm}. These two authors considered a different choice~\cite{Arganaraz:2021fpm}, with the requirement that the null geodesics be orthogonal to the two-sphere of radius $r$ when $r\to\infty$. This is in fact a natural requirement, since it respects the asymptotic spherical symmetry that the Kerr spacetime possesses when $r\to\infty$. Moreover, Arga\~naraz and Moreschi proved that their foliation is regular at the poles $\theta=0,\pi$, as we will see.  
\\

\begin{figure}
\begin{subfigure}{8cm}
\includegraphics[width=\linewidth]{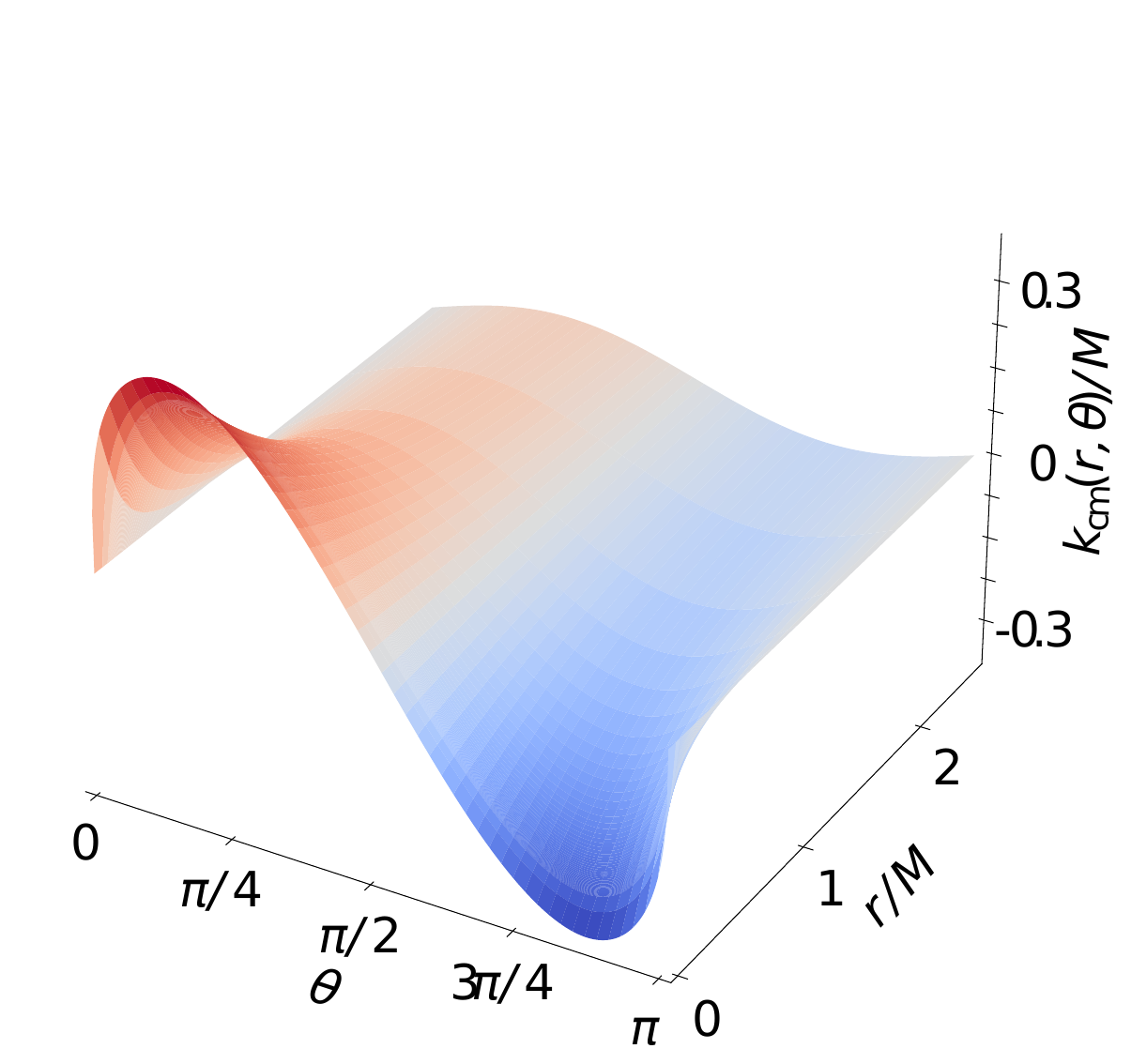}
\end{subfigure}
\begin{subfigure}{8cm}
\includegraphics[width=\linewidth]{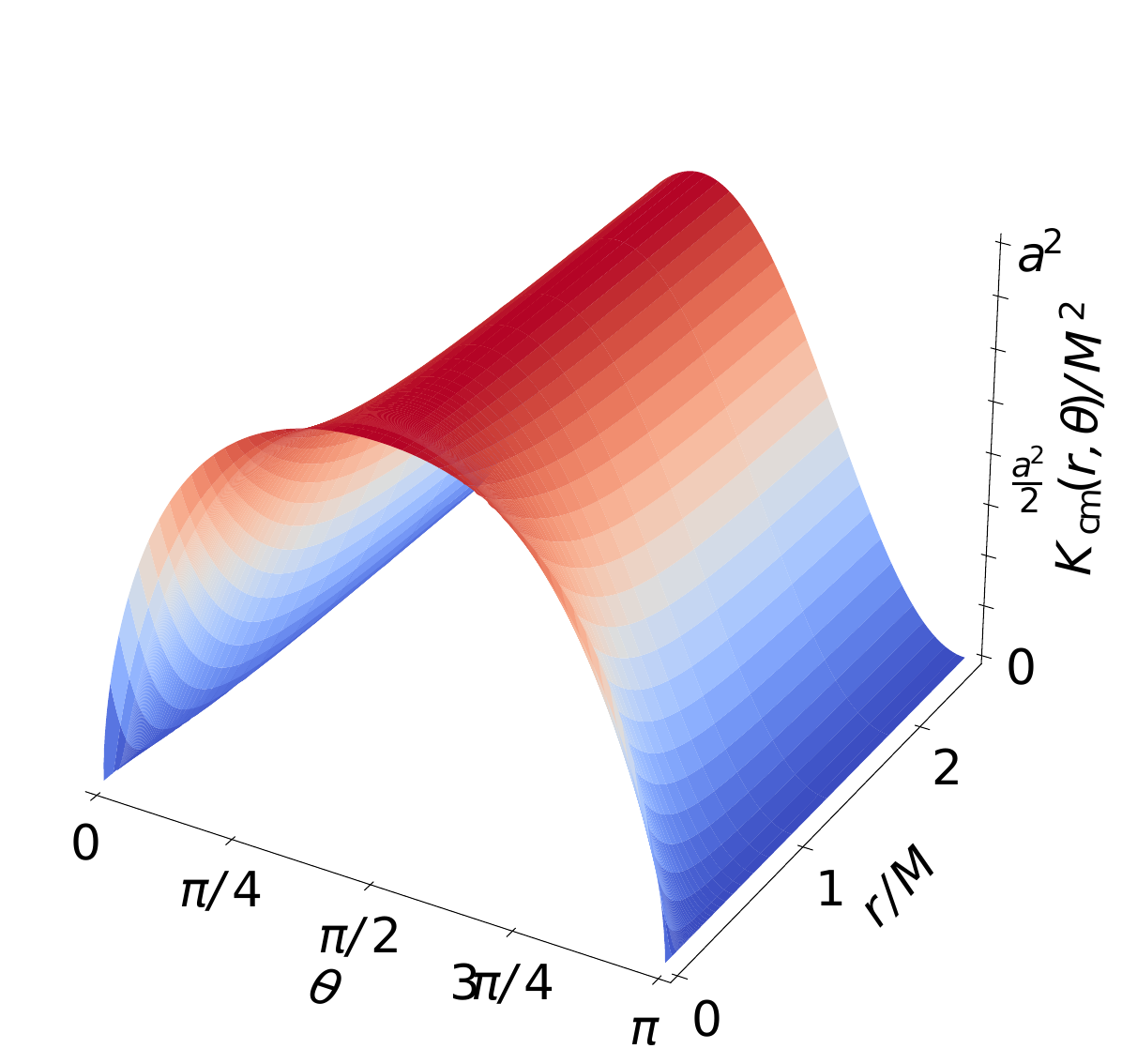}
\end{subfigure}
\caption{Function $k_{\text{cm}}(r,\theta)$ and Carter's function $K_{\text{cm}}(r,\theta)=k^2_\text{cm}\left(r,\theta\right)+a^2\sin^2\theta$ for $a=0.5M$ as a result of numerical integration of Eq. (\ref{eq:pde}).}
\label{fig:7}
\end{figure}

We will follow their vocabulary and refer to this choice of coordinates as \textbf{'center-of-mass' null coordinates} (hence the following subscript 'cm'). Because of Eq. (\ref{eq:nullpairkerr}), the requirement of orthogonality of $L$ and $N$ to $\partial_\theta$ as $r\to\infty$ is equivalent to the following asymptotic boundary condition on $k$,
\begin{equation}
\lim_{r\to\infty}k_{\text{cm}}(r,\theta)=0.\label{eq:arg_choice}
\end{equation}
The PDE (\ref{eq:pde}), along with the boundary condition (\ref{eq:arg_choice}), imply that $k$ is vanishing at the poles (and thus the Carter's function $K$ as well),
\begin{equation}
k_{\text{cm}}\left(r,\theta=0\right)=k_{\text{cm}}\left(r,\theta=\pi\right)=0,\quad K_{\text{cm}}\left(r,\theta=0\right)=K_{\text{cm}}\left(r,\pi\right)=0.\label{eq:vanish_poles}
\end{equation}
The function $k_{\text{cm}}$ is antisymmetric with respect to the equatorial plane, $k_{\text{cm}}(r,\pi-\theta)=-k_{\text{cm}}(r,\theta)$. To prove this, one notices that the function $-k_{\text{cm}}(r,\pi-\theta)$ satisfies the same PDE (\ref{eq:pde}) and boundary condition (\ref{eq:arg_choice}) as $k_{\text{cm}}(r,\theta)$, thus the solutions must coincide. This immediately leads to an unsurprising symmetry of the Carter's function with respect to the equatorial plane, $K_{\text{cm}}(r,\pi-\theta)=K_{\text{cm}}(r,\theta)$. 
This also implies that, at the equatorial plane,
\begin{equation}
k_{\text{cm}}(r,\theta=\pi/2)=0,\quad K_{\text{cm}}(r,\theta=\pi/2)=a^2.\label{eq:kKpi2}
\end{equation}
The functions $k_{\text{cm}}$ and $K_{\text{cm}}$ can be found by numerical integration. This enables to apply the center-of-mass double-null foliation to concrete problems in Kerr spacetime, see e.g.~\cite{Arganaraz:2022mks,Arganaraz:2023eeq}. The numerical integration of the PDE (\ref{eq:pde}) can be carried out for all $(r,\theta)\in\left[0,+\infty\right)\times\left[0,\pi\right]$. We present here our own numerical integration, with a Runge-Kutta 4 algorithm, in Fig.~\ref{fig:7}. All figures from now on are presented for $a=0.5M$, but of course, all inferences we draw from numerical integration have been verified for numerous angular momenta $a\in\left[0,M\right]$.
For instance, we see that $K_\text{cm}\left(r,\theta\neq\pi/2\right)<a^2$ (right panel of Fig.~\ref{fig:7}). Along with Eq. (\ref{eq:kKpi2}), this leads to
\begin{equation}
K_{\text{cm}}(r,\theta)\leq a^2\quad\text{with equality if and only if }\theta=\pi/2,\label{eq:property}
\end{equation}
which implies that
\begin{equation}
\Sigma - k_\text{cm}^2 \geq r^2\quad\text{with equality if and only if }\theta=\pi/2,\label{eq:property_2}
\end{equation}
and that, when $\Delta\geq 0$,
\begin{equation}
\mathcal{R}_\text{cm}\geq r^4+r^2a^2+2Mra^2\quad\text{with equality if and only if }\theta=\pi/2.\label{eq:property_3}
\end{equation}
These results seem technical, but will be useful for the following.
\\

Having established that the choice of boundary condition (\ref{eq:arg_choice}) indeed leads to a well-defined $k(r,\theta)$, let us return to the pair of general null one-forms $\left(L,N\right)$ which satisfy (\ref{eq:nullpairkerr}) and (\ref{eq:pde}). For the moment, we do not assume $k=k_{\text{cm}}(r,\theta)$, rather, we are going to explain why $k=k_{\text{cm}}(r,\theta)$ is the only relevant choice, and only after this will we set $k=k_{\text{cm}}(r,\theta)$.\\

In coordinates $\left(\tau,r,\theta,\varphi_+\right)$ of (\ref{eq:good_coord}), the pair  $\left(L,N\right)$ reads
\begin{align}
L ={}&{} -\mathrm{d}\tau +\frac{\sqrt{\mathcal{R}}+\sqrt{2Mr\left(r^2+a^2\right)}}{\Delta}\mathrm{d}r+k\,\mathrm{d}\theta,\\ N ={}&{} -\mathrm{d}\tau -\frac{\sqrt{\mathcal{R}}-\sqrt{2Mr\left(r^2+a^2\right)}}{\Delta}\mathrm{d}r-k\,\mathrm{d}\theta.
\end{align}
The associated null pair $(l,n)$, with scalar product equal to $-1$ and well-defined in the whole Kerr spacetime (apart from the curvature singularity) is
\begin{align}
l ={}&{} \left(\frac{\Sigma\left(\Sigma-k^2\right)}{2\Upsilon}\right)^{1/2}\frac{L\Delta}{\sqrt{\mathcal{R}}+\sqrt{2Mr\left(r^2+a^2\right)}},\\ n ={}&{} \left(\frac{\Sigma\left(\Sigma-k^2\right)}{2\Upsilon}\right)^{1/2}\frac{N\Delta}{\sqrt{\mathcal{R}}-\sqrt{2Mr\left(r^2+a^2\right)}}.
\end{align}
The well-definedness of $l$ and $n$ follows from the fact that $\sqrt{\mathcal{R}}-\sqrt{2Mr\left(r^2+a^2\right)}$ vanishes when $\Delta$ does, because of Eq. (\ref{eq:beforepde}) and the fact that $\Delta=r^2+a^2-2Mr$. The expansions in the Kerr spacetime are readily computed to be
\begin{equation}
\theta_\pm = \pm\left(\sqrt{\mathcal{R}}\mp \sqrt{2Mr\left(r^2+a^2\right)}\right)\mathfrak{F},\label{eq:thetapluskerr}
\end{equation}
where 
\begin{equation}
\mathfrak{F} \equiv \frac{1}{\sqrt{8\Upsilon\mathcal{R}\Sigma\left(\Sigma-k^2\right)}}\Biggl\{4r\left(r^2+a^2\right)+2\left(M-r\right)K-\Delta\partial_r K+2\sqrt{\mathcal{R}}\left[k\cot\theta+\partial_\theta k\right]\Biggr\}.\label{eq:F}
\end{equation}
For vanishing rotation, $a=0$, one has $K=0=k$, and recovers the expansions of Schwarzschild, Eq. (\ref{eq:exp_sch}). In the case of Schwarzschild, the expansions (\ref{eq:exp_sch}) are diverging only at the curvature singularity $r=0$. We thus expect the expansions of Kerr spacetime to diverge only at the curvature singularity of Kerr, which is at $\Sigma=0$, that is, $r=0$ and $\theta=\pi/2$. However, because of the term $k\cot\theta$ in (\ref{eq:F}), the expansions (\ref{eq:thetapluskerr}) diverge at the poles $\theta=0,\pi$, unless $k$ (and thus $K$) vanish at the poles. 
\\

In particular, for the null foliations considered  by Hayward, Eq. (\ref{eq:hay_choice}), the expansions diverge at the poles, since $K_\text{H}=a^2$ as we noted above. 
More generally, as it was underlined by Arga\~naraz and Moreschi~\cite{Arganaraz:2021fpm}, \textbf{all previously proposed double null coordinates for Kerr spacetime~\cite{SJFletcher2003,PhysRevD.73.084023,Pretorius:1998sf} suffer from the same problem as the ones of Hayward}, due to a conical singularity (along the axis of symmetry of Kerr) of the spacelike surfaces $\mathcal{S}$ (see beginning of paragraph~\ref{subsec:foliations}) induced by the 2+2 foliation.
\\

On the other hand, the expansions $\theta_\pm$ are well-defined at the poles for the center-of-mass foliation of~\cite{Arganaraz:2021fpm}, i.e. when $k=k_\text{cm}$ and $K=K_\text{cm}$, thanks to Eq. (\ref{eq:vanish_poles}). Moreover, in this case, the denominator of $\mathfrak{F}$ vanishes if and only if $\Sigma=0$ ($\mathfrak{F}$ has a double pole there). This can be inferred from the properties (\ref{eq:property_2}-\ref{eq:property_3}) and the definition of $\Upsilon$, Eq. (\ref{eq:notations_new}). 
For the center-of-mass foliation, $\mathfrak{F}$ thus diverges at $r=0$ only for $\theta=\pi/2$. This ensures that, \textbf{for the center-of-mass double-null foliation, the expansions $\theta_\pm$ are finite in all spacetime but the curvature singularity $\Sigma=0$}.
\\

\textbf{In the rest of the discussion, we therefore focus on the center-of-mass double null coordinates, setting from now on $\left(l,n\right)=\left(l_{\text{cm}},n_{\text{cm}}\right)$. 
For brevity, we omit the subscript 'cm' in the following, keeping in mind that we chose the center-of-mass double null coordinates once and for all}.
\\

The property (\ref{eq:property}) implies that $\sqrt{\mathcal{R}}-\sqrt{2Mr\left(r^2+a^2\right)}$ in the expression for the expansions (\ref{eq:thetapluskerr}) has the same sign as $\Delta$ and vanishes if and only if $\Delta$ vanishes\footnote{The expression $\sqrt{\mathcal{R}}-\sqrt{2Mr\left(r^2+a^2\right)}$ also vanishes at the curvature singularity $r=0$ and $\theta=\pi/2$. However $\theta_+$ diverges at the curvature singularity anyway, because $\mathfrak{F}$ has a double pole there.}. 
Furthermore, one can show numerically that the factor $\mathfrak{F}$ is positive throughout spacetime. 
One therefore concludes from Eq. (\ref{eq:thetapluskerr}) that $\theta_+$ vanishes at the two roots of $\Delta$ (i.e. at the outer and inner horizons of Kerr spacetime, $r_+>r_-$), and it is negative for $r_-<r<r_+$ and positive for $r<r_-$ or $r>r_+$. On the other hand, $\theta_-$ is negative in the whole spacetime. Therefore, the 2+2 formalism correctly identifies $r=r_+$ and $r=r_-$ as future trapping horizons.\\
 
\begin{figure}
\begin{subfigure}{7.8cm}
\includegraphics[width=\linewidth]{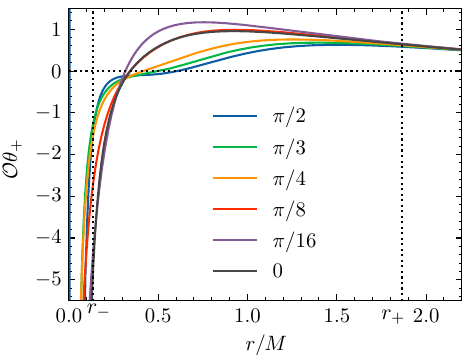}
\end{subfigure}
\begin{subfigure}{8.2cm}
\includegraphics[width=\linewidth]{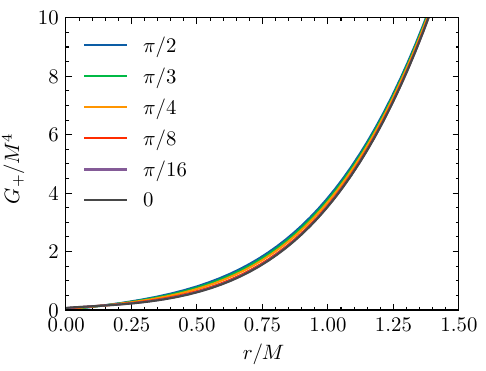}
\end{subfigure}
\caption{Two plots as functions of $r/M$ and for different values of angle $\theta$ indicated by the colors, for the Kerr metric with $a=0.5M$. Left plot: the quantity $\mathcal{O}\theta_+$ of Eqs. (\ref{eq:lmoinsthetaplus}-\ref{eq:op_o}) is seen to be positive at $r_+$ and negative at $r_-$, which are indicated by dotted lines. Right plot: the function $G_+$ of Eq. (\ref{eq:gpm}) is seen to be positive.}
\label{fig:3}
\end{figure}

To see if each of the horizon is outer or inner, we compute the Lie derivative of the expansion,
\begin{equation}
\mathcal{L}_-\theta_+ = -\frac{\sqrt{\mathcal{R}}+\sqrt{2Mr\left(r^2+a^2\right)}}{\sqrt{2\Upsilon\Sigma\left(\Sigma-k^2\right)}}\mathcal{O}\theta_+\label{eq:lmoinsthetaplus}
\end{equation}
where for brevity, we introduce the differential operator $\mathcal{O}$,
\begin{equation}
\mathcal{O} \equiv \sqrt{\mathcal{R}}\partial_r+k\partial_\theta.\label{eq:op_o}
\end{equation}
$\mathcal{L}_-\theta_+$ is seen from (\ref{eq:lmoinsthetaplus}) to have opposite sign to $\mathcal{O}\theta_+$. The plot of $\mathcal{O}\theta_+$, left panel of Fig.~\ref{fig:3}, shows that $\mathcal{O}\theta_+$ is positive at $r=r_+$ and negative at $r=r_-$. This implies that for Kerr, $r=r_+$ is a future outer trapping horizon and $r=r_-$ is a future inner trapping horizon, as it should be.
\\

\begin{figure}
\centering
\includegraphics[width=0.6\textwidth]{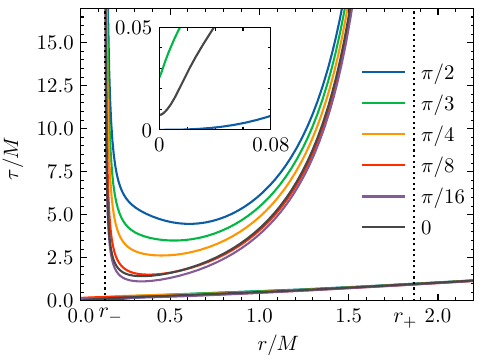}
\caption{Trapping horizons $\tau=\tau_+\left(r,\theta\right)$ and $\tau=\tau_-\left(r,\theta\right)$ for different fixed values of angle $\theta$ indicated by the colors, for the conformal Kerr spacetime with $a=0.5M$, where $r$ is the coordinate radius. $\tau_+$ (curves with a minimum) diverges at $r=r_+$ and $r=r_-$ which are indicated by the dotted lines. $\tau_-$ (cluster of bottom curves) vanishes only at $r=0$ and $\theta=\pi/2$, as is exemplified by the zoom on the curves $\theta=\pi/2$, $\pi/3$ and $0$ near $\tau_-=0$. The plot is obtained for an exponent $\alpha=1$ in the scale factor.}
\label{fig:4}
\end{figure}

In a nutshell, \textbf{for the seed Kerr metric, the center-of-mass double-null foliation yields regular expressions for the expansions, and properly identifies the outer and inner horizons}. We now turn to the study of the conformal Kerr spacetime (\ref{eq:good_coord}). We need the seed contravariant vectors,
\begin{align}
l^\mu\partial_\mu ={}&{} \frac{\sqrt{\mathcal{R}}-\sqrt{2Mr\left(r^2+a^2\right)}}{\sqrt{2\Upsilon\Sigma\left(\Sigma-k^2\right)}}\left\lbrace \frac{G_+}{\Delta}\partial_\tau+\mathcal{O}\right\rbrace+l^{\varphi_+}\partial_{\varphi_+},\\
n^\mu\partial_\mu ={}&{} \frac{\sqrt{\mathcal{R}}+\sqrt{2Mr\left(r^2+a^2\right)}}{\sqrt{2\Upsilon\Sigma\left(\Sigma-k^2\right)}}\left\lbrace \frac{G_-}{\Delta}\partial_\tau-\mathcal{O}\right\rbrace+n^{\varphi_+}\partial_{\varphi_+},
\end{align}
where, for brevity, we used the expression $\mathcal{O}$ of Eq. (\ref{eq:op_o}), and we have not explicitly written the components along $\varphi_+$ since they do not play any role in the following calculations. In the above expressions, we have also introduced
\begin{equation}
G_\pm = \Delta\Sigma\pm\sqrt{2Mr\left(r^2+a^2\right)}\left[\sqrt{\mathcal{R}}\pm\sqrt{2Mr\left(r^2+a^2\right)}\right].\label{eq:gpm}
\end{equation}
It is not difficult to prove analytically that the ratio $G_-/\Delta$ is well-defined and positive. Moreover, numerical analysis, see right plot of Fig.~\ref{fig:3}, shows that $G_+$ is positive. Using Eq. (\ref{eq:tau_plus_minus}), assuming $A=A_0\left\lvert\tau\right\rvert^\alpha$, we find that the expansions $\tilde{\theta}_\pm$ of the conformal Kerr spacetime vanish at $\tau=\tau_\pm$ with
\begin{equation}
\tau_\pm = \frac{-\alpha\sqrt{2}}{\sqrt{\Upsilon\Sigma\left(\Sigma-k^2\right)}}\frac{\left[\sqrt{\mathcal{R}}\mp \sqrt{2Mr\left(r^2+a^2\right)}\right]G_\pm}{\theta_\pm\Delta}. 
\label{eq:tau_conf_kerr}
\end{equation}
We have seen above that $\theta_-$ is negative in the whole spacetime. In addition, $\tau$ has the same sign as $\alpha$ by definition. Consequently, the trapping horizon $\tau=\tau_-\left(r,\theta\right)$ exists for all $r$ and $\theta$. On the other hand, we have seen that $\theta_+$ is negative only for $r_-<r<r_+$. As a consequence, the trapping horizon $\tau=\tau_+\left(r,\theta\right)$ is located between $r_-$ and $r_+$. Fig~\ref{fig:4} shows $\tau_+$ and $\tau_-$ in terms of the comoving radius $r$ for $\alpha=1$. 
The profiles of $\tau_\pm$ for other values of $\alpha$ differ from the one plotted in Fig.~\ref{fig:4} by a factor $\alpha$, as it is clear from (\ref{eq:tau_conf_kerr}).
As we underlined above, for the seed Kerr metric, $\theta_\pm$ diverge only when $\Sigma=0$, i.e. $r=0$ and $\theta=\pi/2$. Therefore, given Eq. (\ref{eq:tau_conf_kerr}), $\tau_+$ does not vanish, while $\tau_-$ vanishes if and only if $\Sigma=0$. 
\\

Note here the difference with the spherically-symmetric case: both Culetu and conformal Kerr spacetimes have a curvature singularity at $r=0$, see Eq. (\ref{eq:ric_conf_kerr}). However, for Culetu metric, $\tau_+$ and $\tau_-$ vanish at $r=0$, while for conformal Kerr metric, $\tau_+$ does not vanish and $\tau_-$ vanishes only at $r=0$ and $\theta=\pi/2$. \\

\begin{figure}
\begin{subfigure}{8.2cm}
\includegraphics[width=\linewidth]{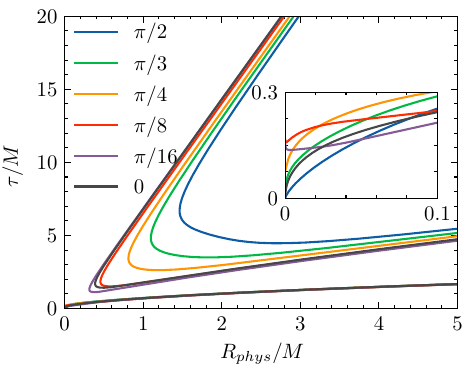}
\end{subfigure}
\begin{subfigure}{7.8cm}
\includegraphics[width=\linewidth]{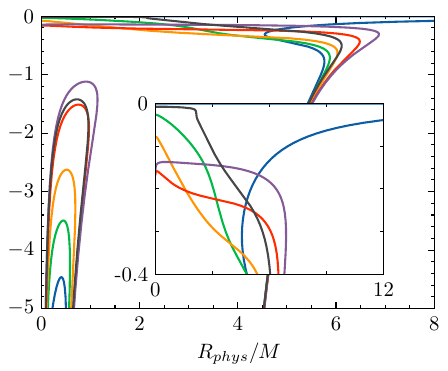}
\end{subfigure}
\caption{Trapping horizons $\tau=\tau_+\left(R_{\text{phys}},\theta\right)$ and $\tau=\tau_-\left(R_{\text{phys}},\theta\right)$ for different values of angle $\theta$, indicated by the colors (same for both panels), for the conformal Kerr spacetime with $a=0.5M$, where $R_{\text{phys}}=A\left(\tau\right)r$ is the physical radius, for $\alpha=1$ (radiation, left plot) and $\alpha=-1$ (cosmological constant, right plot). On the left, $\tau_-$ corresponds to the bottom curves and is zoomed on near $\tau_-=0$, and the factor $A_0$ appearing in $A=A_0\tau^\alpha$ is set to unity. On the right, $\tau_-$ corresponds to the right curves and is zoomed on near $\tau_-=0$, and the Hubble rate is set to $H_0=1/\left(3M\right)$.}
\label{fig:5cul}
\end{figure}

This property of the conformal Kerr metric has an important impact on the behaviour of the curve $\tau_\pm$ close to $R_{\text{phys}}=0$, as demonstrated in Fig~\ref{fig:5cul}. We recall that $R_{\text{phys}}\equiv A(\tau)r$. For positive $\alpha$ (left plot), both trapping horizons start to exist a finite amount of time after the Big Bang $\tau=0$, except the past trapping (cosmological) horizon $\tau=\tau_-$ at $\theta=\pi/2$, which exists at all times. Both horizons expand infinitely: $R_{\text{phys}}\to +\infty$ as $\tau\to+\infty$. \\

In the case of $\alpha=-1$ (right plot), the future trapping (BH) horizon $\tau=\tau_+$ has an upper bound and it ceases to exist at late times for all angles $\theta$, while the past (cosmological) horizon collapses to zero size at late times, unless $\theta=\pi/2$, for which it still extends up to $R_{\text{phys}}\to +\infty$. Let us determine the nature of these trapping horizons, that we have already anticipated in the last few sentences. Eq. (\ref{eq:signs_theta}) yields
\begin{equation}
\tilde{\theta}_-\rvert_{\tau_+} =\frac{\theta_-}{A}\left( 1+\frac{G_-}{G_+}\right),\quad \tilde{\theta}_+\rvert_{\tau_-}= \frac{\theta_+}{A}\left( 1+\frac{G_+}{G_-}\right).
\end{equation}
The ratio $G_-/G_+$ has the same sign as $\Delta$, and one can easily prove that $\left\lvert G_-/G_+\right\rvert<1$. Taking into account the signs of $\theta_\pm$, we conclude that $\tilde{\theta}_-\rvert_{\tau_+}<0$ and $\tilde{\theta}_+\rvert_{\tau_-}>0$, that is, $\tau=\tau_+$ is a future trapping horizon and $\tau=\tau_-$ is a past trapping horizon, for any $\alpha$.
\\

\begin{figure}[!htb]
\begin{subfigure}{5.4cm}
\includegraphics[width=\linewidth]{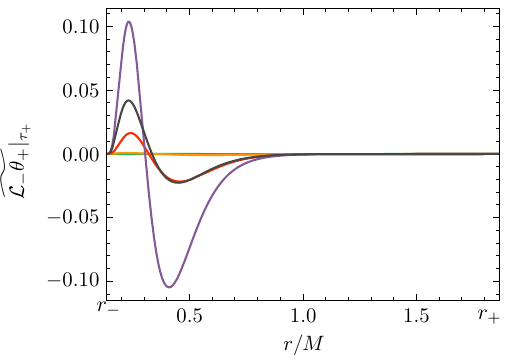}
\end{subfigure}
\begin{subfigure}{4.7cm}
\includegraphics[width=\linewidth]{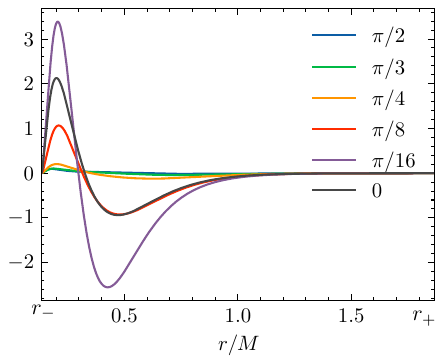}
\end{subfigure}
\begin{subfigure}{4.9cm}
\includegraphics[width=\linewidth]{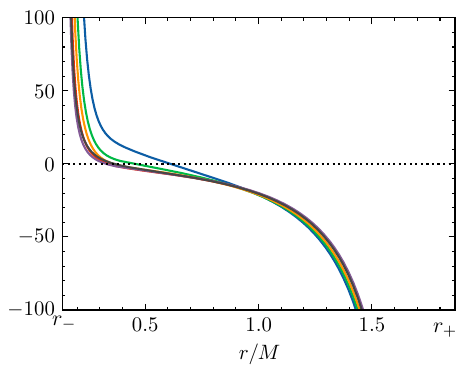}
\end{subfigure}
\caption{$\widetilde{\mathcal{L}_-\theta_+}\rvert_{\tau_+}$ for different values of angle $\theta$ indicated by the colors on the middle plot, for the conformal Kerr spacetime with $a=0.5M$, for $\alpha=2$, $1$ or $-1$ from left to right. The plot is restricted between $r_-$ and $r_+$ precisely because the trapping horizon $\tau=\tau_+$ lies within $r_-<r<r_+$. The magnitudes for $\theta=\pi/2$, $\pi/3$ and $\pi/4$ are barely distinguishable on the left and middle plots, but analysis of the values show that they obey the same pattern as all curves: positive from $r_-$ to some $r_m(\theta)$, then negative from $r_m(\theta)$ to $r_+$.}
\label{fig:6}
\end{figure}

Finally, $\widetilde{\mathcal{L}_-\theta_+}\rvert_{\tau_+}$ and $\widetilde{\mathcal{L}_+\theta_-}\rvert_{\tau_-}$ can be computed from (\ref{eq:signs_lie_1}-\ref{eq:signs_lie}), however the explicit expression is not very instructive. 
Let us focus on the future trapping horizon $\tau=\tau_+$, and on the three interesting cases of cosmology, $\alpha=2,1,-1$. In spherical symmetry, these cases led to $\widetilde{\mathcal{L}_-\theta_+}\rvert_{\tau_+}<0$, that is, the future trapping horizon was outer. Here, since $r=r_+$ and $r=r_-$ are respectively outer and inner for the seed Kerr metric, and given the form of $\tau_+$, see Fig.~\ref{fig:4}, we rather expect to find an outer part close to $r_+$, and an inner part close to $r_-$. 
The plots of Fig.~\ref{fig:6} confirm that this is indeed the case: there exists a radius $r_m(\theta)$ such that for $r_m<r<r_+$, $\widetilde{\mathcal{L}_-\theta_+}\rvert_{\tau_+}<0$, while for $r_-<r<r_m$, $\widetilde{\mathcal{L}_-\theta_+}\rvert_{\tau_+}>0$. In other words, the future trapping horizon $\tau=\tau_+$ is outer for $r_m<r<r_+$ and inner for $r_-<r<r_m$. \textbf{The conformal Kerr spacetime possesses a future outer trapping horizon and is therefore a cosmological BH}. 
\\

Let us conclude this analysis by commenting briefly on the influence of the rotation parameter $a$. The qualitative picture presented above is the same for any $0<a<M$. When $a$ increases towards $M$, the curve $\tau=\tau_+$ of Fig.~\ref{fig:4} moves upwards, and also the range of $r$ for which it exists decrease, because $r_-$ and $r_+$ get closer. The consequence on the plots of Fig.~\ref{fig:5cul} is that, for positive $\alpha$, the BH horizon starts to exist at later times when $a$ increases; while for $\alpha=-1$, the BH horizon disappears earlier when $a$ increases. If $a\geq M$, the expansion $\theta_+$ for the seed Kerr spacetime, see (\ref{eq:thetapluskerr}), is positive everywhere. Therefore, the time $\tau_+$ defined by Eq. (\ref{eq:tau_conf_kerr}) always has opposite sign to $\alpha$, so there is no BH horizon. 

\chapter*{Conclusions}
\label{chap:conclu}
\addcontentsline{toc}{chapter}{\nameref{chap:conclu}}
\chaptermark{CONCLUSIONS}
\sectionmark{CONCLUSIONS}
This thesis presents new advances regarding exact, closed-form BH solutions in scalar-tensor theories. Scalar-tensor theories are a widely studied modified theory of gravity. Finding their BH solutions enables to understand the behaviour of these theories in the strong field regime, and to what extent this behaviour differs from the one of GR. In particular, the no-hair theorem of GR, stating that the final state of gravitational collapse in GR (without Maxwell term) is a BH described only by its mass $M$ and angular momentum $J$, namely the stationary Kerr BH, may be challenged. One may thus wonder if the obtained hairy BHs have only a secondary hair: they are still characterized by $M$ and $J$, but dressed with a non-trivial scalar field inducing possible deviations from the Kerr metric; or a primary hair: the BHs are characterized by another quantity, distinct from $M$ and $J$. On the other hand, one must investigate if the most striking limitation of BHs in GR, that is, the curvature singularity existing at their center, can be avoided in the framework of scalar-tensor theories.
\\

The report starts with Chap.~\ref{chap:modifGR}, which first presents motivations to modify GR, among others, the presence of curvature singularities and the dark energy problem. The possible modifications to GR are constrained by Lovelock's theorem, which finds the most general metric theory of gravity in arbitray spacetime dimension and with second-order field equations. The corresponding action $S\left[g_{\mu\nu}\right]$ is a sum of Lovelock curvature invariants, and reduces to GR in four dimensions. Consequently, modifying GR compels one to violate one of the assumptions of Lovelock's theorem, by for instance allowing higher-order field equations ($f(R)$ gravity), extra dimensions, or additional fields. This latter possibility gives rise to scalar-tensor theories $S\left[g_{\mu\nu},\phi\right]$, where a scalar field $\phi$ is non-minimally coupled to the metric tensor field. The apparent simplicity of scalar-tensor theories does not prevent them to possess direct links with, notably, $f(R)$ gravity and extra-dimensional gravity.
\\

Scalar-tensor theories are then studied in more detail in Chap.~\ref{chap:st}. The most general such theories with second-order field equations, or Horndeski theories, are presented in their modern generalized Galileon formulation. Scalar-tensor theories with field equations of order higher than two propagate in general an additional, unstable degree of freedom. This so-called Ostrogradsky instability can however be evaded if the kinetic matrix of the system is degenerate, leading to the DHOST theories. These DHOST theories of course encompass Horndeski theories, and are related to them by conformal-disformal transformations of the metric field. In particular, a pure disformal transformation maps Horndeski theories to an interesting subclass of DHOST theories, called beyond Horndeski theories.\\

The analysis of scalar-tensor theories simplifies greatly in the presence of symmetries: shift symmetry, under shifts of the scalar by a global additive constant; global conformal symmetry, under a simultaneous Weyl rescaling of the metric and the scalar; and the corresponding local conformal symmetry. The latter can be generalized to the so-called generalized conformal symmetry if the action $S$ is not invariant, but $\delta S/\delta \phi$ is. Interestingly, the most simple Lovelock theory beyond GR, namely EGB gravity, can be compactified, following a KK procedure, down to a sensible four-dimensional scalar-tensor theory, called 4DEGB. This 4DEGB theory reproduces features of the higher-dimensional EGB gravity, and coincides in fact with the most general action with generalized conformal invariance in four dimensions.
\\

Chap.~\ref{sec:state} then moves on to the core topic of this thesis, namely closed-form scalar-tensor BHs, providing a state-of-the-art of solutions which existed before this thesis. To find theories which admit solutions, one must often start by knowing theories which do not, or in other words, no-scalar-hair theorems. The standard such theorems are recalled. They greatly limit the possibility of hairy, asymptotically flat solutions in second-order scalar-tensor theories, i.e. where the Lagrangian is second-order in derivatives. Second-order theories admit up to date a unique such hairy solution, the BBMB BH, for which the scalar field even diverges on the horizon. This justifies the need to look for hairy BHs in the higher-order scalar-tensor theories (from Horndeski to DHOST) introduced in Chap.~\ref{chap:st}. \\

These theories allow quite generically for stealth solutions, that is, for which the metric is as in GR (Schwarzchild or Kerr), but with a non-trivial scalar field, provided the latter has a constant kinetic term $X$. Non-stealth, closed-form solutions, for which the metric differs from GR, are less ubiquitous but also exist. They were first discovered in shift-symmetric theories. However, such non-stealth solutions could also be found more recently in theories without shift symmetry, in the context of the 4DEGB theory. All these spherically-symmetric solutions have a mere secondary hair, i.e., they are all characterized uniquely by their mass $M$.
\\

The new results obtained during this thesis are presented in Chaps.~\ref{chap:4},~\ref{chap:5} and~\ref{chap:6}. Chaps.~\ref{chap:4} and~\ref{chap:5} present new BH solutions, obtained from quite different approaches. In Chap.~\ref{chap:4}, a systematic study of generic beyond Horndeski theories with shift symmetry is performed. Under the assumption of staticity and spherical symmetry, the intricate field equations are rewritten in a very compact way facilitating their analysis. This method leads to many new solutions, but does not provide any physical interpretation of the scalar-tensor theories allowing such solutions. On the other hand, in Chap.~\ref{chap:5}, solutions are obtained in scalar-tensor theories which have interesting theoretical links with conformal invariance, Lovelock theories, and KK compactification of these theories.
\\

More precisely, the analysis of Chap.~\ref{chap:4} enables to find compatibility conditions for the most generic shift-symmetric beyond Horndeski theories to admit BH solutions, Sec.~\ref{sec:nonpar}. These conditions in fact generalize the case of the 4DEGB theory, and the metric solution is obtained as root of a polynomial equation of increasing order, the 4DEGB case corresponding to the lowest order. In a second time, an additional assumption of parity symmetry (under $\phi\to -\phi$) enables to simplify drastically the field equations and to integrate them in various situations. This leads notably to the following solutions: a stealth Schwarzschild with a non-constant scalar kinetic term $X$, which may imply a better behaviour of the perturbations as compared to the usual stealth solutions; non-homogeneous BHs, i.e. with $g_{tt}\neq -g^{rr}$; and BHs with primary hair. \\

These latter BHs with primary hair are characterized by two integration constants, not only their mass $M$, but also the 'primary scalar hair' $q$, which has no equivalent in GR and induces a departure from the Schwarzschild geometry. For a particular relation between $M$ and $q$, the usual curvature singularity disappears: the curvature invariants of the metric, and their derivatives, become regular in the whole spacetime. This latest result thus provides two genuine alternative behaviours as compared to GR: first, a true violation of the no-hair conjecture; second, the avoidance of the BH curvature singularity.
\\

As regards Chap.~\ref{chap:5}, two novel BH solutions are obtained for a theory which generalizes the 4DEGB theory. Remarkably, this theory accomodates these solutions whereas it does not display any of the usual simplifying symmetries, not even the generalized conformal symmetry of the 4DEGB case. Still, the action functional displays Lagrangians which would have local conformal invariance in five dimensions - but not in four. The reason for the appearance of such terms remains unclear for the moment, but is an incentive to study conformal couplings of the scalar field with Lovelock invariants in higher dimensions. The spherically-symmetric BH solutions of these theories are fully described. \\

Then, the diagonal KK compactification of Lovelock theories of arbitrary order is performed. Indeed, such a compactification turns out to yield scalar-tensor theories in four dimensions, where appear the previously analyzed Lagrangians with conformal invariance in other dimensions. The calculations are intricate and, while some results are properly demonstrated for arbitrary order of the Lovelock invariants, other remain at the conjecture stage. Still, the demonstrated results enable to perform the KK reduction of the cubic Lovelock invariant. Interestingly, the obtained theory is a Horndeski theory, without parity symmetry, which fits into the compatibility conditions generalizing the 4DEGB results, as described in Sec.~\ref{sec:nonpar}. Up to the demonstration of the conjecture mentioned above, this paves the way towards an interesting pattern, connecting, on the one hand, the KK compactification of Lovelock invariants of increasing order, and, on the other hand, the Horndeski theories, without parity symmetry, verifying the compatibility conditions presented in Sec.~\ref{sec:nonpar}.
\\

Finally, the stability of DHOST theories under conformal-disformal transformations of the metric field enables to generate new solutions by performing such a transformation on a seed, already existing solution. This principle is illustrated in Chap.~\ref{chap:6} with two examples. In a first time, a non-stealth solution is transformed through a disformal transformation into a wormhole. While the seed spacetime possesses curvature singularities, either naked or covered by a horizon, the obtained wormhole is totally regular. Moreover, as opposed to the wormholes of GR, it does not need to be supported by exotic matter: it is a solution in vacuum of the field equations of a beyond Horndeski theory. In a second time, a conformal transformation is applied onto a stealth Kerr BH. This leads to a non-stationary spacetime which displays the features of both a BH and an expanding FLRW universe.
\\

These new results bring about many perspectives for future investigations. First and foremost, some loose ends mentioned previously deserve precisions. Some of them regard the scalar-tensor theories themselves. For instance, the fact that Lagrangians which would have conformal invariance in other dimensions than four, allow for closed-form solutions in four dimensions, like in Sec.~\ref{sec:selecting}, may need a deeper physical interpretation beyond the mere mathematical result. Also, a proper proof of the undemonstrated conjecture (\ref{eq:kkaction2}), or equivalently (\ref{eq:kkaction3}), would complete the understanding of higher-order Lovelock theories as Horndeski theories. Moreover, and as emphasized above, the integration of field equations performed in Chap.~\ref{chap:4} does only care about the fact that the obtained scalar-tensor theories admit closed-form solutions, but not on the meaning of these theories. One could investigate if the theories permitting BHs with primary scalar hair have any physical interpretation, e.g. any link with higher-dimensional gravity.
\\

Other loose ends rather concern the new BHs themselves, typically, the BHs without curvature singularity obtained in Sec.~\ref{sec:primary}. It would be worth investigating other characterizations of the regularity of these spacetimes, for instance geodesic completeness. Of course, this thesis focused only on the task of discovering new closed-form solutions, while only the study of their perturbative and stability properties would enable to conclude on the physical relevance of these solutions.\\

Finally, the most prominent further investigations which should be pursued, based on the numerous solutions presented in this work, are probably the experimental characterization and constraining of these new BHs. This is one of the advantage of studying exact, closed-form solutions: their observational aspects are easier to study, as compared to numerical or perturbative solutions. There exists as many directions of research as observational constraints: perihelion precession, gravitational lensing, shadows of the BH, and of course the current area undergoing rapid progress, GW astronomy. This includes the study of binary inspirals or of QNMs. These are some of the perspectives which will entrench this work on exact solutions into a more general approach towards understanding better the implications of modifications of gravity.
\appendix

\chapter{Original construction of Horndeski theories}\label{chap:appHornProof}
\chaptermark{Appendix A~~Original construction of Horndeski theories}
\sectionmark{A.1~~Construction of $A^{\mu\nu}$}
This appendix details the construction of Horndeski's scalar-tensor theories, described in his article~\cite{Horndeski:1974wa}, while a very brief overview is given in paragraph~\ref{subsec:horndeski}. Horndeski is looking for the most general four-dimensional scalar-tensor action $S=\int\mathrm{d}^4x\sqrt{-g}\mathcal{L}$ leading to second-order field equations $\mathcal{E}^{\mu\nu}$ and $\mathcal{E}_\phi$, with\footnote{To stick closer to Horndeski's construction, we define this time the metric field equations $\mathcal{E}^{\mu\nu}$ with upper indices, while we define them with lower indices in the rest of the manuscript.}
\begin{equation}
\mathcal{E}^{\mu\nu}\equiv \frac{2}{\sqrt{-g}}\frac{\delta S}{\delta g_{\mu\nu}},\quad \mathcal{E}_\phi\equiv\frac{1}{\sqrt{-g}}\frac{\delta S}{\delta\phi}.
\end{equation}
In general, if $\mathcal{E}^{\mu\nu}$ is second-order in derivatives, $\nabla_\nu\mathcal{E}^{\mu\nu}$ should be third-order. However, as in paragraph~\ref{subsec:horndeski}, diffeomorphism invariance of $S$ leads to the Bianchi identity 
\begin{equation}
\nabla_\nu\mathcal{E}^{\mu\nu}=\mathcal{E}_\phi\nabla^\mu\phi.\label{eq:bianchi_appendix_horn}
\end{equation}
Since $\mathcal{E}_\phi$ is second-order, then $\nabla_\nu\mathcal{E}^{\mu\nu}$ must also be second-order. This places constraints on the structure of $\mathcal{E}^{\mu\nu}$. Horndeski thus starts by constructing the most general symmetric tensor of the form $A^{\mu\nu}=A^{\mu\nu}\left(g_{\mu\nu},g_{\mu\nu,\rho},g_{\mu\nu,\rho\lambda},\phi,\phi_{,\rho},\phi_{,\rho\lambda}\right)$, which is such that $\nabla_\nu A^{\mu\nu}$ is second-order in derivatives. A comma means spacetime derivation:
\begin{equation}
g_{\mu\nu,\rho}\equiv\frac{\partial g_{\mu\nu}}{\partial x^\rho}, \quad \phi_{,\rho\lambda} \equiv \frac{\partial\phi}{\partial x^\lambda\partial x^\rho},
\end{equation} 
and so on. As Horndeski writes, the $\mathcal{E}^{\mu\nu}$ we seek will be contained in such an $A^{\mu\nu}$.
\section{Construction of $A^{\mu\nu}$}
\sectionmark{A.1~~Construction of $A^{\mu\nu}$}
\subsection{Reformulating the conditions on $A^{\mu\nu}$}
Following Horndeski, one may introduce the condensed notations
\begin{equation}
A^{\mu\nu;\alpha\beta,\gamma\lambda}\equiv \frac{\partial A^{\mu\nu}}{\partial g_{\alpha\beta,\gamma\lambda}},\quad A^{\mu\nu;\alpha\beta}\equiv \frac{\partial A^{\mu\nu}}{\partial \phi_{,\alpha\beta}}.\label{eq:notationshornrund}
\end{equation}
The following symmetries hold,
\begin{equation}
A^{\mu\nu;\alpha\beta,\gamma\lambda} = A^{\mu\nu;\beta\alpha,\gamma\lambda}=A^{\mu\nu;\alpha\beta,\lambda\gamma},\quad A^{\mu\nu;\alpha\beta}=A^{\mu\nu;\beta\alpha},\label{eq:symmetryofamunuapp}
\end{equation}
and similarly for any tensor depending on $g_{\mu\nu}$, $\phi$ and their derivatives. \\

First of all, one must establish useful properties~(\ref{eq:Cyclic_result}-\ref{eq:cyclic_result_bis}), first demonstrated by Rund~\cite{Rund_Useful}. Under a change of spacetime coordinates $x^\mu\to \hat{x}^{\hat{\mu}}$, the tensor components $A^{\mu\nu}$ obey the following transformation law:
\begin{equation}
\hat{A}^{\hat{\sigma}\hat{\tau}}(\hat{g}_{\hat{\mu}\hat{\nu}},\hat{g}_{\hat{\mu}\hat{\nu},\hat{\rho}},\hat{g}_{\hat{\mu}\hat{\nu},\hat{\rho}\hat{\lambda}},\hat{\phi},\hat{\phi}_{,\hat{\rho}},\hat{\phi}_{,\hat{\rho}\hat{\lambda}})=J^\sigma_{\hat{\sigma}} J^\tau_{\hat{\tau}} A^{\sigma\tau}\left(g_{\mu\nu},g_{\mu\nu,\rho},g_{\mu\nu,\rho\lambda},\phi,\phi_{,\rho},\phi_{,\rho\lambda}\right).\label{eq:transfo_amunu}
\end{equation}
The inverse Jacobian of the transformation appears,
\begin{equation}
J^\mu_{\hat{\mu}}\equiv\frac{\partial x^{\mu}}{\partial\hat{x}^{\hat{\mu}}}.
\end{equation}
It also appears in the transformation of the metric,
\begin{equation}
\hat{g}_{\hat{\mu}\hat{\nu}} = J^\mu_{\hat{\mu}} J^\nu_{\hat{\nu}} g_{\mu\nu},\quad \hat{\phi}=\phi.
\end{equation}
We introduce the notations
\begin{equation}
J^\mu_{\hat{\mu}\hat{\rho}}\equiv \frac{\partial J^\mu_{\hat{\mu}}}{\partial \hat{x}^{\hat{\rho}}}=\frac{\partial^2 x^\mu}{\partial\hat{x}^{\hat{\rho}}\partial\hat{x}^{\hat{\mu}}},\quad J^\mu_{\hat{\mu}\hat{\rho}\hat{\lambda}}\equiv \frac{\partial J^\mu_{\hat{\mu}\hat{\rho}}}{\partial \hat{x}^{\hat{\lambda}}}=\frac{\partial^3 x^\mu}{\partial\hat{x}^{\hat{\lambda}}\partial\hat{x}^{\hat{\rho}}\partial\hat{x}^{\hat{\mu}}},
\end{equation}
etc. One has
\begin{equation}
\frac{\partial J^\mu_{\hat{\mu}\hat{\rho}\hat{\lambda}}}{\partial J^\alpha_{\hat{\alpha}\hat{\beta}\hat{\gamma}}}=\delta^\mu_\alpha \delta^{\hat{\alpha}}_{(\hat{\mu}}\delta^{\hat{\beta}}_{\hat{\rho}}\delta^{\hat{\gamma}}_{\hat{\lambda})}.\label{eq:partialJpartialJ}
\end{equation}
$J^\mu_{\hat{\mu}\hat{\rho}\hat{\lambda}}$ does not appear on the right hand side of~(\ref{eq:transfo_amunu}), while it appears on the left hand side only through the argument $\hat{g}_{\hat{\mu}\hat{\nu},\hat{\rho}\hat{\lambda}}$. Using~(\ref{eq:partialJpartialJ}), one can directly compute
\begin{equation}
\frac{\partial\hat{g}_{\hat{\mu}\hat{\nu},\hat{\rho}\hat{\lambda}}}{\partial J^\mu_{\hat{\alpha}\hat{\beta}\hat{\gamma}}} =g_{\mu\nu}\left( J^\nu_{\hat{\nu}}\delta^{\hat{\alpha}}_{(\hat{\mu}}\delta^{\hat{\beta}}_{\hat{\rho}}\delta^{\hat{\gamma}}_{\hat{\lambda})}+J^\nu_{\hat{\mu}}\delta^{\hat{\alpha}}_{(\hat{\nu}}\delta^{\hat{\beta}}_{\hat{\rho}}\delta^{\hat{\gamma}}_{\hat{\lambda})}\right).
\end{equation}
Taking the derivative of~(\ref{eq:transfo_amunu}) with respect to $J^\mu_{\hat{\alpha}\hat{\beta}\hat{\gamma}}$ then yields
\begin{equation}
\frac{\partial \hat{A}^{\hat{\sigma}\hat{\tau}}}{\partial \hat{g}_{\hat{\nu}\hat{\mu},\hat{\rho}\hat{\lambda}}}+\frac{\partial \hat{A}^{\hat{\sigma}\hat{\tau}}}{\partial \hat{g}_{\hat{\nu}\hat{\lambda},\hat{\mu}\hat{\rho}}}+\frac{\partial \hat{A}^{\hat{\sigma}\hat{\tau}}}{\partial \hat{g}_{\hat{\nu}\hat{\rho},\hat{\lambda}\hat{\mu}}}=0.
\end{equation}
However, Eq.~(\ref{eq:transfo_amunu}) implies that 
\begin{equation}
\frac{\partial \hat{A}^{\hat{\sigma}\hat{\tau}}}{\partial \hat{g}_{\hat{\nu}\hat{\mu},\hat{\rho}\hat{\lambda}}}= A^{\sigma\tau;\nu\mu,\rho\lambda}J^\sigma_{\hat{\sigma}} J^\tau_{\hat{\tau}}J^\mu_{\hat{\mu}}J^\nu_{\hat{\nu}} J^\rho_{\hat{\rho}}J^\lambda_{\hat{\lambda}},
\end{equation}
where notations~(\ref{eq:notationshornrund}) are used. Therefore, 
\begin{equation}
A^{\sigma\tau;\nu\mu,\rho\lambda}+A^{\sigma\tau;\nu\lambda,\mu\rho}+A^{\sigma\tau;\nu\rho,\lambda\mu}=0.\label{eq:Cyclic_result}
\end{equation}
Repeated use of~(\ref{eq:Cyclic_result}) and~(\ref{eq:symmetryofamunuapp}) gives
\begin{equation}
A^{\mu\nu;\sigma\tau,\rho\lambda}=A^{\mu\nu;\rho\lambda,\sigma\tau}.\label{eq:cyclic_result_bis}
\end{equation}
Now that these useful properties~(\ref{eq:Cyclic_result}-\ref{eq:cyclic_result_bis}) are established, we recall that Horndeski is looking for the most general symmetric $A^{\mu\nu}$ such that 
\begin{align}
A^{\mu\nu}={}&{}A^{\mu\nu}\left(g_{\mu\nu},g_{\mu\nu,\rho},g_{\mu\nu,\rho\lambda},\phi,\phi_{,\rho},\phi_{,\rho\lambda}\right),\label{eq:appgout1}\\
\nabla_\nu A^{\mu\nu}{}&{} \text{ is at most of second-order
in the derivatives of both } g_{\mu\nu} \text{ and } \phi.\label{eq:appgout2}
\end{align}
Condition~(\ref{eq:appgout1}) implies that condition~(\ref{eq:appgout2}) is equivalent to
\begin{equation}
\frac{\partial \nabla_\nu A^{\mu\nu}}{\partial g_{\alpha\beta,\gamma\delta\epsilon}}=0\quad\text{and}\quad \frac{\partial \nabla_\nu A^{\mu\nu}}{\partial \phi_{,\alpha\beta\gamma}}=0.
\end{equation}
Using the chain rule to expand $\nabla_\nu A^{\mu\nu}$, these conditions are in turn equivalent to 
\begin{equation}
A^{\mu\nu;\rho\sigma,\lambda\tau}\frac{\partial g_{\rho\sigma,\lambda\tau\nu}}{\partial g_{\alpha\beta,\gamma\delta\epsilon}}=0\quad\text{and}\quad A^{\mu\nu;\rho\sigma}\frac{\partial \phi_{,\rho\sigma\nu}}{\partial \phi_{,\alpha\beta\gamma}}=0.
\end{equation}
This is finally equivalent to
\begin{align}
A^{\mu\gamma;\rho\sigma,\delta\epsilon}+A^{\mu\delta;\rho\sigma,\epsilon\gamma}+A^{\mu\epsilon;\rho\sigma,\gamma\delta}=0,\label{eq:theohorn1}\\
A^{\mu\alpha;\beta\gamma}+A^{\mu\beta;\gamma\alpha}+A^{\mu\gamma;\alpha\beta}=0.\label{eq:theohorn2}
\end{align}
In a word, \textbf{a symmetric tensor $A^{\mu\nu}=A^{\mu\nu}\left(g_{\mu\nu},g_{\mu\nu,\rho},g_{\mu\nu,\rho\lambda},\phi,\phi_{,\rho},\phi_{,\rho\lambda}\right)$ is such that $\nabla_\nu A^{\mu\nu}$ is at most of second-order
in the derivatives of both $g_{\mu\nu}$ and $\phi$ if and only if it satisfies~(\ref{eq:theohorn1}-\ref{eq:theohorn2}), where the notations~(\ref{eq:notationshornrund}) are used}. Note also that~(\ref{eq:theohorn1}-\ref{eq:theohorn2}), in conjunction with~(\ref{eq:symmetryofamunuapp}), imply
\begin{equation}
A^{\mu\nu;\alpha\beta,\gamma\delta}=A^{\gamma\delta;\alpha\beta,\mu\nu}=A^{\alpha\beta;\mu\nu,\gamma\delta}\quad\text{and}\quad A^{\mu\nu;\alpha\beta}=A^{\alpha\beta;\mu\nu}.\label{eq:lastsymmapp}
\end{equation}
\subsection{Lovelock's property S and its consequences}
In an article~\cite{Lovelock_Property_S}, Lovelock introduces the following definition: \textbf{a quantity with even number of indices $B^{\alpha_1\alpha_2\cdots \alpha_{2h-1}\alpha_{2h}\cdots \alpha_{2p}}$ is said to enjoy property S if it satisfies the following three conditions}:
\begin{itemize}
\item it is symmetric in $\alpha_{2h-1}\alpha_{2h}$ for all $h=1,\cdots,p$,
\item it is symmetric under the exchange of the pair $(\alpha_1\alpha_2)$ with the pair $(\alpha_{2h-1}\alpha_{2h})$ for all $h=2,\cdots,p$,
\item it satisfies the cyclic identity involving any three of the four indices $(\alpha_1\alpha_2)(\alpha_{2h-1}\alpha_{2h})$ for all $h=2,\cdots,p$, for instance for $h=2$,
\begin{equation}
B^{\alpha_1\alpha_2\alpha_3\alpha_4\cdots\alpha_{2p}}+B^{\alpha_3\alpha_1\alpha_2\alpha_4\cdots\alpha_{2p}}+B^{\alpha_2\alpha_3\alpha_1\alpha_4\cdots\alpha_{2p}}=0.
\end{equation}
\end{itemize}
The properties demonstrated in the previous paragraph, namely (\ref{eq:symmetryofamunuapp},\ref{eq:Cyclic_result},\ref{eq:cyclic_result_bis}) and (\ref{eq:theohorn1}-\ref{eq:lastsymmapp}), imply that 
\begin{equation}
A^{\mu\nu;\alpha_1\alpha_2,\alpha_3\alpha_4;\cdots;\alpha_{4h-3}\alpha_{4h-2},\alpha_{4h-1}\alpha_{4h};\alpha_{4h+1}\alpha_{4h+2};\cdots;\alpha_{4h+2k-1}\alpha_{4h+2k}}
\end{equation}
satisfies property S, whenever $h\geq 0$, $k\geq 0$. Note that pair of indices from $\alpha_1$ to $\alpha_{4h}$ are separated by an alternation of commas and semicolons, and thus correspond to derivatives of $A^{\mu\nu}$ with respect to the second derivatives of the metric [first definition in~(\ref{eq:notationshornrund})]; while pair of indices from $\alpha_{4h+1}$ to $\alpha_{4h+2k}$ are separated only by semicolons, and thus correspond to derivatives of $A^{\mu\nu}$ with respect to the second derivatives of the scalar field [second definition in~(\ref{eq:notationshornrund})].
\\

It is easy to prove that if $B^{\alpha_1\cdots \alpha_{4M+2}}$ has property S then it vanishes whenever three (or more) indices are equal~\cite{Lovelock_Property_S}. In particular it vanishes identically in four spacetime dimensions if $M\geq 2$. Consequently, 
\begin{equation}
\text{If }k\text{ even and } h+\frac{k}{2}\geq 2,\,A^{\mu\nu;\alpha_1\alpha_2,\alpha_3\alpha_4;\cdots;\alpha_{4h-3}\alpha_{4h-2},\alpha_{4h-1}\alpha_{4h};\alpha_{4h+1}\alpha_{4h+2};\cdots;\alpha_{4h+2k-1}\alpha_{4h+2k}}=0.\label{eq:lastconsequenceofS}
\end{equation}
\subsection{Expressing $A^{\mu\nu}$ in terms of tensors satisfying property S and depending only on $g_{\alpha\beta}$, $\phi$ and $\phi_{,\alpha}$}
Horndeski then uses~(\ref{eq:lastconsequenceofS}) to construct $A^{\mu\nu}$. Indeed, (\ref{eq:lastconsequenceofS}) with $h=2$ and $k=0$ implies that $A^{\mu\nu}$ is linear in the second derivatives of the metric:
\begin{equation}
A^{\mu\nu}=\beta^{\mu\nu\alpha\beta\gamma\delta}g_{\alpha\beta,\gamma\delta}+\beta^{\mu\nu},\label{eq:equationwithbetas}
\end{equation}
where $\beta^{\mu\nu\alpha\beta\gamma\delta}$ and $\beta^{\mu\nu}$ are concomitants of $g_{\alpha\beta}$, $g_{\alpha\beta,\gamma}$, $\phi$, $\phi_{,\alpha}$ and $\phi_{,\alpha\beta}$ which enjoy property S. Now, the Riemann tensor $R_{\alpha\beta\gamma\delta}$ depends as follows on the second derivatives of the metric:
\begin{equation}
R_{\alpha\beta\gamma\delta} = \frac{1}{2}\left(g_{ad,bc}+g_{bc,ad}-g_{ac,bd}-g_{bd,ac}\right)+\cdots,
\end{equation}
where the dots indicate terms without second derivatives of the metric. Using this, the symmetries of $R_{\alpha\beta\gamma\delta}$ and the property S of $\beta^{\mu\nu\alpha\beta\gamma\delta}$, one can rewrite~(\ref{eq:equationwithbetas}) as
\begin{equation}
A^{\mu\nu}=\hat{\beta}^{\mu\nu\alpha\beta\gamma\delta}R_{\alpha\beta\gamma\delta}+\hat{\beta}^{\mu\nu},\label{eq:hatbeta}
\end{equation}
where $\hat{\beta}^{\mu\nu\alpha\beta\gamma\delta}$ and $\hat{\beta}^{\mu\nu}$ possess the same properties as their counterparts without hats, but in addition, they are tensors.\\

Combining~(\ref{eq:lastconsequenceofS}) (with $h=1$ and $k=2$) and~(\ref{eq:hatbeta}) implies that
\begin{equation}
\hat{\beta}^{\mu\nu\alpha\beta\gamma\delta;\epsilon\eta}=\hat{\beta}^{\mu\nu\alpha\beta\gamma\delta;\epsilon\eta}(g_{\alpha\beta},g_{\alpha\beta,\gamma},\phi,\phi_{,\alpha}),\label{eq:betahatsemicolon}
\end{equation}
where the notation with semicolon means the same as in~(\ref{eq:notationshornrund}). Horndeski then uses a lemma, demonstrated by Du Plessis~\cite{du1968invariance}, and stating that if a tensor $U$ (indices are omitted) is such that $U=U(g_{\alpha\beta},g_{\alpha\beta,\gamma},\phi,\phi_{,\alpha})$, then in fact, it cannot depend on the first derivatives of the metric:
\begin{equation}
U=U(g_{\alpha\beta},g_{\alpha\beta,\gamma},\phi,\phi_{,\alpha})\Rightarrow U=U(g_{\alpha\beta},\phi,\phi_{,\alpha}) \quad \text{(}U\text{ tensor, indices omitted).}\label{eq:horndeskilemmamaster}
\end{equation}
This is demonstrated with similar tools as the ones leading to property~(\ref{eq:Cyclic_result}), that is, by performing a change of spacetime coordinates, writing down the transformation law for the tensor, and differentiating with respect to derivatives of the Jacobian matrix.\\

With this lemma, Eq.~(\ref{eq:betahatsemicolon}) becomes 
\begin{equation}
\hat{\beta}^{\mu\nu\alpha\beta\gamma\delta;\epsilon\eta}=\hat{\beta}^{\mu\nu\alpha\beta\gamma\delta;\epsilon\eta}(g_{\alpha\beta},\phi,\phi_{,\alpha}),
\end{equation}
leading to
\begin{equation}
\hat{\beta}^{\mu\nu\alpha\beta\gamma\delta}=\xi^{\mu\nu\alpha\beta\gamma\delta\epsilon\eta}\phi_{,\epsilon\eta}+\alpha^{\mu\nu\alpha\beta\gamma\delta}.\label{eq:betaxialpha}
\end{equation}
Taking into account the previous results, $\xi^{\mu\nu\alpha\beta\gamma\delta\epsilon\eta}$ depends on $g_{\alpha\beta}$, $\phi$ and $\phi_{,\alpha}$, while $\alpha^{\mu\nu\alpha\beta\gamma\delta}$ depends on $g_{\alpha\beta}$, $g_{\alpha\beta,\gamma}$, $\phi$ and $\phi_{,\alpha}$. They both satisfy property S. Since $\hat{\beta}^{\mu\nu\alpha\beta\gamma\delta;\epsilon\eta}$ is a tensor\footnote{For a tensor $U=U\left(g_{\alpha\beta},g_{\alpha\beta,\gamma},g_{\alpha\beta,\gamma\delta},\phi,\phi_{,\alpha},\phi_{,\alpha\beta}\right)$ (indices omitted), the derivative with respect to the second derivatives of the metric, $U^{;\alpha\beta,\gamma\delta}$, and the derivative with respect to the second derivatives of the scalar, $U^{;\alpha\beta}$, are also tensors~\cite{Rund_Useful}. This is because the Riemann tensor $R_{\mu\nu\rho\sigma}$ depends linearily on $g_{\alpha\beta,\gamma\delta}$, and the second covariant derivative $\phi_{\lvert\alpha\beta}$ depends linearily on $\phi_{,\alpha\beta}$, and dependence of $U$ on $g_{\alpha\beta,\gamma\delta}$ and $\phi_{,\alpha\beta}$ can only come from $R_{\mu\nu\rho\sigma}$ and $\phi_{\lvert\alpha\beta}$ respectively.} , $\xi^{\mu\nu\alpha\beta\gamma\delta\epsilon\eta}$ is also a tensor, but this is not the case for $\alpha^{\mu\nu\alpha\beta\gamma\delta}$. However, $\phi_{,\epsilon\eta}=\phi_{\lvert \epsilon\eta}+\Gamma^\lambda_{\epsilon\eta}\phi_{,\lambda}$, where a vertical bar means a covariant derivative. The notation $\phi_{\epsilon\eta}$, without bars, which is used in all the manuscript to denote covariant differentiation, is given up momentarily, because of the risk of confusion with the partial derivatives $\phi_{,\epsilon\eta}$. This enables to rewrite~(\ref{eq:betaxialpha}) as
\begin{equation}
\hat{\beta}^{\mu\nu\alpha\beta\gamma\delta}=\xi^{\mu\nu\alpha\beta\gamma\delta\epsilon\eta}\phi_{\lvert\epsilon\eta}+\xi^{\mu\nu\alpha\beta\gamma\delta},\label{eq:simplifiedhatbetafirst}
\end{equation}
where $\xi^{\mu\nu\alpha\beta\gamma\delta}$ is a tensor satisfying property S. In addition, thanks again to lemma (\ref{eq:horndeskilemmamaster}), $\xi^{\mu\nu\alpha\beta\gamma\delta}$ depends on $g_{\alpha\beta}$, $\phi$ and $\phi_{,\alpha}$.
\\

The structure of the tensor $\hat{\beta}^{\mu\nu\alpha\beta\gamma\delta}$ has been simplified, and one can perform an analogous work on $\hat{\beta}^{\mu\nu}$, starting again from (\ref{eq:lastconsequenceofS}), with this time $h=0$ and $k=4$, and exploiting also (\ref{eq:simplifiedhatbetafirst}). One finds
\begin{equation}
\hat{\beta}^{\mu\nu} = \psi^{\mu\nu\alpha\beta\gamma\delta\epsilon\eta}\phi_{\lvert \alpha\beta}\phi_{\lvert\gamma\delta}\phi_{\lvert \epsilon\eta}+\psi^{\mu\nu\alpha\beta\gamma\delta}\phi_{\lvert \alpha\beta}\phi_{\lvert\gamma\delta}+\psi^{\mu\nu\alpha\beta}\phi_{\lvert \alpha\beta}+\psi^{\mu\nu},
\end{equation}
where each $\psi$ is a tensor satsifying property S and depending on $g_{\alpha\beta}$, $\phi$ and $\phi_{,\alpha}$. \textbf{In a word, $A^{\mu\nu}$ has been expressed as 
\begin{align}
A^{\mu\nu}=\xi^{\mu\nu\alpha\beta\gamma\delta\epsilon\eta}\phi_{\lvert\epsilon\eta}{}&{}+\xi^{\mu\nu\alpha\beta\gamma\delta}+\psi^{\mu\nu\alpha\beta\gamma\delta\epsilon\eta}\phi_{\lvert \alpha\beta}\phi_{\lvert\gamma\delta}\phi_{\lvert \epsilon\eta}\nonumber\\{}&{}+\psi^{\mu\nu\alpha\beta\gamma\delta}\phi_{\lvert \alpha\beta}\phi_{\lvert\gamma\delta}+\psi^{\mu\nu\alpha\beta}\phi_{\lvert \alpha\beta}+\psi^{\mu\nu},\label{eq:amunuintermexpr}
\end{align}
where each $\xi$ and each $\psi$ is a tensor satsifying property S and depending on $g_{\alpha\beta}$, $\phi$ and $\phi_{,\alpha}$}.
\subsection{Expression of $A^{\mu\nu}$}
The expression of $A^{\mu\nu}$ can thus be obtained, provided one knows the expressions for the tensors $\xi$ and $\psi$ (indices omitted) appearing in (\ref{eq:amunuintermexpr}). In other words, one must construct the most general tensors satsifying property S and depending on $g_{\alpha\beta}$, $\phi$ and $\phi_{,\alpha}$, in a four-dimensional spacetime. \\

Such tensors can be constructed using techniques as the ones used by Lovelock in the appendices of his article~\cite{Lovelock_A14}. There is one tensor with two indices, one with four indices, one with six indices and one with eight indices\footnote{Remember that, in four spacetime dimensions, a tensor with property $S$ and ten indices (or more) vanishes identically.}. The tensors with two, four and six indices each depend on two arbitrary functions of $\phi$ and its kinetic term $X=-(\partial\phi)^2/2$, while the tensor with eight indices depends on only one such arbitrary function. Given the structure of (\ref{eq:amunuintermexpr}), one sees that $A^{\mu\nu}$ will depend on ten arbitrary functions of $\phi$ and $X$. In detail, these tensors read:
\begin{align}
\theta^{\alpha_1\alpha_2}={}&{}C_1g^{\alpha_1\alpha_2}+C_2\phi^{,\alpha_1}\phi^{,\alpha_2},\\
\theta^{\alpha_1\alpha_2\alpha_3\alpha_4}={}&{}C_3\left(g^{\alpha_1\alpha_3}g^{\alpha_2\alpha_4}+g^{\alpha_1\alpha_4}g^{\alpha_2\alpha_3}-2g^{\alpha_1\alpha_2}g^{\alpha_3\alpha_4}\right)\nonumber\\{}&{}+C_4\Bigl[ \phi^{,\alpha_1}\phi^{,\alpha_3}g^{\alpha_2\alpha_4}+\phi^{,\alpha_2}\phi^{,\alpha_4}g^{\alpha_1\alpha_3}+\phi^{,\alpha_1}\phi^{,\alpha_4}g^{\alpha_2\alpha_3}+\phi^{,\alpha_2}\phi^{,\alpha_3}g^{\alpha_1\alpha_4}\nonumber\\{}&{}\hspace{3.8cm}-2\left(\phi^{,\alpha_1}\phi^{,\alpha_2}g^{\alpha_3\alpha_4}+\phi^{,\alpha_3}\phi^{,\alpha_4}g^{\alpha_1\alpha_2}\right)\Bigr],\\
\theta^{\alpha_1\alpha_2\alpha_3\alpha_4\alpha_5\alpha_6}={}&{}\left(C_5\phi_{,\mu}\phi_{,\nu}+C_6 g_{\mu\nu}\right)\Bigl(\epsilon^{\alpha_1\alpha_3\alpha_5\mu}\epsilon^{\alpha_2\alpha_4\alpha_6\nu}+\epsilon^{\alpha_1\alpha_3\alpha_6\mu}\epsilon^{\alpha_2\alpha_4\alpha_5\nu}\nonumber\\{}&{}\hspace{3.4cm}+\epsilon^{\alpha_1\alpha_4\alpha_5\mu}\epsilon^{\alpha_2\alpha_3\alpha_6\nu}+\epsilon^{\alpha_1\alpha_4\alpha_6\mu}\epsilon^{\alpha_2\alpha_3\alpha_5\nu}\Bigr),\\
\theta^{\alpha_1\alpha_2\alpha_3\alpha_4\alpha_5\alpha_6\alpha_7\alpha_8}
={}&{}C_7\Bigl(\epsilon^{\alpha_1\alpha_3\alpha_5\alpha_7}\epsilon^{\alpha_2\alpha_4\alpha_6\alpha_8}+\epsilon^{\alpha_1\alpha_3\alpha_5\alpha_8}\epsilon^{\alpha_2\alpha_4\alpha_6\alpha_7}+\epsilon^{\alpha_1\alpha_3\alpha_6\alpha_7}\epsilon^{\alpha_2\alpha_4\alpha_5\alpha_8}\nonumber\\{}&{}\quad+\epsilon^{\alpha_1\alpha_3\alpha_6\alpha_8}\epsilon^{\alpha_2\alpha_4\alpha_5\alpha_7}+\epsilon^{\alpha_2\alpha_3\alpha_5\alpha_7}\epsilon^{\alpha_1\alpha_4\alpha_6\alpha_8}+\epsilon^{\alpha_2\alpha_3\alpha_5\alpha_8}\epsilon^{\alpha_1\alpha_4\alpha_6\alpha_7}\nonumber\\{}&{}\quad+\epsilon^{\alpha_2\alpha_3\alpha_6\alpha_7}\epsilon^{\alpha_1\alpha_4\alpha_5\alpha_8}+\epsilon^{\alpha_2\alpha_3\alpha_6\alpha_8}\epsilon^{\alpha_1\alpha_4\alpha_5\alpha_7}\Bigr),
\end{align}
where $C_1,\cdots,C_7$ are arbitrary functions of $\phi$ and $X$. Eq.~(\ref{eq:amunuintermexpr}) thus leads, after lengthy calculations, to the following result: \textbf{in a four-dimensional spacetime, the most general symmetric tensor $A^{\mu\nu}=A^{\mu\nu}\left(g_{\mu\nu},g_{\mu\nu,\rho},g_{\mu\nu,\rho\lambda},\phi,\phi_{,\rho},\phi_{,\rho\lambda}\right)$ which is such that $\nabla_\nu A^{\mu\nu}$ is at most of second-order in the derivatives of both $g_{\mu\nu}$ and $\phi$, is given by}
\begin{align}
A^{\mu\nu}={}&{}K_1\delta^{\mu\alpha\beta\gamma}_{\hspace{0.025cm}\delta\hspace{0.02cm}\epsilon\hspace{0.06cm}\eta\hspace{0.02cm}\kappa}g^{\delta\nu}\phi_\alpha^\epsilon R_{\beta\gamma}^{\quad\eta\kappa}+K_2\delta^{\mu\alpha\beta}_{\gamma\hspace{0.03cm}\delta\hspace{0.05cm}\epsilon}g^{\gamma\nu}R_{\alpha\beta}^{\quad\delta\epsilon}+K_3\delta^{\mu\alpha\beta\gamma}_{\hspace{0.025cm}\delta\hspace{0.02cm}\epsilon\hspace{0.06cm}\eta\hspace{0.02cm}\kappa}g^{\delta\nu}\phi_\alpha\phi^\epsilon R_{\beta\gamma}^{\quad\eta\kappa}\nonumber\\{}&{}+K_4\delta^{\mu\alpha\beta\gamma}_{\hspace{0.025cm}\delta\hspace{0.02cm}\epsilon\hspace{0.06cm}\eta\hspace{0.02cm}\kappa}g^{\delta\nu}\phi_\alpha^\epsilon\phi_\beta^\eta\phi_\gamma^\kappa+K_5\delta^{\mu\alpha\beta}_{\gamma\hspace{0.03cm}\delta\hspace{0.05cm}\epsilon}g^{\gamma\nu}\phi_\alpha^\delta \phi_\beta^\epsilon+K_6\delta^{\mu\alpha\beta\gamma}_{\hspace{0.025cm}\delta\hspace{0.02cm}\epsilon\hspace{0.06cm}\eta\hspace{0.02cm}\kappa}g^{\delta\nu}\phi_\alpha\phi^\epsilon\phi_\beta^\eta\phi_\gamma^\kappa\nonumber\\{}&{}+K_7\delta^{\mu\alpha}_{\beta\gamma}g^{\beta\nu}\phi_\alpha^\gamma+K_8\delta^{\mu\alpha\beta}_{\gamma\hspace{0.03cm}\delta\hspace{0.05cm}\epsilon}g^{\gamma\nu}\phi_\alpha\phi^\delta \phi_\beta^\epsilon+K_9g^{\mu\nu}+K_{10}\phi^\mu\phi^\nu,\label{eq:final_expr_amunu}
\end{align}
\textbf{where $K_1,\cdots,K_{10}$ are arbitrary functions of $\phi$ and its kinetic term $X=-\left(\partial\phi\right)^2/2$}. Note that we have reinstated the notation $\phi_\alpha$ and $\phi_{\alpha\beta}$ for the covariant derivatives, since there is no possible confusion here. From now on, we will be dealing only with covariant derivatives and no more with partial derivatives, so this notation $\phi_\alpha$ and $\phi_{\alpha\beta}$ is also used in what follows.
\\

The task is not over yet: the field equations must satisfy the Bianchi identity (\ref{eq:bianchi_appendix_horn}), $\nabla_\nu\mathcal{E}^{\mu\nu}=\mathcal{E}_\phi\nabla^\mu\phi$. Therefore, $\mathcal{E}^{\mu\nu}$ is of the form of $A^{\mu\nu}$, Eq.~(\ref{eq:final_expr_amunu}), but with the additional restriction that \textbf{there must exist a Lorentz scalar $B$ such that}\begin{equation}
\nabla_\nu A^{\mu\nu}=B\,\nabla^\mu\phi \quad\text{with}\quad B=B\left(g_{\mu\nu},g_{\mu\nu,\rho},g_{\mu\nu,\rho\lambda},\phi,\phi_{,\rho},\phi_{,\rho\lambda}\right).\label{eq:divergenceamunu}
\end{equation} 
\section{Restrictions on $A^{\mu\nu}$ implied by~(\ref{eq:divergenceamunu})}
\sectionmark{A.2~~Restrictions on $A^{\mu\nu}$ implied by~(\ref{eq:divergenceamunu})}
The divergence $\nabla_\nu A^{\mu\nu}$ can be computed from Eq.~(\ref{eq:final_expr_amunu}), along with identities valid in four dimensions,
\begin{equation}
\delta^{\alpha\beta\gamma\delta\epsilon}_{\eta\kappa\lambda\mu\nu}\phi^\eta R_{\beta\gamma}^{\quad\kappa\lambda}R_{\delta\epsilon}^{\quad\mu\nu}=0,\quad \delta^{\alpha\beta\gamma\delta\epsilon}_{\eta\kappa\lambda\mu\nu}\phi^\eta \phi_{\beta}^{\kappa}\phi_{\gamma}^{\lambda}R_{\delta\epsilon}^{\quad\mu\nu}=0,\quad \delta^{\alpha\beta\gamma\delta\epsilon}_{\eta\kappa\lambda\mu\nu}\phi^\eta \phi_{\beta}^{\kappa}\phi_{\gamma}^{\lambda}\phi_{\delta}^\mu\phi_{\epsilon}^{\nu}=0.
\end{equation} 
One obtains
\begin{align}
\nabla_\nu A^{\mu\nu}={}&{}Q\,\phi^{\mu}+\alpha \delta^{\mu\alpha\beta}_{\gamma\hspace{0.03cm}\delta\,\epsilon}\phi^\eta\phi_\alpha^\gamma R_{\eta\beta}^{\quad\delta\epsilon}+\beta\delta^{\mu\eta\beta}_{\gamma\hspace{0.01cm}\delta\,\epsilon}\phi^\alpha\phi_\alpha^\gamma R_{\eta\beta}^{\quad\delta\epsilon}-\gamma \delta^{\mu\alpha\beta}_{\lambda\hspace{0.0cm}\delta\,\epsilon}\phi_\rho\phi^{\rho\lambda}\phi_\alpha\phi^\eta R_{\eta\beta}^{\quad\delta\epsilon}\nonumber\\{}&{}+\epsilon \delta^{\mu\alpha\beta\gamma}_{\hspace{0.025cm}\delta\hspace{0.02cm}\epsilon\hspace{0.06cm}\eta\hspace{0.02cm}\kappa}\phi^\rho R_{\rho\alpha}^{\quad \epsilon\delta}\phi_\beta^\eta\phi_\gamma^\kappa+\mu \delta^{\mu\alpha}_{\beta\gamma}\phi^\delta R_{\delta\alpha}^{\quad\gamma\beta}+\nu \delta^{\mu\eta\beta}_{\gamma\hspace{0.01cm}\delta\,\epsilon}\phi^\rho\phi_\rho^\gamma\phi_\eta^\delta\phi_\beta^\epsilon\nonumber\\{}&{}+2\omega\delta^{\mu\alpha}_{\beta\gamma}\phi^\rho\phi_\rho^\beta\phi_\alpha^\gamma+\xi\phi_\rho\phi^{\mu\rho},\label{eq:divergenceofamunuexplicit}
\end{align}
where appear functions of $\phi$ and $X$, related to the $K_1,\cdots,K_{10}$ as follows,
\begin{align}
\alpha ={}&{} 2K_{1\phi}-2K_3+K_5-2XK_6,\quad\beta = K_3+2XK_{3X}-K_{1\phi}-K_{2X},\nonumber\\  \gamma={}&{}K_6-2K_{3X},\quad\epsilon=\frac{3}{2}K_4-K_{1X},\quad \mu=2K_{2\phi}+\frac{K_7}{2}-XK_8,\\\nu={}&{}3K_6+2XK_{6X}-3K_{4\phi}-K_{5X},\quad\omega=K_{8}+X K_{8X}-K_{5\phi}-\frac{K_{7X}}{2}\nonumber\\\xi={}&{}K_{10}-K_{7\phi}-K_{9X}.\nonumber
\end{align}
Finally, $Q=Q\left(g_{\mu\nu},g_{\mu\nu,\rho},g_{\mu\nu,\rho\lambda},\phi,\phi_{,\rho},\phi_{,\rho\lambda}\right)$ is a Lorentz scalar, and its precise expression is not important: one can immediately infer from~(\ref{eq:divergenceofamunuexplicit}) that $A^{\mu\nu}$ will satisfy~(\ref{eq:divergenceamunu}) if and only if there exists a Lorentz scalar $C=C\left(g_{\mu\nu},g_{\mu\nu,\rho},g_{\mu\nu,\rho\lambda},\phi,\phi_{,\rho},\phi_{,\rho\lambda}\right)$ such that
\begin{align}
C\,\phi^\mu ={}&{} \nabla_\nu A^{\mu\nu}-Q\,\phi^{\mu}\\
{}={}&{}\alpha \delta^{\mu\alpha\beta}_{\gamma\hspace{0.03cm}\delta\,\epsilon}\phi^\eta\phi_\alpha^\gamma R_{\eta\beta}^{\quad\delta\epsilon}+\beta\delta^{\mu\eta\beta}_{\gamma\hspace{0.01cm}\delta\,\epsilon}\phi^\alpha\phi_\alpha^\gamma R_{\eta\beta}^{\quad\delta\epsilon}-\gamma \delta^{\mu\alpha\beta}_{\lambda\hspace{0.0cm}\delta\,\epsilon}\phi_\rho\phi^{\rho\lambda}\phi_\alpha\phi^\eta R_{\eta\beta}^{\quad\delta\epsilon}\nonumber\\{}&{}+\epsilon \delta^{\mu\alpha\beta\gamma}_{\hspace{0.025cm}\delta\hspace{0.02cm}\epsilon\hspace{0.06cm}\eta\hspace{0.02cm}\kappa}\phi^\rho R_{\rho\alpha}^{\quad \epsilon\delta}\phi_\beta^\eta\phi_\gamma^\kappa+\mu \delta^{\mu\alpha}_{\beta\gamma}\phi^\delta R_{\delta\alpha}^{\quad\gamma\beta}+\nu \delta^{\mu\eta\beta}_{\gamma\hspace{0.01cm}\delta\,\epsilon}\phi^\rho\phi_\rho^\gamma\phi_\eta^\delta\phi_\beta^\epsilon\nonumber\\{}&{}+2\omega\delta^{\mu\alpha}_{\beta\gamma}\phi^\rho\phi_\rho^\beta\phi_\alpha^\gamma+\xi\phi_\rho\phi^{\mu\rho}.\label{eq:equationforC}
\end{align}
Horndeski then shows that (\ref{eq:equationforC}), as an equation for $C$, admits a solution if and only if $\alpha=\beta=\gamma=\epsilon=\mu=\nu=\omega=\xi=0$, and this solution is then obviously $C=0$. To show the vanishing of the functions $\alpha,\cdots,\xi$, the idea is to differentiate (\ref{eq:equationforC}) with respect to well-chosen arguments, in order to isolate each one of these functions.\\

For instance, on the right hand side of (\ref{eq:equationforC}), the only terms having second derivatives of the metric are the ones with Riemann tensor, that is, $\alpha,\,\beta,\,\gamma,\,\epsilon,\,\mu$. Among these, only the $\epsilon$ term is quadratic in the second derivatives of the scalar field. Therefore, one can isolate $\epsilon$ by differentiating once with respect to $g_{\alpha\beta,\gamma\delta}$ and twice with respect to $\phi_{,\alpha\beta}$. By further contracting with metric symbols (the $g_{\alpha\beta}g_{\gamma\delta}g_{\epsilon\eta}$ in the following equation), one obtains
\begin{equation}
g_{\alpha\beta}g_{\gamma\delta}g_{\epsilon\eta}C^{;\alpha\beta;\gamma\delta;\epsilon\eta,\eta\kappa}\,\phi^\mu = 4 \epsilon(\phi^\mu g^{\eta\kappa}+\phi^\kappa g^{\mu\eta}+\phi^\eta g^{\kappa\mu}).
\end{equation}
Multiplying this last equation by a vector field $X^\nu$ such that $\phi_{,\nu}X^\nu=0$ and $X_\nu X^\nu\neq 0$ (such a vector field exists locally around each point of the manifold), one gets $\epsilon=0$. Similarly, one shows the vanishing of all eight functions $\alpha,\cdots,\xi$. This corresponds to eight partial differential equations on the $K_1,\cdots,K_{10}$. Only six of these partial differential equations are independent, because
\begin{equation}
\gamma-2\epsilon_\phi-\alpha_X=\nu,\quad -\mu_X-\alpha_\phi-2X\gamma_\phi-2\beta_\phi=\omega.
\end{equation}
The remaining six equations imply the following relations between the $K_1,\cdots,K_{10}$:
\begin{align}
K_4={}&{}\frac{2}{3}K_{1X},\quad K_5=2K_3-2K_{1\phi}+4XK_{3X},\quad K_6=2K_{3X},\quad K_2=\frac{1}{2}F,\nonumber\\ K_7={}&{}2XK_8-2F_\phi,\quad K_{10}=2XK_{8\phi}+K_{9X}-2F_{\phi\phi}.
\end{align}
$K_1$, $K_3$, $K_8$ and $K_9$ are arbitrary functions of $\phi$ and $X$, while $F$ is the following function of $\phi$ and $X$,
\begin{equation}
F=2\int\left(K_3+2XK_{3X}-K_{1\phi}\right)\mathrm{d}X,\label{eq:definitionofFappendixHorn}
\end{equation}
defined up to the addition of an arbitrary function of $\phi$. Combining this with~(\ref{eq:final_expr_amunu}) thus leads to the following result: \textbf{in a four-dimensional spacetime, the most general symmetric tensor $A^{\mu\nu}=A^{\mu\nu}\left(g_{\mu\nu},g_{\mu\nu,\rho},g_{\mu\nu,\rho\lambda},\phi,\phi_{,\rho},\phi_{,\rho\lambda}\right)$ verifying that there exists a Lorentz scalar $B=B\left(g_{\mu\nu},g_{\mu\nu,\rho},g_{\mu\nu,\rho\lambda},\phi,\phi_{,\rho},\phi_{,\rho\lambda}\right)$, such that $\nabla_\nu A^{\mu\nu}=B\,\nabla^\mu\phi$, is:
\begin{align}
A^{\mu\nu}={}&{}K_1\delta^{\mu\alpha\beta\gamma}_{\hspace{0.025cm}\delta\hspace{0.02cm}\epsilon\hspace{0.06cm}\eta\hspace{0.02cm}\kappa}g^{\delta\nu}\phi_\alpha^\epsilon R_{\beta\gamma}^{\quad\eta\kappa}+\frac{1}{2}F\delta^{\mu\alpha\beta}_{\gamma\hspace{0.03cm}\delta\hspace{0.05cm}\epsilon}g^{\gamma\nu}R_{\alpha\beta}^{\quad\delta\epsilon}+K_3\delta^{\mu\alpha\beta\gamma}_{\hspace{0.025cm}\delta\hspace{0.02cm}\epsilon\hspace{0.06cm}\eta\hspace{0.02cm}\kappa}g^{\delta\nu}\phi_\alpha\phi^\epsilon R_{\beta\gamma}^{\quad\eta\kappa}\nonumber\\{}&{}+\frac{2}{3}K_{1X}\delta^{\mu\alpha\beta\gamma}_{\hspace{0.025cm}\delta\hspace{0.02cm}\epsilon\hspace{0.06cm}\eta\hspace{0.02cm}\kappa}g^{\delta\nu}\phi_\alpha^\epsilon\phi_\beta^\eta\phi_\gamma^\kappa+2\left(K_3-K_{1\phi}+2XK_{3X}\right)\delta^{\mu\alpha\beta}_{\gamma\hspace{0.03cm}\delta\hspace{0.05cm}\epsilon}g^{\gamma\nu}\phi_\alpha^\delta \phi_\beta^\epsilon\nonumber\\{}&{}+2K_{3X}\delta^{\mu\alpha\beta\gamma}_{\hspace{0.025cm}\delta\hspace{0.02cm}\epsilon\hspace{0.06cm}\eta\hspace{0.02cm}\kappa}g^{\delta\nu}\phi_\alpha\phi^\epsilon\phi_\beta^\eta\phi_\gamma^\kappa+2\left(XK_8-F_\phi\right)\delta^{\mu\alpha}_{\beta\gamma}g^{\beta\nu}\phi_\alpha^\gamma+K_8\delta^{\mu\alpha\beta}_{\gamma\hspace{0.03cm}\delta\hspace{0.05cm}\epsilon}g^{\gamma\nu}\phi_\alpha\phi^\delta \phi_\beta^\epsilon\nonumber\\{}&{}+K_9g^{\mu\nu}+\left(2XK_{8\phi}+K_{9X}-2F_{\phi\phi}\right)\phi^\mu\phi^\nu.\label{eq:final_expr_amunu_taking_div_account}
\end{align}
$K_1$, $K_3$, $K_8$ and $K_9$ are arbitrary functions of $\phi$ and $X$, and $F$ is given by (\ref{eq:definitionofFappendixHorn})}.
\\

The proof of Horndeski is now almost complete: such an $A^{\mu\nu}$ is the most general tensor satisfying the same kind of identity (\ref{eq:bianchi_appendix_horn}) as the field equations $\mathcal{E}^{\mu\nu}$, namely $\nabla_\nu\mathcal{E}^{\mu\nu}=\mathcal{E}_\phi\nabla^\mu\phi$, therefore, $\mathcal{E}^{\mu\nu}$ must be of the form of $A^{\mu\nu}$, Eq. (\ref{eq:final_expr_amunu_taking_div_account}). Still, it remains to be shown that such an $A^{\mu\nu}$ indeed arises as the Euler-Lagrange equations of an action functional.
\section{Action functional associated to $A^{\mu\nu}$}
\sectionmark{A.3~~Action functional associated to $A^{\mu\nu}$}
For this last step of his proof, the intuition of Horndeski is to look for a Lagrangian which resembles the trace $g_{\mu\nu}A^{\mu\nu}$ of the candidate field equations $A^{\mu\nu}$. This can be presented as follows: consider a 'calligraphic version' of Eq. (\ref{eq:final_expr_amunu_taking_div_account}),
\begin{align}
\mathcal{A}^{\mu\nu}={}&{}\mathcal{K}_1\delta^{\mu\alpha\beta\gamma}_{\hspace{0.025cm}\delta\hspace{0.02cm}\epsilon\hspace{0.06cm}\eta\hspace{0.02cm}\kappa}g^{\delta\nu}\phi_\alpha^\epsilon R_{\beta\gamma}^{\quad\eta\kappa}+\frac{1}{2}\mathcal{F}\delta^{\mu\alpha\beta}_{\gamma\hspace{0.03cm}\delta\hspace{0.05cm}\epsilon}g^{\gamma\nu}R_{\alpha\beta}^{\quad\delta\epsilon}+\mathcal{K}_3\delta^{\mu\alpha\beta\gamma}_{\hspace{0.025cm}\delta\hspace{0.02cm}\epsilon\hspace{0.06cm}\eta\hspace{0.02cm}\kappa}g^{\delta\nu}\phi_\alpha\phi^\epsilon R_{\beta\gamma}^{\quad\eta\kappa}\nonumber\\{}&{}+\frac{2}{3}\mathcal{K}_{1X}\delta^{\mu\alpha\beta\gamma}_{\hspace{0.025cm}\delta\hspace{0.02cm}\epsilon\hspace{0.06cm}\eta\hspace{0.02cm}\kappa}g^{\delta\nu}\phi_\alpha^\epsilon\phi_\beta^\eta\phi_\gamma^\kappa+2\left(\mathcal{K}_3-\mathcal{K}_{1\phi}+2X\mathcal{K}_{3X}\right)\delta^{\mu\alpha\beta}_{\gamma\hspace{0.03cm}\delta\hspace{0.05cm}\epsilon}g^{\gamma\nu}\phi_\alpha^\delta \phi_\beta^\epsilon\nonumber\\{}&{}+2\mathcal{K}_{3X}\delta^{\mu\alpha\beta\gamma}_{\hspace{0.025cm}\delta\hspace{0.02cm}\epsilon\hspace{0.06cm}\eta\hspace{0.02cm}\kappa}g^{\delta\nu}\phi_\alpha\phi^\epsilon\phi_\beta^\eta\phi_\gamma^\kappa+2\left(X\mathcal{K}_8-\mathcal{F}_\phi\right)\delta^{\mu\alpha}_{\beta\gamma}g^{\beta\nu}\phi_\alpha^\gamma+\mathcal{K}_8\delta^{\mu\alpha\beta}_{\gamma\hspace{0.03cm}\delta\hspace{0.05cm}\epsilon}g^{\gamma\nu}\phi_\alpha\phi^\delta \phi_\beta^\epsilon\nonumber\\{}&{}+\mathcal{K}_9g^{\mu\nu}+\left(2X\mathcal{K}_{8\phi}+\mathcal{K}_{9X}-2\mathcal{F}_{\phi\phi}\right)\phi^\mu\phi^\nu,
\end{align}
where all functions $K_1$, $K_3$, $K_8$, $K_9$ and $F$ are now written $\mathcal{K}_1$, $\mathcal{K}_3$, $\mathcal{K}_8$, $\mathcal{K}_9$ and $\mathcal{F}$, thus defining a tensor $\mathcal{A}^{\mu\nu}$ distinct from the candidate field equations $A^{\mu\nu}$ of Eq. (\ref{eq:final_expr_amunu_taking_div_account}). Also, $\mathcal{F}$ is given in terms of $\mathcal{K}_1$ and $\mathcal{K}_3$ as $F$ is given in terms of $K_1$ and $K_3$, Eq. (\ref{eq:definitionofFappendixHorn}). Consider now the following action functional,
\begin{align}
S\left[g_{\mu\nu},\phi\right]={}&{}\int\mathrm{d}^4x\sqrt{-g}\,g_{\mu\nu}\mathcal{A}^{\mu\nu}\nonumber\\
{}={}&{}\int\mathrm{d}^4x\sqrt{-g}\Bigl\{\mathcal{K}_1\delta^{\mu\alpha\beta}_{\gamma\hspace{0.03cm}\delta\hspace{0.05cm}\epsilon}\phi_\mu^\gamma R_{\alpha\beta}^{\quad \delta\epsilon}+\frac{2}{3}\mathcal{K}_{1X}\delta^{\mu\alpha\beta}_{\gamma\hspace{0.03cm}\delta\hspace{0.05cm}\epsilon}\phi^\gamma_\mu\phi^\delta_\alpha\phi^\epsilon_\beta+\mathcal{K}_3\delta^{\mu\alpha\beta}_{\gamma\hspace{0.03cm}\delta\hspace{0.05cm}\epsilon}\phi_\mu\phi^\gamma R_{\alpha\beta}^{\quad \delta\epsilon}\nonumber\\{}&{}+2\mathcal{K}_{3X}\delta^{\mu\alpha\beta}_{\gamma\hspace{0.03cm}\delta\hspace{0.05cm}\epsilon}\phi_\mu\phi^\gamma\phi^\delta_\alpha\phi^\epsilon_\beta+\mathcal{F}\delta^{\mu\alpha}_{\beta\gamma}R_{\mu\alpha}^{\quad\beta\gamma}+4\left(\mathcal{K}_3-\mathcal{K}_{1\phi}+2X\mathcal{K}_{3X}\right)\delta^{\mu\alpha}_{\beta\gamma}\phi_\mu^\beta\phi_\alpha^\gamma\nonumber\\{}&{}+2\mathcal{K}_8\delta^{\mu\alpha}_{\beta\gamma}\phi_\mu\phi^\beta\phi_\alpha^\gamma+6\left(X\mathcal{K}_8-\mathcal{F}_\phi\right)\Box\phi+4\mathcal{K}_9\nonumber\\{}&{}+2X\left(2\mathcal{F}_{\phi\phi}-2X\mathcal{K}_{8\phi}-\mathcal{K}_{9X}\right)\Bigr\}.\label{eq:explicitactionappendixhorn}
\end{align}
It is only a matter of laborious calculations to compute the functional derivative of such an action. One finds
\begin{align}
\frac{2}{\sqrt{-g}}\frac{\delta S}{\delta g_{\mu\nu}} ={}&{} 2X\mathcal{K}_{1X}\delta^{\mu\alpha\beta\gamma}_{\hspace{0.025cm}\delta\hspace{0.02cm}\epsilon\hspace{0.06cm}\eta\hspace{0.02cm}\kappa}g^{\delta\nu}\phi_\alpha^\epsilon R_{\beta\gamma}^{\quad\eta\kappa}+\mathcal{J}\delta^{\mu\alpha\beta}_{\gamma\hspace{0.03cm}\delta\hspace{0.05cm}\epsilon}g^{\gamma\nu}R_{\alpha\beta}^{\quad\delta\epsilon}\nonumber\\{}&{}+2X\mathcal{K}_{3X}\delta^{\mu\alpha\beta\gamma}_{\hspace{0.025cm}\delta\hspace{0.02cm}\epsilon\hspace{0.06cm}\eta\hspace{0.02cm}\kappa}g^{\delta\nu}\phi_\alpha\phi^\epsilon R_{\beta\gamma}^{\quad\eta\kappa}+\frac{4}{3}\left(X\mathcal{K}_{1X}\right)_X\delta^{\mu\alpha\beta\gamma}_{\hspace{0.025cm}\delta\hspace{0.02cm}\epsilon\hspace{0.06cm}\eta\hspace{0.02cm}\kappa}g^{\delta\nu}\phi_\alpha^\epsilon\phi_\beta^\eta\phi_\gamma^\kappa\nonumber\\{}&{}+4X\left(2X\mathcal{K}_{3XX}+3\mathcal{K}_{3X}-\mathcal{K}_{1X\phi}\right)\delta^{\mu\alpha\beta}_{\gamma\hspace{0.03cm}\delta\hspace{0.05cm}\epsilon}g^{\gamma\nu}\phi_\alpha^\delta \phi_\beta^\epsilon\nonumber\\{}&{}+4\left(X\mathcal{K}_{3X}\right)_X\delta^{\mu\alpha\beta\gamma}_{\hspace{0.025cm}\delta\hspace{0.02cm}\epsilon\hspace{0.06cm}\eta\hspace{0.02cm}\kappa}g^{\delta\nu}\phi_\alpha\phi^\epsilon\phi_\beta^\eta\phi_\gamma^\kappa+4\left(X^2\mathcal{K}_{8X}-\mathcal{J}_\phi\right)\delta^{\mu\alpha}_{\beta\gamma}g^{\beta\nu}\phi_\alpha^\gamma\nonumber\\{}&{}+2X\mathcal{K}_{8X}\delta^{\mu\alpha\beta}_{\gamma\hspace{0.03cm}\delta\hspace{0.05cm}\epsilon}g^{\gamma\nu}\phi_\alpha\phi^\delta \phi_\beta^\epsilon+2\left(X\mathcal{K}_{9X}-2\mathcal{K}_9\right)g^{\mu\nu}\nonumber\\{}&{}+2\left(X\mathcal{K}_{9XX}-\mathcal{K}_{9X}+2X^2\mathcal{K}_{8X\phi}-2\mathcal{J}_{\phi\phi}\right)\phi^\mu\phi^\nu,\label{eq:functionalderivativeofthetrace}
\end{align}
where 
\begin{equation}
\mathcal{J}=2\int\left[X\mathcal{K}_{3X}+2X\left(X\mathcal{K}_{3X}\right)_X-\left(X\mathcal{K}_{1X}\right)_\phi\right]\mathrm{d}X.
\end{equation}
Comparison of (\ref{eq:functionalderivativeofthetrace}) with (\ref{eq:final_expr_amunu_taking_div_account}) shows that
\begin{equation}
\frac{2}{\sqrt{-g}}\frac{\delta S}{\delta g_{\mu\nu}} = A^{\mu\nu},
\end{equation}
provided
\begin{align}
\mathcal{K}_1={}&{}\int\frac{K_1}{2X}\mathrm{d}X,\quad\mathcal{K}_3=\int\frac{K_3}{2X}\mathrm{d}X,\quad\mathcal{K}_8=\int\frac{K_8}{2X}\mathrm{d}X,\quad\mathcal{K}_9=X^2\int\frac{K_9}{2X^3}\mathrm{d}X,\\
\mathcal{F}={}&{}2\int\left(\mathcal{K}_3+2X\mathcal{K}_{3X}-\mathcal{K}_{1\phi}\right)\mathrm{d}X.
\end{align}
This proves that $A^{\mu\nu}$ as given by~(\ref{eq:final_expr_amunu_taking_div_account}) are indeed field equations arising from a variational principle, the associated action functional being given by Eq.~(\ref{eq:explicitactionappendixhorn}). This action~(\ref{eq:explicitactionappendixhorn}) is of the same form as the one presented in paragraph~\ref{subsec:horndeski}, Eq.~(\ref{eq:original}). This concludes the original construction by Horndeski of the most general four-dimensional scalar-tensor action leading to second-order field equations.
\chapter{Disformal transformations}\label{chap:appdisf}
\chaptermark{Appendix B~~Disformal transformations}
\sectionmark{Appendix B~~Disformal transformations}
Under the disformal transformation
\begin{equation}
g_{\mu\nu}\mapsto \tilde{g}_{\mu\nu} \equiv g_{\mu\nu}+D\left(\phi,X\right)\phi_\mu\phi_\nu,\quad \phi\mapsto\tilde{\phi}\equiv\phi,
\end{equation}
an initial Horndeski action
\begin{align}
S\left[g_{\mu\nu},\phi\right] = \int\mathrm{d}^4 x{}&{}\sqrt{-g}\Bigl\{G_2-G_3\Box\phi+G_4R+G_{4X}\left[\left(\Box\phi\right)^2-\phi_{\mu\nu}\phi^{\mu\nu}\right]\nonumber\\{}&{}+G_5 G^{\mu\nu}\phi_{\mu\nu}-\frac{G_{5X}}{6}\Bigl[\left(\Box\phi\right)^3-3\Box\phi\phi_{\mu\nu}\phi^{\mu\nu}+2\phi_{\mu\nu}\phi^{\nu\rho}\phi_\rho^\mu\Bigr]\Bigr\}
\end{align}
gets mapped to a new action,
\begin{equation}
\tilde{S}\left[\tilde{g}_{\mu\nu},\phi\right]\equiv S\left[g_{\mu\nu},\phi\right].
\end{equation}
The new action $\tilde{S}\left[\tilde{g}_{\mu\nu},\phi\right]$ belongs to the beyond Horndeski class,
\begin{align}
\tilde{S}\left[\tilde{g}_{\mu\nu},\phi\right]{}&{} =\int\mathrm{d}^4 x\sqrt{-\tilde{g}}\Bigl\{\widetilde{G_2}-\widetilde{G_3}\widetilde{\Box}\phi+\widetilde{G_4}\tilde{R}+\widetilde{G_{4}}_{\tilde{X}}\left[\left(\widetilde{\Box}\phi\right)^2-\widetilde{\phi_{\mu\nu}}\widetilde{\phi^{\mu\nu}}\right]\nonumber\\{}&{}+\widetilde{G_5}\tilde{G}^{\mu\nu}\widetilde{\phi_{\mu\nu}}-\frac{\widetilde{G_{5}}_{\tilde{X}}}{6}\Bigl[\left(\widetilde{\Box}\phi\right)^3-3\widetilde{\Box}\phi\widetilde{\phi_{\mu\nu}}\widetilde{\phi^{\mu\nu}}+2\widetilde{\phi_{\mu\nu}}\widetilde{\phi^{\nu\rho}}\widetilde{\phi_\rho^\mu}\Bigr]\nonumber\\{}&{}+ \widetilde{F_4} \tilde{\epsilon}^{\mu\nu\rho\sigma}\tilde{\epsilon}^{\alpha\beta\gamma}_{\,\,\,\,\,\,\,\,\,\,\sigma}\widetilde{\phi_\mu}\widetilde{\phi_\alpha}\widetilde{\phi_{\nu\beta}}\widetilde{\phi_{\rho\gamma}}+\widetilde{F_5}\tilde{\epsilon}^{\mu\nu\rho\sigma}\tilde{\epsilon}^{\alpha\beta\gamma\delta}\widetilde{\phi_\mu}\widetilde{\phi_\alpha}\widetilde{\phi_{\nu\beta}}\widetilde{\phi_{\rho\gamma}}\widetilde{\phi_{\sigma\delta}}\Bigr\}.
\end{align}
The expression of the new functions $\widetilde{G_2}$, $\widetilde{G_3}$, $\widetilde{G_4}$, $\widetilde{G_5}$, $\widetilde{F_4}$, $\widetilde{F_5}$, was computed in all generality in our article~\cite{Babichev:2022awg}, and is given by
\begin{align}
\widetilde{G_2}={}&{}G_2\sqrt{1+2D\tilde{X}}-2\tilde{X}\left(H_3+H_4+H_5\right)_\phi-\frac{2\tilde{X}^2G_3D_\phi}{\left(1+2D\tilde{X}\right)^{3/2}},\\
\widetilde{G_3}={}&{}\frac{G_3}{\sqrt{1+2D\tilde{X}}}-\left(H_3+H_4+H_5\right)\nonumber\\
{}&{}\hspace{-0.5cm}+2\tilde{X}\left\lbrace H_{R,\phi\phi}-H_{\Box,\phi}+\frac{1}{\sqrt{1+2D\tilde{X}}}
\left[2DG_{4\phi}-D_\phi\left(\frac{2\tilde{X}G_{4\tilde{X}}}{1-2\tilde{X}^2D_{\tilde{X}}}-G_4\right)\right]\right\rbrace,\\
\widetilde{G_4}={}&{}G_4\sqrt{1+2D\tilde{X}}+\tilde{X}\left(H_{R,\phi}-\frac{\tilde{X}G_5D_\phi}{\left(1+2D\tilde{X}\right)^{3/2}}\right),\\
\widetilde{G_5}={}&{}\frac{G_5}{\sqrt{1+2D\tilde{X}}}+H_R,
\end{align}
\begin{align}
\widetilde{F_4}={}&{}\frac{D_{\tilde{X}}}{2}\left(\frac{2\tilde{X}G_{4\tilde{X}}\sqrt{1+2D\tilde{X}}}
{1-2\tilde{X}^2D_{\tilde{X}}}-\frac{G_4}{\sqrt{1+2D\tilde{X}}}\right)-
\frac{\tilde{X}^3G_{5\tilde{X}}D_{\tilde{X}}D_\phi}{\left(1-2\tilde{X}^2D_{\tilde{X}}\right)\left(1+2D\tilde{X}\right)^{3/2}}\nonumber\\
{}&{}+\frac{G_5}{2\left(1+2D\tilde{X}\right)^{5/2}}\left\lbrace\tilde{X}
\left(1+2D\tilde{X}\right)D_{\phi\tilde{X}}+D_\phi\left[1-\tilde{X}\left(D+3\tilde{X}D_{\tilde{X}}\right)\right]\right\rbrace\nonumber\\
{}&{}\hspace{7.5cm}+\frac{G_{5\phi}D}{2\left(1+2D\tilde{X}\right)^{3/2}}-\frac{1}{2}H_{R,\phi\tilde{X}},\\ \widetilde{F_5}={}&{}-\frac{\tilde{X}G_{5\tilde{X}}D_{\tilde{X}}}{6\left(1-2\tilde{X}^2D_{\tilde{X}}\right)\sqrt{1+2D\tilde{X}}}.
\end{align}
For conciseness, the following functions were introduced,
\begin{align}
H_\Box ={}&{}
\frac{\tilde{X}G_5D_\phi}{\left(1+2D\tilde{X}\right)^{3/2}},\quad
H_R =
\int\mathrm{d}\tilde{X}\frac{G_5\left(D+\tilde{X}D_{\tilde{X}}\right)}{\left(1+2D\tilde{X}\right)^{3/2}},\nonumber\\
H_5={}&{}\int\mathrm{d}\tilde{X}\left(H_{\Box,\phi}-H_{R,\phi\phi}\right),\quad H_3 =
\int\mathrm{d}\tilde{X}\frac{-G_3\left(D+\tilde{X}D_{\tilde{X}}\right)}{\left(1+2D\tilde{X}\right)^{3/2}},\nonumber\\ H_4={}&{}\int\frac{\mathrm{d}\tilde{X}}{\sqrt{1+2D\tilde{X}}}\left[D_\phi\left(\frac{2\tilde{X}G_{4\tilde{X}}}
{1-2\tilde{X}^2D_{\tilde{X}}}-G_4\right)-2DG_{4\phi}\right],
\end{align}
thus following the notations of~\cite{Bettoni:2013diz}, with the difference that a dependence on the kinetic term $X$ is now included for the disformal function $D\left(\phi,X\right)$. The ingredients used for the computation are the following ones:
\begin{align}
\tilde{g}^{\mu\nu} ={}&{} g^{\mu\nu}-\frac{D}{1-2DX}\phi^\mu\phi^\nu,\\
\tilde{X}={}&\frac{X}{1-2DX},\quad 1+2D\tilde{X}=\frac{1}{1-2DX},\label{eq:xtildeappendixfinal}\\
g^{\mu\nu}={}& \tilde{g}^{\mu\nu}+\frac{D}{1+2D\tilde{X}}\tilde{\phi}^\mu\tilde{\phi}^\nu,\quad \sqrt{-g}=\sqrt{-\tilde{g}}\sqrt{1+2D\tilde{X}}.
\end{align}
One must compute the following tensor:
\begin{align*}
\tilde{C}\indices{^\lambda_\mu_\nu}\equiv{}&\tilde{\Gamma}^\lambda_{\mu\nu}-\Gamma^\lambda_{\mu\nu}\\
{}={}&\frac{D_\phi}{2\left(1+2D\tilde{X}\right)}\tilde{\phi}^\lambda\tilde{\phi}_\mu\tilde{\phi}_\nu-\frac{D_{\tilde{X}}}{1+2D\tilde{X}}\tilde{\phi}^\lambda\tilde{\phi}^\sigma\tilde{\phi}_{\sigma(\mu}\tilde{\phi}_{\nu)}+\frac{D_{\tilde{X}}}{2}\tilde{\phi}^{\lambda\sigma}\tilde{\phi}_\sigma\tilde{\phi}_\mu\tilde{\phi}_\nu\\
{}&\qquad+\frac{DD_{\tilde{X}}}{2\left(1+2D\tilde{X}\right)}\tilde{\phi}^\rho\tilde{\phi}_{\rho\sigma}\tilde{\phi}^\sigma\tilde{\phi}^\lambda\tilde{\phi}_\mu\tilde{\phi}_\nu+\frac{D}{1+2D\tilde{X}}\tilde{\phi}^\lambda\tilde{\phi}_{\mu\nu}.
\end{align*}
Then:
\begin{align}
R_{\sigma\nu}={}&\tilde{R}_{\sigma\nu}+\tilde{\nabla}_\nu\tilde{C}\indices{^\mu_\mu_\sigma}+\tilde{C}\indices{^\lambda_\nu_\sigma}\tilde{C}\indices{^\mu_\mu_\lambda}-\tilde{\nabla}_\mu\tilde{C}\indices{^\mu_\nu_\sigma}-\tilde{C}\indices{^\lambda_\mu_\sigma}\tilde{C}\indices{^\mu_\nu_\lambda},\\
R = {}&\left(\tilde{g}^{\sigma\nu}+\frac{D}{1+2D\tilde{X}}\tilde{\phi}^\sigma\tilde{\phi}^\nu\right)\left(\tilde{R}_{\sigma\nu}+\tilde{\nabla}_\nu\tilde{C}\indices{^\mu_\mu_\sigma}+\tilde{C}\indices{^\lambda_\nu_\sigma}\tilde{C}\indices{^\mu_\mu_\lambda}-\tilde{\nabla}_\mu\tilde{C}\indices{^\mu_\nu_\sigma}-\tilde{C}\indices{^\lambda_\mu_\sigma}\tilde{C}\indices{^\mu_\nu_\lambda}\right),\\
\phi_{\mu\nu}={}&\tilde{\phi}_{\mu\nu}+\tilde{C}\indices{^\lambda_\mu_\nu}\tilde{\phi}_\lambda.
\end{align}
\chapter{Field equations}\label{chap:appfieldeq}
\chaptermark{Appendix C~~Field equations}
\sectionmark{C.1~~Beyond Horndeski field equations}
This appendix details the field equations for both beyond Horndeski and quadratic DHOST theories. As in all this manuscript, we stick to the usual notations for beyond Horndeski and DHOST theories. In particular, the kinetic term $X$ is
\begin{equation}
X=-\frac{1}{2}\partial_\mu\phi\,\partial^\mu\phi\text{ for beyond Horndeski,}\quad X=\partial_\mu\phi\,\partial^\mu\phi\text{ for DHOST.}
\end{equation}
With these notations, the beyond Horndeski action is
\begin{align}
S\left[g_{\mu\nu},\phi\right] = \int\mathrm{d}^4 x{}&{}\sqrt{-g}\Bigl\{G_2-G_3\Box\phi+G_4\,R+G_{4X}\left[\left(\Box\phi\right)^2-\phi_{\mu\nu}\phi^{\mu\nu}\right]\nonumber\\{}&{}+G_5\,G^{\mu\nu}\phi_{\mu\nu}-\frac{G_{5X}}{6}\Bigl[\left(\Box\phi\right)^3-3\Box\phi\phi_{\mu\nu}\phi^{\mu\nu}+2\phi_{\mu\nu}\phi^{\nu\rho}\phi_\rho^\mu\Bigr]\nonumber\\{}&{}+ F_4\,\epsilon^{\mu\nu\rho\sigma}\epsilon^{\alpha\beta\gamma}_{\hspace{0.5cm}\sigma}\phi_\mu\phi_\alpha\phi_{\nu\beta}\phi_{\rho\gamma}+F_5\,\epsilon^{\mu\nu\rho\sigma}\epsilon^{\alpha\beta\gamma\delta}\phi_\mu\phi_\alpha\phi_{\nu\beta}\phi_{\rho\gamma}\phi_{\sigma\delta}\Bigr\},\label{eq:beyond_action_app}
\end{align}
with functions $G_{2,3,4,5}$, $F_{4,5}$ of the scalar field $\phi$ and its kinetic term $X=-\frac{1}{2}\partial_\mu\phi\,\partial^\mu\phi$, while the quadratic DHOST action is
\begin{align}
S\left[g_{\mu\nu},\phi\right]=\int\mathrm{d}^4x\sqrt{-g}\Bigl\{F\,R+P+Q{}&{}\Box\phi+A_1\phi_{\mu\nu}\phi^{\mu\nu}+A_2\left(\Box\phi\right)^2+A_3\Box\phi\phi^\mu\phi_{\mu\nu}\phi^\nu\nonumber \\ {}&{} +A_4\phi^\mu\phi_{\mu\nu}\phi^{\nu\rho}\phi_\rho+A_5\left(\phi^\mu\phi_{\mu\nu}\phi^\nu\right)^2\Bigr\},
\end{align}
with functions $F$, $P$ , $Q$, $A_{1,2,3,4,5}$ of the scalar field $\phi$ and its kinetic term $X=\partial_\mu\phi\,\partial^\mu\phi$. The usual abreviations are used, $\phi_\mu=\partial_\mu\phi=\nabla_\mu\phi$, $\phi_{\mu\nu}=\nabla_\mu\nabla_\nu\phi=\nabla_\nu\nabla_\mu\phi=\phi_{\nu\mu}$. The field equations are
\begin{equation}
\mathcal{E}_{\mu\nu}\equiv\frac{2}{\sqrt{-g}}\frac{\delta S}{\delta g^{\mu\nu}},\quad \mathcal{E}_\phi\equiv \frac{1}{\sqrt{-g}}\frac{\delta S}{\delta \phi}.
\end{equation}
\section{Beyond Horndeski field equations}\label{sec:eq_horn}
\sectionmark{C.1~~Beyond Horndeski field equations}
This section first focuses on the beyond Horndeski field equations. The metric field equations can be decomposed between the contributions of the various terms $G_2$, $G_3$, $G_4$, $G_5$, $F_4$ and $F_5$ as
\begin{equation}
\mathcal{E}_{\mu\nu}=\mathcal{E}_{\mu\nu}^{\left(G_2\right)}+\mathcal{E}_{\mu\nu}^{\left(G_3\right)}+\mathcal{E}_{\mu\nu}^{\left(G_4\right)}+\mathcal{E}_{\mu\nu}^{\left(G_5\right)}+\mathcal{E}_{\mu\nu}^{\left(F_4\right)}+\mathcal{E}_{\mu\nu}^{\left(F_5\right)}.
\end{equation}
The various contributions are computed to be:
\begin{flalign}
\mathcal{E}_{\mu\nu}^{\left(G_2\right)}=-G_2g_{\mu\nu}-G_{2X}\phi_\mu\phi_\nu,&&
\end{flalign}
\begin{flalign}
\mathcal{E}_{\mu\nu}^{\left(G_3\right)}=2G_{3\phi}\Bigl[\phi_\mu\phi_\nu+Xg_{\mu\nu}\Bigr]+G_{3X}\Bigl[\Box\phi\phi_\mu\phi_\nu+\phi^\rho\phi_{\rho\sigma}\phi^\sigma g_{\mu\nu}-2\phi^\rho\phi_{\rho(\mu}\phi_{\nu)}\Bigr],&&
\end{flalign}
\begin{flalign}
\mathcal{E}_{\mu\nu}^{\left(G_4\right)}={}&{}2G_4G_{\mu\nu}+G_{4X}\Bigl\{4\phi^\rho R_{\rho(\mu}\phi_{\nu)}+2\phi_{\mu\rho}\phi^{\rho}_{\nu}+\Bigl[\left(\Box\phi\right)^2-\phi_{\rho\sigma}\phi^{\rho\sigma}-2R_{\rho\sigma}\phi^\rho\phi^\sigma\Bigr]g_{\mu\nu}\nonumber\\{}&{}+2R_{\mu\rho\nu\sigma}\phi^\rho\phi^\sigma-R\phi_\mu\phi_\nu-2\Box\phi\phi_{\mu\nu}\Bigr\}+G_{4XX}\Bigl\{4\Box\phi\phi^\rho\phi_{\rho(\mu}\phi_{\nu)}-4\phi^\rho\phi_{\rho\sigma}\phi^{\sigma}_{(\mu}\phi_{\nu)}\nonumber\\{}&{}+2\phi^\rho\phi_{\rho\sigma}\phi^\sigma\phi_{\mu\nu}+2\Bigl[\phi^\rho\phi_{\rho\sigma}\phi^{\sigma\lambda}\phi_\lambda-\Box\phi\phi^\rho\phi_{\rho\sigma}\phi^\sigma\Bigr]g_{\mu\nu}-\Bigl[\left(\Box\phi\right)^2-\phi_{\rho\sigma}\phi^{\rho\sigma}\Bigr]\phi_\mu\phi_\nu\nonumber\\{}&{}-2\phi^\rho\phi_{\rho\mu}\phi^\sigma\phi_{\sigma\nu} \Bigr\}-2G_{4\phi}\Bigl[\phi_{\mu\nu}-\Box\phi g_{\mu\nu}\Bigr]-2G_{4\phi\phi}\Bigl[\phi_\mu\phi_\nu+2Xg_{\mu\nu}\Bigr]\nonumber\\{}&{}+4G_{4\phi X}\Bigl\{X\Bigl[\phi_{\mu\nu}-\Box\phi g_{\mu\nu}\Bigr]+2\phi^\rho\phi_{\rho(\mu}\phi_{\nu)}-\Box\phi\phi_\mu\phi_\nu-\phi^\rho\phi_{\rho\sigma}\phi^\sigma g_{\mu\nu}\Bigr\},&&
\end{flalign}
\begin{flalign}
\mathcal{E}_{\mu\nu}^{\left(G_5\right)}={}&{}G_{5X}\Bigl\{\frac{1}{2}R\Box\phi\phi_\mu\phi_\nu-R\phi^\rho\phi_{\rho(\mu}\phi_{\nu)}+2R_\rho^\sigma\phi^\rho\phi_{\sigma(\mu}\phi_{\nu)}-R^{\rho\sigma}\phi_{\rho\sigma}\phi_\mu\phi_\nu\nonumber\\{}&{}+2\phi^\rho\phi_{\rho\sigma}R^\sigma_{(\mu}\phi_{\nu)}+\Bigl[\left(\Box\phi\right)^2-\phi_{\rho\sigma}\phi^{\rho\sigma}\Bigr]\phi_{\mu\nu}-2\Box\phi\phi^\rho R_{\rho(\mu}\phi_{\nu)}-2\Box\phi\phi_\mu^\rho\phi_{\rho\nu}\nonumber\\{}&{}-R_{\rho\sigma}\phi^\rho\phi^\sigma\phi_{\mu\nu}+2\phi^\rho\phi^\sigma R_{\rho(\mu}\phi_{\nu)\sigma}-\phi^\rho\phi_{\rho\sigma}\phi^\sigma G_{\mu\nu}+2\phi_{\rho\sigma}\phi^\rho_\mu\phi^\sigma_\nu\nonumber\\{}&{}-2g_{\mu\nu}R_\rho^\sigma\phi^\rho\phi_{\sigma\lambda}\phi^\lambda-\frac{1}{3}g_{\mu\nu}\Bigl[\left(\Box\phi\right)^3\hspace{-0.1cm}-\hspace{-0.1cm}3\Box\phi\phi_{\rho\sigma}\phi^{\rho\sigma}\hspace{-0.1cm}+\hspace{-0.1cm}2\phi_{\rho\sigma}\phi^{\sigma\lambda}\phi_\lambda^\rho\Bigr]+\Box\phi R_{\rho\sigma}\phi^\rho\phi^\sigma g_{\mu\nu}\nonumber\\{}&{}-2\phi^\rho\phi^\sigma\phi^\lambda_{(\mu}R_{\nu)\rho\sigma\lambda}-\Box\phi R_{\mu\rho\nu\sigma}\phi^\rho\phi^\sigma+2R_{(\mu\lvert\rho\rvert\nu)\sigma}\phi^{\rho}\phi^{\sigma}_\lambda\phi^\lambda+2\phi^\sigma\phi^{\rho\lambda}\phi_{(\mu}R_{\nu)\rho\sigma\lambda}\nonumber\\{}&{}-g_{\mu\nu}R_{\rho\sigma\lambda\tau}\phi^\rho\phi^\lambda\phi^{\sigma\tau}\Bigr\}+G_{5XX}\Bigl\{\frac{1}{6}\Bigl[\left(\Box\phi\right)^3-3\Box\phi\phi_{\rho\sigma}\phi^{\rho\sigma}+2\phi_{\rho\sigma}\phi^{\sigma\lambda}\phi_\lambda^\rho\Bigr]\phi_\mu\phi_\nu\nonumber\\{}&{}+\Box\phi\phi^\rho\phi^\sigma\phi_{\rho\mu}\phi_{\sigma\nu}+\Bigl[\left(\Box\phi\right)^2-\phi_{\rho\sigma}\phi^{\rho\sigma}\Bigr]\Bigl[\frac{\phi^\lambda\phi_{\lambda\tau}\phi^\tau}{2}g_{\mu\nu}-\phi^\lambda\phi_{\lambda(\mu}\phi_{\nu)}\Bigr]\nonumber\\{}&{}-2\phi^\rho\phi^\sigma\phi_{\sigma\lambda}\phi_{\rho(\mu}\phi_{\nu)}^\lambda+\Box\phi\phi^\rho\phi_{\rho\sigma}\Bigl[2\phi^\sigma_{(\mu}\phi_{\nu)}-\phi^\sigma\phi_{\mu\nu}\Bigr]+\phi^\rho\phi_{\rho\sigma}\phi^\sigma \phi_\mu^\lambda\phi_{\lambda\nu}\nonumber\\{}&{}+\phi^\rho\phi_{\rho\sigma}\phi^\sigma_\lambda\Bigl[\phi^\lambda\phi_{\mu\nu}-2\phi^\lambda_{(\mu}\phi_{\nu)}\Bigr]+g_{\mu\nu}\phi^\rho\phi^\sigma\phi_{\sigma\lambda}\Bigl[\phi^\lambda_\tau\phi^\tau_\rho-\Box\phi\phi^\lambda_\rho\Bigr]\Bigr\}+G_{5\phi}\Bigl\{R\phi_\mu\phi_\nu\nonumber\\{}&{}-4\phi^\rho R_{\rho(\mu}\phi_{\nu)}-2XG_{\mu\nu}-\Bigl[\left(\Box\phi\right)^2-\phi_{\rho\sigma}\phi^{\rho\sigma}\Bigr]g_{\mu\nu}-2\phi_\mu^\rho\phi_{\rho\nu}\nonumber\\{}&{}+2\Bigl[R_{\rho\sigma}g_{\mu\nu}-R_{\mu\rho\nu\sigma}\Bigr]\phi^\rho\phi^\sigma+2\Box\phi\phi_{\mu\nu}\Bigr\}+G_{5\phi\phi}\Bigl\{\Box\phi\phi_\mu\phi_\nu-2X\Bigl[\phi_{\mu\nu}-\Box\phi g_{\mu\nu}\Bigr]\nonumber\\{}&{}-2\phi^\rho\phi_{\rho(\mu}\phi_{\nu)}+\phi^\rho\phi_{\rho\sigma}\phi^\sigma g_{\mu\nu}\Bigr\} +G_{5\phi X}\Bigl\{\Bigl[\left(\Box\phi\right)^2-\phi_{\rho\sigma}\phi^{\rho\sigma}\Bigr]\Bigl[\phi_\mu\phi_\nu+Xg_{\mu\nu}\Bigr]\nonumber\\{}&{}-4\Box\phi\phi^\rho\phi_{\rho(\mu}\phi_{\nu)}+2\phi^\rho\phi^\sigma\phi_{\rho\mu}\phi_{\sigma\nu}+2X\Bigl[\phi_\mu^\rho\phi_{\rho\nu}-\Box\phi\phi_{\mu\nu}\Bigr]\nonumber\\ {}&{}+2\phi^\rho\phi^\sigma\Bigl[\Bigl(\Box\phi\phi_{\rho\sigma}-\phi_{\rho}^\lambda\phi_{\lambda\sigma}\Bigr)g_{\mu\nu}-\phi_{\rho\sigma}\phi_{\mu\nu}\Bigr]+4\phi^\rho\phi_{\rho\sigma}\phi^\sigma_{(\mu}\phi_{\nu)}\Bigr\}, &&
\end{flalign}
\begin{flalign}
\mathcal{E}_{\mu\nu}^{\left(F_4\right)}={}&{}2XF_{4}\Bigl\{4\phi^\rho R_{\rho(\mu}\phi_{\nu)}+\Bigl[\left(\Box\phi\right)^2+\phi_{\rho\sigma}\phi^{\rho\sigma}-2R_{\rho\sigma}\phi^\rho\phi^\sigma\Bigr]g_{\mu\nu}-2\Box\phi\phi_{\mu\nu}-2\phi^\rho\nabla_\rho\phi_{\mu\nu}\nonumber\\{}&{}+2g_{\mu\nu}\phi^\rho\Box\phi_\rho\Bigr\} + 2F_4\Bigl\{\Bigl[3\phi_{\rho\sigma}\phi^{\rho\sigma}-2\left(\Box\phi\right)^2+\phi^\rho\Box\phi_\rho+R_{\rho\sigma}\phi^\rho\phi^\sigma\Bigr]\phi_\mu\phi_\nu\nonumber\\{}&{}+6\Box\phi\phi^\rho\phi_{\rho(\mu}\phi_{\nu)}-2\phi^\rho\phi^\sigma\phi_{\rho\mu}\phi_{\sigma\nu}+2\phi^\rho\phi_{\rho\sigma}\phi^\sigma\phi_{\mu\nu}-2\phi^\rho\phi^\sigma\nabla_\rho\phi_{\sigma(\mu}\phi_{\nu)}\nonumber\\{}&{}-8\phi^\rho\phi_{\rho\sigma}\phi^\sigma_{(\mu}\phi_{\nu)}+g_{\mu\nu}\phi^\rho\phi^\sigma\Bigl[\phi^\lambda\nabla_\lambda\phi_{\rho\sigma}+3\phi_{\rho}^\lambda\phi_{\lambda\sigma}-2\Box\phi\phi_{\rho\sigma}\Bigr]\Bigr\}\nonumber\\{}&{}+2XF_{4X}\Bigl\{4\Box\phi\phi^\rho\phi_{\rho(\mu}\phi_{\nu)}-4\phi^\rho\phi_{\rho\sigma}\phi^{\sigma}_{(\mu}\phi_{\nu)}+2\phi^\rho\phi_{\rho\sigma}\phi^\sigma\phi_{\mu\nu}-2\Box\phi\phi^\rho\phi_{\rho\sigma}\phi^\sigma g_{\mu\nu}\nonumber\\{}&{}-\Bigl[\left(\Box\phi\right)^2-\phi_{\rho\sigma}\phi^{\rho\sigma}\Bigr]\phi_\mu\phi_\nu\Bigr\}+2F_{4X}\phi^\rho\phi_{\rho\sigma}\Bigl\{2\phi^\sigma\phi^\lambda\phi_{\lambda(\mu}\phi_{\nu)}-\phi^\sigma_\lambda\phi^\lambda\phi_\mu\phi_\nu\nonumber\\{}&{}-g_{\mu\nu}\phi^\sigma\phi^\lambda\phi_{\lambda\tau}\phi^\tau\Bigr\}+4XF_{4\phi}\Bigl\{2X\Bigl[\phi_{\mu\nu}-\Box\phi g_{\mu\nu}\Bigr]-\Box\phi\Bigl[\phi_\mu\phi_\nu+2Xg_{\mu\nu}\Bigr]\nonumber\\{}&{}+2\phi^\rho\phi_{\rho(\mu}\phi_{\nu)}\Bigr\}, &&
\end{flalign}
\begin{flalign}
\mathcal{E}_{\mu\nu}^{\left(F_5\right)}={}&{}-12XF_5\Bigl\{2R_\rho^\sigma\phi^\rho\phi_{\sigma(\mu}\phi_{\nu)}+\left(\Box\phi\right)^2\phi_{\mu\nu}-2\Box\phi\phi^\rho R_{\rho(\mu}\phi_{\nu)}-R_{\rho\sigma}\phi^\rho\phi^\sigma\phi_{\mu\nu}\nonumber\\{}&{}-\frac{1}{3}g_{\mu\nu}\Bigl[\left(\Box\phi\right)^3-\phi_{\rho\sigma}\phi^{\sigma\lambda}\phi_\lambda^\rho\Bigr]+\Box\phi R_{\rho\sigma}\phi^\rho\phi^\sigma g_{\mu\nu}-2\phi^\rho\phi^\sigma\phi^\lambda_{(\mu}R_{\nu)\rho\sigma\lambda}\nonumber\\{}&{}+2\phi^\sigma\phi^{\rho\lambda}\phi_{(\mu}R_{\nu)\rho\sigma\lambda}-g_{\mu\nu}R_{\rho\sigma\lambda\tau}\phi^\rho\phi^\lambda\phi^{\sigma\tau}+\Box\phi\Bigl[\phi^\rho\nabla_\rho\phi_{\mu\nu}-\phi_{\mu}^\rho\phi_{\rho\nu}-\phi^\rho\Box\phi_\rho g_{\mu\nu}\Bigr]\nonumber\\{}&{}+\phi^\rho\Box\phi_\rho\phi_{\mu\nu}-2\phi^\rho\nabla_\sigma\phi_{\rho(\mu}\phi_{\nu)}^\sigma+\phi^\rho\nabla_\sigma\phi_{\rho\lambda}\phi^{\sigma\lambda}g_{\mu\nu}\Bigr\}+12F_{5}\phi^\rho\phi_{\rho\sigma}\Bigl\{\phi^\sigma\phi^\lambda R_{\lambda(\mu}\phi_{\nu)}\nonumber\\{}&{}-4\Box\phi\phi^\sigma_{(\mu}\phi_{\nu)}+2\phi^\lambda\phi_{\lambda(\mu}\phi_{\nu)}^\sigma+\Box\phi\phi^\sigma_\lambda\phi^\lambda-\phi^\sigma_\lambda\phi^\lambda\phi_{\mu\nu}-R^\sigma_\lambda\phi^\lambda\phi_\mu\phi_\nu+\frac{1}{2}\Box\phi\phi^\sigma\phi_{\mu\nu}\nonumber\\{}&{}-\frac{1}{2}\phi^\sigma\phi^\lambda\nabla_\lambda\phi_{\mu\nu}-\phi^\sigma\phi_{\mu}^\rho\phi_{\rho\nu}-\frac{\phi^\sigma}{4}\Bigl[\left(\Box\phi\right)^2-2\phi^\rho\Box\phi_\rho-3\phi_{\lambda\tau}\phi^{\lambda\tau}\Bigr]g_{\mu\nu}\Bigr\}
\nonumber\\{}&{}+3F_5\phi^\rho\Bigl\{\Bigl[3\left(\Box\phi\right)^2-5\phi_{\lambda\tau}\phi^{\lambda\tau}-2\phi^\sigma \Box\phi_{\sigma}\Bigr]\phi_{\rho(\mu}\phi_{\nu)}+2\phi^\sigma\phi^\lambda\nabla_\sigma\phi_{\lambda(\mu}\phi_{\nu)\rho}\nonumber\\{}&{}-2\Box\phi\phi^\sigma\phi_{\sigma\mu}\phi_{\nu\rho}\Bigr\}+F_5\phi_\mu\phi_\nu\Bigl\{6\Box\phi \Bigl[R_{\rho\sigma}\phi^\rho\phi^\sigma+\phi^\rho\Box\phi_\rho+21\phi_{\rho\sigma}\phi^{\rho\sigma}-5\left(\Box\phi\right)^2\Bigr]\nonumber\\{}&{}-16\phi_{\rho\sigma}\phi^{\sigma\lambda}\phi^\rho_\lambda-6\phi^\rho\phi^{\sigma\lambda}\nabla_{\sigma}\phi_{\rho\lambda}-6R_{\rho\sigma\lambda\tau}\phi^\rho\phi^\lambda\phi^{\sigma\tau}\Bigr\}+6F_5\phi^\sigma\phi^\lambda\Bigl\{2\phi^\rho_\sigma\nabla_\rho\phi_{\lambda(\mu}\phi_{\nu)}\nonumber\\{}&{}+\nabla_\rho\phi_{\sigma\lambda}\Bigl[\phi^\rho\Bigl(\Box\phi g_{\mu\nu}-\phi_{\mu\nu}\Bigr)+2\phi^{\rho}_{(\mu}\phi_{\nu)}\Bigr]-2\Box\phi\nabla_{\sigma}\phi_{\lambda(\mu}\phi_{\nu)}-g_{\mu\nu}\Bigl[R_{\sigma\lambda}\phi^\rho\phi_{\rho\tau}\phi^\tau\nonumber\\{}&{}+2\phi^\rho\phi^\tau_{\sigma}\nabla_\tau\phi_{\rho\lambda}+3\phi^\rho_\sigma\phi_{\rho\tau}\phi^\tau_{\lambda}\Bigr]\Bigr\}+60F_5\phi^\rho\phi_{\rho\sigma}\phi^{\sigma\lambda}\phi_{\lambda(\mu}\phi_{\nu)}+2XF_{5X}\Bigl\{-\phi_{\mu}\phi_\nu\nonumber\\{}&{}\Bigl[\left(\Box\phi\right)^3-3\Box\phi\phi_{\rho\sigma}\phi^{\rho\sigma}+2\phi_{\rho\sigma}\phi^{\sigma\lambda}\phi^\rho_\lambda\Bigr]+ 3\Bigl[\left(\Box\phi\right)^2-\phi_{\rho\sigma}\phi^{\rho\sigma}\Bigr]\Bigl[2\phi^\lambda\phi_{\lambda(\mu}\phi_{\nu)}\nonumber\\{}&{}-\phi^\lambda\phi_{\lambda\tau}\phi^\tau g_{\mu\nu}\Bigr]+6\phi^\rho\phi_{\rho\sigma}\Bigl[\phi^\sigma\Box\phi\phi_{\mu\nu}+2\phi^{\sigma\lambda}\phi_{\lambda(\mu}\phi_{\nu)}-2\phi^\sigma\phi_{\mu}^\lambda\phi_{\nu\lambda}-2\Box\phi\phi^\sigma_{(\mu}\phi_{\nu)}\Bigr]\Bigr\}\nonumber\\{}&{}+6F_{5X}\Bigl\{\phi^\rho\phi_{\rho\sigma}\phi^\sigma\Bigl[2\Box\phi\phi^\lambda\phi_{\lambda(\mu}\phi_{\nu)}-\phi^\lambda\phi_{\lambda\mu}\phi^\tau\phi_{\tau\nu}+\phi^\lambda\phi_{\lambda\tau}\phi^\tau\Bigl(\phi_{\mu\nu}-\Box\phi g_{\mu\nu}\Bigr)\nonumber\\{}&{}+\phi^\lambda\phi_{\lambda\tau}\Bigl(\phi^{\tau\gamma}\phi_\gamma g_{\mu\nu}-2\phi^\tau_{(\mu}\phi_{\nu)}\Bigr)\Bigr]+\phi^\rho\phi^\sigma\phi^\lambda_{\rho}\Bigl[\phi_{\lambda\tau}\phi^\tau_\sigma-\Box\phi\phi_{\sigma\lambda}\Bigr]\phi_\mu\phi_\nu\Bigr\}\nonumber\\{}&{}+6XF_{5\phi}\Bigl\{2\Bigl[\phi^\rho\phi_{\rho\sigma}\phi^\sigma+2X\Box\phi\Bigr]\phi_{\mu\nu}-\Bigl[\left(\Box\phi\right)^2-\phi_{\rho\sigma}\phi^{\rho\sigma}\Bigr]\Bigl[\phi_\mu\phi_\nu+2Xg_{\mu\nu}\Bigr]\nonumber\\{}&{}+4\Box\phi\phi^\rho\phi_{\rho(\mu}\phi_{\nu)}-2\phi^\rho\phi_{\rho\mu}\phi^\sigma\phi_{\sigma\nu}-4X\phi_\mu^\rho\phi_{\rho\nu}-4\phi^\rho\phi_{\rho\sigma}\phi^\sigma_{(\mu}\phi_{\nu)}\nonumber\\{}&{}+2g_{\mu\nu}\phi^\rho\phi^\sigma\Bigl[\phi_{\rho\lambda}\phi^\lambda_\sigma-\Box\phi\phi_{\rho\sigma}\Bigr]\Bigr\}. &&
\end{flalign}
On the other hand, taking inspiration from~\cite{Kobayashi:2011nu}, the scalar field equation can be written in a compact way as
\begin{equation}
\mathcal{E}_{\phi}=\sum_{i=2}^5\Bigl(\mathcal{P}^{\left(G_i\right)}-\nabla^\mu \mathcal{J} ^{\left(G_i\right)}_\mu\Bigr)+\sum_{i=4}^5\Bigl(\mathcal{P}^{\left(F_i\right)}-\nabla^\mu \mathcal{J} ^{\left(F_i\right)}_\mu\Bigr)
\end{equation}
The various 'currents' $\mathcal{J}^ \mu$ are obtained by introducing an auxiliary vector field $\chi^\mu$ and making the replacements $\nabla_\mu\phi\to\chi_\mu$, $\nabla_\mu\nabla_\nu\phi\to\nabla_\mu\chi_\nu$ in the action, and computing the functional derivative $\left(\sqrt{-g}\right)^{-1}\delta S/\delta \chi^\mu$. As a consequence, if the action is shift-symmetric, the associated Noether current $\mathcal{J}^\mu_{\text{shift}}$, see Eq. (\ref{eq:shift_current}), is immediately given by
\begin{equation}
\mathcal{J}^\mu_{\text{shift}} = \mathcal{J}^{\left(G_2\right)\,\mu}+\mathcal{J}^{\left(G_3\right)\,\mu}+\mathcal{J}^{\left(G_4\right)\,\mu}+\mathcal{J}^{\left(G_5\right)\,\mu}+\mathcal{J}^{\left(F_4\right)\,\mu}+\mathcal{J}^{\left(F_5\right)\,\mu}\text{ if shift symmetry.} 
\end{equation}
The various currents are given by
\begin{flalign}
\mathcal{J}^{\left(G_2\right)\,\mu}=-G_{2X}\phi^\mu,\quad \mathcal{J}^{\left(G_3\right)\,\mu}=G_{3X}\Bigl[\Box\phi\phi^\mu-\phi_\nu\phi^{\nu\mu}\Bigr]+G_{3\phi}\phi^\mu,&&
\end{flalign}
\begin{flalign}
\mathcal{J}^{\left(G_4\right)\,\mu}={}&{}2G_{4X}G^{\mu\nu}\phi_\nu+G_{4XX}\Bigl\{2\Box\phi\phi_\nu\phi^{\nu\mu}-2\phi^\nu\phi_{\nu\rho}\phi^{\rho\mu}-\Bigl[\left(\Box\phi\right)^2-\phi_{\rho\sigma}\phi^{\rho\sigma}\Bigr]\phi^\mu\Bigr\}\nonumber\\{}&{}-2G_{4\phi X}\Bigl[\Box\phi\phi^\mu-\phi_\nu\phi^{\nu\mu}\Bigr],&&
\end{flalign}
\begin{flalign}
\mathcal{J}^{\left(G_5\right)\,\mu}={}&{}G_{5X}\Bigl\{\frac{R}{2}\Bigl[\Box\phi\phi^\mu-\phi_\nu\phi^{\nu\mu}\Bigr]+R_{\nu\rho}\phi^\nu\phi^{\rho\mu}-R_{\nu\rho}\phi^{\nu\rho}\phi^\mu+\phi^\nu\phi_{\nu\rho}R^{\rho\mu}-\Box\phi\phi_{\nu}R^{\nu\mu}\nonumber\\{}&{}+R^{\mu}_{\,\,\nu\rho\sigma}\phi^\rho\phi^{\nu\sigma}\Bigr\}+G_{5XX}\Bigl\{\frac{\phi^\mu}{6}\Bigl[\left(\Box\phi\right)^3-3\Box\phi\phi_{\nu\rho}\phi^{\nu\rho}+2\phi_{\sigma\nu}\phi^{\nu\rho}\phi_\rho^\sigma\Bigr]-\nonumber\\{}&{}\frac{1}{2}\Bigl[\left(\Box\phi\right)^2-\phi_{\rho\sigma}\phi^{\rho\sigma}\Bigr]\phi_\nu\phi^{\nu\mu}+\Bigl[\Box\phi\phi^\nu\phi_{\nu\rho}-\phi_\nu\phi^{\nu\sigma}\phi_{\sigma\rho}\Bigr]\phi^{\rho\mu}\Bigr\}-G_{5\phi}G^{\mu\nu}\phi_\nu\nonumber\\{}&{}+G_{5\phi X}\Bigl\{\frac{1}{2}\Bigl[\left(\Box\phi\right)^2-\phi_{\rho\sigma}\phi^{\rho\sigma}\Bigr]\phi^\mu+\Bigl[\phi^\nu\phi_{\nu\rho}-\Box\phi\phi_{\rho}\Bigr]\phi^{\rho\mu}\Bigr\},&&
\end{flalign}
\begin{flalign}
\mathcal{J}^{\left(F_4\right)\,\mu}={}&{}2F_4\Bigl\{2XR^{\mu\nu}\phi_\nu-\Bigl[2\left(\Box\phi\right)^2-3\phi_{\rho\sigma}\phi^{\rho\sigma}\Bigr]\phi^\mu+R_{\nu\rho}\phi^\nu\phi^\rho\phi^\mu+\phi^\nu\Box\phi_\nu\phi^\mu\nonumber\\{}&{}-\Bigl[4\phi^\nu\phi_{\nu\rho}-3\Box\phi\phi_{\rho}\Bigr]\phi^{\rho\mu}-\phi^\nu\phi^\rho\nabla_\nu\phi_\rho^\mu\Bigr\}+2F_{4X}\Bigl\{X\Bigl[2\Box\phi\phi_\nu\phi^{\nu\mu}-2\phi^\nu\phi_{\nu\rho}\phi^{\rho\mu}\nonumber\\{}&{}-\Bigl(\left(\Box\phi\right)^2-\phi_{\rho\sigma}\phi^{\rho\sigma}\Bigr)\phi^\mu\Bigr]+\phi^\nu\phi_{\nu\rho}\phi_\sigma\Bigl[\phi^\rho\phi^{\sigma\mu}-\phi^{\rho\sigma}\phi^\mu\Bigr]\Bigr\}\nonumber\\{}&{}+2F_{4\phi}\phi_\nu\Bigl[\phi^{\nu\rho}\phi_\rho\phi^\mu+2X\phi^{\nu\mu}\Bigr],&&
\end{flalign}
\begin{flalign}
\mathcal{J}^{\left(F_5\right)\,\mu}={}&{}F_5\Bigl\{\Bigl[-5\left(\Box\phi\right)^3+21\Box\phi\phi_{\nu\rho}\phi^{\nu\rho}-10\phi_{\sigma\nu}\phi^{\nu\rho}\phi_\rho^\sigma\Bigr]\phi^\mu\nonumber\\{}&{}+3\Bigl[3\left(\Box\phi\right)^2-5\phi_{\rho\sigma}\phi^{\rho\sigma}\Bigr]\phi_\nu\phi^{\nu\mu}+6R_{\nu\rho}\phi^\nu\phi^\mu\Bigl[\Box\phi\phi^\rho-2\phi^{\rho\sigma}\phi_{\sigma}\Bigr]\nonumber\\{}&{}+12X\phi_\nu\Bigl[\Box\phi R^{\mu\nu}-R^{\nu\rho}\phi^\mu_\rho\Bigr]+6\Box\phi_\rho\Bigl[\phi_\nu\phi^{\nu\rho}\phi^\mu+\Box\phi\phi^\rho\phi^\mu-\phi^\rho\phi_\nu\phi^{\nu\mu}\Bigr]\nonumber\\{}&{}-6\Box\phi\phi^\nu\Bigl[\phi^\rho\nabla_\nu\phi_{\rho}^\mu+5\phi_{\nu\rho}\phi^{\rho\mu}\Bigr]+30\phi^\nu\phi_{\nu\rho}\phi^{\rho}_\sigma\phi^{\sigma\mu}+6\phi^\rho\phi_{\rho\sigma}\phi^\sigma R^{\mu\nu}\phi_{\nu}\nonumber\\{}&{}-6\phi^\nu\phi^{\rho\sigma}\Bigl[2XR^{\mu}_{\,\,\rho\nu\sigma}+R^{\mu}_{\,\,\nu\lambda\rho}\phi^\lambda\phi_\sigma+R_{\lambda\rho\nu\sigma}\phi^\lambda\phi^\mu\Bigr]\Bigr\}\nonumber\\{}&{}-12F_{5X}\Bigl\{X\Bigl[\frac{\phi^\mu}{6}\Bigl(\left(\Box\phi\right)^3-3\Box\phi\phi_{\nu\rho}\phi^{\nu\rho}+2\phi_{\sigma\nu}\phi^{\nu\rho}\phi_\rho^\sigma\Bigr)\nonumber\\{}&{}-\frac{1}{2}\Bigl(\left(\Box\phi\right)^2-\phi_{\rho\sigma}\phi^{\rho\sigma}\Bigr)\phi_\nu\phi^{\nu\mu}+\Bigl(\Box\phi\phi^\nu\phi_{\nu\rho}-\phi_\nu\phi^{\nu\sigma}\phi_{\sigma\rho}\Bigr)\phi^{\rho\mu}\Bigr]\nonumber\\{}&{}-\frac{1}{2}\Box\phi\phi^\nu\phi_{\nu\rho}\phi_\sigma\Bigl[\phi^\rho\phi^{\sigma\mu}-\phi^{\rho\sigma}\phi^\mu\Bigr]\Bigr\}+6F_{5\phi}\Box\phi\phi_\nu\Bigl[\phi^{\nu\rho}\phi_\rho\phi^\mu+2X\phi^{\nu\mu}\Bigr].&&
\end{flalign}
The remaining parts $\mathcal{P}$ are nothing but the partial derivatives with respect to $\phi$ of the respective Lagrangians,
\begin{flalign}
\mathcal{P}^{\left(G_2\right)}=G_{2\phi},\quad \mathcal{P}^{\left(G_3\right)}=-G_{3\phi}\Box\phi ,\quad \mathcal{P}^{\left(G_4\right)}=G_{4\phi}R+G_{4\phi X}\Bigl[\left(\Box\phi\right)^2-\phi_{\mu\nu}\phi^{\mu\nu}\Bigr],&&
\end{flalign} 
\begin{flalign}
\mathcal{P}^{\left(G_5\right)}=G_{5\phi}G^{\mu\nu}\phi_{\mu\nu}-\frac{G_{5\phi X}}{6}\Bigl[\left(\Box\phi\right)^3-3\Box\phi\phi_{\mu\nu}\phi^{\mu\nu}+2\phi_{\mu\nu}\phi^{\nu\rho}\phi_\rho^\mu\Bigr],&&
\end{flalign}
\begin{flalign}
\mathcal{P}^{\left(F_4\right)}=2F_{4\phi}\Bigl\{X\Bigl[\left(\Box\phi\right)^2-\phi_{\mu\nu}\phi^{\mu\nu}\Bigr]+\Box\phi\phi^\mu\phi_{\mu\nu}\phi^\nu-\phi^\mu\phi_{\mu\nu}\phi^{\nu\rho}\phi_{\rho}\Bigr\},&&
\end{flalign}
\begin{flalign}
\mathcal{P}^{\left(F_5\right)}={}&{}F_{5\phi}\Bigl\{2X\Bigl[\left(\Box\phi\right)^3 - 3\Box\phi
\phi_{\mu\nu}\phi^{\mu\nu} +2\phi_{\mu\nu}\phi^{\nu\rho}\phi^\mu_\rho\Bigr]+3\Bigl[\left(\Box\phi\right)^2\phi^\mu\phi_{\mu\nu}\phi^\nu\nonumber\\{}&{}-2\Box\phi
\phi_\mu\phi^{\mu\nu}\phi_{\nu\rho}\phi^\rho-\phi_{\mu\nu}\phi^{\mu\nu}
\phi^\rho\phi_{\rho\sigma}\phi^\sigma+2\phi_\mu\phi^{\mu\nu}\phi_{\nu\rho}\phi^{\rho\sigma}\phi_\sigma\Bigr]\Bigr\}.&&
\end{flalign}
\section{Quadratic DHOST field equations}\label{sec:eq_dhost}
\sectionmark{C.2~~Quadratic DHOST field equations}
As regards quadratic DHOST theories, the metric field equations also split as
\begin{equation}
\mathcal{E}_{\mu\nu}=\mathcal{E}_{\mu\nu}^{\left(F\right)}+\mathcal{E}_{\mu\nu}^{\left(P\right)}+\mathcal{E}_{\mu\nu}^{\left(Q\right)}+\mathcal{E}_{\mu\nu}^{\left(A_1\right)}+\mathcal{E}_{\mu\nu}^{\left(A_2\right)}+\mathcal{E}_{\mu\nu}^{\left(A_3\right)}+\mathcal{E}_{\mu\nu}^{\left(A_4\right)}+\mathcal{E}_{\mu\nu}^{\left(A_5\right)}.
\end{equation}
The different contributions are given by
\begin{flalign}
\mathcal{E}_{\mu\nu}^{\left(F\right)}={}&{}2FG_{\mu\nu}+2F_{X}\Bigl\{2\Bigl[\phi^\rho\Box\phi_\rho+\phi_{\rho\sigma}\phi^{\rho\sigma}\Bigr]g_{\mu\nu}-2\phi_{\mu\rho}\phi^{\rho}_{\nu}-2R_{\mu\rho\nu\sigma}\phi^\rho\phi^\sigma+R\phi_\mu\phi_\nu\nonumber\\{}&{}-2\phi^\rho\nabla_\rho\phi_{\mu\nu}\Bigr\}+8F_{XX}\Bigl\{\phi^\rho\phi_{\rho\sigma}\phi^{\sigma\lambda}\phi_\lambda g_{\mu\nu}-\phi^\rho\phi_{\rho\mu}\phi^\sigma\phi_{\sigma\nu} \Bigr\}-2F_{\phi}\Bigl[\phi_{\mu\nu}-\Box\phi g_{\mu\nu}\Bigr]\nonumber\\{}&{}-2F_{\phi\phi}\Bigl[\phi_\mu\phi_\nu-Xg_{\mu\nu}\Bigr]-8F_{\phi X}\Bigl\{\phi^\rho\phi_{\rho(\mu}\phi_{\nu)}-\phi^\rho\phi_{\rho\sigma}\phi^\sigma g_{\mu\nu}\Bigr\},&&
\end{flalign}
\begin{flalign}
\mathcal{E}_{\mu\nu}^{\left(P\right)}=2P_X\phi_\mu\phi_\nu-P\,g_{\mu\nu},&&
\end{flalign}
\begin{flalign}
\mathcal{E}_{\mu\nu}^{\left(Q\right)}=-Q_{\phi}\Bigl[2\phi_\mu\phi_\nu-Xg_{\mu\nu}\Bigr]+2Q_{X}\Bigl[\Box\phi\phi_\mu\phi_\nu+\phi^\rho\phi_{\rho\sigma}\phi^\sigma g_{\mu\nu}-2\phi^\rho\phi_{\rho(\mu}\phi_{\nu)}\Bigr],&&
\end{flalign}
\begin{flalign}
\mathcal{E}_{\mu\nu}^{\left(A_1\right)}={}&{}A_{1}\Bigl\{2\Box\phi\phi_{\mu\nu}-\phi_{\rho\sigma}\phi^{\rho\sigma}g_{\mu\nu}-4\Box\phi_{(\mu}\phi_{\nu)}+2\phi^\rho\nabla_\rho\phi_{\mu\nu}\Bigr\}+2A_{1X}\Bigl\{2\phi^\rho\phi_{\rho\sigma}\phi^\sigma\phi_{\mu\nu}\nonumber\\{}&{}-4\phi^\rho\phi_{\rho\sigma}\phi^{\sigma}_{(\mu}\phi_{\nu)}+\phi_{\rho\sigma}\phi^{\rho\sigma}\phi_\mu\phi_\nu\Bigr\}+2A_{1\phi}\Bigl\{X\phi_{\mu\nu}-2\phi^\rho\phi_{\rho(\mu}\phi_{\nu)}\Bigr\},&&
\end{flalign}
\begin{flalign}
\mathcal{E}_{\mu\nu}^{\left(A_2\right)}={}&{}A_{2}\Bigl\{4\phi^\rho R_{\rho(\mu}\phi_{\nu)}+\Bigl[\left(\Box\phi\right)^2-2R_{\rho\sigma}\phi^\rho\phi^\sigma+2\phi^\rho\Box\phi_{\rho}\Bigr]g_{\mu\nu}-4\Box\phi_{(\mu}\phi_{\nu)}\Bigr\}\nonumber\\{}&{}+2A_{2X}\Bigl\{\left(\Box\phi\right)^2\phi_\mu\phi_\nu+2\Box\phi\phi^\rho\phi_{\rho\sigma}\phi^\sigma g_{\mu\nu}-4\Box\phi\phi^\rho\phi_{\rho(\mu}\phi_{\nu)} \Bigr\}\nonumber\\{}&{}+2A_{2\phi}\Box\phi\Bigl\{Xg_{\mu\nu}-2\phi_\mu\phi_\nu\Bigr\},&&
\end{flalign}
\begin{flalign}
\mathcal{E}_{\mu\nu}^{\left(A_3\right)}={}&{}A_3\Bigl\{\Bigl[R_{\rho\sigma}\phi^\rho\phi^\sigma-\left(\Box\phi\right)^2-\phi^\rho\Box\phi_\rho\Bigr]\phi_\mu\phi_\nu+2\Box\phi\phi^\rho\phi_{\rho(\mu}\phi_{\nu)}-4\phi^\rho\phi_{\rho\sigma}\phi^\sigma_{(\mu}\phi_{\nu)}\nonumber\\{}&{}-2\phi^\rho\phi^\sigma\nabla_\rho\phi_{\sigma(\mu}\phi_{\nu)}+g_{\mu\nu}\phi^\rho\phi^\sigma\Bigl[\phi^\lambda\nabla_\lambda\phi_{\rho\sigma}+2\phi_{\rho}^\lambda\phi_{\lambda\sigma}\Bigr]\Bigr\}\nonumber\\{}&{}+2A_{3X}\phi^\rho\phi_{\rho\sigma}\phi^\sigma\Bigl\{g_{\mu\nu}\phi^\lambda\phi_{\lambda\tau}\phi^\tau-2\phi^\lambda\phi_{\lambda(\mu}\phi_{\nu)}\Bigr\}\nonumber\\{}&{}-A_{3\phi}\Bigl\{\phi^\rho\phi_{\rho\sigma}\phi^\sigma\Bigl[2\phi_\mu\phi_\nu-Xg_{\mu\nu}\Bigr]+X\Box\phi\phi_\mu\phi_\nu\Bigr\}, &&
\end{flalign}
\begin{flalign}
\mathcal{E}_{\mu\nu}^{\left(A_4\right)}={}&{}A_{4}\Bigl\{2\phi^\rho\phi^\sigma\phi_{\rho\mu}\phi_{\sigma\nu}-\phi^\rho\phi_{\rho\sigma}\phi^{\sigma\lambda}\phi_{\lambda}g_{\mu\nu}-2\Bigl[\phi^\rho\Box\phi_\rho+\phi_{\rho\sigma}\phi^{\rho\sigma}\Bigr]\phi_\mu\phi_\nu\Bigr\}\nonumber\\{}&{}-2\phi^\rho\phi^\sigma\Bigl[A_{4X}\phi_{\rho}^\lambda\phi_{\lambda\sigma}+A_{4\phi}\phi_{\rho\sigma}\Bigr]\phi_\mu\phi_\nu, &&
\end{flalign}
\begin{flalign}
\mathcal{E}_{\mu\nu}^{\left(A_5\right)}={}&{}A_{5}\Bigl\{\phi^\rho\phi_{\rho\sigma}\phi^\sigma\Bigl[4\phi^\lambda\phi_{\lambda(\mu}\phi_{\nu)}-\phi^\lambda\phi_{\lambda\tau}\phi^\tau g_{\mu\nu}\Bigr]-2\phi^\rho\phi^\sigma\Bigl[\Box\phi\phi_{\rho\sigma}+\phi^\lambda\nabla_\lambda\phi_{\rho\sigma}\nonumber\\{}&{}+2\phi_{\rho}^\lambda\phi_{\lambda\sigma}\Bigr]\phi_\mu\phi_\nu\Bigr\}-2\phi^\rho\phi_{\rho\sigma}\phi^\sigma\Bigl[A_{5X}\phi^\lambda\phi_{\lambda\tau}\phi^\tau+A_{5\phi}X\Bigr]\phi_\mu\phi_\nu. &&
\end{flalign}
Just as for beyond Horndeski, the scalar field equation can be written as
\begin{equation}
\mathcal{E}_{\phi}=\mathcal{P}^{\left(F\right)}-\nabla^\mu \mathcal{J} ^{\left(F\right)}_\mu+\mathcal{P}^{\left(P\right)}-\nabla^\mu \mathcal{J} ^{\left(P\right)}_\mu+\mathcal{P}^{\left(Q\right)}-\nabla^\mu \mathcal{J} ^{\left(Q\right)}_\mu+\sum_{i=1}^5\Bigl(\mathcal{P}^{\left(A_i\right)}-\nabla^\mu \mathcal{J} ^{\left(A_i\right)}_\mu\Bigr).
\end{equation}
The 'currents' are
\begin{flalign}
\mathcal{J}^{\left(F\right)\,\mu}=2F_{X}R\phi^\mu,\quad \mathcal{J}^{\left(P\right)\,\mu}=2P_{X}\phi^\mu,\quad \mathcal{J}^{\left(Q\right)\,\mu}=2Q_{X}\Bigl[\Box\phi\phi^\mu-\phi_\nu\phi^{\nu\mu}\Bigr]-Q_{\phi}\phi^\mu,&&
\end{flalign}
\begin{flalign}
\mathcal{J}^{\left(A_1\right)\,\mu}=-2A_1\Box\phi^\mu+2A_{1X}\Bigl[\phi_{\rho\sigma}\phi^{\rho\sigma}\phi^\mu-2\phi^\nu\phi_{\nu\rho}\phi^{\rho\mu}\Bigr]-2A_{1\phi}\phi_\nu\phi^{\nu\mu},&&
\end{flalign}
\begin{flalign}
\mathcal{J}^{\left(A_2\right)\,\mu}=2A_2\Bigl[R^{\mu\nu}\phi_{\nu}-\Box\phi^\mu\Bigr]+2A_{2X}\Box\phi\Bigl[\Box\phi\phi^\mu-2\phi_{\nu}\phi^{\nu\mu}\Bigr]-2A_{2\phi}\Box\phi\phi^{\mu},&&
\end{flalign}
\begin{flalign}
\mathcal{J}^{\left(A_3\right)\,\mu}={}&{}A_3\Bigl\{R_{\nu\rho}\phi^\nu\phi^\rho\phi^\mu-\left(\Box\phi\right)^2\phi^\mu-\Bigl[2\phi^\nu\phi_{\nu\rho}-\Box\phi\phi_{\rho}\Bigr]\phi^{\rho\mu}-\phi^\nu\Box\phi_\nu\phi^\mu\nonumber\\{}&{}-\phi^\nu\phi^\rho\nabla_\nu\phi_\rho^\mu\Bigr\}-2A_{3X}\phi^\nu\phi_{\nu\rho}\phi^\rho\phi_\sigma\phi^{\sigma\mu}-A_{3\phi}\Bigl[X\Box\phi+\phi^\nu\phi_{\nu\rho}\phi^\rho\Bigr]\phi^\mu,&&
\end{flalign}
\begin{flalign}
\mathcal{J}^{\left(A_4\right)\,\mu}={}&{}-2A_4\Bigl[\phi_{\rho\sigma}\phi^{\rho\sigma}+\phi^\nu\Box\phi_\nu\Bigr]\phi^\mu-2\phi^\nu\phi^\rho\Bigl[A_{4X}\phi_\nu^\sigma\phi_{\sigma\rho}+A_{4\phi}\phi_{\nu\rho}\Bigr]\phi^\mu,&&
\end{flalign}
\begin{flalign}
\mathcal{J}^{\left(A_5\right)\,\mu}={}&{}2A_5\phi^\nu\phi_{\nu\rho}\Bigl[\phi^{\rho}\phi_\sigma\phi^{\sigma\mu}-\Box\phi\phi^\rho\phi^\mu-2\phi^{\rho\sigma}\phi_\sigma\phi^\mu\Bigr]-2A_5\phi^\nu\phi^\rho\phi^\sigma\nabla_\nu\phi_{\rho\sigma}\phi^\mu\nonumber\\{}&{}-2\phi^\nu\phi_{\nu\rho}\phi^\rho\Bigl[A_{5X}\phi^\sigma\phi_{\sigma\lambda}\phi^\lambda+XA_{5\phi}\Bigr]\phi^\mu,&&
\end{flalign}
and the remaining parts are again the partial derivatives with respect to $\phi$ of the respective Lagrangians,
\begin{flalign}
\mathcal{P}^{\left(F\right)}=F_{\phi}R,\quad \mathcal{P}^{\left(P\right)}=P_\phi ,\quad \mathcal{P}^{\left(Q\right)}=Q_\phi\Box\phi,\quad \mathcal{P}^{\left(A_1\right)}=A_{1\phi}\phi_{\mu\nu}\phi^{\mu\nu},&&
\end{flalign}
\begin{flalign}
\mathcal{P}^{\left(A_2\right)}=A_{2\phi}\left(\Box\phi\right)^2 ,\quad \mathcal{P}^{\left(A_3\right)}=A_{3\phi}\Box\phi\phi^\mu\phi_{\mu\nu}\phi^\nu ,\quad \mathcal{P}^{\left(A_4\right)}=A_{4\phi}\phi^\mu\phi_{\mu\nu}\phi^{\nu\rho}\phi_\rho,&&
\end{flalign}
\begin{flalign}
\mathcal{P}^{\left(A_5\right)}=A_{5\phi}\left(\phi^\mu\phi_{\mu\nu}\phi^\nu\right)^2.&&
\end{flalign}

\chapter{Noether current for the global conformal symmetry}\label{chap:appconfnoeth}
\chaptermark{Appendix D~~Noether current for the global conformal symmetry}
\sectionmark{Appendix D~~Noether current for the global conformal symmetry}
This appendix presents the Noether current associated with the global conformal symmetry presented in paragraph~\ref{subsec:globalconffinal}, under rescalings by a constant parameter $\omega$,
\begin{equation}
g_{\mu\nu}\to \omega^2 g_{\mu\nu},\quad \phi\to\frac{\phi}{\omega}.\label{eq:conf_inv_app}
\end{equation}
The Horndeski theories which are invariant under such a transformation are~\cite{Padilla:2013jza}
\begin{align}
S\left[g_{\mu\nu},\phi\right] = {}&{}\int\mathrm{d}^4 x\sqrt{-g}\Bigl\{\phi^4a_2\left(Y\right)-\phi a_3\left(Y\right)\Box\phi+\phi^2a_4\left(Y\right)R+\frac{a_{4Y}}{\phi^2}\left[\left(\Box\phi\right)^2-\phi_{\mu\nu}\phi^{\mu\nu}\right]\nonumber\\{}&{}+\frac{a_5\left(Y\right)}{\phi}G^{\mu\nu}\phi_{\mu\nu}-\frac{a_{5Y}}{6\phi^5}\Bigl[\left(\Box\phi\right)^3-3\Box\phi\phi_{\mu\nu}\phi^{\mu\nu}+2\phi_{\mu\nu}\phi^{\nu\rho}\phi_\rho^\mu\Bigr]\Bigr\},\label{eq:global_conf_ac_app}
\end{align}
where $a_2$, $a_3$, $a_4$, $a_5$ are arbitrary functions of $Y$, and $Y$ is defined as the combination $Y\equiv X/\phi^4$, invariant under (\ref{eq:conf_inv_app}). \\

The action has the following form,
\begin{equation}
S = \int\mathrm{d}^4x\,L\left(g,\partial g,\partial\partial g,\phi,\partial \phi,\partial\partial\phi\right).\label{eq:sl}
\end{equation}
The Euler-Lagrange equations are
\begin{align}
0={}&{}\frac{\partial L}{\partial g^{\mu\nu}}-\partial_\rho\frac{\partial L}{\partial \left(\partial_\rho g^{\mu\nu}\right)}+\partial_\rho\partial_\sigma\frac{\partial L}{\partial \left(\partial_\rho\partial_\sigma g^{\mu\nu}\right)},\\
0={}&{}\frac{\partial L}{\partial \phi}-\partial_\rho\frac{\partial L}{\partial \left(\partial_\rho \phi\right)}+\partial_\rho\partial_\sigma\frac{\partial L}{\partial \left(\partial_\rho\partial_\sigma\phi\right)}.
\end{align}
The fields are collectively denoted as $\varphi^a$. If under a transformation $\varphi^a\to\varphi^a+\Delta\varphi^a$, the action $S$ is invariant, then $\delta L$ is a total divergence, $\delta L=\partial_\rho j^\rho$. Moreover, one has:
\begin{align}
\delta L ={}&{} \frac{\partial L}{\partial\varphi^a}\Delta\varphi^a+\frac{\partial L}{\partial \left(\partial_\rho\varphi^a\right)}\partial_\rho\Delta\varphi^a+\frac{\partial L}{\partial\left(\partial_\rho\partial_\sigma\varphi^a\right)}\partial_\rho\partial_\sigma\Delta\varphi^a\nonumber\\
{}={}&{}\partial_\rho\frac{\partial L}{\partial \left(\partial_\rho \varphi^a\right)}\Delta\varphi^a-\partial_\rho\partial_\sigma\frac{\partial L}{\partial \left(\partial_\rho\partial_\sigma\varphi^a\right)}\Delta\varphi^a+\frac{\partial L}{\partial \left(\partial_\rho\varphi^a\right)}\partial_\rho\Delta\varphi^a+\frac{\partial L}{\partial\left(\partial_\rho\partial_\sigma\varphi^a\right)}\partial_\rho\partial_\sigma\Delta\varphi^a\nonumber\\
{}={}&{} \partial_\rho\left(\frac{\partial L}{\partial \left(\partial_\rho \varphi^a\right)}\Delta\varphi^a\right)+\partial_\rho\left(\frac{\partial L}{\partial\left(\partial_\rho\partial_\sigma\varphi^a\right)}\partial_\sigma\Delta\varphi^a-\Delta\varphi^a\partial_\sigma\frac{\partial L}{\partial\left(\partial_\rho\partial_\sigma\varphi^a\right)}\right).\label{eq:deltal}
\end{align}
The Noether current $\mathfrak{J}_\text{conf}^\rho$ such that $\partial_\rho \mathfrak{J}_\text{conf}^\rho=0$ is therefore:
\begin{equation}
\mathfrak{J}_\text{conf}^\rho = \left(\frac{\partial L}{\partial \left(\partial_\rho \varphi^a\right)}-\partial_\sigma\frac{\partial L}{\partial\left(\partial_\rho\partial_\sigma\varphi^a\right)}\right)\Delta\varphi^a+\frac{\partial L}{\partial\left(\partial_\rho\partial_\sigma\varphi^a\right)}\partial_\sigma\Delta\varphi^a-j^\rho.
\end{equation}
For the conformal symmetry, the transformation is $g_{\mu\nu}=\omega^2g_{\mu\nu}$ and $\phi\to\phi/\omega$. Its infinitesimal form under $\omega=1-\epsilon$ is $\Delta g^{\mu\nu}=2\epsilon g^{\mu\nu}$ and $\Delta\phi = \epsilon\phi$. Moreover, the invariance is exact at the Lagrangian level itself, i.e. $j^\rho=0$. Finally, the computation of $\mathfrak{J}_\text{conf}^\rho$ yields a result of the form
\begin{equation}
\mathfrak{J}_\text{conf}^\rho = \epsilon\sqrt{-g}\mathcal{J}_\text{conf}^\rho,\label{eq:Jj}
\end{equation}
where $\mathcal{J}_\text{conf}^\rho$ is a Lorentz vector, covariantly conserved,
\begin{equation}
\nabla_\rho \mathcal{J}_\text{conf}^\rho = 0.
\end{equation}
Straightforward computations lead to the following conformal Noether current,
\begin{equation}
\mathcal{J}_\text{conf}^\rho = \sum_{i=2}^5\mathcal{J}_\text{conf,i}^\rho,
\end{equation}
where the contributions of the $G_2$, $G_3$, $G_4$ and $G_5$ terms are respectively:
\begin{align}
\mathcal{J}_\text{conf,2}^\rho ={}&{} -G_{2X}\,\phi\, \partial^\rho\phi,\\
\mathcal{J}_\text{conf,3}^\rho ={}&{}G_3\partial^\rho\phi+\phi\Bigl[\left(G_{3X}\Box\phi+G_{3\phi}\right)\partial^\rho\phi-G_{3X}\partial_\alpha\phi\nabla^\alpha\nabla^\rho\phi\Bigr],\\
\mathcal{J}_\text{conf,4}^\rho ={}&{}-6G_{4\phi}\partial^\rho\phi+\phi\Bigl\{2G_{4X} \,G^{\rho\nu}\partial_\nu\phi+2G_{4X\phi}\Bigl(\partial_\mu\phi\nabla^\mu\nabla^\rho\phi-\Box\phi\partial^\rho\phi\Bigr)\nonumber\\{}&{}+G_{4XX}\Bigl(2\Box\phi\partial_\mu\phi\nabla^\mu\nabla^\rho\phi-2\partial^\mu\phi\nabla_\mu\nabla_\nu\phi\nabla^\nu\nabla^\rho\phi\nonumber\\{}&{}-\left[\left(\Box\phi\right)^2-\left(\phi_{\mu\nu}\right)^2\right]\partial^\rho\phi\Bigr)\Bigr\},\\
\mathcal{J}_\text{conf,5}^\rho ={}&{}G_{5X}\left\lbrace-4XG^{\rho\mu}\phi_\mu+\Box\phi\phi_\mu\phi^{\mu\rho}-\phi^\mu\phi_{\mu\nu}\phi^{\nu\rho}-\frac{1}{2}\left[\left(\Box\phi\right)^2-\phi_{\mu\nu}\phi^{\mu\nu}\right]\phi^\rho\right\rbrace\nonumber\\{}&{}+2G_{5\phi}\left(\Box\phi\phi^\rho-\phi_\mu\phi^{\mu\rho}\right) +\phi\Bigl\{-\mathcal{L}_{5X}\phi^\rho-2G_{5\phi}G^{\rho\mu}\phi_{\mu}-G_{5X}\Bigl[G^{\rho\mu}X_\mu\nonumber\\{}&{}+R^{\rho\mu}\Box\phi\phi_\mu-R_{\mu\nu}\phi^\mu\phi^{\nu\rho}-R^{\alpha\rho\beta\mu}\phi_{\mu}\phi_{\alpha\beta}\Bigr]+G_{5\phi X}\Bigl[\frac{\phi^\rho}{2}\left(\left(\Box\phi\right)^2-\phi_{\mu\nu}\phi^{\mu\nu}\right)\nonumber\\{}&{}+\Box\phi X^\rho-X_\mu \phi^{\mu\rho}\Bigr]+G_{5XX}\Bigl[\frac{X^\rho}{2}\left(\left(\Box\phi\right)^2-\phi_{\mu\nu}\phi^{\mu\nu}\right)\nonumber\\{}&{}-X_\mu\left(\Box\phi\phi^{\mu\rho}-\phi^{\alpha\mu}\phi_\alpha^\rho\right)\Bigr]\Bigr\}.
\end{align}
The following notation is used, $X_\mu \equiv \partial_\mu X$. Remember that in these expressions, the Horndeski functions $G_{2,3,4,5}$ must have the form imposed by the invariant action (\ref{eq:global_conf_ac_app}), for instance $G_2(\phi,X)=\phi^4a_2(Y)$.
\chapter{Expression of $\widetilde{G_2}$, $\widetilde{G_3}$ and $\widetilde{F_4}$ for the wormhole theory}\label{chap:g2g3}
\chaptermark{Appendix E~~Expression of $\widetilde{G_2}$, $\widetilde{G_3}$ and $\widetilde{F_4}$ for the wormhole theory}
\sectionmark{Appendix E~~Expression of $\widetilde{F_4}$, $\widetilde{G_3}$ and $\widetilde{G_2}$ for the wormhole theory}
In paragraph~\ref{subsec:disfbh}, we gave the beyond Horndeski functions for the theory which admits the wormhole (\ref{eq:wormholesolution}-\ref{eq:wormholesolution_bisbis}) as a solution, but the expressions for $\widetilde{G_2}$, $\widetilde{G_3}$ and $\widetilde{F_4}$ are too long and we report it here. $y$ stands for the scalar field $\phi$, $x$ for the disformed kinetic term $\tilde{X}$.
{\small
\begin{flalign}
\widetilde{F_4}\left(y,x\right) ={}&\frac{1}{2 \sqrt{2} \sqrt{a} B \Xi  x \sqrt{\frac{a-a \Xi }{a-1}} (8 (a-1) B x+a)}\Biggl\{\alpha  B_y \Biggr[a^{3/2} (\Xi +1) \left(\ln \left(\frac{\Xi +1}{4B}\right)+2\right)\nonumber\\{}&{}+4 \sqrt{a} B x (a-1) \left((\Xi +4)\ln \left(\frac{\Xi +1}{4B}\right)+2(\Xi -2)+8\right)\Biggr]\Biggr\}\nonumber\\{}&{}-\frac{(a-1) \alpha  B_y(4 (a-1) B x+a \Xi +a-2 \Xi )}{(8 (a-1) B x+a) \sqrt{-\frac{B x}{\Xi +1}} (a (\Xi -1)-4 (a-1) B x)}\nonumber\\{}&{}+\frac{\alpha  \ln \left(\frac{\Xi +1}{4B}\right) B_y (8 (a-1) B x+3 a \Xi -a)}{16 \Xi  (8 (a-1) B x+a) \left(-\frac{B x}{\Xi +1}\right)^{3/2}}\nonumber\\{}&{}+\frac{\sqrt{-\frac{B x}{\Xi +1}} (4 (a-1) B x+a \Xi +a-2 \Xi ) \left(a (\Xi -1) \left(\beta  e^{2 y}-1\right)+8 \alpha  x(1-a)\right)}{4 a \Xi  x^2 (a (\Xi -1)-4 (a-1) B x)}\nonumber\\{}&{}-\frac{\alpha  (a \Xi +a-2) B'(y) (4 (a-1) B x+a \Xi +a)}{16 (a-1) a B^2 \Xi  x^2 \sqrt{-\frac{B x}{\Xi +1}}},&&
\end{flalign}
\begin{flalign}
\widetilde{G_3}\left(y,x\right) ={}&\frac{\alpha  \ln \left(\frac{\Xi +1}{4 B}\right) B_{yy} x^2}{\Xi  \left(-\frac{B x}{\Xi +1}\right)^{3/2}}-\frac{4 (a-1) \alpha  \ln \left(\frac{\Xi +1}{4 B}\right) B_y^2 x^3}{a \Xi ^3 \left(-\frac{B x}{\Xi +1}\right)^{3/2}}+\frac{\alpha  x}{\left(-\frac{B x}{\Xi +1}\right)^{3/2}}\nonumber\\{}&{}+\frac{\alpha  (\Xi  a+a+4 (a-1) B x) B_y^2 x}{B^2 (a+8 (a-1) B x) \sqrt{-\frac{B x}{\Xi +1}}}+\frac{3 \alpha  (\Xi  a+a+4 (a-1) B x) \ln \left(\frac{\Xi +1}{4 B}\right) B_y^2 x}{2 B^2 (a+8 (a-1) B x) \sqrt{-\frac{B x}{\Xi +1}}}\nonumber\\{}&{}-\frac{x}{2 \Xi  \sqrt{-\frac{B x}{\Xi +1}}}\Biggl\{\frac{(\Xi  a+a-2) \left(\alpha  (a \Xi  (\Xi +1)-4 (a-1) B x) B_y-2 a B^2 e^{2 y} \beta  \Xi \right)}{(a-1) a B^2 x}\nonumber\\{}&{}-2 \left(\frac{16 x \alpha  (a-1)^2}{a (4 B x a-\Xi  a+a-4 B x)}+e^{2 y} \beta +\frac{\alpha  (\Xi +1)}{B}-1\right) B_y\Biggr\}\nonumber\\{}&{}+\frac{1}{12 (a-1) B^2 \sqrt{-\frac{B x}{\Xi +1}}}\Biggl\{12 e^{2 y} \beta  \Biggl[2 a (\Xi -2)+\arctan\left(\frac{\sqrt{2}}{\sqrt{\Xi -1}}\right)\nonumber\\{}&{} \sqrt{2} \sqrt{\Xi -1}+2\Biggr] B^2+\Bigl[8 B x \alpha +a \Bigl(-8 B x \alpha -4 (\Xi -1) \alpha +3 B \left(e^{2 y} \beta -1\right) \nonumber\\{}&{}(\Xi -1)\Bigr)\Bigr] B_y\Biggr\}+\frac{1}{a^{3/2} (a+8 (a-1) B x) (\Xi -1) \left(\frac{a (\Xi -1)}{1-a}\right)^{3/2} \Xi  (\Xi +1)^2}\nonumber\\{}&{}\Biggl\{64 \sqrt{2} x^3 \alpha  \Biggl[\Biggl(-8 \Xi  a^{3/2}+3 (a-1) (\Xi +1) \ln \left(\frac{\Xi +1}{4 B}\right) a^{3/2}-8 a^{3/2}+8 \Xi  a^{5/2}\nonumber\\{}&{}+8 a^{5/2}-4 (1-a) \Xi  a^{3/2}+4 (1-a)a^{3/2}+4 (a-1) B x \Biggl(4 (3 \Xi +5) a^{3/2}\nonumber\\{}&{}-4 \sqrt{a} (3 \Xi +5)+(a-1) (5 \Xi +8) \ln \left(\frac{\Xi +1}{4 B}\right) \sqrt{a}+2 (6-5 \Xi ) (1-a)\sqrt{a}\Biggl)\Biggr) B_y^2\nonumber\\{}&{}+2 B (a+8 (a-1) B x) \Biggl(-2 (\Xi +1) a^{3/2}-(a-1) \sqrt{a} (\Xi +1) \ln \left(\frac{\Xi +1}{4 B}\right)\nonumber\\{}&{}+2 (\Xi +1) \sqrt{a}+2 (\Xi -1) (1-a)\sqrt{a}\Biggr)B_{yy}\Biggr]\Biggr\}-\frac{2 \sqrt{2} \alpha  (\Xi -3)}{B \sqrt{\frac{a (\Xi -1)}{1-a}}}-H_5,&&
\end{flalign}
\begin{flalign}
\widetilde{G_2}\left(y,x\right) ={}&\frac{1}{12 (a-1) a^2 \alpha  B^4 (\Xi -1)^{5/2} \Xi }\Biggl\{-2 a^2 B \Xi  (\Xi -1)^2 \sqrt{-\frac{B x}{\Xi +1}} \Biggl[\alpha  (\Xi +1) \sqrt{\Xi -1}\nonumber\\{}&{} B_{yy} \Biggl(a \Biggl(-4 \alpha  (\Xi -1)-8 \alpha  B x+3 B (\Xi -1) \Biggl(\beta  e^{2 y}-1\Biggr)\Biggr)+8 \alpha  B x\Biggr)\nonumber\\{}&{}+6 B \Biggl(\sqrt{\Xi -1} \Biggl(-(a-1) \alpha ^2 (\Xi +1)^2+(a-1) \beta ^2 B^2 e^{4 y}+2 \alpha  \beta  B (\Xi +1) e^{2 y}\nonumber\\{}&{} (4 a \Xi -5 a+1)\Biggr)+4 \sqrt{2} \alpha  \beta  B \Biggl(\Xi ^2-1\Biggr) e^{2 y} \arctan\Biggl(\frac{\sqrt{2}}{\sqrt{\Xi -1}}\Biggr)\Biggr)\Biggr)\nonumber\\{}&{}+a \alpha  (\Xi -1)^{5/2} \sqrt{-\frac{B x}{\Xi +1}} B_y^2 \Biggl(a^2 \Biggl(-4 \alpha  (\Xi -1) (\Xi +1) (4 \Xi +1)-4 B^2 x\nonumber\\{}&{}\Biggl(8 \alpha  x+3 (\Xi +3) \Biggl(\beta  e^{2 y}-1\Biggr)\Biggr)+B \Biggl(3 (\Xi -1) (\Xi +1) (2 \Xi +1) \Biggl(\beta  e^{2 y}-1\Biggr)\nonumber\\{}&{}-8 \alpha  \Biggl(2 \Xi ^2+\Xi -5\Biggr) x\Biggr)\Biggr)+4 a B x \Biggl(2 \alpha  \Biggl(2 \Xi ^2+\Xi -5\Biggr)+16 \alpha  B x+3 B (\Xi +3) \nonumber\\{}&{}\Biggl(\beta  e^{2 y}-1\Biggr)\Biggr)-32 \alpha  B^2 x^2\Biggr)-6 \alpha  B B_y \Biggl(\sqrt{\Xi -1} \Biggl((a-1) a^2 \alpha  (\Xi -1)^2 (\Xi +1)^3 \nonumber\\{}&{}\sqrt{-\frac{B x}{\Xi +1}}-16 (a-1) B^2 x \Biggl(\sqrt{2} (a-1)^2 \alpha  (\Xi +1) x \sqrt{-\frac{a (\Xi -1)}{a-1}}-a \beta  (\Xi -1)^2 \nonumber\\{}&{}e^{2 y} (a (\Xi +4)-2) \sqrt{-\frac{B x}{\Xi +1}}\Biggr)+2 a B (\Xi -1) \Biggl(a \beta  \Biggl(\Xi ^2-1\Biggr) \nonumber\\{}&{}e^{2 y} (a ((\Xi -3) \Xi +4)-2) \sqrt{-\frac{B x}{\Xi +1}}+4 \sqrt{2} \alpha  (a-1)^2 (\Xi -3) \Xi  x \sqrt{-\frac{a (\Xi -1)}{a-1}}\Biggr)\Biggr)\nonumber\\{}&{}+2 \sqrt{2} a \beta  B (\Xi -1)^2 e^{2 y} \arctan\Biggl(\frac{\sqrt{2}}{\sqrt{\Xi -1}}\Biggr) \sqrt{-\frac{B x}{\Xi +1}}\nonumber\\{}&{}\Biggl(8 a B x-a \Xi ^2+a-8 B x\Biggr)\Biggr]\Biggr\}-2xH_{5\phi}.&&
\end{flalign}}The expression for $H_5(y,x)$, which appears in $\widetilde{G_2}$ and $\widetilde{G_3}$, is: 
{\small
\begin{flalign}
H_5\left(y,x\right) ={}&\frac{ \alpha}{4} \int \Biggl\{\frac{8 (a-1) x^2 \ln \left(\frac{\Xi +1}{4 B}\right) B_y^2}{a \Xi ^3 \left(-\frac{B x}{\Xi +1}\right)^{3/2}}+\frac{128 \sqrt{2} x^2 }{\Xi  (\Xi +1)^2 \left(-\frac{a (\Xi -1)}{a-1}\right)^{5/2} (8 (a-1) B x+a)}\nonumber\\{}&{}\Biggl[B_y^2 \Biggl(4 a (2 B (11 \Xi +4) x+3 \Xi +1)+\ln \left(\frac{\Xi +1}{4 B}\right) (4 a B (5 \Xi +8) x\nonumber\\{}&{}+3 a (\Xi +1)-4 B (5 \Xi +8) x)-8 B (11 \Xi +4) x\Biggl)-2 B (8 (a-1) B x+a) \nonumber\\{}&{}\left((\Xi +1) \ln \left(\frac{\Xi +1}{4 B}\right)+4 \Xi \right) B_{yy}\Biggr]-\frac{2 B_y^2 (4 (a-1) B x+a \Xi +a)}{B^2 (8 (a-1) B x+a) \sqrt{-\frac{B x}{\Xi +1}}}\nonumber\\{}&{}-\frac{3 \ln \left(\frac{\Xi +1}{4 B}\right) B_y^2 (4 (a-1) B x+a \Xi +a)}{B^2 (8 (a-1) B x+a) \sqrt{-\frac{B x}{\Xi +1}}}-\frac{2 x \ln \left(\frac{\Xi +1}{4 B}\right) B_{yy}}{\Xi  \left(-\frac{B x}{\Xi +1}\right)^{3/2}}\Biggr\}\mathrm{d}x. &&
\end{flalign}}
\newpage
\chaptermark{BIBLIOGRAPHY}
\sectionmark{BIBLIOGRAPHY}
\addcontentsline{toc}{chapter}{Bibliography}
\bibliography{Refs}{}
\bibliographystyle{utphys}

\end{document}